\RequirePackage[log]{snapshot}
\documentclass[aps,rmp,reprint,amsmath,amssymb,graphicx,longbibliography,floatfix,showpacs]{revtex4-1}
 \pdfoutput=1
\usepackage{bm}
\usepackage{floatrow}

\usepackage{graphicx}

\usepackage{subfig,amssymb}
\usepackage{hyperref} 

\def\R{\ensuremath{\mathbb R}}

\def\Y{\mathcal{Y}}

\def\v{\ensuremath{v}}

\def\1{{\bf 1}}

\newcommand{\diff}{\mathrm{d}}
\newcommand{\A}{\mathbf{A}}
\newcommand{\B}{\mathbf{B}}
\newcommand{\vL}{\mathbf{L}}
\newcommand{\vM}{\mathbf{M}}
\newcommand{\vx}{\mathbf{X}}
\newcommand{\vy}{\mathbf{Y}}
\newcommand{\x}{\mathbf{x}}

\newcommand{\vz}{\mathbf{z}}
\newcommand{\Vr}{\mathbf{r}}

\newcommand{\vm}{\boldsymbol{\mu}}

\usepackage{color}
\newcommand{\vH}{\mathbf{H}}
\definecolor{Brown}{rgb}{0.7,0.6,0}
\newcommand{\br}{\color{Brown}}
\newcommand{\bb}{\color{blue}}
\newcommand{\re}{\color{red}}
\newcommand{\gr}{\color{green}}

\begin{document}

\title{The Physics of Climate Variability and Climate Change}

\author{Michael Ghil}
\affiliation{Geosciences Department and Laboratoire de M\'et\'eorologie Dynamique (CNRS and IPSL), 
Ecole Normale Sup\'erieure and PSL University, F-75231 Paris Cedex 05, France}
\affiliation{Department of Atmospheric and Oceanic Sciences, University of California,Los Angeles, CA 90095-1565, USA}
\author{Valerio Lucarini} 
\affiliation{Department of Mathematics and Statistics, University of Reading, Reading, RG66AX, UK}
\affiliation{Centre for the Mathematics of Planet Earth, University of Reading, Reading, RG66AX, UK}
\affiliation{CEN - Institute of Meteorology, University of Hamburg, Hamburg, 20144, Germany}

\date{\today{}}

\begin{abstract}
{The climate system is a forced, dissipative, nonlinear, complex and heterogeneous system that is out of thermodynamic equilibrium. The system exhibits natural variability on many scales of motion, in time as well as space, and it is subject to various external forcings, natural as well as anthropogenic.
This paper reviews the observational evidence on climate phenomena and the governing equations of planetary-scale flow, as well as presenting the key concept of a hierarchy of models as used in the climate sciences. Recent advances in the application of dynamical systems theory, on the one hand, and of nonequilibrium statistical physics, on the other, are brought together for the first time and shown to complement each other in helping understand and predict the system's behavior. These complementary points of view permit a self-consistent handling of subgrid-scale phenomena as stochastic processes, as well as a unified handling of natural climate variability and forced climate change, along with a treatment of the crucial issues of climate sensitivity, response, and predictability.
}
\end{abstract}


\maketitle

\tableofcontents{}

\section{Introduction and Motivation}
\label{intro}

\subsection{Basic Facts of the Climate Sciences}
\label{intro1}
The climate system is a forced, dissipative, chaotic system that is out of equilibrium and whose complex natural variability arises from the interplay of positive and negative feedbacks, instabilities and saturation mechanisms. These processes span a broad range of spatial and temporal scales and include many chemical species and all physical phases. The system's heterogeneous phenomenology includes the mycrophysics of clouds, cloud--radiation interactions, atmospheric and oceanic boundary layers, and several scales of turbulence \cite{Ghil.2019}; it evolves, furthermore, under the action of large-scale agents that drive and modulate its evolution, mainly differential solar heating and the Earth's rotation and gravitation. 

As is often the case, the complexity of the physics is interwoven with the chaotic character of the dynamics. 
Moreover, the climate system's large natural variability on different time scales is strongly affected by relatively small changes in the forcing, anthropogenic as well as natural \cite{Peixoto1992, Ghil1987, Lucarini.ea.2014}

\begin{figure}
\centering
\includegraphics[angle=0,width=0.9\columnwidth]{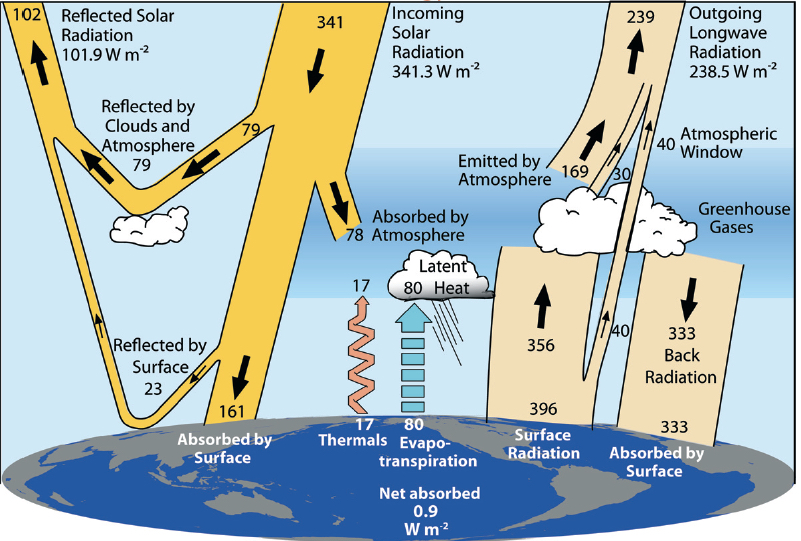}
\caption{\small Globally averaged energy fluxes in the Earth system [W m$^{-2}$]. The fluxes on the left represent solar radiation in the visible and the ultraviolet, those on the right terrestrial radiation in the infrared, and those in the middle nonradiative fluxes. Reproduced from \citet{Trenberth2009} \copyright American Meteorological Society; used with permission.}
\label{Trenberth1}
\end{figure}

On the macroscopic level, climate is driven by differences in the absorption of solar radiation throughout the depth of the atmosphere, as well as in a narrow surface layer of the ocean and of the soil; the system's actual governing equations are given in Sect.~\ref{conservation} below. The prevalence of absorption at the surface and in the atmosphere's lower levels leads, through several processes, to compensating vertical energy fluxes --- most notably, fluxes of infrared radiation throughout the atmosphere and convective motions in the troposphere; see Fig.~\ref{Trenberth1}. 

\begin{figure}[ht]
\subfloat[Net radiative fluxes]{
    \label{fig:rad_flux} 
    \centering
   \includegraphics [angle=270,width=0.9\columnwidth]{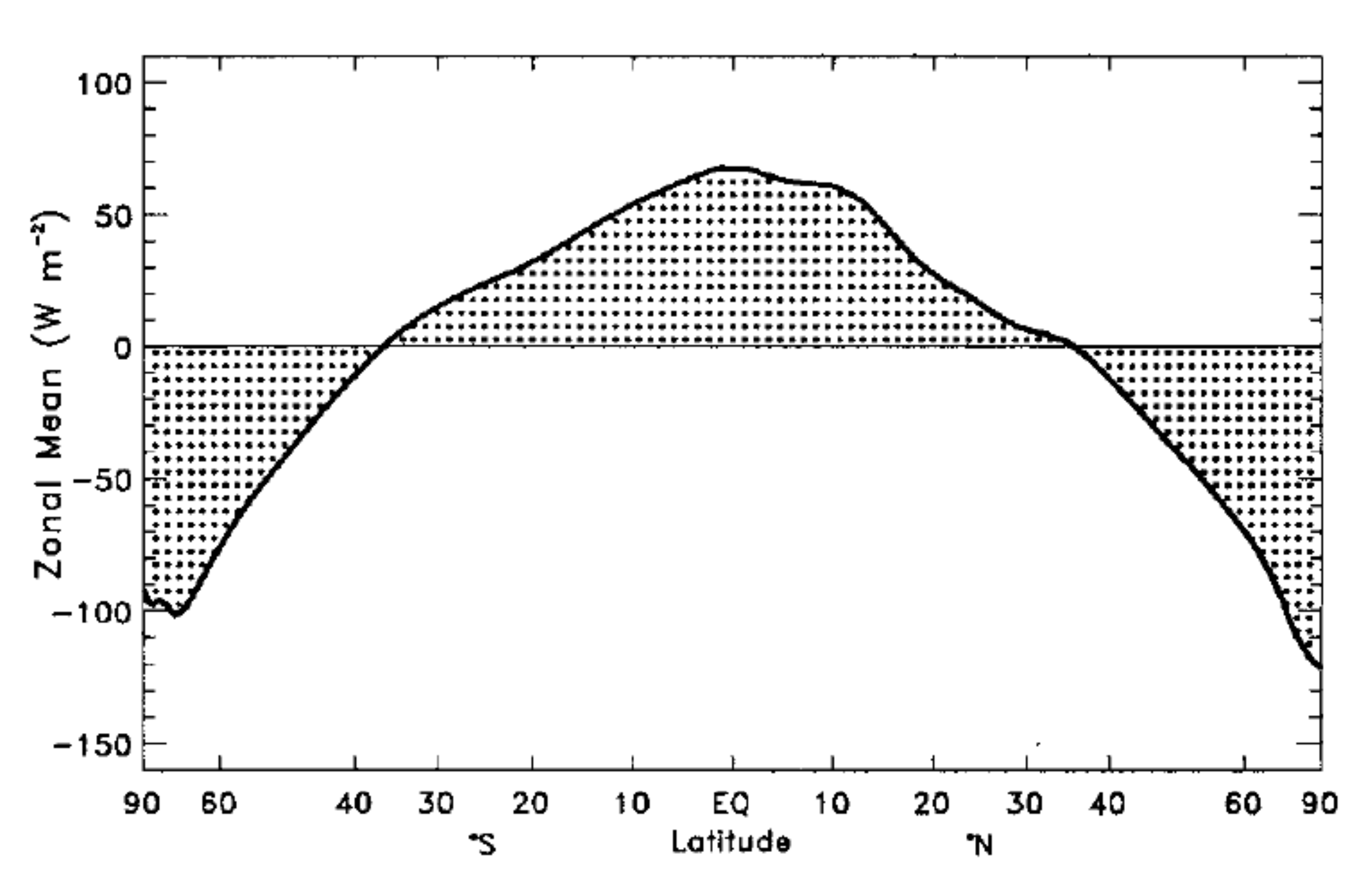}
}

\subfloat[Inferred horizontal fluxes]{
    \label{fig:horiz_flux} 
    \centering
   \includegraphics [angle=270,width=0.9\columnwidth]{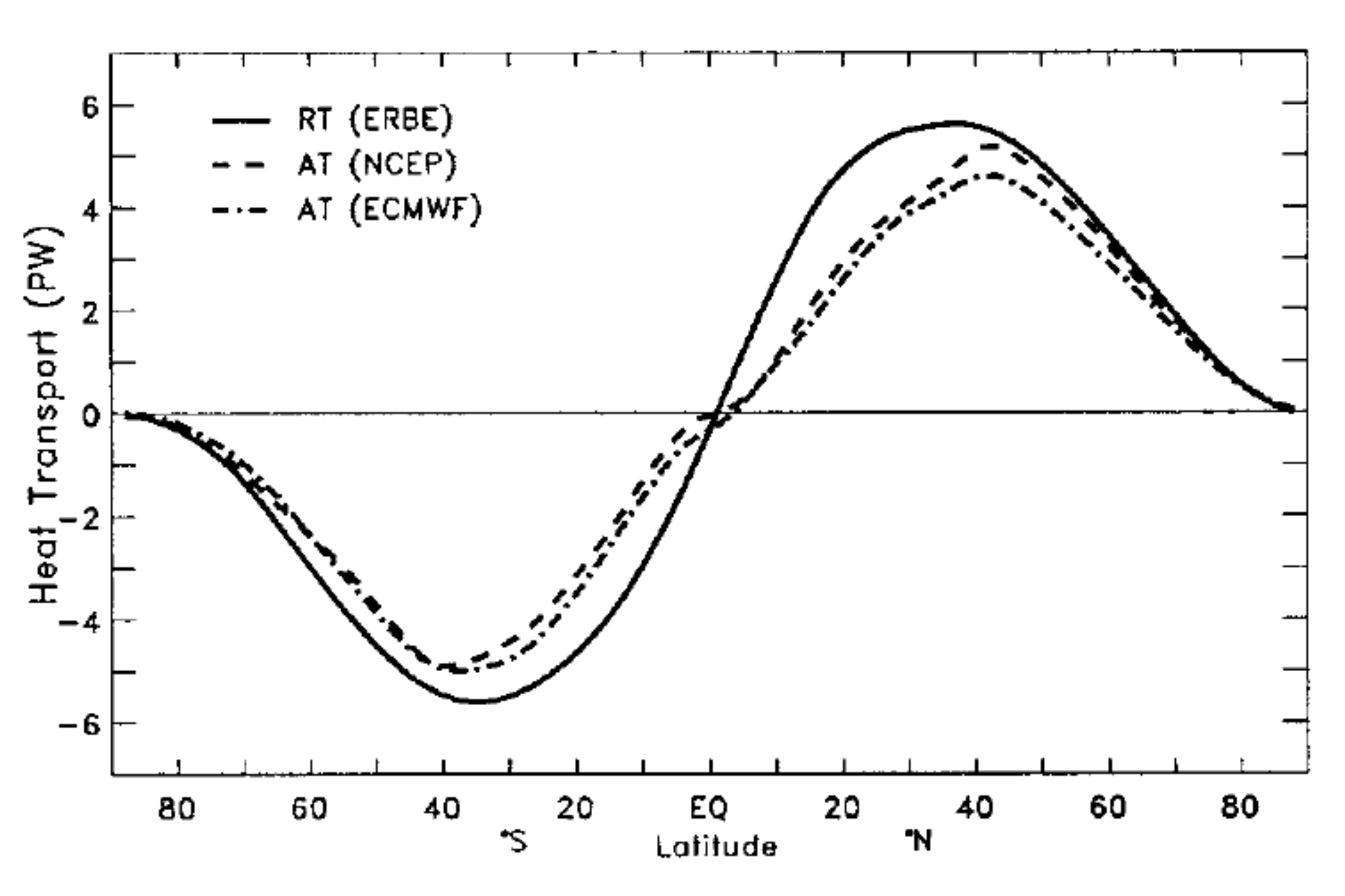}
   
}
 \caption{\small Meridional distribution of net radiative fluxes and of horizontal enthalpy fluxes. (a) Observed zonally averaged radiative imbalance at the top of the atmosphere from the ERBE experiment (1985--1989). (b) Inferred meridional enthalpy transport from ERBE observations (solid line) and estimate of the atmospheric enthalpy transport from two reanalysis datasets (ECMWF and NCEP). 
 Reproduced from \citet{TreCar} \copyright American Meteorological Society; used with permission.}
 \label{Trenberth2} 
\end{figure}

More solar radiation is absorbed in the low latitudes, leading to horizontal energy fluxes as well. The atmosphere's large-scale circulation is, to first order, a result of these horizontal and vertical fluxes arising from the gradients in solar radiation absorption, in which the hydrological cycle plays a key role as well.  The ocean circulation, in turn, is set into motion by surface or near-surface exchanges of mass, momentum and energy with the atmosphere: the so-called wind-driven component of the circulation is due mainly to the wind stress and the thermohaline one is due mainly to buoyancy fluxes 
\cite{DijkstraB2000, DG05}. The coupled atmospheric and oceanic circulation reduces the temperature differences between tropics and polar regions with respect to that on an otherwise similar planet with no horizontal energy transfers \cite{Lor67,Peixoto1992,Held.transport.01,LucariniRagone}.  At steady state, the convergence of enthalpy transported by the atmosphere and by the oceans compensates the radiative imbalance at the top of the atmosphere; see Fig.~\ref{Trenberth2}.


The classical theory of the general circulation of the atmosphere \cite{Lor67} describes in further detail how the mechanisms of energy generation, conversion and dissipation produce the observed circulation, which deviates substantially from the highly idealized, zonally symmetric picture sketched so far. According to \citet{Lor55}, atmospheric large-scale flows result from the conversion of available potential energy --- which is produced by the atmosphere's differential heating --- into kinetic energy, and the \citet{Lor67} energy cycle is completed by energy cascading to smaller scales to be eventually dissipated. \citet{JCM.2019} provides an up-to-date criticism of and further perspective on this theory.

Overall, the climate system can be seen as a thermal engine capable of transforming radiative heat into mechanical energy with a given, highly suboptimal efficiency, given the many irreversible processes that make it less than ideal \cite{Pauluis,Kleidon05,Lucarini:2009_PRE,Lucarini.ea.2014}. This conversion occurs through genuinely three-dimensional (3-D) baroclinic instabilities \cite{char48, Eady1949} that are triggered by large temperature gradients and would break zonal symmetry even on a so-called aqua-planet, with no topographic or thermal asymmetries at its surface. These instabilities give rise to a negative feedback, as they tend to reduce the temperature gradients they feed upon by favoring the mixing between masses of fluids at different temperatures. 

\begin{figure}[ht]
\begin{center}
\includegraphics[angle=0,width=0.9\columnwidth]{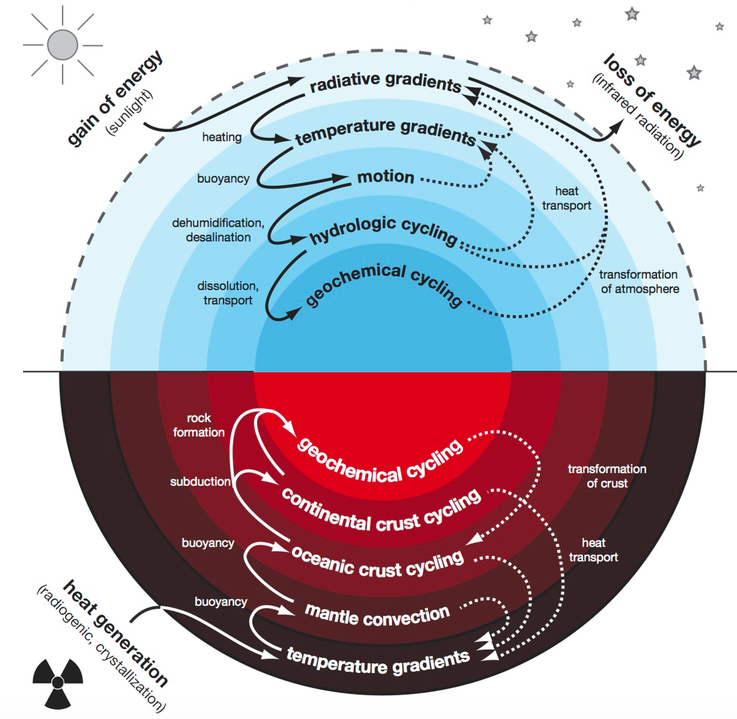}
\caption{\small Schematic diagram representing forcings, dissipative and mixing processes, gradients of temperature and chemical species, and coupling mechanisms across the Earth system. Bluish (reddish) colors refer to the fluid (solid) Earth. Reproduced from \citet{Kleidon10} with permission.}
\label{kleidon}
\end{center}
\end{figure}

Note that, while these baroclinic and other large-scale instabilities do act as negative feedbacks, they cannot be treated as diffusive, \citet{Onsager.1931}-like processes. Faced with the Earth system's complexity discussed herein and illustrated by Fig.~\ref{kleidon}, the closure of the coupled thermodynamical equations governing the general circulation of the atmosphere and oceans would provide a self-consistent theory of climate. Such a theory should able to connect instabilities and large-scale stabilizing processes on longer spatial and temporal scales, and to predict its response to a variety of forcings, both natural and anthropogenic \cite{Ghil1987,Lucarini:2009_PRE,Lucarini.ea.2014,Ghil2015}. This goal is being actively pursued but is still out of reach at the time of this writing \cite[e.g.,][and references]{Ghil.2019}; see also Sects. \ref{sensitivity} and \ref{critical} herein. The observed persistence of spatial gradients in chemical concentrations and temperatures, as well as the associated  mass and energy fluxes, are a basic signature of the climate system's intrinsic disequilibrium.


Figure~\ref{kleidon} emphasizes, moreover, that the fluid and the solid parts of the Earth system are coupled on even longer time scales, on which geochemical processes become of paramount importance \citep{Kleidon09, Rothman.ea.2003}. In contrast, closed, isolated systems cannot maintain disequilibrium and have to evolve towards homogenous thermodynamical equilibrium, as a result of the second law of thermodynamics \cite{Prigogine61}.

Studying the climate system's entropy budget provides a good global perspective on this system. The Earth as a whole absorbs shortwave radiation carried by low-entropy solar photons at $T_{\rm{Sun}} \simeq 6000$~K and emits infrared radiation to space via high-entropy thermal photons at $T_{\rm{Earth}}\simeq 255$~K  \cite{Peixoto1992,Lucarini.ea.2014}. Besides the viscous dissipation of kinetic energy, many other irreversible processes---such as turbulent diffusion of heat and chemical species, irreversible phase transitions associated with various hydrological processes, and  chemical reactions involved in the biogeochemistry of the planet---contribute to the total material entropy production \citep{Goody00, Kleidon09}. 

These and other important processes appear in the schematic diagram of Fig.~\ref{kleidon}. In general, in a forced dissipative system, 
entropy is continuously produced by irreversible processes and at steady state, this production is balanced by a net outgoing flux of entropy at the system's boundaries \citep{Prigogine61, deGroot84}; in the case at hand, this flux leaves mainly through the top of the atmosphere \cite{Goody00,Lucarini:2009_PRE}. Thus, on average, the climate system's entropy budget is balanced, just like its energy budget.


The phenomenology of the climate system is commonly approached by focusing on distinct and complementary aspects that include:
\begin{itemize}
\item wave-like features such as Rossby waves or equatorially trapped waves \cite[e.g.,][]{Gill1982}, which play a key role in the transport of energy, momentum, and water vapor, as well as in the study of atmospheric, oceanic and coupled-system predictability;
\item particle-like features such as hurricanes, extratropical cyclones or oceanic vortices \cite[e.g.,][]{JCM.2019, SalmonBook}, which strongly affect the local properties of the climate system and of its subsystems and subdomains; 
\item turbulent cascades, which are of crucial importance in the development of large eddies through the mechanism of geostrophic turbulence \cite{Charney_QG}, as well as in mixing and dissipation within the planetary boundary layer \cite{Zil75}.
\end{itemize}
 
Each of these points of view is useful and they do overlap and complement each other \cite{Ghil.Rob.2002,Lucarini.ea.2014} but neither by itself provides a comprehensive understanding of the properties of the climate system. { It is a key objective of this review paper to provide the interested and motivated reader with the tools for achieving such a comprehensive understanding with predictive potential.}


While much progress has been achieved \cite{Ghil.2019}, understanding and predicting the dynamics of the climate system faces --- on top of all the difficulties that are intrinsic to any nonlinear, complex system out of equilibrium --- the following additional obstacles that make it especially hard to grasp fully:
\begin{itemize} 
\item the presence of well-defined subsystems --- the atmosphere, the oceans, the cryosphere --- characterized by distinct physical and chemical properties and widely differing time and space scales;
\item the complex processes coupling these subsystems;
\item the continuously varying set of forcings that result from fluctuations in the incoming solar radiation and in the processes, both natural and anthropogenic, that alter the atmospheric composition;
\item the lack of scale separation between different processes, which requires a profound revision of the standard methods for model reduction, 
and calls for unavoidably complex parametrization of subgrid-scale processes in numerical models;  
\item the  lack of detailed, homogeneous, high-resolution and long-lasting observations of climatic fields that leads to the need for combining direct and indirect measurements when trying to reconstruct past climate states preceding the industrial era; and, last but not least,
\item the fact that we only have one realization of the processes that give rise to climate evolution in time.
\end{itemize}  

For all these reasons, it is far from trivial to separate the climate system's response to various forcings from its natural variability in the absence of time-dependent forcings. In simpler words, and as noted already by \citet{lorenz79}, it is hard to separate forced and free  climatic fluctuations \cite{Lucarini2011,Lucarini.ea.2014,Lucarini2017}. This difficulty is a major stumbling block on the road to a unified theory of climate evolution  \cite{Ghil2015, Ghil2016} but some promising ideas for overcoming it are emerging and will be addressed in Sects.~\ref{sensitivity} and \ref{critical} herein; see also \citet{Ghil.2019}. 

\subsection{More Than ``Just'' Science}

\subsubsection{The Intergovernmental Panel on Climate Change}
\label{sssec:IPCC}
Besides the strictly scientific aspects of climate research, much of the recent interest in it has been driven by the accumulated observational and modeling evidence on the ways humans influence the climate system. In order to review and coordinate the research activities carried out by the scientific community in this respect, the United Nations Environment Programme (UNEP) and the World Meteorological Organization (WMO) established in 1988 the Intergovernmental Panel on Climate Change (IPCC); its assessments reports (ARs) are issued every 4--6 years. By compiling systematic reviews of the scientific literature relevant to climate change, the ARs summarize the scientific progress, the open questions, and the bottlenecks regarding our ability to observe, model, understand and predict the climate system's evolution. 

More specifically, it is the IPCC Working Group I that focuses on the physical basis of climate change;  see \citet{IPCC01,IPCC07,IPCC13} for the three latest reports --- AR3, AR4 and AR5 --- in this area. Working Groups II and III are responsible for the reports that cover the advances in the interdisciplinary areas of adapting to climate change and of mitigating its impacts; see \citet{IPCC14b, IPCC14c} for the contributions of Working Groups II and III, respectively, to AR5. AR6 is currently in preparation, cf.~\url{https://www.ipcc.ch/assessment-report/ar6/}. 

Moreover, the IPCC supports the preparation of special reports on themes which are of interest across two of the working groups, e.g. climatic extremes \cite{IPCC12}, or across all three of them. The IPCC experience and working group structure is being replicated for addressing climate change at regional level, as in the case for the Hindu Kush Himalayan region, called sometimes the ``third pole'' \cite{HIMAP2019}. 

The IPCC reports are based on the best science available and are policy-relevant but not policy-prescriptive. Their multi-stage review is supposed to guarantee neutrality but they are still  inherently official, UN-sanctioned documents and have to bear the imprimatur of the IPCC's 195 member countries. Their release thus leads to considerable and often adversarial debates involving a variety of stakeholders from science, politics, civil society, and business; they also affect media production, cinema, videogames, and art at large, and are reflected more and more by them. 

Climate change has thus become an increasingly central topic of discussion in the public arena, involving all levels of decision making from local through regional and on to global. 
In recent years, climate services have emerged as a new area at the intersection of science, technology, policy making and business. They emphasize tools to enable climate change adaptation and mitigation strategies, and have benefited from large public investments like the European Union's Copernicus Programme, \url{https://climate.copernicus.eu}.

The lack of substantial progress in national governments' and international bodies' tackling climate change has recently led to the rapid growth of global, young-people--driven grassroot movements like Extinction Rebellion, \url{https://rebellion.earth/}, and FridaysForFuture, \url{https://www.fridaysforfuture.org}. Some countries, like the United Kingdom, have declared a state of climate emergency, cf.~\url{https://www.bbc.co.uk/news/uk-politics-48126677}, and some influential media outlets have started to use the expression climate crisis instead of climate change, \url{https://tinyurl.com/y2v2jwzy}. While of great consequence,
the present paper does not further dwell on such socio-economic and political issues.

\subsubsection{Hockey Stick Controversy and Climate Blogs}
\label{sssec:hockey}
\begin{figure}
\begin{center}
\includegraphics[width=0.9\columnwidth]{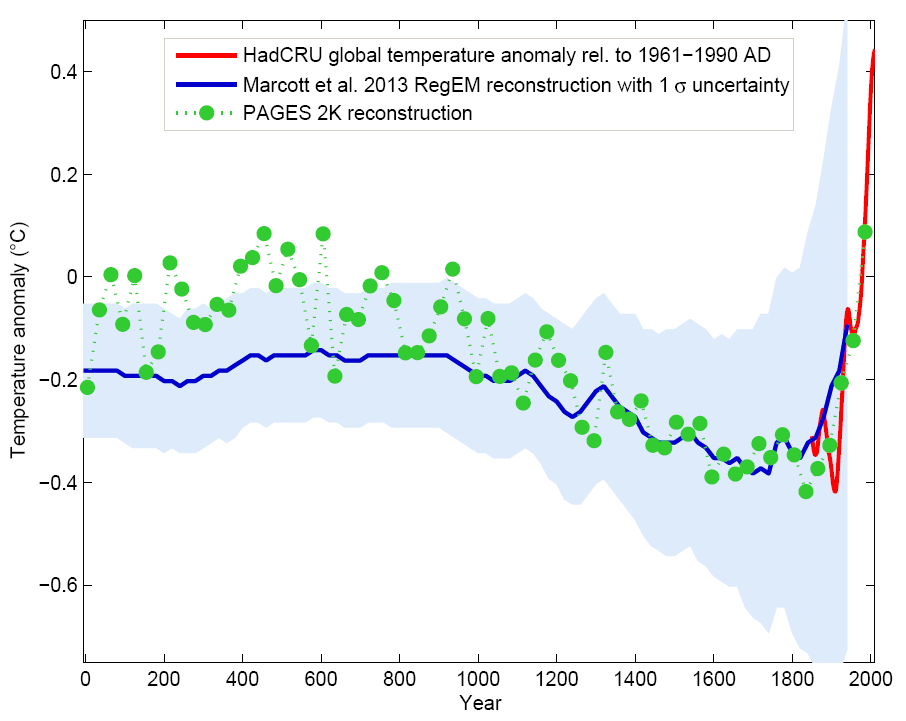}
\caption{\small Surface air temperature record for the last two millennia. Green dots show the 30-year average of the latest PAGES 2k reconstruction \cite{pages2013}, while the red curve shows the global mean temperature, according to HadCRUT4 data from 1850 onwards; the original ``hockey stick'' of \citet{mann99} is plotted in dark blue and its uncertainty range in light blue. Graph by Klaus Bitterman.} 
\label{obs4}
\end{center}
\end{figure}

\citet[Fig.~3(a)]{mann99} produced a temperature reconstruction from proxy data, cf. Sect. \ref{ssec:observations}, for the last 1000~yr, shown as part of the blue curve in Fig.~\ref{obs4}. This curve was arguably the most striking, and hence controversial, scientific result contained in AR3 \cite{IPCC01} and it was dubbed, for obvious reasons, the hockey stick. The AR3 report combined into one figure --- Fig.~1(b) of the Summary for Policy Makers (SPM) --- this blue curve and the red curve shown in Fig.~\ref{obs4} here, which was based on instrumental data over the last century-and-one-half; see Fig.~1(a) of the SPM. This superposition purported to demonstrate that recent temperature increase was unprecedented over the last two millennia, in both values attained and rate of change. 

Figure~1(b) of the AR3's SPM received an enormous deal of attention from the social and political forces wishing to underscore the urgency of tackling anthropogenic climate change. For the opposite reasons, the paper and its authors were the subject of intense political and judicial scrutiny and attack by other actors in the controversy, claiming that the paper was both politically motivated and scientifically unsound. 

\citet{McMc05}, among others, strongly criticized the results of \citet{mann99}, claiming that the statistical procedures used for smoothly combining the diverse proxy records used --- including tree rings, coral records, ice cores, and long historical records, with their diverse sources and ranges of uncertainty --- into a single multiproxy record, and the latter with instrumental records, were marred by biases and underestimation of the actual statistical uncertainty. Later papers criticized in turn the statistical methods of \citet{McMc05} and confirmed the overall correctness of the hockey stick reconstruction \cite[e.g.,][]{Huybers05, mann08, Taricco.ea.09, pages2013}. \citet{NRC.06} provided an excellent review of the state of our knowledge concerning the last two millennia of climate change and variability.

This controversy included the notorious ``Climategate'' incident, in which data hacked from the computer of a well-known UK scientist were used to support the thesis that scientific misconduct and data falsification had been routinely used to support the hockey stick reconstruction. These claims were later dismissed but they did lead to an important change in the relationship between the climate sciences, society, and politics, and in the way climate scientists interact among themselves and with the public. In certain countries --- e.g., the UK  and the USA ---  stringent rules have been imposed to ascertain that scientists working in governmental institutions have to publicly reveal the data they use in the preparation of scientific work, if formally requested to do so. 

Faced with the confusion generated by the polemics, several leading scientists started blogs --- e.g., \texttt{www.realclimate.org}, \texttt{www.ClimateAudit.org}, \texttt{www.climate-lab-book.ac.uk}, and \texttt{judithcurry.com} --- in which scientific literature and key ideas are presented for a broader audience and debated outside the traditional media of peer-reviewed journals or public events, such as conferences and workshops. Most contributions are of high quality, but sometimes arguments appear to sink to the level of a bitter strife between those in favor and those against the reality of climate change and of the anthropogenic contribution to its causes. 
 
\subsection{This Review}
The main purpose of this review paper is to bring together a substantial body of literature published over the last few decades in the geosciences, as well as in mathematical and physical journals, and provide a comprehensive picture of climate dynamics. Moreover, this picture should appeal to a readership of physicists, and help stimulate interdisciplinary research activities. 

For decades meteorology and oceanography, on the one side, and physics, on the other, have had a relatively low level of interaction, with by-and-large separate scientific gatherings and scholarly journals. Recent developments in dynamical systems theory, both finite- and infinite-dimensional, as well as in random processes and statistical mechanics, have created a common language that makes it possible, at this time, to achieve a higher level of communication and mutual stimulation.

The key aspects of the field that we want to tackle here are the natural variability of the climate system, the deterministic and random processes that contribute to this variability, its response to perturbations, and the relations between internal and external causes of observed changes in the system. Moreover, we will present tools for the study of critical transitions in the climate system, which can help understand and possibly predict the potential for catastrophic climate change. 

In Sect.~\ref{climatedynamics},  we provide an overview for non-specialists of the way climate researchers collect and process information on the state of the atmosphere, the land surface, and the oceans. Next, the conservation laws and the equations that govern climatic processes are introduced.

An important characteristic of the climate system is the already mentioned coexistence and nonlinear interaction of multiple subsystems, processes and scales of motion. This state of affairs entails two important consequences that are also addressed in Sect.~\ref{climatedynamics}. 
First is the need for scale-dependent filtering: on the positive side, this filtering leads to simplified equations; on the negative one, it calls for so-called parametrization of unresolved processes, i.e. for the representation of subgrid-scale processes in terms of the resolved, larger-scale ones.
Second is the fact that no single model can encompass all the subsystems, processes and scales, hence the need for resorting to a hierarchy of models. The section ends by discussing 
present-day standard protocols for climate modeling and the associated problem of evaluating the models' performance in a coherent way.


Section~\ref{climatevariability} treats in greater depth climate variability. We describe the most important modes of climate variability and provide an overview of the coexistence of several equilibria in the climate system, and of their dependence on parameter values. While the study of bifurcations and exchange of stability in the climate system goes back to the work of E. N. Lorenz, H. M. Stommel and G. Veronis in the 1960s \citep[e.g.,][]{Ghil1987, dijkstra2013, Ghil.2019}, a strikingly broadened interest in these matters has been stimulated by the borrowing from the social sciences of the term ``tipping points'' \cite{Gladwell00, Lenton.tip.08}.

Proceeding beyond multiple equilibria, we show next how complex processes give rise to the system's internal variability by successive instabilities setting in, competing and eventually leading to the quintessentially chaotic nature of the evolution of climate. This section concludes by addressing the need for using random processes to model the faster and smaller scales of motion in multiscale systems, and by discussing markovian and non-markovian approximations for the representation of the neglected degrees of freedom. We will also discuss top-down vs. data-driven approaches. 

Section~\ref{sensitivity} delves into the analysis of climate response. The response to external forcing of a physico-chemical system out of equilibrium is the overarching concept we use in clarifying the mathematical and physical bases of climate change. We critically appraise climate models as numerical laboratories and review ways to test their skill at simulating past and present changes, as well as at predicting future ones. The classical concept of equilibrium climate sensitivity is critically presented first, and we discuss its merits and limitations. 

We present next the key concepts and methods of nonautonomous and random dynamical systems, as a framework for the unified understanding of intrinsic climate variability and forced climate change, and emphasize the key role of pullback attractors in this framework. These concepts have been introduced only quite recently into the climate sciences, and we show how pullback attractors and the associated dynamical-systems machinery provide an excellent setting for studying the statistical mechanics of the climate system as an open system.

This system is subject to variations in the forcing and in its boundary conditions on all time scales. Such variations include, on different time scales, the incoming solar radiation, the position of the continents, and the sources of aerosols and greenhouse gases. We further introduce time-dependent invariant measures on a parameter-dependent pullback attractor, and the Wasserstein distance between such measures, as the main ingredients for a more geometrical treatment of climate sensitivity in the presence of large and sudden changes in the forcings. 

We then outline, in the context of non-equilibrium statistical mechanics, Ruelle's response theory as an efficient and flexible tool for calculating climate response to small and moderate natural and anthropogenic forcings, and reconstruct the properties of the pullback attractor from a suitably defined reference background state. 
The response of a system near a tipping point is studied, and we emphasize the link between properties of the autocorrelation of the unperturbed system and its vicinity to the critical transition, along with their implications in terms of tell-tale properties of associated time series. 

Section~\ref{critical} is devoted to discussing multistability in the climate system, and the critical transitions that occur in the vicinity of tipping points in systems possessing multiple steady states. The corresponding methodology is then applied to the transitions between a fully frozen, so-called snowball state of our planet and its warmer states. These transitions have played a crucial role in modulating the appearance of complex life forms. We  introduce the concept of an edge state, a dynamical object that has helped explain bistability in fluid mechanical systems, and argue that such states will also yield a more complete picture of tipping points in the climatic context. Finally, we present an example of a more exotic chaos-to-chaos critical transition that occurs in a delay differential equation model for the Tropical Pacific.

In Section~\ref{conclusions}, we briefly summarize this review paper's main ideas and introduce complementary research lines that have not been discussed herein, as well as a couple of the many still open questions. Appendix \ref{app:acronyms} contains a table of scientific acronyms and one of institutional acronyms that are used throughout the paper. 

We have started this section by characterizing the climate system and giving a broad-brush description of its behavior. But we have not defined the concept of climate as such, since we do not have as yet a consensual definition of what the climate, as opposed to weather, really is. An old adage states that ``Climate is what you expect, weather is what you get.'' 

Clearly this implies that stochastic and ergodic approaches must play a role in disentangling the proper types of averaging on the multiple time and space scales involved. A fuller understanding of the climate system's behavior should eventually lead to a proper definition of climate. Mathematically rigorous work aimed at such a definition  is being undertaken but is far from complete (e.g., F. Flandoli, Lectures at the Institut Henri Poincar\'e, Paris, October 2019; cf.~\url{http://users.dma.unipi.it/flandoli/IHP_contribution_Flandoli_Tonello_v3.pdf}).



\section{A Brief Introduction to Climate Dynamics}
\label{climatedynamics}

\subsection{{Climate Observations: Direct and Indirect}} 
\label{ssec:observations}

A fundamental difficulty in the climate sciences arises from humanity's insufficient ability to collect data of standardized quality, with sufficient spatial detail and of sufficient temporal coverage. Instrumental data sets have substantial issues of both synchronic and diachronic coherence. Moreover, such data sets only extend, at best, for about one-to-two centuries into the past. In this section, we cover first instrumental data sets and then so-called historical and proxy data sets, which use indirect evidence on the value of meteorological observables before the industrial era. 

\vspace{-0.35cm}

\subsubsection{Instrumental Data and Reanalyses}
\label{sssec:instrumental}

Since the establishment of the first meteorological stations in Europe and in North America in the $19^{\rm{th}}$ century, the extent and quality of the network of observations and the technology supporting the collection and storage of data have rapidly evolved. Still, at any given time, the spatial density of data changes dramatically across the globe, with much sparser observations over the oceans and over land areas characterized by low population density or a low degree of technological development \citep[e.g.,][Fig.~1]{Ghil.Mal.1991}. 

\begin{figure}[ht]
\centering
\includegraphics[width=0.9\columnwidth]{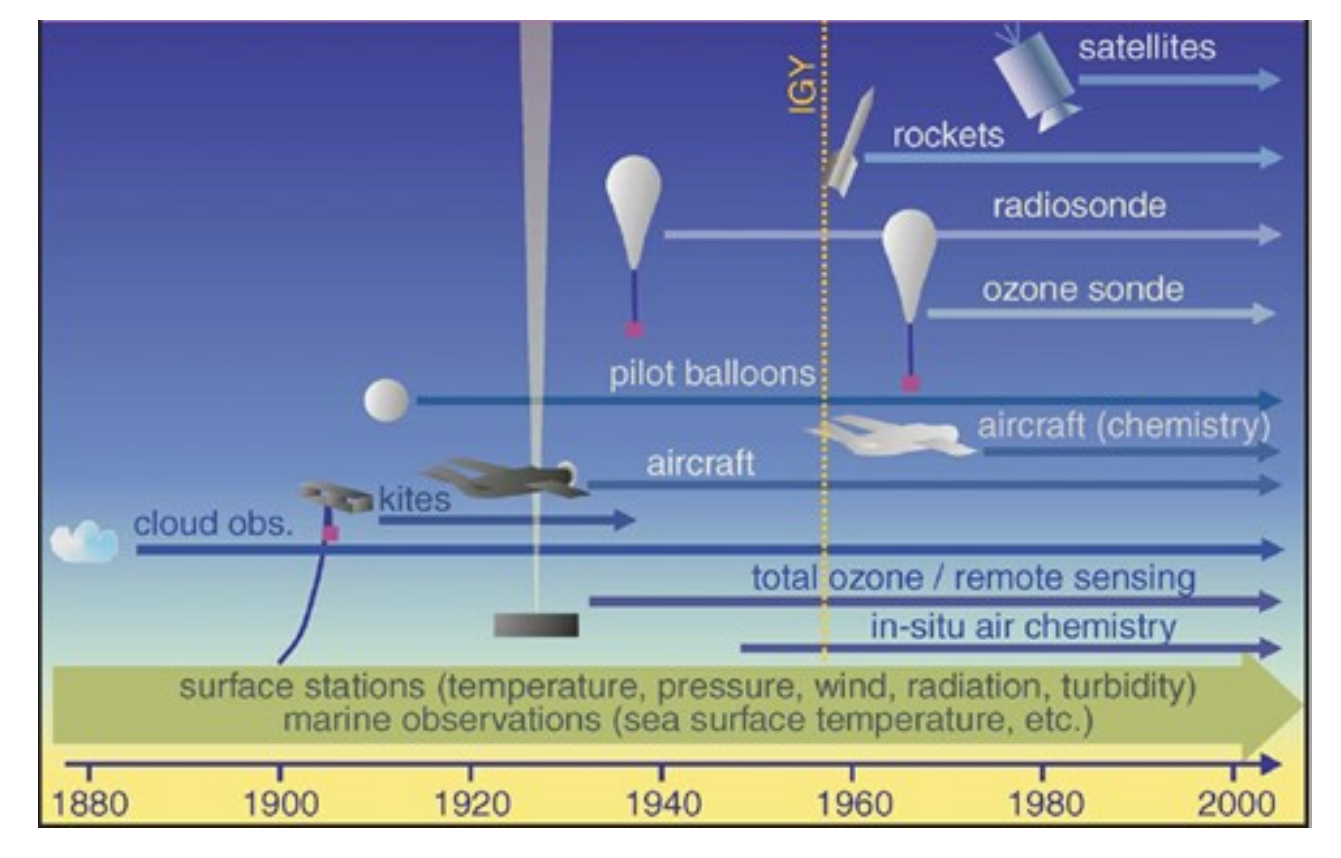}
\caption{{\small Schematic diagram representing the evolution of the observing network for weather and climate data. The dotted vertical line corresponds to the International Geophysical Year (IGY).} Courtesy of Dick Dee.} 
\label{obs0}
\end{figure}

Starting in the late 1960s, polar orbiting and geostationary satellites have led to a revolution in collecting weather, land surface and ocean surface data. Space-borne instruments are now remotely sensing many climatic variables from the most remote areas of the globe; for instance, they measure the overall intensity and spectral features of emitted infrared and reflected visible and ultraviolet radiation, and complex algorithms relate their raw measurements to the actual properties of the atmosphere, such as temperature and cloud cover.

\begin{figure}[ht]
\centering
\includegraphics[width=0.9\columnwidth]{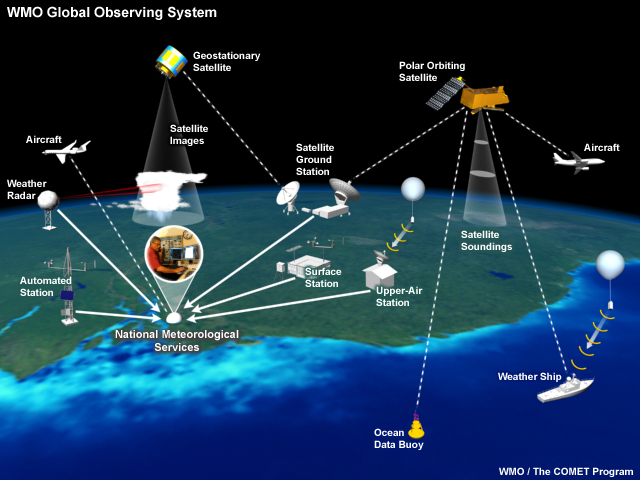}
\caption{{\small An illustration of the instruments and platforms that comprise the World Meteorological Organization's (WMO's) Global Observing System (GOS). From the COMET\copyright~website at \texttt{http://meted.ucar.edu/} of the University Corporation for Atmospheric Research (UCAR), sponsored in part through a cooperative agreements with the National Oceanic and Atmospheric Administration (NOAA), U.S. Department of Commerce (DOC). \copyright 1997--2016 UCAR; all rights reserved.}\label{obs1}}
\end{figure}

Figure~\ref{obs0} represents schematically the evolution of the observational network for climatic data, while Fig.~\ref{obs1} portrays the instruments that today comprise the Global Observing System of the Word Meteorological Organization (WMO), the United Nations agency that coordinates the collection and quality check of weather and climate data over the entire globe.

\begin{figure*}[ht]
\centering
\includegraphics[width=0.8\textwidth]{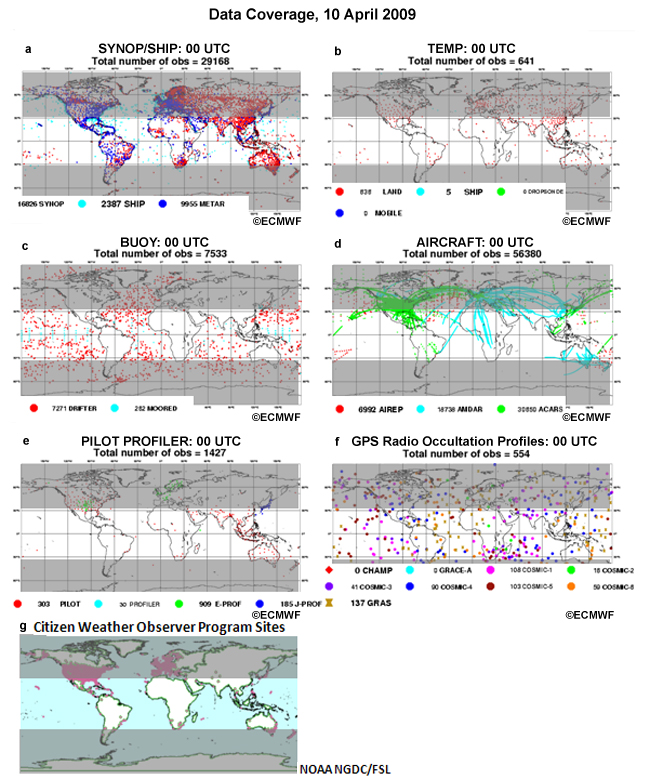}
\caption{{\small Maps of point observations from the World Meteorological Organization's (WMO's) Global Observing System (GOS) on 10 April 2009: (a) synoptic weather station and ship reports; (b) upper-air station reports; (c) buoy observations; (d) aircraft wind and temperature; (e) wind profiler reports; (f) temperature and humidity profiles from Global Positioning System (GPS) radio occultation; and (g) observations from citizen weather observers. The tropics are the bright areas bordered by $\pm30^\circ$ latitude. From the COMET\copyright~website at \texttt{http://meted.ucar.edu/} of the University Corporation for Atmospheric Research (UCAR), sponsored in part through cooperative agreements with the National Oceanic and Atmospheric Administration (NOAA), U.S. Department of Commerce (DOC). \copyright 1997-2016 UCAR; all rights reserved.}}\label{obs2}
\end{figure*}

\begin{figure*}[ht]
\centering
\includegraphics[width=0.8\textwidth]{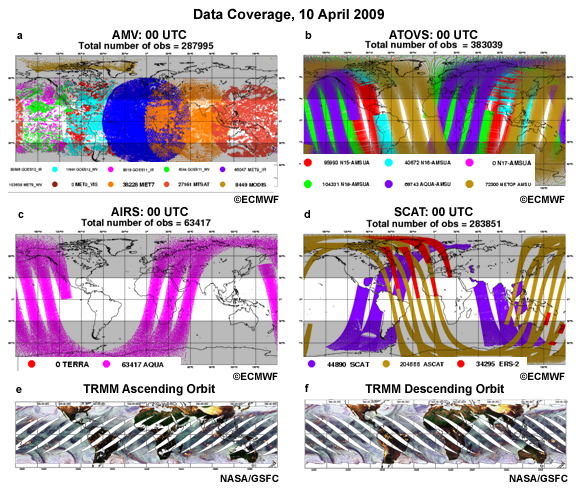}
\caption{{\small Geographic distribution of observing systems: (a) geostationary satellite observations; (b) and (c) polar-orbiting satellite soundings; (d) ocean surface scatterometer-derived winds; and (e, f) TRMM Microwave Imager (TMI) orbits. Each color represents the coverage of a single satellite. Observations in (b) and (c) represent vertical layers and area-averaged values. The tropics are marked by the lighter areas bordered by $\pm30^\circ$ latitude. 
Same source as Fig.~\ref{obs2}. \copyright 1997-2016 UCAR; all rights reserved.}}
\label{obs3}
\end{figure*}

Since the early $20^{\rm{th}}$ century, the daily 
measurements have grown in number by many orders of magnitude and cover now more regularly the entire globe, even though large swaths of the Earth still feature relatively sparse observations. Figures~\ref{obs2} and~\ref{obs3} illustrate the coverage and variety of the observing systems available at present to individual researchers and practitioners, as well as to environmental and civil-protection agencies. Note that the so-called conventional network of ground-based weather stations and related observations has evolved since the Global Weather Experiment in the late 1970s, cf. Fig.~1 in \citet{bengtsson81} but only marginally so: it is the remote-sensing observations that have increased tremendously in number, variety and quality.

The number and quality of oceanographic observations was several orders of magnitude smaller than that of meteorological ones in the 1980s \cite{Munk.Wunsch.82, Ghil.Mal.1991}. Here also the advent of space-borne altimetry for sea surface heights (SSHs), scatterometry for surface winds, and other remote-sensing methods has revolutionized the field \cite[e.g.,][]{Rob.10}.  This number, however, is still smaller by at least one order of magnitude than that of atmospheric observations since, as pointed out by \citet{Munk.Wunsch.82}, the oceans' interior is not permeable to exploration by electromagnetic waves. This is a fundamental barrier hindering our ability to directly observe the deep ocean.

Observational data for the atmosphere and oceans are at any rate sparse, irregular, and of different degrees of accuracy, while in many applications one has to obtain the best estimate, with known error bars, of the state of the atmosphere or oceans at a given time and with a given, uniform spatial resolution. More often than not, this estimate also needs to include meteorological, oceanographic or coupled-system variables, such as vertical wind velocity or surface heat fluxes, that can only be observed very poorly or not at all.

The very active field of {\it data assimilation} has developed to bridge the gap between the observations that are, typically, discrete in both time and space, and the continuum of the atmospheric and oceanic fields. Data assimilation --- as distinct from polynomial interpolation, statistical regression or the inverse methods used in solid-earth geophysics --- first arose in the late 1960s from the needs of numerical weather prediction (NWP), on the one hand, and the appearance of time-continuous data streams from satellites, on the other \cite{Charney.ea.69, Ghil.ea.79}. NWP is, essentially, an initial-value problem for the partial differential equations (PDEs) governing large-scale atmospheric flows \cite{Richardson.1922} that needed a complete and accurate initial state every 12 or 24 hours.
 
\begin{figure}
\centering
\includegraphics[width=0.9\columnwidth]{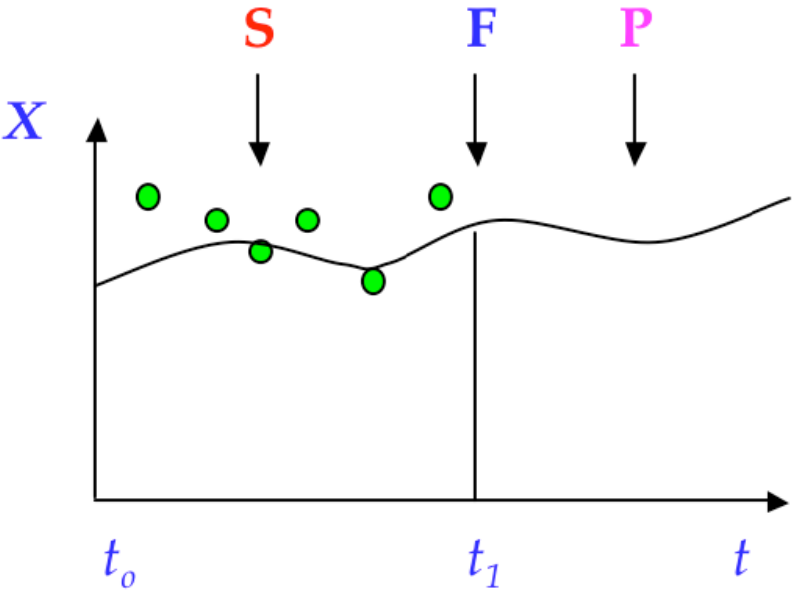}
\caption{{\small Schematic diagram of filtering ({\bf F}), smoothing ({\bf S}) and prediction ({\bf P}); green filled circles are observations. Based on \citet{Wiener.49}.}} 
\label{fig:smooth}
\end{figure}

Data assimilation combines partial and inaccurate observational data with a dynamic model, based on physical laws, that governs the evolution of the continuous medium under study in order to provide the best estimates of the state of the medium. 
This model also is subject to errors, due to incomplete knowledge of the smaller-scale processes, numerical discretization errors and other factors. Given these two sources of information, observational and physico-mathematical, there are three types of problems that can be formulated and solved, given measurements over a time interval $\{t_0 \le t \le t_1\}$: filtering, smoothing and prediction; see Fig.~\ref{fig:smooth}. 

{\it Filtering} involves obtaining a best-possible estimate of the state $\vx(t)$ at $t = t_1$, {\it smoothing} at all times $t_0 \le t \le t_1$, and {\it prediction} at times $t > t_1$. Filtering and prediction are typically used in NWP, and can be considered as the generation of short ``video loops,'' while smoothing is typically used in climate studies, and resembles the generation of long ``feature movies'' \cite{Ghil.Mal.1991}.

\begin{figure}
\centering
\includegraphics[width=0.9\columnwidth]{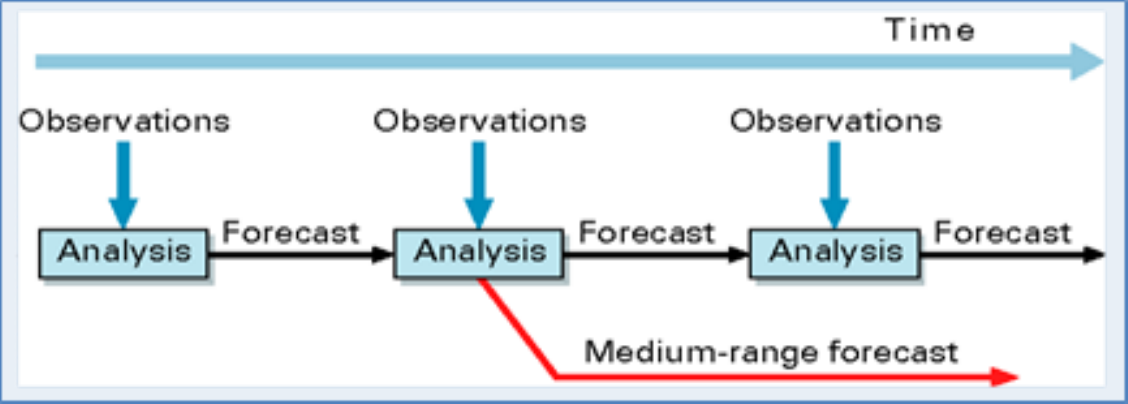}
\caption{{\small Schematic diagram of a forecast--assimilation cycle that is used for constructing the best estimates of the state of the atmosphere, oceans or both through the procedure of data assimilation: observational data are dynamically interpolated using the a meteorological, oceanographic or coupled model to yield the analysis products. The red arrow corresponds to a longer forecast, made only from time to time. Greater detail for the case of operational weather prediction appears in \citet[Fig.~1]{Ghil.89}.}} 
\label{analysis}
\end{figure}

Figure~\ref{analysis} illustrates a so-called {\it forecast--assimilation cycle}, as used originally in NWP: at evenly spaced, pre-selected times $\{t_k: k = 1, 2, \ldots, K\}$, one obtains an \textit{analysis} of the state $\vx(t_k)$ by combining the observations over some interval preceding the time $t_k$ with the forecast from the previous state $\vx(t_{k-1})$ \cite{bengtsson81, kalnay2003}. Many variations on this relatively simple scheme have been introduced in adapting it to oceanographic data \cite[and references therein]{Ghil.Mal.1991} and to space plasmas \cite[e.g.,][]{Merkin.ea.16} or to actually using the time-continuous stream of remote-sensing data \cite{Ghil.ea.79}. \citet{Carrassi2018} have recently provided a  comprehensive review of data assimilation methodology and applications.

Analyses are routinely used for numerical weather forecasts and take advantage of the continuous improvements of models and observations. But climate studies require data of consistent spatial resolution and accuracy over long time intervals, over which an operational NWP center might have changed its numerical model or its data assimilation scheme, as well as its raw data sources. To satisfy this need, several NWP centers have started in the 1990s to produce so-called {\it reanalyses} that use the archived data over multidecadal time intervals, typically since World War II, as well as the best model and data assimilation method available at the time of the reanalysis project. For obvious reasons of computational cost, reanalyses are often run at a lower spatial resolution than the latest version in operational use at that time.

Some leading examples of such diachronically coherent reanalyses for the atmosphere are those produced by the European Centre for Medium-range Weather Forecasts \cite[ECMWF:][]{Dee2011}, the NCEP-NCAR Reanalysis produced in collaboration by the U.S. National Centers for Environmental Prediction (NCEP) and the National Center for Atmospheric Research \cite[NCAR:][]{Kistler01}, and the JRA-25 reanalysis produced by the Japan Meteorological Agency \cite[JMA:][]{Onogi07}. While these reanalyses agree pretty well for fields that are relatively well observed, such as the geopotential field (see Sect.~\ref{conservation} 
herein) over the continents of the Northern Hemisphere, substantial differences persist in their fields over the Southern Hemisphere or those that are poorly observed or not at all \cite{KZZ05,Dellaquila05,Dellaquila07,Marques09,Marques10,Kim13}. 

Recently, \citet{Compo2011}  produced a centennial reanalysis from 1871 to the present by assimilating only surface pressure reports and using observed monthly sea-surface temperature and sea-ice distributions as boundary conditions, while \citet{Poli2016} provided a similar product for the time interval 1899--2010, where the surface pressure and the surface winds were assimilated. These enterprises are motivated by the need to provide a benchmark for testing the performance of climate models for the late $19^{\rm{th}}$ and the $20^{\rm{th}}$ century. 

A similar need has arisen for the oceans: on the one hand, several much more sizeable data sources have become available through remote sensing, and led to detailed ocean modeling; on the other, the study of the coupled climate system requires a more uniform data set, albeit less accurate than for the atmosphere alone. Thus the equivalent of a reanalysis for the oceans had to be produced, in spite of the fact that the equivalent of NWP for the oceans did not exist. A good example of a diachronically coherent data set for the global ocean is the Simple Ocean Data Assimilation \cite[SODA:][]{Carton.Giese.2008}. More recently, the community of ocean modelers and observationalists has delivered several ocean reanalyses able to provide a robust estimate of the state of the ocean \cite{Lee2009, Balmaseda2015}.

Finally, by relying on recent advances in numerical methods and in the increased availability of observational data, as well as of increased performance of computing and storage capabilities, 
coupled atmosphere--ocean data assimilation systems have been constructed \cite[e.g.,][]{Penny2017}. \citet{Vannitsem2016} provide a theoretical rationale for the need of coupled data assimilation schemes to be able to deal effectively with the climate system's multiscale instabilities. These coupled systems play a key role in efforts to produce seamless weather, subseasonal-to-seasonal (S2S), seasonal, and interannual climate predictions \cite{palmer_2008, S2S.book}; they have already been used for constructing climate reanalyses \cite[e.g.,][]{Karspeck2018,Laloyaux2018}.

\subsubsection{Proxy Data}
\label{sssec:proxy}
As mentioned already repeatedly, and discussed in greater detail in Sect.~\ref{ssec:time_scales} 
below, climate variability covers a vast range of time scales, and the information we can garner from the instrumental record is limited to the last century or two. Even so-called historical records only extend to the few millennia of a literate humanity \cite{Lamb1972}. In order to extend our reach beyond this eyeblink of the planet's life, it is necessary to resort to indirect measures of past climatic conditions able to inform us on its state thousands or even millions of years ago. 

Climate proxies are physical, chemical or biological characteristics of the past that have been preserved in various natural repositories and that can be correlated with the local or global state of the atmosphere, oceans or cryosphere at that time.  Paleoclimatologists and geochemists currently take into consideration multiple proxy records, including coral records \cite{Boiseau1999, Kara.ea.15} and tree rings \cite{Esper.ea.02} for the last few millennia, as well as  marine-sediment \cite{Duplessy1985, Taricco.ea.09} and ice-core \cite{Jouzel1993, Andersen2004} records for the last two million years of Earth history, the Quaternary. Glaciation cycles, i.e. an alternation of warmer and colder climatic episodes,  dominated  the latter era. The chemical and physical characteristics, along with the accumulation rate of the samples, require suitable calibration and dating, and are used thereafter to infer some of the properties of the climate of the past \cite[e.g.,][and references therein]{Cronin2010, Ghil1994b}.

Proxies differ enormously in terms of precision, uncertainties in the values and dating, and spatio-temporal extent, nor do they cover homogeneously the Earth. It is common practice to combine and cross-check many different sources of data to have a more precise picture of the past \cite{Imbrie1986,Cronin2010}. Recently, data assimilation methods have started to be applied to this problem as well, using simple models and addressing the dating uncertainties in particular \cite[e.g.,][]{Roques.ea.14}. 
Combining the instrumental and proxy data --- with their very different characteristics of resolution and accuracy --- is a complex, and sometimes controversial, exercise in applied statistics. An important example is that of estimating the globally averaged surface air temperature record well before the industrial era, cf. Fig.~\ref{obs4} and previous discussion on the hockey stick controversy in Sect.~\ref{sssec:hockey}. 

\subsection{Climate Variability on Multiple Time Scales}
\label{ssec:time_scales}
\begin{figure}
\subfloat[Stommel diagram of ocean variability]{
    \label{fig:Stommel} 
    \centering
   \includegraphics [angle=270,width=0.8\columnwidth]{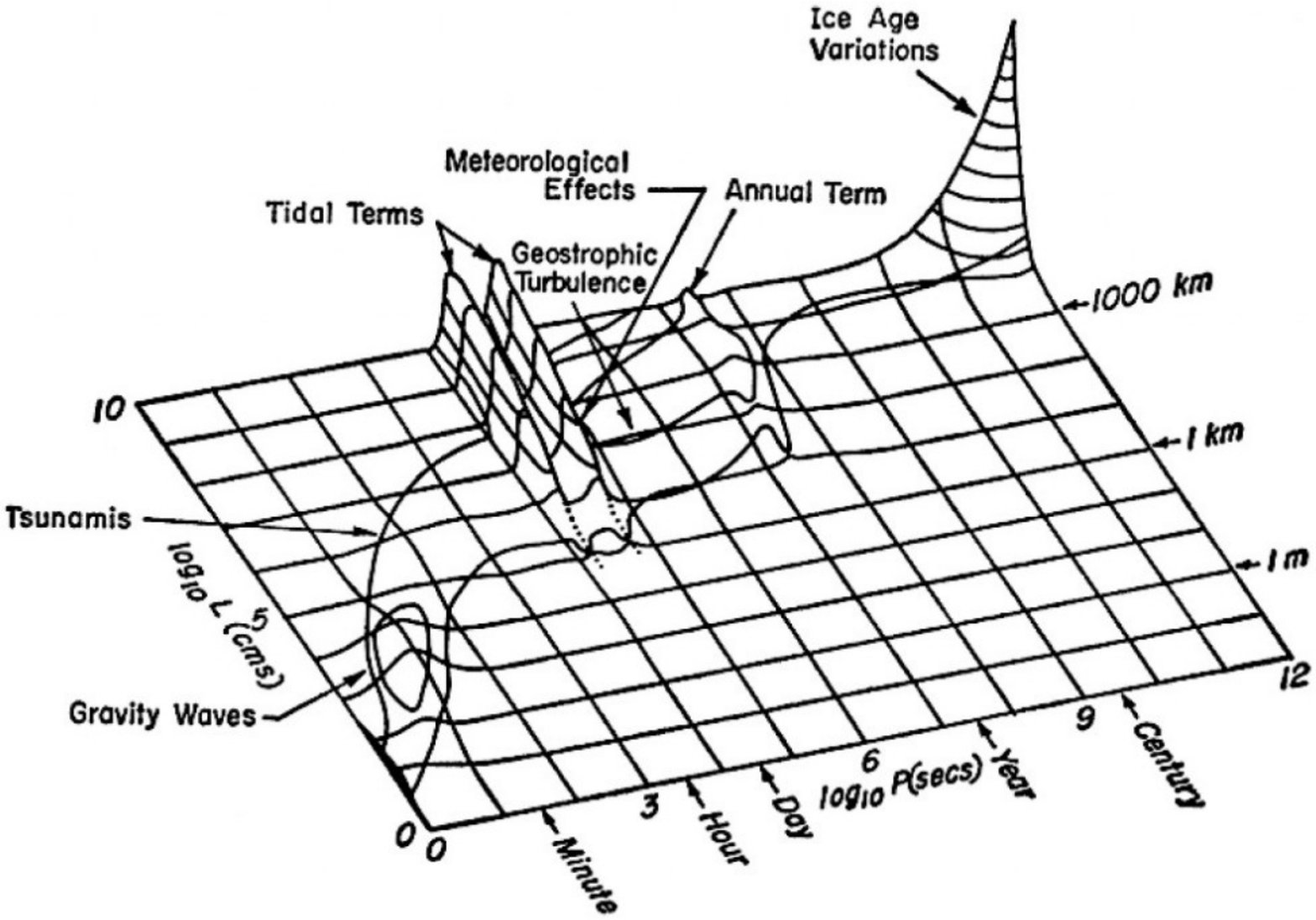}
}

\subfloat[Chelton diagram of ocean variability]{
    \label{fig:Chelton} 
    \centering
   \includegraphics [angle=0,width=0.8\columnwidth]{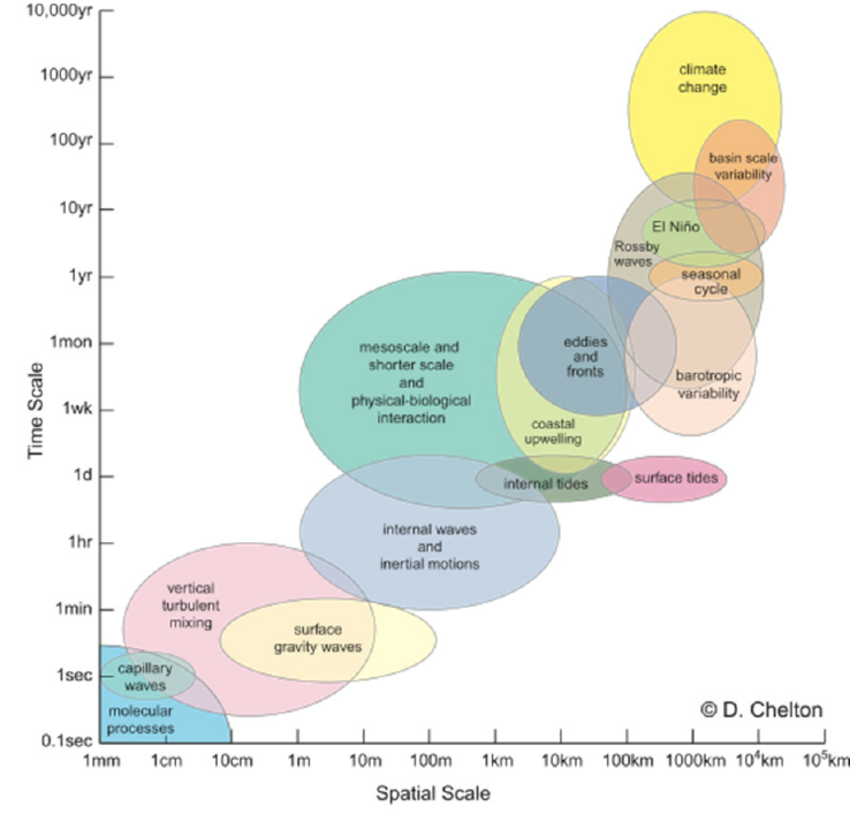}   
}

\subfloat[COMET diagram of atmospheric variability]{
    \label{fig:COMET} 
    \centering
   \includegraphics [angle=0, width=.9\columnwidth]{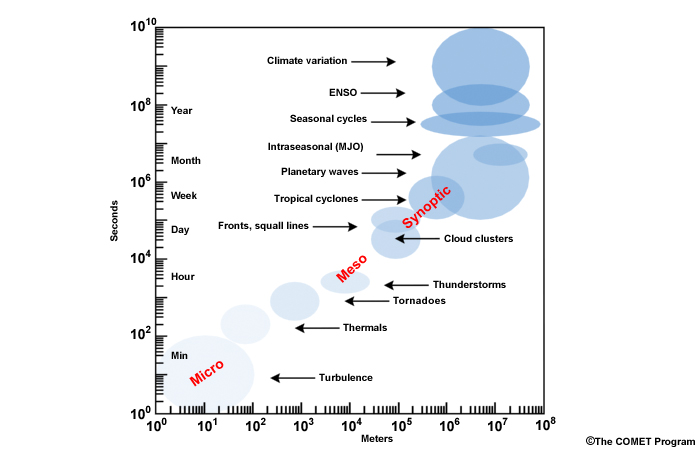}
}
\caption{\small Idealized wavelength-and-frequency power spectra for the climate system. (a) The original \citet{Stommel63} diagram representing the spectral density (vertical coordinate) of the oceans' variability as a function of the spatial and temporal scale. Reproduced with permission. (b) Diagram representing qualitatively the main features of the ocean variability; courtesy of D. Chelton.(c) Same as b), describing here the variability of the atmosphere, \copyright{The COMET program}.}
\label{stommel}
\end{figure}

The presence of multiple scales of motions --- in space and in time --- in the climate system can be summarized through so-called Stommel diagrams. Figure~\ref{fig:Stommel} presents the original \citet{Stommel63} diagram, in which a somewhat idealized spectral density associated with the oceans' variability was plotted in logarithmic space and time scales, while identifying characteristic oceanic phenomena whose variance exceeds the background level. Stommel diagrams describe the spatial-temporal variability in a climatic subdomain by associating different, phenomenologically well-defined dynamical features --- such as cyclones and long waves in the atmosphere or meanders and eddies in the ocean --- with specific ranges of scales; they emphasize relationships between spatial and temporal scales of motion. Usually, specific dynamical features are associated with specific approximate balances governing the properties of the evolution equations of the geophysical fluids, cf. Sect.~\ref{balance}.

In Figs.~\ref{fig:Chelton} and \ref{fig:COMET}, a qualitative Stommel diagram portrays today's estimates of the main range of spatial and temporal scales in which variability is observed for the oceans and the atmosphere, respectively. One immediately notices that larger spatial scales are typically associated with longer temporal scales, in both the atmosphere and oceans. The two plots clearly show that, for both geophysical fluids,  a ``diagonal'' of high spectral density in the wavelength--frequency plane predominates. As the diagonal reaches the size of the planet in space, the variability can no longer maintain this proportionality of scales and it keeps increasing in time scales, which are not bounded, while the spatial ones are.

Both extratropical cyclones in the atmosphere and eddies in the ocean are due to baroclinic instability, but their characteristic spatial extent in the ocean is ten times smaller and their characteristic duration is one hundred times longer than in the atmosphere. In Fig.~\ref{fig:COMET}, three important meteorological scales are explicitly mentioned: the microscale (small-scale turbulence), the mesoscale (e.g. thunderstorms and frontal structures), and the synoptic scale (e.g. extratropical cyclones).
 
Given the different dynamical variability ranges in space and time, different classes of numerical models, based on different dynamical balances, can  simulate explicitly only one or a few such dynamical ranges. The standard way of modeling processes associated with a particular range of scales is to ``freeze'' processes on slower time scales or to prescribe their slow, quasi-adiabatic effect on the variability being modeled,  in order to handle the processes that are too large or too slow in scale to be included in the model. 

As for the faster processes, these are ``parametrized,'' i.e. one attempts to model their net effect on the variability of interest. 
Such parametrizations have been, until fairly recently, purely deterministic but have started over the last decade or so to be increasingly stochastic \cite[and references therein]{palmer_stochastic_2009}. We will discuss the mathematics behind parametrizations and provide a few examples in Sect.~\ref{multiplescales}.

To summarize, there are about 15 orders of magnitude in space and in time that are active in the climate system, from continental drift at millions of years to cloud processes at hours and shorter. The presence of such a wide range of scales in the system provides a formidable challenge for its direct numerical simulation. There is no numerical model that can include all the processes that are active on the various spatial and temporal 
 scales and can run for 10$^7$ simulated years. Ockham's razor and its successors, including Poincar\'e's parsimony principle \cite{Poincare.1902}, suggest that, if we had one, it wouldn't necessarily be such a good tool for developing scientific insight, rather than just a gigantic simulator not helping scientists to distinguish the forest from the trees.
 
Still, there are increasing efforts for achieving ``seemless prediction'' across time and space scales \citep[e.g.,][]{palmer_2008, S2S.book}. \citet{Merryfield.ea.2020} provide a comprehensive review of the most recent efforts for bridging the gap between S2S and seasonal-to-decadal prediction.

\subsubsection{A Survey of Climatic Time Scales}\label{sssc:survey}

\begin{figure}[ht]
\subfloat[Composite power spectrum of climate]{
    \label{fig:Mitchell} 
    \centering
   \includegraphics [angle=0,width=0.7\columnwidth]{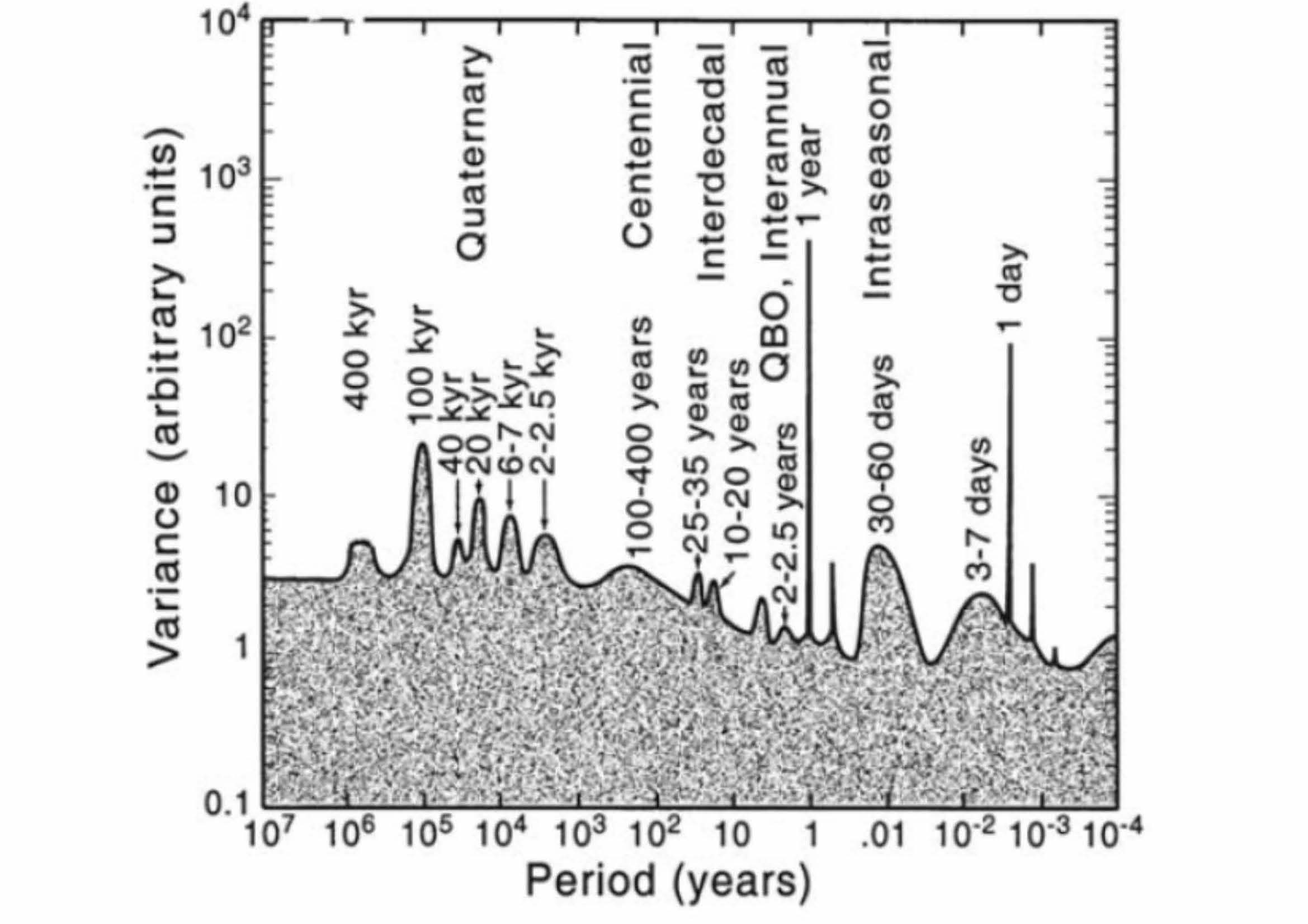}
}\\
\subfloat[Spectrum of Central England Temperatures]{
    \label{fig:CET} 
    \centering
   \includegraphics [width=0.7\columnwidth]{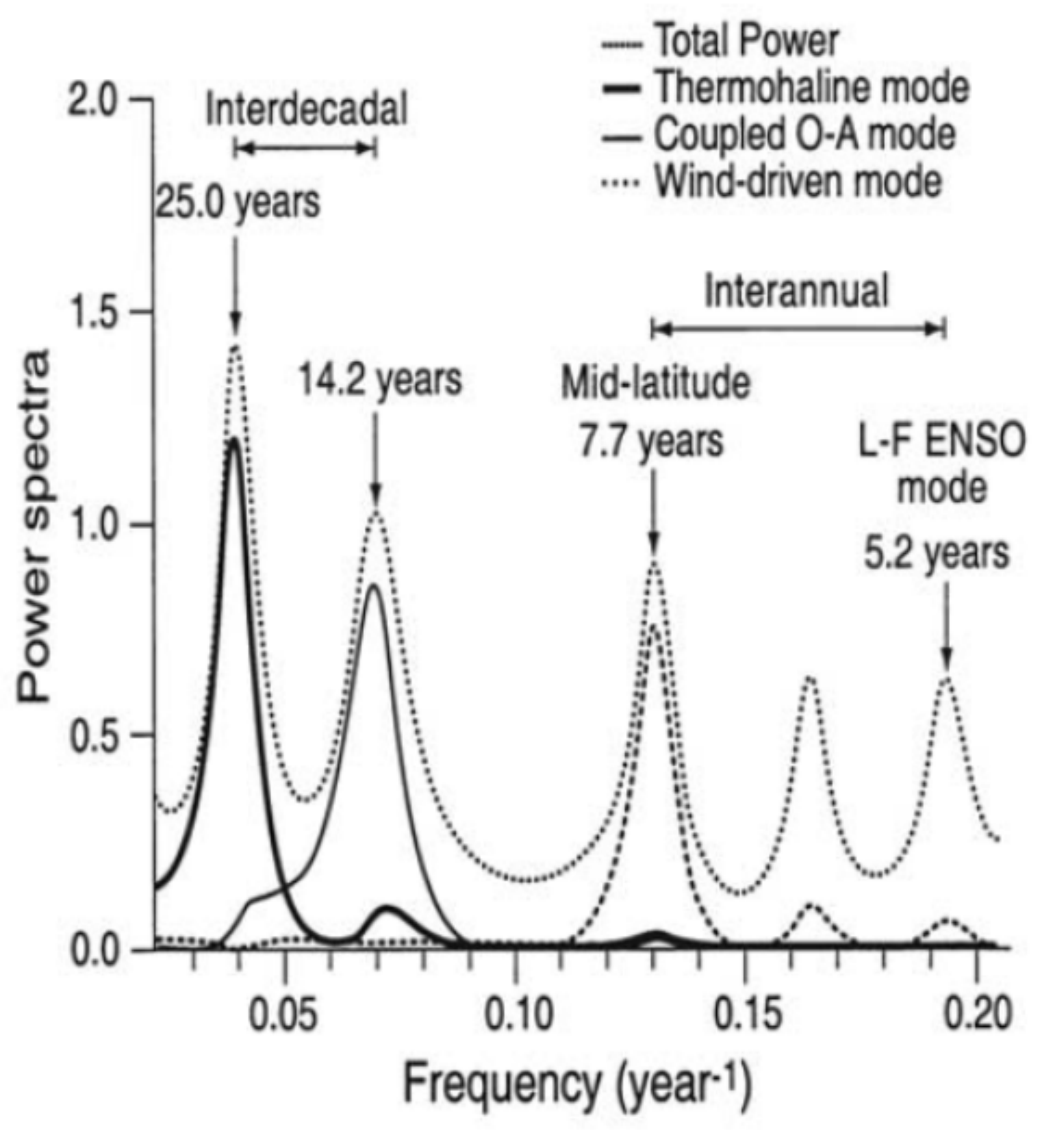}
}
\caption{\small Power spectra of climate variability across time scales.
(a) An artist's rendering of the composite power spectrum of climate variability for a generic climatic variable, from hours to millions of years; it shows the amount of variance in each frequency range. 
(b) Spectrum of the Central England Temperature time series from 1650 to the present. Each peak in the
spectrum is tentatively attributed to a physical mechanism; see \citet{Plaut1995} for details.
From \citet{Ghil2002}. \copyright John Wiley and Sons, Ltd. Reproduced with permission. } 
\label{f:MG1}
\end{figure}

Combining proxy and instrumental data allows one to gather information not only on the mean state of the climate system, but also on its variability on many different scales. An artist's rendering of climate variability on all time scales is provided in Fig.~\ref{f:MG1}a.  The first version of this figure was produced by \citet{Mitchell1976} and many versions thereof have circulated since. The figure is meant to provide semi-quantitative information on the spectral  power $S = S(\omega)$, where the angular frequency $\omega$ is $2\pi$ times the inverse of the oscillation period; $S(\omega)$ is supposed to give the amount of variability in a given frequency band for a  generic climatic variable, although one has typically in mind the globally averaged surface air temperature.  Unlike in the Stommel diagrams of Fig.~\ref{stommel}, there is no information on the spatial scales of interest.

This power spectrum is not computed directly by spectral analysis from a time series of a given climatic quantity, such as (local or global) temperature; indeed, there is no single time series that is 10$^7$ years long and has a sampling interval of hours, as the figure would suggest.  Figure~\ref{f:MG1}a  includes, instead, information obtained by analyzing the spectral content of many different time series, for example, the spectrum  of the 335-year long record of Central England temperatures in Fig.~\ref{f:MG1}b. This time series is the longest instrumentally measured record of temperatures; see, though, \citet{Nile2005} for Nile River water levels. Given the lack of earlier instrumental records, one can imagine, but cannot easily confirm, that the higher-frequency spectral features might have changed, in amplitude, frequency or both, over the course of climatic history.  

With all due caution in its interpretation, Fig.~\ref{f:MG1}a reflects three types of variability: (i) sharp lines that correspond to periodically forced variations, at one day and one year; (ii) broader peaks that arise from internal modes of variability; and (iii)  a continuous portion of the spectrum that reflects stochastically forced variations, as well as deterministic chaos \cite{Ghil2002}. 

Between the two sharp lines at 1 day and 1 year lies the synoptic variability of mid-latitude weather systems, concentrated at 3--7 days, as well as intraseasonal variability, i.e. variability that occurs on the time scale of 1--3 months. The latter is also called low-frequency atmospheric variability, a name that refers to the fact that this variability has lower frequency, or longer periods, than the life cycle of weather systems.   Intraseasonal variability comprises phenomena such as the Madden--Julian oscillation of winds and cloudiness in the tropics or the alternation between episodes of zonal and blocked flow in mid-latitudes \cite[]{Ghil1987, Ghil1991, Ghil1991b, Haines1994, Molteni2002}. 

Immediately to the left of the seasonal cycle in Fig.~\ref{f:MG1}a lies interannual, i.e. year-to-year, variability.  An important component of this variability is the El Ni\~no phenomenon in the Tropical Pacific: once about every 2--7 years, the sea-surface  temperatures (SSTs) in the Eastern Tropical Pacific increase by one or more degrees over a time interval of about one year.  This SST variation is associated with changes  in the trade winds over the Tropical Pacific and in sea level pressures \cite[]{Bjerknes1969,Philander1990}; an East--West seesaw in the latter is called the Southern Oscillation. The combined El Ni\~no--Southern Oscillation (ENSO) phenomenon arises through large-scale interaction between the equatorial Pacific and the atmosphere above.  Equatorial wave dynamics in the ocean plays a key role in setting ENSO's  time scale \cite[]{Cane1985, Neelin1994, Neelin1998, Dijkstra2002_ARFM}.  

The greatest excitement among climate scientists, as well as the public, is more recently being generated by interdecadal variability, i.e. climate variability on the time scale of a few decades, the time scale of an individual human's life cycle \cite[]{NRC1995}.  
Figure~\ref{f:MG1}b  represents a ``blow-up" of the interannual-to-interdecadal portion of Fig.~\ref{f:MG1}a. The broad peaks are due to the climate system's internal processes: each spectral component can be associated, at least tentatively, with a mode of interannual 
or interdecadal variability \cite[]{Plaut1995}.  Thus the rightmost peak, with a period of 5.5 years,  can be attributed 
to the remote effect of ENSO's low-frequency mode \cite{Ghil2000, Ghil.SSA.2002}, while the 7.7-year peak captures a North Atlantic 
mode of variability that arises from the Gulf Stream's interannual cycle of meandering and intensification \cite[and references therein]{DG05}. The two interdecadal peaks, near 14 and 25 years,  are  also present in global records, instrumental as well as paleoclimatic \cite[]{Kushnir1994, Mann1998, Moron1998, Delworth2000a, Ghil.SSA.2002}. 

Finally, the leftmost part of Fig.~\ref{f:MG1}a represents paleoclimatic variability. The information summarized here comes exclusively from 
proxy indicators of climate; see Sect. \ref{sssec:proxy}. 

The presence of near-cyclicity is manifest, in this range,  in the broad peaks  present in Fig.~\ref{f:MG1}a between roughly 1~kyr and 1~Myr.  The two peaks at about 20~kyr and 40~kyr reflect variations in Earth's orbit, while the dominant peak at 100~kyr remains to be convincingly explained \cite[]{Imbrie1986, Ghil1994b, Gildor2001}. Quaternary glaciation cycles provide a fertile  testing ground for theories of climate variability  for two reasons: (i) they represent a wide range of climatic conditions;  and (ii) they are much better documented than earlier parts of paleoclimatic history. 

Within these glaciation cycles, there is higher-frequency variability prominent in North Atlantic paleoclimatic records. These are the \citet{Heinrich1988} events, marked by a sediment layer that is rich in ice-rafted  debris, whose near-periodicity is of 6--7 kyr, and the Dansgaard-Oeschger cycles \cite{Dansgaard1993complete}  that provide the peak at around 1--2.5  kyr in Fig.~\ref{f:MG1}a. Rapid changes in temperature, of up to one half of the amplitude of a typical glacial--interglacial temperature difference,  occurred  during Heinrich events and somewhat smaller ones over a Dansgaard-Oeschger  cycle. Progressive cooling through several of the latter cycles, followed by an abrupt warming, defines a Bond cycle \cite{Bond1995}. 
None of these higher-frequency phenomena can be directly connected to orbital or other cyclic forcings. 

In summary,  climate variations  range  from the  large-amplitude climate  excursions of past millennia to  smaller-amplitude fluctuations  on shorter time  scales. Several spectral peaks of variability can be  clearly related to forcing mechanisms; others cannot. In fact, even if the external  forcing were  constant in time --- i.e., if no systematic changes in insolation or atmospheric composition, such as trace gas or  aerosol concentrations,  would occur ---   the climate system would still display variability on many time scales. This statement is clearly true for interannual ENSO variability in the equatorial Pacific, as discussed above.  We will understand multiscale climate variability better in Sects.~\ref{climatevariability} and \ref{sensitivity} 
, where we will look in greater detail at natural climate variability and climate response to forcings, respectively. 

 
\subsubsection{Atmospheric Variability in Mid-latitudes}
\label{ssec:midlat}

Mid-latitude atmospheric variability during boreal winter --- when the winds are stronger and the variability is larger in the Northern Hemisphere --- has long been a major focus of dynamic meteorology. The intent of this section is to motivate the reader to appreciate the complexity of large-scale atmospheric dynamics by focusing on a relatively well understood aspect thereof. We shall see that fairly diverse processes contribute to the spectral features discussed in connection with Fig.~\ref{f:MG1}.

The synoptic disturbances that are most closely associated with mid-latitude weather have characteristic time scales of the order of 3--10 days, with a corresponding spatial scale of the order of 1000--2000~km \cite{Holton}. They roughly correspond to the familiar eastward-propagating cyclones and anticyclones, and emerge as a result of the process of baroclinic instability, which converts available potential energy of the zonal flow into eddy kinetic energy. 

This conversion is a crucial part of the \citet{Lor55,Lor67} energy cycle 
and it occurs through the lowering of the center of mass of the atmospheric system undergoing an unstable development. 
Baroclinic instability \cite{char48, Holton,Vallis_atmospheric_2006} is active when the meridional temperature gradient or, equivalently (see below), the vertical wind shear are strong enough. These conditions are more readily verified in the winter season, which features a large equator-to-pole temperature difference and a strong mid-latitude jet \cite{Speranza83, Holton}. 

The space--time spectral analysis introduced by \citet{Haya71} and refined by \citet{Pratt76} and by \citet{KF78} builds upon the idea of the Stommel diagrams in Fig.~\ref{stommel}. In addition, it provides information about the direction and speed at which the atmospheric eddies move and associates to each range of spatial and temporal scales a corresponding weight in terms of spectral power. This information may be obtained, in the first instance, by Fourier analysis of a one-dimensional spatial field and it allows one to reconstruct the propagation of atmospheric waves. This analysis is usually carried out in the so-called zonal, i.e. west-to-east direction; see Sect.~\ref{conservation} 
below.

Next, one can compute the power spectrum in the frequency domain for each spatial Fourier component, and then average the results across consecutive winters to derive a climatology of winter atmospheric waves. The difficulty here lies in the fact that straightforward space--time decomposition will not distinguish between standing and traveling waves: a standing wave will give two spectral peaks corresponding to waves that travel east- and westward at the same speed and with the same phase. This problem can only be circumvented by making assumptions regarding a given wave's nature. For instance, we may assume complete coherence between the eastward and westward components of standing waves and attribute the incoherent part of the spectrum to actual traveling waves. 

\begin{figure*}[ht]
\centering
\includegraphics[angle=270,width=0.9\textwidth]{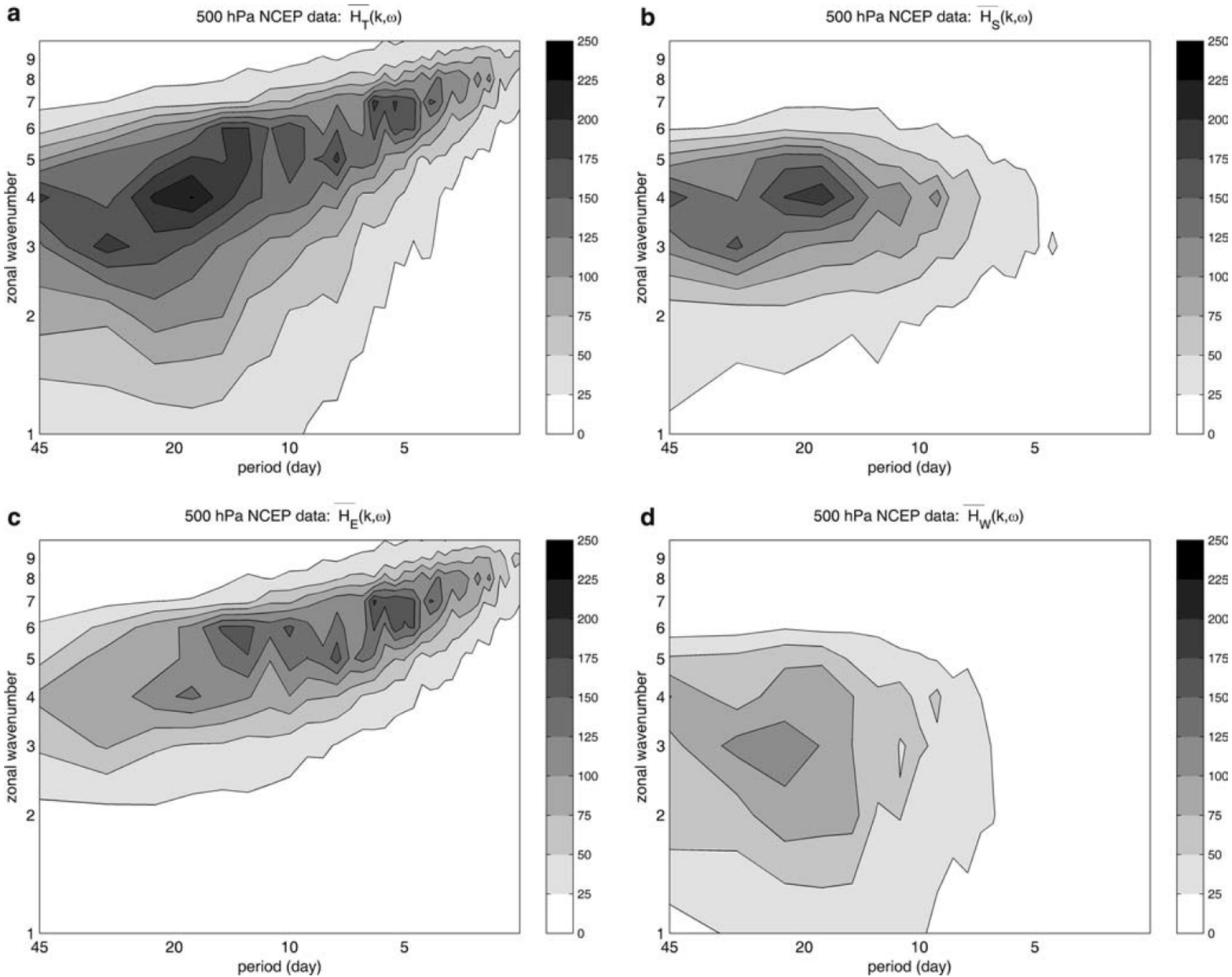}
\caption{\small Variance $\overline{\bf H}$ of the winter (December-January-February) atmospheric fields in the mid-latitudes of the Northern Hemisphere. (a) Total variance $\overline{\bf H}_{\rm T}$; (b) variance associated with standing waves $\overline{\bf H}_{\rm S}$; (c) variance associated with eastward-propagating waves $\overline{\bf H}_{\rm E}$; and (d) variance associated with westward-propagating waves $\overline{\bf H}_{\rm W}$. Based on NCEP-NCAR reanalysis data  \cite{Kistler01}. See text for details. From \citet{Dellaquila05}, with permission from the American Geophysical Union.}
\label{Dellaquila}
\end{figure*}

Figure~\ref{Dellaquila} 
shows the spectral properties of the winter 500-hPa geopotential height field meridionally averaged across the mid-latitudes of the northern hemisphere (specifically, between 30$^\circ$ and 75$^\circ$ N) for the time interval 1957--2002. The properties of all the waves, as well as the standing, eastward- and westward-traveling waves appear in panels (a)--(d), respectively. As discussed later, the 500-hPa height field provides a synthetic yet comprehensive picture of the atmosphere's synoptic to large-scale dynamics. 

Figure~\ref{Dellaquila}c clearly shows that the eastward-propagating waves are dominated by synoptic variability, concentrated between 3--12-day periods and zonal wavenumbers 5--8; note that a single cyclone or anticyclone counts for half-a-wavelength. In addition, the slanting high-variability ridge in $\overline{\bf H}_{\rm E}$ indicates the existence of a statistically defined dispersion relation that relates frequency and wavenumber, in agreement with the basic tenets of baroclinic instability theory \cite{Holton}. 

When looking at the westward propagating variance in Fig.~\ref{Dellaquila}d, one finds that  the dominant portion of the variability is associated with low-frequency, planetary-scale Rossby waves. 
Finally, Fig.~\ref{Dellaquila}b shows the contribution to the variance given by standing waves, which correspond to large-scale, geographically locked and persistent phenomena like blocking events. Note that westward propagating and stationary waves provide the lion's share of the overall variability of the atmospheric field; see also \citet[Fig.~7]{Kimoto1993a}.

The dynamics and energetics of planetary waves are still under intensive scrutiny. Descriptions and explanations of several highly nonlinear aspects thereof are closely interwoven with those of blocking events, identified as persistent, large-scale deviations from the zonally symmetric general circulation \cite{Benzi.ea.1986, Benzi1989, Charney1979, Ghil1987, Kimoto1993a, Legras1985}. \citet[Fig.~1]{Weeks.ea.1997} provide a fine example of contrast between a blocking event and the climatologically more prevalent zonal flow.

Persistent blocking events affect strongly the weather for up to a month over continental-size areas. Such persistence offers some hope for extended predictability of large-scale flows, and of the associated synoptic-scale weather, beyond the usual predictability of the latter, which is believed not to exceed 10--15 days \cite[and references therein]{Lorenz_predictability_1996}. 

Today, the most highly resolved and physically detailed NWP models are reasonably good at predicting the persistence of a blocking event, once the model's initial state is within that event, but not at predicting the onset of such an event or its collapse \cite{Ferranti2015}. Likewise, the capability of necessarily lower-resolution climate models to simulate the spatio-temporal statistics of such events is far from perfect; in fact, relatively limited improvement has been realized in the last two decades \cite{Davini2016}. \citet{Weeks.ea.1997} reproduced successfully in the laboratory key features of the dynamics and statistics of blocking events.
 
 \citet{Ghil.Rob.2002} reviewed several schematic descriptions of the mid-latitude atmosphere's low-frequency variability (LFV) as jumping between a zonal regime and a blocked one or, more generally, a small number of such regimes. This coarse-graining of the LFV's phase space and Markov chain representation of the dynamics continues to inform current efforts at understanding what atmospheric phenomena can be predicted beyond 10--15 days and how. Recently, analyses based on dynamical systems theory have associated blocked flow configurations with higher instability of the atmosphere as a whole \cite{Schubert2016,Faranda2017,LucariniG2019}, as predicted by \citet{Legras1985}.
 
The effect of global warming on the statistics of blocking events has become recently a matter of considerable controversy. The sharper increase of near-surface temperatures in the Arctic than near the Equator is fairly well understood \cite[e.g.,][Fig.~7]{Ghil1976} and has been abundantly documented in recent observations \cite[e.g.,][Fig.~8]{Walsh.2014}. This decrease of pole-to-equator temperature difference $\Delta T$ is referred to as polar amplification of global warming.

\citet{Francis.2012} and \citet{Liu.Curry.2012} have suggested that reduced $\Delta T$ slows down the prevailing westerlies and increases the north--south meandering of the subtropical jet stream, resulting in more frequent blocking events. This suggestion seems to agree with fairly straightforward arguments of several authors on the nature of blocking \cite{Charney1979, Ghil1987, Legras1985}, as illustrated, for instance, in \citet[Fig.~2]{Ghil.Rob.2002}: The strentgh $\psi^{\ast}_{\rm A}$ of the driving jet in the figure is proportional to $\Delta T$, in accordance with standard quasi-geostrophic flow theory,\footnote{See Sects.~\ref{balance} and \ref{QGdynamics} for geostrophic balance, quasi-geostrophy and their consequences.}
and lower jet speeds $\psi_{\rm A}$ favor blocking. \citet{Ruti2006} have also presented observational evidence for a nonlinear relation between the strength of the subtropical jet and the probability of occurrence of blocking events, in agreement with the bent-resonance theory of mid-latitude LFV proposed by \citet{Benzi.ea.1986}.

Considerable evidence against the apparently straightforward argument for global warming as the cause of an increase in midlatitude blocking has accumulated, too, from both observational and modeling studies \cite[and references therein]{Barnes.AA.2015, Hassanz.ea.2014}. The issue is far from settled, as are many questions about other regional effects of global warming.


\subsection{Basic Properties and Fundamental Equations}
\label{conservation}


\subsubsection{Governing Equations}
\label{equations}
The evolution of the atmosphere, the oceans, the soil, and the ice masses can be described by using the continuum approximation, in which these subsystems are represented by field variables that depend on three spatial dimensions and time. For each of the climatic subdomains, we consider the following field variables: the density $\rho$ and the heat capacity at constant volume $C$, with the specific expression for $\rho$ and $C$ defining the thermodynamics of the medium; the concentration of the chemical species $\{\xi_k: 1 \le k \le K\}$ contained in the medium, and present in different phases, e.g. the salt dissolved in the oceans or the water vapor in the atmosphere; the three components $\{v_i: 1 \le i \le 3\}$ of the velocity vector; the temperature $T$; the pressure $p$; the heating rate $J$; and the gravitational potential $\Phi$. 

Note that, by making the thin-shell approximation $H/R\ll1$, where $H$ is the vertical extent of the geophysical fluid and $R$ is the radius of the Earth, we can assume that the gravitational potential at the local sea level is zero and we can thus safely use the approximation $\Phi=gz$, where $\Phi$ is then called the geopotential, $g$ is the gravity at the surface of the Earth, and $z$ is the geometric height above sea level. Moreover, one has to take into account that the climate system is embedded in a noninertial frame of reference that rotates with an angular velocity $\mathbf{\Omega}$ with components $\{\Omega_i: 1 \le i \le 3\}$.  

The PDEs that govern the evolution of the field variables are based on the budget of mass, momentum and energy. When the fluid contains several chemical species, their separate budgets also have to be accounted for \cite{Vallis_atmospheric_2006}. 
In order to have a complete picture of the Earth system, one should in principle also consider the evolution of biological species. Doing so, however, is well beyond our scope here, even though present Earth system models do attempt to represent biological processes, albeit in a simplified way; see later discussion in Sect.~\ref{climatemodelprediction}.

The mass budget equations for the constituent {species} can be written as follows: 
\begin{equation}
\partial_t (\rho \xi_k)= -\partial_i  (\rho \xi_k v_i) +D_{\xi_k}+L_{\xi_k}+S_{\xi_k}.  \label{speciesb}
 \end{equation}
Here $\partial_t$ is the partial derivative in time and $\partial_i$ in the $x_i$-direction; 
$D_{\xi_k}$, $L_{\xi_k}$ and  $S_{\xi_k}$ are the diffusion operator, phase changes, and local mass budget due to other chemical reactions that are associated with $k$.  

The momentum budget's $i^{\rm{th}}$ component is written:
\begin{align}
\partial_t (\rho v_i) = & -\partial_j (\rho v_j v_i) -\partial_i p +\rho \partial_i \Phi \nonumber\\
& - 2\rho \epsilon _{ijk}\Omega_j v_k +T_i+ F_i. \label{NSE}
\end{align}
Here the Levi-Civita antisymmetric tensor $\epsilon_{ijk}$ has been used to write the Coriolis force; $T_i$ indicates direct mechanical forcings, e.g. those resulting from luni-solar tidal effects; $F_i=-\partial_j \tau_{ij}$ corresponds to friction, with $\{\tau_{ij}\}$ the stress tensor; and summation over equal indices is used. Equation~\eqref{NSE} is just a forced version of the momentum equation in the Navier-Stokes equations (NSEs), written in a rotating frame of reference. 

A general state equation valid for both fluid envelopes of the Earth, i.e., the atmosphere and the oceans, is 
\begin{equation}\label{eq:state}
\rho = g(T, p, \xi_1,.\ldots,\xi_K).
\end{equation}
As a first approximation, one can take $K=1$, where $\xi_1=\xi$ is, respectively, moisture for the atmosphere and salinity for the oceans. In the interests of brevity, we will restrict ourselves here to the atmosphere.

In general, one can write the specific energy of 
the climate system as the sum of the specific internal, potential, and kinetic energies, taking into account also the contributions coming from chemical species in different phases. In order to get manageable formulas, some approximations are necessary \cite[e.g.,][]{Peixoto1992}.

Neglecting reactions other than the phase changes between the liquid and gas phases of water, the expression of the specific energy in the atmosphere is 
$$e=c_v T+\Phi+ v_jv_j/2+Lq,$$
where $c_v$ is the specific heat at constant volume for the gaseous atmospheric mixture, $L$ is the latent heat of evaporation, and $q=\rho\xi$ is the specific humidity. In this formula, we neglect the heat content of the liquid and solid water and the heat associated with the phase transition between solid and liquid water. Instead, the approximate expression for the specific energy in the oceans is 
$$e=c_W T+\Phi+v_jv_j/2,$$ 
where $c_W$ is the specific heat at constant volume of water, while neglecting the effects of salinity and of pressure. Finally, for the specific energy of soil or ice, we can take $e=c_{\{S, I\}} T+\Phi$, respectively. 

After some nontrivial calculations, one derives the following general equation for the local energy budget:
\begin{align}
\partial_t (\rho e )&= -\partial_j (\rho \varepsilon v_j) -\partial_j Q^{\rm{SW}}_{j} -\partial_j Q^{\rm{LW}}_{j}\nonumber\\
			   & -\partial_j J^{\rm{SH}}_{j}-\partial_j J^{\rm{LH}}_j -\partial_j ( v_i \tau_{ij} ) +v_iT_i, \label{energyb}
\end{align}
where $e$ is the energy per unit mass and $\varepsilon=e+p/\rho$ is the enthalpy per unit mass. The energy source and sink terms can be written as the sum of the work done by the mechanical forcing $v_i T_i$ and of the respective divergences of the shortwave (solar) and longwave (terrestrial) components of the Poynting vector, $Q_j^{\rm{SW}}$ and $Q_j^{\rm{LW}}$; of the turbulent sensible and latent heat fluxes, $J_j^{\rm{SH}}$ and $J_j^{\rm{LH}}$; and of the scalar product of the velocity field with the stress tensor $v_i \tau_{ij}$. 

   
Equation~\eqref{energyb} is written in a conservative form, with the right-hand side containing only the sum of flux divergences, except for the last term, which is negligible. By taking suitable volume integrals of Eq.~\eqref{energyb} and assuming steady-state conditions, one derives meridional enthalpy transports from the zonal budgets of energy fluxes \cite{Ghil1987,Peixoto1992,LucariniRagone,Lucarini.ea.2014}; recall Figs.~\ref{Trenberth2}(a,b). 
The presence of inhomogeneous absorption of shortwave radiation due to the geometry of the Sun--Earth system and of the physico-chemical properties of the climatic subdomains determines the presence of non-equilibrium conditions for the climate system, as already discussed in Sect.~\ref{intro1}. 

Various versions of Eqs.~\eqref{speciesb}--\eqref{energyb} have been studied for over a century, and well-established thermodynamical and chemical laws describe accurately the phase transitions and reactions of the climate system's constituents. Finally, quantum mechanics allows one to calculate in detail the interaction between matter and radiation. 

Still, despite the fact that climate dynamics is governed by well-known evolution equations, it is way beyond our scientific abilities to gain a complete picture of the mathematical properties of the solutions. In fact, many fundamental questions are still open regarding the basic NSEs in a homogeneous fluid, without phase transitions and rotation \cite{Temam1984, Temam1997}. In particular, even for the basic NSEs, providing analytical,  closed-form solutions is only possible in some highly simplified cases that are either linear or otherwise separable \cite{Batchelor1974}.

As already discussed above, Eqs.~\eqref{speciesb}--\eqref{energyb} can simulate a range of phenomena that spans many orders of magnitude in spatial and temporal scales. Hence, it is virtually impossible to construct numerical codes able to represents explicitly all the ongoing processes, at the needed resolution in space and time. It is thus necessary to parametrize the processes that occur at subgrid scales and cannot, therewith, be directly represented. Among the most important processes of this type are cloud--radiation interactions and turbulent diffusion. We will discuss the theoretical framework behind the formulation of parametrizations in Sect.~\ref{multiplescales}. 


\subsubsection{Approximate Balances and Filtering}
\label{balance}
Equations~\eqref{speciesb}--\eqref{energyb} are too general and contain too many wave propagation speeds and instabilities in order to properly focus on certain classes of phenomena of interest. For instance, the fluid's compressibility only plays a role in the propagation of very fast acoustic waves, whose energy is negligible, compared to that of the winds and currents we are interested in here.

Therefore, starting with \citet{char48}, more or less systematic approximations have been introduced to filter out undesirable waves from the equations of motion and to study particular classes of phenomena and processes. Depending on the climate subsystem and the range of scales of interest, different approximations can be adopted. For example, if one considers ice sheets and bedrock on the time scale of millennia, it is reasonable to assume that $v_j\simeq 0$, \textit{i.e.}, to remove the flow field from the evolution equation \cite[and references therein]{Ghil1994b,saltzman_dynamical}. Obviously, this approximation is not valid if one wants to describe explicitly the motion of the ice sheets on shorter time scales \cite{Pat81}.

More precisely, the filtering process consists in applying a set of mathematical approximations into the fundamental governing equations --- usually considered to be a suitable generalization of the classical NSEs, like Eqs.~\eqref{speciesb}--\eqref{energyb}. The purpose of this filtering is to exclude from the filtered system certain physical processes that are heuristically assumed to play only a minor role in its dynamics at the time and space scale under investigation. The magnitudes of all the terms in the governing equations for a particular type of motion are estimated by dimensional analysis \cite{Barenblatt1987}, and various classes of simplified equations have been derived by considering distinct asymptotic regimes in the associated scales \cite{Ghil1987, Pedlosky1987, McW06, klein2010}. 

The approximations adopted rely on assuming that the continuous medium, whether fluid or solid, obeys suitably defined time-independent relations --- or undergoes only small departures from such relations --- and that these relations result from the balance of two or more dominating forces. Imposing such balances leads to reducing the number of independent field variables of the system that obey a set of evolution equations. 

In meteorological terminology, a \textit{prognostic} variable whose tendency, i.e. time derivative, appears in the full equations, may thus become a \textit{diagnostic} variable, which only appears in a nondifferentiated form in the filtered equations. Additionally, the imposition of dynamical balances leads to removing specific wave motions from the range of allowed dynamical processes. Below we shall see two of the most important examples of filtering, which, among other things, are essential for the practical implementation of numerical models of the atmosphere and oceans \cite{CRB11, Holton, Vallis_atmospheric_2006}. 

\paragraph{Hydrostatic Balance.}
A classical example of filtering is the hydrostatic approximation. In a local Cartesian coordinate system $(x,y,z)$, we define by \^z$=\nabla \Phi/|\nabla \Phi|$ the direction perpendicular to a geopotential isosurface $\Phi \equiv \mathrm{const.}$ An obvious stationary solution of the NSEs is given in these coordinates by the time-independent \textit{hydrostatic balance} equation:
\begin{equation}
\rho_{\rm h} g = -\partial_z p_{\rm h}, 
\label{hydro}
\end{equation}
where the subscript `h' denotes this particular solution. 

On sufficiently large spatial and temporal scales, many geophysical flows --- e.g., atmospheric and oceanic, as well as continental surface and ground water --- are near hydrostatic equilibrium. In general, stable hydrostatic equilibrium is achieved when fluid with lower specific entropy lies below fluid with higher specific entropy; in this case, Eq.~\eqref{hydro} is obeyed for all $(x,y,z)$ within the domain occupied by the fluid. 

When this condition is broken because of external forcing or internal heating, say, the stratification is not stable and the fluid readjusts so as to reestablish the hydrostatic equilibrium. This readjustment is achieved by vertical convective motions, in which available potential energy is transformed into kinetic energy responsible for vertical motions that can be locally much faster than the large-scale flow.  Violations of hydrostatic balance thus exist only on short time and space scales. 

 A large class of  models used in studying, simulating and attempting to predict climate variability and climate change is based  on a particular simplification of the full set of Eqs.~\eqref{speciesb}--\eqref{energyb}. This simplification leads to the so-called \textit{primitive equations}, which filter out nonhydrostatic motions. As a result, sound waves 
are excluded from the solutions, a fact that greatly facilitates the numerical implementation of such models \cite{CRB11, Washington2005}.

The primitive equations are derived by assuming that the time-independent hydrostatic balance given in Eq.~\eqref{hydro} does apply at all times, even when the flow is time dependent. One thus assumes that the vertical acceleration of the fluid is everywhere much smaller than gravity. Neglecting the vertical acceleration altogether leads to a monotonic relationship between the vertical coordinate $z$ and pressure $p$.\footnote{The validity of this approximation, even locally in space and at most times, explains Pascal's experimental observation during the famed 1648 Puy-de-D\^ome ascension that a barometer's reading decreases monotonically with altitude.} 

One can hence replace $z$ by $p$ as a vertical coordinate, and rewrite the modified Eqs.~\eqref{speciesb}--\eqref{energyb} using a new coordinate system in which one also replaces, typically, the horizontal coordinates $(x,y)$ by the longitude $\lambda$ and the latitude $\phi$, respectively. In such a coordinate system, $v_3 \equiv \omega \equiv \partial_t p$ is the change of pressure with time, the three-dimensional (3-D) velocity field is nondivergent, $\partial_j v_j=0$, and the density is automatically constant and set to $1$ \cite{Holton,Vallis_atmospheric_2006}.

Despite the use of primitive equations, climate models aim at simulating a system that nonetheless features nonhydrostatic motions. Additionally, geophysical flows obeying primitive equations can lead to unstable vertical configurations of the fluids. Typical climate models are formulated so as to quickly eliminate local nonhydrostatic conditions and to parametrize the balanced condition that is thus recovered in terms of variables that are explicitly represented in the model, following the  idea of \textit{convective adjustment} originally proposed by \citet{Manabe1964}. Finding optimal ways to parametrize the effect of convection on hydrostatic climate models has become a key research area in climate dynamics; see \citet{Emanuel.1994} for a classical treatise on the problem of atmospheric convection and \citet{plant2016parameterization} for a recent overview. 

The primitive equations were at the core of both climate models and weather prediction models for three decades, from about the 1970s through the 1990s. Nonhydrostatic models --- constructed through discretization of the full NSEs with forcing and rotation --- have become increasingly available over the last decade, first as limited-area models and, more recently, even as global ones \cite[e.g.,][]{Marshall1997a}. 

These models require extremely high resolution in space and time and are computationally quite expensive; as they start to enter operational use, they require state-of-the-art hard- and software, in terms of both computing and storage, as well as sophisticated post-processing of the output. Moreover, for any practical purpose, sufficiently detailed initial and boundary data are not available and methodological problems reappear in connection with formulating such data. Still, the use of nonhydrostatic models allows, in principle, for bypassing the key problem of parametrizing convection. 

\begin{figure}
\centering
\subfloat[Geostrophic balance in Northern Hemisphere]{
    \label{fig:balance} 
    \centering
   \includegraphics [width=0.9\columnwidth]{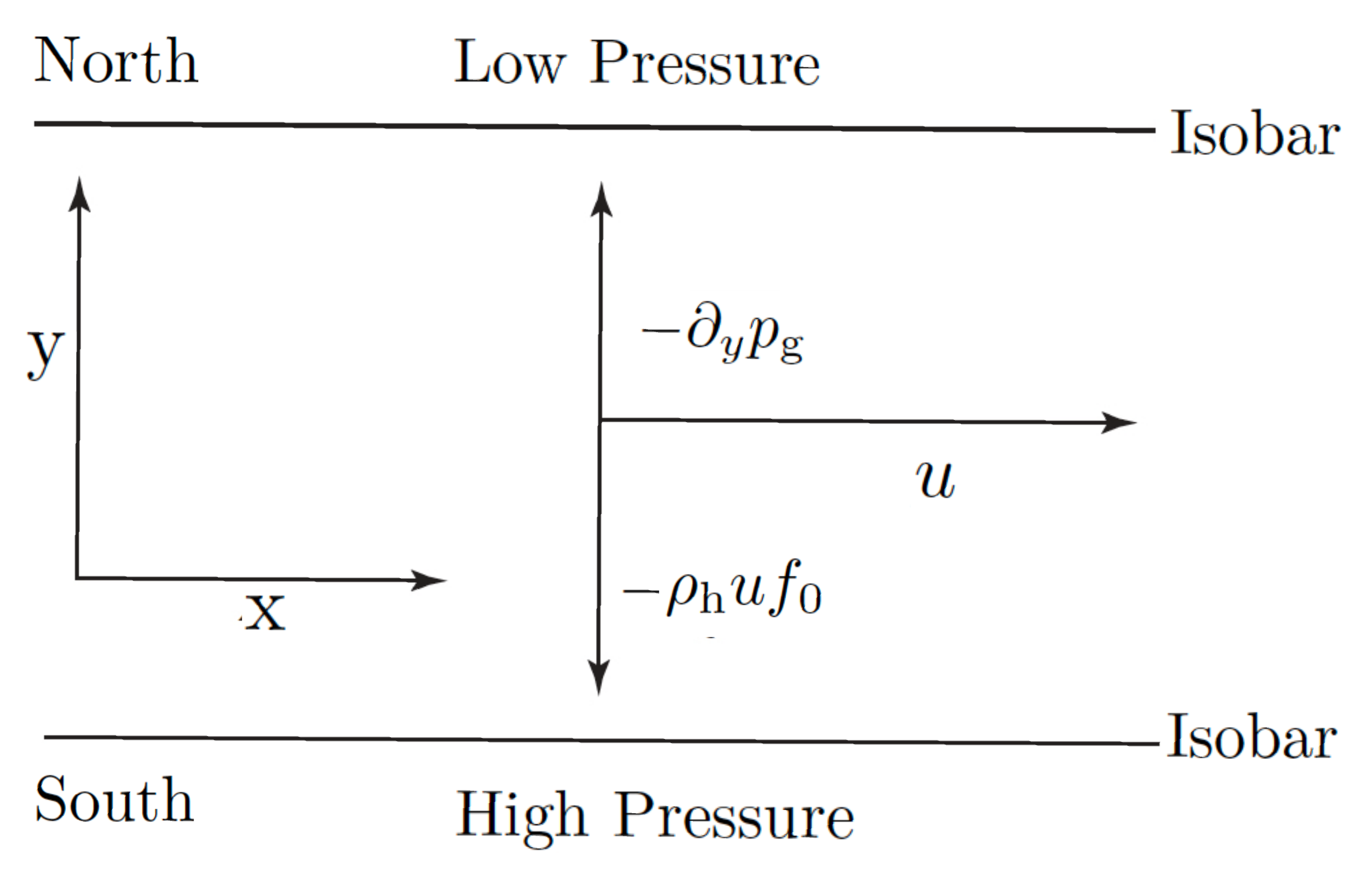}
}

\subfloat[Synoptic conditions at 500~hPa over U.S.]{
    \label{fig:synoptic} 
    \centering
   \includegraphics [angle=270, width=0.9\columnwidth]{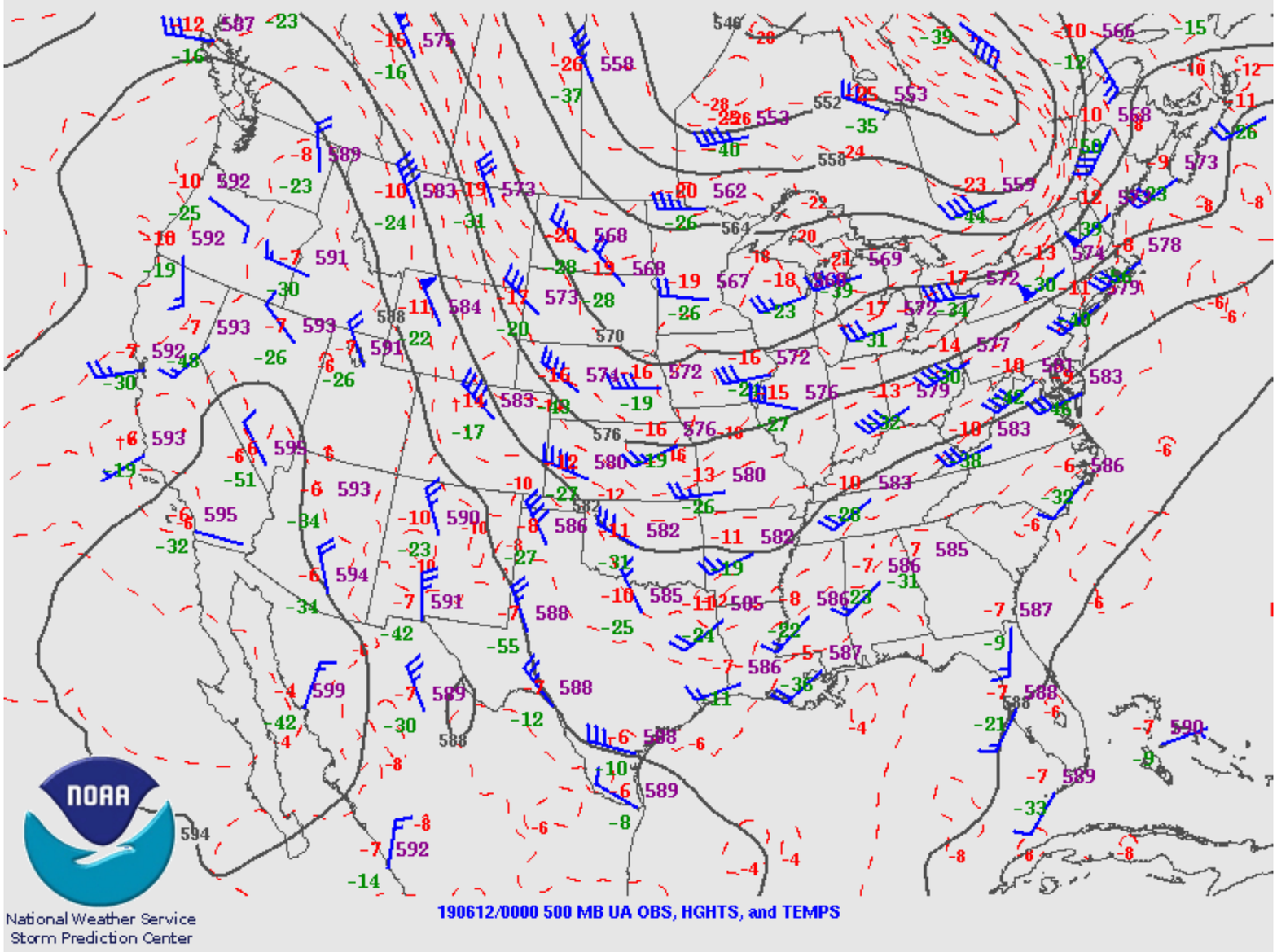} 
}
\caption{\small Geostrophic balance. (a) Schematics of geostrophically balanced flow in the Northern Hemisphere: at a given $z$-level, the pressure gradient force (upward-pointing arrow), and Coriolis force (downward-pointing arrow) cancel out and the flow (horizontal arrow) is parallel to the isobars. (b) Synoptic conditions for the 500-hPa level over the United States at 0000 Greenwich mean time (GMT) on 12 July 2019. The dark gray lines indicate the isolines of geopotential height $z=\Phi/g$ (in units of 10 m), and the barbed blue arrows indicate the direction of the wind; the barbs indicate the wind speed, each short (long) barb corresponding to a speed of 5 (10) knots, with 2 knots $ =1.03$~ms$^{-1}$. Reproduced with permission from the NOAA-National Weather Service \url {https://www.spc.noaa.gov/obswx/maps/}.} 
\label{geostrophic}
\end{figure}

\paragraph{Geostrophic Balance.}
Another important example of filtering is the one associated with time-independent purely horizontal balanced flows, in which the horizontal components of the pressure gradient force and the Coriolis force cancel out in the NSEs. Such flows are termed {\it geostrophic}; etymologically, this means Earth-turning. The nondimensional parameter that determines the validity of this approximation is the Rossby number $Ro=U/f_0 L$, where $U$ is a characteristic
horizontal velocity, $L$ a characteristic horizontal extent, and $f_0=2\Omega\sin\phi_0$; here $\Omega$ is the modulus of Earth's angular frequency of rotation around its axis, and $\phi_0$ is a reference latitude. 


The geostrophic approximation provides a rather good diagnostic description of the flow when $Ro\ll1$. Such small $Ro$-values prevail for the atmosphere on synoptic and planetary scales, such as those of extratropical weather
systems --- say poleward of $30^\circ$, where $f_0$ is large enough --- and in the free atmosphere, i.e. above the planetary boundary layer, where frictional forces can be comparable to the Coriolis force. In the oceans, this approximation is extremely accurate everywhere except very close to the equator. The smallness of Rossby number in the oceans arises from oceanic currents being orders of magnitude slower than atmospheric winds, so that $Ro$ nears unity only where $f_0$ is extremely small.

Purely geostrophic flow fields are obtained via a zeroth-order expansion in $Ro$ of the NSEs. Fluid motion is introduced by considering small perturbations $\rho_{\rm g}$ and $p_{\rm g}$ that break the translation symmetry of the basic, purely hydrostatic density and pressure fields, $\rho_{\rm h}$ and $p_{\rm h}$, but preserve the geostrophic balance, i.e.
one sets $\rho = \rho_{\rm h} + \rho_{\rm g}$ and $p = p_{\rm h} + p_{\rm g}$, respectively. 

Letting \^x be locally the zonal direction and \^y be locally the  meridional direction, one can write \cite{Holton} geostrophic balance at each height $z$ as:
\begin{subequations}
\label{geostrophiceq}
\begin{eqnarray}
\rho_{\rm h} u f_0&=-\partial_y p_{\rm g},\\
\rho_{\rm h} v f_0&=\partial_x p_{\rm g},
\end{eqnarray}
\end{subequations}
where the derivatives are taken at constant $z$, $(u, v) = (v_1, v_2)$ and, furthermore, the geostrophic perturbation itself is in hydrostatic equilibrium, i.e. $g\rho_{\rm g}=-\partial_z p_{\rm g}$.  The combined hydrostatic and geostrophic balance constrain atmospheric and oceanic motions so that the velocity field is uniquely determined by the pressure field. In such doubly balanced flows, the winds or currents are parallel  to the isobars at a given geopotential height $z$, as shown in Fig.~\ref{fig:balance}, rather than perpendicular to the isobars, as in nonrotating fluids. Moreover, the vertical component of the velocity field vanishes, $w=v_3 \equiv 0$. 

Using the pressure coordinate system $(x,y,p)$ described above, where $p$ plays the role of the vertical coordinate, it is possible to express the geostrophic balance as follows \cite{Holton}:
\begin{subequations}
\label{geostrophiceqb}
\begin{eqnarray}
f_0 u_{\rm g} &=-\partial_y \Phi,\\
f_0 v_{\rm g} &=\partial_x \Phi, 
\end{eqnarray}
\end{subequations}
where, this time, derivatives are taken at constant $p$. The vertical derivative $\partial_p \Phi$ can, furthermore, be expressed in terms of $\rho_{\rm g}$. For the atmosphere, one can make the simplifying assumption of dry conditions, for which the equation of state is described by the gas constant $R$ and the heat capacity at constant pressure $C_p$. 

Assuming, moreover, a horizontally homogeneous background state with $T_{\rm h} = T_{\rm h}(p)$, introducing the potential temperature $\Theta_{\rm h} = T_{\rm h}(p/p_{\rm s})^{R/C_p}$, with $p_{\rm s} = \rm{const.}$ the reference surface pressure, and letting $\sigma := -RT_{\rm h} \mathrm{d}\ln \Theta_h/\mathrm{d}p$, which defines the stratification of the background state, one obtains: 
\begin{equation}
\partial_p \Phi=-\frac{Rp}{\sigma}T_{\rm g}.\label{vertical}
\end{equation}

Thus, in the geostrophic approximation, the scalar field $\Phi$ provides complete diagnostic information on the state of the fluid: its horizontal derivatives give us the velocity field, while the vertical derivative gives us the geopotential, or mass field, via perturbations with respect to the background temperature field \cite{Holton}.

Essentially, geophysical fluid dynamics (GFD) is the study of large-scale flows dominated by the combined hydrostatic and geostrophic balances. The former is a consequence of the shallowness of planetary flows, the latter of their rotation \cite{CRB11, Ghil1987, McW06, Pedlosky1987}. The next subsection will introduce the more complex situation of finite-but-small perturbations from this combined balance.

Although modern computers allow a fully ageostrophic (beyond geostrophic approximation) description of most geophysical fluids, geostrophic balance remains a fundamental tool of theoretical research, as well as being used for practical applications in everyday weather prediction. Figure~\ref{fig:synoptic} illustrates a typical  mid-latitude synoptic situation: at a given pressure level, the winds blow, to a very good approximation, parallel to the isolines of geopotential height, and the speed of the wind is higher where the gradient of geopotential height is stronger. An appendix in \citet{Ghil.ea.79} describes the extent to which diagnostic relations based on the geostrophic approximation are still used in the analysis and prediction of mid-latitude weather, as simulated by advanced numerical models. Indeed, geostrophic approximation is  implicit in the day-to-day \textit{reading} of weather maps on synoptic-to-planetary scales.

\subsubsection{Quasi-Geostrophy and Weather Forecasting}\label{QGdynamics}
The diagnostic, i.e. time-independent nature of the geostrophic balance implies that the ageostrophic terms, although relatively small, are important for the time evolution of the flow. A planetary flow that evolves slowly in time compared to $1/f_0$ can be described using \textit{quasi-geostrophic} theory, namely a perturbative theory that expands the NSEs in $Ro$ and truncates at first order.  

The use of the quasi-geostrophic approximation effectively filters out solutions that correspond to higher-speed atmospheric or oceanic inertia-gravity waves; the latter, also called Poincar\'e waves, are gravity waves modified by the presence of rotation \cite{CRB11, Ghil1987, Holton, McW06, Pedlosky1987, Vallis_atmospheric_2006}.  This approximation breaks down, though, near frontal discontinuities and in other situations in which the ageostrophic component of the velocity field plays an important advective role, and it has to be improved upon by retaining higher-order terms.

In quasi-geostrophic theory, a fundamental role is played by the quasi-geostrophic potential vorticity\footnote{More general formulations of potential vorticity are also quite important in GFD \cite[e.g.,][]{Hoskins1985}.}:
\begin{equation}
q = \frac{1}{f_0}\Delta_2 \Phi + f+\partial_p\left(\frac{f_0}{\sigma} \partial_p \Phi\right).  \label{diagnostic0}
\end{equation}
Here $\Delta_2$ is the horizontal Laplacian in $p$-coordinates, the first term $\zeta_{\rm g} = \Delta_2 \Psi/f_0$ is the relative vorticity of the geostrophic flow, $f$ is the planetary vorticity, and the last term is the stretching vorticity; the static stability $\sigma$, which measures the stratification of the fluid was introduced in Eq.~\eqref{vertical}.

The potential vorticity $q$ thus combines dynamical and thermodynamical information on the fluid flow. In more abstract terms, one can write:
\begin{equation}
q-f=\mathcal{L}\Phi,   \label{diagnostic}
\end{equation}
where $\mathcal{L}$ is a modified three-dimensional Laplacian operator. Using suitable boundary conditions, one can invert the expression above and derive the geopotential field from the vorticity field,
\begin{equation}
\Phi=\mathcal{L}^{-1}(q-f).\label{diagnostic2}
\end{equation}

In the absence of forcing and dissipation, $q$ is conserved by geostrophic motions, so that:
\begin{equation}
\partial_t q+J\left(\frac{\Phi}{f_0}, q\right)=0,  \label{prognostic}
\end{equation}
where $J(A,B)=\partial_x A \partial_y B - \partial_y A \partial_x B$ is the Jacobian operator, so that $J(\Phi/f_0,q) = (1/f_0)(\partial_x \Phi/ \partial_y q - \partial_y \Phi \partial_x q)$ describes vorticity advection by the geostrophic velocity field $(u_{\rm g}, v_{\rm g})$, cf. Eq.~\eqref{geostrophiceq}. Note that, in the limit $\sigma\rightarrow\infty$, i.e. if one assumes infinitely stable stratification, the third term on the right-hand side of Eq.~\eqref{diagnostic0} drops out, and Eq.~\eqref{prognostic} describes the conservation of the absolute vorticity in a barotropic, two-dimensional flow.

Quasi-geostrophic theory, as introduced by \citet{char48}, arguably provided a crucial advance for the understanding of the dynamics of planetary flows and provided the foundation for the successful start of NWP \cite{charney50}. The filtering associated with this theory was, in particular, instrumental in eliminating the numerical instability that marred the pioneering weather forecast experiment of \citet{Richardson.1922} 

In fact, L.~F.~Richardson had used a more accurate set of equations of motion for the atmosphere than the set of Eqs.~\eqref{diagnostic0}--\eqref{prognostic}. Somewhat counterintuitively, the quasi-geostrophic model eventually provided a more accurate prediction  tool, because it represented more robustly the dynamical processes at the scales of interest. The inconclusive result of Richardson's one-step, 6-hour numerical experiment was received by the meteorological community at the time as proof that NWP, as proposed by \citet{Bjerknes.1904}, was not possible. 

We outline herewith how the quasi-geostrophic potential-vorticity equations can be used to perform stable numerical weather forecasts. Assume that, at time $t=t_0$, we have information on the field $\Phi_0=\Phi(t_0)$. 
\begin{itemize}
\item Step 0: Using Eq.~\eqref{diagnostic}, one computes the corresponding potential vorticity field $q_0=q(t_0)$;
\item Step A: Equation~\eqref{prognostic} is used next to predict the potential vorticity field at time $t_{1}$ as follows: $q_{1}=q(t_{1}) = q(t_0) + J(\Phi_0/f_0, q_0) \Delta t$, where $\Delta t = t_1 - t_0$;
\item Step B: Equation~\eqref{diagnostic2} is used now to infer ${\Phi_1}$;
\item Step C: Go to Step A and predict $q_{2}$, and so on. 
\end{itemize}


Three decades after \citet{Richardson.1922}, J.~G. Charney, R. Fj{\o}rtoft and J. von Neumann performed the first successful numerical weather forecasts on the ENIAC computer in Princeton by using essentially the procedure detailed above for the simpler case of barotropic, two-dimensional flow \citep{charney50}. In this case, the standard vorticity is given by the sum of relative and planetary vorticity and it plays a role analogous to potential vorticity in quasi-geostrophic three-dimensional flow. The procedure was adapted to describe the evolution of the 500-hPa field, which deviates the least from the behavior of an idealized atmosphere of homogeneous density with no vertical velocities. The simulation was performed on a limited domain that covered North America and the adjacent ocean areas. 

The authors investigated the issues associated with horizontal boundary conditions, as well as with the numerical stability of the integration. Their succes \cite{charney50} paved the way for the theory's successive applications to physical oceanography and climate dynamics as a whole \cite{DijkstraB2000, Ghil1987, Pedlosky1987}. 
The saga of this scientific and technological breakthrough was told many times; see \citet{Lynch08} for a good recent account.

\subsection{Climate Prediction and Climate Model Performance}\label{climatemodelprediction}


A key area of interest in the climate sciences is the development and testing of numerical models used for simulating the past, present, and future of the climate system. As discussed later in the paper, climate models differ enormously in terms of scientific scope, computational cost, and flexibility, so that one has to consider systematically a \textit{hierarchy of climate models}, rather than one model that could incorporate all subsystems, processes, and scales of motion \cite{Ghil2000, Ghil2001, Lucarini2002, Held.gap.05}. 

\begin{figure*}
\begin{center}
\subfloat[Bretherton horrendogram]{
    \label{fig:horrendogram} 
    \centering
   \includegraphics [width=0.8\textwidth]{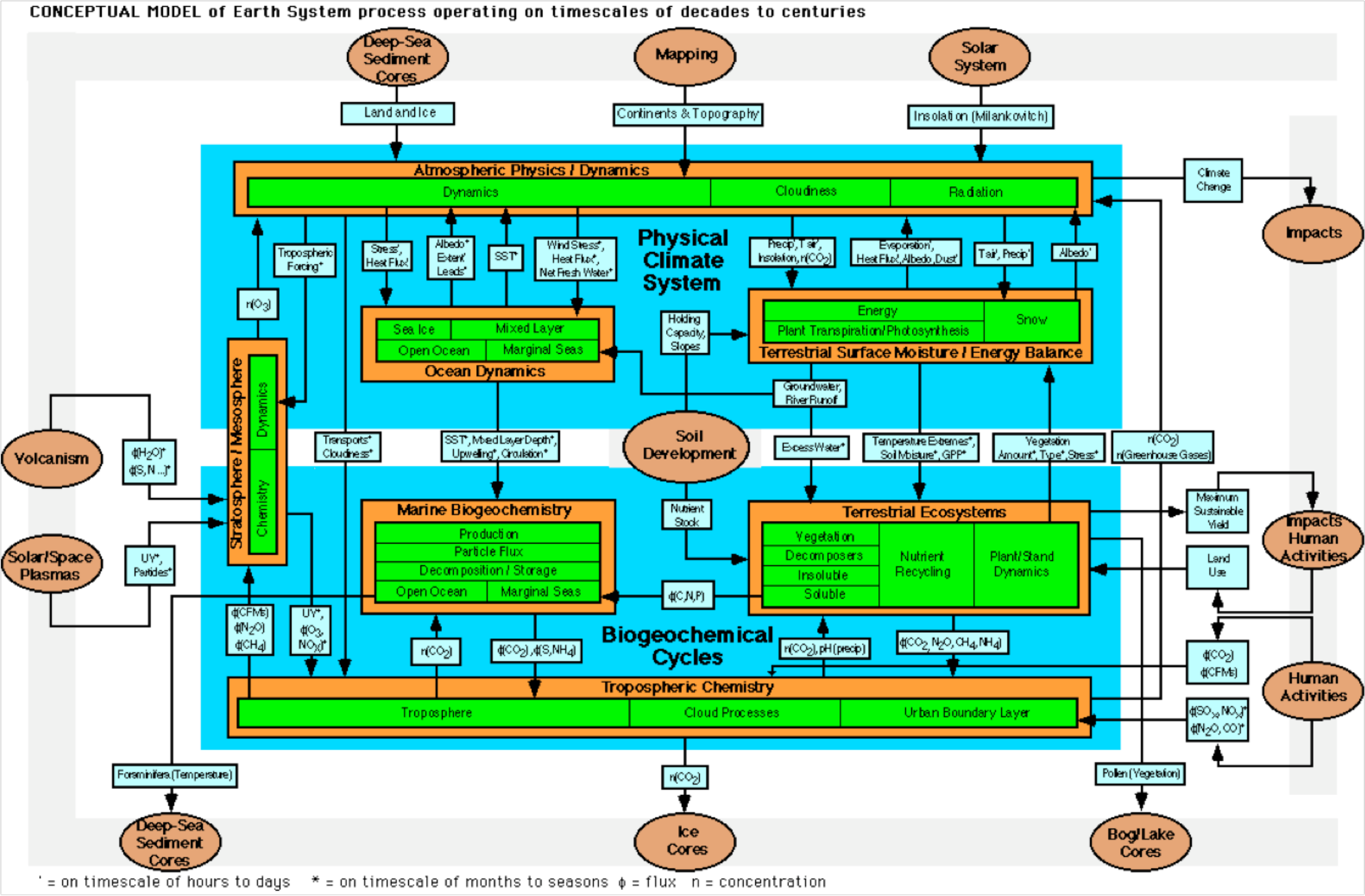}
}

\subfloat[Climate model evolution]{
    \label{fig:ESMs} 
    \centering
   \includegraphics[width=0.8\textwidth]{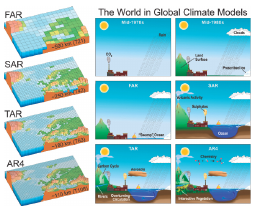} 
}
\caption{\small The Earth system, its components and its modelling. (a) The \citet{Bretherton86} horrendogram that illustrates the main components of the Earth system and the interactions amongst them. (b) Evolution of climate models across the first four IPCC assessment reports, ranging from the early 1990s to the mid-2000s \cite{IPCC07}.} 
\label{modelstoday}
\end{center}
\end{figure*}

Figure~\ref{fig:horrendogram} shows the so-called Bretherton \textit{horrendogram} \cite{Bretherton86}, which displays the full range of subsystems one needs to deal with when addressing the Earth system as a whole, along with some of the many  interactions among these subsystems. The climate modeling community has slowly caught up with the complexities illustrated by F. Bretherton and his colleagues in this diagram, three decades ago. 

Models of different levels of complexity and detail are suited for addressing different kinds of questions, according, in particular, to the main spatial and temporal scales of interest. At the top of the hierarchy, one finds global climate models, known in the 1970s and `80s as general circulation models. Both these phrases share the acronym GCMs, with the change in name reflecting a change in emphasis from understanding the planetary-scale circulation --- first of the atmosphere, then of the oceans --- to simulating and predicting the global climate. GCMs aim to represent, at the highest computationally achievable resolution, the largest number of physical, chemical, and biological processes of the Earth system;  see an early overview in \citet{Randall2000}. 

The atmosphere and oceans, whose modeling relies on the equations discussed in Sect.~\ref{equations} 
above, are still at the core of the Earth system models being developed today. Following improvements in basic scientific knowledge, as well as in computing and data storage capabilities, the latter models 
include an increasing number of physical, chemical and biological processes. They also rely on a much higher spatial and temporal resolution of the fields of interest. 

Figure~\ref{fig:ESMs} illustrates schematically the process of model development across the last three decades, from the first to the fourth assessment report: FAR--SAR--TAR--AR4. The graph does not cover the development of the last 5--10 years, which has largely dealt with the inclusion of eco-biological modules and
the above-mentioned nonhydrostatic effects. Currently available state-of-the-art models result from the cumulative efforts of generations of climate scientists and coding experts, with hundreds of individual researchers contributing to different parts of the code. As in the early stages of post-WWII NWP, climate modelling exercises now are some of the heaviest users of civilian High Performance Computing.

\subsubsection{Predicting the State of the System}\label{firstkind}
In spite of its remarkable progress, climate modeling and prediction face several kinds of uncertainties: First, uncertainties in predicting the state of the system at a certain lead time, given the uncertain knowledge of its state at the present time.  Geophysical flows are typically chaotic, as are other processes in the system. Hence, as discussed in the next two sections, the climate system depends sensitively on its initial data, as suggested already by \citet{Poincare.1908} and recognized more fully by \citet{Lorenz1963a}. Following up on terminology introduced by \citet{Lorenz76}, these are \textit{uncertainties of the first kind}, namely small errors in the initial data that can lead at later times to large errors in the flow pattern.
 
Assume that two model runs start from nearby initial states.  In a nearly linear regime, in which the phase space distance between the system's orbits is small, their divergence rate can be studied by considering the spectrum of Lyapunov exponents and, in particular, the algebraically largest ones. A chaotic system has at least one positive Lyapunov exponent, and physical instabilities that act on distinct spatial and temporal scales are related to distinct positive exponents \cite{ER85}. These instabilities can be described by using the formalism of covariant Lyapunov vectors 
\cite{GPTCLP07}. 

Addressing the uncertainties of the first kind and providing good estimates for the future state of the system is the classical goal of NWP. After the initially exponential increase of errors, nonlinear effects kick in, and the orbits in the ensemble populate the attractor of the system, so that any predictive skill is lost. In such an ensemble, several simulations using the same model are started with slightly perturbed initial states, and the ensemble  of orbits produced is used to provide a probabilistic estimate of how the system will actually evolve \cite{Palmer2017}. 

\begin{figure}
\begin{center}
\subfloat[Evolution in phase space]{
    \label{fig:ball} 
    \centering
   \includegraphics [width=0.9\columnwidth]{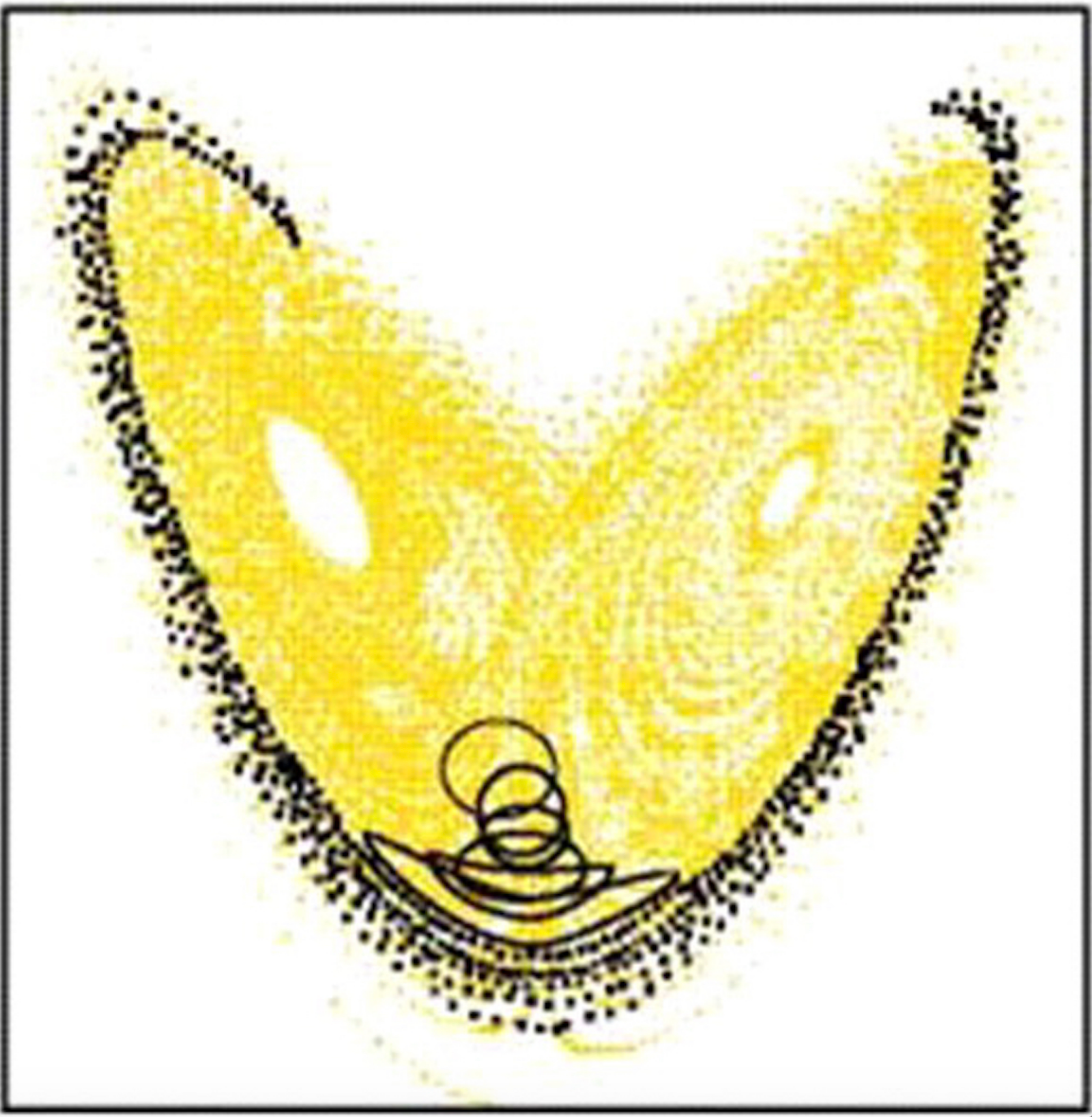}
}

\subfloat[Evolution in time]{
    \label{fig:spread} 
    \centering
   \includegraphics[width=0.9\columnwidth]{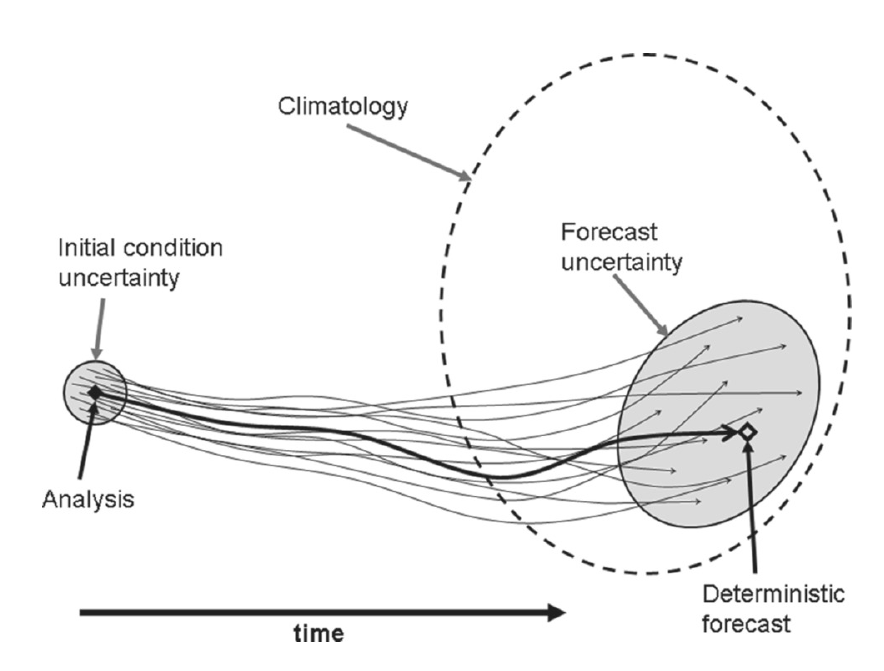}
   
}
\caption{\small Schematic diagram of the evolution of an ensemble of initial states in a chaotic system. (a) A small ball of initial states in the \citet{Lorenz1963a} convection model evolves initially in phase space according to the stretching and contracting directions associated with positive and negative Lyapunov exponents, until nonlinear effects become important and the set of initial states propagated by the system's vector field populates its attractor. Reproduced with permission from \citet{Slingo2011}. (b) Evolution in time of such an initial ball, in a generic chaotic system. Reproduced with permission from \citet{Friedrichs2016}.} \label{errorgrowth} 
\end{center}
\end{figure}

Figure~\ref{fig:ball} illustrates the main phases of error growth from an ensemble of initial states in the \citet{Lorenz1963a} model and Fig.~\ref{fig:spread} outlines how an ensemble forecast system actually works.  Obvious limitations are related to the computational difficulties of running a sufficient number of ensemble members in order to obtain an accurate estimate, as well as with the fact that an initial ellipsoid of states tends to become flattened out in time, making the error estimate more laborious. Thus, care has to be exercised in the choice of the initial states, which is done in practice by taking advantage of the reanalyses described earlier in Sect.~\ref{sssec:instrumental}.

Recently, similar methods are being used on longer time scales to perform experimental climate predictions, seen as an initial value problem \cite{Eade2014}. These predictions are performed on seasonal \cite{Doblas2013} to decadal \cite{Meehl2014} time scales; they aim not only to reach a better understanding of the multiscale dynamics of the climate system, but also to help achieve the many potential benefits of skillful medium-to-long--term predictions.  One interesting example of such an application is the wind energy market \cite{Torralba2017}.

The World Climate Research Programme (WCRP) has created an encompassing project focusing on the challenges of S2S prediction; see \url{https://www.wcrp-climate.org/s2s-overview} and \citet{S2S.book}. We will delve more specifically into extended-range forecasting in Sect.~\ref{sssec:Ext_pred} 
and see how its performance depends on a model's ability to capture the natural modes of variability of the climate system on different time scales.

\subsubsection{Predicting the System's Statistical Properties}\label{secondkind}

A key goal of climate modeling is to capture the system's statistical properties --- i.e., its mean state and its variability --- and its response to forcings of different nature. These problems will be treated thoroughly in the next two sections but are rendered quite difficult by a second set of uncertainties. Uncertainties in model formulation, as well as an unavoidably limited knowledge of the external forcings are referred to as \textit{uncertainties of the second kind} \cite{Lor67, Lorenz76, Peixoto1992} and limit intrinsically the possibility of providing realistic simulations of the statistical properties of the climate system; they affect, in particular, severely the modeling of abrupt climate changes and of the processes that may lead to such changes. 

These deficiencies are related to uncertainties in many key parameters of the climate system, as well as to the fact that each model may represent incorrectly certain processes that are relevant on the temporal and spatial scales of interest or that it may miss them altogether. These types of uncertainty are termed parametric and structural uncertainty, respectively \cite{lucarini_modelling_2013}. 

\begin{figure}
\centering
\includegraphics[width=0.9\columnwidth]{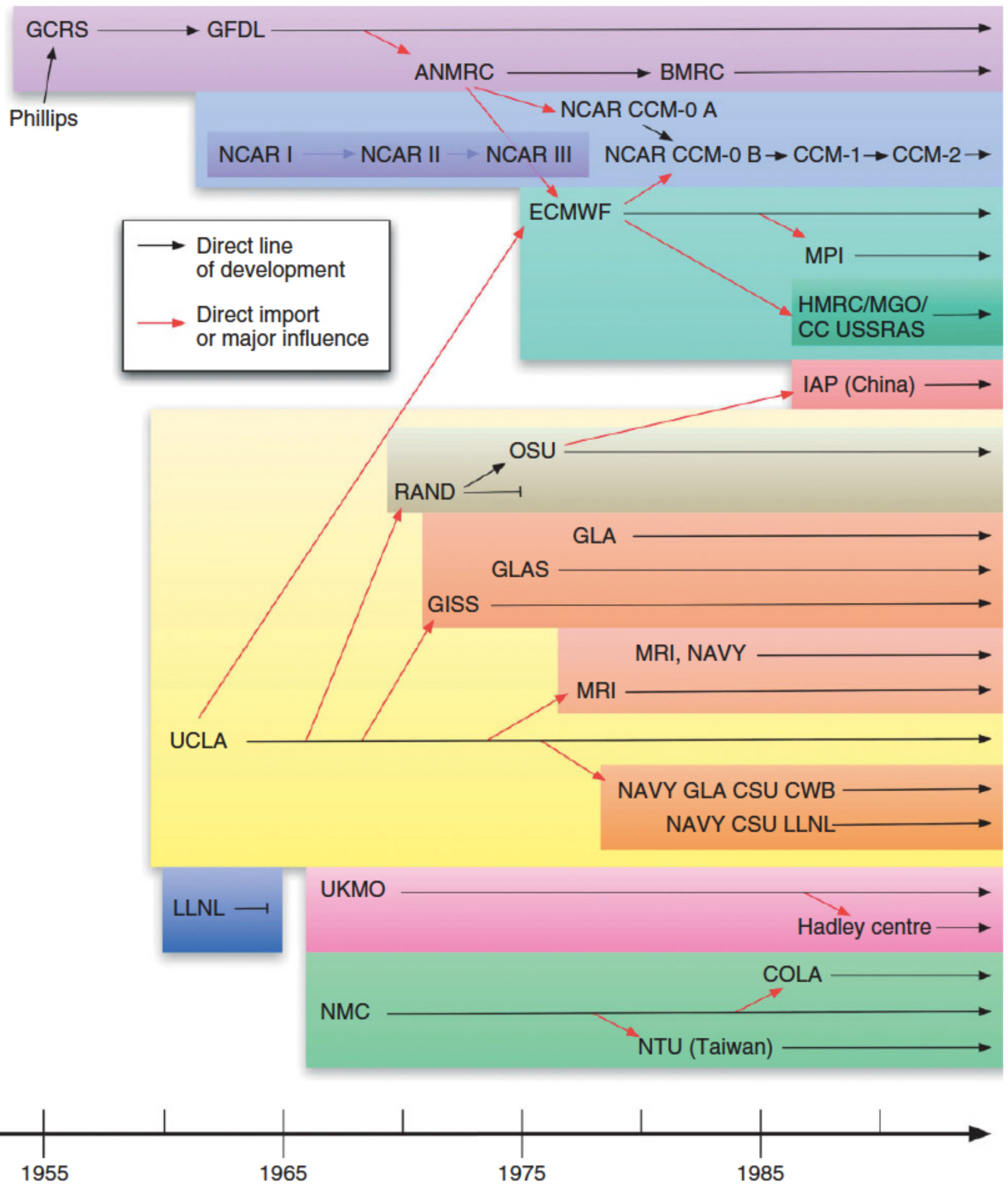}
\caption{\small Family tree of the main atmospheric general circulation models. 
The tree clearly shows that many of today's state-of-the-art models share a --- smaller or larger --- portion of their \textit{genes}, i.e., of the basic ideas and parameter values that went into the coding. Reproduced with permission from \citet{edwards10}. \label{familytree}}. 
\end{figure}

A growing number of  comprehensive climate models are available to the international scientific community for studying the properties of the climate system and predicting climate change; currently there are about 50 such models \cite{IPCC13}. Still, many of these models have in common a substantial part of their numerical code, as they originate from a relatively small number of models, atmospheric and oceanic, that were originally developed in the 1960s and 1970s. Figure~\ref{familytree} illustrates this family tree for the case of atmospheric GCMs. Hence the widely noted fact that the climate simulations produced by this large number of models often fall into classes that bear marked similarities and do not necessarily resemble to a desired extent the observed climate evolution over the last century or so.

IPCC's evolution over three decades has led to a coordination and restructuring of modeling activities around the world. In order to improve comparisons among distinct models and the replicability of investigations aimed at climate change, the Program for Climate Model Diagnostics and Intercomparison (PCMDI), through its climate model intercomparison projects (CMIPs), defines standards for the modeling exercises to be performed by research groups that wish to participate in a given assessment report (AR) and provide projections of future climate change; see \url{http://cmip-pcmdi.llnl.gov/cmip5/data\_portal.html} for CMIP5 and \url{https://www.wcrp-climate.org/wgcm-cmip/wgcm-cmip6} for CMIP6, respectively. 

The PCMDI's CMIPs have also supported a single website that gathers climate model outputs contributing to the IPCC-initiated activities, thereby providing a unique opportunity to evaluate the state-of-the-art capabilities of climate models in simulating the climate system's past, present and future behavior.  A typical IPCC-style package of experiments includes simulating the climate system under various conditions, such as:
\begin{itemize}
\item a reference state, e.g., a statistically stationary preindustrial state with fixed parameters;
\item industrial era and present-day conditions, including known natural and anthropogenic forcings;
\item future climate projections, performed by using a set of future scenarios of greenhouse gas and aerosol emissions and land-use change with some degree of realism, as well as idealized ones (e.g. instantaneous doubling of CO$_2$ concentration).
\end{itemize}


The latter changes of  greenhouse gas, aerosol concentrations and land use follow a prescribed evolution over a given time window, and are then fixed at a certain value to observe the relaxation of the system to a new stationary state. Each such evolution of the forcing is defined a ``scenario'' in ARs 1--4 \cite{IPCC01,IPCC07}, and a Representative Concentration Pathway (RCP) in AR5 \cite{IPCC13}. Note that each AR has involved an increasing number of models, people --- reaching many hundreds by the latest AR --- and hence acronyms. 

\begin{figure*}[ht]
	\centering
    \includegraphics[clip=true,trim=0cm 2cm 0cm 2cm, angle=0,width=0.8\textwidth]{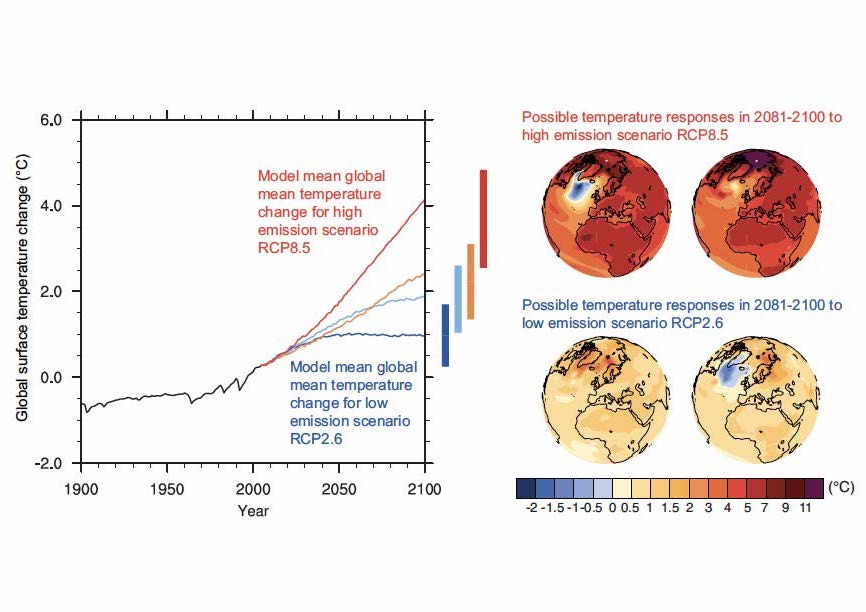}
    \caption{\small State-of-the-art climate model outputs for various climate change scenarios. Left panel: Change in the globally averaged surface temperature as simulated by climate models included in \citet{IPCC13}. Vertical bands indicate the range of model outputs, and the colors correspond to different { Representative Concentration Pathways (RCPs).} Right panel: Spatial patterns of temperature change --- i.e., 2081--2100 average with respect to the present --- for the two most extreme { RCPs}. Reproduced with permission from \cite{IPCC13}}
    \label{Fig_IPCC}
\end{figure*}

Each scenario or RCP is a representation of the expected greenhouse gas and aerosol concentrations resulting from a specific path of industrialization and change in land use, as provided by Working Groups II and III to Working Group I; see Fig.~\ref{Fig_IPCC} for AR5's simulations of the $20^{\rm th}$ and $21^{\rm st}$ century with two extreme RCPs. At the same time, the attribution of unusual climatic conditions to specific climate forcing is far from being a trivial matter  \cite{Allen2003,Hannart2016}. 

In the most recent CMIP exercise, CMIP6, the scope of model intercomparisons has grown beyond just testing future climate response to specific forcing scenarios or RCPs. In order to test more comprehensively the performance of climate models, various standardized intercomparison projects focus on the analysis of specific subsystems, processes or time scales; 
see \url{https://search.es-doc.org}. \citet{Eyring2016} provide a useful summary of the strategy for PCMDI's CMIP6 project and of the scientific questions it addresses.

\begin{figure*}[ht]
\centering
\includegraphics[angle=270,width=0.8\textwidth]{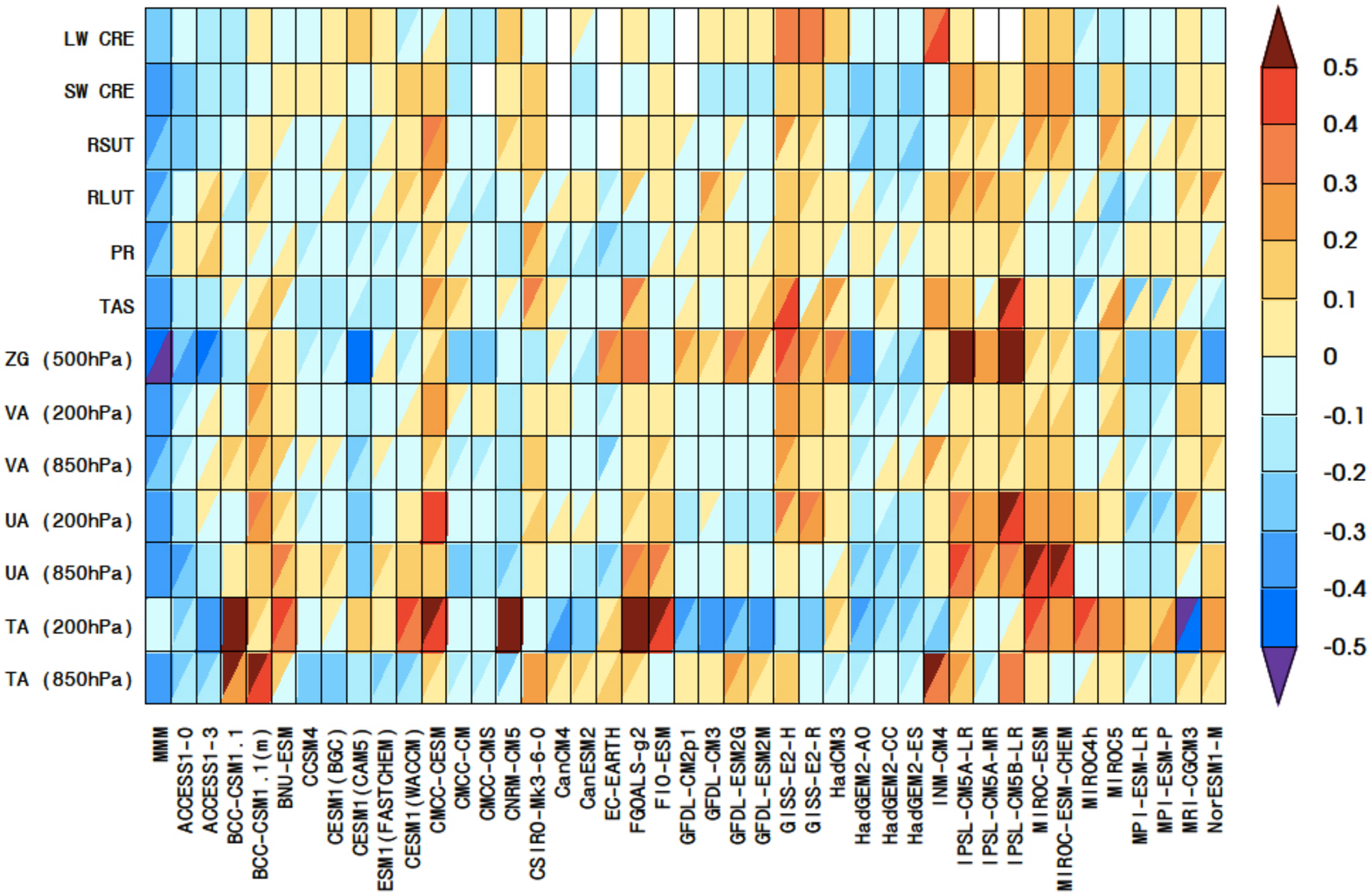}
\caption{\small Example of model performance evaluation. This diagram shows how 42 models participating in the Fifth Climate Model Intercomparison Project (CMIP5) fare in terms of representation of the seasonal cycle during the time interval 1980--2005 for 13 climate variables. Values are normalized and perfect agreement with the observations is given by 0. 
See text for details. Reproduced with permission from \citet{IPCC13}. \
} 
\label{modelstaudit}. 
\end{figure*}

\subsubsection{Metrics for Model Validation}
The standardization of climate model outputs promoted by the CMIPs has also helped address a third major uncertainty in climate modeling: what are the best metrics for analyzing a model's outputs and evaluating its skill? Note that, in this context, the term metric does not have its usual mathematical meaning of a distance in function space but refers instead to a statistical estimator of model performance, whether quadratic or not.

The validation, or auditing --- i.e., the overall evaluation of the faithfulness --- of a set of climate models, is a delicate operation, which can be decomposed into two distinct but related procedures. The first one is model intercomparison, which aims to assess the consistency of different models in the simulation of certain physical phenomena in a certain time frame. The second procedure is model verification, whose goal is to compare a model's outputs with corresponding observed or reconstructed scalar quantities or fields \cite{Lucarini08enc}. 

In principle, there are numerous ways to construct a metric by taking any reasonable function of a climate model's variables. Nevertheless, even if several observables are mathematically well defined, their physical relevance and robustness can differ widely. There is no a priori valid criterion for selecting a good climatic observable, even though taking into account basic physical properties of the climate system can provide useful guidance. 

There is thus no unique recipe for testing  climate models, a situation that is in stark contrast with more traditional areas of physics. For instance, in high-energy physics, the variables mass, transition probability or cross-section are suggested by the very equations that one tries to solve or to study experimentally. 

In the absence of dissipation and of sources and sinks, certain scalars---such as the atmosphere's or the oceans' total mass, energy and momentum---are also conserved by the fundamental equations of Sect.~\ref{equations}. 
Local quantities, though, like total rain over India during the summer monsoon or the intensity of the subtropical jet over North America during a given winter month, often matter in evaluating a climate model's skill: current models still have substantial difficulties in simulating the statistics of major regional processes, such as ENSO in the Tropical Pacific \cite{Bellenger2014,Lu2018}, the Indian Monsoon \cite{Turner12,Boos13,Hasson13}, and mid-latitude LFV associated with blocking \cite{Davini2016,Woollings2018}. 

From the end user's point of view, it is important to check how realistic the modeled fields of practical interest are. But if the aim is to define strategies for the radical improvement of model performance --- beyond incremental advances often obtained at the price of large increases in computer power --- it is crucial to fully understand the differences among models in their representation of the climate system, and to decide whether specific physical processes are correctly simulated by a given model. 

Additional issues, practical as well as epistemological, emerge when we consider the actual process of comparing theoretical and numerical investigations with observational or reanalysis data. Model results and approximate theories can often be tested only against observational data from a sufficiently long past history, which may pose problems of both accuracy and coverage, as mentioned in Sect.~\ref{sssec:instrumental}.

To summarize, difficulties emerge in evaluating climate model performance because (i) we always have to deal with three different kinds of attractors: the attractor of the real climate system, its reconstruction from observations, and the attractors of the climate models; and (ii) because of the high dimensionality of both the phase space and the parameter space of these attractors.

In order to address these issues, multi-variable metrics are currently used to try to assess the skill of available climate models. Thus, Fig.~\ref{modelstaudit} shows a diagram describing the performance of the 42 models participating in PCMDI's CMIP5. Even superficial analysis of the diagram indicates that no model is the best for all the variables under consideration. An improved assessment package is given by PCMDI's Metrics Package \cite{Gleckler.ea.2016} and by the ESMValTool package \cite{Eyring2016b}. Recently, \citet{Lembo2019} released TheDiato, a flexible diagnostic tool able to evaluate comprehensively the energy, entropy and water budgets, and their transports for climate models. This package will become part of the second generation of the ESMValTool package \cite{Eyring2020}.
 
Finally, in order to describe synthetically and comprehensively the outputs of a growing number of climate models, it has become common to consider multi-model ensembles and focus the attention on the ensemble mean and the ensemble spread. Mean and spread have been taken as the, possibly weighted, first two moments of the models' outputs for the  metric under study \cite{Tebaldi2007}; see Fig. \ref{Fig_IPCC}. 

This approach merges information from different attractors and the resulting statistical estimators cannot be interpreted in the standard way, with the mean approximating the true field and the standard deviation describing its uncertainty. Such a naive interpretation relies on an implicit assumption that the set is a probabilistic ensemble formed by equivalent realizations of a given process, and that the underlying probability distribution is unimodal; see \citet{Parker2010} for a broader epistemological discussion of the issues.

While the models in such an ``ensemble of opportunity'' may be related to each other, as shown in Fig.~\ref{familytree}, they are by no means drawn from the same distribution. A number of alternative approaches for uncertainty quantification in climate modeling have been proposed but they go beyond the scope of the present review.



\section{Climate Variability and the Modeling Hierarchy}
\label{climatevariability}

\subsection{Radiation Balance and Energy-Balance Models}
\label{ssec:radiation}

The concepts and methods of the theory of deterministic dynamical systems \citep{Andro.Pont.37, V.Arnold.83, Guckenheimer1983} have been applied first to simple models of atmospheric and oceanic flows, starting about half-a-century ago \citep{Lorenz1963a, Stommel1961, Veronis1963}. More powerful computers now allow their application to fairly realistic and detailed models of the atmosphere, ocean, and the coupled atmosphere--ocean system. We start herewith by presenting such a hierarchy of models.

This presentation is interwoven with that of the successive bifurcations that lead from simple to more complex solution behavior for each climate model. Useful tools for comparing model behavior across the hierarchy and with observations are provided by ergodic theory \citep{ER85, GCS08}. Among these tools, advanced methods for the analysis and prediction of uni- and multivariate time series play an important role \citep[and references therein]{Ghil.SSA.2002}.

The concept of a modeling hierarchy in climate dynamics was introduced by \citet{Schneider1974}. Several authors have discussed  lately in greater detail the role of such a hierarchy in the understanding and prediction of climate variability \cite{Ghil2000, Held.gap.05, Lucarini2002, saltzman_dynamical, DG05}. At present, the best-developed hierarchy is for atmospheric models. These models were originally developed for weather simulation and prediction on the time scale of hours to days. Currently they serve --- in a stand-alone mode or coupled to oceanic and other models --- to address climate variability on all time scales.

The first rung of the modeling hierarchy for the atmosphere is formed by zero-dimensional (0-D) models, where the number of dimensions, from zero to three, refers to the number of independent space variables used to describe the model domain, i.e. to physical-space dimensions. Such 0-D models essentially attempt to follow the evolution of the globally averaged  air temperature at surface   as a result of changes in global radiative balance:
\noindent
\begin{subequations}
\label{eq:rad_bal_0}
\begin{eqnarray}
c\frac{\diff \bar{T}}{\diff t} & = & R_{\mathrm i} - R_{\mathrm o}, \label{0DEBM}\\
R_{\mathrm{i}} & = & \mu Q_0 \{1 - \alpha(\bar T)\}, \label{eq:R_i}\\
R_{\mathrm o} & = & \sigma m(\bar{T}) (\bar{T})^4. \label{eq:R_o}
\end{eqnarray}
\end{subequations}
	 								
Here $R_{\rm i}$ and $R_{\rm o}$  are incoming solar radiation and outgoing terrestrial radiation. The heat capacity $c$ is that of the global atmosphere, plus that of the global ocean or some fraction thereof, depending on the time scale of interest: one might only include in $c$ the ocean mixed layer when interested in subannual time scales but the entire ocean when studying paleoclimate \cite[e.g.,][]{saltzman_dynamical}.

The rate of change of $\bar T$ with time $t$ is given by ${\diff \bar{T}}/{\diff t}$, while $Q_0$ is the solar radiation received at the top of the atmosphere, also called the ``solar constant,'' $\sigma$ is the Stefan-Boltzmann constant, and $\mu$ is an insolation parameter equal to unity for present-day conditions. To have a closed, self-consistent model, the planetary reflectivity or albedo $\alpha$ and grayness factor $m$ have to be expressed as functions of  $\bar{T}$; $m = 1$ for a perfectly black body and $0 < m < 1$ for a grey body like planet Earth.


There are two kinds of one-dimensional (1-D) atmospheric models, for which the single spatial variable is latitude or height, respectively. The former are so-called energy-balance models (EBMs), which consider the generalization of the model (2.1) for the evolution of surface-air temperature $T = T(x,t)$, say,
\begin{equation}
\label{eq:rad_bal_1}
c(x)\frac{\partial T}{\partial t} = R_{\mathrm i} - R_{\mathrm o} + D.
\end{equation}

The terms $R_{\mathrm{i}}$ and $R_{\mathrm{o}}$ are similar to those given in Eqs.~\eqref{eq:R_i} and \eqref{eq:R_o} for the 0-D case above, but can now be functions of the meridional coordinate $x$ --- latitude, co-latitude, or sine of latitude --- as well as of time $t$ and temperature $T$.  The horizontal heat-flux term $D$ describes the convergence of the heat transport across latitude belts; it typically contains first and second partial derivatives of $T$ with respect to $x$, while $c(x)$ represents the system's space-dependent heat capacity.

Thus Eq.~\eqref{eq:rad_bal_1} corresponds physically to a nonlinear heat or reaction--diffusion equation, and mathematically to a nonlinear parabolic PDE. Hence the rate of change of local temperature $T$ with respect to time also becomes a partial derivative, ${\partial T (x,t)}/{\partial t}$. Two such models were introduced independently by \citet{Budyko} in the then Soviet Union and by \citet{Sellers} in the United States. 

\begin{figure}[ht] 
\centering
	\includegraphics[width=0.9\columnwidth]{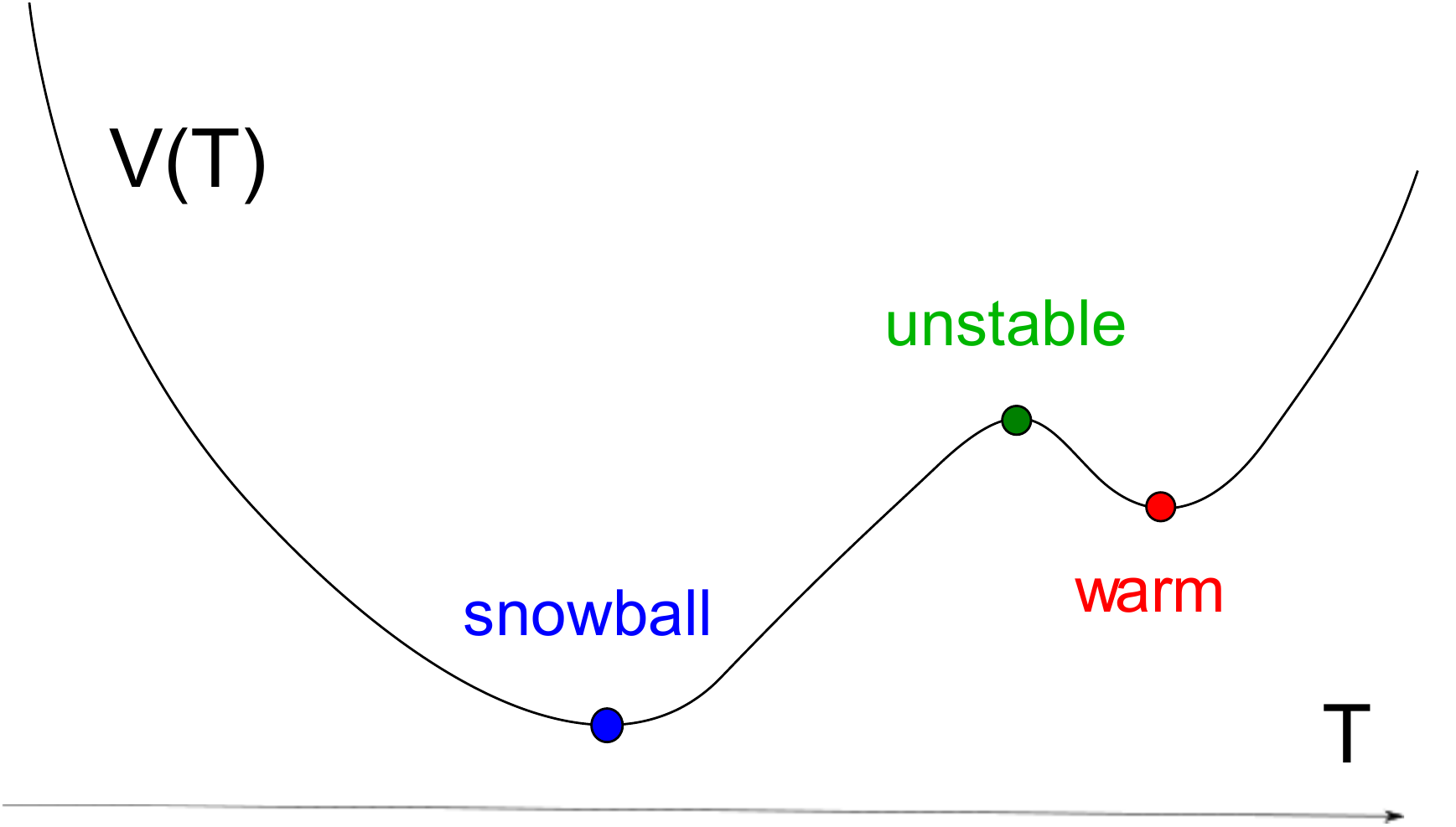}
        \caption{\label{fig:double_well} 
        \small Scalar double-well potential function $V(T)$; the {\re warm} and the ``deep-freeze'' or {\bb snowball} states 
        correspond to the system's two stable fixed points, separated by an {\gr unstable} one.}
\end{figure}

The first striking results of theoretical climate dynamics were obtained in showing that Eq.~\eqref{eq:rad_bal_1} 
could have two stable steady-state solutions, depending on the value of the insolation parameter $\mu$, cf. Eq.~\eqref{eq:R_i}\footnote{After the publication of the \citet{Budyko} and \citet{Sellers} models, it became apparent that massive nuclear explosions could induce a \textit{nuclear winter}: namely, reducing the incoming solar radiation by a dramatic increase in atmospheric particulate matter could potentially trigger an even greater disaster for life on Earth than nuclear war itself \citep[e.g.,][]{TTAPS.1983}. Studies to this effect have been influential in reducing the size of the nuclear arsenals at the end of the Cold War. Strikingly, the two contributions came almost simultaneously from scientists belonging to the Cold War's two opposing geopolitical blocks. }. 

This multiplicity of stable steady states, or physically possible stationary climates of our planet, can be explained, in its simplest form, already by the 0-D model of Eq.~\eqref{eq:rad_bal_0}. Note that the time derivative of the global temperature $\bar T$ in Eq.~\eqref{0DEBM} can be written as minus the derivative of a potential { $V(\bar T) = - \int\{R_{\mathrm i}(\bar T) - R_{\mathrm o}(\bar T)\}\diff \bar T$, viz.  ${\diff \bar{T}}/{\diff t} = - {\diff {V(\bar{T})}/{\diff \bar{T}}}.$} In the case of bistability, the two local minima of $V$ correspond to the stable solutions, or fixed points, and the local maximum of $V$ corresponds to the unstable solution; see the sketch in Fig.~\ref{fig:double_well}. 

The physical explanation lies in the fact that --- for a fairly broad range of $\mu$-values around $\mu = 1.0$ --- the curves for $R_{\mathrm i}$ and $R_{\mathrm o}$ as a function of  $\bar T$ intersect in three points.  One of these points corresponds to the present climate (highest  $\bar T$-value), and another one to an ice-covered  planet (lowest $\bar T$-value); both of these are stable, while the third one, i.e. the intermediate $\bar T$-value, is unstable. 

To obtain this result, it suffices to make two assumptions: (i) that $\alpha = \alpha(\bar T)$ in Eq.~\eqref{eq:R_i} is a piecewise-linear function of $\bar T$ --- or, more generally, a monotonically increasing one with a single inflection point --- with high albedo at low temperature, due to the presence of snow and ice, and low albedo at high  $\bar T$, due to their absence; and (ii) that $m = m(\bar T)$ in Eq.~\eqref{eq:R_o} is a smooth, increasing function of $\bar T$  that captures in its simplest from the ``greenhouse effect'' of trace gases and water vapor.	

The EBM modelers \cite{Ghil1976, Held1974, North1975} called the ice-covered state a ``deep freeze.'' The possibility of such a state in Earth history was met with considerable incredulity by much of the climate community, as incompatible with existing paleoclimatic evidence at the time. Geochemical evidence led in the early 1990s to the discovery of a snowball or, at least, slushball Earth prior to the emergence of multicellular life, 600 Myr B.P. \cite{Hoffman, HoffmanSchrag}. This discovery did not lead, however, to more enthusiasm for the theoretical prediction of such a state, almost two decades earlier. 

\begin{figure}
\centerline{
\includegraphics[width = .9\columnwidth]{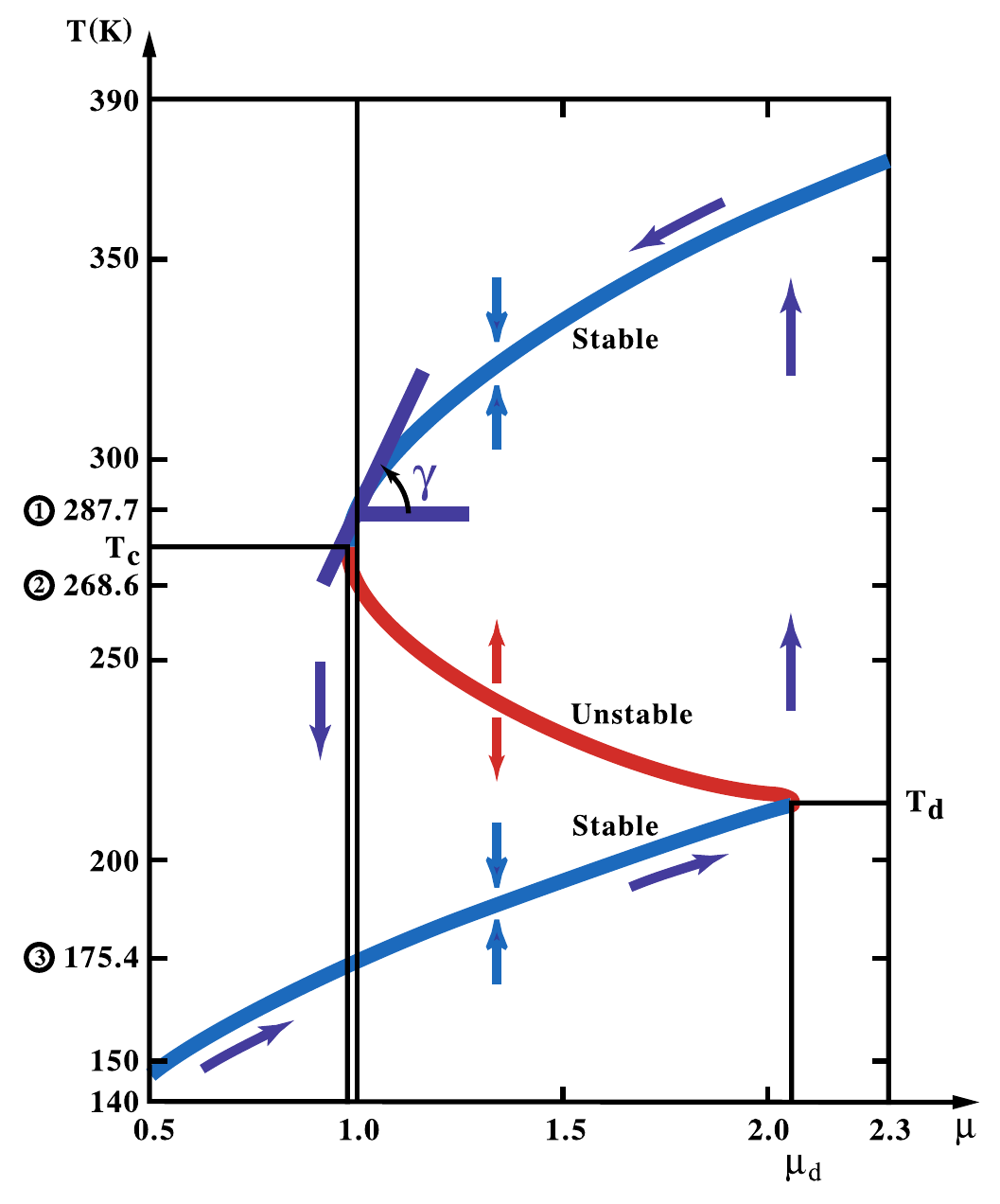}}
\caption{\small Bifurcation diagram for the solutions of an energy-balance model (EBM), 
     showing the global-mean temperature $\bar T$ vs. the fractional change $\mu$ of insolation 
     at the top of the atmosphere. The arrows pointing up and down at about $\mu = 1.4$ 
     indicate the stability of the branches: towards a given branch if it is stable and away if it is unstable. 
     The other arrows show the hysteresis cycle that global temperatures would 
     have to undergo for transition from the upper stable branch to the lower one and back. 
     The angle $\gamma$ gives the measure of the present climate's sensitivity to changes in insolation. 
     [After \citet{Ghil1987} with permission from Springer Science+Business Media.]}
\label{Fig_1}
\end{figure}

The bifurcation diagram of an 1-D EBM, like the one of Eq.~\eqref{eq:rad_bal_1}, is shown in Fig.~\ref{Fig_1}.  It displays the model's mean temperature $\bar T$  as a function of the fractional change $\mu$ in the insolation $Q_0 = Q_0(x)$ at the top of the atmosphere. The `$S$'-shaped curve in the figure arises from two back-to-back {\it saddle-node bifurcations}. \citet{Bensid.2019} have recently provided a mathematically rigorous treatment of the bifurcation diagram of a Budyko-type 1-D EBM.

The normal form of the first saddle-node bifurcation is
\begin{equation}
\label{eq:saddle-node}
\dot X = \mu - X^2.
\end{equation}
Here $X$ stands for a suitably normalized form of  $\bar T$ and $\dot X$ is the rate of change of $X$, while $\mu$ is a parameter that measures the stress on the system, in particular a normalized form of the insolation parameter in Eq.~\eqref{eq:R_i}.

The uppermost branch corresponds to the steady-state solution $X = + \mu^{1/2}$ of Eq.~\eqref{eq:saddle-node} and it is stable. This branch matches rather well Earth's present-day climate for $\mu = 1.0$; more precisely the steady-state solution $T = T(x; \mu)$  of the full 1-D EBM (not shown) matches closely the annual mean temperature profile from instrumental data over the last century \citep{Ghil1976}.

The intermediate branch starts out at the left as the second solution, $X = - \mu^{1/2}$, of Eq.~\eqref{eq:saddle-node} and it is unstable. It blends smoothly into the upper branch of a coordinate-shifted and mirror-reflected version of Eq.~\eqref{eq:saddle-node}, say
\begin{equation}
\label{eq:back-saddle-node}
\dot X = (\mu - \mu_0) + (X - X_0)^2.
\end{equation}

This branch, $X = X_0 + (\mu_0 - \mu)^{1/2}$, is also unstable. Finally, the lowermost branch in Fig.~\ref{Fig_1} is the second steady-state solution of Eq.~\eqref{eq:back-saddle-node}, $X = X_0 - (\mu_0 - \mu)^{1/2}$, and it is stable, like the uppermost branch.  The lowermost branch corresponds to an ice-covered planet at the same distance from the Sun as Earth. 

The fact that the upper-left bifurcation point $(\mu_{\rm c}, T_{\rm c})$ in Fig.~\ref{Fig_1} is so close to present-day insolation values created great concern in the climate dynamics community in the mid-1970s, when these results were obtained.  Indeed, much more detailed computations (see below) confirmed that a reduction of about 2--5\% of insolation values would suffice to precipitate Earth into a ``deep freeze.'' The great distance of the lower-right bifurcation point $(\mu_{\rm d}, T_{\rm d})$ from present-day insolation values, on the other hand, suggests that one would have to nearly double atmospheric opacity, say, for the Earth's climate to jump back to more comfortable temperatures.

The results above follow \citet{Ghil1976}. \citet{Held1974} and  \citet{North1975} obtained similar results, and a detailed comparison between EBMs appears in Chapter 10 of \citet{Ghil1987}. \citet{Ghil1976}, rigorously, and then \citet{North.ea.1979}, numerically, pointed out that a double-well potential, like the one sketched in Fig.~\ref{fig:double_well}, does exist for higher-dimensional \citep{North.ea.1979} and even infinite-dimensional \citep{Ghil1976} versions of an EBM. In higher dimensional cases, the maximum of the potential shown in  Fig. \ref{fig:double_well} is replaced by a saddle, or ``mountain pass'' \cite[e.g.,][Sec.~10.4]{Ghil1987}. In this case, dimensionality refers to phase space, rather than physical space; see a more detailed discussion in Sect.~\ref{critical}. 


\subsection{Other Atmospheric Processes and Models}
\label{ssec:processes}
The 1-D atmospheric models in which the details of radiative equilibrium are investigated with respect to a height coordinate $z$ (geometric height, pressure, etc.) are often called radiative-convective models \citep{Ram.Coak.78}. This name emphasizes the key role that convection plays in vertical heat transfer. While these models preceded historically EBMs as rungs on the modeling hierarchy, it was only recently shown that they, too, could exhibit multiple equilibria \citep{Li.ea.1997}. The word ``equilibrium,'' here and in the rest of this article, refers simply to a steady state of the model, rather than a to true thermodynamic equilibrium.

Two-dimensional (2-D) atmospheric models are also of two kinds, according to the third space coordinate that is not explicitly included. Models that resolve explicitly two horizontal coordinates, on the sphere or on a plane tangent to it, tend to emphasize the study of the dynamics of large-scale atmospheric motions. They often have a single layer or two. Those that resolve explicitly a meridional coordinate and height are essentially combinations of EBMs and radiative-convective models and emphasize therewith the thermodynamic state of the system, rather than its dynamics.

Yet another class of ``horizontal'' 2-D models is the extension of EBMs to resolve zonal, as well as meridional surface features, in particular land-sea contrasts. We shall see in Sec.~\ref{ssec:THC_GCMs} 
how such a 2-D EBM is used, when coupled to an oceanic model.

\citet{Schneider1974} and \citet{Ghil2000} discuss additional types of 1-D and 2-D atmospheric models, 
along with some of their main applications. Finally, to encompass and resolve the main atmospheric phenomena with respect to all three spatial coordinates, as discussed in Sect.~\ref{climatemodelprediction}, GCMs occupy the pinnacle of the modeling hierarchy \cite[e.g.,][]{Randall2000}.

The mean zonal temperature's dependence on the insolation parameter $\mu$ 
--- as obtained for 1-D EBMs and shown in Fig.~\ref{Fig_1} here --- was confirmed, to the extent possible, by using a simplified GCM, coupled to a ``swamp'' ocean model \citep{Wether.Manabe.1975}. More precisely, forward integrations with a GCM cannot confirm the presence of the intermediate, unstable branch. Nor was it possible in the mid-70s, when this numerical experiment was carried out, to reach the deep-freeze stable branch, {as it was called at the time,} because of the GCM's computational limitations. 

Still, the normal form of the saddle-node bifurcation, given by Eq.~\eqref{eq:saddle-node}, suggests a parabolic shape of the upper, present-day--like branch near the upper-left bifurcation point in our figure, namely $(\mu_{\rm c}, T_{\rm c})$.  This parabolic shape is characteristic of the dependence of a variable that represents the model solution on a parameter that represents the intensity of the forcing in several types of bifurcations; moreover, this shape is not limited to the bifurcation's normal form, like Eqs.~\eqref{eq:saddle-node} or \eqref{eq:back-saddle-node}, but is much more general. The GCM simulations supported quite well a similar shape for the globally averaged temperature profiles of the GCM's five vertical layers, cf. \citet[Fig.~5]{Wether.Manabe.1975}. See discussion below in Sects.~\ref{tantet} and \ref{critical}. 

\citet{Ghil2000} also describe the separate hierarchies that have grown over the last quarter-century in modeling the ocean and the coupled ocean--atmosphere system. More recently, an overarching hierarchy of Earth system models --- that encompass all the subsystems of interest, atmosphere, biosphere, cryosphere, hydrosphere and lithosphere --- has been developing. Eventually, the partial results about each subsystem's variability, outlined in this section and the next one, will have to be verified from one rung to the next of the full Earth system modeling hierarchy.

\subsection{Oscillations in the Oceans' Thermohaline Circulation}
\label{sec:THC}
\subsubsection{Theory and Simple Models}
\label{ssec:THC_theory}
Historically, the thermohaline circulation (THC) \cite[e.g.,][]{DG05, Kuhlbrodt2007} was first among the climate system's major processes to be studied using a very simple mathematical model. 
\citet{Stommel1961} formulated a two-box model and showed that it possesses multiple equilibria.

\begin{figure}
\centerline{
\includegraphics[width = .95\columnwidth]{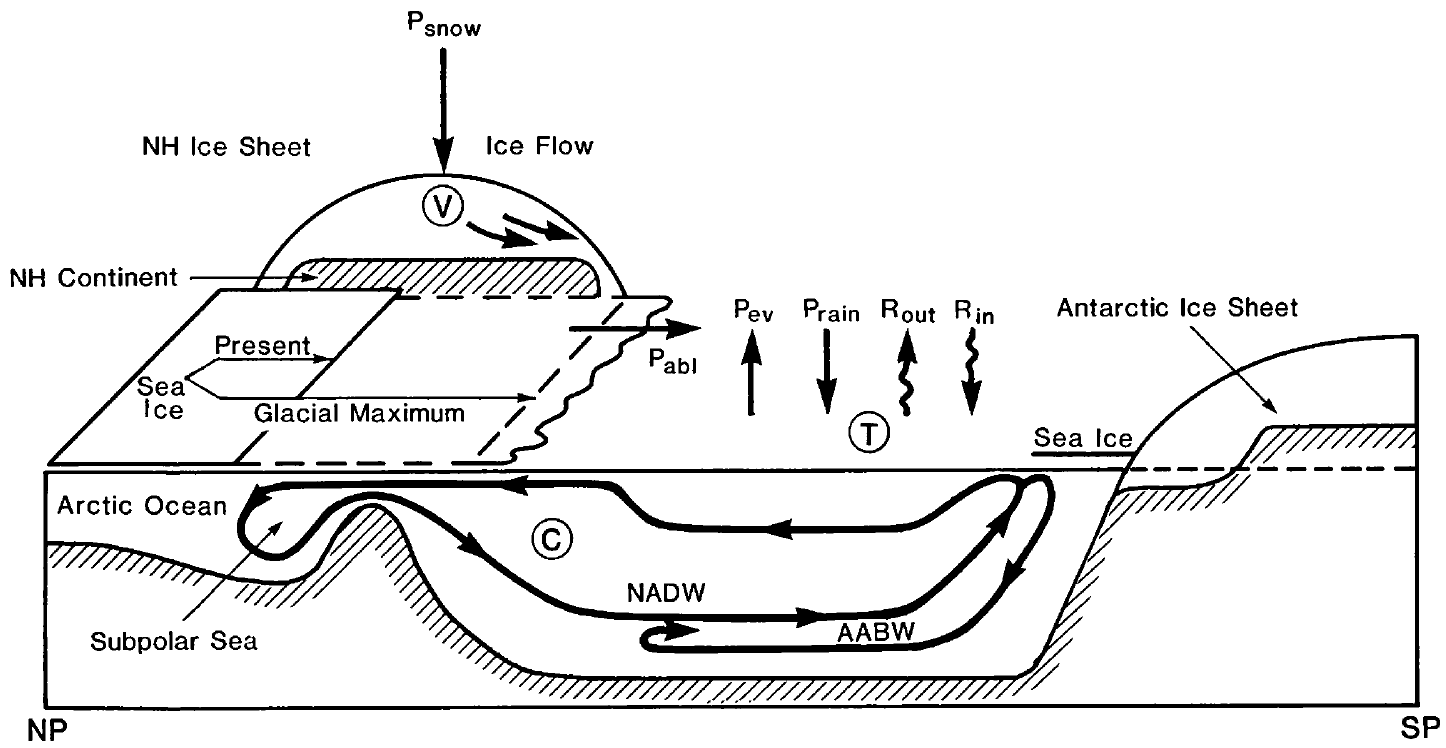}}
\caption{Schematic diagram of an Atlantic meridional cross section from North Pole (NP) to South Pole (SP), 
        showing mechanisms likely to affect the thermohaline circulation (THC) on various time-scales. 
        Changes in the radiation balance $R_{\rm in} - R_{\rm out}$ are due, at least in part, to changes in extent of 
        Northern Hemisphere (NH) snow and ice cover $V$, and to how these changes affect the global temperature $T$; 
        the extent of Southern Hemisphere ice is assumed constant, to a first approximation. 
        The change in hydrologic cycle expressed in the terms $P_{\rm rain} - P_{\rm evap}$ for the ocean and $P_{\rm snow} - P_{\rm abl}$ 
        for the snow and ice is due to changes in ocean temperature.  Deep-water formation in the North Atlantic Subpolar Sea 
        (North Atlantic Deep Water: NADW) is affected by changes in ice volume and extent, and regulates the intensity $C$ of the THC; 
        changes in Antarctic Bottom Water (AABW) formation are neglected in this approximation.  
        The THC intensity $C$ in turn affects the system's temperature, and is also affected by it. 
        [After \citet{Ghil.ea.1987} with permission from Springer Nature.]}
\label{Fig_2}
\end{figure}

A sketch of the Atlantic Ocean's THC and its interactions with the atmosphere and cryosphere on long time scales is shown in Fig.~\ref{Fig_2}.  These interactions can lead to climate oscillations with multi-millennial periods, such as the Heinrich events \citep[e.g.,][and references therein]{Ghil1994b}, and are summarized in the figure's caption. An equally schematic view of the global THC is provided by the widely known ``conveyor belt" \citep{Broecker1991} diagram. The latter diagram captures greater horizontal, 2-D detail but it does not commonly include the THC's interactions with water in both its gaseous and solid phases, which our Fig.~\ref{Fig_2} here does include.
  
Basically, the THC is due to denser water sinking, lighter water rising, and water mass continuity closing the circuit through near-horizontal flow between the areas of rising and sinking\footnote{A complementary point of view suggests, instead, that surface winds and tides play a major role in the driving and maintenance of the large-scale ocean circulation \cite{Wunsch2002, Wunsch2013}.} effects of temperature $T$ and salinity $S$ on the ocean water's density, $\rho = \rho(T, S)$, oppose each other: the density $\rho$ {\it decreases} as $T$ increases and it {\it increases} as $S$ increases. It is these two effects that give the {\it thermohaline} circulation its name, from the Greek words for $T$ and $S$.  In high latitudes, $\rho$ increases as the water loses heat to the air above and, if sea ice is formed, as the water underneath is enriched in brine.  In low latitudes, $\rho$ increases due to evaporation but decreases due to sensible heat flux into the ocean.

For the present climate, the temperature effect is stronger than the salinity effect, and ocean water is observed to sink in certain areas of the high-latitude North Atlantic and Southern Ocean --- with very few and limited areas of deep-water formation elsewhere --- and to rise everywhere else. Thus, in a {\it thermohaline} regime, $T$ is more important than $S$ and hence comes before it.  

During certain remote geological times, deep water may have formed in the global ocean near the equator; such an overturning circulation of opposite sign to that prevailing today has been dubbed {\it halothermal}, $S$ before $T$. The quantification of the relative effects of $T$ and $S$ on the oceanic water masses' buoyancy in high and low latitudes is far from complete, especially for paleocirculations; the association of the latter with salinity effects that exceed the thermal ones \citep[e.g., during the Palaeocene, $\simeq 57$~Myr ago, cf. ][]{Kennett.Stott.1991} is thus rather tentative.

To study the reversal of the abyssal circulation, due to the opposite effects of $T$ and $S$, \citet{Stommel1961} considered a model with two pipes connecting two boxes. He showed that the system of two nonlinear, coupled ordinary differential equations that govern the temperature and salinity differences between the two well-mixed boxes has two stable steady-state solutions; these two steady states are distinguished by the direction of flow in the upper and the lower pipe.  

Stommel's paper was primarily concerned with distinct local convection regimes, and hence vertical stratifications, in the North Atlantic and the Mediterranean or the Red Sea, say. Today, we mainly think of one box as representing the low latitudes and the other one the high latitudes in the global THC \citep[and references therein]{Marotzke2000}. Subsequently, many other THC models that used more complex box-and-pipe geometries have been proposed and studied \cite[e.g.,][]{Rooth1982, Scott1999, Titz2001, LucariniStone2005}. 

The next step in the hierarchical modeling of the THC is that of 2-D meridional-plane models, in which the temperature and salinity fields are governed by coupled nonlinear PDEs with two independent space variables, say latitude and depth \cite[e.g.,][]{Quon.Ghil.1992,Cessi1992,Lucariniea2005,Lucariniea2007}. Given boundary conditions for such a model that are symmetric about the Equator, like the equations themselves, one expects a symmetric solution, in which water either sinks near the poles and rises everywhere else (thermohaline) or sinks near the Equator and rises everywhere else (halothermal). These two symmetric solutions would correspond to the two equilibria of the \citet{Stommel1961} two-box model; see \citet{Thual1992} for a discussion of the relationship between 2-D and box models. 

In fact, {\it symmetry breaking} can occur, leading gradually from a symmetric two-cell circulation to an antisymmetric one-cell circulation. In between, all degrees of dominance of one cell over the other are possible. A situation lying somewhere between the two seems to resemble most closely the meridional overturning diagram of the Atlantic Ocean in Fig.~\ref{Fig_2}. 

\begin{figure}
\centerline{
\includegraphics[width = 0.9\columnwidth]{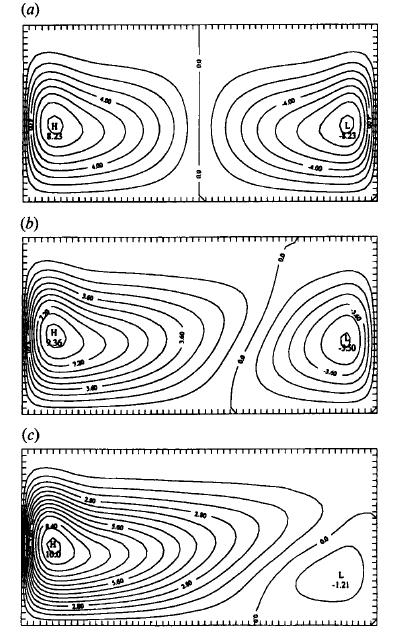}}
\caption{Transition from a symmetric to an increasingly asymmetric meridional ocean circulation. 
      The streamfunction plots represent results from an idealized 2-D model of thermosolutal convection in a rectangular domain, 
      for a prescribed value of the Rayleigh number and increasing values of the nondimensional intensity $\gamma$ of the salinity flux at the surface. 
      (a) $\gamma = 0.40$; (b) $\gamma = 0.50$: and (c) $\gamma = 0.55$.
      [After \citet{Quon.Ghil.1992} with permission from Cambridge University Press.]}
\label{fig:THC_Quon}
\end{figure}

The gradual transition is illustrated by Fig.~\ref{fig:THC_Quon} and it can be described by a {\it pitchfork bifurcation}, cf.~\citet{DG05}:
\begin{equation}
\label{eq:pitchfork}
\dot X = f(X; \mu) = \mu X - X^3.
\end{equation}
Here $X$ measures how asymmetric the  solution is, so that $X = 0$ is the symmetric branch, and $\mu$ measures the stress on the system, in particular a normalized form of the buoyancy flux at the surface.  For $\mu < 0$ the symmetric branch is stable, while for $\mu > 0$ the two branches $X = \pm \mu^{1/2}$ inherit its stability.

In the 2-D THC problem, the left cell dominates on one of the two branches, while the right cell dominates on the other: for a given value of $\mu$, the two stable steady-state solutions --- on the $\{X = + \mu^{1/2}\}$ branch and on the $\{X = - \mu^{1/2}\}$ branch, respectively --- 
are mirror images of each other. The idealized THC in Fig.~\ref{Fig_2}, with the North Atlantic Deep Water extending to the Southern Ocean's polar front, corresponds to one of these two branches. In theory, therefore, a mirror-image circulation, with the Antarctic Bottom Water extending to the North Atlantic's polar front, is equally possible.

More recently, \citet[and references therein]{Cessi.2019} have argued that the meridional overturning is actually powered by momentum fluxes and not by buoyancy fluxes. Because of the ongoing arguments on this topic, the THC is termed increasingly the meridional overturning circulation (MOC).

\subsubsection{\hbox{Bifurcation Diagrams for GCMs}}
\label{ssec:THC_GCMs}

\begin{table*}
{\begin{tabular}{p{0.7in}p{2.4in}p{2.4in} }
\toprule
Time scale & Phenomena & Mechanism \\
\colrule
Decadal & $\cdot$ Local migration of surface anomalies.
     in the NW corner of the ocean basin &  $\cdot$ Localized surface density anomalies due to surface coupling \\
& $\cdot$ Gyre advection in mid-latitudes & $\cdot$ Gyre advection\\
Centennial & Loop-type, meridional circulation  & Conveyor-belt advection of density anomalies \\
Millennial & Relaxation oscillation, with ``flushes" and superimposed decadal fluctuations & Bottom water warming, due to strong braking effect of salinity forcing \\
\botrule
\end{tabular}
\caption{Oscillations in the oceans' thermohaline circulation. See also \citet{Ghil1994b}. \label{tbl_1}}
}
\end{table*}

\citet{Bryan1986} was the first to document transition from a two-cell to a one-cell circulation in a simplified ocean GCM with idealized, symmetric forcing. In Sect.~\ref{ssec:processes} 
atmospheric GCMs confirmed essentially the EBM results. Results of coupled ocean--atmosphere GCMs, however, have led to questions about whether the presence of more than one stable THC equilibrium is actually realistic. The situation with respect to the THC's pitchfork bifurcation \eqref{eq:pitchfork} is thus subtler than it was with respect to Fig.~\ref{Fig_1} for radiative equilibria. 

\begin{figure}
\centering
\subfloat[Intermediate-complexity models]{
    \label{fig:intermediate} 
    \centering
   \includegraphics [angle=0,width=0.9\columnwidth]{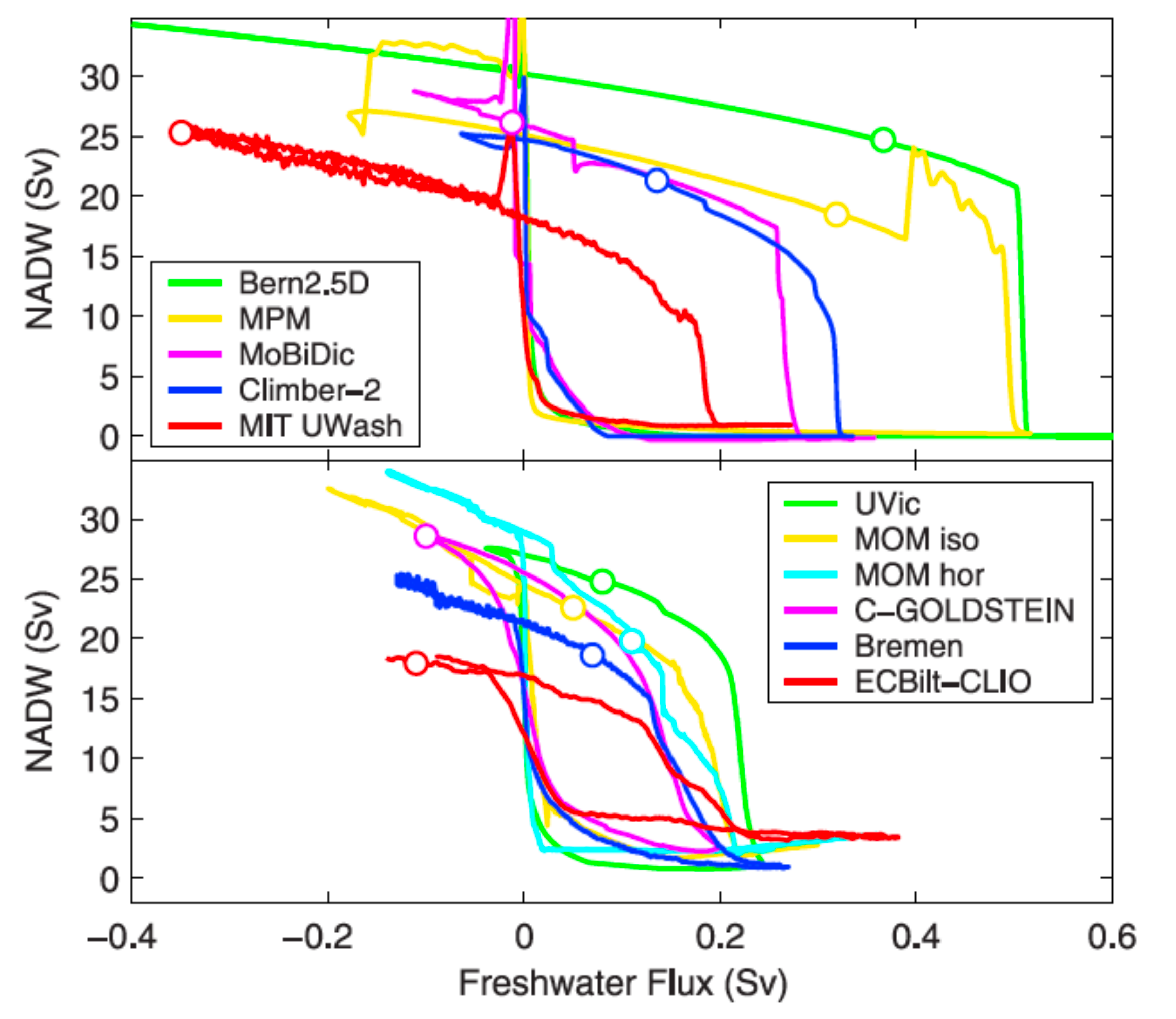}
}

\subfloat[IPCC-class climate model]{
    \label{fig:IPCC_class} 
    \centering
   \includegraphics [width=0.8\columnwidth]{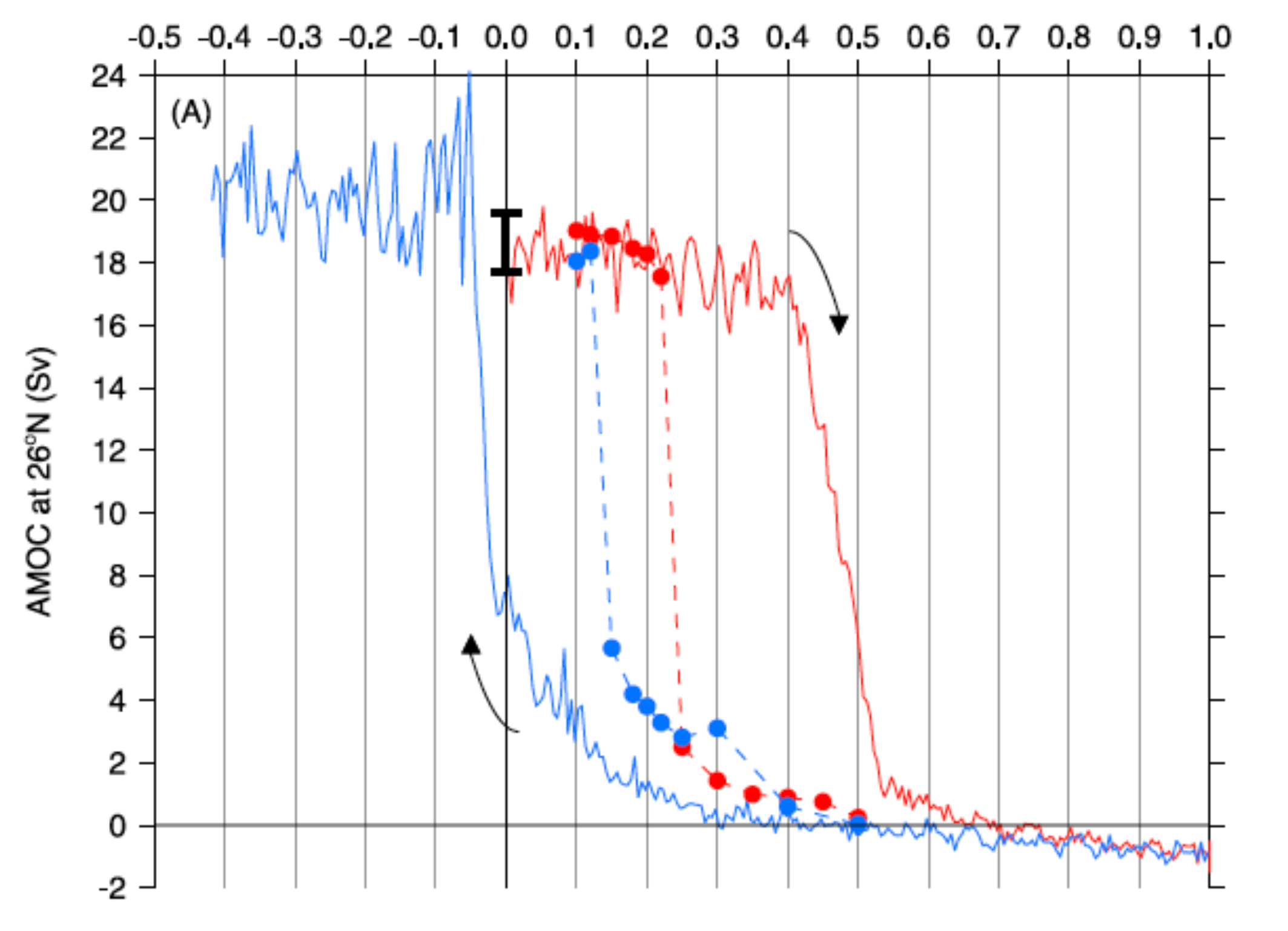} 
}
\caption{\small Multistability of the THC for (a) Earth models of intermediate complexity, reproduced with permission from \citet{Rahmstorf2005}; and (b) for an IPCC-class climate model, reproduced with permission from \citet{Hawkins2011}. Parameter controlling the freshwater input in the North Atlantic basin on the abscissa and  the THC's intensity on the ordinate.}
\label{THCbistable}
\end{figure}

\citet{Dijkstra.multiple.2007} showed, however, by comparing observational and reanalysis data with high-end ocean GCMs, that the current Atlantic Ocean's THC is actually in its multiple-equilibria regime. Climate models of intermediate complexity did indeed find the Atlantic Ocean to be multistable \cite[e.g.,][]{Rahmstorf2005}, while several of the GCMs used in recent IPCC ARs did not show such a behavior, with an exception coming from the study by \citet{Hawkins2011}; see Fig. \ref{THCbistable}. Thus a failure of high-end models in the hierarchy to confirm results obtained on the hierarchy's lower rungs does not necessarily imply it is the simpler models that are wrong. To the contrary, such a failure might well indicate that the high-end models, no matter how detailed, may still be rather imperfect; see also \citet{Ghil2015} for a summary.


Internal variability of the THC --- with smaller and more regular excursions than the huge and totally irregular jumps associated with bistability --- was studied intensively in the late 1980s and the 1990s. These studies placed themselves on various rungs of the modeling hierarchy, from box models through 2-D models and all the way to ocean GCMs.  A summary of the different kinds of oscillatory variability found in the latter appears in Table~\ref{tbl_1}\footnote{Anomalies in the table are defined as the difference between the monthly mean value of a variable and its climatological mean.}.
Such oscillatory behavior seems to match more closely the instrumentally recorded THC variability, as well as the paleoclimatic records for the recent geological past, than bistability.

\begin{figure*}
{
\includegraphics[width = 0.47\columnwidth]{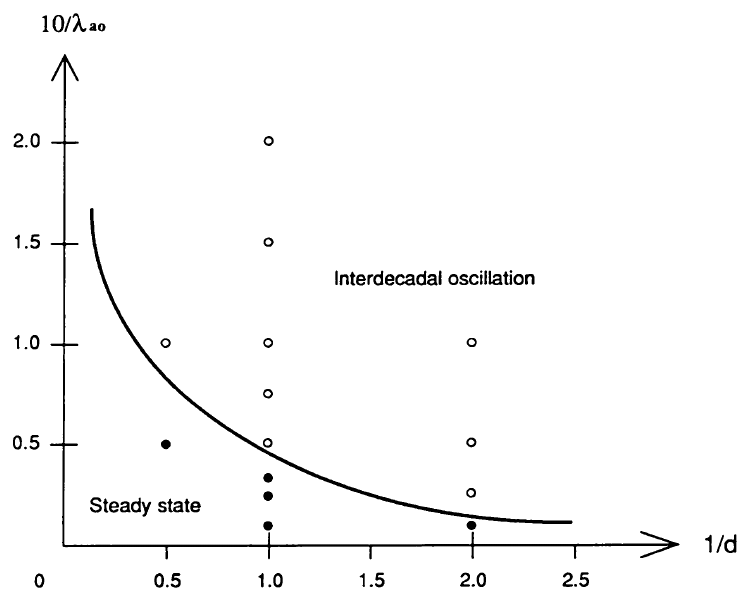}\hfill
\includegraphics[width = 0.47\columnwidth]{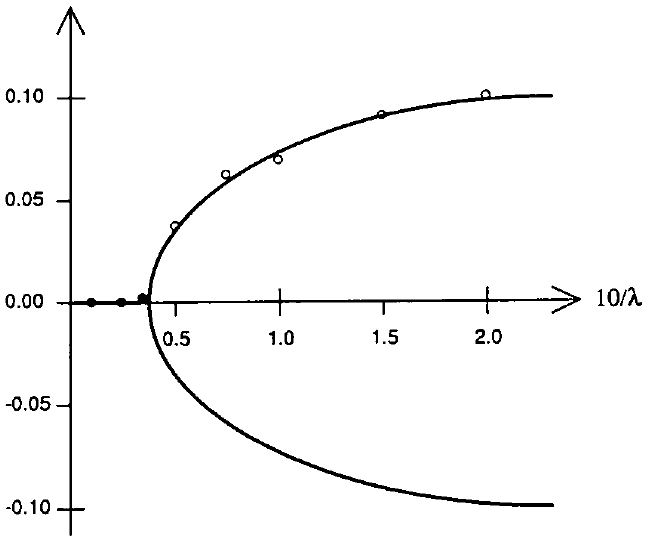}
\par
\hfill (a) Regime diagram\hfill (b) Bifurcation diagram\hfill}
\caption{\small Dependence of THC solutions on two parameters in a hybrid coupled model; the two parameters are the atmosphere--ocean coupling coefficient $\lambda_{\rm ao}$ and the atmospheric thermal diffusion coefficient $d$. (a) Schematic regime diagram.  The full circles stand for the model's stable steady states, the open circles for stable limit cycles, and the solid curve is the estimated neutral stability curve between the former and the latter. (b) Hopf bifurcation curve at fixed $d$  = 1.0 and varying $\lambda_{\rm ao}$; this curve was obtained by fitting a parabola to the model's numerical-simulation results, shown as full and open circles. [From \citet{Chen1996} with permission from the American Meteorological Society.]}
\label{Fig_3}
\end{figure*}

The (multi)millennial oscillations interact with variability in the surface features and processes shown in Fig.~\ref{Fig_2}. \citet{Chen1996}, in particular, studied some of the interactions between atmospheric processes and the THC.  They used a so-called hybrid coupled model, namely a 2-D EBM, coupled to a rectangular-box version of the North Atlantic rendered by a low-resolution ocean GCM.  This hybrid model's regime diagram is shown in Fig.~\ref{Fig_3}(a).  A steady state is stable for higher values of the coupling parameter $\lambda_{\rm ao}$ or of the EBM's diffusion parameter $d$.  Interdecadal oscillations with a period of 40--50 years are self-sustained and stable for lower
 values of these parameters.

The self-sustained THC oscillations in question are characterized by a pair of vortices of opposite sign that grow and decay in quadrature with each other in the ocean's upper layers.
Their centers follow each other anti-clockwise through the northwestern quadrant of the model's rectangular domain. Both the period and the spatio-temporal characteristics of the oscillation are thus rather similar to those seen in a fully coupled GCM with realistic geometry \citep{Delworth.ea.1993}. The transition from a stable equilibrium to a stable limit cycle, via {\it Hopf bifurcation,} in this hybrid coupled model, is shown in Fig.~\ref{Fig_3}(b), and we will further elaborate upon it in Sect.~\ref{sssec:Hopf} below. 

\subsection{Bistability, {Oscillations and} Bifurcations}
\label{ssec:bifurcations}

In Sects.~\ref{ssec:radiation}, \ref{ssec:processes} and \ref{ssec:THC_theory}, 
we have introduced bistability of steady-state solutions via saddle-node and pitchfork bifurcations, while in Sect.~\ref{ssec:THC_GCMs}, 
we mentioned oscillatory solutions and Hopf bifurcation as the typical way the latter are reached as a model parameter changes. We start this subsection by summarizing the bifurcations of steady-state solutions, often referred to as equilibria, and will then introduce and discuss the normal forms of Hopf bifurcation, both supercritical and subcritical.

\subsubsection{Bistability and Steady-State Bifurcations}
\label{sssec:Bistability}

Section~\ref{ssec:processes} 
introduced EBMs and explained how the present climate and a totally ice-covered Earth can result as coexisting stable steady states over a certain parameter range. The normal forms of a {\it supercritical} and a {\it subcritical} saddle-node bifurcation were given as Eqs.~\eqref{eq:saddle-node} and \eqref{eq:back-saddle-node}, respectively. Here the criticality is defined as the existence of the two equilibria, stable an unstable, to the right or the left of the {\it critical} or bifurcation point.

An $S$-shaped bifurcation curve, such as the one that appears in Fig.~\ref{Fig_1} for the 1-D EBM of Eq.~\eqref{eq:rad_bal_1}, can be obtained easily by soldering smoothly together the back-to-back saddle-node bifurcations of Eqs.~\eqref{eq:saddle-node} and \eqref{eq:back-saddle-node}. While there is no generic normal form for such a curve, here is a simple example:
\noindent
\begin{subequations}
\label{eq:bifu}
\begin{eqnarray}
\dot X & = & \mu - X^2, \label{eq:saddle}\\
\dot X & = & (\mu - \mu_0) + (X - X_0)^2, \label{eq:back-saddle}
\end{eqnarray}
\end{subequations}
with $\mu_0 = 1$ and $X_0 = - 1/2$.

Note that both the sub- and supercritical saddle-node bifurcations are {\it structurally stable}, i.e. they persist as the system of evolution equations, be it a system of ordinary or partial differential equations, is smoothly perturbed \citep{Andro.Pont.37, Guckenheimer1983, Arnold.2003, Temam1997} by a small amount. This robustness is the reason why saddle-node bifurcations --- and other elementary bifurcations called of {\it co-dimension one}, i.e., depending on a single parameter, like the Hopf bifurcation --- are so important and can, in practice, be a guide through the hierarchy of models, in the climate sciences and elsewhere. A striking example was provided in Sect.~\ref{ssec:processes} 
for the supercritical saddle-node bifurcation that can be found in the 3-D GCM of \citet[Fig.~5]{Wether.Manabe.1975}, as well as the 1-D EBM reproduced in Fig.~\ref{Fig_1} herein.

The next kind of bifurcation that leads to bistability of stationary states is the pitchfork bifurcation introduced in Sect.~\ref{ssec:THC_theory}, 
in connection with the THC, and whose normal form is given by Eq.~\eqref{eq:pitchfork}. This bifurcation, however, only arises in the presence of a mirror symmetry in the model under study, such as seen in Fig.~\ref{fig:THC_Quon}a. One suspects, therefore, that it is nor structurally stable with respect to perturbations of the dynamics that do not preserve this symmetry.

\begin{figure}
\centerline{
\includegraphics[width = 0.95\columnwidth]{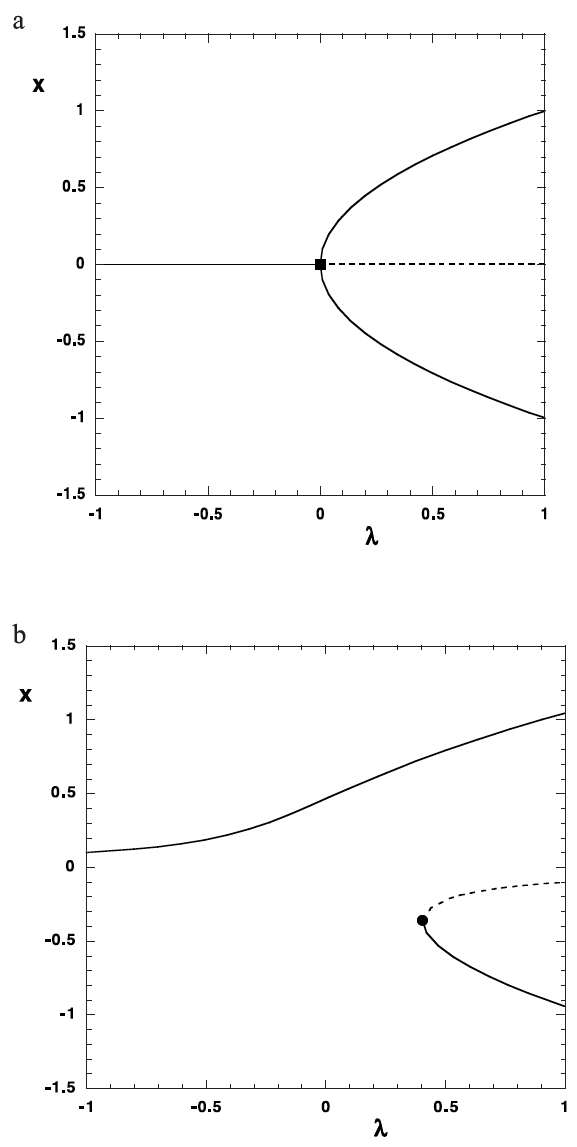}}
\caption{\small Bifurcation diagram of the pitchfork bifurcation in Eq.~\eqref{eq:pitch-perturb},
     (a) for $\epsilon = 0$; and (b) for $\epsilon = 0.1.$ Solid lines indicate stable solutions, 
     and dashed lines indicate unstable ones. [After \citet{DG05} with permission from Elsevier.]}
\label{fig:pitch}
\end{figure}

A simple example is given by the following perturbed pitchfork bifurcation:
\begin{equation}
\label{eq:pitch-perturb}
\dot X = f(X; \mu, \epsilon) = X(\mu - X^2) + \epsilon,
\end{equation}
where $\epsilon$ is a small parameter. The bifurcation diagrams for $\epsilon = 0$ and $\epsilon = + 0.1$ are given in Figs.~\ref{fig:pitch}(a,b), respectively. Clearly, a nonzero value of $\epsilon$ breaks the $X \rightarrow -X$ symmetry of Eq.~\eqref{eq:pitchfork}, since it is no longer the case that $f(X) = f(-X)$. Hence, the pitchfork breaks up into a continuous uppermost branch that is stable for all $X$-values, and a supercritical saddle-node bifurcation, whose lower branch is stable. If $\epsilon < 0$, it will be the lowermost branch that is stable for all $X$-values (not shown), and the upper branch of the saddle-node bifurcation above it that is stable. In both cases, above a critical value associated with the saddle-node bifurcation, three solution branches exist, with the middle one that is unstable and the other two that are stable. 

An obvious way that the symmetric pitchfork bifurcation present in 2-D models of the THC can be broken, as illustrated in Fig.~\ref{fig:pitch}, is simply by considering 3-D models with zonally asymmetric basin boundaries.
Perturbed pitchfork bifurcations were also found in shallow-water models of the wind-driven ocean circulation \citep[e.g.,][]{Jiang1995,Speich1995,Ghil2016}.

\subsubsection{{Oscillatory Instabilities and Hopf Bifurcations}}
\label{sssec:Hopf}

In spite of the considerable detail and 3-D character of the ocean GCM involved in the hybrid coupled model of \citet{Chen1996}, it is clear that the numerically obtained bifurcation diagram in Fig.~\ref{Fig_3}b is of a very simple, fundamental type. The normal form of such a Hopf bifurcation is given by 
\begin{equation}
\label{eq:Hopf_z}
\dot z = (\mu + i \omega)z + c(z \bar z)z,
\end{equation}
where $z = x + iy$ is a complex variable, while $c$ and $\omega$ are nonnegative and $\mu$ is a real parameter. Clearly, for small $z$, this equation describes a rotation in the complex plane with an increasing radius $|z|$ when $\mu > 0$, i.e., it contains the possibility of an oscillatory instability, while the cubic term corresponds to an increasing modification of this simple rotation away from the origin.

The parsimonious complex notation above follows \citet{V.Arnold.83} and \citet{Ghil1987} and is, we believe, more suggestive and transparent than the more common one that uses separately the two real variables $x$ and $y$ \citep[e.g.,][]{Guckenheimer1983}. The advantages of the former are apparent when introducing polar variables via $z = \rho^{1/2}e^{i\theta},$ with $\rho = z \bar z \geq 0.$ One can then separate the flow induced by Eq.~\eqref{eq:Hopf_z} into a constant rotation with angular velocity $\omega$ and a radius $r = \rho^{1/2}$ that either increases or decreases as $\dot \rho \gtrless 0$, according to
\noindent
\begin{subequations}
\label{eq:Hopf_r}
\begin{eqnarray}
\dot \rho & = & 2 \rho(\mu + c \rho), \label{eq:rho}\\
\dot \theta & = & \omega. \label{eq:theta}
\end{eqnarray}
\end{subequations}

Equation~\eqref{eq:rho} is quadratic in $\rho$ and has the two roots $\rho = 0$ and $\rho = - \mu/c.$ The former corresponds to a fixed point at the origin $z = 0$ in Eq.~\eqref{eq:Hopf_z}, while the latter only exists when $c \mu < 0$. 

\begin{figure*}
{
\includegraphics[width = 0.4\textwidth]{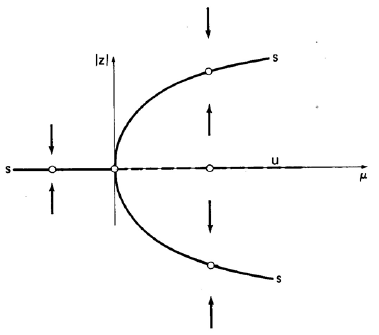}\hfill
\includegraphics[width = 0.5\textwidth]{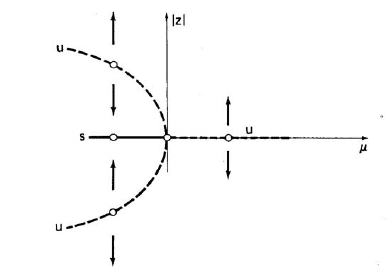}
\par
\hfill (a) supercritical Hopf\hfill (b) subcritical Hopf\hfill}
\caption{\small Bifurcation diagram of the Hopf bifurcation in Eq.~\eqref{eq:Hopf_z}:
     (a) supercritical Hopf bifurcation for $c = -1$; and (b) subcritical Hopf bifurcation for $c = +1.$
     Solid lines indicate stable solutions (s), and dashed lines indicate unstable ones (u); it is common to plot $|z| = - \mu^{1/2}$ along with $|z| = \mu^{1/2},$ in order to emphasize 
     that one is actually projecting onto the $(|z|, \mu)$-plane a paraboloid-shaped, one-parameter family of limit cycles with given $r = |z| = \mu^{1/2}$ and $0 \leq \theta < 2\pi.$
     [After \citet{Ghil1987} with permission from Springer Science+Business Media.]}
\label{fig:Hopf}
\end{figure*}

We thus anticipate a solution that is a circle with radius $r = (- \mu/c)^{1/2}$ when $c \mu \neq 0$ and the two parameters have opposite signs. The simpler case is that of the nonzero radii given by Eq.~\eqref{eq:rho} for positive stability parameter $\mu$  and negative saturation parameter $c$: it corresponds to the {\it supercritical} Hopf bifurcation sketched in Fig.~\ref{fig:Hopf}a. In this case, the stable fixed point at the origin loses its stability as $\mu$ changes sign, and transmits it to a limit cycle of parabolically increasing radius.

The opposite case, of $c > 0$, is plotted in Fig.~\ref{fig:Hopf}b: it corresponds to {\it subcritical} Hopf bifurcation. Here, in fact, the stability of the fixed point is also lost at $\mu = 0$ but there is no oscillatory solution for $\mu > 0$ at all, since the higher-order terms in either Eq.~\eqref{eq:Hopf_z} or \eqref{eq:rho} cannot stabilize the linearly unstable oscillatory solution, which spirals out to infinity. It is not just that the parabolas in Figs.~\ref{fig:Hopf}a and \ref{fig:Hopf}b point in opposite directions, as for the saddle-node bifurcations in Eqs.~\eqref{eq:saddle} and \eqref{eq:back-saddle} or Eqs.~\eqref{eq:saddle-node} and \eqref{eq:back-saddle-node}, but the two figures are topologically distinct.


\subsection{Main Modes of Variability}
\label{ssec:modes}


The atmosphere, ocean, and the coupled ocean--atmosphere climate system have many modes of variability, as initially discussed in Sects.~\ref{ssec:observations}-\ref{ssec:time_scales}-\ref{conservation}. 
We review here some of the most important ones, and comment on the general features of such modes. 

\subsubsection{Modes of Variability and Extended-Range Prediction}
\label{sssec:Ext_pred}

Several large-scale spatial patterns of atmospheric covariability have been studied, starting in the second half of the 19$^{\rm{th}}$ century. \citet{Lor67} and \citet{Wallace.Gutzler.1981} provided good reviews of the earlier studies. The earliest work tended to emphasize ``centers of action,'' where the variability is strongest \cite{TeissBort1881}, while more recently, it is so-called {\it teleconnections} between such centers of action that have been emphasized. J. Namias (1910--1997) played a key role in developing the interest in such teleconnections, by applying systematically the use of their spatio-temporal properties to the development of operational extended-range weather forecasting \cite[e.g.,][]{ Namias1968}.

Returning to the discussion of prediction in Sect.~\ref{climatemodelprediction}, 
it is important to recall John von Neumann's (1903--1957) important distinction \citep{JvN.predict.1960} between weather and climate prediction. To wit, short-term NWP is the easiest --- i.e., it is a pure initial-value problem; long-term climate prediction is next easiest --- it corresponds to studying the system's asymptotic behavior; while intermediate-term prediction is hardest --- both initial and boundary values are important. Von Neumann's role in solving the NWP problem by integrating the discretized equations that govern large-scale atmospheric motions  \citep{Bjerknes.1904, Richardson.1922, CFN.1950} is well known. In fact, he also played a key role in convening the conference on the ``Dynamics of Climate'' that was held at Princeton's Institute for Advanced Studies in October 1955, and whose proceedings \cite{Pfeffer.1960} were finally  published three years after Von Neumann's untimely death.

Today, routine NWP is quite skillful for several days, but we also know that detailed prediction of the weather is limited in theory by the exponential growth of small errors \citep{Lorenz1963a} and by their turbulent cascading from small to large scales \citep{Thompson.1957, Lorenz1969a, Lorenz1969b, Leith.Kraichnan.1972}. The theoretical limit of detailed prediction --- in the sense of predicting future values of temperature, wind and precipitation at a certain point, or within a small volume, in time and space --- is of the order of 10--15~days \citep{Epstein.1988, Tribbia.1987}.

In the sense of \citet{JvN.predict.1960}, short-term prediction is being improved by meteorologists and oceanographers through more-and-more accurate discretization of the governing equations, increased horizontal and vertical resolution of the numerical models, improved observations and data assimilation methodologies, and improved parametrization of subgrid-scale processes. 

Important strides in solving the theoretical problems of the climate system's asymptotic behavior are being taken by the use of idealized models --- either by simplifying the governing equations in terms of the number of subsystems and of active physical processes, e.g., by eliminating phase transitions or chemical processes in the atmosphere --- or by systematic model reduction to small or intermediate-size sets of ordinary differential or stochastic differential equations \citep[e.g.,][]{Chang.ea.2015, palmer_stochastic_2009}. We shall return to the latter in Sects.~\ref{sensitivity} and \ref{critical}. 

What, then, if anything, and how, and how accurately can climate-related scalars or fields be predicted beyond the limits of NWP? In other words, what can be done about the gap between short-term and asymptotic prediction of climate? These issues have been actively pursued for the last three decades \cite[e.g.,][and many others]{Epstein.1988, Tribbia.1987, Ghil.Rob.2002}. Atmospheric, oceanic and coupled modes of variability play an important role in extended- and long-range forecasting. 

The key idea is that a mode that is stationary or oscillatory can, by its persistence or periodicity, contribute a fraction of variance that is predictable for the mode's characteristic time or, at least, for a substantial fraction thereof.  We start therefore with a brief review of some of the most promising modes of variability. 


\subsubsection{Coupled Atmosphere--Ocean Modes of Variability}\label{coupledmode}

\begin{figure}
\centerline{
\includegraphics[width = .95\columnwidth]{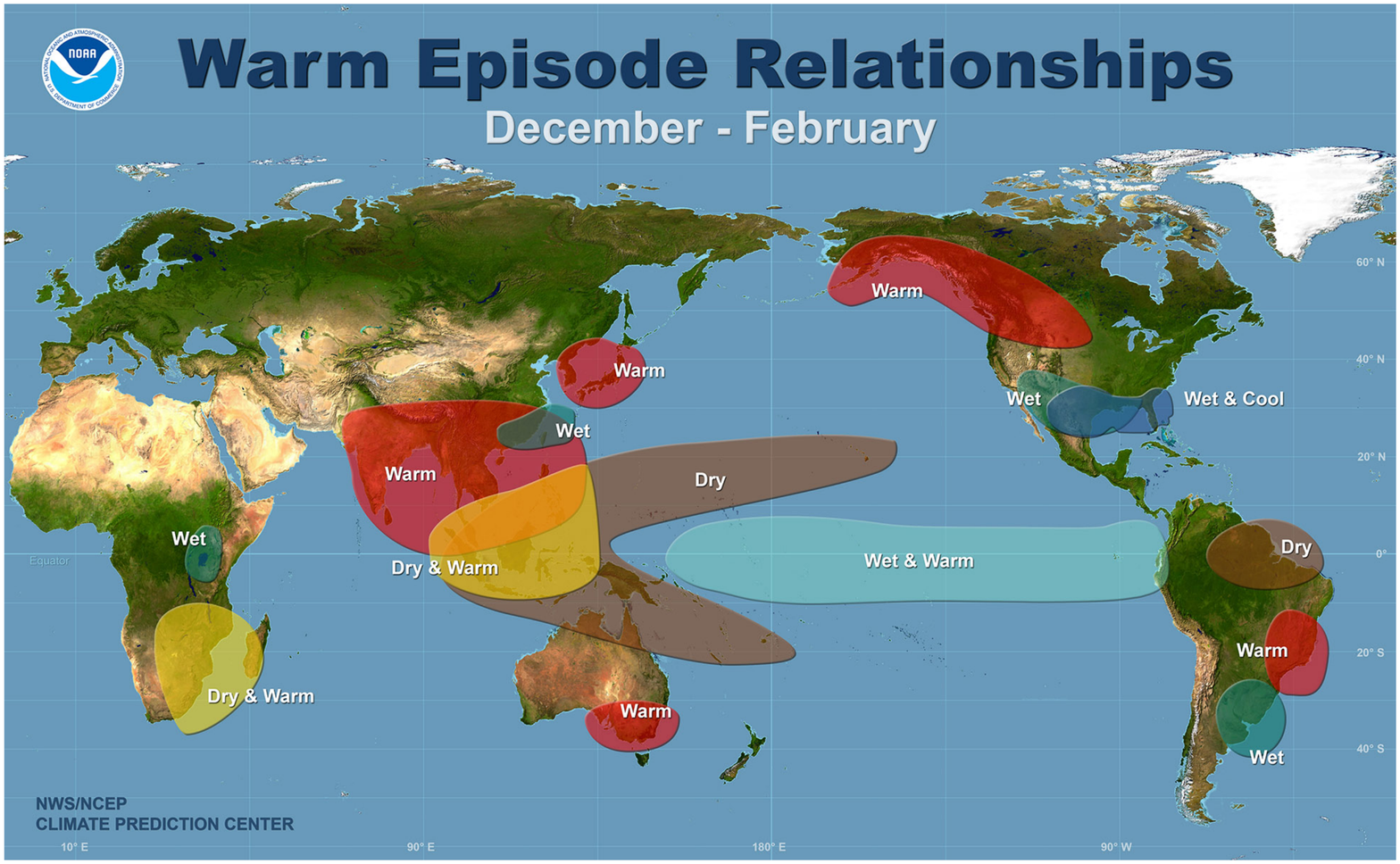}}
\caption{Teleconnection pattern for a warm ENSO episode, during the boreal winter months 
     December-January-February. Both colors and labels indicate {\re warm}-vs.-{\bb cold} 
     and {\gr wet}-vs.-{\br dry} anomalies,
     with anomalies being defined as the difference between a 
     monthly mean value of a variable and its climatological mean.
     [Reproduced, with permission, from \url{https://www.meted.ucar.edu/ams/wim_2014/9b.html}.]
     }
\label{ENSO-tel}
\end{figure}

The best known of all these modes is the El Ni\~no--Southern Oscillation \citep[ENSO:][and references therein]{Philander1990, DijkstraB2000}, mentioned earlier in Sect.~\ref{ssec:time_scales}. 
The ENSO phenomenon is particularly strong over the Tropical Pacific, but it affects temperatures and precipitations far away and over a large area of the globe. Some of the best documented and statistically most significant ones of these teleconnections are illustrated in Fig.~\ref{ENSO-tel}. Just as an example, destructive droughts over Northeast Brazil, Southeast Africa or cold spells over Florida are often associated with particularly strong warm ENSO episodes. 

Such strong episodes recurred every 2--7 years during the instrumental record of roughly 150 years, and the ENSO phenomenon, with it alternation of warm episodes (El Ni\~nos) and cold ones (La Ni\~nas), is quite irregular. Still, there is a marked tendency for year-to-year alternation of, not necessarily strong, El Ni\~nos and La Ni\~nas; this alternation  is associated with a well-known quasi-biennial near-periodicity \citep[e.g.,][]{Rasmusson1990, Ghil.SSA.2002}.

Even larger variance, accompanied by lesser regularity, is associated with a quasi-quadrennial mode \cite{JiangN1995, Ghil.SSA.2002}, sometimes just called the low-frequency ENSO mode. The positive interference of these two modes generates large ENSO events that visually coincide with the instrumentally recorded ones \cite[Fig.~9]{JiangN1995}. 

These ENSO features have been used since 1992 for real-time forecasting that essentially relies on predicting the oscillatory modes of two scalar indices that capture much of the ENSO variability, namely the Southern Oscillation Index (SOI) and the so-called  Ni\~no-3 index, obtained by averaging mean-monthly sea surface temperatures (SSTs) over an area of the Eastern Tropical Pacific. For the time being, such a data-driven forecast appears to still be quite competitive with those made by high-end, detailed GCMs \cite{iri12}.

\subsubsection{Atmospheric Low-Frequency Variability}
\label{sssec:LFV}
As discussed in Sect.~\ref{ssec:observations}, 
datasets for the atmosphere are both longer and more plentiful than for the oceans. Thus, there are quite a few modes of variability that have been detected and described, especially for the Northern Hemisphere, where both the human population and the data sets are denser.

As discussed in Sect.~\ref{ssec:time_scales}, 
in connection with Fig. \ref{f:MG1}, 
atmospheric phenomena are designated as having low, or intraseasonal, frequency if their characteristic time is longer than the life cycle of a mid-latitude weather system but still shorter than a season. We are thus talking here of intrinsic variability, as opposed to the externally forced seasonal cycle, the latter being easier to understand. More recently, one is also referring to this variability as \emph{subseasonal} \cite[e.g.,][]{S2S.book}.

There are essentially two approaches for describing this intrinsic, subseasonal variability: (i) as being episodic or intermittent, and (ii) as being oscillatory. The two approaches are complementary, as we shall see, and they have been dubbed the `particle' vs. the `wave' description, by a crude analogy with quantum mechanics \cite{Ghil.Rob.2002}. 

The key ingredient of the particle approach is provided by so-called \emph{persistent anomalies} or regimes. The best known among these are blocking vs. zonal flow, which were already mentioned in Sect. \ref{ssec:time_scales}. 
Other well-known persistent features, especially during boreal winter, are the positive and negative phase of the North Atlantic Oscillation (NAO) and the Pacific North American (PNA) pattern. A rich literature exists on the reliable identification, description and modeling of these patterns; see \citet[and references therein]{Ghil.ea.S2S}. In the Southern Hemisphere, there has also been some interest in an approximate counterpart of the PNA, called the Pacific South American (PSA) pattern \cite{Mo.Ghil.1987}.

The key ingredient of the wave approach is provided by oscillatory modes that do not necessarily possess exact periodicities, but rather the broad spectral peaks that were illustrated in Fig. \ref{f:MG1} 
and discussed already to some extent in Sect.~\ref{ssec:time_scales}. 
Probably the best known example of this type in the subseasonal band is the Madden-Julian Oscillation \cite[MJO:][]{Mad.Jul.71}. It has a near-periodicity of roughly 50 days and affects winds and precipitation in the Tropics, being strongest in the Indo-Pacific sector. Like the much lower-frequency ENSO phenomenon, important extratropical effects have been documented \cite[e.g.,][]{Mal.Hart.00}.

In spite of the considerable amount of observational, theoretical and modeling work dedicated to the MJO, it is still incompletely understood and not very well simulated or predicted. Some of the reasons for these difficulties include the key role played in its mechanism by transitions among liquid and gaseous phases of water in tropical clouds, its multiscale character, and the pronounced interactions with the oceans \cite{Zhang.2005}.

Two extratropical modes in the subseasonal band are the \citet{Branstator.1987}--\citet{Kushnir.1987} wave, and the 40-day mode associated with the topographic instability first described by \citet{Charney1979} in a low-order model. The Branstator-Kushnir wave is, like the MJO, an eastward traveling wave, while the 40-day mode is a standing wave anchored by the topography. 

\citet{Charney1979} emphasized the bimodality of the solutions of a model version with only three Fourier modes --- one stable steady state being zonal and the other being blocked --- and did not pursue further the fact that, in a more highly resolved version of the model, oscillatory solutions did appear. M. Ghil and associates did clarify the role of the higher meridional modes in the Hopf bifurcation that gives rise to the oscillatory solutions \cite{Ghil1987, Legras1985, Jin.Ghil.1990}. The potential role of these results in the controversy surrounding the effect of anthropogenic polar amplification on blocking frequency was mentioned in Sect. \ref{ssec:midlat}. 

While clearly mid-latitude weather systems, like the tropical ones, also involve precipitation, their large-scale properties seem to be much less affected by wet processes; the former arise essentially from purely dynamical, as opposed to largely thermodynamical mechanisms. For the sake of simplicity, we will thus try to illustrate in further detail the complementarity of the wave and the particle approach for the extratropical topographic oscillation.

Given the recent interest in the physical literature for synchronization in continuous media, cf.~\citet{Duane.ea.2017} and references therein, it might be challenging to this readership that one can accomodate in a fairly narrow frequency band, between roughly $0.1$ and $0.01$ day$^{-1}$, three distinct oscillatory modes that do not seem to synchronize: in the Tropics the MJO, with a period $\simeq 50$~days, and in mid-latitudes the Branstator-Kushnir wave, with a period $\simeq 30$~days, along with the topographic oscillatory mode, with a period $\simeq 40$~days; see, for instance, \citet[Fig.~11]{Dickey.ea.1991}. Not only are these three frequencies fairly close. i.e., the detuning fairly small, but the characteristic wave lengths of all three of these oscillatory modes are quite large with respect to the radius of the Earth, and several teleconnections mentioned so far extend across continents and oceans, cf. Fig.~\ref{ENSO-tel} herein.

Be that as it may, let us now concentrate on the topographic `wave' mode and its relationship with the blocking and zonal `particles'. \citet{Ghil.ea.S2S} recently reviewed the evidence provided by a hierarchy of models for the presence of a Hopf bifurcation that arises from the interaction of the large-scale westerly flow in mid-latitudes with the topography of the Northern Hemisphere. 

Certain spatial features of the phases of this mode do present striking similarities to the blocked and zonal flows that appear not just as the two stable equilibria in the highly idealized model of \citet{Charney1979}, but also as unstable, though long-lived, patterns in much more realistic models, like the three-level, quasi-geostrophic (QG3) model originally formulated by \citet{Mar.Mol.1993}. The latter model is still widely used to study the nonlinear dynamics of large-scale, mid-latitude atmospheric flows \cite[e.g.,][]{Kondras.ea.2004,LucariniG2019}. 

\begin{figure}[h]
	\centering
	\includegraphics[width=0.9\linewidth]{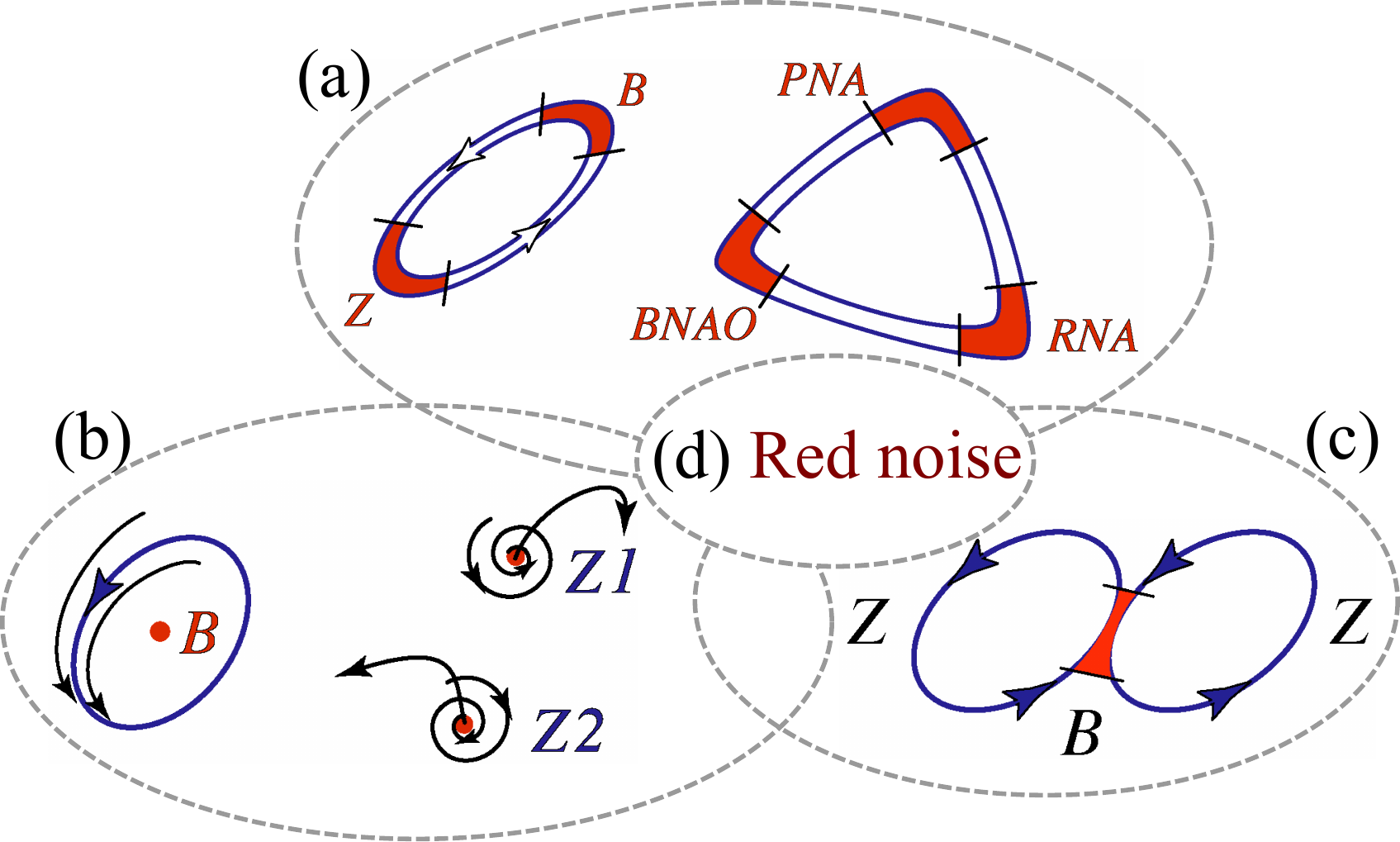}
	\caption{Schematic overview of atmospheric low-frequency variability (LFV) mechanisms; see text for details. 
	Reproduced from \citet{Ghil.ea.S2S}, with permission from Elsevier.}
	\label{fig:part_wave}
\end{figure}

More broadly, we summarize here the relevant conclusions of the \citet{Ghil.ea.S2S} review paper on the observational, theoretical and modeling literature of multiple regimes and oscillatory modes of subseasonal variability. Four complementary approaches to explaining this variability are illustrated in Fig.~\ref{fig:part_wave}.

One approach to persistent anomalies in mid-latitude atmospheric flows on subseasonal time scales is to consider them simply as due to slowing down of Rossby waves or to their linear interference \citep{Lindzen.1986}. This approach is illustrated in the sketch labeled (c) within the figure: zonal flow $Z$ and blocked flow $B$ are simply slow phases of an harmonic oscillation, like the neighborhood of $t = \pi/2$ or $t = 3 \pi/2$ for a sine wave $\sin(t)$; or else they are due to an interference of two or more linear waves, like the one occurring for a sum $A\sin(t) + B\sin(3t)$ near $t = (2k + 1) \pi/2$. A more ambitious, quasi-linear version of this approach is to study long-lived resonant wave triads between a topographic Rossby wave and two free Rossby waves \citep{Egger.1978, Trevi.Buzzi.1980, Ghil1987}. Neither version of this line of thought, though, explains the  organization of the persistent anomalies into distinct flow regimes. 

\citet{Rossby.ea.1939} initiated a different, genuinely nonlinear approach by suggesting that multiple equilibria may explain preferred atmospheric flow patterns. These authors drew an analogy between such equilibria and hydraulic jumps, and formulated simple models in which similar transitions between faster and slower atmospheric flows could occur. This multiple-equilibria approach was then pursued vigorously in the 1980s \citep{Charney1979, Charney.ea.1981, Legras1985, Ghil1987} and it is illustrated in Fig.~\ref{fig:part_wave} by the sketch labeled (a): one version of the sketch illustrates models that concentrated on the $B$--$Z$ dichotomy \citep{Charney1979, Charney.ea.1981, Benzi.ea.1986}, the other on models \citep[e.g.,][]{Legras1985} that allowed for the presence of additional clusters, found by \citet{Kimoto1993a} and \citet{Smyth1999}, among others, in observations. The latter include opposite phases of the NAO and PNA anomalies ($PNA, RNA$ and $BNAO$ in sketch (a) of Fig.~\ref{fig:part_wave}). The LFV dynamics in this approach is given by the preferred transition paths between the two or more regimes; see again Table~1 in \citet{Ghil.ea.S2S} and references therein.

A third approach is associated with the idea of oscillatory instabilities of one or more of the multiple fixed points that can play the role of regime centroids. Thus, \cite{Legras1985} found a 40-day oscillation arising by Hopf bifurcation off their blocked regime $B$, as illustrated in sketch (b) of the figure. 

An ambiguity arises, though, between this point of view and the complementary possibility that the regimes are just slow phases of such an oscillation, caused itself by the interaction of the mid-latitude jet with topography that gives rise to a supercritical Hopf bifurcation. Thus, \citet{Kimoto1993a,Kimoto1993b} 
found, in their observational data, closed paths within a Markov chain whose states resemble well-known phases of an intraseasonal oscillation. Such a possibility was confirmed in the QG3 model by \citet{Kondras.ea.2004}. Furthermore, multiple regimes and intraseasonal oscillations can coexist in a two-layer model on the sphere within the scenario of ``chaotic itinerancy'' \citep{Itoh.Kimoto.1996, Itoh.Kimoto.1997}. 

\citet{LucariniG2019} observed that blockings occur when the system's trajectory is in the neighborhood of a specific class of unstable periodic orbits (UPOs). UPOs in general are natural modes of variability that cover a chaotic system's attractor \cite{svita88, Cvitanovic1991}. Here, the UPOs that correspond to blockings have a higher degree of instability compared to UPOs associated with zonal flow; thus blockings are associated with anomalously unstable atmospheric states, as suggested theoretically by \citet{Legras1985} and confirmed experimentally by \citet{Weeks.ea.1997}. Different regimes may be associated with different bundles of UPOs, a conjecture that could also explain the efficacy of Markov chains fin describing the transitions between qualitatively different regimes.  

Sketch (d) in the figure refers to the role of stochastic processes in S2S variability and prediction, whether it be noise that is white in time --- as in \citet{Hasselmann1976} or in linear inverse models \citep[LIMs:][]{Penland_MWR89, Penland_PD96, PenlandGhil_MWR93, PenlandSardeshmukh_JCL95} --- or red in time, as in certain nonlinear data-driven models \citep{KravtsovKondrashovGhil_JCL05, KravtsovGhilKondrashov_09, Kondrashov.Kravtsov.ea.2006, Kondras.MJO.2013, MSM2015} or even non-Gaussian \citep{Sard.Pen.2015}. Stochastic processes may enter into models situated on various rungs of the modeling hierarchy, from the simplest conceptual models to high-resolution GCMs. In the former, they may enter via stochastic forcing, whether additive or multiplicative, Gaussian or not \citep[e.g.,][]{MSM2015}, while in the latter, they may enter via stochastic parametrizations of subgrid-scale processes \citep[e.g.,][and references therein]{palmer_stochastic_2009}.

Figure~\ref{fig:part_wave} summarizes some of the key dynamical mechanisms of mid-latitude subseasonal variability, as discussed in this section and in \citet{Ghil.ea.S2S}, without attempting to provide a definitive answer as to which approach to modeling and prediction of this variability is or will be most productive in the near future.


\subsection{Internal Variability and Routes to Chaos}
\label{routes}

In the present section, we illustrate on hand of the wind-driven ocean circulation a sequence of successive bifurcations that lead from a highly symmetric, steady-state circulation to much more finely structured, irregular, possibly chaotic oscillations. Various issues arise in pursuing such a bifurcation sequence across a hierarchy of models and on to the observational data.

\begin{figure*} 
	\centering
    \includegraphics[width=0.9\textwidth, scale=1]{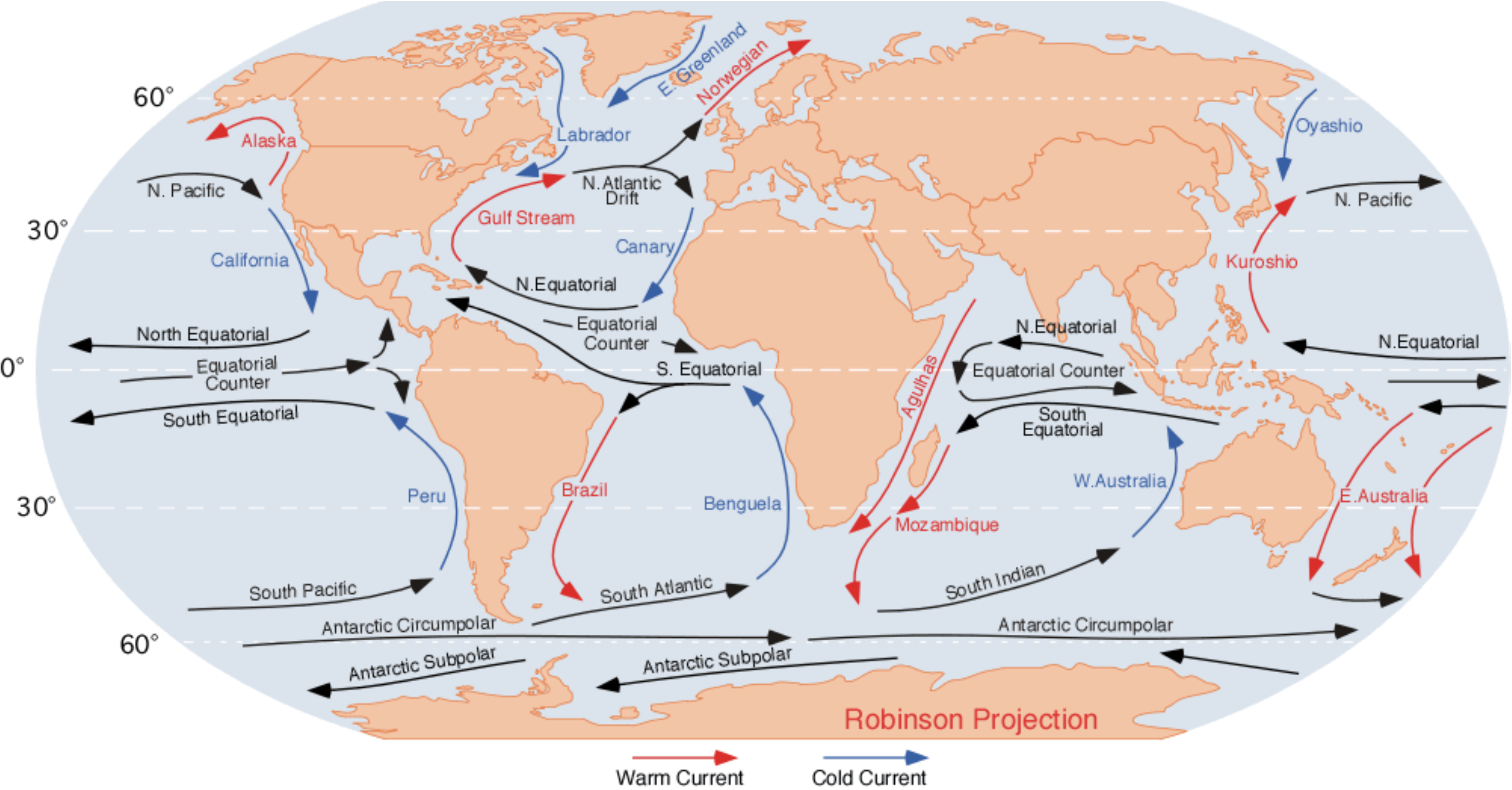}
    \caption{A map of the main oceanic currents: warm currents in {\re red}
    and cold ones in {\bb blue}. Reproduced from \citet{GCS08}, with permission from Elsevier.}
    \label{fig:currents}
\end{figure*}

Mid-latitude oceanic gyres appear clearly in Fig.~\ref{fig:currents} below, in the four major extratropical ocean basins, namely the North and South Atlantic, and the North and South Pacific. The large, subtropical ocean gyres are formed by a poleward-flowing western boundary cur{\color{red}r}ent, an equatorward-flowing eastern boundary current, and the roughly zonally flowing currents that connect these two coastal currents off the equator and on the poleward basin side, respectively. These gyres are characterized by so-called anticyclonic rotation, clockwise in the Northern and anticlockwise in the Southern Hemisphere. In the North Pacific and the North Atlantic, they are accompanied by smaller, cyclonically rotating gyres, while in the Southern Hemisphere, such subpolar gyres are missing and replaced by the Antarctic Circumpolar Current.

The basic phenomenology of these gyres and the detailed physical mechanisms that give rise to it are described in several books \cite[e.g.,][]{CRB2011, Ghil1987, Gill1982, Pedlosky1996, Sverdrup1946, Vallis_atmospheric_2006} and review papers \cite{DG05, Ghil2016}. Clearly the sharp western boundary currents --- like the Gulf Stream in the North Atlantic. the Kuroshio and its cross-basin extension in the North Pacific, and the Brazil Current in the South Atlantic --- as well as the more diffuse eastern boundary currents --- like the Canaries Current in the North Atlantic and the California and Peru Currents in the North and South Pacific, respectively --- play a major role in carrying heat poleward and colder waters equatorward. Hence these gyres' interannual and interdecadal variability is a major contributor to climate variability.

To illustrate the bifurcation sequence that might lead to this oceanic LFV, we use a highly idealized model of the wind-driven double-gyre circulation in a rectangular geometry. Note that the counterparts of synoptic weather systems in the ocean are eddies and meanders that have much shorter spatial scales than in the atmosphere, but considerably longer time scales: $\mathcal{O}(100)$~km in the ocean vs. $\mathcal{O}( 1000)$~km in the atmosphere, but order of several months in time vs. order of merely several days. Thus the definition we used for LFV in the atmosphere, cf. Sect.~\ref{sssec:LFV}, 
does correspond in the ocean to a time scale of years to decades.

\subsubsection{A simple model of the double-gyre circulation}
The simplest model that includes several of the most pertinent mechanisms described in Sect. \ref{ssec:time_scales} 
is governed by the barotropic quasi-geostrophic equations. We consider an idealized, rectangular basin geometry and simplified forcing that mimics the distribution of vorticity due to the wind stress, as sketched in \citet[Fig.~2]{Simonnet.ea.2005}. In this idealized model, the amounts of subpolar and subtropical vorticity injected into the basin are equal and the rectangular domain $\Omega = (0,L_x) \times (0,L_y)$ is symmetric about the axis of zero wind stress curl $y = L_y/2$. 

The barotropic $2$-D quasi-geostrophic equations in this idealized setting are:
\begin{subequations}\label{QGE}
\begin{eqnarray}
\partial_t q & + & J(\psi,q) -\nu \Delta^2 \psi + \mu \Delta \psi = -\tau \sin\big(\frac{2\pi y}{L_y}\big), \label{eq:PV_t}\\
q & = & \Delta \psi - \lambda_{\rm R}^{-2} \psi + \beta y. \label{eq:PV}
\end{eqnarray}
\end{subequations}
Here  $x$ points east and $y$ points north, while $q$ and $\psi$ are the potential vorticity and the streamfunction, respectively, and the Jacobian $J$ gives the advection of potential vorticity by the flow, as already discussed in Sect.~\ref{QGdynamics} 
, so that $J(\psi,q) = \psi_x q_y - \psi_y q_x = {\bf u} \cdot \nabla q$.

The physical parameters are the strength of the planetary vorticity gradient $\beta = \partial f/\partial y$, the Rossby radius of deformation $\lambda_{\mathrm R}^{-2}$, the eddy-viscosity coefficient $\nu$, the bottom friction coefficient $\mu$, and the wind-stress intensity $\tau$. One considers  
here free-slip boundary conditions  $\psi = \Delta^2 \psi = 0$; the qualitative results described below do not depend on the choice of homogeneous boundary conditions \cite{Jiang1995, DG05}.

The nonlinear system of PDEs \eqref{QGE} is an infinite-dimensional dynamical system and one can thus study its bifurcations as the parameters change. Two key parameters  are  the wind stress intensity $\tau$ and the eddy viscosity $\nu$: as $\tau$ increases the solutions become rougher, while an increase in $\nu$ renders them smoother. 

An important property of (\ref{QGE}) is its mirror symmetry in the $y=L_y/2$ axis. This symmetry can be expressed as invariance with respect to the discrete $\mathbb{Z}_2$ group ${\mathcal S}$, 
$$ {\mathcal S}\left[ \psi(x,y) \right] = -\psi(x,L_y-y); $$
any solution of \eqref{QGE} is thus accompanied by its mirror-conjugated solution. Hence the prevailing bifurcations are of either the symmetry-breaking or the Hopf type. 


\subsubsection{Bifurcations in the  double-gyre problem}
\label{sssec:dble_gyre}
\begin{figure}[htpb]
\includegraphics[width=.9\columnwidth,scale=1]{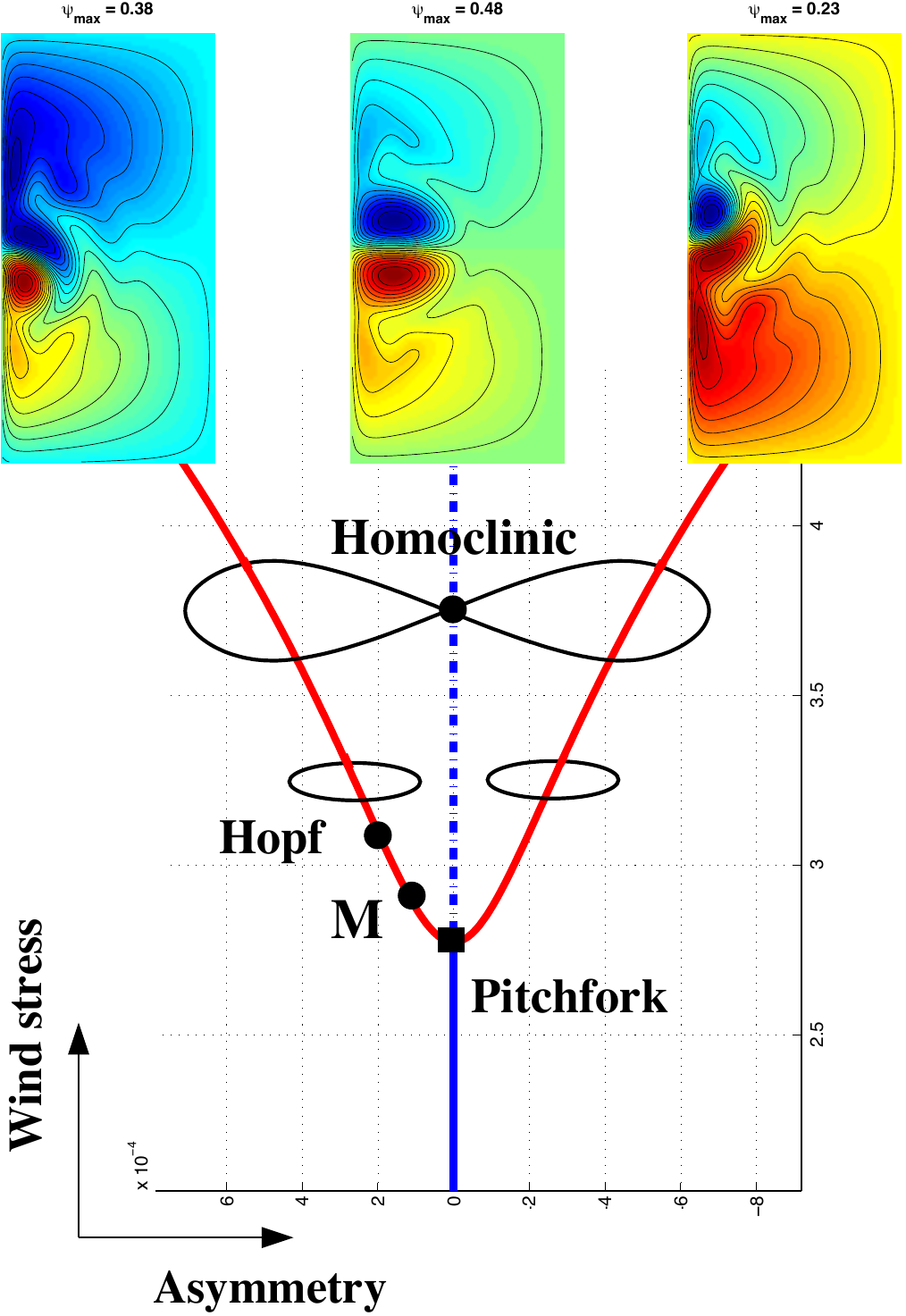}
\caption{Generic bifurcation diagram for the barotropic quasi-geostrophic model of
   the double-gyre problem: the asymmetry of the solution is plotted versus
   the intensity of the wind stress $\tau$. The streamfunction field is
   plotted for a steady-state solution associated with each of
   the three branches; positive values in red and negative ones in blue.
   After \citet{Simonnet.ea.2005}.} \label{fig:bif_tree}
\end{figure}

The development of a comprehensive nonlinear theory of
the double-gyre circulation over the last two decades has
gone through four main steps. These four steps can be
followed through the bifurcation tree in Fig.~\ref{fig:bif_tree}.

\paragraph{Symmetry-breaking bifurcation.}
The ``trunk'' of the bifurcation tree is plotted as the solid blue line in the lower part of the figure. When the forcing $\tau$ is weak or the dissipation $\nu$ is large, there is only one steady solution, which is antisymmetric with respect to the mid-axis of the basin. This solution exhibits two large gyres, along with their $\beta$-induced western boundary currents.  Away from the western boundary, such a near-linear solution (not shown) is dominated by so-called Sverdrup balance between wind stress curl and the meridional mass transport \cite{Gill1982, Sverdrup1947}.

The first generic bifurcation of this quasi-geostrophic model was found to be a genuine pitchfork bifurcation that breaks the system's symmetry as the nonlinearity becomes large enough with increasing wind stress intensity $\tau$ \cite{Jiang1995, Cessi1995}. As $\tau$ increases, the near-linear Sverdrup solution that lies along the solid blue line in the figure develops an eastward jet along the mid-axis, which penetrates farther into the domain and also forms two intense recirculation vortices, on either side of the jet and near the western boundary of the domain.

The resulting more intense, and hence more nonlinear solution is still antisymmetric about the mid-axis, but loses its stability for some critical value of the wind-stress intensity, $\tau = \tau_{\rm P}$. This value is indicated by the filled square on the symmetry axis of Fig.~\ref{fig:bif_tree} and is labeled ``{\bf Pitchfork}" in the figure.

A pair of mirror-symmetric solutions emerges and it is plotted as the two red solid lines in the figure's middle part. The streamfunction fields associated with the two stable steady-state branches have a rather different vorticity distribution and they are plotted in the two small panels to the upper-left and upper-right of Fig.~\ref{fig:bif_tree}. In particular, the jet in such a solution exhibits a large, stationary meander, reminiscent of the semi-permanent one that  occurs in the Gulf Stream, just downstream of Cape Hatteras. These asymmetric flows are characterized by one recirculation vortex being stronger in intensity than the other; accordingly the eastward jet is deflected either to the southeast, as is the case in the observations for the North Atlantic, or to the northeast.

\paragraph{Gyre modes.}
The next step in the theoretical treatment of the problem was taken in part concurrently with the first one above \cite{Jiang1995} and in part shortly thereafter \cite{SDG1995, Dijkstra1997b, Sheremet1997}. It involved the study of time-periodic instabilities that arise through Hopf bifurcation from either an antisymmetric or an asymmetric steady flow. Some of these studies treated wind-driven circulation models limited to a stand-alone, single gyre \cite{Pedlosky1996, Sheremet1997}; such a model concentrates on the larger, subtropical gyre while neglecting the smaller, subpolar one.

The overall idea was to develop a full, generic picture of the time-dependent behavior of the solutions in more turbulent regimes, by classifying the various instabilities in a comprehensive way. However, it quickly appeared that a particular kind of instability leads to so-called {\it gyre modes} \cite{Jiang1995, SDG1995}, and was prevalent across the full hierarchy of models of the double-gyre circulation; furthermore, this instability triggers the lowest nonzero frequency present in all such models \cite{DijkstraB2000, DG05}.

These gyre modes always appear after the first pitchfork bifurcation, and it took several years to understand their genesis: gyre modes arise as two eigenvalues merge --- one of the two is associated with a symmetric eigenfunction and responsible for the pitchfork bifurcation, the other one with an antisymmetric eigenfunction \cite{Simonnet2002}. This merging is marked by a filled circle on the left branch of antisymmetric stationary solutions and is labeled as $\bf M$ in Fig.~\ref{fig:bif_tree}.

Such a merging phenomenon is not a bifurcation in the term's usual meaning: Although it corresponds to a topological change in phase space, the oscillatory behavior at and near $\bf M$ is damped. Nevertheless, this oscillatory eigenmode is eventually destabilized through a Hopf bifurcation, which is indicated in Fig.~\ref{fig:bif_tree} by a heavy dot marked ``{\bf Hopf}," from which a stylized limit cycle emerges. A mirror-symmetric $\bf M$ and {\bf Hopf} bifurcation also occur on the right branch of stationary solutions, but have been omitted for visual clarity. This merging is as generic as the $\bf Pitchfork$ bifurcation in the figure, and arises in much more complex situations and models \cite[e.g.,][]{Simonnet2003b, Simonnet2005}.

In fact, such merging is quite common in small-dimensional dynamical systems with symmetry, as exemplified by the unfolding of codimension-2 bifurcations of Bogdanov-Takens type \cite{Guckenheimer1983}. In particular, the fact that gyre modes trigger the longest, multi-annual periodicity of the model is due to the frequency of this mode growing quadratically with the control parameter from zero --- i.e., from infinite period --- until nonlinear saturation sets in \cite[e.g.,][]{Simonnet2002, SDG.Hdbk.2009}.

More generally, Hopf bifurcations give rise to features that recur more-or-less periodically in fully turbulent planetary-scale flows, atmospheric, oceanic and coupled \cite{Ghil1987, DijkstraB2000, DG05, Ghil2015, Ghil2016}. It is precisely this kind of near-periodic recurrence that is identified in the climate sciences as LFV.

\paragraph{Global bifurcations.}
The bifurcations studied so far --- in this subsection, as well as in the preceding ones --- are collectively known as {\em local bifurcations}: they result from an instability of a specific solution that arises at a particular value of a control parameter. This term is meant to distinguish them from the {\em global bifurcations} that will be studied forthwith.

The importance of the gyre modes was further confirmed through an even more puzzling discovery. Several authors realized, independently of each other, that the low-frequency dynamics of their respective double-gyre models was driven by intense relaxation oscillations of the jet \cite{SGITW1995, Meacham2000, Chang2001, Nadiga2001, Simonnet2003a, Simonnet2003b, Simonnet.ea.2005}. These relaxation oscillations, already described by \citet{Jiang1995} and \citet{Speich1995}, were now attributed to a homoclinic bifurcation, which is no longer due to a linear instability of an existing solution but to a so-called homoclinic reconnection, whose character is global in phase space \cite{Guckenheimer1983, Ghil1987}. In effect, the quasi-geostrophic model reviewed herein undergoes a genuine homoclinic bifurcation that is generic across the full hierarchy of double-gyre models. 

This bifurcation is due to the growth and eventual merging of the two limit cycles, each of which arises from either one of the two mutually symmetric Hopf bifurcations. The corresponding bifurcation is marked in Fig.~\ref{fig:bif_tree} by a filled circle and labeled ``{\bf Homoclinic}''; the reconnecting orbit itself is illustrated in the figure by a stylized lemniscate, and plotted accurately in \citet[Fig.~2]{Simonnet.ea.2005}. This global bifurcation is associated with chaotic behavior of the flow due to the Shilnikov phenomenon \cite{Nadiga2001,Simonnet.ea.2005}, which induces Smale horseshoes in phase space.

The connection between such homoclinic bifurcations and gyre modes was not immediately obvious, but \citet{Simonnet.ea.2005} emphasized that the two were part of a single, global dynamical phenomenon. The homoclinic bifurcation indeed results from the unfolding of the gyre modes' limit cycles. This familiar dynamical scenario is again well illustrated by the unfolding of a codimension-$2$ Bogdanov-Takens bifurcation, where the homoclinic orbits emerge naturally.

Since homoclinic orbits have an infinite period, it was natural to hypothesize that the gyre-mode mechanism, in this broader, global-bifurcation context, gave rise to the observed $7$-yr and $14$-yr North Atlantic oscillations. Although this hypothesis may appear a little farfetched --- given the simplicity of the double-gyre models analyzed so far --- it is reinforced by results with much more detailed models in the hierarchy \cite[e.g.,][]{DG05, DijkstraB2000, Vann.ea.2015}.

The successive-bifurcation theory appears therewith to be fairly complete for barotropic, single-layer models of the double-gyre circulation. This theory also provides a self-consistent, plausible explanation for the climatically important 7-year and 14-year oscillations of the oceanic circulation and the related atmospheric phenomena in and around the North Atlantic basin \cite{DG05,DijkstraB2000,Moron1998,Wunsch1999,Costa2004, Plaut1995, FGS2004, FGS2007, FGR2010, FGR2011, Nile2005, Simonnet2003b, Simonnet.ea.2005}. The dominant 7- and 14-year modes of this theory survive, moreover, perturbation by seasonal-cycle changes in the intensity and meridional position of the westerly winds \cite{Laxmi.ea.2007}.

In baroclinic models, with two or more active layers of different density, baroclinic instabilities \cite{DG05, FGS2007, Gill1982, Ghil1987, Pedlosky1987, Stommel1965, Pedlosky1996, Simonnet2003b, Berloff.ea.2007, Kravtsov.ea.2007} surely play a fundamental role, as they do in the observed dynamics of the oceans. However, it is not known to what extent baroclinic instabilities can destroy gyre-mode dynamics. The difficulty lies in a deeper understanding of the so-called rectification process \cite{Katsman1998}, which arises from the nonzero mean effect of the baroclinic eddying and meandering of the flow on its barotropic component.

Roughly speaking, rectification drives the dynamics farther away from any stationary solutions. In this situation, dynamical systems theory by itself cannot be used as a full explanation of complex, observed behavior resulting from successive bifurcations that are rooted in simple stationary or periodic solutions. 

Other tools from statistical mechanics and nonequilibrium thermodynamics have, therefore, to be considered \cite{Bouchet2002, Farrell1996, chavanis_sommeria_1996, Lucarini2011, Lucarini.ea.2014, MajdaWangBook, Robert1991, Trefethen1993}, and will be discussed in Sects.~\ref{sensitivity} and \ref{critical}. 
Combining these tools with those of the successive-bifurcation approach could lead to a more complete physical characterization of gyre modes in realistic models. Preparing the ground for combining dynamical-systems tools and statistical-physics tools in this way is one of the main purposes of our review paper.

\subsection{Multiple Scales: Stochastic and Memory Effects} 
\label{multiplescales}

In Sect.~\ref{ssec:time_scales}, 
we have illustrated in Fig.~\ref{f:MG1} 
the multiplicity of time scales present in the climate spectrum.  We also pointed out that this multiplicity of scales gives rise to the need for a hierarchy of models that enable the study of separate scales of motion and of the phenomena associated with each, as well as of the interactions between two or more scales, cf. Fig.~\ref{stommel}. 
In this subsection, we discuss several ways in which one can address these issues, using the theory of stochastic processes and taking into account non-markovian effects.

\subsubsection{Fast Scales and Their Deterministic Parametrization}
\label{ssec:fast_stoch}

Let us concentrate, for the sake of definiteness, on variability within a particular range of frequencies $f$ in the climatic power spectrum of Fig.~\ref{f:MG1}a, 
say seasonal to centennial, i.e., $10^{-2}$~yr$^{-1} =f_1 \le f \le f_2 = 10^0$~yr$^{-1}$, and write a model of this variability as 
\begin{equation}\label{eq:model}
\dot \vz = \vH(\vz; \vm).
\end{equation}
We have seen that oscillatory modes of both the THC, cf. Sect.~\ref{ssec:bifurcations} 
, and of the wind-driven circulation, cf. Sec.~\ref{routes}, 
lie in this range. How should one then take into account the slower time scales to the left of this range, $f_0 < f  < f_1$, and the faster ones to the right, $f_2 < f < f_3$, where $0 \le f_0 < 10^{-2}$~yr$^{-1}$ and $1 < f_3 < \infty$? 

A time-honored approach in physical modeling is to describe variability to the left as a prescribed (slow) evolution of parameters, 
\begin{equation}\label{eq:slow-fast}
\vm = \vm(\epsilon t), \quad 0 < \epsilon \ll 1,
\end{equation} 
where  $\epsilon$ is small and $\epsilon t$ is, therefore, a slow time. One might, for instance, consider a particular $\mu = \mu(\epsilon t)$ in Eq.~\eqref{eq:slow-fast} above to represent a slow change in the solar constant or in the height of the topography.

To the right, one might approximate the more rapid fluctuations as being very fast with respect to those of main interest or even infinitely fast. There are two distinct approaches based on these ideas: the first one is purely deterministic, the second one introduces a noise process and thus stochastic considerations.

The standard slow--fast formulation of a system of differential equations --- when assuming a large but finite separation of time scales --- is given, for  $\epsilon \neq 0$, by
\noindent
\begin{subequations}
\label{eq:fast}
\begin{eqnarray}
x' & = & F(x,y; \epsilon), \label{eq:x}\\
y' & = & \epsilon G(x,y; \epsilon). \label{eq:y}
\end{eqnarray}
\end{subequations}
Here $\vz = (x,y)^{\rm T}$, with $x$ the fast and $y$ the slow variable, and $\vH = (F,G)^{\rm T}$, while $(\cdot)^{\rm T}$ designates the transpose and $(\cdot)' = \diff (\cdot)/\diff t$ stands for differentiation with respect to the fast time $t$.

As long as $\epsilon \neq 0$, system \eqref{eq:fast} is equivalent to the so-called slow system
\noindent
\begin{subequations}
\label{eq:slow}
\begin{eqnarray}
\epsilon \dot x & = & F(x,y; \epsilon), \label{eq:x_slow}\\
\dot y & = & G(x,y; \epsilon), \label{eq:y_slow}
\end{eqnarray}
\end{subequations}
in which the dot stands for differentiation $\diff (\cdot)/\diff \tau$ with respect to the slow time $\tau = \epsilon t$.

The classical way of dealing with such problems has been matched asymptotic expansions \cite[e.g.,][]{Grasman.1987, Lagerstrom.1988}. This methodology arose originally from dealing with boundary layers in fluid dynamics, with the inner problem refering to the fast variations in the boundary layer, while the outer problem refers to the more slowly varying free flow outside this layer.

More recently, a point of view inspired by dynamical systems theory \cite{Fenichel.1979} has led to geometric singular perturbation theory \cite[e.g.,][]{Jones.1995}. In this approach, one considers the invariant manifolds that arise in the two complementary limits obtained by letting $\epsilon \to 0$ in the fast and the slow system, respectively.

In the fast system \eqref{eq:fast}, the limit is given by 
\begin{subequations}
\label{eq:faster}
\begin{eqnarray}
x' & = & F(x,y; 0), \\
y' & = & 0,  
\end{eqnarray}
\end{subequations}
while in the slow one, it only makes sense if the right-hand side of \eqref{eq:x_slow} is identically zero; if so, the latter limit is given by
\noindent
\begin{subequations}
\label{eq:slower}
\begin{eqnarray}
0 & = & F(x,y; 0), \label{eq:crit}\\
\dot y & = & G(x,y; 0). 
\end{eqnarray}
\end{subequations}
The algebraic equation $F(x,y; 0) = 0$ defines the critical manifold $\cal{M}_{\rm c}$ on which the solutions of the reduced problem $\dot y  = G(x,y; 0)$ evolve; here $x = X_F(y)$ are the explicit solutions of the implicit equation \eqref{eq:crit}. 


The splitting of the full, slow--fast system given by either Eqs.~\eqref{eq:fast} or \eqref{eq:slow} into the two systems \eqref{eq:faster} and \eqref{eq:slower} has proven very helpful in the study of relaxation oscillations \citep[e.g.,][]{Grasman.1987}. We saw such sawtooth-shaped, slow--fast oscillations arise in either the THC (Sect.~\ref{ssec:THC_GCMs}) 
or the wind-driven circulation of the oceans (Sect.~\ref{sssec:dble_gyre}). 

Another important application of this methodology is in the reduction of large multiscale problems to much smaller ones. In the systems \eqref{eq:fast} and \eqref{eq:slow}, we considered both $x$ and $y$ to be scalar variables. We saw in Fig.~\ref{stommel}, 
though, that the characteristic spatial and temporal scales of atmospheric, oceanic and coupled climate phenomena are highly correlated with each other; large-scale motions tend to be slow and the smaller-scale ones faster. Thus, it is much more judicious to consider $\vz = 
(\vx^{\rm T}, \vy^{\rm T})^{\rm T}$, with $\vx \in \mathbb{R}^m$, $\vy \in \mathbb{R}^n$, and $m \gg n$, i.e., the number of small and fast degrees of freedom much larger than that of the large and slow ones.

This set-up corresponds conceptually to the parametrization problem, which we defined in Sect.~\ref{ssec:time_scales} 
as finding a representation of the unresolved subgridscale processes described by $\vx \in \mathbb{R}^m$ in terms of the resolved, larger-scale ones described by $\vy \in \mathbb{R}^n$. A paradigmatic example is that of parametrizing cloud processes, with spatial scales of 1~km and smaller and with temporal scales of one hour and less, given the large-scale fields characterized by lengths of tens and hundreds of kilometers and by durations of substantial fractions of a day and longer. In this case, the critical manifold appears to be $S$-shaped, as for the periodically forced Van der Pol oscillator \citep[e.g.,][Fig.~2.1]{Guck.ea.2003}, with jumps that occur between the branch on which convection, and hence rain, is prevalent, and the one on which the mean vertical stratification is stable, and thus no rain is possible. Next, we discuss specifically the parametrization of convective processes and of clouds in this perspective.

\subsubsection{An Example: Convective Parametrization}
Clouds have a dramatic role in climate modeling and in determining the climate's sensitivity to natural and anthropogenic perturbations \citep[e.g.,][]{IPCC01, IPCC13}. A substantial literature exists, therefore, on cloud observations, modeling and simulation \citep[e.g.,][and references therein]{Emanuel.1994}. See our earlier discussion in Sects.~\ref{balance} and \ref{ssec:fast_stoch}. 
One of the oldest, best known and most widely used cumulus parametrizations is the \citet{AS.74} (AS) one. Cumulus convection occurs due to moist convective instability, which converts the potential energy of the large-scale mean state into the kinetic energy of the cumulus clouds, A fundamental parameter in this process is the fractional entrainment rate $\lambda$ of a cumulus updraft. In the AS parametrization of moist atmospheric convection, the key idea is that an ensemble of cumulus clouds is in quasi-equilibrium with the large-scale environment. 

The cloud work function $A(\lambda)$ changes in time according to
\begin{equation}\label{eq:work}
\dot A(\lambda) = J \otimes M_{\rm B}(\lambda) + F(\lambda);
\end{equation}
here $M_{\rm B}(\lambda)$ is the nonnegative mass flux through the cloud base, and $J \otimes M_{\rm B}$ is a weighted average over cloud types, with $J$ standing for the weights in the averaging integral, while $F$ is the large-scale forcing.

The dot stands for differentiation with respect to the slow time $\tau = \epsilon t$, as in Eq.~\eqref{eq:x_slow}. The quasi-equilibrium assumption in the AS parametrization corresponds simply to the critical manifold equation Eq.~\eqref{eq:crit} above, which becomes
\begin{equation}\label{eq:QE}
0 = J \otimes M_{\rm B}(\lambda) + F(\lambda).
\end{equation}
In this case, the small parameter that corresponds to the $\epsilon$ of the general slow--fast formulation above is the reciprocal of the adjustment time $\tau_{\rm{adj}}$ of a cloud ensemble to the mean state, $\epsilon \sim 1/\tau_{\mathrm{adj}}$.

\citet{Pan.Randall.1998} proposed an equation that corresponds to the behavior of a cumulus ensemble off the critical manifold given by \eqref{eq:QE}, which they termed a prognostic closure. In their formulation, one computes a cumulus kinetic energy $K$ from
\begin{equation}\label{eq:prog}
\dot K = B + S - D;
\end{equation}
here $B$ is the buoyancy production term, $S$ the shear production term, and $D$ the vertically integrated dissipation rate. The main parameters on which the behavior of the slow--fast system given by Eqs.~(\ref{eq:QE}, \ref{eq:prog}) depends are 
\begin{equation}\label{eq:param}
\alpha = M_{\rm B}^2/K, \quad  \tau_{\rm D} = K/D.
\end{equation}
While \citet{Pan.Randall.1998} do not determine $\alpha$, $\tau_{\rm D}$ or $\tau_{\rm{adj}}$ explicitly, they provide qualitative arguments based on the physics of cumulus convection that make the quasi-equilibrium limit plausible.


\subsubsection{Stochastic Parametrizations}
\label{ssec:stoch_param}

It is of broader interest, though, to consider now the slow--fast formulation of a system of differential equations in the case of infinite separation of time scales, i.e., when the fast motions have infinite frequency or, more precisely, zero decorrelation time. In this case, one has to introduce a white-noise process and the associated stochastic considerations.

The basic idea relies on the \citet{Einstein.1905} explanation of Brownian motion, in which a large particle is immersed in a fluid formed of many small ones. Let the large particle move along a straight line with velocity $u = u(t)$, subject to a random force $\eta(t)$ and to linear friction $- \lambda u$, with coefficient $\lambda$. The equation of motion is 
\begin{equation}\label{eq:Langevin}
{\rm d}u = - \lambda u {\rm d}t + \eta(t). 
\end{equation}
The random force $\eta(t)$ is assumed to be a ``white noise,'' i.e., it has mean zero ${\mathcal E}[\eta(t; \omega)] = 0$ and autocorrelation ${\mathcal E}[\eta(t; \omega)\eta(t+s; \omega)] = \sigma^2 \delta(s)$, where $\delta(s)$ is a Dirac function, $\sigma^2$ is the variance of the white-noise process, $\omega$ labels the realization of the random process, and ${\mathcal E}$ is the expectation operator, which averages over the realizations $\omega$. Alternative notations for the latter are the overbar, in climate sciences, and the angle brackets, in quantum mechanics, ${\mathcal E}[F] := {\bar F} := \langle F \rangle$.

Equation~\eqref{eq:Langevin}, with $\eta = \sigma {\rm d}W$, is a linear stochastic differential equation (SDE) of a form that is now referred to as a Langevin equation, where $W(t)$ is a normalized  Wiener process, also called Brownian motion. It was introduced into climate dynamics by \citet{Hasselmann1976}, who identified slow, ``climate'' changes with the motion of the large particle and fast, ``weather'' fluctuations with the motions of the small fluid particles. He also thought of weather as associated with the atmosphere and climate with the ocean, cryosphere and land vegetation.

Specifically, \citet{Hasselmann1976} assumed that, in a system like \eqref{eq:model}, and without formally introducing the time-scale separation parameter $\epsilon$, one would have $\tau_x \ll \tau_y$, where $\tau_x$ and $\tau_y$ are the characteristic times of the fast $x$- and slow $y$-variables, respectively. From this assumption, and relying also upon the results of \citet{Taylor.1921}, he derived then a linear SDE for the deviations $y_j'$of the slow variables $\vy$ from a reference state $\vy_0$, and the properties of the corresponding covariance matrix and spectral densities. In particular, the red-noise character of the spectrum $S=S(f)$, with $S \sim f^{-2}$ for many oceanic observed time series, gave considerable credence to the thermal-flywheel role that \citet{Hasselmann1976} attributed to the ocean in the climate system.

In the light of recent mathematical results on very large time scale separation in slow--fast deterministic systems, let us consider a relatively simple --- but still sufficiently relevant and instructive --- case, in which the reduction to an SDE can be rigorously derived, cf. \citet{Pavliotis2008} and \citet{Mel.Stuart.2011}. Their deterministic system of ordinary differential equations (ODEs) is a small modification of Eq.~\eqref{eq:slow}, namely
\noindent
\begin{subequations}
\label{eq:skew}
\begin{eqnarray}
\dot x & = &\epsilon^{-1} f_0 (y^{(\epsilon)}) + f_1(x^{(\epsilon)}, y^{(\epsilon)}), 
                  \; x^{(\epsilon)}(0) = x_0, \label{eq:x_skew}\\
\dot y^{(\epsilon)} & = & \epsilon^{-2} g(y^{(\epsilon)}),  
                  \;  y^{(\epsilon)}(0) = y_0,  \label{eq:y_skew}
\end{eqnarray}
\end{subequations}
where $x^{(\epsilon)} \in \mathbb{R}^d$ and $y^{(\epsilon)} \in \mathbb{R}^{d'}$. 

The formal difference with respect to the situation studied in the previous subsection is that $F(x,y; \epsilon)$ of \eqref{eq:x_slow} has been expanded in $\epsilon$ as $F(x,y; \epsilon) = f_0 (y^{(\epsilon)}) + \epsilon f_1(x^{(\epsilon)}, y^{(\epsilon)})$, while $G(x,y; \epsilon)$ of \eqref{eq:y_slow} has been both simplified, in becoming $x$-independent, and ``accelerated,'' to read $G(x,y; \epsilon) = \epsilon^{-2} g(y^{(\epsilon)})$. The basic idea is that the the chaotic and fast $y^{(\epsilon)}$ induces, as $\epsilon \to 0$, a white-noise driving of the slow $x$. Note that one needs $d' \ge 3$ for the autonomous Eq.~\eqref{eq:y_skew} to have chaotic solutions.

\citet{Mel.Stuart.2011} assume merely that the fast equation \eqref{eq:y_skew} has a compact attractor $\Lambda \in \mathbb{R}^{d'}$, which supports an invariant measure $\mu$, and that $\mathcal{E}_\mu f_0(x) = 0$, along with certain boundedness and regularity conditions on $f(x,y)$. They then show rigorously that $x^{(\epsilon)}(t) \rightarrow_p X(t)$ as $\epsilon \to 0$, where the convergence is with respect to the appropriate probability measure, and $X(t)$ is the solution of the SDE 
\begin{equation}\label{eq:Ito}
X(t) = x_0 + \int_0^t {\overline F}(X(s))\diff s + \sigma W(t).
\end{equation}
Here $W$ is the Brownian motion with variance $\sigma ^2$, such that the white noise in Eq.~\eqref{eq:Langevin} be given by $\eta = \sigma \diff W$. Moreover, ${\overline F}(X) = \mathcal{E}_\mu F(x, \cdot)$ with respect to $\mu$.

Fundamental mathematical issues associated with the above diffusive limit of slow--fast systems were explored by \citet{Papa.K.1974} and early results in the physical literature include \citet{Beck.1990} and \citet{Kantz.ea.2001}. Many aspects of the applications to climate modeling are covered in \citet{palmer_stochastic_2009}. More specifically, \citet{Berner.ea.2017} and \citet{Franzke.ea.2015} discuss issues of stochastic parametrization of subgrid-scale processes.

To conclude this subsection, it is of interest to consider, in a broader perspective, the potential for a unified theory of nonautonomous dynamical systems, in which the fast processes may be modeled as either deterministic or stochastic. The theory of purely deterministic, skew-product flows was laid on a solid basis by \citet[and references therein]{Sell.1971} and, more recently, by \citet{KR11}. Random dynamical systems are extensively covered by \citet{Arnold.1998}, with many recent results in an active field.

\citet{Berger.Sieg.2003} point out that ``Quite often, results for random dynamical systems and continuous skew product flows are structurally similar,'' and thus open the way to a unified theory. They outline both commonalities and distinctions between the two broad classes of nonautonomous dynamical systems, in order to shed further light on existing results, as well as stimulate the development of common concepts and methods. 

\citet{Caraballo.Han.2017} provide a solid and accessible introduction to random dynamical systems, as well as to deterministically nonautonomous ones, along with several interesting applications. They also consider the two distinct types of formulation of the deterministic ones, the so-called process formulation and the skew-product flow one. System \eqref{eq:skew} above, for instance, is a particular case of a master--slave system 
\begin{equation}\label{eq:master}
\dot x = f(x,y), \quad \dot y = g(y), \quad \mathrm{with} \; x \in \mathbb{R}^{d}, \; y \in \mathbb{R}^{d'}
\end{equation}
that induces a skew-product flow, where $y(t)$ is the driving force for $x(t)$.

These developments bear following, since the climate sciences offer a rich source of relevant problems and could thus lead to novel and powerful applications of the unified theory. \citet{Caraballo.Han.2017} already studied the \citet{L84} model, in which seasonal forcing acting on deterministic subseasonal variability can induce interannual variability. Other applications will be discussed in the next sections, especially in Sect. \ref{ssec:CR}. 

\subsubsection{Modeling Memory Effects}
\label{ssec:memory}
While the study of differential equations goes back to Isaac Newton, the interest for including explicitly delays into evolution equations that govern physical and biological processes is relatively recent. A mid-20th--century reference is \citet{Bellman.Cooke.1963}, followed by \citet{Driver.1977} and \citet{Hale.1977}. Delay-differential equations (DDEs) were introduced into the climate sciences by \citet{Bhat.ea.1982} and have been used widely in studying ENSO \citep[e.g.,][and references therein]{Tziperman1994b, Ghil.ea.2015}.

As we shall see, memory effects can play a key role when there is little separation between scales, in contrast to the assumptions of \citet{Hasselmann1976} and of other authors mentioned in the two immediately preceding subsections. Moreover, when properly incorporated in the mathematical formulation of the climate problem at hand, relying on such effects can lead to highly efficient and accurate model reduction methods.

\paragraph{The Mori-Zwanzig formalism.} 
In statistical physics, the \citet{mori_transport_1965}-\citet{zwanzig_memory_1961} (MZ) formalism arose from describing the interaction of a Hamiltonian many-particle system with a heat bath. Today, though, it is being used in a large number of applications that include dissipative systems. 

The fundamental idea is illustrated by the following very simple, linear system of two ODEs \cite{E.Lu.2011}:
\noindent
\begin{subequations}
\label{eq:MZ}
\begin{eqnarray}
\dot x & = & a_{11}x + a_{12}y, \label{eq:x_MZ}\\
\dot y & = & a_{21}x + a_{22}y.  \label{eq:y_MZ}
\end{eqnarray}
\end{subequations}
The sole assumption is that we are interested in the details of the behavior of $x(t)$ but only in the statistics of $y$; naturally, one thinks of $y$ as fluctuating faster than $x$ but this is not actually required for the formalism outlined below to work. One considers $x$ as a (slowly varying) parameter in solving Eq.~\eqref{eq:y_MZ} for $y$ by using the variation-of-constants formula
$$y(t) = e^{a_{22}t} y(0)+\int_0^t e^{a_{22}(t - s)} a_{21} x(s)\diff s,$$
and plugs this result back into Eq.~\eqref{eq:x_MZ}, to yield
\begin{equation}\label{eq:GLE_1}
\dot x = a_{11}x + \int_0^t K(t-s)x(s) \diff s + f(t).
\end{equation}

Equation~\eqref{eq:GLE_1} is a generalized Langevin equation (GLE), in which $K(t) = a_{12} \exp({a_{22}t})a_{21}$ is the {\it memory kernel} and $f(t) = a_{12}\exp(a_{22}t)y(0)$ is the {\it noise term}, since one thinks of $y(0)$ as randomly drawn from the rapidly fluctuating $y(t)$. The essential difference with respect to Eq.~\eqref{eq:Langevin} is the convolution integral in \eqref{eq:GLE_1}, which expresses the delayed action of the slow variable $x$ on the fast variable $y$.

The MZ formalism consists --- in a general, nonlinear set of Markovian evolution equations with a large or even infinite number of degrees of freedom --- in selecting the variables one is interested in via a projection operator, and deriving the generalized form of the GLE above. Examples of Markovian evolution equations are systems of ODEs or PDEs for which an instantaneous initial state carries all the information from the past.

In this general case, the memory term involves repeated convolutions between decaying memory kernels and the resolved modes, and the GLE is therewith a non-Markovian,  stochastic integro-differential system that is very difficult to solve without further simplifications. Among the latter, the short-term memory approximation \cite[e.g.,][]{Chorin.Stinis.2007} posits rapidly decaying memory and is equivalent to assuming a relatively large separation of scales, as in Sects.~\ref{ssec:fast_stoch} and \ref{ssec:stoch_param} above. 

Fortunately, \citet{MSM2015} have shown that there is a way to approximate the GLE in a very broad setting, efficiently and accurately, by a set of Markovian SDEs without the need for pronounced scale separation and in the presence of so-called intermediate-range memory. This way relies on a methodology that was developed at first quite independently of the MZ formalism, namely empirical model reduction \cite[EMR:][]{Kondrashov.Kravtsov.ea.2005, KravtsovKondrashovGhil_JCL05, KravtsovGhilKondrashov_09}. 

\paragraph{Empirical model reduction (EMR) methodology.} 
The purpose of EMR development was deriving relatively simple nonlinear, stochastic-dynamic models from time series of observations or of long simulations with high-end models, such as GCMs. An EMR model can be compactly written as
\begin{equation}\label{eq:EMR}
\dot \x = - \A \x + \B(\x, \x) + \vL(\x, \Vr_t^l, \xi_t, t), \; 0 \le l \le \vL - 1.
\end{equation}

Here $\x$ typically represents the resolved and most energetic modes. Most often, these are chosen by first selecting a suitable basis of empirical orthogonal functions \citep[EOFs:][]{Preisendorfer1988} or other data-adaptive basis \citep[e.g.,][and references therein]{KravtsovGhilKondrashov_09}, and retaining a set of principal components that capture a satisfactory fraction of the total variance in the data set.

The terms $- \A \x$ and $\B(\x, \x)$ represent, respectively, the linear dissipation and the quadratic self-interactions of these modes, while $\vL$ is a time-dependent operator that is bilinear in the resolved variables $\x$ and the unresolved ones $\Vr_t^l$. These interactions take a prescribed form in the EMR formulation of Eq.~\eqref{eq:EMR}, and arise by
integrating recursively---from the lowest level $L$ to the top level, $l = 0$---the ``matrioshka'' of linear SDEs 
$$ \diff \Vr_t^l = \vM_l (\x, \Vr_t^0, \ldots,  \Vr_t^l) \diff +  \Vr_t^{l+1} \diff t.$$
At each level $l$, the coupling between the variable  $\Vr_t^l$ and the previous-level variables $(\x, \Vr_t^0, \ldots,  \Vr_t^{l-1})$ is modeled by $L - 1$ rectangular matrices $\vM_l$ of increasing order.

In practice, the matrices $\A, \vM_l$, and the quadratic terms $\B$ are estimated by a recursive least-square procedure, which is stopped when the $L$th-level residual noise $\Vr^{L-1} = \xi_t$ has a lag$-1$ vanishing autocorrelation. The stochastic residuals $\Vr_t^l$, obtained in this recursive minimization procedure, are ordered in decreasing order of decorrelation time, from $\Vr_t^0$ to $\xi_t$. 

Note that the integral terms arising in the $\vL$ operator are convolution integrals between the macro-state variables $\x$ and memory kernels that decay according to the dissipative properties of the matrices $\vM_l$. These decay times are not necessarily short and one can thus treat the case of intermediate-range memory, in the MZ terminology. Furthermore, cf.~\citet{Kondrashov.Kravtsov.ea.2005}, note that any external forcing, such as the seasonal cycle, can be typically introduced as a time dependence in the linear part $\A$ of the main level of Eq.~\eqref{eq:EMR} .

\paragraph{Role of memory effects in EMR.}
We propose here a simple analytic example that should help understand the general description of EMR in the previous paragraphs, as well as the connection to the MZ formalism. The model is given by
\noindent
\begin{subequations}
\label{eq:MSM}
\begin{eqnarray}
\diff x & = & (f(x) + r) \diff t, \label{eq:x_MSM}\\
\diff r & = & (\gamma x - \alpha r) \diff t + \diff W_t,  \label{eq:y_MSM}
\end{eqnarray}
\end{subequations}
where $f$ is a nonlinear function, $W_t = W_t(\omega)$ is a standard Wiener process, as in Eq.~\eqref{eq:Langevin}, $\alpha > 0$ and $\gamma$ is real. We are interested in $x$, which is slow, and want to parameterize $r$, which is fast. 

Proceeding for \eqref{eq:MSM} as we did for the system \eqref{eq:MZ}, but now for a fixed realization $\omega \in \Omega$, we get
$$r(t; \omega) = e^{- \alpha t} r_0 + \gamma \int_0^t e^{- \alpha (t - s)} x(s)\diff s + W_t(\omega),$$
where $r(t_0; \omega) = r_0$. Substituting this back into \eqref{eq:x_MSM} yields the following randomly forced integro-differential
equation,
\begin{equation}\label{eq:RDDE}
\dot x = f(x) + e^{- \alpha t} r_0 + \gamma \int_0^t e^{- \alpha (t - s)} x(s)\diff s + W_t(\omega),
\end{equation}
which is the analog of Eq.~\eqref{eq:GLE_1} in the stochastic-dynamic context of \eqref{eq:MSM}.

\citet{MSM2015} proved rigorously that the EMR \eqref{eq:EMR} is equivalent to a suitable generalization of the GLE \eqref{eq:RDDE}. \citet{Pavliotis2014} defines as \textit{quasi-Markovian} a stochastic process where the noise in the GLE can be described by adding a finite number of auxiliary variables, as in the case of the EMR approach. Thus, the EMR methodology can be seen as an efficient implementation of the MZ formalism, i.e., an efficient solution of the associated GLE, even in the absence of large-scale separation. This result explains the remarkable success of EMR in producing reduced models that capture the multimodality as well as the nontrivial power spectrum of phenomena merely known from time series of observations or of high-end model simulations. 

In the remainder of this subsection, we give two examples of this success and further references to many more. Alternative approaches to efficient solutions of the GLE can be found, for instance, in \citet{chorin_optimal_2002}. 

\paragraph{EMR applications.} We choose here an EMR model to simulate Northern Hemisphere mid-latitude flow \citep{KravtsovKondrashovGhil_JCL05} and a real-time ENSO prediction model \citep{Kondrashov.Kravtsov.ea.2005}. Further examples of successful application of the methodology appear in \citet{KravtsovGhilKondrashov_09, kravtsov2011empirical} and elsewhere.

\citet{KravtsovKondrashovGhil_JCL05} introduced the EMR methodology and illustrated it at first with quadratically nonlinear models of the general form given here in Eq.~\eqref{eq:EMR}. The applications were to the \citet{Lorenz1963a} convection model, the classical double-well potential in one space dimension, and a triple-well potential in two dimensions with an exponential shape for the wells. More challenging was a real-data application to geopotential height data for 44 boreal winters (1 December 1949--31 March 1993). The data set consisted of $44 \times 90 = 3960$ daily maps of winter data, defined as 90-day sequences starting on 1 December of each year. The best EMR fit for the data required the use of nine principal components and of $L = 3$ levels.

\begin{figure}
	\centering
    \includegraphics[width=0.9\columnwidth, scale=1]{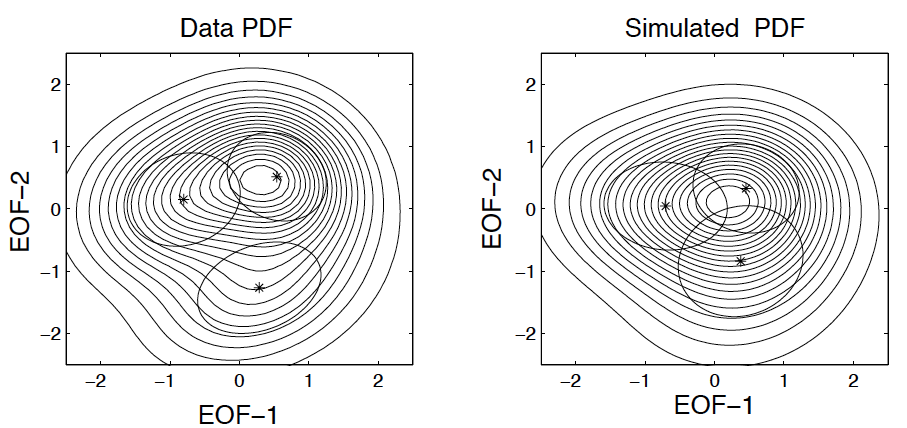}
    \caption{Multimodal probability density function (PDF) for the Northern Hemisphere's 
    geopotential height anomalies of 44 boreal winters; see \citet{Smyth1999} for 
    details of the data set and of the mixture model methodology for computing the PDFs. 
    (a) PDF of the observed height anomalies; and 
    (b) of the anomalies given by the EMR model.
    Modified from \citet{KravtsovKondrashovGhil_JCL05}, 
    with the permission of the American Meteorological Society.}
    \label{fig:NH_PDFs}
\end{figure}

The probability density functions (PDFs) for the observed and the EMR model--generated data sets are plotted in Fig.~\ref{fig:NH_PDFs}. The EMR clearly captures quite well the three modes obtained with a Gaussian mixture model, cf.~\citet{Smyth1999} and \citet{Ghil.Rob.2002}. These three modes correspond to three clusters found by  by \citet{Cheng1993} using very different methods on a somewhat different data set. The maps of the corresponding centroids appear as Fig.~1 in \citet{Ghil.Rob.2002} and are discussed therein; they agree quite well with those of \citet{Cheng1993}, cf.~\citet[Fig.~9]{Smyth1999}.

\citet{Kondrashov.Kravtsov.ea.2005} fitted the global SST field between $30^{\circ}$S--$60^{\circ}$N over the time interval January 1950--September 2003 by using linear and quadratic EMR models with one and two noise levels, $L = 1, 2$, based on monthly SST anomaly maps and allowing a seasonal dependence of the dissipative terms in Eq.~\eqref{eq:EMR}. Their results when using $L = 2$ were much better, for either a linear or a quadratic model, which clearly shows the role of memory effects in EMR modeling and the connection with the MZ formalism that was explained above.

\begin{figure}[ht]
	\centering
    \includegraphics[width=0.95\columnwidth, scale=1]{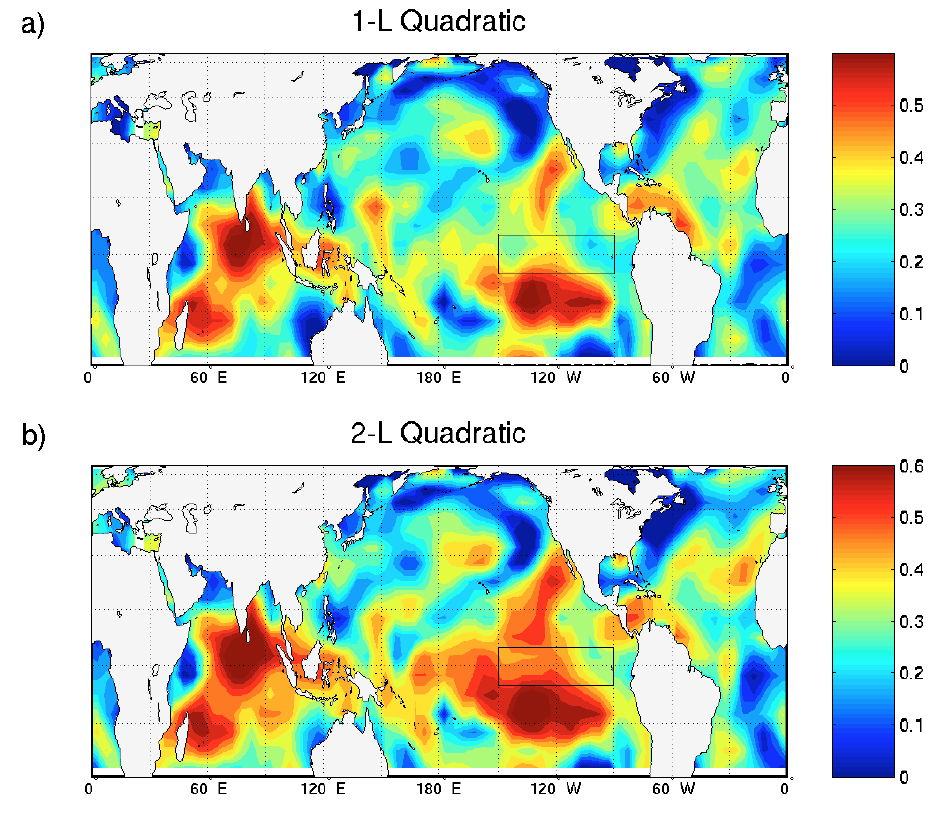}
    \caption{Validation of forecast skill of the EMR-based ENSO model of \citet{Kondrashov.Kravtsov.ea.2005} via
    anomaly correlation maps of SSTs at 9-month lead time. (a) Using a quadratic EMR model with $L = 1$; and 
    (b) a quadratic model with $L = 2$.
    Modified from \citet{Kondrashov.Kravtsov.ea.2005},
    with the permission of the American Meteorological Society.}
    \label{fig:ENSO_EMR}
\end{figure}

The use of the EMR models in prediction was tested by so-called hindcasting or retrospective forecasting, i.e., a protocol --- also called ``no look-ahead'' --- in which the data available past a certain time instant are eliminated when constructing the model to be used in the forecast to be evaluated. The results of these tests are plotted in Fig.~\ref{fig:ENSO_EMR}, for $L = 2$ and a linear vs. a quadratic model. The light-black rectangle in the Eastern Tropical Pacific corresponds to the region $(5^{\circ}$S--$5^{\circ}$N, $150^{\circ}$--$90^{\circ}$W) over which SST anomalies are averaged to obtain the Ni\~no-3 index, already mentioned in Sect.~\ref{coupledmode}.

Climate forecast skill is measured mainly via root-mean-square errors and anomaly correlations. The former skill scores are given in \citet[Fig.~2d]{Kondrashov.Kravtsov.ea.2005} and clearly indicate the superiority of the quadratic model. The latter apeear in Fig.~\ref{fig:ENSO_EMR} here, too. 

Anomaly correlations of roughly 0.5--0.6 (red in the figure) are considered quite useful in climate prediction, and the red area is substantially larger in panel (b) here, covering most of the Tropical Pacific and Indian Oceans. Concerning actual real-time forecasts, \citet{iri12} have found that, over the 2002--2011 interval of their evaluation, the EMR-based forecasts of UCLA's Theoretical Climate Dynamics (TCD) group were at the top of the eight statistical models being evaluated and exceeded in skill by only a few of the 12 dynamical, high-end models in the group that participated in the IRI ENSO Forecast plume \url{https://iri.columbia.edu/our-expertise/climate/forecasts/enso/current/?enso_tab=enso-sst_table}.

\paragraph{Explicit derivation of the parametrized equations.}
The EMR methodology is data-driven, so that it allows to construct bottom-up an effective dynamics able to account for the observed data. \citet{wouters_disentangling_2012,wouters_multi-level_2013}, showed that the MZ formalism can be used also to derive the parametrizations in a top-down manner in a rather general way. Assume the system of interest is described by the following evolution equations:
\begin{subequations}
\label{eq:sWL}
\begin{eqnarray}
 \dot \vx & = & f_\vx(\vx)+\epsilon\Psi_\vx(\vx,\vy), \label{eq:sWLx} \\
\dot \vy & = & f_\vy(\vy)+\epsilon\Psi_\vy(\vx,\vy); \label{eq:sWLy}
\end{eqnarray}
\end{subequations}
here, again,  $\vx \in \mathbb{R}^m$, $\vy \in \mathbb{R}^n$, $\vx$ is the set of large-scale, energetic variables of interest and, typically, $m \gg n$.

We assume, furthermore, that $\epsilon$ is a small parameter describing the strength of the coupling between the two sets of variables, and that, if $\epsilon=0$, the dynamics is chaotic for both the $\vx$ and $\vy$ variables. By expanding the MZ projection operator, it is possible to derive the following expression for the projected dynamics on the $\vx$ variables, which is valid up to order $O(\epsilon^3)$:
\begin{equation}
\dot \vx  =  f_\vx(\vx)+\epsilon M(\vx)+ \epsilon S(\vx)+\epsilon^2\int K(\vx,t-s)\diff s.  \label{eq:sWLsolution} 
\end{equation}
This equation provides the explicit expression of the mean-field, deterministic term $M$; the time-correlation properties of the stochastic term $S$ that is, in general, multiplicative; and of the integration kernel $K$, which defines the non-Markovian contribution. 

\begin{figure}[ht]
	\centering
    \includegraphics[clip=true,trim=3cm 3cm 3cm 5cm, angle=270,width=0.95\columnwidth]{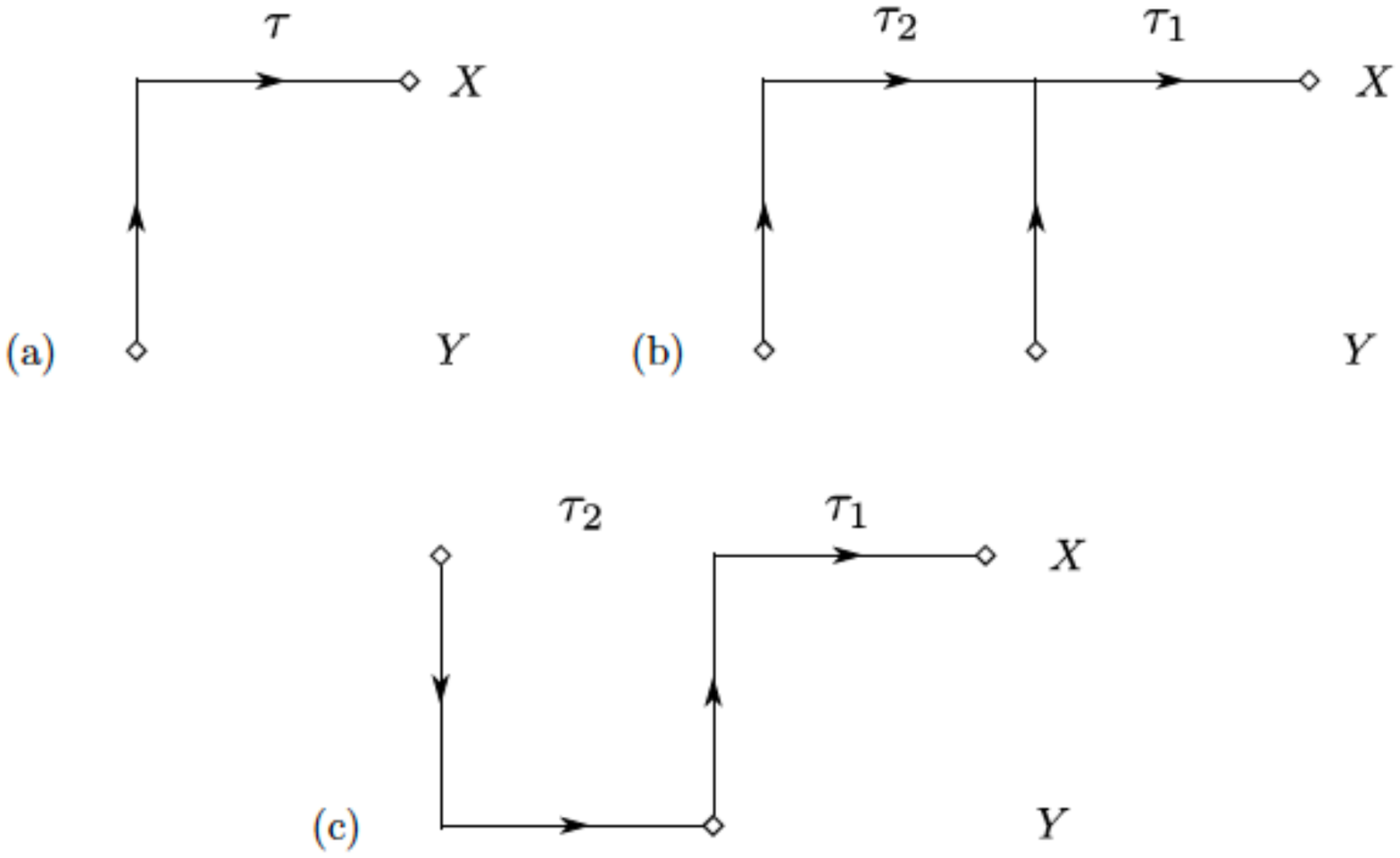}
    \caption{Schematic representation of the three terms contributing to the parametrization 
    of the fast variables $\vx$ in Eq.~\eqref{eq:sWLsolution}.
    (a) Mean-field term $M$; (b) stochastic term $S$; and (c) memory term with kernel $K$. 
    {The symbols $(\tau, \tau_1, \tau_2)$ denote the delays involved in these effects.}
    Reproduced with permission from \citet{wouters_multi-level_2013}.}
    \label{fig:WL}
\end{figure}

Figure~\ref{fig:WL} provides a schematic diagram of the three terms of the parametrization: the $M$-term comes from time averaging of the effects of the $\vy$ variables on the $\vx$ variables; the $S$-term results from the fluctuations of the forcing of the $\vy$ variables on the $\vx$ variables; and the non-Markovian contribution represents the self-interaction of the $\vx$ variables on themselves at a later time, mediated by the $\vy$ variables.

These terms are derived using the statistical properties of the uncoupled dynamics of the $\vy$ variables, at $\epsilon=0$. In the limit of infinite time scale separation between the  $\vx$ and $\vy$ variables, the non-Markovian term drops out and the stochastic term becomes a white-noise contribution, where one needs to use the Stratonovich definition of the stochastic integral \cite{Pavliotis2008}. If, instead, $\Psi_\vy=0$ in the master--slave system \eqref{eq:sWL}, the non-Markovian term is identically zero, as expected.

A surprising property of the surrogate dynamics given by Eq.~\eqref{eq:sWLsolution} is that the expectation value of any observable $\phi(\vx)$ is the same as for the full dynamics governed by Eq.~\eqref{eq:sWL}, up to third order in $\epsilon$ \cite{wouters_multi-level_2013}. This property is due to the fact that one can derive Eq.~\eqref{eq:sWLsolution} by treating the weak coupling using Ruelle response theory \cite{Ruelle:1998,ruelle_smooth_1999,ruelle2009}; see Sect.~\ref{ruellert} below. 
As a result, the average model error due to the use of parametrized dynamics is well under control; see discussion in \citet{Hu2019}. Moreover, \citet{Vissio2018, Vissio2018a} showed by explicit examples that this top-down approach can, in some cases, help derive scale-adaptive parametrizations, while \citet{Demaeyer2017} demonstrated its effectiveness in an intermediate complexity climate model. 

%
%



\section{Climate Sensitivity and Response}
\label{sensitivity}

A central goal of the climate sciences is to predict the impact of changes 
in the system's internal or external parameters --- such as the greenhouse gas (GHG) concentration or the solar constant --- on its statistical properties. A key concept in doing so is \textit{climate sensitivity}, which aims to measure the response of the climate system to external perturbations of Earth's radiative balance. As we shall see below, this measure is being used for projecting, for instance, mean temperature changes over the coming century as a response to increasing concentrations of atmospheric GHGs. While a good start, accurate and flexible predictions of climate changes require, though, more sophisticated concepts and methods.

\subsection{A Simple Framework for Climate Sensitivity}
\label{ssec:ECS}
In order to illustrate the main ideas, let us consider the simple energy balance model (EBM) introduced in Eqs.~\eqref{eq:rad_bal_0}  of Sect. \ref{ssec:radiation} 
where the net radiation $R = R_{\rm i} - R_{\rm o}$ at the top of the atmosphere is related to the corresponding average temperature $T$ near the Earth's surface by $R=R(T)$. This simple, 0-D EBM includes both longwave and shortwave processes, so that $c{\diff T}/{\diff t} = R(T)$, as in Eq.~\ref{0DEBM}. 

Following \citet{Peixoto1992} and \citet{ghil2010}, we assume, furthermore, that there are $N$ climatic variables $\{\alpha_k=\alpha_k(T)$, $k=1,\ldots,N\}$ that are, to a first approximation, directly affected by the temperature change only and which can, in turn, affect the radiative balance. Hence, one can write $R=R\left(T,\alpha_1(T),\ldots,\alpha_N(T)\right)$. Let us assume, furthermore, that, for a certain reference temperature $T=T_0$ one has $R(T_0)=0$, which corresponds to steady-state conditions. 

The simplest framework for climate sensitivity is to think of the difference in global annual mean surface air temperature $\Delta T$ between two statistical steady states, which have distinct $CO_2$ concentration levels. We then assume that changing the CO$_2$ concentration corresponds to applying an extra net radiative forcing $\Delta \tilde R$ to the system, and look for the corresponding change $\Delta T$ in the average temperature, so that  $R(T_0+\Delta T)+\Delta \tilde R=0$. 

For small $\Delta T$ and smooth $R=R(T)$, the Taylor expansion yields 
\begin{eqnarray}
\Delta \tilde R & = - {\displaystyle \frac{\diff R}{\diff T}}\bigg|_{T=T_0} \Delta T+\mathcal{O}\left((\Delta\,T)^2\right) \nonumber\\
                       & = {\displaystyle \frac{\partial R}{\partial T}}\bigg|_{T=T_0} \Delta\,T - \sum_{k=1}^N {\displaystyle\frac{\partial R}{\partial \alpha_k}}{\displaystyle \frac{\partial \alpha_k}{\partial T} \bigg|_{T=T_0}\Delta T} \nonumber\\
			&+\mathcal{O}\left((\Delta\,T)^2\right).
\label{Taylor}
\end{eqnarray}
Here, $\mathcal{O}(x)$ is a function such that $\mathcal{O}(x)\le C\,x$ as 
soon as $0<x<\epsilon$ for some positive constants $C$ and $\epsilon$. While the higher-order terms in $\Delta T$ are usually small, they can become important where the smooth dependence of $R$ on $T$ breaks down. Specifically, rapid climate change may ensue when the system crosses a tipping point, 
as will be explained further below.

Introducing the notations
\begin{subequations}
\label{eq:ECS}
\begin{eqnarray}
\frac{1}{\lambda_0(T_0)} = & {\displaystyle -\frac{\partial R}{\partial T}}\bigg|_{T=T_0}, \label{eq:ref_sens}\\
f_k(T_0) = & -\lambda_0(T_0) {\displaystyle \frac{\partial R}{\partial \alpha_k}\frac{\partial \alpha_k}{\partial T}} \bigg|_{T=T_0}, \label{eq:feedback}
\end{eqnarray}
\end{subequations}
for the ``reference sensitivity'' $\lambda_0$ and the ``feedback factors'' $f_k$ at the reference state $T=T_0$, we obtain
\begin{equation}
\Delta \tilde R=\frac{1-\sum_{k=1}^Nf_k(T_0)}{\lambda_0(T_0)}\,\Delta\,T+\mathcal{O}\left((\Delta\,T)^2\right),
\label{eq0}
\end{equation}
which readily leads to
\begin{equation}
\Delta T=\Lambda(T_0)  \Delta \tilde R 
+\mathcal{O}\left((\Delta\,T)^2\right).
\label{eq}
\end{equation}
Here  
\begin{equation}\label{eq:gain}
\Lambda(T_0) = - \frac{\diff R}{\diff T}\bigg|_{T=T_0} = \frac{\lambda_0((T_0))}{1-\sum_{k=1}^N f_k(T_0)}
\end{equation}
is the linear gain factor, which can be defined as long as $\sum_{k=1}^Nf_k((T_0))\ne 1$; note that, if the sum of the feedback factors exceeds unity, i.e. $\sum_{k=1}^Nf_k((T_0))> 1$, the system is unstable. 

Referring again to Eqs.~\eqref{eq:rad_bal_0}, the feedback associated with the dependence of $\alpha$ on the temperature in Eq.~\eqref{eq:R_i} 
is usually taken to be the ice-albedo feedback, cf. Sect.~\ref{ssec:radiation}, while the dependence of the emissivity $m$ on the temperature in Eq.~\eqref{eq:R_o} 
is associated with the changes in the atmospheric opacity. For the latter one, the standard, reference sensitivity, associated with deviations from Planck's law for black-body radiation, is $\lambda_0(T_0)=-\partial R/\partial T|_{T=T_0} = 4\sigma m(T_0) T_0^3$. 

More specifically, feedbacks that can contribute to changes in reflectivity in Eq.~\eqref{eq:R_i} 
include, aside from the incremental presence of snow and ice, also the climate-vegetation feedback \cite[e.g.,][and references therein]{watson_biological_1983, zeng_role_2000, Rombouts.2015}. The feedbacks that can affect the sensitivity of emitted radiation in Eq.~\eqref{eq:R_o} 
include atmospheric alteration in water vapor content and cloud cover, as well as in GHGs and aerosol concentration. 

In climate studies, different measures of climate sensitivity are used. The so-called equilibrium climate sensitivity (ECS) denotes the globally and annually averaged surface air temperature increase that would result from sustained doubling of the concentration of carbon dioxide in Earth's atmosphere vs. that of the reference state, after the climate system had reached a new steady-state equilibrium \cite{Charney.ea.1979}. The ECS was used extensively by the IPCC's first three assessment reports, up to \citet{ipcc2001}. 

\begin{figure}
\includegraphics[width = .9\columnwidth]{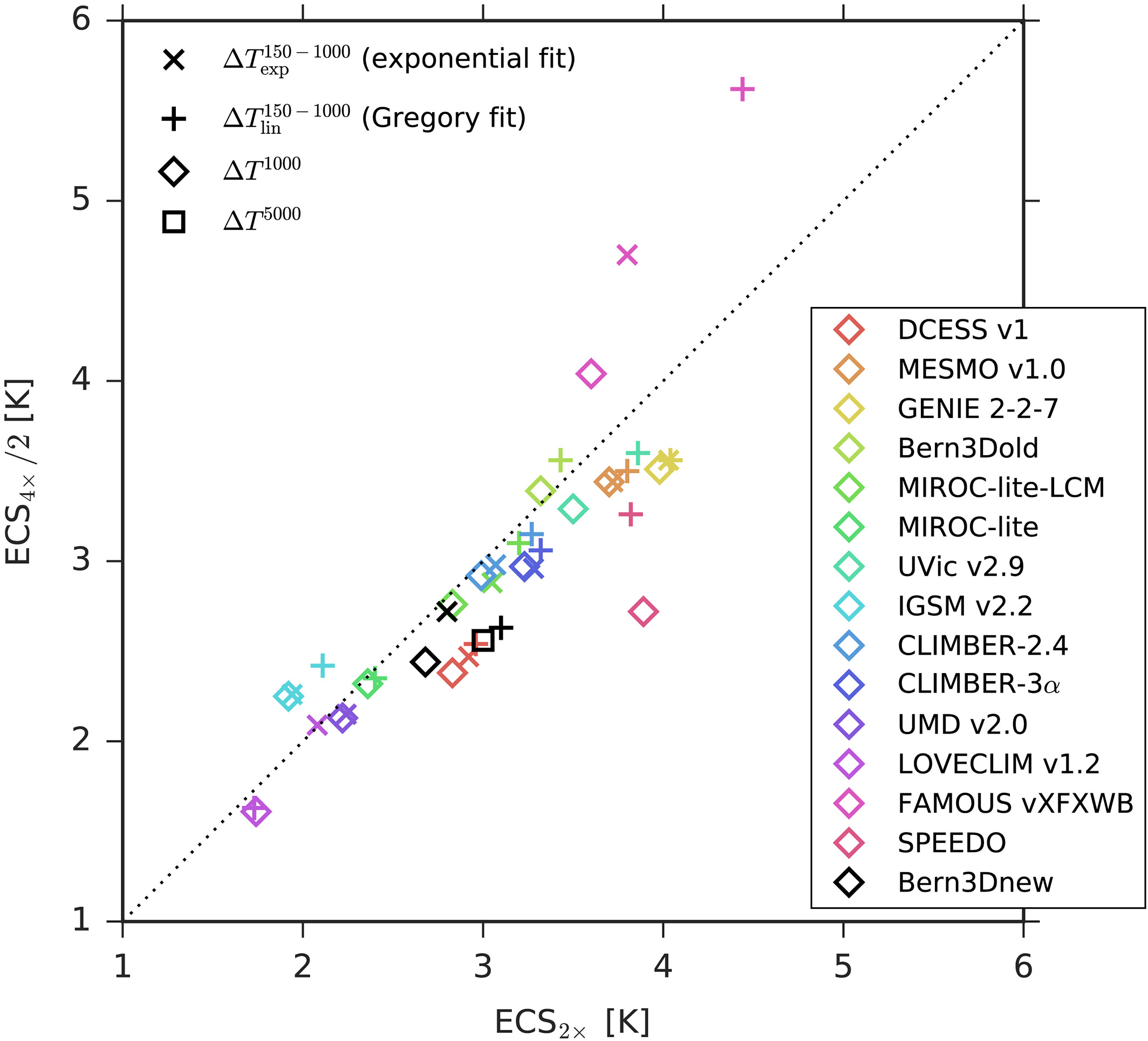}
\caption{Test of the linear scaling of the long-term climate response with respect to the CO$_2$ concentration increase in 15 GCMs. On the abscissa: standard $ECS$, as in Eq.~\eqref{eqECS}; on the ordinate: long-term response of the globally averaged surface air temperature to quadrupling of the CO$_2$ concentration. Reproduced with permission from \citet{Pfister17}.}
\label{Linearity}
\end{figure}

Taking the linear approximation in Eqs.~\eqref{Taylor}--\eqref{eq} above, one has:
\begin{eqnarray}
ECS & = \Lambda(T_0)\Delta \tilde R_{2\times {\rm{CO}_2}} \nonumber\\
& = {\displaystyle \frac{\lambda_0((T_0))}{1-\sum_{k=1}^Nf_k(T_0)}}
 \Delta \tilde R_{2\times {\rm CO}_2}.
\label{eqECS}
\end{eqnarray}
Note that  the radiative forcing is, to a good approximation, proportional to the logarithm of the CO$_2$ concentration. Hence, in the linear-response regime, and for a given reference state $T_0$, the long-term globally averaged surface air temperature change resulting from a quadrupling of the CO$_2$ concentration is twice as large as the ECS; see Fig.~\ref{Linearity}.

\begin{figure}
\includegraphics[trim=2cm 0cm 2cm 10cm, clip=true, width = .9\columnwidth]{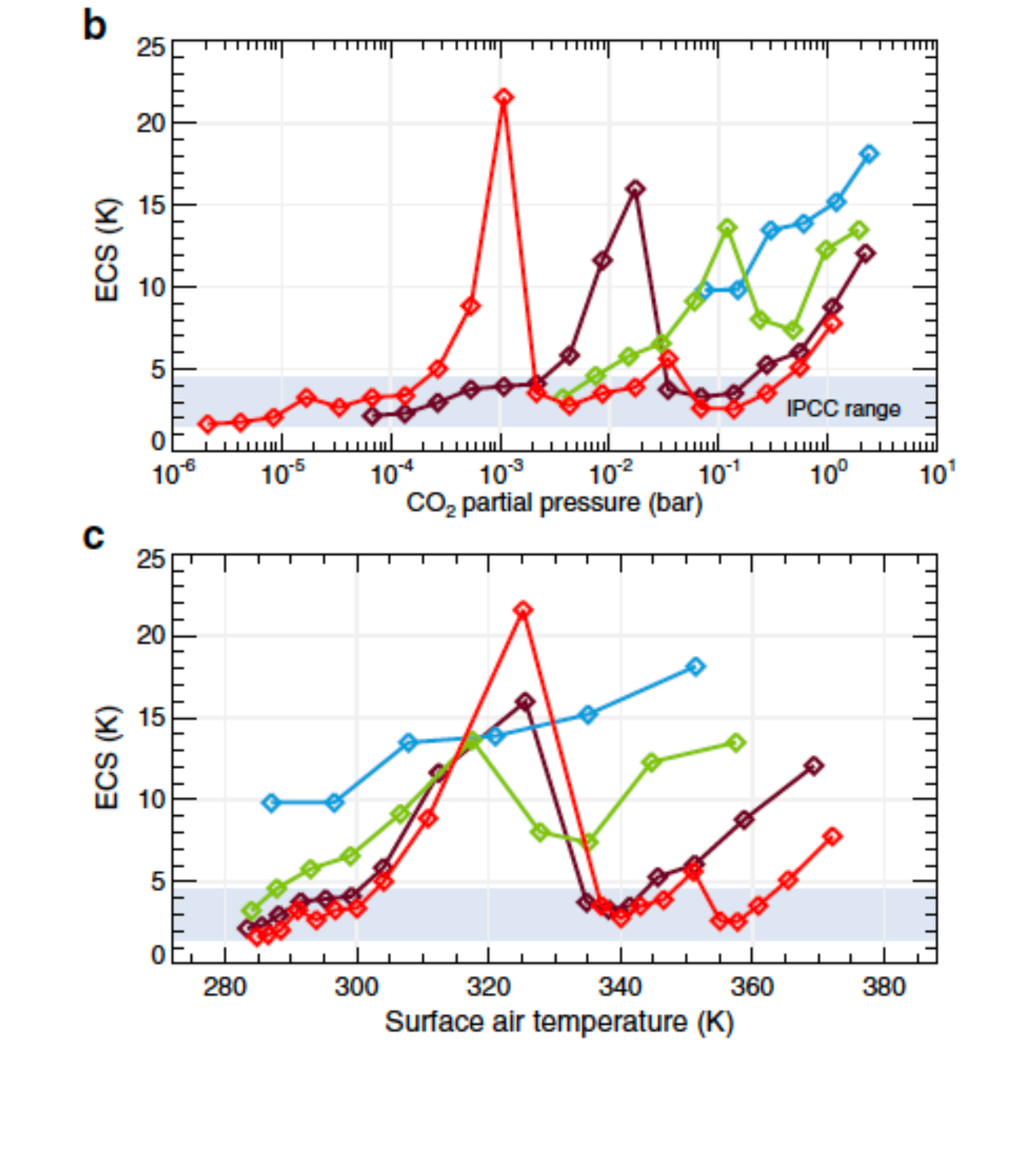}
\caption{Estimates of the state dependence of the ECS using the Community Earth System Model (CESM). 
    The four curves correspond to CESM simulations with four multiples of the solar constant $Q_0$: in the notation 
    of Eqs.~\eqref{eq:rad_bal_0} 
    here, $\mu = 0.75, 0.875, 1.0$ and $1.1$ 
    for the blue, green, brown and red curves. (b) ECS as a function of CO$_2$ partial pressure; 
    and (c) ECS as a function of global mean surface air temperature.
    The temperatures shown in (c) are an average between the base and doubled-CO$_2$ state. 
    The shaded region in (b) and (c) indicates the IPCC estimated range for ECS. 
    Reproduced with permission from Figs.~1(b,c) of \citet{Wolf2018}; see also \citet{Gomez_2018}.}
\label{statedependence1}
\end{figure}

The concept of climate sensitivity can be generalized to describe the linear dependence of the long-term average of any climatic observable with respect to the radiative forcing due to changes in CO$_2$ or in other GHGs, as well as to changes in solar radiation, aerosol concentration or any other sudden changes in the forcing \cite{Ghil1976, Ghil2015, vonderHeydt2016, Lucarini2010b}.

\subsection{Climate Sensitivity: Uncertainties and Ambiguities}
\label{ssec:UCS}

The ECS is widely considered to be the most important indicator in understanding climate response to natural and anthropogenic forcings. It is usually estimated from instrumental data coming from the industrial age, from proxy paleoclimatic data, and from climate models of different levels of complexity. In climate models, the ECS results from a nontrivial combination of several model parameters that enter the feedback factors $\{f_k\}$ in Eq.~\eqref{eq:feedback} above, and it requires careful tuning.  Despite many years of intense research, major uncertainties still exist in estimating it from past climatic data, as well as substantial discrepancies among different climate models \cite{IPCC01,IPCC07,IPCC13}. In fact, \citet{Charney.ea.1979} estimated the ECS uncertainty as $1.5$--$4.5$~K for CO$_2$ doubling and this range of uncertainties has increased rather than decreased over the four intervening decades.

The basic reason for these uncertainties lies in the high sensitivity of the ECS to the strength of the feedbacks $f_k$, as a result of the factor $1/\{1-\sum_{k=1}^Nf_k(T_0)\}$ in Eq.~\eqref{eq:gain}. Efforts to reduce the uncertainty in ECS values for the current climate include adopting ultra-high resolution GCMs \cite{Satoh2018}, in which one may better account for feedbacks that act on a larger range of scales, to applying relations that are rigorously valid for simple stochastic models to the climate data \cite{Cox2018}, where one hopes to take advantage of general, and possibly universal, relationships between climatic variables.

In particular, the largest uncertainty in defining the ECS for the current climate state is associated with the difficulty in estimating correctly the strength of the two main feedbacks associated with clouds \cite{Bony2015,Schneider2017}. A warmer climate leads to increased presence of water vapor in the atmosphere, and, in turn, to more clouds. An increased cloud cover leads, on the one side, to an increase in the climate system's albedo (cooling effect) and, on the other side, to a more efficient trapping of longwave radiation near the surface (warming effect). The balance between the two feedbacks changes substantially according to the type of cloud, with the cooling effect dominant for low-lying clouds, while the warming effect is dominant for high-altitude ones. This is a striking example of the multiscale nature of the climate system: an extremely small-scale, short-lived  dynamical process --- cloud formation --- has a substantial effect on the planet's global and long-term energy budget.

\begin{figure}
\includegraphics[width = .95\columnwidth]{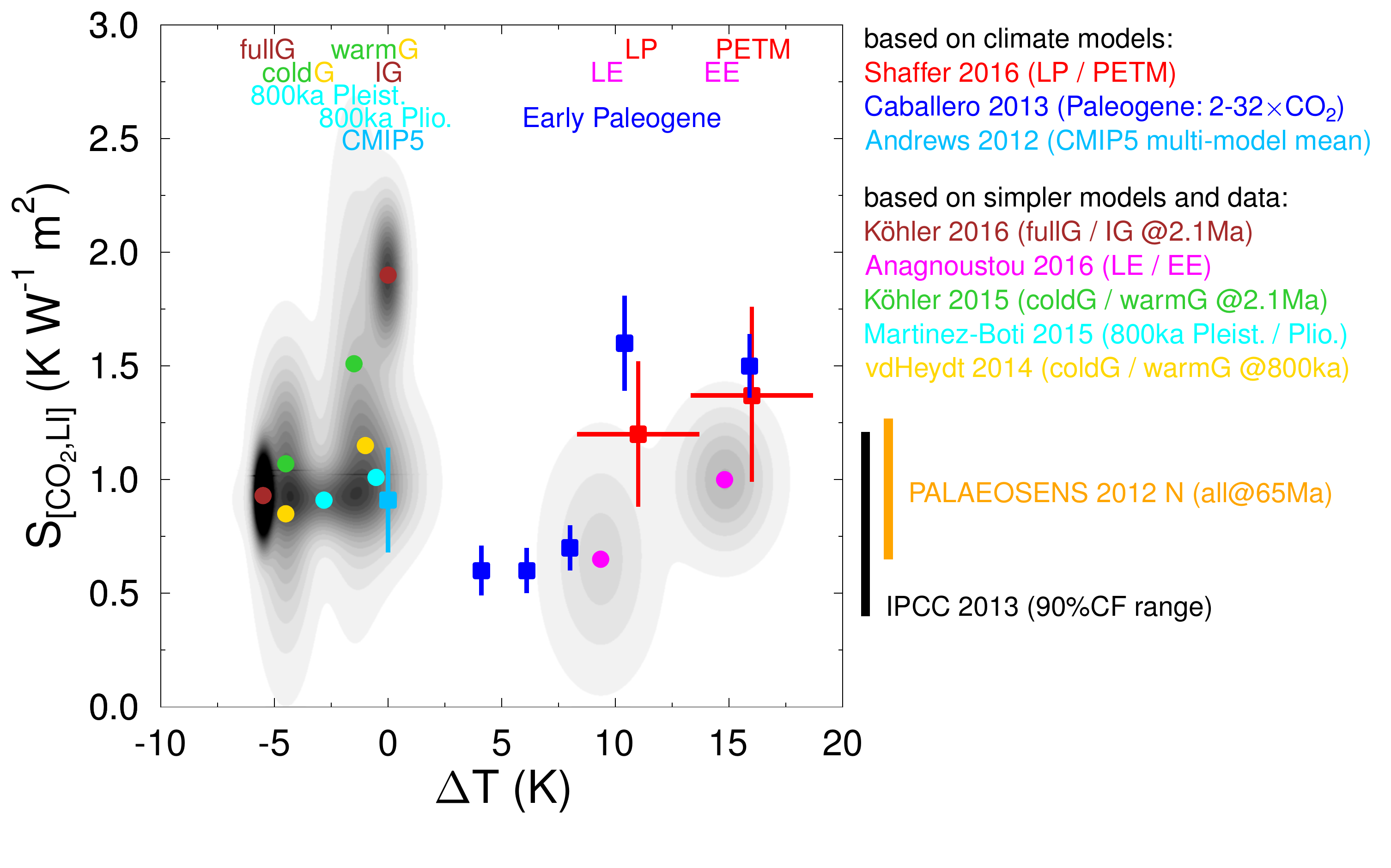}
\caption{State dependence of the ECS: estimates from proxy data and from climate models. The large number of data sets, models and acronyms is detailed in the \citet{vonderHeydt2016} review paper. The $\Delta T$ on the $x$-axis refers to the difference of the given $T_0$ from the pre-industrial value $T_{00}$, i.e., $T_0 = T_{00}+ \Delta T$; the $S_{[{\rm CO}_2, {\rm LI}]}$ on the $y$-axis refers to sensitivity with respect to CO$_2$ concentration corrected for land-ice albedo feedback. Mean values: for data -- color-coded circles with shaded probability density functions; for models -- squares with error bars. Reproduced with permission from \citet{vonderHeydt2016}.}
\label{statedependence2}
\end{figure}

Despite the highly simplified description above, it should be clear that the ECS is a state-dependent indicator. This state dependence is further supported by the evidence in Figs.~\ref{statedependence1} and \ref{statedependence2}. In particular:
\begin{itemize}

\item Both the Planck response and the strength of the feedbacks that determine the gain factor $\Lambda$ in Eq.~\eqref{eq:gain} depend on the reference state $T_0$. As an example, in warmer climates where sea ice is absent, the positive ice-albedo feedback is greatly reduced, thus contributing to a smaller climate sensitivity. On the other hand, in warmer climates the atmosphere is more opaque as a result of the presence of more water vapor, leading to a strong enhancement of the greenhouse effect. 

\item The radiative forcing is only approximately linear with the logarithm of the CO$_2$ concentration, so that $\Delta \tilde R_{2\times {\rm CO}_2}$ depends on the concentration's reference value. In fact, this dependence is weak across a large range of CO$_2$ concentrations, but it is greatly strengthened by optical saturation effects in the CO$_2$ absorption bands.

\item  Near the moist greenhouse threshold, which corresponds to a tipping point of the Earth system, the ECS is greatly strengthened. Figure \ref{statedependence1} shows that for a solar irradiance comparable or stronger than the present one, the peak in the value of the ECS is realised at a surface temperature of about 320 K, which corresponds to a lower CO$_2$ concentration in the case of weaker irradiance; see the discussion in \citet{Gomez_2018}. Note that reaching the moist greenhouse threshold for lower values of the solar irrandiance requires exceedingly high CO$_2$ concentrations.

\item Recently, \citet{Schneider2019}, using very high-resolution simulations that represent explicitly convective processes, proposed a mechanism of instability of stratocumulus clouds occurring at high CO$_2$ concentrations that greatly enhances the ECS and eventually leads to an abrupt transition to a much warmer climatic regime.

\end{itemize}

\begin{figure}
\includegraphics[trim=0cm 0cm 0cm 0cm, clip=true, width = 0.95\columnwidth]{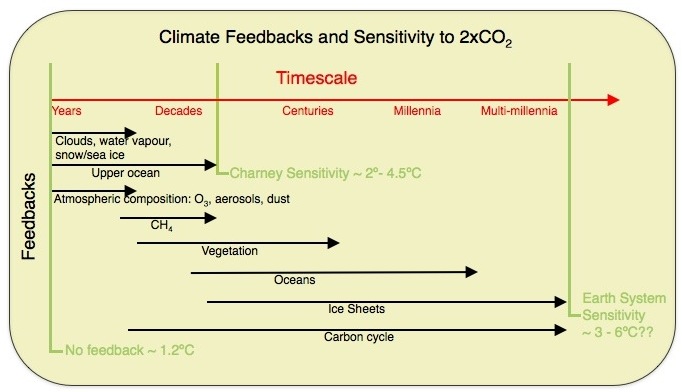}
\caption{Dependence of the effective ECS on the reference time scale. Consideration of longer time scales entails taking into account a larger set of slow climate processes. Reproduced with permission from \url{http://www.realclimate.org/index.php/archives/2013/01/on-sensitivity-part-i/}.}
\label{sensitivitymultiscale}
\end{figure}

While useful, the ECS concept faces practical difficulties because its definition assumes that, after the forcing is applied, the climate reaches a new steady state after all transients have died out. Since the climate is multiscale in both time and space, it is extremely non-trivial to define an effective cut-off time scale able to include all transient behavior. Thus, a time scale of 100 years is long compared to atmospheric processes, but short with respect to oceanic ones that involve the deep ocean. While a time scale of 5~000 years is long compared to oceanic processes, but short with respect to cryospheric ones that involve the dynamics of the Antarctic ice sheets. Therefore, one needs to associate each ECS estimate from observational or model data to a reference time scale; see Fig.~\ref{sensitivitymultiscale} for an illustrative cartoon, and its discussion by \citet{PALEOSENS} and \citet{vonderHeydt2016}. 

\subsection{Transient Climate Response (TCR)}
\label{ssec:TCR}

Transient climate response (TCR) has recently gained popularity in the study of climate change and climate variability because of its ability to help capture the evolution in time of climate change by addressing the transient, rather than asymptotic, response of the climate system to perturbations in the CO$_2$ concentration. TCR is defined as the change in the globally averaged surface air temperature recorded at the time at which CO$_2$ has doubled due to an increase at a 1\% annual rate, i.e. roughly after 70 years, having started at a given reference value $T_0$ \cite{Otto2013}. 
This operational definition agrees rather well with the standard IPCC-like simulation protocols; 
and TCR is therewith better suited than ECS to test model outputs against observational data sets from the industrial era. 

\begin{figure}
\includegraphics[angle=270,width = .9\columnwidth]{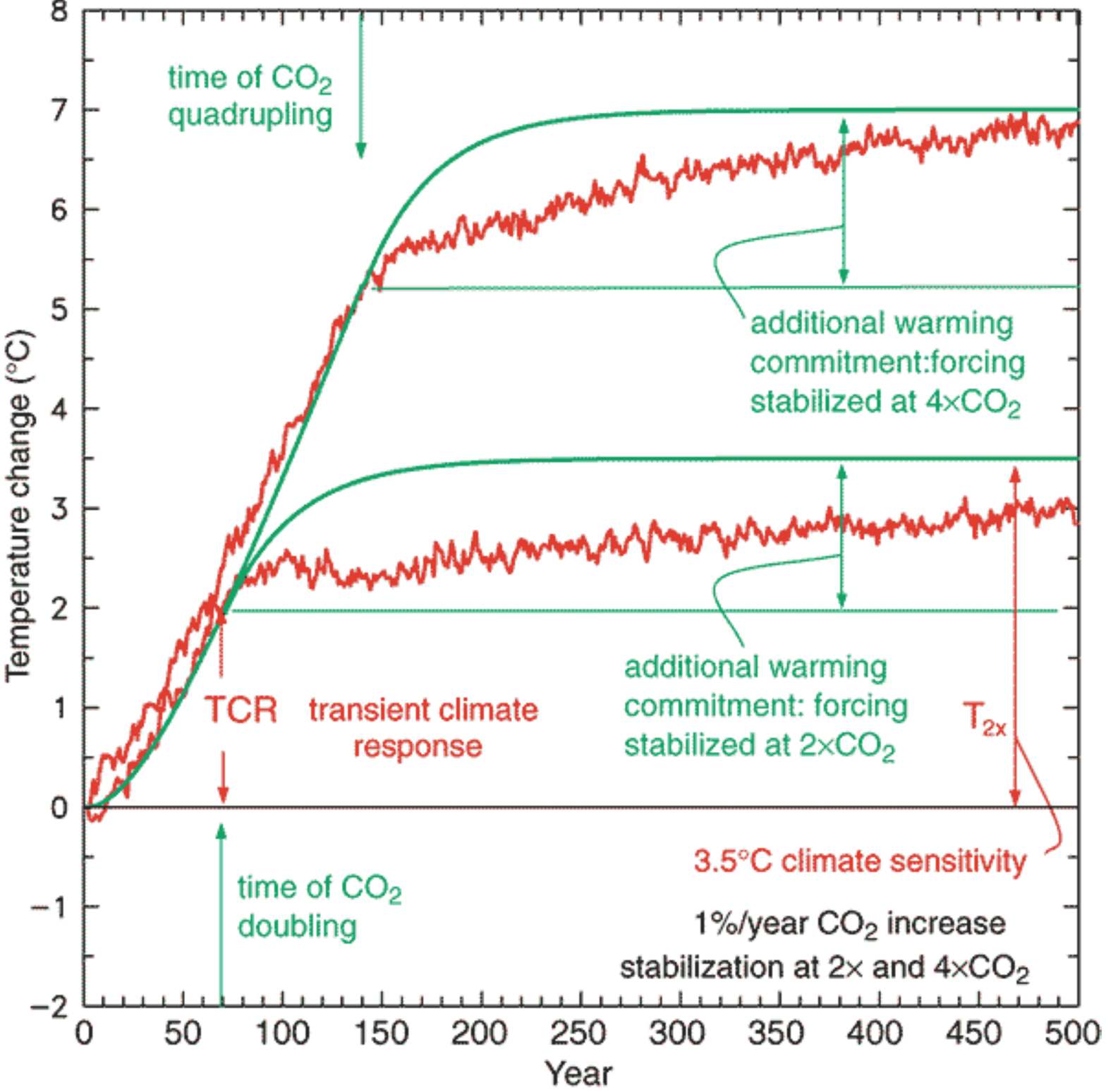}
\caption{Transient climate response (TCR) estimates from two climate models: (i) a coupled atmosphere--ocean
    GCM (red curve) and (ii) a simple illustrative model with no energy exchange with the deep ocean (green curve).
    Time on abscissa from start of CO$_2$ concentration increase at pre-industrial levels, 
    with change in global mean temperature on the ordinate. The ``additional warming commitment'' 
    corresponds to temperature stabilization at a given CO$_2$ level, i.e., at $2 \times {\rm CO}_2$ 
    or at $4 \times {\rm CO}_2$.
    Reproduced with permission from \citet[Fig.~9.1]{IPCC01}.}
\label{TCR}
\end{figure}

As shown in Fig.~\ref{TCR}, the TCR is found to be lower than the ECS for a long time, because of the climate system's thermal inertia, which is dominated by the oceans' heat capacity. A smaller effective heat capacity $c$  in Eq.~\eqref{0DEBM}, and hence a shorter relaxation time, would result in the TCR catching up much faster with the ECS, as is the case in regular diffusion processes; see Fig.~\ref{Fig_6}a. Assuming linearity in the response, a relationship must clearly exist between ECS and TCR. So far, this inference has been based, by-and-large, on heuristic arguments of time scale separation between climate feedbacks rather than being rigorously derived \cite{Otto2013}; it will be derived more systematically in Sect. \ref{ssec:CR} herein. 

\subsection{Beyond Climate Sensitivity}\label{ssec:wasserstein}
The standard viewpoint on climate sensitivity discussed above is associated with the idea that the climate is in equilibrium, in the absence of external perturbations. In the setting of deterministic, autonomous dynamical systems, this view can be described by the change in the position of a fixed point, $\vx_0 = \vx_0(\mu)$, as a function of a parameter $\mu$.

\begin{figure*}[ht]
\centering
    \includegraphics[angle=0,width=0.8\columnwidth]{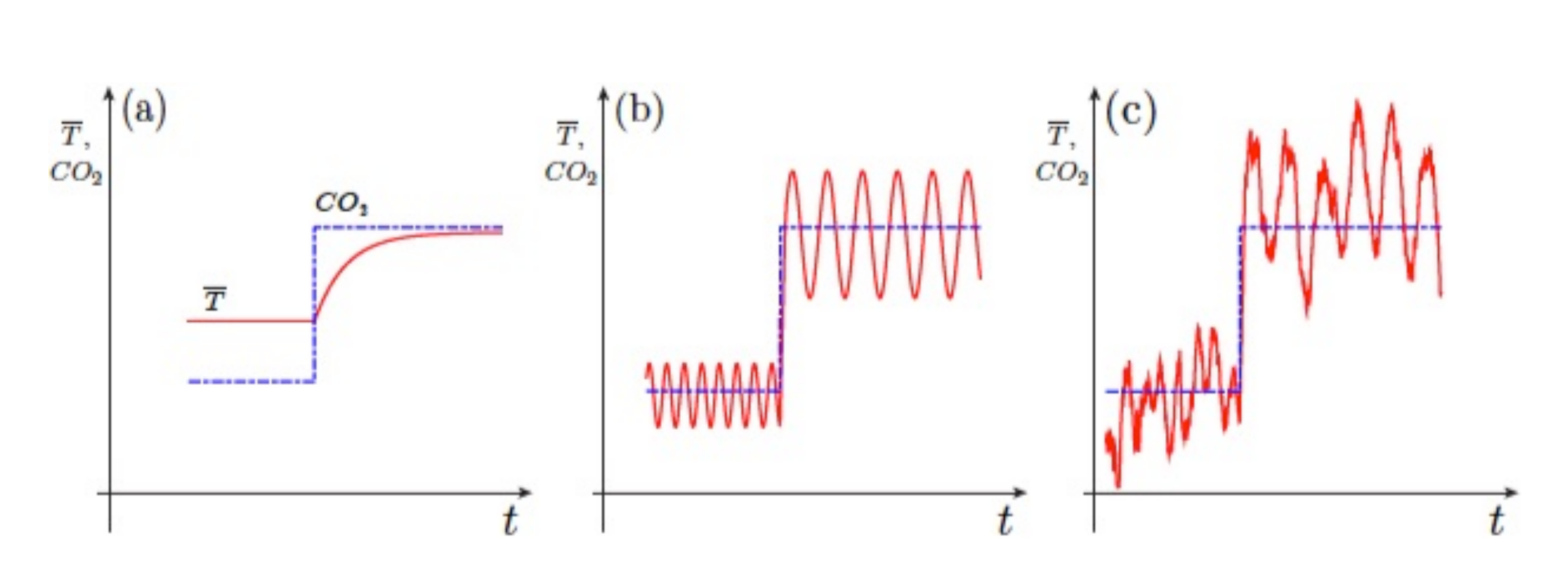}
    \caption{Climate sensitivity (a) for an equilibrium model; (b) for a nonequilibrium, oscillatory model; 
    and (c) for a nonequilibrium model featuring chaotic dynamics and stochastic perturbations. 
    As a forcing (atmospheric CO$_2$ concentration, say, blue dash-dotted line) changes suddenly, 
    global temperature (red heavy solid) undergoes a transition.
    (a) Only the mean temperature $\bar T$ changes; (b) the amplitude, frequency, and phase of 
    the oscillation change, too;
    and (c) all the details of the invariant measure, as well as the correlations at all orders, are affected.
    Reproduced from \citet{Ghil2016}, with permission from the American Institute of Mathematical Sciences.} 
\label{Fig_6}
\end{figure*}

We illustrate in Fig.~\ref{Fig_6} the difference between the ways that a change in a parameter can affect a climate model's behavior in the case of equilibrium solutions, cf.~ panel (a), vs. more complex dynamical behavior, cf. panels (b) and (c). Assume that the climate state is periodic, i.e., lies on a limit cycle, rather than being a fixed point, as in panel (a). In this case, climate sensitivity can no longer be defined by a single scalar, ${\partial \bar T}/{\partial \mu}$, but needs four scalars --- the sensitivity of the mean temperature along with that of the limit cycle's frequency, amplitude, and phase --- or more, e.g., the orbit's ellipticity, too.

But the internal climate variability can be better described in terms of strange attractors than by fixed points or limit cycles. Moreover, the presence of time-dependent forcing, deterministic as well as stochastic, introduces additional complexities into the proper definition of climate sensitivity. It is thus apparent that a rigorous definition of climate sensitivity requires considerably more effort. 

\citet{Ghil2015} proposed to measure the change in the overall properties of the attractor before and after the change in forcing by computing the {\it Wasserstein distance} $d_{\rm W}$ between the two invariant measures. The Wasserstein distance or ``earth mover's distance" $d_{\rm W}$ quantifies the minimum ``effort'' in morphing one measure into another one of equal mass on a metric space, 
like an $n$-dimensional Euclidean space \cite{Dobrushin1970}. 

\citet{Monge1781} originally introduced this distance to study a problem of military relevance. Roughly speaking, $d_{\rm W}$ represents the total work needed to move the ``dirt" (i.e., the measure) from a trench you are digging to another one you are filling, over the distance between the two trenches. In general, the shape of the two trenches and the depth along the trench --- i.e., the support of the measure and its density --- can differ. \citet{Robin2017} and \citet{Vissio2018} recently showed the effectiveness of applying this idea to climate problems.

\subsection{A General Framework for Climate Response}\label{ssec:CR}
A major use of state-of-the-art climate models is to produce projections of climate change taking into account different possible future scenarios of emission of greenhouse gases and pollutants like aerosols, as well as changes in land use, which has substantial impacts on the terrestrial carbon cycle. Projections are needed not only for quantities like the global average surface air temperature, but on spatially and temporally detailed information is needed for a multitude of practical needs; see, for instance, Fig.~\ref{Fig_IPCC}.

The ECS concept is well suited for describing the properties of equilibrium solutions of heuristically simplified equations of the climate system, like Eq.~\ref{0DEBM}, and has clear intuitive appeal, as in Fig.~\ref{Fig_6}a. But it also has basic scientific limitations:
\begin{itemize}
\item it only addresses long term-climatic changes and no detailed temporal information, an issue only partially addressed by TCR information;
\item it only addresses changes in the globally averaged surface air temperature and no spatial information at the regional scale and at different levels of the atmosphere, of the ocean, and of the soil;
\item it cannot discriminate between radiative forcings resulting from different physical and chemical processes, e.g. differences resulting from changes in aerosol vs. GHG concentration; the two impact quite differently shortwave and longwave radiation, and different atmospheric levels.
\end{itemize}

We will thus try to address these shortcomings by taking the complementary points of view of nonequilibrium statistical mechanics and dynamical systems theory. The setting of nonautonomous and of stochastically forced dynamical systems allows one to examine the interaction of internal climate variability with the forcing, whether natural or anthropogenic; it also helps provide a general definition of climate response that takes into account the climate system's nonequilibrium behavior, its time-dependent forcing, and its spatial patterns.   

\subsubsection{Pullback Attractors (PBAs)}\label{sssec:PBA}
The climate system experiences forcings that vary on many different time scales \cite[e.g.,][]{saltzman_dynamical}, and its feedbacks also act on multiple time scales \cite[e.g.,][]{Ghil1987}. Hence, defining rigorously what climate response to forcing, vs. intrinsic variability, actually is requires some care: Observed variations can be related to the presence of natural periodicities --- such as the daily and the seasonal cycle, and orbital forcings; to rapid, impulsive forcings, such as volcanic eruptions; or to slow modulations to the parameters of the system, as in the case of anthropogenic climate change.  

For starters, consider a dynamical system in continuous time,  
\begin{equation}
\dot{x}=F(x,t)\label{eq1}
\end{equation}
on a compact manifold $\Y \subset \R^d$; here $x(t)=\phi(t,t_0)x(t_0)$, with initial state $x(t_0)=x_0 \in \Y$. The evolution operator $\phi(t,t_0)$ is assumed to be defined for all $t\geq t_0$, with $\phi(s,s) = \mathbf 1$, and it thus generates a two-parameter semi-group. In the autonomous case, time-translational invariance reduces the latter to a one-parameter semigroup since, $\forall t\geq s$, $\phi(t,s)=\phi(t-s)$. In the nonautonomous case, in other terms, there is an absolute clock. 

We are interested in forced and dissipative systems such that, with probability one, initial states in the remote past are attracted at time $t$ towards ${\A}(t)$, a time-dependent family of geometrical sets that define the system's pullback attractor (PBA). In the autonomous case, 
${\A}(t) \equiv \A_0$ is the time-independent attractor of the system, and it is known to support, under suitable conditions, a physical measure $\mu(\textrm{d}x)$ \cite{ER85, LY.1988}.\footnote{Among invariant measures, a natural measure is one obtained by flowing a volume forward in time, and a physical measure is one for which the time average equals the ensemble average almost surely with respect to Lebesgue measure; moreover physical implies natural. A particular class of invariant measures of interest are Sinai-Ruelle-Bowen (SRB) measures \cite[e.g.,][]{young_what_2002}, and an important result is that --- for a system with no null Lyapunov exponents, except the one corresponding to the flow --- an ergodic SRB measure is physical.}

Such a PBA can also be constructed when random forcing is present \citep[e.g.,][and references therein]{arnold1988},
\begin{equation}
\label{eq:RDS}
\diff {x}=F(x,t) \diff t+g(x) \diff \eta,
\end{equation}			
where $\eta = \eta(t; \omega)$ is a Wiener process, while $\omega$ labels the particular realization of this random process, and $\diff \eta(t)$ is commonly referred to as ``white noise.'' The noise can be multiplicative, and one then uses the It\^o calculus for the integration of Eq.~\eqref{eq:RDS}. 

In the random case, the PBA $\A(t;\omega)$ is commonly referred to as a random attractor.   A more detailed and mathematically rigorous discussion of these concepts appears in \citet{Chekroun2011,CLR13} and \citet[Appendix A]{GCS08}.  Careful numerical applications of PBAs to explain the wind-driven circulation and the THC are now available \cite{Sevellec2015, Pierini2016, Pierini.ea.2018}.  

In the purely deterministic case, the theory of nonautonomous dynamical systems goes back to the skew-product flows of \citet{Sell.1967}. A concept that is closely related to PBAs is that of snapshot attractors; it was introduced in a more intuitive and less rigorous manner into the physical literature by \citet{Ott.ea.1990}. Snapshot attractors have also been used for studying  time-dependent problems of climatic relevance \cite{BKT11,BT12,BKT13,DBT15}.

\begin{figure}
\includegraphics[clip=true, trim= 0cm .5cm 0cm 0cm, angle=0, width = .9\columnwidth]{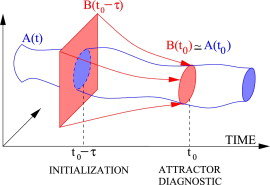}
\caption{A cartoon depicting the pullback attractor (PBA) for a nonautonomous system as in Eq.~\eqref{eq1}. The set $\B$ of finite Lebesgue measure, initialized at time $t=t_0-\tau$, evolves towards the 
set $\A(t_0)$. The construction is exact in the limit $\tau\rightarrow \infty$; for a random attractor, see \citet[Figs.~A.1 and A.2]{GCS08}. Adapted from \citet{Sevellec2015}, with permission.}
\label{pullba}
\end{figure}

Note that, in the most recent IPCC reports  \cite{IPCC01,IPCC07,IPCC13} and according to the standard protocols described in Sect.~\ref{secondkind} 
, future climate projections are virtually always performed using as initial states the final states of sufficiently long simulations of historical climate conditions. As a result, it is reasonable to assume that the pullback time $\tau$, as defined in Fig.~\ref{pullba}, is large enough, and that the covariance properties of the associated ${\A}(t)$ sets are therefore well approximated.

\subsubsection{Fluctuation Dissipation and Climate Change}
\label{sssec:FDT}
The fluctuation-dissipation theorem (FDT) has its roots in the classical theory of many-particle systems in thermodynamic equilibrium. The idea is very simple: the system's return to equilibrium will be the same whether the perturbation that modified its state is due to a small external force or to an internal, random fluctuation. The FDT thus relates  natural and forced fluctuations of a system \cite[e,g.,][]{K57, Kubo.1966};
it is a cornerstone of statistical mechanics and has applications in many areas \cite[and references therein]{marconi2008}. 


We have emphasized already in Sect. \ref{ssec:processes} 
that, even when the climate system is in a steady state, it is not at all in thermodynamic equilibrium \cite[e.g.,][]{Ghil.2019, LucariniRagone}. Still, \citet{Leith75} showed that FDT applies to a 2-D or QG turbulent flow with two integral invariants, kinetic energy $E$ and enstrophy $Z$, under some additional assumptions of normal distribution of the realizations and stationarity. Soon thereafter, \citet{Bell1980} showed that the FDT still seemed to work for a highly truncated version of such a model, even in the presence of dissipative terms that invalidate the thermodynamic equilibrium assumption.

The FDT has been applied to the output of climate models to predict the climate response to a step-like  increase of the solar irradiance \cite{North1993} as well as to increases in atmospheric CO$_2$ concentration \cite{Cionni2004,langen_estimating_2005}, while \citet{gritsun2007} and \citet{gritsun2008b} used it to predict the response of an atmospheric model to localised heating anomalies. Most recently, \citet{Cox2018} tried to reduce the uncertainty in the ECS discussed in Sect. \ref{ssec:UCS} 
by a systematic FDT application to an ensemble of model outputs, as well as to the observed instrumental climate variability.

The FDT-based response of the system to perturbations in the above examples reproduces the actual changes at a good qualitative, rather than strictly quantitative level. An important limitation of these insightful studies is the use of a severely simplified version of the FDT that is heuristically constructed by 
taking a gaussian approximation for the invariant measure of the unperturbed system. This approximation amounts to treating the climate as being in thermodynamic equilibrium; see also \citet{majda2010b}, who specifically address the applicability of the FDT in a reduced phase space. 

In order to address the problem of gaussian approximation, \citet{cooper_climate_2011} proposed to construct a kernel-based approximation of the actual invariant measure of the unperturbed system and then use Eq.~ \eqref{correlation} below to construct the system's Green function. This approach has been applied successfully in a very low-dimensional system, but  its robustness may be limited by the kernel's arbitrary cut-off not being able to account for the smaller scales of the invariant measure's fine structure.

\subsubsection{Ruelle Response Theory}\label{ruellert}
FDT generalizations to systems out of equilibrium have been developed since the early 1950s \cite[e.g.,][]{Kubo.1966}. But a particularly fruitful change in point of view was provided by D. Ruelle \cite{Ruelle:1998, ruelle_smooth_1999, ruelle2009}, who considered the problem in the setting of dynamical systems theory, rather than that of statistical mechanics. The former point of view is justified in this context by the so-called chaotic hypothesis \cite{gallavotti_dynamical_1995}, which states, roughly, that chaotic systems with many degrees of freedom possess a physically relevant invariant measure, as discussed in Sect.~\ref{sssec:PBA}. 

It is common in practice to assume that a time-dependent measure $\mu_t(\diff x)$ associated with the evolution of the dynamical system given by Eq.~\eqref{eq1} does exist. Still, computing the expectation value of measurable observables with respect to this measure is in general far from trivial and requires setting up a large ensemble of initial states in the Lebesgue measurable set $\B$ mentioned before. Moreover, PBAs and the physical measures they support only set the stage for predicting the system's sensitivity to small changes in the forcing or the dynamics. 

Ruelle's response theory allows one to compute the change in the measure $\mu(\diff x)$ of an autonomous Axiom A system due to weak perturbations of intensity $\epsilon$ applied to the dynamics, in terms of the unperturbed system's properties. The basic idea behind it is that the invariant measure of  
such a system, even though supported on a strange attractor, is differentiable with respect to $\epsilon$. \citet{B08} reviewed extensions of response theory and provided a complementary approach to it, while \citet{Lucarini2016} estimated its radius of convergence in $\epsilon$.

The nonautonomous version of the Ruelle response theory allows one to calculate the time-dependent 
measure $\mu_t(\mathrm{d}x)$ on the PBA by computing the time-dependent corrections to it with respect to a reference state $x(t) = \tilde x(t)$. This provides a fairly general formulation for the climate system's response to perturbations. 

Let us assume that we can write 
\begin{equation}
\dot{x}=F(x,t)=F(x)+\epsilon X(x,t)\label{eqpertu}
\end{equation}
where,  $\forall t \in \R$ and $\forall x \in \Y \subset \R^d$, $|\epsilon X(x,t)|\ll |F(x)|$. Hence, we can take $F(x)$ as the background dynamics and $\epsilon X(x,t)$ as a perturbation. We only treat here the case of deterministic dynamics; see  \citet{lucarini2012} for stochastic perturbations. As shown by \citet{Lucarini2017}, we can restrict our analysis without loss of generality to the separable case of $F(x,t)=F(x)+\epsilon X(x)T(t)$.

To evaluate the expectation value $\langle \Psi \rangle^\epsilon(t)$ of a measurable observable $\Psi(x)$ with respect to the measure $\mu_t(\diff x)$ of the system governed by Eq.~\eqref{eq1}, one writes: 
\begin{equation}
\langle \Psi \rangle^\epsilon(t) = \int \Psi(x) \mu_t(\diff x) = \langle \Psi \rangle_0+\sum_{j=1}^\infty \epsilon^j \langle \Psi\rangle_0^{(j)}(t)\label{mut};
\end{equation}
here $\langle \Psi \rangle_0 = \int \Psi(x) \bar{\mu}(\diff x)$ is the expectation value of $\Psi$ with respect to the SRB invariant measure $\bar{\mu}(\diff x)$ of the autonomous dynamical system $\dot{x} = F(x)$. 
We will restrict ourselves here to the linear correction term, which can be written as:
\begin{eqnarray}
\langle \Psi\rangle_0^{(1)}(t) & = {\displaystyle \int \int_0^\infty}  \Lambda S_0^{\tau}  \Psi(x) T(t-\tau) \diff \tau \bar{\mu}(\diff x); \nonumber \\
& = {\displaystyle \int_0^\infty} G^{(1)}_{\Psi,X}(\tau) T(t-\tau) \diff \tau. \label{lrruelle}
\end{eqnarray}


The Green's function $G^{(1)}_{\Psi,X}(\tau)$ above is given by
\begin{equation}
G^{(1)}_{\Psi,X}(\tau) = \int \Theta(\tau)\Lambda S_0^{\tau} \Psi(x)\bar{\mu}(\diff x), \label{Greenf}
\end{equation}
where $\Lambda(\bullet) = {X} \cdot \nabla (\bullet)$ and $S_0^t(\bullet)=\exp(t {F} \cdot \nabla )(\bullet)$ is the semigroup of unperturbed Koopman operators\footnote{In discrete-time dynamics given by $x_{k+1} = g(x_k)$, the Koopman operator is a linear operator acting on observables $h: \Y \to \R$ via $Uh(x) = h(f(x))$. In continuous-time dynamics, like Eq.~\eqref{eq1} herein, this operator is replaced by a semigroup of operators, as in Eq.~\eqref{Greenf} above; see \citet{Mezic.ea.2012} for a good review.}, $(\cdot)$ denotes the inner product in $\Y$, and the Heaviside distribution $\Theta(\tau)$  enforces causality. If the unperturbed invariant measure $\diff \bar \mu(x)$ is smooth with respect to the standard Lebesgue measure, 
one has $\bar{\mu}(\diff {x}) = \tilde {\mu}(x)\diff {x}$, with $\tilde {\mu}(x)$ the density, and the Green's function can be writen as follows:
\begin{eqnarray}
G^{(1)}_{\Psi,X}(\tau) & = \Theta(\tau) {\displaystyle \int  \frac{-\nabla \cdot ( \tilde{\mu}(x) X) }{ \tilde{\mu}(x)}} S_0^{\tau} \Psi(x) \tilde{\mu}(x) \diff x \nonumber\\
& = \Theta(\tau)\mathcal{C}(\Phi,S_0^{\tau} \Psi); \label{correlation}
\end{eqnarray}
here $\Phi = {- \left( \nabla \cdot ( \tilde{\mu}(x) X) \right) }/{ \tilde{\mu}(x)}$ and $\mathcal{C}(A,S_0^{\tau} B)$ is the $\tau$-lagged correlation between the variables $A$ and $B$, and the average of $\Phi$ vanishes. 
Note that Eq.~\eqref{correlation} is the appropriate generalization of the FDT for the nonautonomous, out-of-equilibrium system \eqref{eqpertu}.

Given any specific choice of the forcing's time dependency $T(t)$ in Eqs.~\eqref{eqpertu}--\eqref{lrruelle} and measuring the linear correction term $\langle \Psi\rangle_0^{(1)}(t)$ from a set of experiments, the same equations allow one to derive the appropriate Green's function. Therefore, using the output of a specific set of experiments or of GCM simulations, we achieve predictive power for any temporal pattern of the forcing $X(x)$.


Consider now the Fourier transform of Eq.~\eqref{lrruelle}: 
\begin{equation}
\langle \Psi\rangle_0^{(1)}(\omega) = \chi^{(1)}_{\Psi,X}(\omega) T(\omega),\label{lrruelle2}
\end{equation}
where we have introduced the susceptibility $\chi^{(1)}_{\Psi,X}(\omega)=\mathcal{F}[G^{(1)}_{\Psi,X}]$, defined as the Fourier transform of the Green's function $G^{(1)}_{\Psi,X}(t)$. Under suitable integrability conditions, the fact that the Green's function $G_{\Psi,X}^{(t)}$ is causal is equivalent to saying that its susceptibility obeys the so-called Kramers-Kronig relations \cite{ruelle2009,Lucarini2011}; these provide integral constraints that link the real and imaginary parts, so that $\chi^{(1)}_{\Psi,X}(\omega) = i \mathcal{P}(1/\omega)\star\chi_{\Psi,X}^{(1)}(\omega)$, where $i=\sqrt{-1}$, $\star$ indicates the convolution product, and $\mathcal{P}$ stands for integration by parts. Extensions to the case of higher-order susceptibilities are also available \cite{lucarini2005,lucarini08,L09,LC12}.

Instead of studying merely individual trajectories, one can study the evolution of ensembles of trajectories to obtain probabilistic estimates. The evolution of the measure $\rho$ driven by the dynamical system $\dot{x}=F(x)$ is described by the transfer or Perron-Frobenius operator $\mathcal{L}_0^t$ \cite{B00, Chekroun2014, Villani.2009}, which is the adjoint of the Koopman or composition operator $S_0^t$; it is defined as follows:
\begin{equation}
 \int \mathcal{L}_0^t \rho(x)\Psi(x) \diff x=\int \rho(x)S_0^t\Psi(x) \diff x. \label{Perron}
\end{equation}
Assuming that no degenaries are present, the generator of the semigroup $\{\mathcal{L}_0^t\}$ is the so-called Liouville operator $L$ and it satisfies $ \partial_t\rho=-\nabla (\rho F)=L\rho$. 

Let $\{(\sigma_k, \rho_k): k=1,\ldots,\infty\}$ be the eigenpairs of $L$. Then the eigenvalues of $\mathcal{L}_0^t$ are given by $\{\exp(\sigma_k t)\}$, with the same eigenvectors $\{\rho_k\}$. Correspondingly, the invariant measure $\bar \mu$ is the eigenvector having the null eigenvalue $\sigma_1=0$ for $L$ or, $\forall t\geq0$, unit eigenvalue of $\mathcal{L}_0^t$. Note that $\forall k\geq2$, $\Re(\sigma_k)<0$, which implies exponential decay of correlations for all the system's smooth observables, with an asymptotic rate that is given by the largest value of $\Re(\sigma_k)$, $k\geq2$. Moreover, the presence of a small spectral gap between the unit eigenvalue of $\mathcal{L}_0^t$ and its other eigenvalues within the unit disk leads to a small radius of expansion for Ruelle's perturbative approach \cite{Chekroun2014,Lucarini2016}. 

It turns out that, if one neglects the essential part of the spectrum, the susceptibility can be written as
\begin{equation}
\chi^{(1)}_{\Psi,X}(\omega)=\sum_{k=1}^\infty \frac{\alpha_k\{\Psi,X\}}{\omega-\sigma_k}, \label{chispectral}
\end{equation}
where the factor $\alpha_k$  evaluates how the response of the system to the forcing $X$ for the observable $\Psi$ projects on the  eigenvector $\{\rho_k\}$. 
The constants $\pi_k=i\sigma_k$  are usually referred to as Ruelle-Pollicott poles \cite{Ruelle1986,Pollicott1985}. Note that Eq.~\eqref{chispectral} implies that the Green's function corresponding to the susceptibility can be written as a weighted sum of exponential functions, if one neglects possible degeneracies.\footnote{Note that Eq.~\eqref{chispectral} mirrors the quantum-mechanical expressions for the electric susceptibility of atoms or molecules. In the latter case, the summation involves all the pairs of eigenstates of the system's unperturbed Hamiltonian operator; in each term, the poles' imaginary part corresponds to the energy difference between the pair of considered eigenstates; and the real part is the so-called line width of the transition, whose inverse is the life time; finally, the numerator is its so-called dipole strength \cite{CohenT1997}.} 

\subsubsection{Climate Change Prediction via Ruelle Response Theory}

A somewhat different approach to constructing the climate response to forcings focuses on computing it directly from Eq.~\eqref{Greenf}, without relying on the applicability of the FDT, which fails in certain cases of geophysical relevance \cite[e.g.,][]{gritsun2017}. The difficulty in applying this direct approach lies in the fact that the formula contains contributions from both stable and unstable directions in the tangent space \cite{ruelle2009}. 

Evaluating the contribution of the unstable directions is especially hard; hence, \citet{AM07a} proposed a blended approach that also uses the FDT. Using adjoint methods in the direct approach has also yielded promising results \cite{W13}. Faced with the so-called cold-start problem of climate simulations, \citet{hasselmann.ea.1993} suggested a heuristic approach to computing a climate model's Green's function and applied it to study the relaxation to steady state of a coupled GCM's globally averaged surface temperatures.


Lucarini and associates \cite{Lucarini2011, Lucarini2017, Ragone2016} proposed to evaluate the Green's function using an experimental but rigorous  approach, suggested by standard optics laboratory practice \cite{lucarini2005}. The idea is to use a set of carefully selected probe experiments --- typically, step-like increases of the parameter of interest --- to construct the Green's function and then, exploiting Eq.~\eqref{Greenf}, use this operator to predict the response of the system to a temporal pattern of interest for the forcing. 

Given a set of forced climate simulations and a background unperturbed one, such 
an approach 
allows one to construct the Green's functions response operators for as many observables as desired, global as well as local in space. Such a toolkit   
allows one to treat a continuum of scenarios of temporal patterns forcings, thus providing a general framework for improving climate change projections given in the form shown in Fig.~\ref{Fig_IPCC}. 

The idea is to consider the set of equations describing an unperturbed climate evolution in the form $\dot{x}=F(x)$, with the vector field $X = X(x)$ in Eq.~\eqref{eqpertu} as the 3-D radiative forcing associated with the increase of CO$_2$ concentration, and $\epsilon T(t)$ its time modulation. By plugging $T(t)=\epsilon\Theta(t)$ into Eq.~\eqref{lrruelle}, we have, for any climatic observable $\Psi$,
\begin{equation}
\frac{\diff}{\diff t}\langle \Psi \rangle_0^{(1)}(t)=\epsilon G_{\Psi,[{\rm CO}_2]}^{(1)}(t). \label{computeG}
\end{equation}
We estimate $\langle \Psi \rangle_0^{(1)}(t)$ by taking the system's average of response over an ensemble of initial states and use the previous equation to derive our estimate of $ G_{\Psi,[CO_2]}^{(1)}(t)$, by assuming linearity in the response. 

Note that these Green's functions are specifically related to changes in the atmospheric CO$_2$ concentration, rather than to a generic 
radiative forcing. Indeed, for each climatic variable, 
the response to changes in the solar irradiance --- e.g., via modulation of the parameter $\mu$  in 
Eqs.~\eqref{eq:rad_bal_0} 
%
--- is different from the impact of changes in the CO$_2$ concentration, because the details of the radiative forcing are very different in the two cases. Thus, geoengineering proposals that aim to reduce solar irradiance by injecting sulphate aerosols in the stratosphere 
\cite[e.g.,][]{Smith_2018} rely on flawed scientific reasoning \cite{lucarini_modelling_2013,Bodai2018}, as well as being of dubious practical help \cite{Proctor2018} and hardly defensible from an ethical perspective \cite{Lawrence2018}.
 

\citet{Ragone2016} derived, moreover, a closed formula relating the TCR and the ECS. Indeed, with the definition of $TCR(\tau)$ given 
at the beginning of Sect.~\ref{ssec:TCR}, 
one gets
\begin{eqnarray}
& ECS-TCR(\tau) = INR(\tau) \nonumber\\ & = \Delta \tilde R_{2\times {\rm CO}_2}\times 
\mathcal{P} {\displaystyle \int_{-\infty}^{\infty}}  \chi^{(1)}_{T_S,[{\rm CO}_2]}(\omega') \nonumber\\
& {\displaystyle \frac{1 + {\displaystyle \frac{\sin(\omega'\tau/2)}{\omega'\tau/2} } \exp[-i\omega'\tau/2]}
{\displaystyle {2\pi i\omega} } } \diff \omega'. \label{TCRb}
\end{eqnarray}
The difference between ECS and TCR is a weighted integral of the susceptibility, accounting for the contribution of processes and feedbacks occurring at different time scales. 
The integral in Eq.~\eqref{TCRb} yields $\chi^{(1)}_{T_S,[{\rm CO}_2]}(0)$ if $\tau\rightarrow 0$ (i.e., $TCR(0)=0$), decreases with $\tau$, and vanishes in the limit $\tau\rightarrow\infty$. $INR(\tau)$ is a measure of the system's inertia at the time scale $\tau$, due to the overall contribution of the internal physical processes and characteristic time scales of the relevant climatic subsystems.

\begin{figure}
\includegraphics[clip=true, trim= 2cm 2cm 2cm 2cm, angle=270, width = .9\columnwidth]{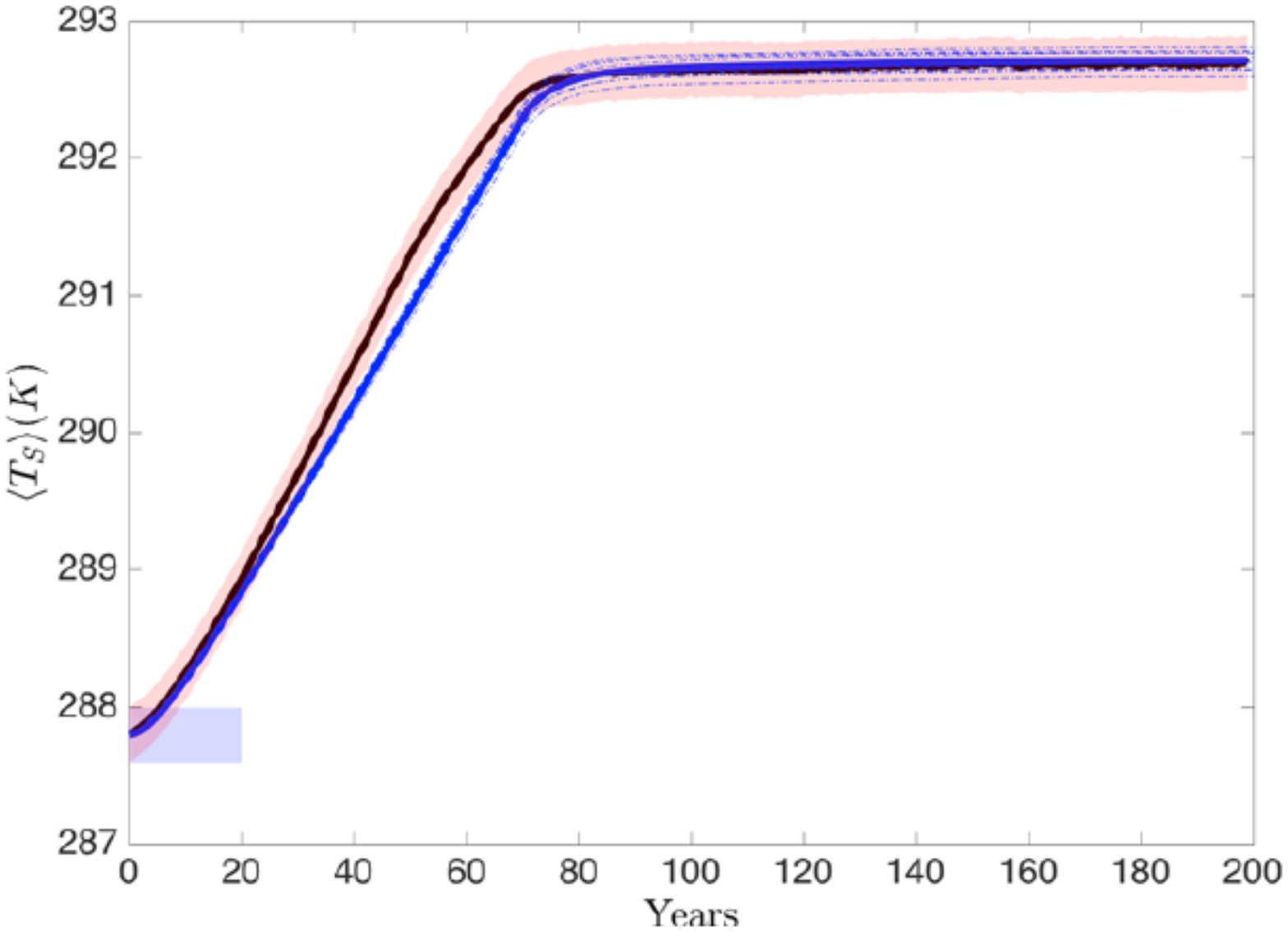}
\caption{Comparison between the climate model simulation (black) and response theory prediction (blue) for a TCR experiment using PLASIM, a GCM of intermediate complexity. The CO$_2$ concentration was ramped up by 1\% per year to double its initial value. The upper and lower limit of the light-shaded bands are computed as two standard deviations of the ensemble distribution. Reproduced with permission from \cite{Lucarini2017}.}
\label{ClimatePrediction1}
\end{figure}

Figure~\ref{ClimatePrediction1} illustrates a 
climate prediction obtained by applying the response theory above to the open-source model PLASIM (see \url{http://tiny.cc/zgk0bz}), an atmospheric GCM of intermediate complexity coupled with a mixed-layer ocean model \cite{Fraedrich.ea.2005, Fraedrich.e.2012}. 
The figure shows a good agreement between (a) the ensemble average of 200 simulations where the CO$_2$ concentration is increased at the rate of 1\% until doubling (black curve), and then kept constant; and (b) the prediction done by convolving the Green's function of Eq.~\eqref{computeG} with the time pattern of the forcing (blue curve). Note that the temperature increase predicted at year 70, when doubling of the CO$_2$ is reached, gives as estimate the $TCR \simeq 4.1$~K (blue curve). The true value of the $TCR \simeq 4.3$~K is given by the black line. The TCR  is indeed smaller than the $ECS \simeq 4.8$~K, which corresponds to a good approximation to the  temperature increase predicted at year 200. 

The power of response theory lies in the fact that --- once the Green's function of interest has been computed --- one can predict the future evolution of bespoke observables, defined as needed. Furthermore, \citet{Lucarini2018JSP} has shown that it is possible to use certain classes of observables as surrogate forcings of other observables, in the spirit of the linear feedback analysis of Eq.~\eqref{Taylor}. Given the notoriously more difficult prediction of precipitation, \citet{Lucarini2017} have found that predictions of changes in the globally averaged precipitation can be as good as those of the temperature. 

\begin{figure}
\includegraphics[clip=true, trim= 0cm .5cm 0cm 0cm, angle=0, width = .9\columnwidth]{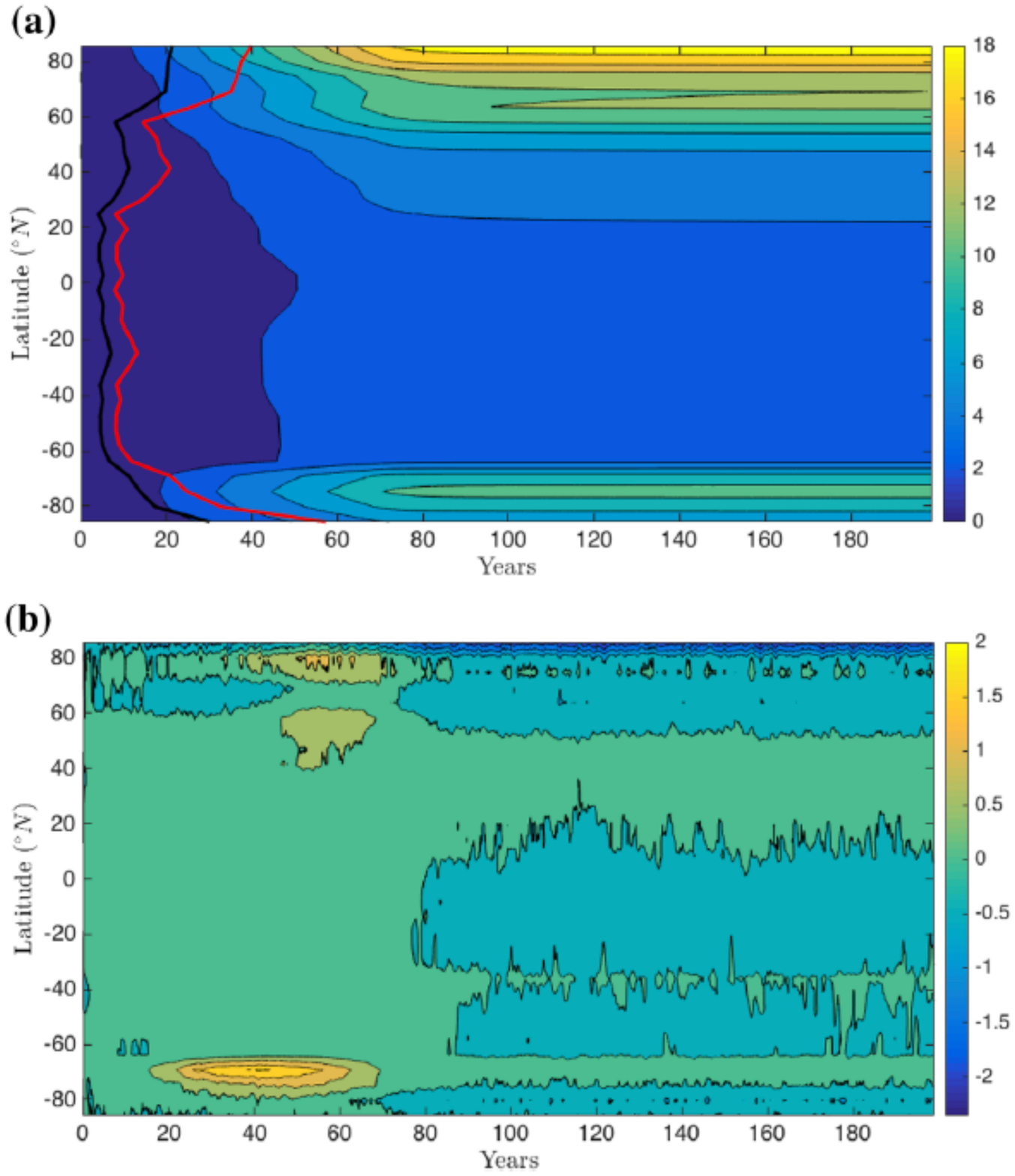}
\caption{Patterns of climate response for the zonal average of the surface temperature. (a) Prediction via response theory (red and black lines not relevant in this context). b) Difference between the ensemble average of the direct numerical simulations and the predictions obtained using the response theory. Reproduced with permission from \cite{Lucarini2017}.}
\label{ClimatePrediction2}
\end{figure}

Proceeding from global predictions to more localized ones, Fig.~\ref{ClimatePrediction2} shows the outcome of predicting the change in the zonal averages of the surface temperature. It is clear that response theory does a good job in reproducing the spatial patterns of temperature change, except for an underestimate of the temperature change in the high latitudes on the scale of few decades. Indeed, such mishap is due the strong polar amplification of the warming due to the ice-albedo feedback, which can only qualitatively captured by a linear approach like the one used by \citet{Lucarini2017}. 

\subsubsection{Slow Correlation Decay and Sensitive Parameter Dependence} 
\label{tantet}

Equation~\eqref{chispectral} implies that resonant amplification occurs if the forcing acts a a frequency $\omega$ that is close to the imaginary part of a subdominant Ruelle-Pollicott pole $\pi_{k}, \; k\geq2$ that has a small real part, i.e., for some $k \geq 2$, 
\begin{equation}
\label{eq:resonant}
|\omega - \Im(\pi_{k})| \ll \omega, \quad |\Re(\pi_{k})| \ll 1,
\end{equation}
because the system's susceptibility is greatly enhanced at such an $\omega$. Conditions \eqref{eq:resonant} are easily satisfied for a broadband forcing and a system that has a small spectral gap. Conversely, as discussed after Eq.~\eqref{Perron}, the presence of a small spectral gap is associated with a slow decay of correlation for smooth observables. We now provide two examples that show how linear response breaks down for forced systems possessing slow decay of correlations due to a small spectral gap in the unperturbed dynamics. 

The first example is due to \citet{Chekroun2014}, who investigated the response of the highly simplified ENSO model of \citet{Jin1993p1}, to which \citet{Jin1994} added periodic forcing that led to chaotic behavior. Chekroun and coauthors focussed on how the response changes when one varies the model parameter $\delta$ that controls the travel time of the equatorially trapped waves, which play an essential role in the ENSO mechanism.  \citet{Chekroun2014} found that when the spectral gap is small, the system exhibits rough dependence of its properties with respect to small modulations of the parameters, since Ruelle's perturbative expansion  breaks down even for a very small intensity of the forcing. 

\begin{figure}
\includegraphics[trim=0cm 0cm 0cm 0cm, clip=true,width = .9\columnwidth]{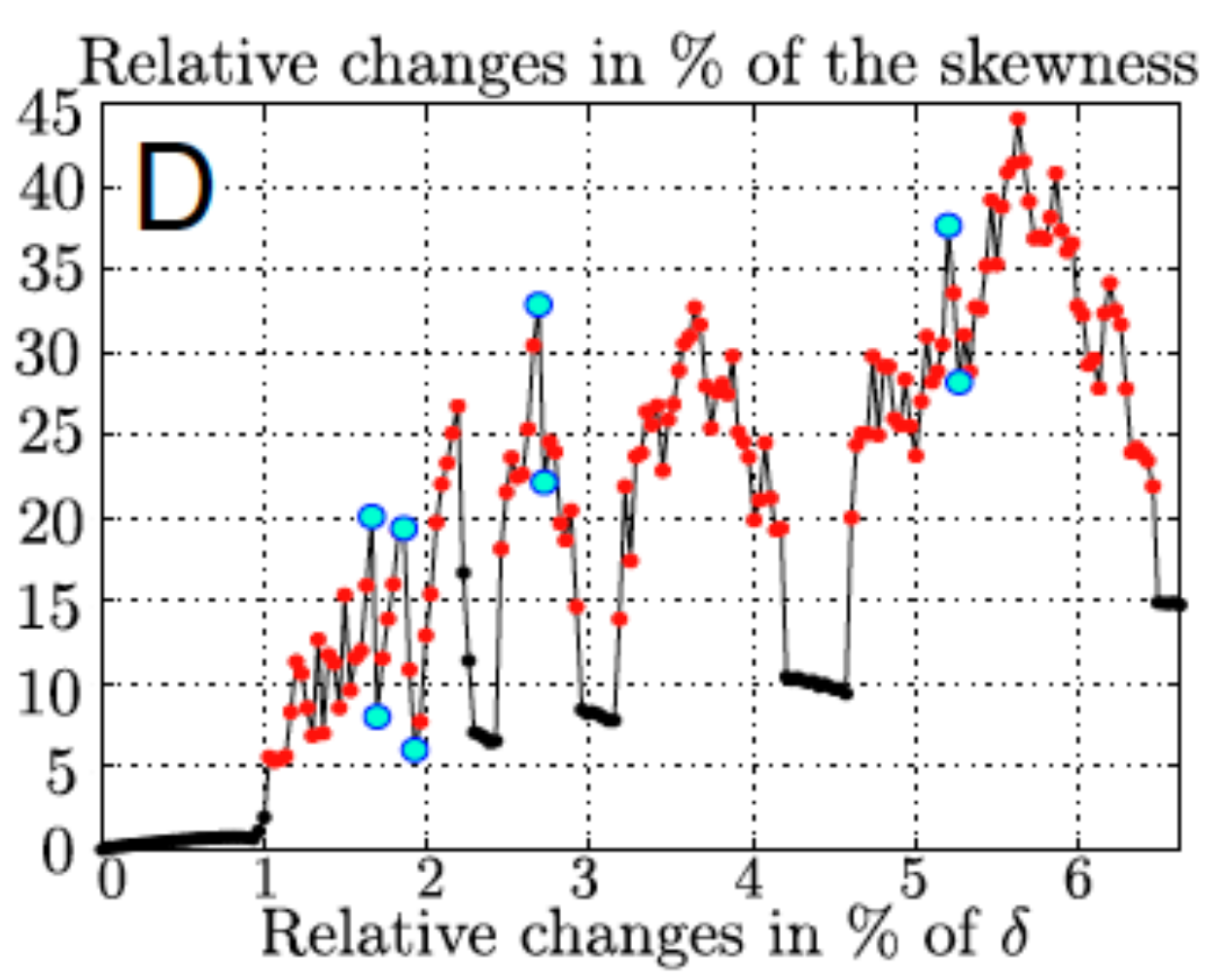}
\caption{Rough dependence (red dots) of the skewness of the equatorial ocean temperature's distribution with respect to changes in the parameter $\delta$ that controls the travel time of equatorially trapped waves in a simplified ENSO model; see text for details. Black points refer to nonchaotic windows where the parameter dependence is smooth; blue points indicate particularly sharp jumps in the parameter dependence. Reproduced with permission from \citet{Chekroun2014}.}
\label{ENSO_Chekrun}
\end{figure}

\begin{figure}
a)\includegraphics[angle=270,trim=0cm 0cm 0cm 0cm, clip=true, width = 0.9\columnwidth]{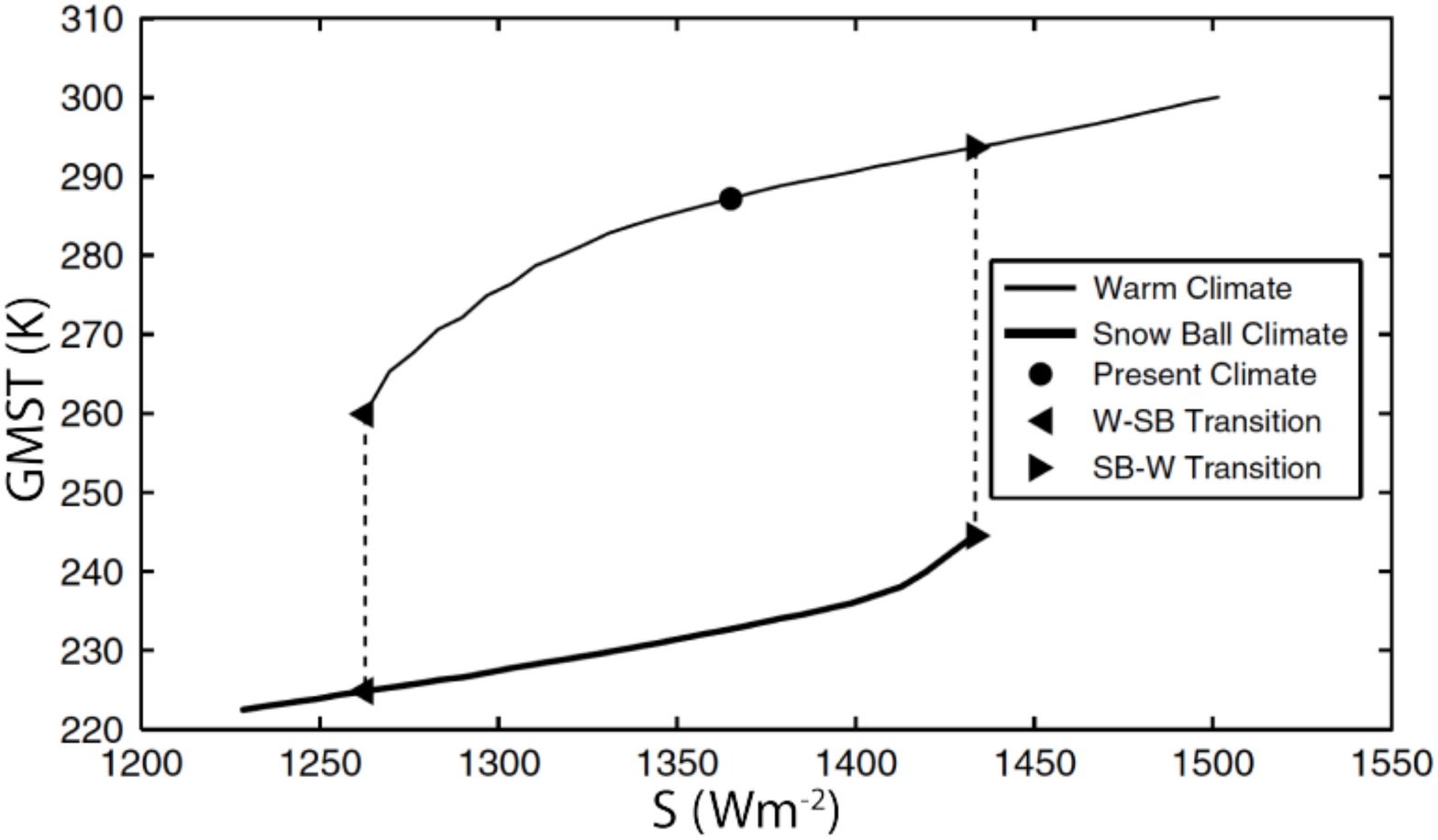}\\
b)\includegraphics[trim=0cm 0cm 0cm 0cm, clip=true, width =0.9\columnwidth]{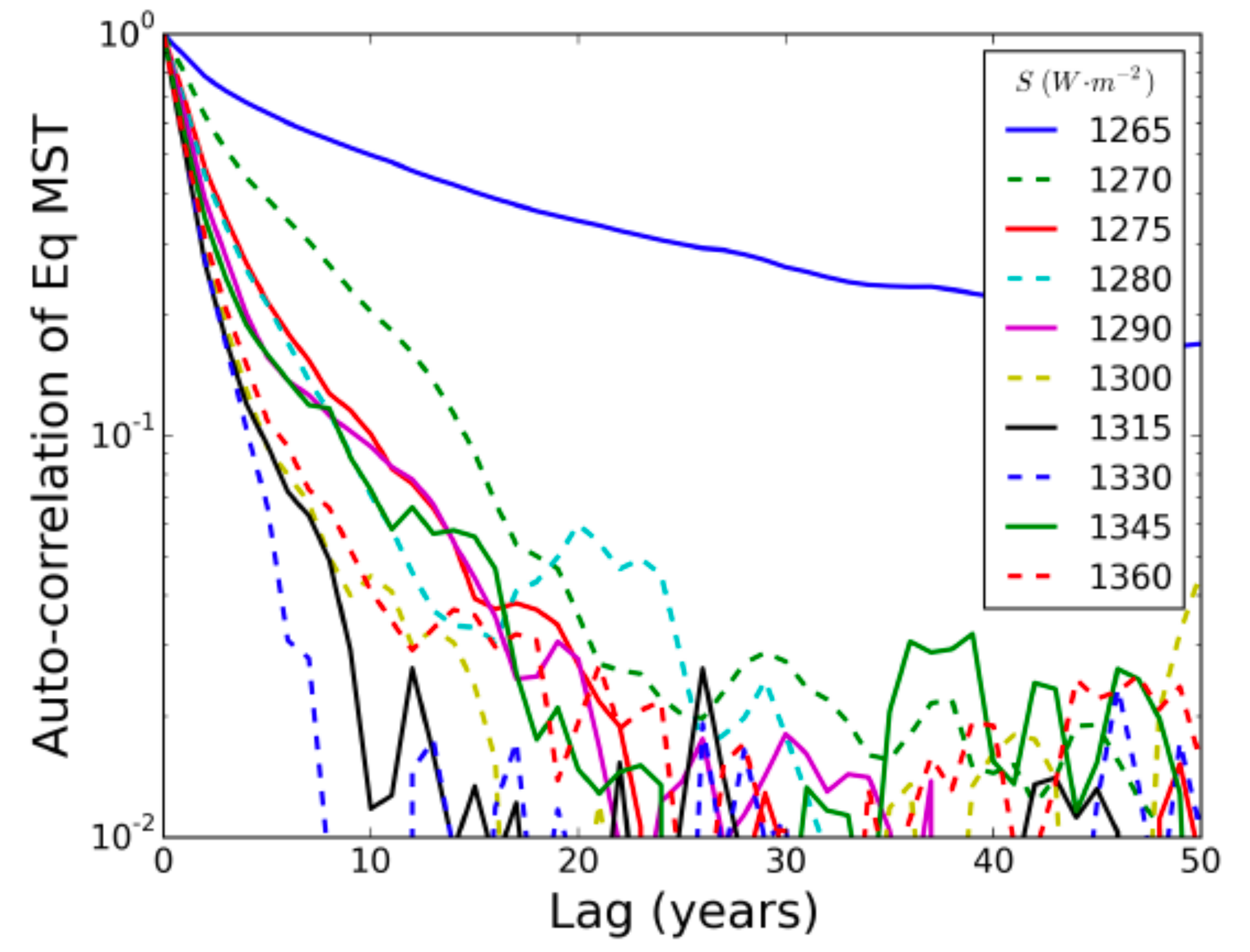}
\caption{Slowdown of correlations near a saddle-node bifurcation point in the PLASIM model. (a) The model's bifurcation diagram as a function of the value of the solar irradiance $S$. The critical transitions Warm-to-Snowball and Snowball-to-Warm as well as the present-day conditions are indicated. Reproduced with permission from \citet{Luchyst}. (b) Decay of correlation of the equatorial mean surface temperature for various values of the solar irradiance $S$. The decay becomes considerably slower as $S$ approaches the critical value of $S \simeq 1265$~Wm$^{-2}$. Reproduced with permission from \citet{Tantet2018}.}
\label{correlation_tantet}
\end{figure}

The second example shows the importance of correlation slowdown near saddle-node bifurcation points in the PLASIM model discussed already in the previous subsection. This intermediate-complexity 
model is multistable: its bifurcation diagram with respect to the solar insolation parameter $\mu$ is very similar to Fig.~\ref{Fig_1} 
of Sect.~\ref{ssec:radiation} 
see \citet{Luchyst,Lucarini2013a}. PLASIM's bifurcation diagram is reproduced in Fig.~\ref{correlation_tantet}a here, in which the present climate $S=S_0=1360$~Wm$^{-2}$ is denoted by a bullet $\bullet$ and the bifurcation points associated with the transition from the warm to the snowball state ($W-SB$) and from the snowball to the warm state ($SB-W$) are indicated by solid arrowheads $\blacktriangle$. 
For comparison with  Fig.~\ref{Fig_1}, 
note that $\mu = S/S_0$ there.

\citet{Tantet2018} investigated PLASIM's response to changes in the solar irradiance for $S<S_0$. They found that as the $W-SB$ transition nears, the lag correlation for a large-scale observable --- namely the average equatorial near surface air temperature --- decays more and more slowly, as a result of a narrowing spectral gap. Indeed, near the saddle-node bifurcations, the response of the system to perturbations is greatly amplified, and, by definition, becomes singular exactly at the bifurcation point.\footnote{\citet{scheffer2009early} have proposed a set of ``warning signals'' when a dynamical system approaches a critical transition. Correlation slowdown is one of the main ones and it is verified in the present case for the back-to-back saddle-node bifurcations that make up the backbone of the PLASIM model's hysteresis cycle. \citet{Colon.ea.2015}, though, have found that --- in a predator--prey system, modeled by an ODE system as well as by an agent-based mode --- more complex critical transitions can behave rather differently than in the by-and-large saddle-node--bound early warning literature.}

This second example leads us naturally to the investigation of the  climate system's global stability properties and of its critical transitions, which are the the subject of the next section. 

%
%


\section{Critical Transitions and Edge States}
\label{critical}

In Sect.~\ref{climatevariability}, 
we introduced several types of bifurcations that involve transitions between two or more regimes: saddle-node bifurcations whose pairing can lead to coexistence of two stable steady states (fixed points); pitchfork bifurcations that, in the presence of a mirror symmetry, can lead from one stable symmetric steady state to the coexistence of two steady states that are mirror images of each other; Hopf bifurcations that lead from a stable steady state to a stable periodic solution (limit cycle) and then again from the stable limit cycle to a stable torus on which an infinity of quasi-periodic solutions live; and nonlocal bifurcations associated with the existence of homoclinic and heteroclinic orbits that lead on to chaotic regimes. 

Successive bifurcation scenarios \citep{Eckmann1981, Ghil1987, DG05,  dijkstra2013} that involve several of these bifurcations and additional ones lead from solutions with high symmetry in space and time to successively more complex and chaotic ones. We saw in Sect.~\ref{climatevariability} 
that such bifurcation scenarios shed considerable light on the phenomenology of large-scale atmospheric, oceanic and coupled ocean--atmosphere flows. Finally, in Sects. \ref{multiplescales} and \ref{ssec:CR}, 
we introduced stochastic effects into the nonlinearly deterministic setting recalled above, and outlined the role of pullback attractors (PBAs) and of the invariant measures they support in the theory of nonautonomous and random dynamical systems (NDSs and RDSs).

Recently, the interest in bifurcations in the climate and environmental sciences has greatly increased due to the introduction of the concept of tipping points from the social sciences \citep{Gladwell00, Lenton.tip.08}. Clearly, a tipping point sounds a lot more threatening than a bifurcation point, especially when dealing with a hurricane or a dramatic and irreversible climate change. 

Beyond the linguistic effectiveness of the term, it does also generalize the bifurcation concept in the context of open systems that are modeled mathematically by NDSs or RDSs. As we saw in Sects.~\ref{climatevariability} and \ref{climatedynamics}, 
the climate system  --- as well as its subsystems, namely the atmosphere, oceans, biosphere and cryosphere ---  are open and exchange time-dependent fluxes of mass, energy and momentum with each other and with outer space. Hence it is quite appropriate to consider this generalization. Following \citet{Ashwin2012}, one distinguishes among three kinds of tipping points, as shown in Table~\ref{table:TPs}.

The investigation of systems possessing multiple attractors --- which may include fixed points, limit cycles, strange attractors as well as PBAs or random attractors --- is an active area of research, encompassing mathematics, as well as the natural and socio-economic sciences. \citet{Feudel2018}, for instance, provide a brief review that introduces a special issue of the journal {\it Chaos} on the topic 
and sketch several interesting examples. 

This section is devoted to providing a general framework for the study of multistability in the Earth system, to forced transitions between different regimes when multistability is present, and to the critical transitions taking place at classical bifurcation and more general tipping points. Recall, for motivation, two cases of multistablility in the climate system. 

First, planet Earth as a whole has boundary conditions that arise from its space environment 
in the solar system, which are compatible with at least two competing, stable regimes: today's relatively ice-free and warm climate and a deep freeze or snowball climate; see Sect.~\ref{ssec:radiation}. 
Second, the oceans' THC is --- over a rather broad range of the parameters that control its heat and freshwater budget --- bistable, with the current active state coexisting with a greatly reduced or even reversed circulation; see Sect.~\ref{sec:THC}. 

In order to illustrate the main mathematical and physical aspects of the problem, we choose at first the specific example of coexistence between the warmer, Holocene-like, and the much colder, snowball state of planet Earth, as discussed in Sects. \ref{ssec:radiation} and \ref{tantet}; 
see, in particular, Figs. \ref{Fig_1} and \ref{correlation_tantet}a.

\begin{table*}
{\begin{tabular}{p{0.7in}p{1.5in}p{3.2in} }
\toprule
Type & Cause & Mechanism \\
\colrule
B-Tipping & Bifurcation-due tipping & Slow change in a parameter leads to the system's passage through a classical bifurcation\\

N-Tipping & Noise-induced tipping & Random fluctuations lead to the system's crossing an attractor basin boundary\\

R-Tipping & Rate-induced tipping & Due to rapid changes in the forcing, the system loses track of a slow change in its attractors\\

\botrule
\end{tabular}
\caption{Tipping points in open systems; see also \citet{Ashwin2012} and Fig.~\ref{fig:TPs}. \label{table:TPs}}
}
\end{table*}

\subsection{Bistability for Gradient Flows and EBMs}\label{ssec:gradient}

The theory of SDEs and of RDSs provides a comprehensive framework for deriving the probability of occurrence of coexisting regimes in multistable systems and for estimating the probability of stochastic-forcing--triggered transitions between them. A good starting point is gradient flow in the presence of additive white noise. 

The governing SDE in this case is:
\begin{equation}
\diff x =-\nabla_x V(x) \diff t +\epsilon \diff W, \label{ito}
\end{equation}
where $x\in\mathbb{R}^d$, the potential $V:\mathbb{R}^d\rightarrow \mathbb{R}$ is sufficiently smooth, and $\mathrm{d}W$ is a vector of $d$ independent increments of Brownian motion, while $\epsilon$ determines the strength of the noise. 
The Fokker-Planck equation associated with Eq.~\eqref{ito} describes the evolution of the PDF $p_\epsilon(x,t)$ of  an ensemble of trajectories obeying the SDE. 

The stationary solution corresponding to the latter equation's invariant measure $\lim_{t\rightarrow \infty} p_\epsilon(x,t)=p_\epsilon(x)$ is given by a large-deviation law \citep{Varadhan.1966}: 
\begin{equation}
p_\epsilon(x)=\frac{1}{Z}\exp\left(-\frac{2V(x)}{\epsilon^2}\right) \label{invariant}
\end{equation}
where $Z$ is a normalization constant.
The local minima of the potential $V$ are the stable fixed points of the deterministic dynamics. One obtains them by setting $\epsilon=0$ in the SDE~\eqref{ito}, and they  correspond to the local maxima of $p$. Thus, for instance, the double-well potential of Fig.~\ref{fig:double_well} corresponds to a bimodal PDF. 

\begin{figure}
\includegraphics[angle=0,trim=0cm 0cm 0cm 0cm, clip=true, width = 0.9\columnwidth]{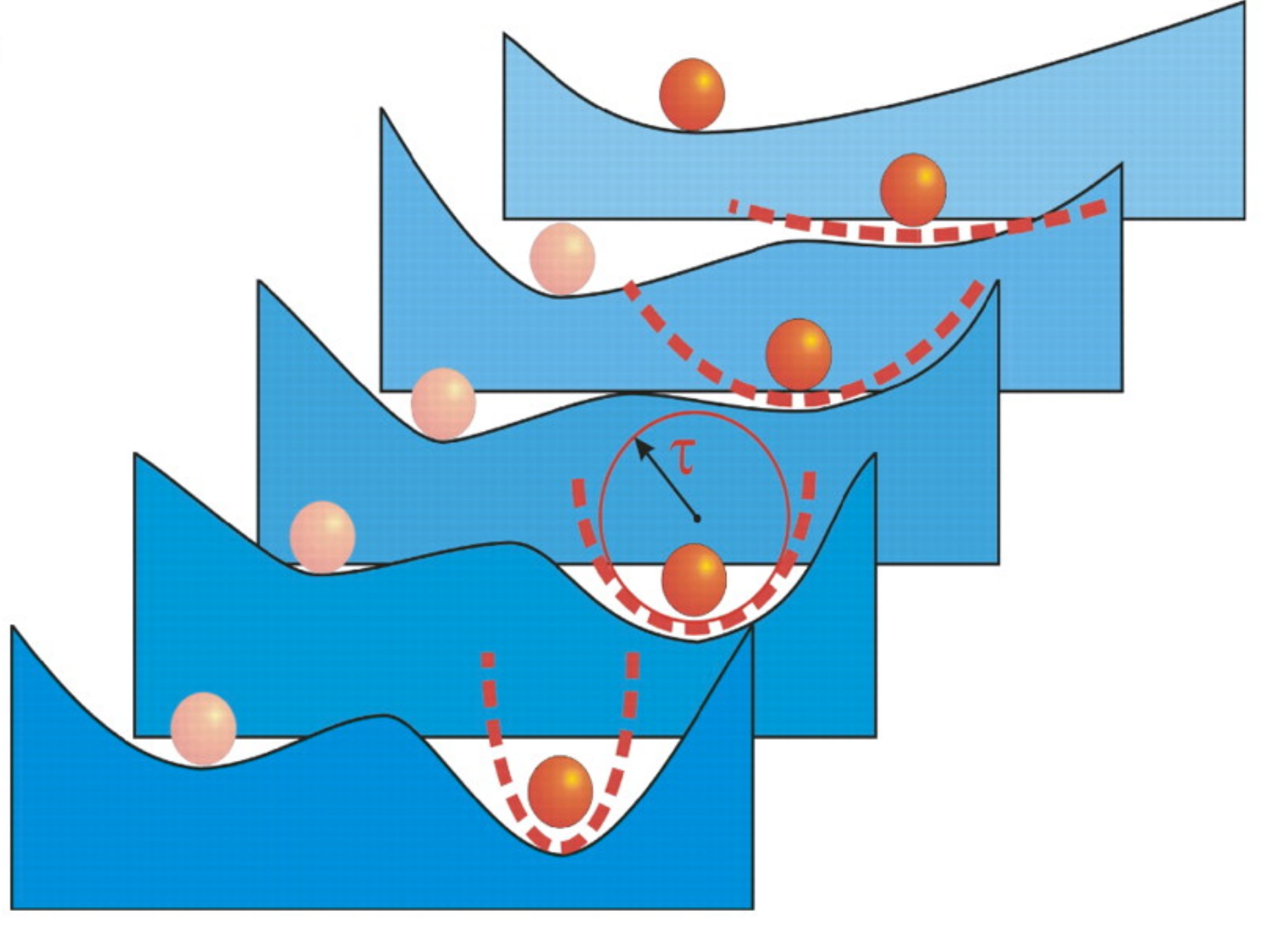}
\caption{Schematic diagram of B-Tipping: as a control parameter is being changed slowly, the number of stable equilibria is reduced from two to one. Reproduced with permission from \citet{Lenton.tip.08}.}
\label{fig:TPs}
\end{figure}

The mountain pass lemma \cite[e.g.,][]{Bisgard2015} states that the two minima of $V$ that give rise to the two stable fixed points have to be separated by an unstable one of the saddle type, which is a maximum of $V$ in one direction and a minimum in all the other ones. Such a saddle looks like a mountain pass on a topographic map, hence the name of the lemma. \citet[Sec.~10.4]{Ghil1987} discussed its application to 1-D EBMs. 

In the limit of weak noise, trajectories starting near a local minimum of $V$ located at $x=x_1$ typically wait a long time before moving to the neighborhood of a different local minimum of $V$. The transitions are most likely to occur through the lowest energy saddle $x=x_{\rm s}$ that links the initial basin of attraction to any other basin. Let us call the corresponding local minimum $x=x_2$. 

The optimal path linking $x_1$ to $x_{\rm s}$ minimizes the \citet{Freidlin1984} action and is called an instanton. In the weak-noise limit $\epsilon\rightarrow 0$, the persistence time inside the basin of attraction of $x=x_1$ is 
\begin{equation}\label{eq:tau}
\bar{\tau}_\epsilon \propto \exp\left(\frac{2(V(x_{\rm s})-V(x_1))}{\epsilon^2}\right),
\end{equation}
which is referred to as the \citet{Kramers1940} formula.


Assume now, for simplicity, that we deal with a bistable system, as in Fig.~\ref{fig:double_well}  
of Sect.~\ref{ssec:radiation}. 
If the potential $V$ depends on a slowly varying control parameter $\phi$, a B-tipping would involve, for instance, the number of local minima decreasing from $2$ to $1$, as a local minimum merges with a saddle point at $\phi=\phi_{\rm c}$. This merging is illustrated in Fig.~\ref{fig:TPs} for a simple 1-D case, like that of the 0-D EBM governed by Eqs.~\eqref{eq:rad_bal_0}\footnote{Note that the number of dimensions in phase space is 1, while that of the model in physical space is 0.}. 

When the system nears the tipping point, the persistence time in the shallow, metastable minimum is reduced, because $\Delta V \equiv V_{\rm s} - V_1 \rightarrow 0$. Another flag anticipating the tipping point is the increase of the autocorrelation time $\tau$ in the metastable state, which is proportional to the inverse of the second derivative of the potential evaluated at the minimum, $\bar \tau_{\epsilon} \propto 1/V''_{x_1}$. 

These two easily observable phenomena are the simplest manifestation of the slowing-down process \cite{scheffer2009early}, which we have discussed already in Sect.~ \ref{tantet} when describing the findings of \citet{Tantet2018}. As indicated in the footnote there, the warning signals of tipping-point approach have been mostly studied in the B-tipping case for saddle-node bifurcations. Much remains to be done in this context for more complex situations that may involve N-tipping, R-tipping or global bifurcations rather than local ones. For instance, \citet{Ditlevsen2010} analyzed a high-resolution ice core record and excluded the possibility of interpreting Dansgaard-Oeschger events as B-tipping by noting the lack of the two early warning signals mentioned above. Their results leave us with the alternative of interpreting the associated irregular oscillations \cite[e.g.,][]{Boers.ea.2018} either as the result of N-tipping for an S-shaped back-to-back saddle-node bifurcation or as B-tipping for a Hopf bifurcation; see Fig.~\ref{fig:part_wave} and its discussion in Sect.~\ref{sssec:LFV} for similar ambiguities in interpreting changes in atmospheric LFV.

\begin{figure}
\centering
a)\includegraphics[angle=270,trim=0cm 0cm 0cm 0cm, clip=true, width = 0.8\columnwidth]{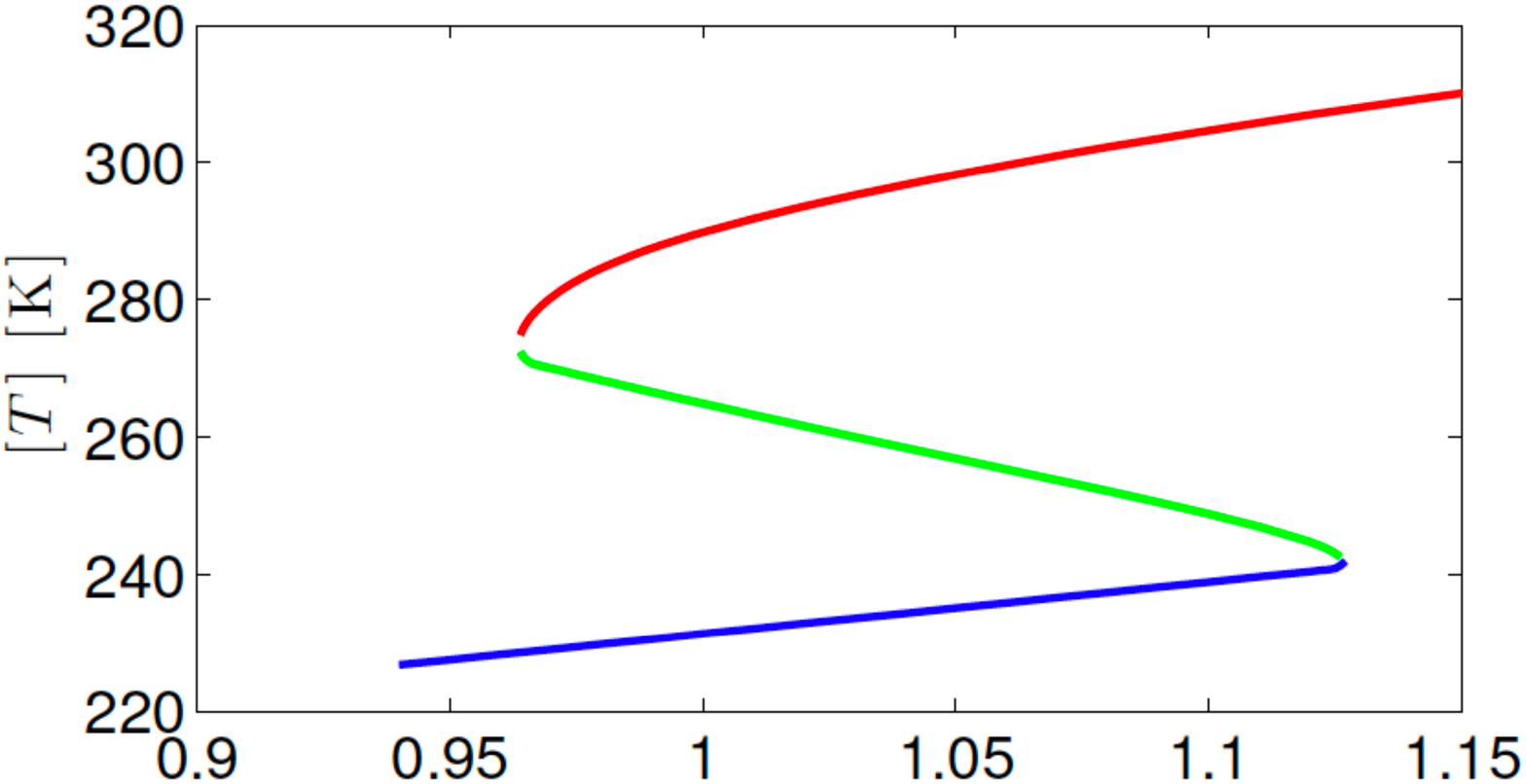}\\
b)\includegraphics[angle=0,trim=0cm 15cm 0cm 0cm, clip=true, width = 0.8\columnwidth]{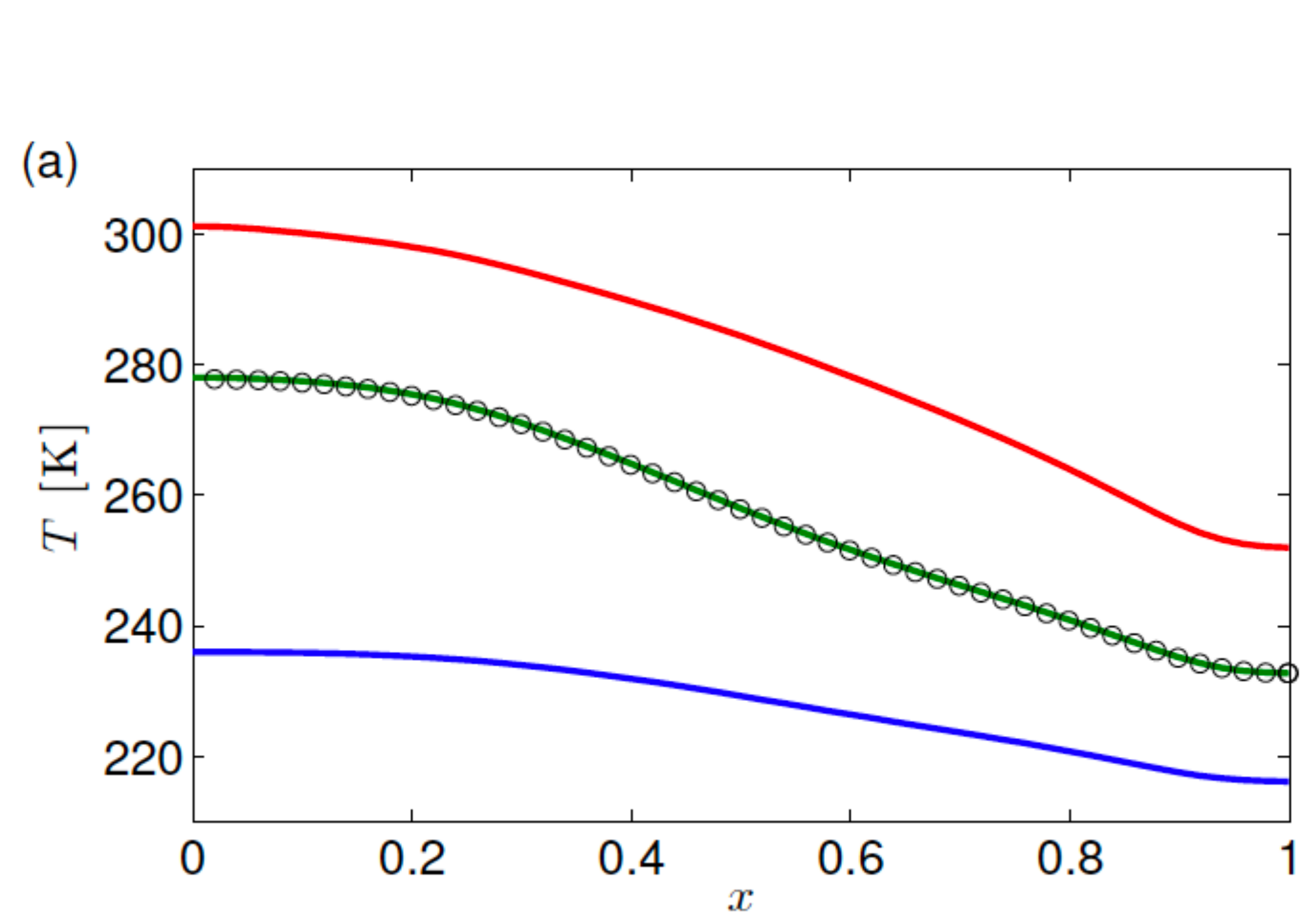}
\caption{Bistability for the Ghil-Sellers 1-D EBM. (a) Simplified bifurcation diagram for the \citet{Ghil1976} 1-D EBM; compare Figs.~\ref{Fig_1} and \ref{correlation_tantet}.  The red, blue, and green segments correspond to the globally averaged surface air temperature for the warm, snowball, and Melancholia state, respectively, as a function of the relative solar irradiance $\mu$; see also \citet{Diaz.ea.1998}. (b) Meridional temperature profile for the warm (red), snowball (blue), and $M$-state (green) for this model under present-day conditions ($\mu=1$), with the sine of latitude on the $x$-axis; compare \citet[Fig.~3a]{Ghil1976}. The circles indicate the $M$-state estimate obtained using the edge tracking method. Adapted with permission from \citet{Bodai2015}.}
\label{GSmodel1}
\end{figure}

Returning to the right-hand side of the 0-D EBM of Eq.~\eqref{0DEBM}, it can be written as minus the derivative of a potential $V(T)$, cf.~\citet{Ghil1976}. The gain factor $\Lambda$ given in Eq.~\eqref{eq}, as well as the ECS, is proportional to the inverse of the second derivative of the potential $V$ evaluated at the local minimum defining the reference climate $T_0$, $ECS \propto \Lambda \propto 1/V''_{T_0}$. Hence, near the critical transition, a small forcing in the right direction can lead the system to jump to the other basin of attraction, even in the absence of noise. Note, though, that, near such a transition, the linear stability analysis performed in Eqs.~(\ref{eq0}, \ref{eq}) is no longer valid, and an accurate quantitative analysis of climate response requires taking into account the essential nonlinearity in the problem; see the \citet{ghil2010} comments on \citet{Roe2007}. 
 
Determining the saddle point that potentially connects, via instantons, two local minima is not at all trivial when looking at high- or even infinite-dimensional systems. Figure~\ref{GSmodel1} reports the findings of \citet{Bodai2015} on the \citet{Ghil1976} 1-D EBM.  Panel (a) shows the bifurcation diagram of the model, in which the control variable is the globally averaged surface air temperature $\bar T$, and the tuning parameter is the insolation parameter $\mu = S/S_0$ introduced in Eq.~\eqref{eq:R_i}.

Panel (b) shows the zonally averaged temperatures of the 1-D model's steady states, for the reference conditions $T = T(x)$ at $\mu = 1$. The red and blue lines represent the two stable solutions --- warm ($W$) and snowball ($SB$), respectively --- while the green line is the unstable solution lying in-between. Note that, at each tipping point, the unstable state comes in contact with a stable one, and then both disappear, in accordance with the scenario of basin boundary crisis \cite{Ott2002}. In fact, the lack of stability of the steady-state solution described by the green line is apparent by observing that $\bar T$ in Fig.~\ref{GSmodel1}a decreases with increasing solar radiation --- as opposed to the $W$ and $SB$ states --- which is clearly not physical \citep{Ghil1976, North1981}
and can be loosely interpreted as the result of a negative heat capacity; see discussion in \citet{Bodai2015}.

\subsection{Finding the Edge States}\label{ssec:melancholia}

In one phase space dimension, every first-order system $\dot x = f(x; \mu)$ is a gradient system, since one can always write $V(x; \mu) = -\int f(\xi, \mu) \diff \xi$. This is not so in two or more dimensions and one might wonder whether it is possible to  compute unstable solutions for more general multistable systems, with no gradient property. \citet{Scott1999} presented an example of such an unstable stationary solution in a simple box model of the THC featuring competing stable solutions and no gradient structure. 
Moreover, the actual computation of the unstable steady state in \citet[Fig.~3a]{Ghil1976} did not rely on the system's potential $V(T(x))$ but on its 1-D character in physical space and on a shooting method to solve the Sturm-Liouville equation that results when setting $\partial T(x,t)/\partial t \equiv 0$ in the Ghil-Sellers model. Neither of these approaches \cite{Ghil1976, Scott1999} can easily be extended to more general systems.

In the remainder of this section, we discuss a more general paradigm of multistability for determininistic systems and, subsequently, how this paradigm may be useful for studying 
the problem of noise-induced transitions between competing states, without the simplifying assumption of a gradient, as in Eq.~\eqref{ito}.

\begin{figure}
\includegraphics[angle=270,trim=0cm 0cm 0cm 0cm, clip=true, width = 0.9\columnwidth]{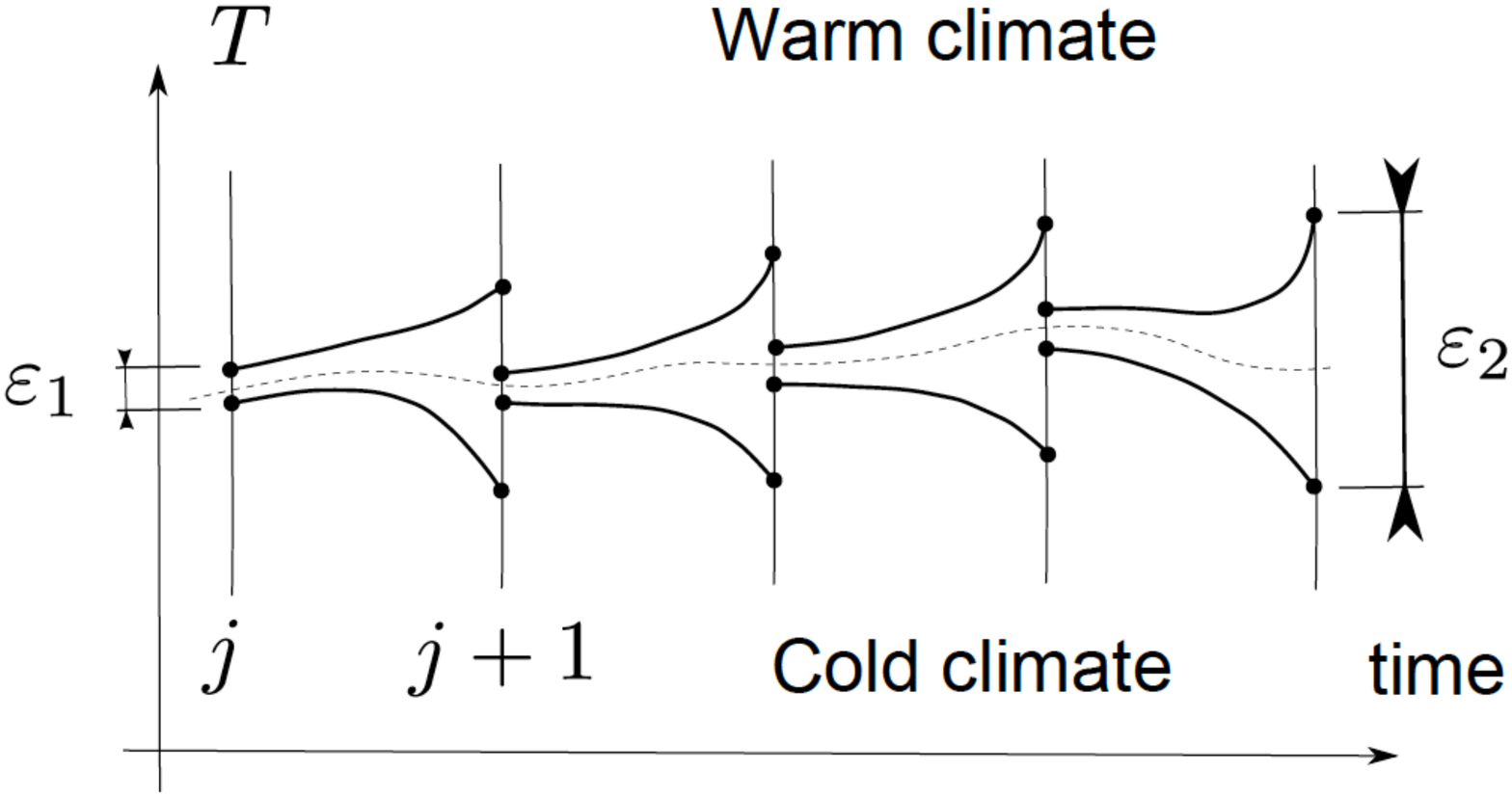}
\caption{Schematic diagram of the edge tracking algorithm proposed by \citet{Skufca2006} and \citet{Schneider2007}, as applied to the 1-D EBM of Fig.~\ref{GSmodel1}. Reproduced with permission from \citet{Bodai2015}.}
\label{edgetracking}
\end{figure}

One can formally introduce general multistable systems as follows.   We consider a smooth autonomous continuous-time dynamical system acting on a smooth finite-dimensional compact manifold $\mathcal{M}$, described by a ODE system of the form ${\dot{{x}}}={F}({x})$. 
The system is multistable if it possesses more than one asymptotically stable state, defined by the attractors $\{\Omega_j: j = 1, \ldots, J\}$. The  asymptotic state of an orbit is determined by its initial state, and the phase space is partitioned between the basins of attraction $\{B_j\}$ of the attractors $\{\Omega_j\}$ and the boundaries $\{\partial B_\ell$: $\ell = 1,\ldots,L\}$ separating these basins. 

The basin boundaries can be strange geometrical objects with codimension smaller than one. Orbits initialized on the basin boundaries $\{\partial B_\ell\}$ are attracted towards invariant saddles $\{\Pi_\ell{_k}$: $\ell = 1,\ldots,L$, $k = 1,\ldots,k_\ell\}$, where we allow for the existence of  $k_\ell$ separate saddles in the basin boundary $\partial B_\ell$; these saddles are often referred to as edge states \cite[cf.~][]{Skufca2006,Schneider2007}.\footnote{Note that, strictly speaking, edge states refer to solutions separating long-lived turbulence and laminar flow, rather than describing the interface between truly separated attractor basins.} For nongradient flows, the edge states, like the asymptotic states, can feature chaotic dynamics \cite{Grebogi1983,Robert2000,Ott2002,Vollmer2009}. \citet{Lucarini2017N,Lucarini2019,LucariniBodai2019arxiv} refer to the chaotic edge states as Melancholia ($M$) states. Hence, we cannot expect to find easily edge states and, a fortiori,  $M$-states for such general multistable systems. 

The edge tracking algorithm of B.~Eckhart, J.~A.~Yorke and associates \cite{Skufca2006,Schneider2007} allows one to do so by constructing a shadowing trajectory that leads an orbit starting on the basin boundary toward the corresponding edge state, cf. Fig.~\ref{edgetracking}; therein one uses a bisection method to control the instability associated with the trajectories' diverging away from the basin boundary. \citet{Bodai2015} first used edge tracking in a geophysical context to reproduce the unstable solution of \citet{Ghil1976} with this more easily generalizable approach; compare green line and circles in Fig.~\ref{GSmodel1}b. 

\begin{figure}
\includegraphics[trim=1cm 0cm 0cm 1.22cm, clip=true, width=0.9\columnwidth]{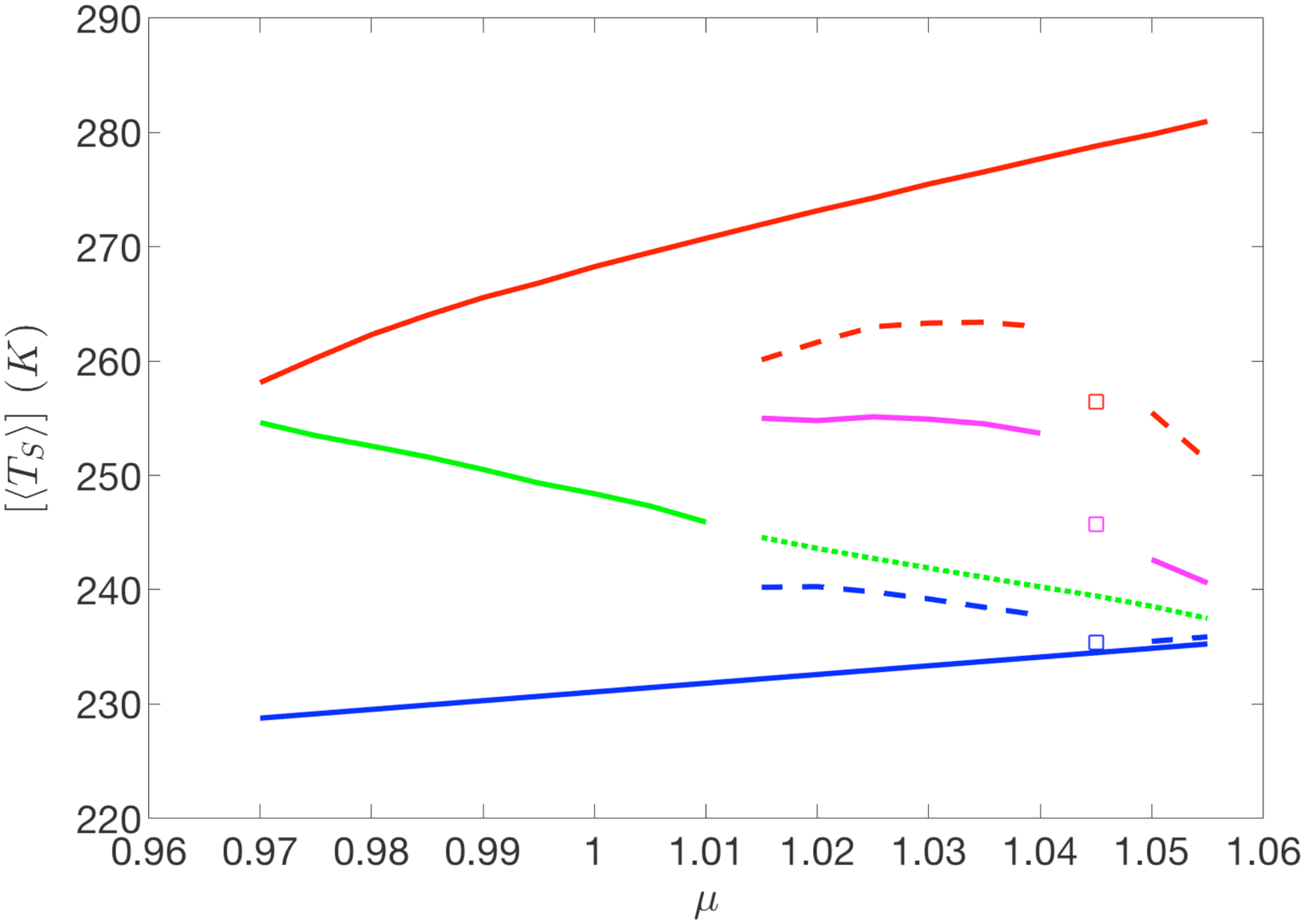}%
\caption{Bifurcation diagram for the intermediate-complexity climate model in \citet{Lucarini2017N}, drawn for the long-term, globally averaged ocean temperatures $[\langle T_S\rangle]$.  
Bistability is found for a large range of values of the control parameter $\mu = S/S_0$. 
The relevant solid lines are red for the Warm $W$-states, blue for the Snowball $SB$-states, and green for the Melancholia $M$-states, as in Fig.~\ref{GSmodel1}. 
Reproduced with permission from \citet{Lucarini2019}.}\label{fig1}
\end{figure}

\citet{Lucarini2017N} computed $M$-states for an intermediate complexity climate model with O($10^5$) degrees of freedom that couples PUMA, an atmospheric primitive equations model \cite{Frisius1998} with a modified version of the 1-D EBM of \citet{Ghil1976} that acts as a surrogate ocean and contributes to meridional heat transport. Figure~\ref{fig1} shows the bifurcation diagram of this climate model, where the $W \rightarrow SB$ tipping point is located near $\mu=0.98$, while the $SB \rightarrow W$ tipping point is located near $\mu=1.06$. Just as in the 1-D EBM of Fig.~\ref{GSmodel1}, the tipping points are associated with basin boundary crises \cite{Ott2002}.

Over a wide range of $\mu$-values, the edge state features chaotic dynamics that arises from the atmospheric model's baroclinic instability; it leads to weather variability and to a limited predictability horizon. Since this instability is much faster than the climatic one due to the ice-albedo feedback, the basin boundary is a fractal set with near-zero codimension, in agreement with results obtained in low-dimensional cases \cite{Grebogi1983,LT:2011}. In other words, the basin boundary has almost full Lebesgue measure. As a result, near the basin boundary there is virtually no predictability on the asymptotic state of the system, because infinitesimally close initial states have  a high probability of belonging to separate basins of attraction. 

\subsection{Invariant Measures and Noise-induced Transitions}\label{stochastic}

As mentioned above, transitions across the basin boundaries of competing attractors are possible and quite likely in the presence of stochastic perturbations. Noise-induced escape from an attractor has, in fact, long been studied in the natural sciences \cite[e.g.,][]{Hanggi1986,Kautz1987,Grassberger1989}.  


Let us generalize Eq.~\eqref{ito} as follows: 
\begin{equation}\label{eqapp}
\diff x = {F}({x})\diff t+\epsilon{s}({x})\diff W;
\end{equation}
here $x\in \R^d$, ${F}({x})\diff t$ is the drift term given by a vector flow field that admits multiple steady states, as discussed in the previous two subsections, $\diff W$ are the increments of a $d$-dimensional Brownian motion. The volatility matrix ${s}({x}) \in \mathbb{R}^{d\times d}$ is such that ${s}({x})^T{s}({x})$ is the covariance matrix of the noise. Finally, the parameter $\epsilon\geq 0 $ controls the noise intensity. 

Recent extensions of the classical Freidlin-Wentzell~\cite{Freidlin1984} theory \cite{Graham1991, Hamm1994,LT:2011} yield results that mirror closely those summarized before in the case of gradient flows with additive noise, cf. Eqs.~\eqref{invariant} and \eqref{eq:tau}, given suitable assumptions on the drift term and the volatility \cite[e.g.,][]{Lucarini2019,LucariniBodai2019arxiv}. In the weak-noise limit given by $\epsilon\rightarrow 0$, the invariant measure  of the system can be written as a large-deviation law. Formally, one just has to replace $V(x)$ by a general pseudo-potential $\Phi({x})$, which depends in a nontrivial way on ${F}(x)$ and ${s}({x})$. Moreover, the constant $Z$ in Eq.~\eqref{invariant} is replaced by a function $\zeta(x)$, which is relatively unimportant because the behavior of the system depends mostly on the properties of $\Phi(x)$. 

Certain general properties of $\Phi({x})$ apply to all choices of the noise law and, once a noise law is chosen, one can derive how the properties of the system change as a function of the parameter $\epsilon$. In general, regardless of the noise law, $\Phi({x})$ has local minima supported on the attractors $\{\Omega_j: j=1,\ldots, J\}$ of the deterministic dynamics. Correspondingly, the invariant measure has local maxima on the attractors. Moreover, $\Phi({x})$ has a constant value on the support of each edge state $\Pi_\ell$ and each attractor $\Omega_j$, respectively. 

In the simplest case of just $J=2$ attractors and $L=1$ edge states then, in the weak-noise limit, transitions away from either attractor basin take place exponentially more likely along the instanton connecting the corresponding attractor with the edge state $\Pi_1$. Just as for the gradient flows in Sect.~\ref{ssec:gradient}, instantons can be calculated as minimizers of a suitably defined action \cite{Kautz1987, Grassberger1989, Kraut2002, Beri2005}.

Consider now the case that more than one edge state separate a given $\Omega_{j}$ from the other attractors $\{\Omega_k : k\neq j\}$. Let us denote then by $\Pi_j$ the edge state for which the value of $\Phi$ is lowest. In this case, the most probable exit path connects $\Omega_{j}$ with $\Pi_j$, while the other escape channels are basically switched off in the weak-noise limit.

Note that, if the attractors and the edge states are more complex sets than isolated points, the instantons connecting them are not unique, because in principle any point of the attractor can be linked by an instanton to any point of the edge state, because the pseudo-potential is constant on $\Omega_j$ and $\Pi_j$. Indeed, the edge states are the gateways for the noise-induced escapes from the deterministic basins of attraction, as illustrated in the bistable example with nongradient flow given below.

\begin{figure}
\includegraphics[trim=2cm .5cm 1cm 1.9cm, clip=true, width=0.9\columnwidth]{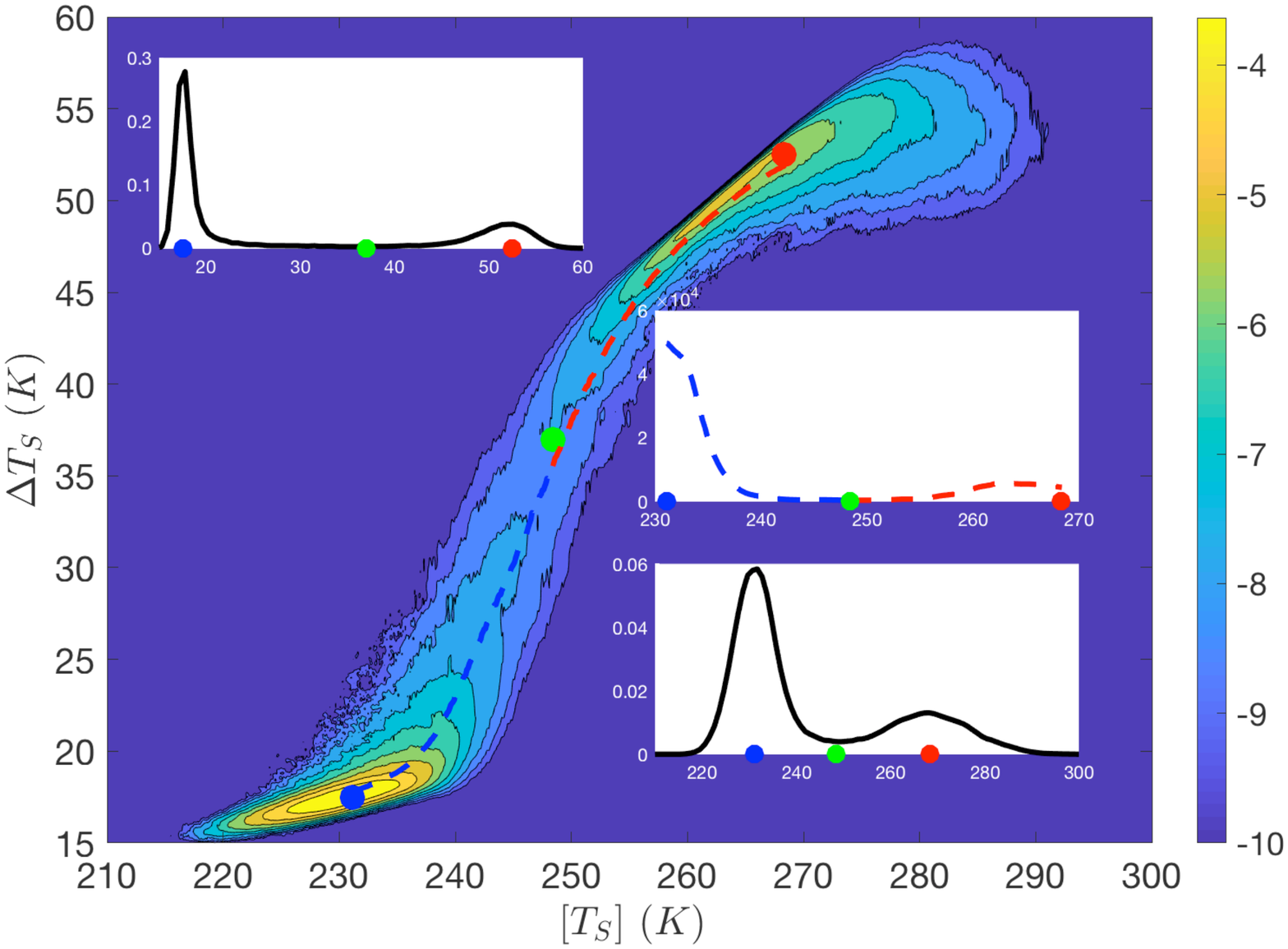}
\caption{Invariant measure of the transitions between the two stable regimes of the stochastically perturbed climate model of \citet{Lucarini2017N} via the $M$-state separating them, for $\mu=1$. 
Main graph: projection of this invariant measure on the reduced space $(T_S,[\Delta T_S])$; $W$ attractor -- red dot; $SB$ attractor -- blue dot; and $M$-state -- green dot. Red and blue dashed lines plot the $W\rightarrow SB$ and the $SB\rightarrow W$ instanton, respectively. Top left inset: marginal PDF with respect to $\Delta T_S$; bottom right inset: marginal PDF with respect to $[T_S]$; and center right inset: probability along the two instantons. Reproduced with permission from \citet{Lucarini2019}. 
} \label{lownoise}
\end{figure}

\citet{Lucarini2019} introduced stochastic forcing into the climate model studied by \citet{Lucarini2017N} as a fluctuating factor of the form $1+\epsilon \diff W/\diff t$, 
where $\diff W$ is a Brownian motion that modulates the solar insolation parameter $\mu$. This forcing yields a multiplicative noise law because the energy input into the system depends on the product of $\mu$ times the co-albedo $(1-\alpha)$, where the albedo $\alpha = \alpha(T)$ depends explicitly on the surface air temperature $T = T(x,t))$, which is a state variable of the model; compare with Eq.~\eqref{eq:R_i}. 

In Fig.~\ref{lownoise} we reproduce the estimate by \citet{Lucarini2019} of a 2-D projection of the invariant measure for $\mu=1$ from a climate simulation of  $\simeq 6.0\times10^4$~yr with fluctuations in the solar insolation of $1.5\%$ on the scale of $100$~yr. The simulation features 92 $SB\rightarrow W$ and $W\rightarrow SB$ transitions, and the measure is projected onto the $([T_S],\Delta T_S)$-plane; here $[T_S]$ is the globally averaged surface air 
temperature, and $\Delta T_S$ measures the temperature difference between the surface temperature in the low- and high-latitude regions. 

The peaks of the PDFs are very close to the $W$ and $SB$ attractors, and the agreement improves even further when considering the two marginal PDFs (top left and bottom right insets). 
Both the $W\rightarrow SB$ and the $SB\rightarrow W$ instantons can be estimated: their starting and final points agree remarkably well with the attractors and the $M$-state. The instantons follow a path of monotonic descent that tracks closely the crests of the PDF, with the minimum occurring at the $M$-state. 

These features are all in excellent agreement with the theoretical predictions for multistable systems with a generalized pseudo-potential.

\begin{figure}
\includegraphics[trim=1cm .5cm 1cm 1.9cm, clip=true, width=0.9\columnwidth]{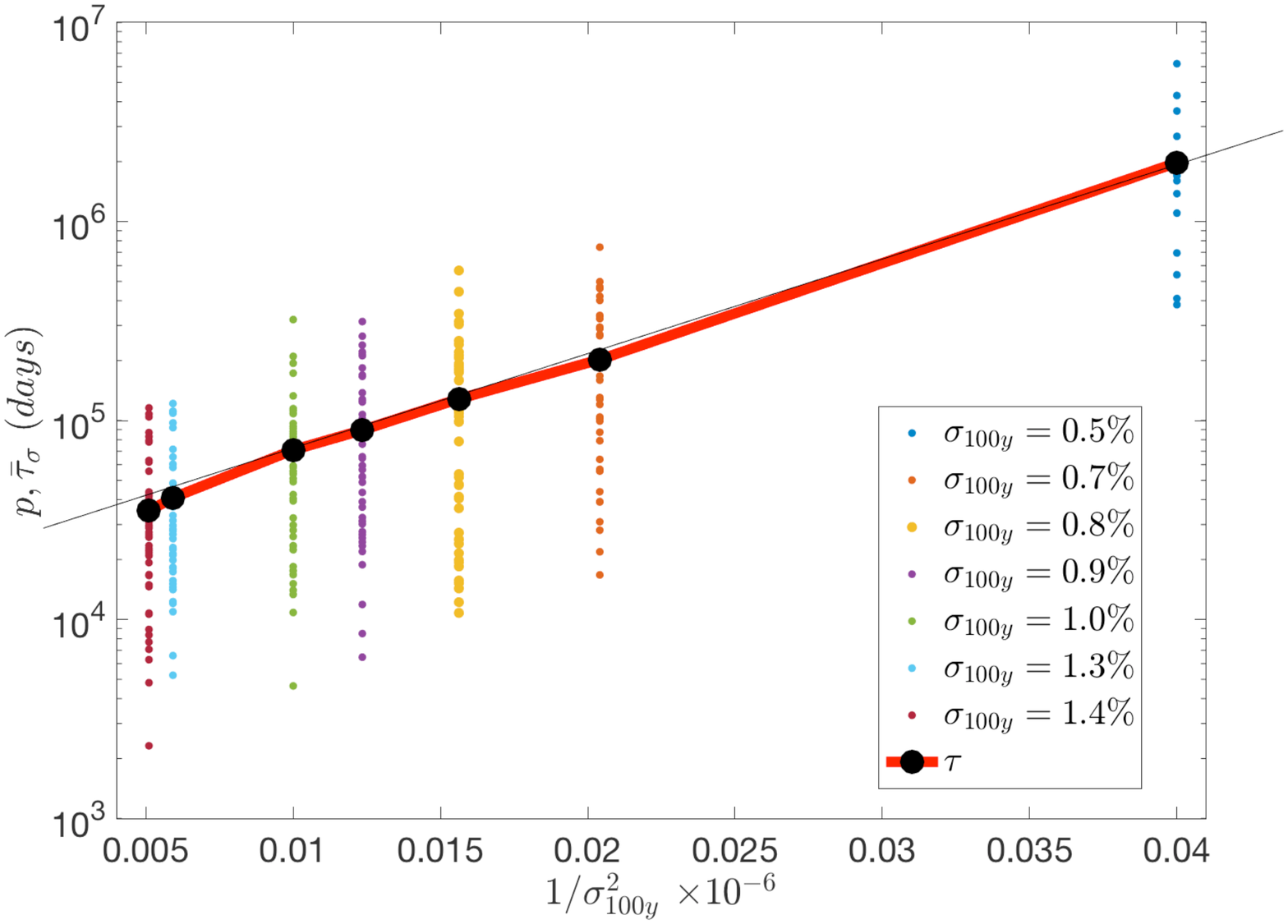}\\
\includegraphics[trim=1cm .5cm 1cm 1.9cm, clip=true, width=0.9\columnwidth]{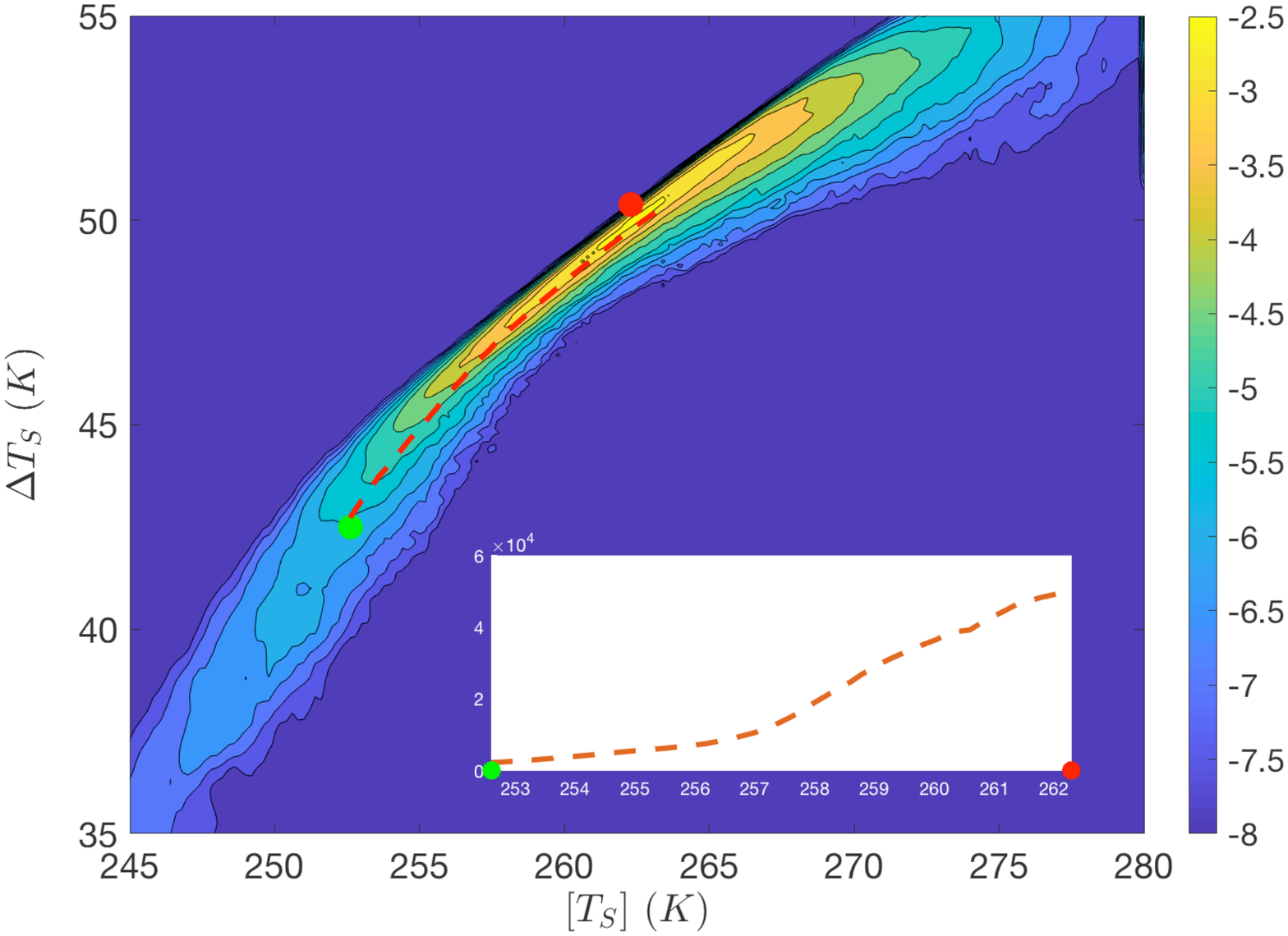}
\caption{Escape from the $W$ attractor near the $W \rightarrow SB$ critical transition of the stochastically perturbed climate model of \citet{Lucarini2017N}, for $\mu=0.98$. (a) The expected value of the transition time obeys Eq.~\eqref{eq:tau} in the weak-noise regime; the diagram is in log--linear coordinates and $\sigma$ on the $x$-axis is the relative fluctuation of the solar insolation $\mu$. (b) Estimate of the $W \rightarrow SB$ instanton and, in the inset, empirical density; both used 50 trajectories that escape to the $SB$ state. Reproduced with permission from \citet{Lucarini2019} \label{mu098tau}}
\end{figure}

\subsection{Nearing Critical Transitions}\label{nearcritical}
The general framework outlined in the previous subsection is also quite useful for studying the properties of such systems near a critical transition \cite[e.g.,][]{Lucarini2019}. In Fig.~\ref{fig1}, it is clear that $\mu=0.98$ is close to the $W\rightarrow SB$ tipping point.  One finds that at low noise intensities --- i.e., at a relative $\mu$-fluctuation smaller or equal to 1.4\% on a centennial scale --- it is extremely hard to escape from the $SB$ state. This finding happens to agree with snowball state simulations that used more detailed models as well \cite[e.g.,][]{Crowley.ea.2001, Pierrehumbert.2004, Ghil.2019}

Let us then focus on the escape from the $W$-state. \citet{Lucarini2019} estimated the expected value of the transition time from the $W$ to the $SB$ state using 50 simulations per noise strength value. These values grow exponentially with the inverse of the square of the parameter $\epsilon$, as predicted by the theory; see Fig. ~\ref{mu098tau}a. Note that the difference between the pseudo-potential value $\Phi$ at the $M$-state and at the $W$-state is equal to half the slope of the straight line, in agreement with Eq.~\eqref{eq:tau}. 

Finally, the escape transition paths from the $W$-state to the $M$-state are plotted in Fig. ~\ref{mu098tau}b. In the weak-noise case of relative $\mu$ fluctuation smaller than 1\% on a centennial scale, the highest densities of these paths lie quite close to the instanton connecting the $W$ attractor to the $M$-state, and follow a path of decreasing probability.

These encouraging results suggest that the underlying methodology of edge tracking and instanton estimation could be applied to observational and reanalysis data sets. As discussed in Sect. \ref{sssec:LFV} 
the episodic, or `particle', approach to atmospheric LFV results in a Markov chain of regimes $\{R_j: j = 1, \ldots,J\}$ that involves preferential transition paths $\{\Theta_{j,k}: j = 1, \ldots,J\;  k = 1, \ldots,J\}$ between them. Heretofore, these transition paths, as well as the regimes themselves, were estimated by purely statistical methods; see \citet{Ghil.ea.S2S} and Fig. \ref{fig:part_wave}a 
herein. \citet{Deloncle.ea.2007} and \citet{Kondrashov.ea.2007} applied a random-forest algorithm \cite{Breiman.2001} to find the best real-time predictors of the next regime in a Markov chain, conditional upon the one currently occupied. The results were quite satisfying for the QG3 model of \citet{Mar.Mol.1993} and encouraging for an observational data set of 55 winters of Northern Hemisphere 700-hPa geopotential height anomalies \cite[respectively]{Deloncle.ea.2007, Kondrashov.ea.2007}.

\subsection{Chaos-to-Chaos Transition}\label{ssec:Chaos2Chaos}

So far, we have seen that critical transitions associated with saddle-node bifurcations and mild generalizations thereof are fairly well understood by now. We conclude this section and the main part of the text with a somewhat more exotic example of chaos-to-chaos transition for a delay differential equation (DDE) system with and without stochastic perturbations.


We saw in Sect. \ref{sssc:survey} and \ref{coupledmode}
that ENSO plays a key role in the global climate on interannual-to-interdecadal scales. Hence, a large number of relatively simple models thereof exist to better understand its main features. Important mechanisms involved are air--sea interaction, equatorial wave dynamics, and radiative forcing by the seasonal cycle \cite{Bjerknes1969}. In particular, the role of the wave dynamics has been captured by introducing one or two delays into the governing equations of some of the simpler models \cite[e.g.,][and references therein]{Ghil.ea.2008a, Tziperman.ea.1994}.

\citet{Chekroun.ea.2018} studied the PBA of the seasonally forced \citet{Tziperman.ea.1994} model both with and without stochastic perturbations. The model has two delays, associated with a positive and a negative feedback; these delays are based on the basin-crossing times of the eastward-traveling Kelvin waves and the westward-traveling Rossby waves. The control parameter $a$ is the intensity of the positive feedback and the PBA undergoes a crisis that consists of a chaos-to-chaos transition; the authors refer to it as a strange PBA since model  behavior is chaotic within it in the purely deterministic case.

\begin{figure*}
\subfloat[High-variance PBA; $\delta = 0$]{
    \label{fig:noisy_PBA} 
    \centering
   \includegraphics [height=0.4\columnwidth, width=0.45\columnwidth]{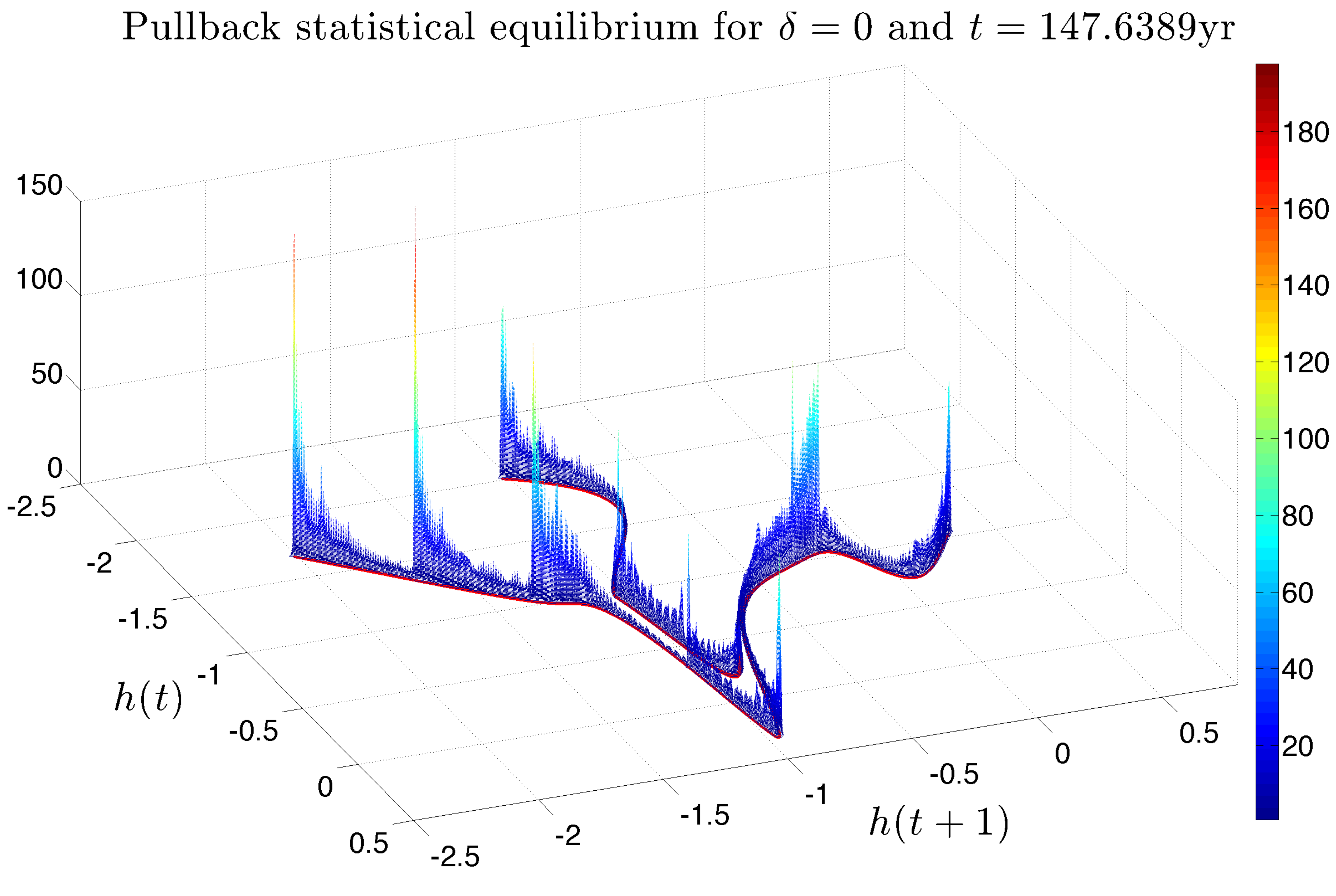}
}
\subfloat[$\delta = 0.01500$]{
    \label{fig:UR_PBA} 
    \centering
   \includegraphics [height=0.4\columnwidth, width=0.45\columnwidth]{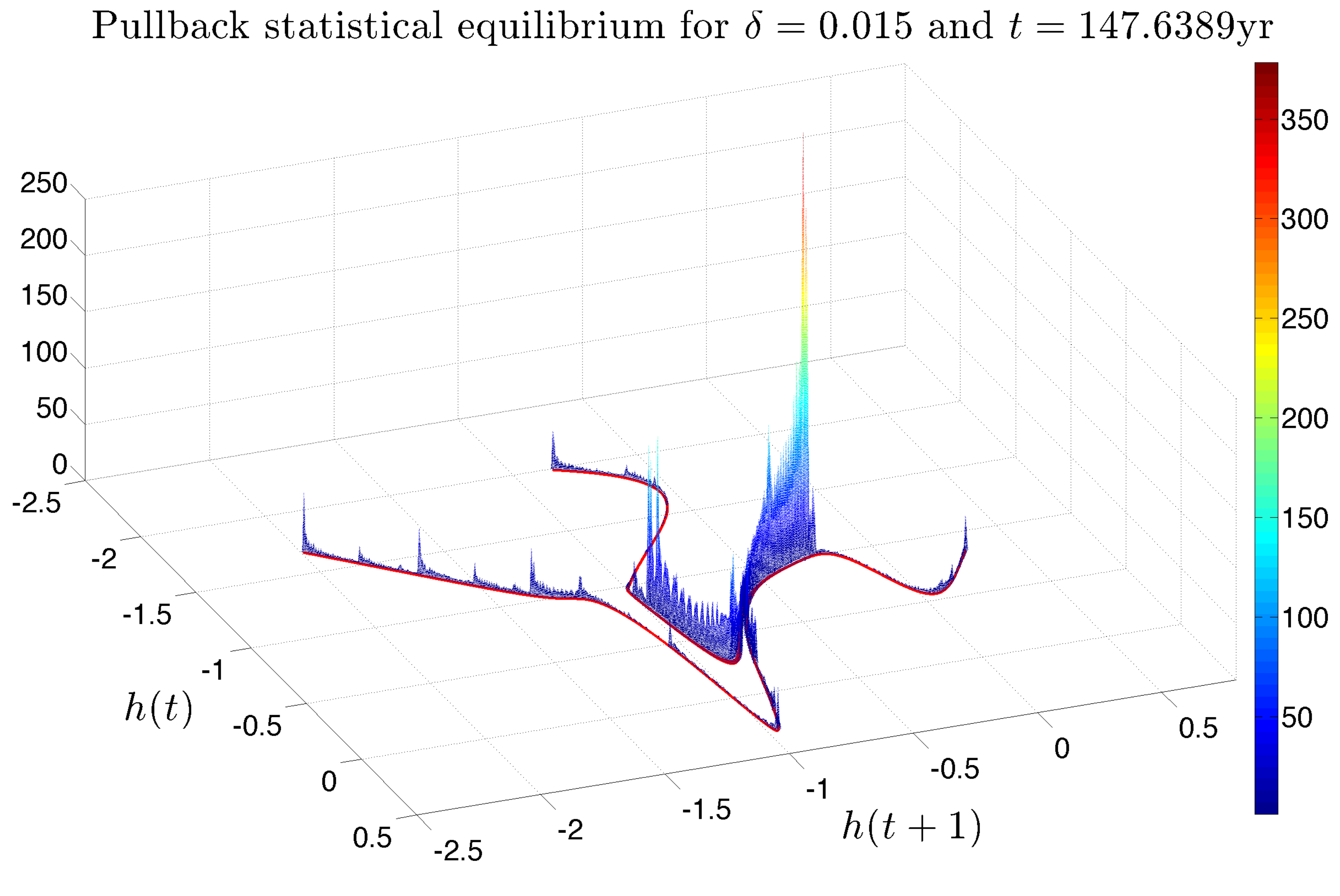}
}\\
\subfloat[$\delta = 0.0157$]{
    \label{fig:LL_PBA} 
    \centering
   \includegraphics [height=0.4\columnwidth, width=0.45\columnwidth]{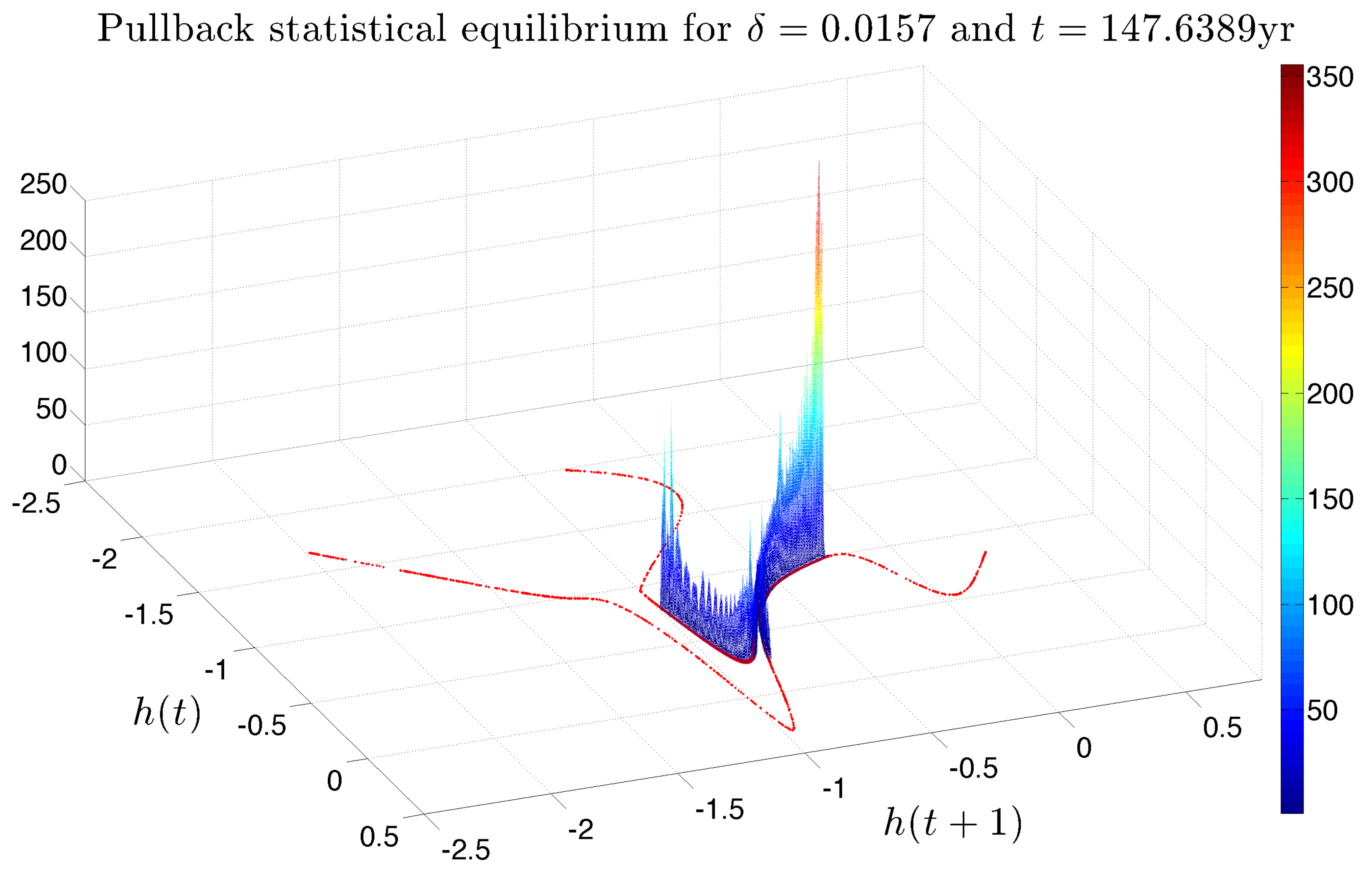}
}
\subfloat[Low-variance PBA; $\delta = 0.015707$]{
    \label{fig:quiet_PBA} 
    \centering
   \includegraphics [height=0.4\columnwidth, width=0.45\columnwidth]{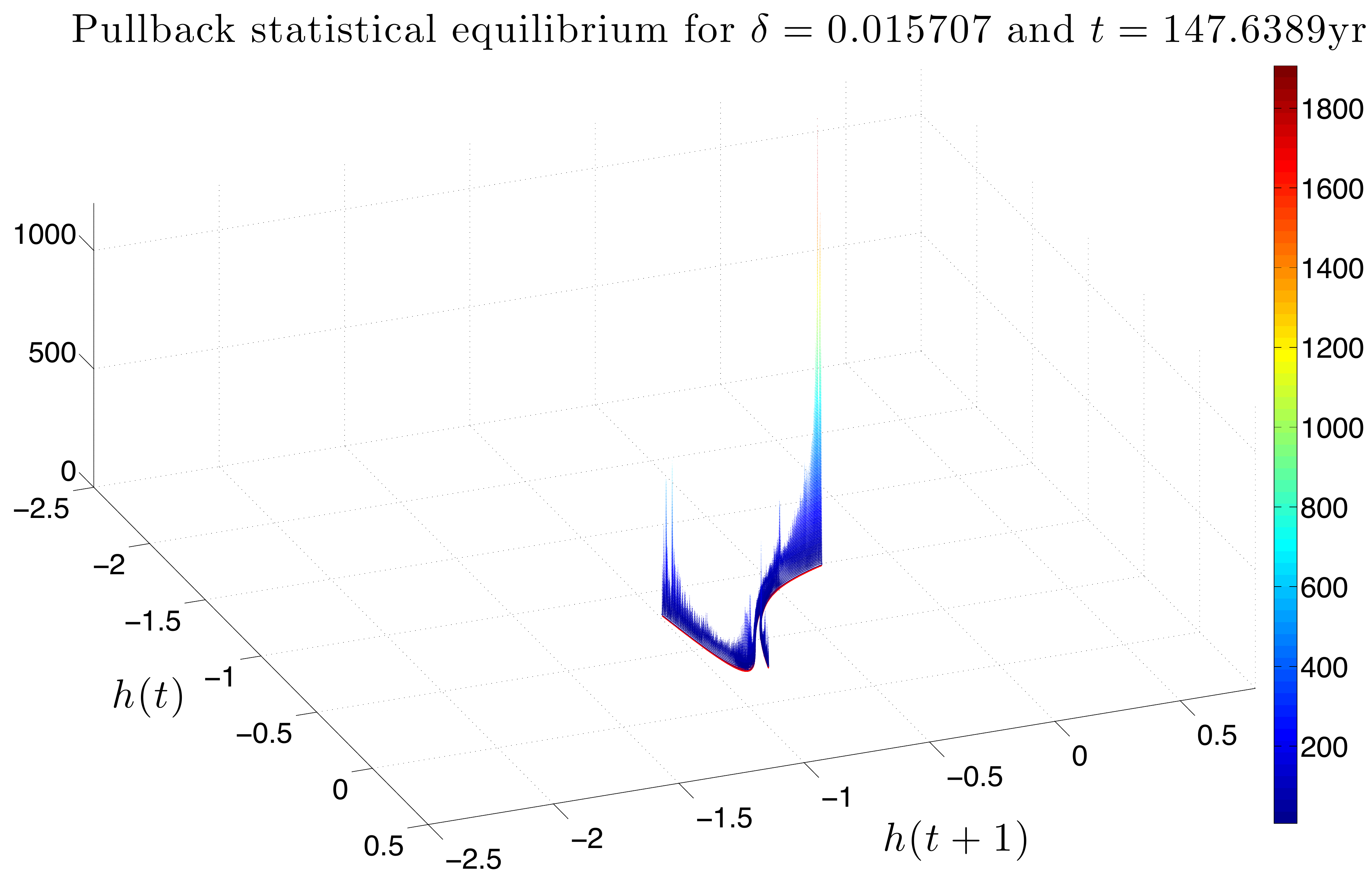}
}
 \caption{Embedding of the invariant, time-dependent measure $\mu_t$ supported on the PBA
      associated with the highly idealized ENSO DDE model of \citet{Tziperman.ea.1994}. 
      The embedding is shown within the $(h(t),h(t+1))$-plane for 
      $a = (1.12 + \delta))/180$, $t \simeq 147.64$~yr and, respectively, 
      (a) $\delta = 0.0$; and (b) $\delta = 0.01500$;
      (c) $\delta = 0.0157$; and (d) $\delta = 0.015707$.
      The red curves in the four panels represent the singular support of the measure.
      Reproduced with permission from \citet{Chekroun.ea.2018}.}
\label{fig:ENSO_PBA} 
\end{figure*}

The changes in the invariant, time-dependent measure $\mu_t$ supported on this ENSO model's PBA are illustrated in Figs.~\ref{fig:noisy_PBA}--\ref{fig:quiet_PBA}, as a function of the control parameter $a$. The PBA experiences a critical transition at a value $a_{\ast}$, as illustrated in Fig.~\ref{fig:ENSO_PBA}, where $h(t)$ is the thermocline depth anomaly from seasonal depth values at the domain's eastern boundary, with $t$ in years. Here $a = (1.12 + \delta))/180$ and $0.015700 < \delta_{\ast} < 0.015707$. 

The transition in the Kolmogorov-Smirnov metric of the invariant measure's dependence on the parameter $a$, i.e. in $\mu_t = \mu_t(a)$, is quite sharp, according to \citet[Fig.~3; not shown here]{Chekroun.ea.2018}. The singular support of the measure is in full agreement with rigorous mathematical results, as well as with the numerical results \cite{Chekroun2011, Ghil2016} on the random attractor of the stochastically perturbed \citet{Lorenz1963a} convection model.

The change in the PBA is clearly associated with the population lying towards the ends of the elongated  filaments apparent in Figs.~\ref{fig:noisy_PBA}--\ref{fig:quiet_PBA}. This population is due to the occurrence of large warm, El Ni\~no and cold, La Ni\~a events. Thus, $\mu_t(a)$ encripts faithfully the disappearance of such extreme events as $a \nearrow a_{\ast}$.

Note that \citet{Mukhin2015a,Mukhin2015b} have used neural-network methodology to predict such transitions between two types of chaotic regimes in inverse models learned from time series simulated by ENSO models of  increasing complexity. The ENSO models were subjected to a linear change in a parameter meant to represent gradual global warming. Strikingly, transitions between regimes in the inverse model did occur out of sample, even when the training interval did not contain both types of behavior.

Returning to the results of Chekroun et al. (2018), they found that perturbing the \citet{Tziperman.ea.1994} model by small additive noise eliminates the crisis. The explanation of this numerical observation is tied to the role played in ENSO dynamics by the interaction between the intrinsic frequency of the coupled ocean--atmosphere system \cite[e.g.,][]{Neelin1998} and the seasonal forcing \cite{Ghil.ea.2008a, Jin1994, Jin1996a, Tziperman.ea.1994}. This interaction induces a Devil's staircase in model frequency, which has plausible counterparts in observations \cite{Ghil2000}.

As shown by \citet[Appendix B]{GCS08}, a Devil's staircase step that corresponds to a rational rotation number can be smoothed out by a sufficiently intense noise. In fact, the narrower a Devil's staircase step is, the less robust is it to noise perturbations, while the wider ones are the most robust. The effect of noise on the paradigmatic example of such a staircase, the standard circle map, has been examined in greater depth by \citet{Galatolo.ea.2019}.

This example is just one step on the long road of using the tools of nonautonomous and random dynamical systems for a better understanding of major climatic phenomena and processes. Note, however, that the model, while relatively simple, is actually infinite-dimensional because of the dependence of a DDE solution on a full interval of initial values on the real axis.
%
%
%
%


\section{Concluding Remarks}
\label{conclusions}

The goal of this review was to highlight some of the key physical and mathematical ingredients that can help address the description, understanding and prediction of climate variability and climate change. Complementary aspects of observations, theory and numerics have been taken into consideration. We have leaned heavily on dynamical systems theory and nonequilibrium statistical mechanics, and have tried to present a coherent picture of the time dependence of the climate system, its multiscale nature, and its multistability. 

We have {emphasized} the complex interplay between intrinsic climate variability and the climate's response to perturbations. The topic is, in fact, relevant for three problems of great scientific relevance: (a) anthropogenic climate change; (b) coevolution of the Earth's climate and of the biosphere, and (c) the quest for life on other planets, along with the habitability of our own planet.

The presentation { also aimed to show} the extent to which basic mathematical and physical tools { can help solve} the main challenges inherent to the climate sciences.  These challenges cannot be overcome merely by increasing the resolution of numerical models and including in them more and more physical and biogeochemical processes. { In addition,} a balanced interplay of observations, modeling and theory is definitely { needed to achieve the necessary} progress. 

This review is far from exhaustive: the authors had to make hard and, obviously, personal choices on the topics to be covered. We would like to briefly mention here {several additional} approaches to the { problems at hand that have been} developed in recent years. 
\begin{itemize}
\item {\it Network Theory} \cite[e.g.][]{Barrat2008, newman_networks_2010} has provided a novel viewpoint for constructing a parsimonious yet efficient representation of many complex processes taking place in the Earth system \cite[e.g.,][]{Tsonis2004,Tsonis2006,Donges2009}. \citet{Gozolchiani2011} and \citet{Wang2014} provide examples of use of networks for capturing specific climatic processes, while \citet{Boers2014} an example of applying network theory to climate prediction.

\item {\it Extreme Value Theory (EVT)} \cite[e.g.,][]{EKM99,Coles.2001} has been used extensively in studying the statistical properties of rare hydrometeorological events, such as the occurrence of very intense rain or of extreme temperatures \cite[and references therein]{KPN02,ghil2011}. Yet this classical methodology has been almost entirely neglected in IPCC reports dedicated to the study of extremes, cf.~\citet{IPCC12}. Recently, though, rapid advances in EVT theory  for general observables of chaotic deterministic dynamical systems \cite{HVR12,LKFW14,lucarini2016extremes} have led to the derivation of indicators of weather regimes based on extreme value statistics in the recurrence of atmospheric fields. In addition to the mere classification of such regimes, as mentioned at the end of Sect. \ref{nearcritical}, 
this approach allows one to infer higher or lower instability of these regimes and, hence, their lower or higher predictability \cite{Faranda2017,Hochman2019}.
 
\item {\it Large Deviation Theory} \cite[e.g.,][]{T09, V84} has been first used in the context of geophysical fluid dynamics for studying the self-organization and the multistability of turbulent flows and, specifically, of jet structures \cite{Bouchet:2012,Bouchet2014}; see also Sects.~\ref{ssec:gradient} and \ref{stochastic}
herein. More recently, it has been applied successfully in studying weather and climate extremes. First, it has helped formulate rare events algorithms able to nudge a climate model towards representing preferentially the class of extreme events of interest \cite{Ragone2017}. Second, it has provided a solid theoretical and numerical basis for the study of spatially extended or temporally persistent temperature extremes \cite{Galfi2019}. Large Deviation Theory has also been recently used to study multiscale  and coupled atmosphere-ocean  instabilities in a hierarchy of climate models \cite{Vannitsem2016,decruz2018}.

\item {\it Detection and Attribution Studies.} In a complex system like the climate, inferring causal relationships among events and phenomena is far from obvious. Nonetheless, doing so for well-defined weather and climate events is essential for building simplified models and, at a practical level, for causal attribution of weather- and climate-related events. This is quite important in the case of detection and attribution studies of anthropogenic climate change, especially when aiming to go beyond changes in mean climate properties, such as globally averaged temperatures, and on to determining to what extent the occurrence of an individual extreme event --- e.g., of a hurricane or an extended drought --- can be attributed to climate change \cite{Allen2003, Adam.2011}.

Methodologies based on the causal counterfactual theory of \citet{Pearl2009} are being increasingly recognised as a key instrument for providing a more rigorous basis for detection and attribution studies \cite[e.g.,][]{Hannart2016}. They also seem better suited for defining reliably the link between anthropogenic climate forcing and individual events \cite{Hannart2018}. Combining these methods with those of data assimilation, discussed herein in Sect. \ref{ssec:observations}, 
appears to be well suited to refine the distinction between the factual and counterfactual world that separates causation from the lack thereof \cite[e.g.,][]{Hannart.ea.2016}.

\item {\it Beyond Linear Response Theory.} As shown in Sect.~\ref{ssec:CR}, 
linear response theory can provide a systematic improvement upon the standard methodology of forward integration of model ensembles with perturbed initial states and parameters. While it does apply to systems out of thermodynamic equilibrium, it is still limited by its linearity to fairly small perturbations in parameters. In Sect.~\ref{ssec:wasserstein}, we mentioned that one can use the Wasserstein distance to measure arbitrary changes between two invariant measures, whether supported on a time-independent, classical attractor or a time-dependent PBA. One might thus wish to apply the PBA-based methodologies herein directly to observational data sets or to the simulations of IPCC-class models, either instead of or in combination with previously tested statistical methods. 

In Sect. \ref{ssec:Chaos2Chaos}, we showed that the PBA of an intermediate, but still infinite-dimensional ENSO model can undergo a chaos-to-chaos transition that involves major changes in its invariant measure. Moreover, these changes could be connected to a physically quite significant change in model behavior, namely in the number and size of extreme events, i.e. of the largest warm and cold events. Thus exploring similarly interesting changes in model PBAs and in the time-dependent invariant measures supported by them appears to be a very promising road toward a deeper understanding of climate variability and its interaction with both natural and anthropogenic forcing.

\end{itemize}

In the text above, we have pointed to several open scientific problems; see also \citet{Ghil.2019} for a historical perspective. An important matter that was left practically untouched, aside from the brief discussion of ENSO and its prediction in Sec. \ref{coupledmode},
is the tropical atmosphere. Half the surface of the Earth lies in the Tropics, between $30$~degrees N  and $30$~degrees S, and much of humanity lives there. As one gets closer to the Equator, the quasi-geostrophic approximation collapses and moist processes take on a much greater importance. Four of the classical references are \citet{Palmer.1951, Riehl.1954, Krish.ea.1979} and \citet{Yanai.1975}. The current literature is too plethoric to marshall here.

More broadly, we have already alluded at the end of Sec. \ref{intro} 
to the lack of a proper definition of climate and to the efforts to provide a mathematically rigorous definition that would clearly distinguish it from weather. We have also mentioned in Sec. \ref{sssec:Ext_pred} 
the distinction made by \citet{JvN.predict.1960} between the short-, medium- and long-term prediction problem in the climate sciences. 

At an even higher level of abstraction, one might wish to recall the parallel drawn by H. Poincar\'e between the deterministic unpredictability of weather and that of the position of the planets on the ecliptic \citep[pp. 69-70]{Poincare.1902}. Harking back to these two giants of both physics and mathematics, we could say that: (i) the astronomical ephemerides are the analog of weather and of its short-term prediction; (ii) the problem of climate corresponds to that of the long-term stability of the solar system; and (iii) the secular variations of the orbital configurations in the solar system are the appropriate counterpart of the slow evolution of the climate system's low-frequency modes of variability \cite{Chown.2004}.

Making these analogies stick at a level that might help determine more reliably climate sensitivity to human activities is clearly a worthwhile effort for the best mathematicians, physicists and climate scientists.

%
%

\section*{Acknowledgments}

We have benefited immensely from discussions with many colleagues and friends in the last few years, but we would like to thank especially T. B\'odai, N. Boers, F. Bouchet, M.~D. Chekroun, J.~I. D\'iaz, H.~A. Dijkstra, D. Faranda, A. Feigin, U. Feudel, K. Fraedrich, G. Gallavotti, A. Gritsun, A. Groth, D. Holm, F.-F. Jin, D. Kondrashov, S. Kravtsov, J. Kurths, J.~C. McWilliams, J.~D. Neelin, S. Pierini, F. Ragone, D. Ruelle, A. Speranza, H. Swinney, O. Talagrand, A. Tantet, S. Vannitsem, J. Yorke, and J. Wouters. Two anonymous reviewers provided constructive comments. The authors acknowledge the financial support provided by  Horizon 2020 through the project TiPES (grant No. 820970). This activity has been supported by the EIT Climate-KIC. EIT Climate-KIC is supported by EIT, a body of the European Union. VL  also acknowledges the financial support provided by Horizon 2020 through the projects CRESCENDO (grant No. 641816) and Blue-Action (grant No. 727852) and by the Royal Society (grant No. IEC R2 170001).

\newpage

\appendix

\section{Acronyms}{\label{app:acronyms}

This Appendix contains two tables of acronyms used throughout the paper: Table~\ref{tab:glossary1} contains the scientific acronyms and Table~\ref{tab:glossary2} the institutional ones.

\begin{table}[!hpb]

\caption{Scientific acronyms} 

\centering

\begin{tabular}{l c}

\hline

Acronym & Meaning \\

\hline

$AS$ & Arakawa-Schubert (parametrization) \\

$CN$s & Complex networks \\

$DDE$ & Delay differential equation \\

$DNS$ & {Direct numerical simulation} \\

$EBM$ & Energy Balance Model \\

$ECS$ & Equilibrium Climate Sensitivity\\

$ENSO$ & El Ni\~no--Southern Oscillation \\

$EMS$ & Empirical Mode Reduction \\

$FDT$ & Fluctuation--Dissipation Theorem \\

$GCM$ & General Circulation Model \\

$GCM$ & Global Climate Model \\

$GFD$ & Geophysical fluid dynamics \\

$GHGs$ & Greenhouse Gases\\

$GLE$ & Generalized Langevin equation \\

$LFV$ & Low Frequency Variability \\

$MJO$ & Madden-Julian Oscillation \\

$MOC$ & Meridional overturning circulation \\

$MZ$ & Mori-Zwanzig \\

$NAO$ & North Atlantic Oscillation \\

$NDS$ & Nonautonomous dynamical system \\

$NWP$ & Numerical Weather Prediction \\

$NSEs$ & Navier-Stokes Equations \\

$ODE$ & Ordinary differential equation \\

$PBA$ & Pullback attractor \\

$PDE$ & Partial differential equation \\

$PNA$ & Pacific North American (pattern) \\

$PSA$ & Pacific South American (pattern) \\

$QG$ & Quasi-geostrophic (flow, model) \\

$RDS$ & Random dynamical system \\ 

$SDE$ & Stochastic differential equation \\

$TCR$ & Transient Climate Response\\

$THC$ & Thermohaline Circulation \\

\hline

\multicolumn{2}{l} 

\end{tabular}\label{tab:glossary1}

\end{table}

\begin{table}[!hpb]

\caption{Institutional acronyms}  

\centering

\begin{tabular}{l c}

\hline

Acronym & Meaning \\

\hline

$AR$ & Assessment Report \\

$CMIP$ & Climate Model Intercomparison Project\\

$ECMWF$ & European Centre for \\
 & Mid-range Weather Forecast\\

$ENSO$ & El-Ni\~no-Southern Oscillation \\

$IPCC $ & Intergovernmental Panel on Climate Change \\

$NCAR$ & National Center for Atmospheric Research\\

$NCEP$ & National Center for Environmental Prediction \\


$PCMDI$ & Program for Climate Model \\
& Diagnostics and Intercomparison\\

$SPM$ & Summary for Policy Makers \\

$UNEP$ & United Nations Environmental Programme \\

$WCRP$ & World Climate Research Programme \\

$WMO$ & World Meteorological Organization \\

\hline

\multicolumn{2}{l} 

\end{tabular}\label{tab:glossary2}

\end{table}


\clearpage


\begin{thebibliography}{537}%
\makeatletter
\providecommand \@ifxundefined [1]{%
 \@ifx{#1\undefined}
}%
\providecommand \@ifnum [1]{%
 \ifnum #1\expandafter \@firstoftwo
 \else \expandafter \@secondoftwo
 \fi
}%
\providecommand \@ifx [1]{%
 \ifx #1\expandafter \@firstoftwo
 \else \expandafter \@secondoftwo
 \fi
}%
\providecommand \natexlab [1]{#1}%
\providecommand \enquote  [1]{``#1''}%
\providecommand \bibnamefont  [1]{#1}%
\providecommand \bibfnamefont [1]{#1}%
\providecommand \citenamefont [1]{#1}%
\providecommand \href@noop [0]{\@secondoftwo}%
\providecommand \href [0]{\begingroup \@sanitize@url \@href}%
\providecommand \@href[1]{\@@startlink{#1}\@@href}%
\providecommand \@@href[1]{\endgroup#1\@@endlink}%
\providecommand \@sanitize@url [0]{\catcode `\\12\catcode `\$12\catcode
  `\&12\catcode `\#12\catcode `\^12\catcode `\_12\catcode `\%12\relax}%
\providecommand \@@startlink[1]{}%
\providecommand \@@endlink[0]{}%
\providecommand \url  [0]{\begingroup\@sanitize@url \@url }%
\providecommand \@url [1]{\endgroup\@href {#1}{\urlprefix }}%
\providecommand \urlprefix  [0]{URL }%
\providecommand \Eprint [0]{\href }%
\providecommand \doibase [0]{http://dx.doi.org/}%
\providecommand \selectlanguage [0]{\@gobble}%
\providecommand \bibinfo  [0]{\@secondoftwo}%
\providecommand \bibfield  [0]{\@secondoftwo}%
\providecommand \translation [1]{[#1]}%
\providecommand \BibitemOpen [0]{}%
\providecommand \bibitemStop [0]{}%
\providecommand \bibitemNoStop [0]{.\EOS\space}%
\providecommand \EOS [0]{\spacefactor3000\relax}%
\providecommand \BibitemShut  [1]{\csname bibitem#1\endcsname}%
\let\auto@bib@innerbib\@empty
\bibitem [{\citenamefont {Abramov}\ and\ \citenamefont {Majda}(2007)}]{AM07a}%
  \BibitemOpen
  \bibfield  {author} {\bibinfo {author} {\bibnamefont {Abramov}, \bibfnamefont
  {R~V}}, \ and\ \bibinfo {author} {\bibfnamefont {A.~J.}\ \bibnamefont
  {Majda}}} (\bibinfo {year} {2007}),\ \bibfield  {title} {\enquote {\bibinfo
  {title} {Blended response algorithms for linear fluctuation-dissipation for
  complex nonlinear dynamical systems},}\ }\href@noop {} {\bibfield  {journal}
  {\bibinfo  {journal} {Nonlinearity}\ }\textbf {\bibinfo {volume}
  {20}}~(\bibinfo {number} {12}),\ \bibinfo {pages} {2793--2821}}\BibitemShut
  {NoStop}%
\bibitem [{\citenamefont {Adam}(2011)}]{Adam.2011}%
  \BibitemOpen
  \bibfield  {author} {\bibinfo {author} {\bibnamefont {Adam}, \bibfnamefont
  {D}}} (\bibinfo {year} {2011}),\ \bibfield  {title} {\enquote {\bibinfo
  {title} {Climate change in court},}\ }\href {\doibase 10.1038/nclimate1131}
  {\bibfield  {journal} {\bibinfo  {journal} {Nature Climate Change}\ }\textbf
  {\bibinfo {volume} {1}}~(\bibinfo {number} {3}),\ \bibinfo {pages}
  {127--130}}\BibitemShut {NoStop}%
\bibitem [{\citenamefont {Allen}(2003)}]{Allen2003}%
  \BibitemOpen
  \bibfield  {author} {\bibinfo {author} {\bibnamefont {Allen}, \bibfnamefont
  {M}}} (\bibinfo {year} {2003}),\ \bibfield  {title} {\enquote {\bibinfo
  {title} {Liability for climate change},}\ }\href {\doibase 10.1038/421891a}
  {\bibfield  {journal} {\bibinfo  {journal} {Nature}\ }\textbf {\bibinfo
  {volume} {421}}~(\bibinfo {number} {6926}),\ \bibinfo {pages}
  {891--892}}\BibitemShut {NoStop}%
\bibitem [{\citenamefont {Andersen}\ and\ \citenamefont
  {Coauthors}(2004)}]{Andersen2004}%
  \BibitemOpen
  \bibfield  {author} {\bibinfo {author} {\bibnamefont {Andersen},
  \bibfnamefont {K~K}}, \ and\ \bibinfo {author} {\bibnamefont {Coauthors}}}
  (\bibinfo {year} {2004}),\ \bibfield  {title} {\enquote {\bibinfo {title}
  {{High-resolution record of Northern Hemisphere climate extending into the
  last interglacial period}},}\ }\href {\doibase 10.1038/nature02805}
  {\bibfield  {journal} {\bibinfo  {journal} {Nature}\ }\textbf {\bibinfo
  {volume} {431}}~(\bibinfo {number} {7005}),\ \bibinfo {pages}
  {147--151}}\BibitemShut {NoStop}%
\bibitem [{\citenamefont {Andronov}\ and\ \citenamefont
  {Pontryagin}(1937)}]{Andro.Pont.37}%
  \BibitemOpen
  \bibfield  {author} {\bibinfo {author} {\bibnamefont {Andronov},
  \bibfnamefont {A~A}}, \ and\ \bibinfo {author} {\bibfnamefont {L.~S.}\
  \bibnamefont {Pontryagin}}} (\bibinfo {year} {1937}),\ \bibfield  {title}
  {\enquote {\bibinfo {title} {Syst\`emes grossiers},}\ }\href@noop {}
  {\bibfield  {journal} {\bibinfo  {journal} {Dokl. Akad. Nauk SSSR}\ }\textbf
  {\bibinfo {volume} {14}}~(\bibinfo {number} {5}),\ \bibinfo {pages}
  {247--250}}\BibitemShut {NoStop}%
\bibitem [{\citenamefont {Arakawa}\ and\ \citenamefont
  {Schubert}(1974)}]{AS.74}%
  \BibitemOpen
  \bibfield  {author} {\bibinfo {author} {\bibnamefont {Arakawa}, \bibfnamefont
  {A}}, \ and\ \bibinfo {author} {\bibfnamefont {W.~H.}\ \bibnamefont
  {Schubert}}} (\bibinfo {year} {1974}),\ \bibfield  {title} {\enquote
  {\bibinfo {title} {{Interaction of a cumulus cloud ensemble with the
  large-scale environment, Part I}},}\ }\href@noop {} {\bibfield  {journal}
  {\bibinfo  {journal} {Journal of the Atmospheric Sciences}\ }\textbf
  {\bibinfo {volume} {31}}~(\bibinfo {number} {3}),\ \bibinfo {pages}
  {674--701}}\BibitemShut {NoStop}%
\bibitem [{\citenamefont {Arcoya}\ \emph {et~al.}(1998)\citenamefont {Arcoya},
  \citenamefont {Diaz},\ and\ \citenamefont {Tello}}]{Diaz.ea.1998}%
  \BibitemOpen
  \bibfield  {author} {\bibinfo {author} {\bibnamefont {Arcoya}, \bibfnamefont
  {D}}, \bibinfo {author} {\bibfnamefont {J.~I.}\ \bibnamefont {Diaz}}, \ and\
  \bibinfo {author} {\bibfnamefont {L.}~\bibnamefont {Tello}}} (\bibinfo {year}
  {1998}),\ \bibfield  {title} {\enquote {\bibinfo {title} {S-shaped
  bifurcation branch in a quasilinear multivalued model arising in
  climatology},}\ }\href@noop {} {\bibfield  {journal} {\bibinfo  {journal}
  {Journal of Differential Equations}\ }\textbf {\bibinfo {volume}
  {150}}~(\bibinfo {number} {1}),\ \bibinfo {pages} {215--225}}\BibitemShut
  {NoStop}%
\bibitem [{\citenamefont {Arnold}(1998)}]{Arnold.1998}%
  \BibitemOpen
  \bibfield  {author} {\bibinfo {author} {\bibnamefont {Arnold}, \bibfnamefont
  {L}}} (\bibinfo {year} {1998}),\ \href@noop {} {\emph {\bibinfo {title}
  {{Random Dynamical Systems}}}}\ (\bibinfo  {publisher} {Springer-Verlag},\
  \bibinfo {address} {New York/Berlin})\BibitemShut {NoStop}%
\bibitem [{\citenamefont {Arnold}(1988)}]{arnold1988}%
  \BibitemOpen
  \bibfield  {author} {\bibinfo {author} {\bibnamefont {Arnold}, \bibfnamefont
  {Ludwig}}} (\bibinfo {year} {1988}),\ \href@noop {} {\emph {\bibinfo {title}
  {{Random Dynamical Systems}}}}\ (\bibinfo  {publisher} {Springer},\ \bibinfo
  {address} {New York})\BibitemShut {NoStop}%
\bibitem [{\citenamefont {Arnold}(1983)}]{V.Arnold.83}%
  \BibitemOpen
  \bibfield  {author} {\bibinfo {author} {\bibnamefont {Arnold}, \bibfnamefont
  {V}}} (\bibinfo {year} {1983}),\ \href@noop {} {\emph {\bibinfo {title}
  {Geometric Methods in the Theory of Ordinary Differential Equations}}}\
  (\bibinfo  {publisher} {Springer},\ \bibinfo {address} {New
  York/Berlin})\BibitemShut {NoStop}%
\bibitem [{\citenamefont {Arnold}(2003)}]{Arnold.2003}%
  \BibitemOpen
  \bibfield  {author} {\bibinfo {author} {\bibnamefont {Arnold}, \bibfnamefont
  {VI}}} (\bibinfo {year} {2003}),\ \href@noop {} {\emph {\bibinfo {title}
  {Catastrophe Theory}}}\ (\bibinfo  {publisher} {Springer Nature},\ \bibinfo
  {address} {Berlin})\BibitemShut {NoStop}%
\bibitem [{\citenamefont {Ashwin}\ \emph {et~al.}(2012)\citenamefont {Ashwin},
  \citenamefont {Wieczorek}, \citenamefont {Vitolo},\ and\ \citenamefont
  {Cox}}]{Ashwin2012}%
  \BibitemOpen
  \bibfield  {author} {\bibinfo {author} {\bibnamefont {Ashwin}, \bibfnamefont
  {P}}, \bibinfo {author} {\bibfnamefont {S.}~\bibnamefont {Wieczorek}},
  \bibinfo {author} {\bibfnamefont {R.}~\bibnamefont {Vitolo}}, \ and\ \bibinfo
  {author} {\bibfnamefont {P.}~\bibnamefont {Cox}}} (\bibinfo {year} {2012}),\
  \bibfield  {title} {\enquote {\bibinfo {title} {Tipping points in open
  systems: bifurcation, noise-induced and rate-dependent examples in the
  climate system},}\ }\href {\doibase 10.1098/rsta.2011.0306} {\bibfield
  {journal} {\bibinfo  {journal} {Philosophical Transactions of the Royal
  Society A: Mathematical, Physical and Engineering Sciences}\ }\textbf
  {\bibinfo {volume} {370}}~(\bibinfo {number} {1962}),\ \bibinfo {pages}
  {1166--1184}}\BibitemShut {NoStop}%
\bibitem [{\citenamefont {Baladi}(2000)}]{B00}%
  \BibitemOpen
  \bibfield  {author} {\bibinfo {author} {\bibnamefont {Baladi}, \bibfnamefont
  {V}}} (\bibinfo {year} {2000}),\ \href@noop {} {\emph {\bibinfo {title}
  {Positive Transfer Operators and Decay of Correlations}}}\ (\bibinfo
  {publisher} {World Scientific},\ \bibinfo {address} {Singapore})\BibitemShut
  {NoStop}%
\bibitem [{\citenamefont {Baladi}(2008)}]{B08}%
  \BibitemOpen
  \bibfield  {author} {\bibinfo {author} {\bibnamefont {Baladi}, \bibfnamefont
  {V}}} (\bibinfo {year} {2008}),\ \bibfield  {title} {\enquote {\bibinfo
  {title} {Linear response despite critical points},}\ }\href@noop {}
  {\bibfield  {journal} {\bibinfo  {journal} {Nonlinearity}\ }\textbf {\bibinfo
  {volume} {21}}~(\bibinfo {number} {6}),\ \bibinfo {pages} {T81}}\BibitemShut
  {NoStop}%
\bibitem [{\citenamefont {Balmaseda}\ \emph {et~al.}(2015)\citenamefont
  {Balmaseda}, \citenamefont {Hernandez}, \citenamefont {Storto}, \citenamefont
  {Palmer}, \citenamefont {Alves}, \citenamefont {Shi}, \citenamefont {Smith},
  \citenamefont {Toyoda}, \citenamefont {Valdivieso}, \citenamefont {Barnier},
  \citenamefont {Behringer}, \citenamefont {Boyer}, \citenamefont {Chang},
  \citenamefont {Chepurin}, \citenamefont {Ferry}, \citenamefont {Forget},
  \citenamefont {Fujii}, \citenamefont {Good}, \citenamefont {Guinehut},
  \citenamefont {Haines}, \citenamefont {Ishikawa}, \citenamefont {Keeley},
  \citenamefont {KÃÂ¶hl}, \citenamefont {Lee}, \citenamefont {Martin},
  \citenamefont {Masina}, \citenamefont {Masuda}, \citenamefont {Meyssignac},
  \citenamefont {Mogensen}, \citenamefont {Parent}, \citenamefont {Peterson},
  \citenamefont {Tang}, \citenamefont {Yin}, \citenamefont {Vernieres},
  \citenamefont {Wang}, \citenamefont {Waters}, \citenamefont {Wedd},
  \citenamefont {Wang}, \citenamefont {Xue}, \citenamefont {Chevallier},
  \citenamefont {Lemieux}, \citenamefont {Dupont}, \citenamefont {Kuragano},
  \citenamefont {Kamachi}, \citenamefont {Awaji}, \citenamefont {Caltabiano},
  \citenamefont {Wilmer-Becker},\ and\ \citenamefont
  {Gaillard}}]{Balmaseda2015}%
  \BibitemOpen
  \bibfield  {author} {\bibinfo {author} {\bibnamefont {Balmaseda},
  \bibfnamefont {MA}}, \bibinfo {author} {\bibfnamefont {F.}~\bibnamefont
  {Hernandez}}, \bibinfo {author} {\bibfnamefont {A.}~\bibnamefont {Storto}},
  \bibinfo {author} {\bibfnamefont {M.D.}\ \bibnamefont {Palmer}}, \bibinfo
  {author} {\bibfnamefont {O.}~\bibnamefont {Alves}}, \bibinfo {author}
  {\bibfnamefont {L.}~\bibnamefont {Shi}}, \bibinfo {author} {\bibfnamefont
  {G.C.}\ \bibnamefont {Smith}}, \bibinfo {author} {\bibfnamefont
  {T.}~\bibnamefont {Toyoda}}, \bibinfo {author} {\bibfnamefont
  {M.}~\bibnamefont {Valdivieso}}, \bibinfo {author} {\bibfnamefont
  {B.}~\bibnamefont {Barnier}}, \bibinfo {author} {\bibfnamefont
  {D.}~\bibnamefont {Behringer}}, \bibinfo {author} {\bibfnamefont
  {T.}~\bibnamefont {Boyer}}, \bibinfo {author} {\bibfnamefont {Y-S.}\
  \bibnamefont {Chang}}, \bibinfo {author} {\bibfnamefont {G.A.}\ \bibnamefont
  {Chepurin}}, \bibinfo {author} {\bibfnamefont {N.}~\bibnamefont {Ferry}},
  \bibinfo {author} {\bibfnamefont {G.}~\bibnamefont {Forget}}, \bibinfo
  {author} {\bibfnamefont {Y.}~\bibnamefont {Fujii}}, \bibinfo {author}
  {\bibfnamefont {S.}~\bibnamefont {Good}}, \bibinfo {author} {\bibfnamefont
  {S.}~\bibnamefont {Guinehut}}, \bibinfo {author} {\bibfnamefont
  {K.}~\bibnamefont {Haines}}, \bibinfo {author} {\bibfnamefont
  {Y.}~\bibnamefont {Ishikawa}}, \bibinfo {author} {\bibfnamefont
  {S.}~\bibnamefont {Keeley}}, \bibinfo {author} {\bibfnamefont
  {A.}~\bibnamefont {KÃÂ¶hl}}, \bibinfo {author} {\bibfnamefont
  {T.}~\bibnamefont {Lee}}, \bibinfo {author} {\bibfnamefont {M.J.}\
  \bibnamefont {Martin}}, \bibinfo {author} {\bibfnamefont {S.}~\bibnamefont
  {Masina}}, \bibinfo {author} {\bibfnamefont {S.}~\bibnamefont {Masuda}},
  \bibinfo {author} {\bibfnamefont {B.}~\bibnamefont {Meyssignac}}, \bibinfo
  {author} {\bibfnamefont {K.}~\bibnamefont {Mogensen}}, \bibinfo {author}
  {\bibfnamefont {L.}~\bibnamefont {Parent}}, \bibinfo {author} {\bibfnamefont
  {K.A.}\ \bibnamefont {Peterson}}, \bibinfo {author} {\bibfnamefont {Y.M.}\
  \bibnamefont {Tang}}, \bibinfo {author} {\bibfnamefont {Y.}~\bibnamefont
  {Yin}}, \bibinfo {author} {\bibfnamefont {G.}~\bibnamefont {Vernieres}},
  \bibinfo {author} {\bibfnamefont {X.}~\bibnamefont {Wang}}, \bibinfo {author}
  {\bibfnamefont {J.}~\bibnamefont {Waters}}, \bibinfo {author} {\bibfnamefont
  {R.}~\bibnamefont {Wedd}}, \bibinfo {author} {\bibfnamefont {O.}~\bibnamefont
  {Wang}}, \bibinfo {author} {\bibfnamefont {Y.}~\bibnamefont {Xue}}, \bibinfo
  {author} {\bibfnamefont {M.}~\bibnamefont {Chevallier}}, \bibinfo {author}
  {\bibfnamefont {J-F.}\ \bibnamefont {Lemieux}}, \bibinfo {author}
  {\bibfnamefont {F.}~\bibnamefont {Dupont}}, \bibinfo {author} {\bibfnamefont
  {T.}~\bibnamefont {Kuragano}}, \bibinfo {author} {\bibfnamefont
  {M.}~\bibnamefont {Kamachi}}, \bibinfo {author} {\bibfnamefont
  {T.}~\bibnamefont {Awaji}}, \bibinfo {author} {\bibfnamefont
  {A.}~\bibnamefont {Caltabiano}}, \bibinfo {author} {\bibfnamefont
  {K.}~\bibnamefont {Wilmer-Becker}}, \ and\ \bibinfo {author} {\bibfnamefont
  {F.}~\bibnamefont {Gaillard}}} (\bibinfo {year} {2015}),\ \bibfield  {title}
  {\enquote {\bibinfo {title} {The ocean reanalyses intercomparison project
  (ora-ip)},}\ }\href {\doibase 10.1080/1755876X.2015.1022329} {\bibfield
  {journal} {\bibinfo  {journal} {Journal of Operational Oceanography}\
  }\textbf {\bibinfo {volume} {8}}~(\bibinfo {number} {sup1}),\ \bibinfo
  {pages} {s80--s97}}\BibitemShut {NoStop}%
\bibitem [{\citenamefont {Barenblatt}(1987)}]{Barenblatt1987}%
  \BibitemOpen
  \bibfield  {author} {\bibinfo {author} {\bibnamefont {Barenblatt},
  \bibfnamefont {G~L}}} (\bibinfo {year} {1987}),\ \href@noop {} {\emph
  {\bibinfo {title} {Dimensional Analysis}}}\ (\bibinfo  {publisher} {Gordon
  and Breach, New York})\BibitemShut {NoStop}%
\bibitem [{\citenamefont {Barnes}\ and\ \citenamefont
  {Screen}(2015)}]{Barnes.AA.2015}%
  \BibitemOpen
  \bibfield  {author} {\bibinfo {author} {\bibnamefont {Barnes}, \bibfnamefont
  {E~A}}, \ and\ \bibinfo {author} {\bibfnamefont {J.~A.}\ \bibnamefont
  {Screen}}} (\bibinfo {year} {2015}),\ \bibfield  {title} {\enquote {\bibinfo
  {title} {{The impact of Arctic warming on the midlatitude jetstream: Can it?
  Has it? Will it?}}}\ }\href {\doibase 10.1002/wcc.337} {\bibfield  {journal}
  {\bibinfo  {journal} {WIREs Climate Change}\ }\textbf {\bibinfo {volume}
  {6}},\ \bibinfo {pages} {277--286}}\BibitemShut {NoStop}%
\bibitem [{\citenamefont {Barnston}\ \emph {et~al.}(2012)\citenamefont
  {Barnston}, \citenamefont {Tippett}, \citenamefont {Heureux}, \citenamefont
  {Li},\ and\ \citenamefont {DeWitt}}]{iri12}%
  \BibitemOpen
  \bibfield  {author} {\bibinfo {author} {\bibnamefont {Barnston},
  \bibfnamefont {A~G}}, \bibinfo {author} {\bibfnamefont {M.~K.}\ \bibnamefont
  {Tippett}}, \bibinfo {author} {\bibfnamefont {M.~L.}\ \bibnamefont
  {Heureux}}, \bibinfo {author} {\bibfnamefont {S.}~\bibnamefont {Li}}, \ and\
  \bibinfo {author} {\bibfnamefont {D.~G.}\ \bibnamefont {DeWitt}}} (\bibinfo
  {year} {2012}),\ \bibfield  {title} {\enquote {\bibinfo {title} {Skill of
  real-time seasonal {ENSO} model predictions during 2002--2011 --- is our
  capability improving?}}\ }\href {\doibase 10.1175/BAMS-D-11-00111.1}
  {\bibfield  {journal} {\bibinfo  {journal} {Bull. Am. Meteorol. Soc.}\
  }\textbf {\bibinfo {volume} {93}}~(\bibinfo {number} {5}),\ \bibinfo {pages}
  {631--651}}\BibitemShut {NoStop}%
\bibitem [{\citenamefont {Barrat}\ \emph {et~al.}(2008)\citenamefont {Barrat},
  \citenamefont {Barthelemy},\ and\ \citenamefont {Vespignany}}]{Barrat2008}%
  \BibitemOpen
  \bibfield  {author} {\bibinfo {author} {\bibnamefont {Barrat}, \bibfnamefont
  {A}}, \bibinfo {author} {\bibfnamefont {M.}~\bibnamefont {Barthelemy}}, \
  and\ \bibinfo {author} {\bibfnamefont {A.}~\bibnamefont {Vespignany}}}
  (\bibinfo {year} {2008}),\ \href@noop {} {\emph {\bibinfo {title} {Dynamical
  Processes in Complex Networks}}}\ (\bibinfo  {publisher} {Cambridge
  University Press})\BibitemShut {NoStop}%
\bibitem [{\citenamefont {Batchelor}(1974)}]{Batchelor1974}%
  \BibitemOpen
  \bibfield  {author} {\bibinfo {author} {\bibnamefont {Batchelor},
  \bibfnamefont {G}}} (\bibinfo {year} {1974}),\ \href@noop {} {\emph {\bibinfo
  {title} {Introduction to Fluid Dynamics}}}\ (\bibinfo  {publisher} {Cambridge
  University Press},\ \bibinfo {address} {Cambridge, UK})\BibitemShut {NoStop}%
\bibitem [{\citenamefont {Beck}(1990)}]{Beck.1990}%
  \BibitemOpen
  \bibfield  {author} {\bibinfo {author} {\bibnamefont {Beck}, \bibfnamefont
  {C}}} (\bibinfo {year} {1990}),\ \bibfield  {title} {\enquote {\bibinfo
  {title} {Brownian motion from deterministic dynamics},}\ }\href@noop {}
  {\bibfield  {journal} {\bibinfo  {journal} {{Physica A: Statistical Mechanics
  and its Applications}}\ }\textbf {\bibinfo {volume} {169}}~(\bibinfo {number}
  {2}),\ \bibinfo {pages} {324--336}}\BibitemShut {NoStop}%
\bibitem [{\citenamefont {Bell}(1980)}]{Bell1980}%
  \BibitemOpen
  \bibfield  {author} {\bibinfo {author} {\bibnamefont {Bell}, \bibfnamefont
  {T~L}}} (\bibinfo {year} {1980}),\ \bibfield  {title} {\enquote {\bibinfo
  {title} {Climate sensitivity from fluctuation dissipation: Some simple model
  tests},}\ }\href {\doibase 10.1175/1520-0469(1980)037<1700:CSFFDS>2.0.CO;2}
  {\bibfield  {journal} {\bibinfo  {journal} {Journal of the Atmospheric
  Sciences}\ }\textbf {\bibinfo {volume} {37}}~(\bibinfo {number} {8}),\
  \bibinfo {pages} {1700--1707}}\BibitemShut {NoStop}%
\bibitem [{\citenamefont {Bellenger}\ \emph {et~al.}(2014)\citenamefont
  {Bellenger}, \citenamefont {Guilyardi}, \citenamefont {Leloup}, \citenamefont
  {Lengaigne},\ and\ \citenamefont {Vialard}}]{Bellenger2014}%
  \BibitemOpen
  \bibfield  {author} {\bibinfo {author} {\bibnamefont {Bellenger},
  \bibfnamefont {H}}, \bibinfo {author} {\bibfnamefont {E.}~\bibnamefont
  {Guilyardi}}, \bibinfo {author} {\bibfnamefont {J.}~\bibnamefont {Leloup}},
  \bibinfo {author} {\bibfnamefont {M.}~\bibnamefont {Lengaigne}}, \ and\
  \bibinfo {author} {\bibfnamefont {J.}~\bibnamefont {Vialard}}} (\bibinfo
  {year} {2014}),\ \bibfield  {title} {\enquote {\bibinfo {title} {Enso
  representation in climate models: from cmip3 to cmip5},}\ }\href {\doibase
  10.1007/s00382-013-1783-z} {\bibfield  {journal} {\bibinfo  {journal}
  {Climate Dynamics}\ }\textbf {\bibinfo {volume} {42}}~(\bibinfo {number}
  {7}),\ \bibinfo {pages} {1999--2018}}\BibitemShut {NoStop}%
\bibitem [{\citenamefont {Bellman}\ and\ \citenamefont
  {Cooke}(1963)}]{Bellman.Cooke.1963}%
  \BibitemOpen
  \bibfield  {author} {\bibinfo {author} {\bibnamefont {Bellman}, \bibfnamefont
  {R~E}}, \ and\ \bibinfo {author} {\bibfnamefont {K.~L.}\ \bibnamefont
  {Cooke}}} (\bibinfo {year} {1963}),\ \href@noop {} {\emph {\bibinfo {title}
  {{Differential-Difference Equations}}}}\ (\bibinfo  {publisher} {Rand
  Corporation},\ \bibinfo {address} {Santa Monica, CA})\BibitemShut {NoStop}%
\bibitem [{\citenamefont {Bengtsson}\ \emph {et~al.}(1981)\citenamefont
  {Bengtsson}, \citenamefont {Ghil},\ and\ \citenamefont
  {K{\"a}ll{\'e}n}}]{bengtsson81}%
  \BibitemOpen
  \bibfield  {author} {\bibinfo {author} {\bibnamefont {Bengtsson},
  \bibfnamefont {L}}, \bibinfo {author} {\bibfnamefont {M.}~\bibnamefont
  {Ghil}}, \ and\ \bibinfo {author} {\bibfnamefont {E.}~\bibnamefont
  {K{\"a}ll{\'e}n}}} (\bibinfo {year} {1981}),\ \href@noop {} {\emph {\bibinfo
  {title} {Dynamic Meteorology: Data Assimilation Methods}}}\ (\bibinfo
  {publisher} {Springer})\BibitemShut {NoStop}%
\bibitem [{\citenamefont {Bensid}\ and\ \citenamefont
  {Diaz}(2019)}]{Bensid.2019}%
  \BibitemOpen
  \bibfield  {author} {\bibinfo {author} {\bibnamefont {Bensid}, \bibfnamefont
  {S}}, \ and\ \bibinfo {author} {\bibfnamefont {J.~I.}\ \bibnamefont {Diaz}}}
  (\bibinfo {year} {2019}),\ \bibfield  {title} {\enquote {\bibinfo {title} {On
  the exact number of monotone solutions of a simplified {Budyko climate model
  and their different stability}},}\ }\href@noop {} {\bibfield  {journal}
  {\bibinfo  {journal} {{Discrete and Continuous Dynamical Systems - B}}\
  }\textbf {\bibinfo {volume} {24}},\ \bibinfo {pages} {1033}}\BibitemShut
  {NoStop}%
\bibitem [{\citenamefont {Benzi}\ \emph {et~al.}(1986)\citenamefont {Benzi},
  \citenamefont {Malguzzi}, \citenamefont {Speranza},\ and\ \citenamefont
  {Sutera}}]{Benzi.ea.1986}%
  \BibitemOpen
  \bibfield  {author} {\bibinfo {author} {\bibnamefont {Benzi}, \bibfnamefont
  {R}}, \bibinfo {author} {\bibfnamefont {P.}~\bibnamefont {Malguzzi}},
  \bibinfo {author} {\bibfnamefont {A.}~\bibnamefont {Speranza}}, \ and\
  \bibinfo {author} {\bibfnamefont {A.}~\bibnamefont {Sutera}}} (\bibinfo
  {year} {1986}),\ \bibfield  {title} {\enquote {\bibinfo {title} {The
  statistical properties of general atmospheric circulation: observational
  evidence and a minimal theory of bimodality},}\ }\href {\doibase
  10.1256/smsqj.47305} {\bibfield  {journal} {\bibinfo  {journal} {{Q. J. R.
  Meteorol. Soc.}}\ }\textbf {\bibinfo {volume} {112}}~(\bibinfo {number}
  {473}),\ \bibinfo {pages} {661--674}}\BibitemShut {NoStop}%
\bibitem [{\citenamefont {Benzi}\ and\ \citenamefont
  {Speranza}(1989)}]{Benzi1989}%
  \BibitemOpen
  \bibfield  {author} {\bibinfo {author} {\bibnamefont {Benzi}, \bibfnamefont
  {R}}, \ and\ \bibinfo {author} {\bibfnamefont {A.}~\bibnamefont {Speranza}}}
  (\bibinfo {year} {1989}),\ \bibfield  {title} {\enquote {\bibinfo {title}
  {Statistical properties of low-frequency variability in the northern
  hemisphere},}\ }\href {\doibase
  10.1175/1520-0442(1989)002<0367:SPOLFV>2.0.CO;2} {\bibfield  {journal}
  {\bibinfo  {journal} {Journal of Climate}\ }\textbf {\bibinfo {volume}
  {2}}~(\bibinfo {number} {4}),\ \bibinfo {pages} {367--379}}\BibitemShut
  {NoStop}%
\bibitem [{\citenamefont {Berger}\ and\ \citenamefont
  {Siegmund}(2003)}]{Berger.Sieg.2003}%
  \BibitemOpen
  \bibfield  {author} {\bibinfo {author} {\bibnamefont {Berger}, \bibfnamefont
  {A}}, \ and\ \bibinfo {author} {\bibfnamefont {S.}~\bibnamefont {Siegmund}}}
  (\bibinfo {year} {2003}),\ \bibfield  {title} {\enquote {\bibinfo {title} {On
  the gap between random dynamical systems and continuous skew products},}\
  }\href@noop {} {\bibfield  {journal} {\bibinfo  {journal} {{Journal of
  Dynamics and Differential Equations}}\ }\textbf {\bibinfo {volume}
  {15}}~(\bibinfo {number} {2-3}),\ \bibinfo {pages} {237--279}}\BibitemShut
  {NoStop}%
\bibitem [{\citenamefont {Beri}\ \emph {et~al.}(2005)\citenamefont {Beri},
  \citenamefont {Mannella}, \citenamefont {Luchinsky}, \citenamefont
  {Silchenko},\ and\ \citenamefont {McClintock}}]{Beri2005}%
  \BibitemOpen
  \bibfield  {author} {\bibinfo {author} {\bibnamefont {Beri}, \bibfnamefont
  {S}}, \bibinfo {author} {\bibfnamefont {R.}~\bibnamefont {Mannella}},
  \bibinfo {author} {\bibfnamefont {D.~G.}\ \bibnamefont {Luchinsky}}, \bibinfo
  {author} {\bibfnamefont {A.~N.}\ \bibnamefont {Silchenko}}, \ and\ \bibinfo
  {author} {\bibfnamefont {P.~V.~E.}\ \bibnamefont {McClintock}}} (\bibinfo
  {year} {2005}),\ \bibfield  {title} {\enquote {\bibinfo {title} {Solution of
  the boundary value problem for optimal escape in continuous stochastic
  systems and maps},}\ }\href {\doibase 10.1103/PhysRevE.72.036131} {\bibfield
  {journal} {\bibinfo  {journal} {Phys. Rev. E}\ }\textbf {\bibinfo {volume}
  {72}},\ \bibinfo {pages} {036131}}\BibitemShut {NoStop}%
\bibitem [{\citenamefont {Berloff}\ \emph {et~al.}(2007)\citenamefont
  {Berloff}, \citenamefont {Hogg},\ and\ \citenamefont
  {Dewar}}]{Berloff.ea.2007}%
  \BibitemOpen
  \bibfield  {author} {\bibinfo {author} {\bibnamefont {Berloff}, \bibfnamefont
  {P}}, \bibinfo {author} {\bibfnamefont {A.}~\bibnamefont {Hogg}}, \ and\
  \bibinfo {author} {\bibfnamefont {W.~K.}\ \bibnamefont {Dewar}}} (\bibinfo
  {year} {2007}),\ \bibfield  {title} {\enquote {\bibinfo {title} {The
  turbulent oscillator: {A mechanism of low-frequency variability of the
  wind-driven ocean gyres}},}\ }\href@noop {} {\bibfield  {journal} {\bibinfo
  {journal} {J. Phys. Oceanogr.}\ }\textbf {\bibinfo {volume} {37}},\ \bibinfo
  {pages} {2363--2386}}\BibitemShut {NoStop}%
\bibitem [{\citenamefont {Berner}\ \emph {et~al.}(2017)\citenamefont {Berner},
  \citenamefont {Achatz}, \citenamefont {Batt{\'e}}, \citenamefont {Bengtsson},
  \citenamefont {C{\'a}mara}, \citenamefont {Christensen}, \citenamefont
  {Colangeli}, \citenamefont {Coleman}, \citenamefont {Crommelin},
  \citenamefont {Dolaptchiev},\ and\ \citenamefont {Franzke}}]{Berner.ea.2017}%
  \BibitemOpen
  \bibfield  {author} {\bibinfo {author} {\bibnamefont {Berner}, \bibfnamefont
  {J}}, \bibinfo {author} {\bibfnamefont {U.}~\bibnamefont {Achatz}}, \bibinfo
  {author} {\bibfnamefont {L.}~\bibnamefont {Batt{\'e}}}, \bibinfo {author}
  {\bibfnamefont {L.}~\bibnamefont {Bengtsson}}, \bibinfo {author}
  {\bibfnamefont {A.~de~la}\ \bibnamefont {C{\'a}mara}}, \bibinfo {author}
  {\bibfnamefont {H.~M.}\ \bibnamefont {Christensen}}, \bibinfo {author}
  {\bibfnamefont {M.}~\bibnamefont {Colangeli}}, \bibinfo {author}
  {\bibfnamefont {D.~R.~B.}\ \bibnamefont {Coleman}}, \bibinfo {author}
  {\bibfnamefont {D.}~\bibnamefont {Crommelin}}, \bibinfo {author}
  {\bibfnamefont {S.~I.}\ \bibnamefont {Dolaptchiev}}, \ and\ \bibinfo {author}
  {\bibfnamefont {C.~L.}\ \bibnamefont {Franzke}}} (\bibinfo {year} {2017}),\
  \bibfield  {title} {\enquote {\bibinfo {title} {Stochastic parameterization:
  {Toward a new view of weather and climate models}},}\ }\href@noop {}
  {\bibfield  {journal} {\bibinfo  {journal} {Bulletin of the American
  Meteorological Society}\ }\textbf {\bibinfo {volume} {98}}~(\bibinfo {number}
  {3}),\ \bibinfo {pages} {565--588}}\BibitemShut {NoStop}%
\bibitem [{\citenamefont {Bhattacharya}\ \emph {et~al.}(1982)\citenamefont
  {Bhattacharya}, \citenamefont {Ghil},\ and\ \citenamefont
  {Vulis}}]{Bhat.ea.1982}%
  \BibitemOpen
  \bibfield  {author} {\bibinfo {author} {\bibnamefont {Bhattacharya},
  \bibfnamefont {K}}, \bibinfo {author} {\bibfnamefont {M.}~\bibnamefont
  {Ghil}}, \ and\ \bibinfo {author} {\bibfnamefont {I.~L.}\ \bibnamefont
  {Vulis}}} (\bibinfo {year} {1982}),\ \bibfield  {title} {\enquote {\bibinfo
  {title} {Internal variability of an energy-balance model with delayed albedo
  effects},}\ }\href@noop {} {\bibfield  {journal} {\bibinfo  {journal}
  {Journal of the Atmospheric Sciences}\ }\textbf {\bibinfo {volume} {39}},\
  \bibinfo {pages} {1747--1773}}\BibitemShut {NoStop}%
\bibitem [{\citenamefont {Bisgard}(2015)}]{Bisgard2015}%
  \BibitemOpen
  \bibfield  {author} {\bibinfo {author} {\bibnamefont {Bisgard}, \bibfnamefont
  {J}}} (\bibinfo {year} {2015}),\ \bibfield  {title} {\enquote {\bibinfo
  {title} {Mountain passes and saddle points},}\ }\href {\doibase
  10.1137/140963510} {\bibfield  {journal} {\bibinfo  {journal} {SIAM Review}\
  }\textbf {\bibinfo {volume} {57}}~(\bibinfo {number} {2}),\ \bibinfo {pages}
  {275--292}}\BibitemShut {NoStop}%
\bibitem [{\citenamefont {Bjerknes}(1969)}]{Bjerknes1969}%
  \BibitemOpen
  \bibfield  {author} {\bibinfo {author} {\bibnamefont {Bjerknes},
  \bibfnamefont {J~P}}} (\bibinfo {year} {1969}),\ \bibfield  {title} {\enquote
  {\bibinfo {title} {{Atmospheric teleconnections from the equatorial
  Pacific}},}\ }\href@noop {} {\bibfield  {journal} {\bibinfo  {journal} {Mon.\
  Weather\ Rev.}\ }\textbf {\bibinfo {volume} {97}},\ \bibinfo {pages}
  {163--172}}\BibitemShut {NoStop}%
\bibitem [{\citenamefont {Bjerknes}(1904)}]{Bjerknes.1904}%
  \BibitemOpen
  \bibfield  {author} {\bibinfo {author} {\bibnamefont {Bjerknes},
  \bibfnamefont {V}}} (\bibinfo {year} {1904}),\ \bibfield  {title} {\enquote
  {\bibinfo {title} {{Das Problem der Wettervorhersage, betrachtet vom
  Standpunkte der Mechanik und der Physik}},}\ }\href@noop {} {\bibfield
  {journal} {\bibinfo  {journal} {Meteorologische Zeitschrift}\ }\textbf
  {\bibinfo {volume} {21}},\ \bibinfo {pages} {1--7}}\BibitemShut {NoStop}%
\bibitem [{\citenamefont {B\'odai}\ \emph {et~al.}(2013)\citenamefont
  {B\'odai}, \citenamefont {K\'arolyi},\ and\ \citenamefont {T\'el}}]{BKT13}%
  \BibitemOpen
  \bibfield  {author} {\bibinfo {author} {\bibnamefont {B\'odai}, \bibfnamefont
  {T}}, \bibinfo {author} {\bibfnamefont {G.}~\bibnamefont {K\'arolyi}}, \ and\
  \bibinfo {author} {\bibfnamefont {T.}~\bibnamefont {T\'el}}} (\bibinfo {year}
  {2013}),\ \bibfield  {title} {\enquote {\bibinfo {title} {Driving a
  conceptual model climate by different processes: Snapshot attractors and
  extreme events},}\ }\href {\doibase 10.1103/PhysRevE.87.022822} {\bibfield
  {journal} {\bibinfo  {journal} {Phys. Rev. E}\ }\textbf {\bibinfo {volume}
  {87}},\ \bibinfo {pages} {022822}}\BibitemShut {NoStop}%
\bibitem [{\citenamefont {B\'odai}\ \emph {et~al.}(2011)\citenamefont
  {B\'odai}, \citenamefont {K\'arolyi},\ and\ \citenamefont {T\'el}}]{BKT11}%
  \BibitemOpen
  \bibfield  {author} {\bibinfo {author} {\bibnamefont {B\'odai}, \bibfnamefont
  {T}}, \bibinfo {author} {\bibfnamefont {Gy.}\ \bibnamefont {K\'arolyi}}, \
  and\ \bibinfo {author} {\bibfnamefont {T.}~\bibnamefont {T\'el}}} (\bibinfo
  {year} {2011}),\ \bibfield  {title} {\enquote {\bibinfo {title} {A
  chaotically driven model climate: extreme events and snapshot attractors},}\
  }\href {\doibase 10.5194/npg-18-573-2011} {\bibfield  {journal} {\bibinfo
  {journal} {Nonlinear Processes in Geophysics}\ }\textbf {\bibinfo {volume}
  {18}}~(\bibinfo {number} {5}),\ \bibinfo {pages} {573--580}}\BibitemShut
  {NoStop}%
\bibitem [{\citenamefont {B\'odai}\ \emph {et~al.}(2020)\citenamefont
  {B\'odai}, \citenamefont {Lucarini},\ and\ \citenamefont
  {Lunkeit}}]{Bodai2018}%
  \BibitemOpen
  \bibfield  {author} {\bibinfo {author} {\bibnamefont {B\'odai}, \bibfnamefont
  {T}}, \bibinfo {author} {\bibfnamefont {V.}~\bibnamefont {Lucarini}}, \ and\
  \bibinfo {author} {\bibfnamefont {F.}~\bibnamefont {Lunkeit}}} (\bibinfo
  {year} {2020}),\ \bibfield  {title} {\enquote {\bibinfo {title} {Can we use
  linear response theory to assess geoengineering strategies?}}\ }\href
  {\doibase 10.1063/1.5122255} {\bibfield  {journal} {\bibinfo  {journal}
  {Chaos: An Interdisciplinary Journal of Nonlinear Science}\ }\textbf
  {\bibinfo {volume} {30}}~(\bibinfo {number} {2}),\ \bibinfo {pages}
  {023124}},\ \Eprint {http://arxiv.org/abs/https://doi.org/10.1063/1.5122255}
  {https://doi.org/10.1063/1.5122255} \BibitemShut {NoStop}%
\bibitem [{\citenamefont {B{\'o}dai}\ \emph {et~al.}(2015)\citenamefont
  {B{\'o}dai}, \citenamefont {Lucarini}, \citenamefont {Lunkeit},\ and\
  \citenamefont {Boschi}}]{Bodai2015}%
  \BibitemOpen
  \bibfield  {author} {\bibinfo {author} {\bibnamefont {B{\'o}dai},
  \bibfnamefont {T}}, \bibinfo {author} {\bibfnamefont {V.}~\bibnamefont
  {Lucarini}}, \bibinfo {author} {\bibfnamefont {F.}~\bibnamefont {Lunkeit}}, \
  and\ \bibinfo {author} {\bibfnamefont {R.}~\bibnamefont {Boschi}}} (\bibinfo
  {year} {2015}),\ \bibfield  {title} {\enquote {\bibinfo {title} {Global
  instability in the {Ghil--Sellers model}},}\ }\href {\doibase
  10.1007/s00382-014-2206-5} {\bibfield  {journal} {\bibinfo  {journal} {Clim.
  Dyn.}\ }\textbf {\bibinfo {volume} {44}}~(\bibinfo {number} {11}),\ \bibinfo
  {pages} {3361--3381}}\BibitemShut {NoStop}%
\bibitem [{\citenamefont {B\'odai}\ and\ \citenamefont {T\'el}(2012)}]{BT12}%
  \BibitemOpen
  \bibfield  {author} {\bibinfo {author} {\bibnamefont {B\'odai}, \bibfnamefont
  {T}}, \ and\ \bibinfo {author} {\bibfnamefont {T.}~\bibnamefont {T\'el}}}
  (\bibinfo {year} {2012}),\ \bibfield  {title} {\enquote {\bibinfo {title}
  {Annual variability in a conceptual climate model: Snapshot attractors,
  hysteresis in extreme events, and climate sensitivity},}\ }\href {\doibase
  10.1063/1.3697984} {\bibfield  {journal} {\bibinfo  {journal} {Chaos: An
  Interdisciplinary Journal of Nonlinear Science}\ }\textbf {\bibinfo {volume}
  {22}}~(\bibinfo {number} {2}),\ \bibinfo {eid} {023110}}\BibitemShut
  {NoStop}%
\bibitem [{\citenamefont {Boers}\ \emph {et~al.}(2014)\citenamefont {Boers},
  \citenamefont {Bookhagen}, \citenamefont {Barbosa}, \citenamefont {Marwan},
  \citenamefont {Kurths},\ and\ \citenamefont {Marengo}}]{Boers2014}%
  \BibitemOpen
  \bibfield  {author} {\bibinfo {author} {\bibnamefont {Boers}, \bibfnamefont
  {N}}, \bibinfo {author} {\bibfnamefont {B.}~\bibnamefont {Bookhagen}},
  \bibinfo {author} {\bibfnamefont {H.~M.~J.}\ \bibnamefont {Barbosa}},
  \bibinfo {author} {\bibfnamefont {N.}~\bibnamefont {Marwan}}, \bibinfo
  {author} {\bibfnamefont {J.}~\bibnamefont {Kurths}}, \ and\ \bibinfo {author}
  {\bibfnamefont {J.~A.}\ \bibnamefont {Marengo}}} (\bibinfo {year} {2014}),\
  \bibfield  {title} {\enquote {\bibinfo {title} {Prediction of extreme floods
  in the eastern central andes based on a complex networks approach},}\ }\href
  {https://doi.org/10.1038/ncomms6199} {\bibfield  {journal} {\bibinfo
  {journal} {Nature Communications}\ }\textbf {\bibinfo {volume} {5}},\
  \bibinfo {pages} {5199 EP --}}\BibitemShut {NoStop}%
\bibitem [{\citenamefont {Boers}\ \emph {et~al.}(2018)\citenamefont {Boers},
  \citenamefont {Ghil},\ and\ \citenamefont {Rousseau}}]{Boers.ea.2018}%
  \BibitemOpen
  \bibfield  {author} {\bibinfo {author} {\bibnamefont {Boers}, \bibfnamefont
  {N}}, \bibinfo {author} {\bibfnamefont {M.}~\bibnamefont {Ghil}}, \ and\
  \bibinfo {author} {\bibfnamefont {D.-D.}\ \bibnamefont {Rousseau}}} (\bibinfo
  {year} {2018}),\ \bibfield  {title} {\enquote {\bibinfo {title} {Ocean
  circulation, ice shelf, and sea ice interactions explain {Dansgaard--Oeschger
  cycles}},}\ }\href@noop {} {\bibfield  {journal} {\bibinfo  {journal}
  {Proceedings of the National Academy of Sciences}\ }\textbf {\bibinfo
  {volume} {115}}~(\bibinfo {number} {47}),\ \bibinfo {pages}
  {E11005--E11014}}\BibitemShut {NoStop}%
\bibitem [{\citenamefont {Boiseau}\ \emph {et~al.}(1999)\citenamefont
  {Boiseau}, \citenamefont {Ghil},\ and\ \citenamefont
  {Juillet-Leclerc}}]{Boiseau1999}%
  \BibitemOpen
  \bibfield  {author} {\bibinfo {author} {\bibnamefont {Boiseau}, \bibfnamefont
  {M}}, \bibinfo {author} {\bibfnamefont {M.}~\bibnamefont {Ghil}}, \ and\
  \bibinfo {author} {\bibfnamefont {A.}~\bibnamefont {Juillet-Leclerc}}}
  (\bibinfo {year} {1999}),\ \bibfield  {title} {\enquote {\bibinfo {title}
  {{Trends and interdecadal variability from South-Central Pacific coral
  records}},}\ }\href@noop {} {\bibfield  {journal} {\bibinfo  {journal}
  {Geophys. Res. Lett.}\ }\textbf {\bibinfo {volume} {26}},\ \bibinfo {pages}
  {2881--2884}}\BibitemShut {NoStop}%
\bibitem [{\citenamefont {Bond}\ \emph {et~al.}(1995)\citenamefont {Bond},
  \citenamefont {Showers}, \citenamefont {Cheseby}, \citenamefont {Lotti},
  \citenamefont {Almasi}, \citenamefont {deMenocal}, \citenamefont {Priori},
  \citenamefont {Cullen}, \citenamefont {Hajdas},\ and\ \citenamefont
  {Bonani}}]{Bond1995}%
  \BibitemOpen
  \bibfield  {author} {\bibinfo {author} {\bibnamefont {Bond}, \bibfnamefont
  {G}}, \bibinfo {author} {\bibfnamefont {W.}~\bibnamefont {Showers}}, \bibinfo
  {author} {\bibfnamefont {M.}~\bibnamefont {Cheseby}}, \bibinfo {author}
  {\bibfnamefont {R.}~\bibnamefont {Lotti}}, \bibinfo {author} {\bibfnamefont
  {P.}~\bibnamefont {Almasi}}, \bibinfo {author} {\bibfnamefont
  {P.}~\bibnamefont {deMenocal}}, \bibinfo {author} {\bibfnamefont
  {P.}~\bibnamefont {Priori}}, \bibinfo {author} {\bibfnamefont
  {H.}~\bibnamefont {Cullen}}, \bibinfo {author} {\bibfnamefont
  {I.}~\bibnamefont {Hajdas}}, \ and\ \bibinfo {author} {\bibfnamefont
  {G.}~\bibnamefont {Bonani}}} (\bibinfo {year} {1995}),\ \bibfield  {title}
  {\enquote {\bibinfo {title} {{A pervasive millennial-scale cycle in the North
  Atlantic Holocene and glacial climates}},}\ }\href@noop {} {\bibfield
  {journal} {\bibinfo  {journal} {Science}\ }\textbf {\bibinfo {volume}
  {278}},\ \bibinfo {pages} {1257--1265}}\BibitemShut {NoStop}%
\bibitem [{\citenamefont {Bony}\ \emph {et~al.}(2015)\citenamefont {Bony},
  \citenamefont {Stevens}, \citenamefont {Frierson}, \citenamefont {Jakob},
  \citenamefont {Kageyama}, \citenamefont {Pincus}, \citenamefont {Shepherd},
  \citenamefont {Sherwood}, \citenamefont {Siebesma}, \citenamefont {Sobel},
  \citenamefont {Watanabe},\ and\ \citenamefont {Webb}}]{Bony2015}%
  \BibitemOpen
  \bibfield  {author} {\bibinfo {author} {\bibnamefont {Bony}, \bibfnamefont
  {S}}, \bibinfo {author} {\bibfnamefont {B.}~\bibnamefont {Stevens}}, \bibinfo
  {author} {\bibfnamefont {D.~M.~W.}\ \bibnamefont {Frierson}}, \bibinfo
  {author} {\bibfnamefont {C.}~\bibnamefont {Jakob}}, \bibinfo {author}
  {\bibfnamefont {M.}~\bibnamefont {Kageyama}}, \bibinfo {author}
  {\bibfnamefont {R.}~\bibnamefont {Pincus}}, \bibinfo {author} {\bibfnamefont
  {T.~G.}\ \bibnamefont {Shepherd}}, \bibinfo {author} {\bibfnamefont {S.~C.}\
  \bibnamefont {Sherwood}}, \bibinfo {author} {\bibfnamefont {A.~P.}\
  \bibnamefont {Siebesma}}, \bibinfo {author} {\bibfnamefont {A.~H.}\
  \bibnamefont {Sobel}}, \bibinfo {author} {\bibfnamefont {M.}~\bibnamefont
  {Watanabe}}, \ and\ \bibinfo {author} {\bibfnamefont {M.~J.}\ \bibnamefont
  {Webb}}} (\bibinfo {year} {2015}),\ \bibfield  {title} {\enquote {\bibinfo
  {title} {Clouds, circulation and climate sensitivity},}\ }\href {\doibase
  10.1038/ngeo2398} {\bibfield  {journal} {\bibinfo  {journal} {Nat Geosci}\
  }\textbf {\bibinfo {volume} {8}},\ 10.1038/ngeo2398}\BibitemShut {NoStop}%
\bibitem [{\citenamefont {Boos}\ and\ \citenamefont {Hurley}(2013)}]{Boos13}%
  \BibitemOpen
  \bibfield  {author} {\bibinfo {author} {\bibnamefont {Boos}, \bibfnamefont
  {W~R}}, \ and\ \bibinfo {author} {\bibfnamefont {J.~V.}\ \bibnamefont
  {Hurley}}} (\bibinfo {year} {2013}),\ \bibfield  {title} {\enquote {\bibinfo
  {title} {Thermodynamic bias in the multimodel mean boreal summer monsoon},}\
  }\href@noop {} {\bibfield  {journal} {\bibinfo  {journal} {J. Climate}\
  }\textbf {\bibinfo {volume} {26}}}\BibitemShut {NoStop}%
\bibitem [{\citenamefont {Bouchet}\ \emph {et~al.}(2014)\citenamefont
  {Bouchet}, \citenamefont {Laurie},\ and\ \citenamefont
  {Zaboronski}}]{Bouchet2014}%
  \BibitemOpen
  \bibfield  {author} {\bibinfo {author} {\bibnamefont {Bouchet}, \bibfnamefont
  {F}}, \bibinfo {author} {\bibfnamefont {J.}~\bibnamefont {Laurie}}, \ and\
  \bibinfo {author} {\bibfnamefont {O.}~\bibnamefont {Zaboronski}}} (\bibinfo
  {year} {2014}),\ \bibfield  {title} {\enquote {\bibinfo {title} {Langevin
  dynamics, large deviations and instantons for the quasi-geostrophic model and
  two-dimensional euler equations},}\ }\href {\doibase
  10.1007/s10955-014-1052-5} {\bibfield  {journal} {\bibinfo  {journal}
  {Journal of Statistical Physics}\ }\textbf {\bibinfo {volume}
  {156}}~(\bibinfo {number} {6}),\ \bibinfo {pages} {1066--1092}}\BibitemShut
  {NoStop}%
\bibitem [{\citenamefont {Bouchet}\ and\ \citenamefont
  {Sommeria}(2002)}]{Bouchet2002}%
  \BibitemOpen
  \bibfield  {author} {\bibinfo {author} {\bibnamefont {Bouchet}, \bibfnamefont
  {F}}, \ and\ \bibinfo {author} {\bibfnamefont {J.}~\bibnamefont {Sommeria}}}
  (\bibinfo {year} {2002}),\ \bibfield  {title} {\enquote {\bibinfo {title}
  {Emergence of intense jets and {Jupiter's Great Red Spot} as maximum-entropy
  structures},}\ }\href {\doibase 10.1017/S0022112002008789} {\bibfield
  {journal} {\bibinfo  {journal} {J. Fluid Mech.}\ }\textbf {\bibinfo {volume}
  {464}},\ \bibinfo {pages} {165--207}}\BibitemShut {NoStop}%
\bibitem [{\citenamefont {Bouchet}\ and\ \citenamefont
  {Venaille}(2012)}]{Bouchet:2012}%
  \BibitemOpen
  \bibfield  {author} {\bibinfo {author} {\bibnamefont {Bouchet}, \bibfnamefont
  {F}}, \ and\ \bibinfo {author} {\bibfnamefont {A.}~\bibnamefont {Venaille}}}
  (\bibinfo {year} {2012}),\ \bibfield  {title} {\enquote {\bibinfo {title}
  {Statistical mechanics of two-dimensional and geophysical flows},}\ }\href
  {\doibase 10.1016/j.physrep.2012.02.001} {\bibfield  {journal} {\bibinfo
  {journal} {Physics Reports}\ }\textbf {\bibinfo {volume} {515}}~(\bibinfo
  {number} {5}),\ \bibinfo {pages} {227--295}}\BibitemShut {NoStop}%
\bibitem [{\citenamefont {Branstator}(1987)}]{Branstator.1987}%
  \BibitemOpen
  \bibfield  {author} {\bibinfo {author} {\bibnamefont {Branstator},
  \bibfnamefont {G}}} (\bibinfo {year} {1987}),\ \bibfield  {title} {\enquote
  {\bibinfo {title} {A striking example of the atmosphere's leading traveling
  pattern},}\ }\href {\doibase 10.1175/1520-0469(1987)044<2310:aseota>2.0.co;2}
  {\bibfield  {journal} {\bibinfo  {journal} {J. Atmos. Sci.}\ }\textbf
  {\bibinfo {volume} {44}}~(\bibinfo {number} {16}),\ \bibinfo {pages}
  {2310--2323}}\BibitemShut {NoStop}%
\bibitem [{\citenamefont {Breiman}(2001)}]{Breiman.2001}%
  \BibitemOpen
  \bibfield  {author} {\bibinfo {author} {\bibnamefont {Breiman}, \bibfnamefont
  {L}}} (\bibinfo {year} {2001}),\ \bibfield  {title} {\enquote {\bibinfo
  {title} {Random forests},}\ }\href@noop {} {\bibfield  {journal} {\bibinfo
  {journal} {Machine Learning}\ }\textbf {\bibinfo {volume} {45}}~(\bibinfo
  {number} {1}),\ \bibinfo {pages} {5--32}}\BibitemShut {NoStop}%
\bibitem [{\citenamefont {Broecker}(1991)}]{Broecker1991}%
  \BibitemOpen
  \bibfield  {author} {\bibinfo {author} {\bibnamefont {Broecker},
  \bibfnamefont {W~S}}} (\bibinfo {year} {1991}),\ \bibfield  {title} {\enquote
  {\bibinfo {title} {{The great ocean conveyor}},}\ }\href@noop {} {\bibfield
  {journal} {\bibinfo  {journal} {Oceanography}\ }\textbf {\bibinfo {volume}
  {4}},\ \bibinfo {pages} {79--89}}\BibitemShut {NoStop}%
\bibitem [{\citenamefont {Bryan}(1986)}]{Bryan1986}%
  \BibitemOpen
  \bibfield  {author} {\bibinfo {author} {\bibnamefont {Bryan}, \bibfnamefont
  {F~O}}} (\bibinfo {year} {1986}),\ \bibfield  {title} {\enquote {\bibinfo
  {title} {High-latitude salinity effects and interhemispheric thermohaline
  circulations},}\ }\href@noop {} {\bibfield  {journal} {\bibinfo  {journal}
  {Nature}\ }\textbf {\bibinfo {volume} {323}},\ \bibinfo {pages}
  {301--304}}\BibitemShut {NoStop}%
\bibitem [{\citenamefont {Budi{\v{s}}i{\'c}}\ \emph {et~al.}(2012)\citenamefont
  {Budi{\v{s}}i{\'c}}, \citenamefont {Mohr},\ and\ \citenamefont
  {Mezi{\'c}}}]{Mezic.ea.2012}%
  \BibitemOpen
  \bibfield  {author} {\bibinfo {author} {\bibnamefont {Budi{\v{s}}i{\'c}},
  \bibfnamefont {M}}, \bibinfo {author} {\bibfnamefont {R.}~\bibnamefont
  {Mohr}}, \ and\ \bibinfo {author} {\bibfnamefont {I.}~\bibnamefont
  {Mezi{\'c}}}} (\bibinfo {year} {2012}),\ \bibfield  {title} {\enquote
  {\bibinfo {title} {Applied {Koopmanism}},}\ }\href@noop {} {\bibfield
  {journal} {\bibinfo  {journal} {Chaos: An Interdisciplinary Journal of
  Nonlinear Science}\ }\textbf {\bibinfo {volume} {22}}~(\bibinfo {number}
  {4}),\ \bibinfo {pages} {047510}}\BibitemShut {NoStop}%
\bibitem [{\citenamefont {Budyko}(1969)}]{Budyko}%
  \BibitemOpen
  \bibfield  {author} {\bibinfo {author} {\bibnamefont {Budyko}, \bibfnamefont
  {M~I}}} (\bibinfo {year} {1969}),\ \bibfield  {title} {\enquote {\bibinfo
  {title} {The effect of solar radiation variations on the climate of the
  {Earth}},}\ }\href@noop {} {\bibfield  {journal} {\bibinfo  {journal}
  {Tellus}\ }\textbf {\bibinfo {volume} {21}},\ \bibinfo {pages}
  {611--619}}\BibitemShut {NoStop}%
\bibitem [{\citenamefont {Cane}\ and\ \citenamefont {Zebiak}(1985)}]{Cane1985}%
  \BibitemOpen
  \bibfield  {author} {\bibinfo {author} {\bibnamefont {Cane}, \bibfnamefont
  {M~A}}, \ and\ \bibinfo {author} {\bibfnamefont {S.~E.}\ \bibnamefont
  {Zebiak}}} (\bibinfo {year} {1985}),\ \bibfield  {title} {\enquote {\bibinfo
  {title} {{A theory for El Ni\~{n}o and the Southern Oscillation}},}\
  }\href@noop {} {\bibfield  {journal} {\bibinfo  {journal} {Science}\ }\textbf
  {\bibinfo {volume} {228}},\ \bibinfo {pages} {1084--1087}}\BibitemShut
  {NoStop}%
\bibitem [{\citenamefont {Caraballo}\ and\ \citenamefont
  {Han}(2017)}]{Caraballo.Han.2017}%
  \BibitemOpen
  \bibfield  {author} {\bibinfo {author} {\bibnamefont {Caraballo},
  \bibfnamefont {T}}, \ and\ \bibinfo {author} {\bibfnamefont {X.}~\bibnamefont
  {Han}}} (\bibinfo {year} {2017}),\ \href@noop {} {\emph {\bibinfo {title}
  {{Applied Nonautonomous and Random Dynamical Systems: Applied Dynamical
  Systems}}}}\ (\bibinfo  {publisher} {Springer Science + Business
  Media})\BibitemShut {NoStop}%
\bibitem [{\citenamefont {Carrassi}\ \emph {et~al.}(2018)\citenamefont
  {Carrassi}, \citenamefont {Bocquet}, \citenamefont {Bertino},\ and\
  \citenamefont {Evensen}}]{Carrassi2018}%
  \BibitemOpen
  \bibfield  {author} {\bibinfo {author} {\bibnamefont {Carrassi},
  \bibfnamefont {A}}, \bibinfo {author} {\bibfnamefont {M.}~\bibnamefont
  {Bocquet}}, \bibinfo {author} {\bibfnamefont {L.}~\bibnamefont {Bertino}}, \
  and\ \bibinfo {author} {\bibfnamefont {G.}~\bibnamefont {Evensen}}} (\bibinfo
  {year} {2018}),\ \bibfield  {title} {\enquote {\bibinfo {title} {Data
  assimilation in the geosciences: {An overview of methods, issues, and
  perspectives}},}\ }\href {\doibase 10.1002/wcc.535} {\bibfield  {journal}
  {\bibinfo  {journal} {WIREs Climate Change}\ }\textbf {\bibinfo {volume}
  {9}}~(\bibinfo {number} {5}),\ \bibinfo {pages} {{e535}}},\ \Eprint
  {http://arxiv.org/abs/https://onlinelibrary.wiley.com/doi/pdf/10.1002/wcc.535}
  {https://onlinelibrary.wiley.com/doi/pdf/10.1002/wcc.535} \BibitemShut
  {NoStop}%
\bibitem [{\citenamefont {Carton}\ and\ \citenamefont
  {Giese}(2008)}]{Carton.Giese.2008}%
  \BibitemOpen
  \bibfield  {author} {\bibinfo {author} {\bibnamefont {Carton}, \bibfnamefont
  {J~A}}, \ and\ \bibinfo {author} {\bibfnamefont {B.~S.}\ \bibnamefont
  {Giese}}} (\bibinfo {year} {2008}),\ \bibfield  {title} {\enquote {\bibinfo
  {title} {A reanalysis of ocean climate using {Simple Ocean Data Assimilation
  (SODA)}},}\ }\href {\doibase 10.1175/2007MWR1978.1} {\bibfield  {journal}
  {\bibinfo  {journal} {Mon. Wea. Rev.}\ }\textbf {\bibinfo {volume}
  {136}}~(\bibinfo {number} {8}),\ \bibinfo {pages} {2999--3017}}\BibitemShut
  {NoStop}%
\bibitem [{\citenamefont {Carvalho}\ \emph {et~al.}(2013)\citenamefont
  {Carvalho}, \citenamefont {Langa},\ and\ \citenamefont {Robinson}}]{CLR13}%
  \BibitemOpen
  \bibfield  {author} {\bibinfo {author} {\bibnamefont {Carvalho},
  \bibfnamefont {A~N}}, \bibinfo {author} {\bibfnamefont {J.}~\bibnamefont
  {Langa}}, \ and\ \bibinfo {author} {\bibfnamefont {J.~C.}\ \bibnamefont
  {Robinson}}} (\bibinfo {year} {2013}),\ \bibfield  {title} {\enquote
  {\bibinfo {title} {The pullback attractor},}\ }in\ \href {\doibase
  10.1007/978-1-4614-4581-4\_1} {\emph {\bibinfo {booktitle} {Attractors for
  infinite-dimensional non-autonomous dynamical systems}}},\ \bibinfo {series}
  {Applied Mathematical Sciences}, Vol.\ \bibinfo {volume} {182}\ (\bibinfo
  {publisher} {Springer New York})\ pp.\ \bibinfo {pages} {3--22}\BibitemShut
  {NoStop}%
\bibitem [{\citenamefont {Cessi}(2019)}]{Cessi.2019}%
  \BibitemOpen
  \bibfield  {author} {\bibinfo {author} {\bibnamefont {Cessi}, \bibfnamefont
  {P}}} (\bibinfo {year} {2019}),\ \bibfield  {title} {\enquote {\bibinfo
  {title} {The global overturning circulation},}\ }\href@noop {} {\bibfield
  {journal} {\bibinfo  {journal} {Annual Review of Marine Science}\ }\textbf
  {\bibinfo {volume} {11}},\ \bibinfo {pages} {249--270}}\BibitemShut {NoStop}%
\bibitem [{\citenamefont {Cessi}\ and\ \citenamefont
  {Ierley}(1995)}]{Cessi1995}%
  \BibitemOpen
  \bibfield  {author} {\bibinfo {author} {\bibnamefont {Cessi}, \bibfnamefont
  {P}}, \ and\ \bibinfo {author} {\bibfnamefont {G.~R.}\ \bibnamefont
  {Ierley}}} (\bibinfo {year} {1995}),\ \bibfield  {title} {\enquote {\bibinfo
  {title} {Symmetry-breaking multiple equilibria in quasi-geostrophic,
  wind-driven flows},}\ }\href@noop {} {\bibfield  {journal} {\bibinfo
  {journal} {J. Phys. Oceanogr.}\ }\textbf {\bibinfo {volume} {25}},\ \bibinfo
  {pages} {1196--1205}}\BibitemShut {NoStop}%
\bibitem [{\citenamefont {Cessi}\ and\ \citenamefont
  {Young}(1992)}]{Cessi1992}%
  \BibitemOpen
  \bibfield  {author} {\bibinfo {author} {\bibnamefont {Cessi}, \bibfnamefont
  {P}}, \ and\ \bibinfo {author} {\bibfnamefont {W.~R.}\ \bibnamefont {Young}}}
  (\bibinfo {year} {1992}),\ \bibfield  {title} {\enquote {\bibinfo {title}
  {Multiple equilibria in two-dimensional thermohaline circulation},}\
  }\href@noop {} {\bibfield  {journal} {\bibinfo  {journal} {J.\ Fluid Mech.}\
  }\textbf {\bibinfo {volume} {241}},\ \bibinfo {pages} {291--309}}\BibitemShut
  {NoStop}%
\bibitem [{\citenamefont {Cvitanovi\ifmmode~\acute{c}\else
  \'{c}\fi{}}(1988)}]{svita88}%
  \BibitemOpen
  \bibfield  {author} {\bibinfo {author} {\bibnamefont
  {Cvitanovi\ifmmode~\acute{c}\else \'{c}\fi{}}, \bibfnamefont {Predrag}}}
  (\bibinfo {year} {1988}),\ \bibfield  {title} {\enquote {\bibinfo {title}
  {Invariant measurement of strange sets in terms of cycles},}\ }\href
  {\doibase 10.1103/PhysRevLett.61.2729} {\bibfield  {journal} {\bibinfo
  {journal} {Phys. Rev. Lett.}\ }\textbf {\bibinfo {volume} {61}},\ \bibinfo
  {pages} {2729--2732}}\BibitemShut {NoStop}%
\bibitem [{\citenamefont {Chang}\ \emph {et~al.}(2015)\citenamefont {Chang},
  \citenamefont {Ghil}, \citenamefont {Latif},\ and\ \citenamefont
  {Wallace}}]{Chang.ea.2015}%
  \BibitemOpen
  \bibinfo {editor} {\bibnamefont {Chang}, \bibfnamefont {C~P}}, \bibinfo
  {editor} {\bibfnamefont {M.}~\bibnamefont {Ghil}}, \bibinfo {editor}
  {\bibfnamefont {M.}~\bibnamefont {Latif}}, \ and\ \bibinfo {editor}
  {\bibfnamefont {J.~M.}\ \bibnamefont {Wallace}},\ Eds. (\bibinfo {year}
  {2015}),\ \href@noop {} {\emph {\bibinfo {title} {{Climate Change :
  Multidecadal and Beyond}}}}\ (\bibinfo  {publisher} {World Scientific
  Publishing Co./Imperial College Press})\BibitemShut {NoStop}%
\bibitem [{\citenamefont {Chang}\ \emph {et~al.}(2001)\citenamefont {Chang},
  \citenamefont {Ghil}, \citenamefont {Ide},\ and\ \citenamefont
  {Lai}}]{Chang2001}%
  \BibitemOpen
  \bibfield  {author} {\bibinfo {author} {\bibnamefont {Chang}, \bibfnamefont
  {K-Il}}, \bibinfo {author} {\bibfnamefont {M.}~\bibnamefont {Ghil}}, \bibinfo
  {author} {\bibfnamefont {K.}~\bibnamefont {Ide}}, \ and\ \bibinfo {author}
  {\bibfnamefont {C-C.~A.}\ \bibnamefont {Lai}}} (\bibinfo {year} {2001}),\
  \bibfield  {title} {\enquote {\bibinfo {title} {{Transition to aperiodic
  variability in a wind-driven double-gyre circulation model}},}\ }\href@noop
  {} {\bibfield  {journal} {\bibinfo  {journal} {J.\ Phys.\ Oceanogr.}\
  }\textbf {\bibinfo {volume} {31}},\ \bibinfo {pages}
  {1260--1286}}\BibitemShut {NoStop}%
\bibitem [{\citenamefont {Charney}(1947)}]{char48}%
  \BibitemOpen
  \bibfield  {author} {\bibinfo {author} {\bibnamefont {Charney}, \bibfnamefont
  {J~G}}} (\bibinfo {year} {1947}),\ \bibfield  {title} {\enquote {\bibinfo
  {title} {The dynamics of long waves in a baroclinic westerly current},}\
  }\href {\doibase 10.1175/1520-0469(1947)004<0136:TDOLWI>2.0.CO;2} {\bibfield
  {journal} {\bibinfo  {journal} {J. Meteorol.}\ }\textbf {\bibinfo {volume}
  {4}}~(\bibinfo {number} {5}),\ \bibinfo {pages} {136--162}}\BibitemShut
  {NoStop}%
\bibitem [{\citenamefont {Charney}({1971})}]{Charney_QG}%
  \BibitemOpen
  \bibfield  {author} {\bibinfo {author} {\bibnamefont {Charney}, \bibfnamefont
  {J~G}}} (\bibinfo {year} {{1971}}),\ \bibfield  {title} {\enquote {\bibinfo
  {title} {Geostrophic turbulence},}\ }\href {\doibase
  {10.1175/1520-0469(1971)028<1087:GT>2.0.CO;2}} {\bibfield  {journal}
  {\bibinfo  {journal} {J.\ Atmos.\ Sci.}\ }\textbf {\bibinfo {volume}
  {{28}}}~(\bibinfo {number} {{6}}),\ \bibinfo {pages}
  {1087--1095}}\BibitemShut {NoStop}%
\bibitem [{\citenamefont {Charney}\ \emph {et~al.}(1979)\citenamefont
  {Charney}, \citenamefont {Arakawa}, \citenamefont {Baker}, \citenamefont
  {Bolin}, \citenamefont {Dickinson}, \citenamefont {Goody}, \citenamefont
  {Leith}, \citenamefont {Stommel},\ and\ \citenamefont
  {Wunsch}}]{Charney.ea.1979}%
  \BibitemOpen
  \bibfield  {author} {\bibinfo {author} {\bibnamefont {Charney}, \bibfnamefont
  {J~G}}, \bibinfo {author} {\bibfnamefont {A.}~\bibnamefont {Arakawa}},
  \bibinfo {author} {\bibfnamefont {D.~J.}\ \bibnamefont {Baker}}, \bibinfo
  {author} {\bibfnamefont {B.}~\bibnamefont {Bolin}}, \bibinfo {author}
  {\bibfnamefont {R.~E.}\ \bibnamefont {Dickinson}}, \bibinfo {author}
  {\bibfnamefont {R.~M.}\ \bibnamefont {Goody}}, \bibinfo {author}
  {\bibfnamefont {C.~E.}\ \bibnamefont {Leith}}, \bibinfo {author}
  {\bibfnamefont {H.~M.}\ \bibnamefont {Stommel}}, \ and\ \bibinfo {author}
  {\bibfnamefont {C.~I.}\ \bibnamefont {Wunsch}}} (\bibinfo {year} {1979}),\
  \href@noop {} {\emph {\bibinfo {title} {{Carbon Dioxide and Climate: A
  Scientific Assessment}}}}\ (\bibinfo  {publisher} {National Academy of
  Sciences},\ \bibinfo {address} {Washington, DC})\BibitemShut {NoStop}%
\bibitem [{\citenamefont {Charney}\ and\ \citenamefont
  {DeVore}(1979)}]{Charney1979}%
  \BibitemOpen
  \bibfield  {author} {\bibinfo {author} {\bibnamefont {Charney}, \bibfnamefont
  {J~G}}, \ and\ \bibinfo {author} {\bibfnamefont {J.~G.}\ \bibnamefont
  {DeVore}}} (\bibinfo {year} {1979}),\ \bibfield  {title} {\enquote {\bibinfo
  {title} {Multiple flow equilibria in the atmosphere and blocking},}\
  }\href@noop {} {\bibfield  {journal} {\bibinfo  {journal} {J. Atmos. Sci.}\
  }\textbf {\bibinfo {volume} {36}},\ \bibinfo {pages}
  {1205--1216}}\BibitemShut {NoStop}%
\bibitem [{\citenamefont {Charney}\ \emph
  {et~al.}(1950{\natexlab{a}})\citenamefont {Charney}, \citenamefont
  {Fj{\o}rtoft},\ and\ \citenamefont {von Neumann}}]{charney50}%
  \BibitemOpen
  \bibfield  {author} {\bibinfo {author} {\bibnamefont {Charney}, \bibfnamefont
  {J~G}}, \bibinfo {author} {\bibfnamefont {R.}~\bibnamefont {Fj{\o}rtoft}}, \
  and\ \bibinfo {author} {\bibfnamefont {J.}~\bibnamefont {von Neumann}}}
  (\bibinfo {year} {1950}{\natexlab{a}}),\ \bibfield  {title} {\enquote
  {\bibinfo {title} {Numerical integration of the barotropic vorticity
  equation},}\ }\href@noop {} {\bibfield  {journal} {\bibinfo  {journal}
  {Tellus}\ }\textbf {\bibinfo {volume} {2}},\ \bibinfo {pages}
  {237--254}}\BibitemShut {NoStop}%
\bibitem [{\citenamefont {Charney}\ \emph
  {et~al.}(1950{\natexlab{b}})\citenamefont {Charney}, \citenamefont
  {Fj{\o}rtoft},\ and\ \citenamefont {Von~Neumann}}]{CFN.1950}%
  \BibitemOpen
  \bibfield  {author} {\bibinfo {author} {\bibnamefont {Charney}, \bibfnamefont
  {J~G}}, \bibinfo {author} {\bibfnamefont {R.}~\bibnamefont {Fj{\o}rtoft}}, \
  and\ \bibinfo {author} {\bibfnamefont {J.}~\bibnamefont {Von~Neumann}}}
  (\bibinfo {year} {1950}{\natexlab{b}}),\ \bibfield  {title} {\enquote
  {\bibinfo {title} {Numerical integration of the barotropic vorticity
  equation},}\ }\href@noop {} {\bibfield  {journal} {\bibinfo  {journal}
  {Tellus}\ }\textbf {\bibinfo {volume} {2}},\ \bibinfo {pages}
  {38--54}}\BibitemShut {NoStop}%
\bibitem [{\citenamefont {Charney}\ \emph {et~al.}(1969)\citenamefont
  {Charney}, \citenamefont {Halem},\ and\ \citenamefont
  {Jastrow}}]{Charney.ea.69}%
  \BibitemOpen
  \bibfield  {author} {\bibinfo {author} {\bibnamefont {Charney}, \bibfnamefont
  {J~G}}, \bibinfo {author} {\bibfnamefont {M.}~\bibnamefont {Halem}}, \ and\
  \bibinfo {author} {\bibfnamefont {R.}~\bibnamefont {Jastrow}}} (\bibinfo
  {year} {1969}),\ \bibfield  {title} {\enquote {\bibinfo {title} {Use of
  incomplete historical data to infer the present state of the atmosphere},}\
  }\href@noop {} {\bibfield  {journal} {\bibinfo  {journal} {J. Atmos. Sci.}\
  }\textbf {\bibinfo {volume} {26}},\ \bibinfo {pages}
  {1160--1163}}\BibitemShut {NoStop}%
\bibitem [{\citenamefont {Charney}\ \emph {et~al.}(1981)\citenamefont
  {Charney}, \citenamefont {Shukla},\ and\ \citenamefont
  {Mo}}]{Charney.ea.1981}%
  \BibitemOpen
  \bibfield  {author} {\bibinfo {author} {\bibnamefont {Charney}, \bibfnamefont
  {J~G}}, \bibinfo {author} {\bibfnamefont {J.}~\bibnamefont {Shukla}}, \ and\
  \bibinfo {author} {\bibfnamefont {K.~C.}\ \bibnamefont {Mo}}} (\bibinfo
  {year} {1981}),\ \bibfield  {title} {\enquote {\bibinfo {title} {Comparison
  of a barotropic blocking theory with observation},}\ }\href {\doibase
  10.1175/1520-0469(1981)038<0762:coabbt>2.0.co;2} {\bibfield  {journal}
  {\bibinfo  {journal} {{J. Atmos. Sci.}}\ }\textbf {\bibinfo {volume}
  {38}}~(\bibinfo {number} {4}),\ \bibinfo {pages} {762--779}}\BibitemShut
  {NoStop}%
\bibitem [{\citenamefont {Chavanis}\ and\ \citenamefont
  {Sommeria}(1996)}]{chavanis_sommeria_1996}%
  \BibitemOpen
  \bibfield  {author} {\bibinfo {author} {\bibnamefont {Chavanis},
  \bibfnamefont {P~H}}, \ and\ \bibinfo {author} {\bibfnamefont
  {J.}~\bibnamefont {Sommeria}}} (\bibinfo {year} {1996}),\ \bibfield  {title}
  {\enquote {\bibinfo {title} {Classification of self-organized vortices in
  two-dimensional turbulence: the case of a bounded domain},}\ }\href {\doibase
  10.1017/S0022112096000316} {\bibfield  {journal} {\bibinfo  {journal}
  {Journal of Fluid Mechanics}\ }\textbf {\bibinfo {volume} {314}},\ \bibinfo
  {pages} {267--297}}\BibitemShut {NoStop}%
\bibitem [{\citenamefont {Chekroun}\ \emph {et~al.}(2018)\citenamefont
  {Chekroun}, \citenamefont {Ghil},\ and\ \citenamefont
  {Neelin}}]{Chekroun.ea.2018}%
  \BibitemOpen
  \bibfield  {author} {\bibinfo {author} {\bibnamefont {Chekroun},
  \bibfnamefont {M~D}}, \bibinfo {author} {\bibfnamefont {M.}~\bibnamefont
  {Ghil}}, \ and\ \bibinfo {author} {\bibfnamefont {J.~D.}\ \bibnamefont
  {Neelin}}} (\bibinfo {year} {2018}),\ \bibfield  {title} {\enquote {\bibinfo
  {title} {Pullback attractor crisis in a delay differential {ENSO model}},}\
  }in\ \href@noop {} {\emph {\bibinfo {booktitle} {Advances in Nonlinear
  Geosciences}}},\ \bibinfo {editor} {edited by\ \bibinfo {editor}
  {\bibfnamefont {A.}~\bibnamefont {Tsonis}}}\ (\bibinfo  {publisher}
  {Springer})\ pp.\ \bibinfo {pages} {1--33}\BibitemShut {NoStop}%
\bibitem [{\citenamefont {Chekroun}\ \emph {et~al.}(2014)\citenamefont
  {Chekroun}, \citenamefont {Neelin}, \citenamefont {Kondrashov}, \citenamefont
  {McWilliams},\ and\ \citenamefont {Ghil}}]{Chekroun2014}%
  \BibitemOpen
  \bibfield  {author} {\bibinfo {author} {\bibnamefont {Chekroun},
  \bibfnamefont {M~D}}, \bibinfo {author} {\bibfnamefont {J.~D.}\ \bibnamefont
  {Neelin}}, \bibinfo {author} {\bibfnamefont {D.}~\bibnamefont {Kondrashov}},
  \bibinfo {author} {\bibfnamefont {J.~C.}\ \bibnamefont {McWilliams}}, \ and\
  \bibinfo {author} {\bibfnamefont {M.}~\bibnamefont {Ghil}}} (\bibinfo {year}
  {2014}),\ \bibfield  {title} {\enquote {\bibinfo {title} {Rough parameter
  dependence in climate models and the role of {Ruelle-Pollicott
  resonances}},}\ }\href {\doibase 10.1073/pnas.1321816111} {\bibfield
  {journal} {\bibinfo  {journal} {Proceedings of the National Academy of
  Sciences}\ }\textbf {\bibinfo {volume} {111}}~(\bibinfo {number} {5}),\
  \bibinfo {pages} {1684--1690}}\BibitemShut {NoStop}%
\bibitem [{\citenamefont {Chekroun}\ \emph {et~al.}(2011)\citenamefont
  {Chekroun}, \citenamefont {Simonnet},\ and\ \citenamefont
  {Ghil}}]{Chekroun2011}%
  \BibitemOpen
  \bibfield  {author} {\bibinfo {author} {\bibnamefont {Chekroun},
  \bibfnamefont {M~D}}, \bibinfo {author} {\bibfnamefont {E.}~\bibnamefont
  {Simonnet}}, \ and\ \bibinfo {author} {\bibfnamefont {M.}~\bibnamefont
  {Ghil}}} (\bibinfo {year} {2011}),\ \bibfield  {title} {\enquote {\bibinfo
  {title} {Stochastic climate dynamics: {Random attractors and time-dependent
  invariant measures}},}\ }\href {\doibase 10.1016/j.physd.2011.06.005}
  {\bibfield  {journal} {\bibinfo  {journal} {Physica D: Nonlinear Phenomena}\
  }\textbf {\bibinfo {volume} {240}}~(\bibinfo {number} {21}),\ \bibinfo
  {pages} {1685--1700}}\BibitemShut {NoStop}%
\bibitem [{\citenamefont {Chen}\ and\ \citenamefont {Ghil}(1996)}]{Chen1996}%
  \BibitemOpen
  \bibfield  {author} {\bibinfo {author} {\bibnamefont {Chen}, \bibfnamefont
  {F}}, \ and\ \bibinfo {author} {\bibfnamefont {M.}~\bibnamefont {Ghil}}}
  (\bibinfo {year} {1996}),\ \bibfield  {title} {\enquote {\bibinfo {title}
  {Interdecadal variability in a hybrid coupled ocean-atmosphere model},}\
  }\href@noop {} {\bibfield  {journal} {\bibinfo  {journal} {J.\ Phys.\
  Oceanogr.}\ }\textbf {\bibinfo {volume} {26}},\ \bibinfo {pages}
  {1561--1578}}\BibitemShut {NoStop}%
\bibitem [{\citenamefont {Cheng}\ and\ \citenamefont
  {Wallace}(1993)}]{Cheng1993}%
  \BibitemOpen
  \bibfield  {author} {\bibinfo {author} {\bibnamefont {Cheng}, \bibfnamefont
  {X}}, \ and\ \bibinfo {author} {\bibfnamefont {J.~M.}\ \bibnamefont
  {Wallace}}} (\bibinfo {year} {1993}),\ \bibfield  {title} {\enquote {\bibinfo
  {title} {{Cluster analysis of the Northern Hemisphere wintertime 500 pHa
  heigh field: Spatial patterns}},}\ }\href@noop {} {\bibfield  {journal}
  {\bibinfo  {journal} {J.\ Atmos.\ Sci.}\ }\textbf {\bibinfo {volume} {50}},\
  \bibinfo {pages} {2674--2696}}\BibitemShut {NoStop}%
\bibitem [{\citenamefont {Chorin}\ and\ \citenamefont
  {Stinis}(2007)}]{Chorin.Stinis.2007}%
  \BibitemOpen
  \bibfield  {author} {\bibinfo {author} {\bibnamefont {Chorin}, \bibfnamefont
  {A}}, \ and\ \bibinfo {author} {\bibfnamefont {P.}~\bibnamefont {Stinis}}}
  (\bibinfo {year} {2007}),\ \bibfield  {title} {\enquote {\bibinfo {title}
  {Problem reduction, renormalization, and memory},}\ }\href@noop {} {\bibfield
   {journal} {\bibinfo  {journal} {Communications in Applied Mathematics and
  Computational Science}\ }\textbf {\bibinfo {volume} {1}}~(\bibinfo {number}
  {1}),\ \bibinfo {pages} {1--27}}\BibitemShut {NoStop}%
\bibitem [{\citenamefont {Chorin}\ \emph {et~al.}(2002)\citenamefont {Chorin},
  \citenamefont {Hald},\ and\ \citenamefont {Kupferman}}]{chorin_optimal_2002}%
  \BibitemOpen
  \bibfield  {author} {\bibinfo {author} {\bibnamefont {Chorin}, \bibfnamefont
  {A~J}}, \bibinfo {author} {\bibfnamefont {O.~H.}\ \bibnamefont {Hald}}, \
  and\ \bibinfo {author} {\bibfnamefont {R.}~\bibnamefont {Kupferman}}}
  (\bibinfo {year} {2002}),\ \bibfield  {title} {\enquote {\bibinfo {title}
  {Optimal prediction with memory},}\ }\href {\doibase
  10.1016/S0167-2789(02)00446-3} {\bibfield  {journal} {\bibinfo  {journal}
  {Physica D: Nonlinear Phenomena}\ }\textbf {\bibinfo {volume} {166}},\
  \bibinfo {pages} {239--257}}\BibitemShut {NoStop}%
\bibitem [{\citenamefont {Chown}(2004)}]{Chown.2004}%
  \BibitemOpen
  \bibfield  {author} {\bibinfo {author} {\bibnamefont {Chown}, \bibfnamefont
  {M}}} (\bibinfo {year} {2004}),\ \bibfield  {title} {\enquote {\bibinfo
  {title} {Chaotic heavens},}\ }\href@noop {} {\bibfield  {journal} {\bibinfo
  {journal} {New Scientist}\ }\textbf {\bibinfo {volume} {181}}~(\bibinfo
  {number} {2436}),\ \bibinfo {pages} {32--35}}\BibitemShut {NoStop}%
\bibitem [{\citenamefont {Cionni}\ \emph {et~al.}(2004)\citenamefont {Cionni},
  \citenamefont {Visconti},\ and\ \citenamefont {Sassi}}]{Cionni2004}%
  \BibitemOpen
  \bibfield  {author} {\bibinfo {author} {\bibnamefont {Cionni}, \bibfnamefont
  {I}}, \bibinfo {author} {\bibfnamefont {G.}~\bibnamefont {Visconti}}, \ and\
  \bibinfo {author} {\bibfnamefont {F.}~\bibnamefont {Sassi}}} (\bibinfo {year}
  {2004}),\ \bibfield  {title} {\enquote {\bibinfo {title} {Fluctuation
  dissipation theorem in a general circulation model},}\ }\href {\doibase
  10.1029/2004GL019739} {\bibfield  {journal} {\bibinfo  {journal} {Geophysical
  Research Letters}\ }\textbf {\bibinfo {volume} {31}}~(\bibinfo {number}
  {9}),\ 10.1029/2004GL019739}\BibitemShut {NoStop}%
\bibitem [{\citenamefont {Cohen-Tannoudji}\ \emph {et~al.}(2007)\citenamefont
  {Cohen-Tannoudji}, \citenamefont {Dupont-Roc},\ and\ \citenamefont
  {Grynberg}}]{CohenT1997}%
  \BibitemOpen
  \bibfield  {author} {\bibinfo {author} {\bibnamefont {Cohen-Tannoudji},
  \bibfnamefont {C}}, \bibinfo {author} {\bibfnamefont {J.}~\bibnamefont
  {Dupont-Roc}}, \ and\ \bibinfo {author} {\bibfnamefont {G.}~\bibnamefont
  {Grynberg}}} (\bibinfo {year} {2007}),\ \href@noop {} {\emph {\bibinfo
  {title} {Photons and Atoms : Introduction to Quantum Electrodynamics}}}\
  (\bibinfo  {publisher} {Wiley},\ \bibinfo {address} {New York})\BibitemShut
  {NoStop}%
\bibitem [{\citenamefont {Coles}(2001)}]{Coles.2001}%
  \BibitemOpen
  \bibfield  {author} {\bibinfo {author} {\bibnamefont {Coles}, \bibfnamefont
  {S}}} (\bibinfo {year} {2001}),\ \href@noop {} {\emph {\bibinfo {title} {An
  Introduction to Statistical Modeling of Extreme Values}}}\ (\bibinfo
  {publisher} {Springer Science \& Business Media})\BibitemShut {NoStop}%
\bibitem [{\citenamefont {Colon}\ \emph {et~al.}(2015)\citenamefont {Colon},
  \citenamefont {Claessen},\ and\ \citenamefont {Ghil}}]{Colon.ea.2015}%
  \BibitemOpen
  \bibfield  {author} {\bibinfo {author} {\bibnamefont {Colon}, \bibfnamefont
  {C}}, \bibinfo {author} {\bibfnamefont {D.}~\bibnamefont {Claessen}}, \ and\
  \bibinfo {author} {\bibfnamefont {M.}~\bibnamefont {Ghil}}} (\bibinfo {year}
  {2015}),\ \bibfield  {title} {\enquote {\bibinfo {title} {Bifurcation
  analysis of an agent-based model for predator--prey interactions},}\
  }\href@noop {} {\bibfield  {journal} {\bibinfo  {journal} {Ecological
  Modelling}\ }\textbf {\bibinfo {volume} {317}},\ \bibinfo {pages}
  {93--106}}\BibitemShut {NoStop}%
\bibitem [{\citenamefont {Compo}\ \emph {et~al.}(2011)\citenamefont {Compo},
  \citenamefont {Whitaker}, \citenamefont {Sardeshmukh}, \citenamefont
  {Matsui}, \citenamefont {Allan}, \citenamefont {Yin}, \citenamefont
  {Gleason}, \citenamefont {Vose}, \citenamefont {Rutledge}, \citenamefont
  {Bessemoulin}, \citenamefont {Br\"onnimann}, \citenamefont {Brunet},
  \citenamefont {Crouthamel}, \citenamefont {Grant}, \citenamefont {Groisman},
  \citenamefont {Jones}, \citenamefont {Kruk}, \citenamefont {Kruger},
  \citenamefont {Marshall}, \citenamefont {Maugeri}, \citenamefont {Mok},
  \citenamefont {Nordli}, \citenamefont {Ross}, \citenamefont {Trigo},
  \citenamefont {Wang}, \citenamefont {Woodruff},\ and\ \citenamefont
  {Worley}}]{Compo2011}%
  \BibitemOpen
  \bibfield  {author} {\bibinfo {author} {\bibnamefont {Compo}, \bibfnamefont
  {G~P}}, \bibinfo {author} {\bibfnamefont {J.~S.}\ \bibnamefont {Whitaker}},
  \bibinfo {author} {\bibfnamefont {P.~D.}\ \bibnamefont {Sardeshmukh}},
  \bibinfo {author} {\bibfnamefont {N.}~\bibnamefont {Matsui}}, \bibinfo
  {author} {\bibfnamefont {R.~J.}\ \bibnamefont {Allan}}, \bibinfo {author}
  {\bibfnamefont {X.}~\bibnamefont {Yin}}, \bibinfo {author} {\bibfnamefont
  {B.~E.}\ \bibnamefont {Gleason}}, \bibinfo {author} {\bibfnamefont {R.~S.}\
  \bibnamefont {Vose}}, \bibinfo {author} {\bibfnamefont {G.}~\bibnamefont
  {Rutledge}}, \bibinfo {author} {\bibfnamefont {P.}~\bibnamefont
  {Bessemoulin}}, \bibinfo {author} {\bibfnamefont {S.}~\bibnamefont
  {Br\"onnimann}}, \bibinfo {author} {\bibfnamefont {M.}~\bibnamefont
  {Brunet}}, \bibinfo {author} {\bibfnamefont {R.~I.}\ \bibnamefont
  {Crouthamel}}, \bibinfo {author} {\bibfnamefont {A.~N.}\ \bibnamefont
  {Grant}}, \bibinfo {author} {\bibfnamefont {P.~Y.}\ \bibnamefont {Groisman}},
  \bibinfo {author} {\bibfnamefont {P.~D.}\ \bibnamefont {Jones}}, \bibinfo
  {author} {\bibfnamefont {M.~C.}\ \bibnamefont {Kruk}}, \bibinfo {author}
  {\bibfnamefont {A.~C.}\ \bibnamefont {Kruger}}, \bibinfo {author}
  {\bibfnamefont {G.~J.}\ \bibnamefont {Marshall}}, \bibinfo {author}
  {\bibfnamefont {M.}~\bibnamefont {Maugeri}}, \bibinfo {author} {\bibfnamefont
  {H.~Y.}\ \bibnamefont {Mok}}, \bibinfo {author} {\bibfnamefont
  {O.}~\bibnamefont {Nordli}}, \bibinfo {author} {\bibfnamefont {T.~F.}\
  \bibnamefont {Ross}}, \bibinfo {author} {\bibfnamefont {R.~M.}\ \bibnamefont
  {Trigo}}, \bibinfo {author} {\bibfnamefont {X.~L.}\ \bibnamefont {Wang}},
  \bibinfo {author} {\bibfnamefont {S.~D.}\ \bibnamefont {Woodruff}}, \ and\
  \bibinfo {author} {\bibfnamefont {S.~J.}\ \bibnamefont {Worley}}} (\bibinfo
  {year} {2011}),\ \bibfield  {title} {\enquote {\bibinfo {title} {The
  twentieth century reanalysis project},}\ }\href {\doibase 10.1002/qj.776}
  {\bibfield  {journal} {\bibinfo  {journal} {Quarterly Journal of the Royal
  Meteorological Society}\ }\textbf {\bibinfo {volume} {137}}~(\bibinfo
  {number} {654}),\ \bibinfo {pages} {1--28}}\BibitemShut {NoStop}%
\bibitem [{\citenamefont {Cooper}\ and\ \citenamefont
  {Haynes}(2011)}]{cooper_climate_2011}%
  \BibitemOpen
  \bibfield  {author} {\bibinfo {author} {\bibnamefont {Cooper}, \bibfnamefont
  {F~C}}, \ and\ \bibinfo {author} {\bibfnamefont {P.~H.}\ \bibnamefont
  {Haynes}}} (\bibinfo {year} {2011}),\ \bibfield  {title} {\enquote {\bibinfo
  {title} {Climate sensitivity via a nonparametric fluctuation-dissipation
  theorem},}\ }\href@noop {} {\bibfield  {journal} {\bibinfo  {journal}
  {Journal of the Atmospheric Sciences}\ }\textbf {\bibinfo {volume}
  {68}}~(\bibinfo {number} {5}),\ \bibinfo {pages} {937--953}}\BibitemShut
  {NoStop}%
\bibitem [{\citenamefont {Cox}\ \emph {et~al.}(2018)\citenamefont {Cox},
  \citenamefont {Huntingford},\ and\ \citenamefont {Williamson}}]{Cox2018}%
  \BibitemOpen
  \bibfield  {author} {\bibinfo {author} {\bibnamefont {Cox}, \bibfnamefont
  {P~M}}, \bibinfo {author} {\bibfnamefont {C.}~\bibnamefont {Huntingford}}, \
  and\ \bibinfo {author} {\bibfnamefont {M.~S.}\ \bibnamefont {Williamson}}}
  (\bibinfo {year} {2018}),\ \bibfield  {title} {\enquote {\bibinfo {title}
  {Emergent constraint on equilibrium climate sensitivity from global
  temperature variability},}\ }\href {\doibase 10.1038/nature25450} {\bibfield
  {journal} {\bibinfo  {journal} {Nature}\ }\textbf {\bibinfo {volume} {553}},\
  \bibinfo {pages} {319--322}}\BibitemShut {NoStop}%
\bibitem [{\citenamefont {Cronin}(2010)}]{Cronin2010}%
  \BibitemOpen
  \bibfield  {author} {\bibinfo {author} {\bibnamefont {Cronin}, \bibfnamefont
  {T~N}}} (\bibinfo {year} {2010}),\ \href@noop {} {\emph {\bibinfo {title}
  {Paleoclimates: {Understanding Climate Change Past and Present}}}}\ (\bibinfo
   {publisher} {Columbia University Press},\ \bibinfo {address} {New
  York})\BibitemShut {NoStop}%
\bibitem [{\citenamefont {Crowley}\ \emph {et~al.}(2001)\citenamefont
  {Crowley}, \citenamefont {Hyde},\ and\ \citenamefont
  {Peltier}}]{Crowley.ea.2001}%
  \BibitemOpen
  \bibfield  {author} {\bibinfo {author} {\bibnamefont {Crowley}, \bibfnamefont
  {T~J}}, \bibinfo {author} {\bibfnamefont {W.~T.}\ \bibnamefont {Hyde}}, \
  and\ \bibinfo {author} {\bibfnamefont {W.~R.}\ \bibnamefont {Peltier}}}
  (\bibinfo {year} {2001}),\ \bibfield  {title} {\enquote {\bibinfo {title}
  {{CO}$_2$ levels required for deglaciation of a {"near-snowball" Earth}},}\
  }\href@noop {} {\bibfield  {journal} {\bibinfo  {journal} {Geophysical
  Research Letters}\ }\textbf {\bibinfo {volume} {28}}~(\bibinfo {number}
  {2}),\ \bibinfo {pages} {283--286}}\BibitemShut {NoStop}%
\bibitem [{\citenamefont {Cushman-Roisin}\ and\ \citenamefont
  {Beckers}(2011{\natexlab{a}})}]{CRB11}%
  \BibitemOpen
  \bibfield  {author} {\bibinfo {author} {\bibnamefont {Cushman-Roisin},
  \bibfnamefont {B}}, \ and\ \bibinfo {author} {\bibfnamefont {J.-M.}\
  \bibnamefont {Beckers}}} (\bibinfo {year} {2011}{\natexlab{a}}),\ \href@noop
  {} {\emph {\bibinfo {title} {Introduction to Geophysical Fluid Dynamics:
  Physical and Numerical Aspects}}}\ (\bibinfo  {publisher} {Academic
  Press})\BibitemShut {NoStop}%
\bibitem [{\citenamefont {Cushman-Roisin}\ and\ \citenamefont
  {Beckers}(2011{\natexlab{b}})}]{CRB2011}%
  \BibitemOpen
  \bibfield  {author} {\bibinfo {author} {\bibnamefont {Cushman-Roisin},
  \bibfnamefont {B}}, \ and\ \bibinfo {author} {\bibfnamefont {J-M.}\
  \bibnamefont {Beckers}}} (\bibinfo {year} {2011}{\natexlab{b}}),\ \href@noop
  {} {\emph {\bibinfo {title} {Introduction to Geophysical Fluid Dynamics:
  Physical and Numerical Aspects}}}\ (\bibinfo  {publisher} {Academic Press,
  2nd Edition})\ \bibinfo {note} {875 pp.}\BibitemShut {Stop}%
\bibitem [{\citenamefont {{Cvitanov\'ic}}\ and\ \citenamefont
  {Eckhardt}(1991)}]{Cvitanovic1991}%
  \BibitemOpen
  \bibfield  {author} {\bibinfo {author} {\bibnamefont {{Cvitanov\'ic}},
  \bibfnamefont {P}}, \ and\ \bibinfo {author} {\bibfnamefont {B}~\bibnamefont
  {Eckhardt}}} (\bibinfo {year} {1991}),\ \bibfield  {title} {\enquote
  {\bibinfo {title} {{Periodic orbit expansions for classical smooth flows}},}\
  }\href@noop {} {\bibfield  {journal} {\bibinfo  {journal} {J. Phys. Math.
  Gen.}\ }\textbf {\bibinfo {volume} {24}}}\BibitemShut {NoStop}%
\bibitem [{\citenamefont {Da~Costa}\ and\ \citenamefont {Colin~de
  Verdi\'ere}(2004)}]{Costa2004}%
  \BibitemOpen
  \bibfield  {author} {\bibinfo {author} {\bibnamefont {Da~Costa},
  \bibfnamefont {E~D}}, \ and\ \bibinfo {author} {\bibfnamefont {A.~C.}\
  \bibnamefont {Colin~de Verdi\'ere}}} (\bibinfo {year} {2004}),\ \bibfield
  {title} {\enquote {\bibinfo {title} {{The 7.7 year North Atlantic
  oscillation}},}\ }\href@noop {} {\bibfield  {journal} {\bibinfo  {journal}
  {{Q. J. R. Meteorol. Soc.}}\ }\textbf {\bibinfo {volume} {128A}},\ \bibinfo
  {pages} {797--817}}\BibitemShut {NoStop}%
\bibitem [{\citenamefont {Dansgaard}\ \emph {et~al.}(1993)\citenamefont
  {Dansgaard}, \citenamefont {Johnsen}, \citenamefont {Clausen}, \citenamefont
  {Dahl-Jensen}, \citenamefont {Gundestrup}, \citenamefont {Hammer},
  \citenamefont {Hvidberg}, \citenamefont {Steffensen}, \citenamefont
  {Sveinbj{\"o}rnsdottir}, \citenamefont {Jouzel},\ and\ \citenamefont
  {Bond}}]{Dansgaard1993complete}%
  \BibitemOpen
  \bibfield  {author} {\bibinfo {author} {\bibnamefont {Dansgaard},
  \bibfnamefont {W}}, \bibinfo {author} {\bibfnamefont {S.~J.}\ \bibnamefont
  {Johnsen}}, \bibinfo {author} {\bibfnamefont {H.~B.}\ \bibnamefont
  {Clausen}}, \bibinfo {author} {\bibfnamefont {D.}~\bibnamefont
  {Dahl-Jensen}}, \bibinfo {author} {\bibfnamefont {N.~S.}\ \bibnamefont
  {Gundestrup}}, \bibinfo {author} {\bibfnamefont {C.~U.}\ \bibnamefont
  {Hammer}}, \bibinfo {author} {\bibfnamefont {C.~S.}\ \bibnamefont
  {Hvidberg}}, \bibinfo {author} {\bibfnamefont {J.~P.}\ \bibnamefont
  {Steffensen}}, \bibinfo {author} {\bibfnamefont {A.~E.}\ \bibnamefont
  {Sveinbj{\"o}rnsdottir}}, \bibinfo {author} {\bibfnamefont {J.}~\bibnamefont
  {Jouzel}}, \ and\ \bibinfo {author} {\bibfnamefont {G.}~\bibnamefont {Bond}}}
  (\bibinfo {year} {1993}),\ \bibfield  {title} {\enquote {\bibinfo {title}
  {Evidence for general instability of past climate from a 250-kyr ice-core
  record},}\ }\href {\doibase 10.1038/364218a0} {\bibfield  {journal} {\bibinfo
   {journal} {Nature}\ }\textbf {\bibinfo {volume} {364}}~(\bibinfo {number}
  {6434}),\ \bibinfo {pages} {218--220}}\BibitemShut {NoStop}%
\bibitem [{\citenamefont {Davini}\ and\ \citenamefont
  {D'Andrea}(2016)}]{Davini2016}%
  \BibitemOpen
  \bibfield  {author} {\bibinfo {author} {\bibnamefont {Davini}, \bibfnamefont
  {P}}, \ and\ \bibinfo {author} {\bibfnamefont {F.}~\bibnamefont {D'Andrea}}}
  (\bibinfo {year} {2016}),\ \bibfield  {title} {\enquote {\bibinfo {title}
  {Northern {Hemisphere atmospheric blocking representation in global climate
  models: Twenty years of improvements?}}}\ }\href {\doibase
  10.1175/JCLI-D-16-0242.1} {\bibfield  {journal} {\bibinfo  {journal} {Journal
  of Climate}\ }\textbf {\bibinfo {volume} {29}}~(\bibinfo {number} {24}),\
  \bibinfo {pages} {8823--8840}}\BibitemShut {NoStop}%
\bibitem [{\citenamefont {De~Cruz}\ \emph {et~al.}(2018)\citenamefont
  {De~Cruz}, \citenamefont {Schubert}, \citenamefont {Demaeyer}, \citenamefont
  {Lucarini},\ and\ \citenamefont {Vannitsem}}]{decruz2018}%
  \BibitemOpen
  \bibfield  {author} {\bibinfo {author} {\bibnamefont {De~Cruz}, \bibfnamefont
  {L}}, \bibinfo {author} {\bibfnamefont {S.}~\bibnamefont {Schubert}},
  \bibinfo {author} {\bibfnamefont {J.}~\bibnamefont {Demaeyer}}, \bibinfo
  {author} {\bibfnamefont {V.}~\bibnamefont {Lucarini}}, \ and\ \bibinfo
  {author} {\bibfnamefont {S.}~\bibnamefont {Vannitsem}}} (\bibinfo {year}
  {2018}),\ \bibfield  {title} {\enquote {\bibinfo {title} {Exploring the
  {Lyapunov instability properties of high-dimensional atmospheric and climate
  models}},}\ }\href {\doibase 10.5194/npg-25-387-2018} {\bibfield  {journal}
  {\bibinfo  {journal} {Nonlinear Processes in Geophysics}\ }\textbf {\bibinfo
  {volume} {25}}~(\bibinfo {number} {2}),\ \bibinfo {pages}
  {387--412}}\BibitemShut {NoStop}%
\bibitem [{\citenamefont {Dee}\ and\ \citenamefont
  {Coauthors}(2011)}]{Dee2011}%
  \BibitemOpen
  \bibfield  {author} {\bibinfo {author} {\bibnamefont {Dee}, \bibfnamefont
  {D~P}}, \ and\ \bibinfo {author} {\bibnamefont {Coauthors}}} (\bibinfo {year}
  {2011}),\ \bibfield  {title} {\enquote {\bibinfo {title} {{The ERA-Interim
  reanalysis: configuration and performance of the data assimilation
  system}},}\ }\href@noop {} {\bibfield  {journal} {\bibinfo  {journal} {Q. J.
  R. Meteorol. Soc.}\ }\textbf {\bibinfo {volume} {137}}~(\bibinfo {number}
  {656}),\ \bibinfo {pages} {553--597}}\BibitemShut {NoStop}%
\bibitem [{\citenamefont {Dell'Aquila}\ \emph {et~al.}(2005)\citenamefont
  {Dell'Aquila}, \citenamefont {Lucarini}, \citenamefont {Ruti},\ and\
  \citenamefont {Calmanti}}]{Dellaquila05}%
  \BibitemOpen
  \bibfield  {author} {\bibinfo {author} {\bibnamefont {Dell'Aquila},
  \bibfnamefont {A}}, \bibinfo {author} {\bibfnamefont {V.}~\bibnamefont
  {Lucarini}}, \bibinfo {author} {\bibfnamefont {P.~M.}\ \bibnamefont {Ruti}},
  \ and\ \bibinfo {author} {\bibfnamefont {S.}~\bibnamefont {Calmanti}}}
  (\bibinfo {year} {2005}),\ \bibfield  {title} {\enquote {\bibinfo {title}
  {{Hayashi spectra of the Northern Hemisphere mid-latitude atmospheric
  variability in the NCEP and ERA-40 reanalyses}},}\ }\href@noop {} {\bibfield
  {journal} {\bibinfo  {journal} {Climate Dynamics}\ }\textbf {\bibinfo
  {volume} {25}},\ \bibinfo {pages} {639--652}}\BibitemShut {NoStop}%
\bibitem [{\citenamefont {Dell'Aquila}\ \emph {et~al.}(2007)\citenamefont
  {Dell'Aquila}, \citenamefont {Ruti}, \citenamefont {Calmanti},\ and\
  \citenamefont {Lucarini}}]{Dellaquila07}%
  \BibitemOpen
  \bibfield  {author} {\bibinfo {author} {\bibnamefont {Dell'Aquila},
  \bibfnamefont {A}}, \bibinfo {author} {\bibfnamefont {P.~M.}\ \bibnamefont
  {Ruti}}, \bibinfo {author} {\bibfnamefont {S.}~\bibnamefont {Calmanti}}, \
  and\ \bibinfo {author} {\bibfnamefont {V.}~\bibnamefont {Lucarini}}}
  (\bibinfo {year} {2007}),\ \bibfield  {title} {\enquote {\bibinfo {title} {{
  Southern Hemisphere Mid-latitude Atmospheric Variability of the NCEP-NCAR and
  ECMWF Reanalyses}},}\ }\href@noop {} {\bibfield  {journal} {\bibinfo
  {journal} {J. Geophys. Res.}\ }\textbf {\bibinfo {volume} {112}},\ \bibinfo
  {pages} {D08106}}\BibitemShut {NoStop}%
\bibitem [{\citenamefont {Deloncle}\ \emph {et~al.}(2007)\citenamefont
  {Deloncle}, \citenamefont {Berk}, \citenamefont {D'Andrea},\ and\
  \citenamefont {Ghil}}]{Deloncle.ea.2007}%
  \BibitemOpen
  \bibfield  {author} {\bibinfo {author} {\bibnamefont {Deloncle},
  \bibfnamefont {A}}, \bibinfo {author} {\bibfnamefont {R.}~\bibnamefont
  {Berk}}, \bibinfo {author} {\bibfnamefont {F.}~\bibnamefont {D'Andrea}}, \
  and\ \bibinfo {author} {\bibfnamefont {M.}~\bibnamefont {Ghil}}} (\bibinfo
  {year} {2007}),\ \bibfield  {title} {\enquote {\bibinfo {title} {Weather
  regime prediction using statistical learning},}\ }\href@noop {} {\bibfield
  {journal} {\bibinfo  {journal} {Journal of the Atmospheric Sciences}\
  }\textbf {\bibinfo {volume} {64}}~(\bibinfo {number} {5}),\ \bibinfo {pages}
  {1619--1635}}\BibitemShut {NoStop}%
\bibitem [{\citenamefont {Delworth}\ \emph {et~al.}(1993)\citenamefont
  {Delworth}, \citenamefont {Manabe},\ and\ \citenamefont
  {Stouffer}}]{Delworth.ea.1993}%
  \BibitemOpen
  \bibfield  {author} {\bibinfo {author} {\bibnamefont {Delworth},
  \bibfnamefont {T}}, \bibinfo {author} {\bibfnamefont {S.}~\bibnamefont
  {Manabe}}, \ and\ \bibinfo {author} {\bibfnamefont {R.~J.}\ \bibnamefont
  {Stouffer}}} (\bibinfo {year} {1993}),\ \bibfield  {title} {\enquote
  {\bibinfo {title} {Interdecadal variations of the thermohaline circulation in
  a coupled ocean-atmosphere model},}\ }\href@noop {} {\bibfield  {journal}
  {\bibinfo  {journal} {J. Climate}\ }\textbf {\bibinfo {volume} {6}},\
  \bibinfo {pages} {1993--2011}}\BibitemShut {NoStop}%
\bibitem [{\citenamefont {Delworth}\ and\ \citenamefont
  {Mann}(2000)}]{Delworth2000a}%
  \BibitemOpen
  \bibfield  {author} {\bibinfo {author} {\bibnamefont {Delworth},
  \bibfnamefont {T~L}}, \ and\ \bibinfo {author} {\bibfnamefont {M.~E.}\
  \bibnamefont {Mann}}} (\bibinfo {year} {2000}),\ \bibfield  {title} {\enquote
  {\bibinfo {title} {{Observed and simulated multidecadal variability in the
  Northern Hemisphere}},}\ }\href@noop {} {\bibfield  {journal} {\bibinfo
  {journal} {Clim.\ Dyn.}\ }\textbf {\bibinfo {volume} {16}},\ \bibinfo {pages}
  {661--676}}\BibitemShut {NoStop}%
\bibitem [{\citenamefont {Demaeyer}\ and\ \citenamefont
  {Vannitsem}(2017)}]{Demaeyer2017}%
  \BibitemOpen
  \bibfield  {author} {\bibinfo {author} {\bibnamefont {Demaeyer},
  \bibfnamefont {J}}, \ and\ \bibinfo {author} {\bibfnamefont {S.}~\bibnamefont
  {Vannitsem}}} (\bibinfo {year} {2017}),\ \bibfield  {title} {\enquote
  {\bibinfo {title} {Stochastic parametrization of subgrid-scale processes in
  coupled ocean-atmosphere systems: benefits and limitations of response
  theory},}\ }\href {\doibase 10.1002/qj.2973} {\bibfield  {journal} {\bibinfo
  {journal} {Quarterly Journal of the Royal Meteorological Society}\ }\textbf
  {\bibinfo {volume} {143}}~(\bibinfo {number} {703}),\ \bibinfo {pages}
  {881--896}}\BibitemShut {NoStop}%
\bibitem [{\citenamefont {Dickey}\ \emph {et~al.}(1991)\citenamefont {Dickey},
  \citenamefont {Ghil},\ and\ \citenamefont {Marcus}}]{Dickey.ea.1991}%
  \BibitemOpen
  \bibfield  {author} {\bibinfo {author} {\bibnamefont {Dickey}, \bibfnamefont
  {J~O}}, \bibinfo {author} {\bibfnamefont {M.}~\bibnamefont {Ghil}}, \ and\
  \bibinfo {author} {\bibfnamefont {S.~L.}\ \bibnamefont {Marcus}}} (\bibinfo
  {year} {1991}),\ \bibfield  {title} {\enquote {\bibinfo {title}
  {Extratropical aspects of the 40--50 day oscillation in length-of-day and
  atmospheric angular momentum},}\ }\href {\doibase 10.1029/91jd02339}
  {\bibfield  {journal} {\bibinfo  {journal} {J. Geophys. Res.: Atmospheres}\
  }\textbf {\bibinfo {volume} {96}}~(\bibinfo {number} {D12}),\ \bibinfo
  {pages} {22643--22658}}\BibitemShut {NoStop}%
\bibitem [{\citenamefont {Dijkstra}(2005)}]{DijkstraB2000}%
  \BibitemOpen
  \bibfield  {author} {\bibinfo {author} {\bibnamefont {Dijkstra},
  \bibfnamefont {H~A}}} (\bibinfo {year} {2005}),\ \href@noop {} {\emph
  {\bibinfo {title} {{Nonlinear Physical Oceanography: A Dynamical Systems
  Approach to the Large Scale Ocean Circulation and El Ni\~no.}}}},\ \bibinfo
  {edition} {2nd}\ ed.\ (\bibinfo  {publisher} {Springer Science+Business
  Media.},\ \bibinfo {address} {Berlin-Heidelberg, Germany})\BibitemShut
  {NoStop}%
\bibitem [{\citenamefont {Dijkstra}(2007)}]{Dijkstra.multiple.2007}%
  \BibitemOpen
  \bibfield  {author} {\bibinfo {author} {\bibnamefont {Dijkstra},
  \bibfnamefont {H~A}}} (\bibinfo {year} {2007}),\ \bibfield  {title} {\enquote
  {\bibinfo {title} {Characterization of the multiple equilibria regime in a
  global ocean model},}\ }\href@noop {} {\bibfield  {journal} {\bibinfo
  {journal} {Tellus A}\ }\textbf {\bibinfo {volume} {59}},\ \bibinfo {pages}
  {695--705}}\BibitemShut {NoStop}%
\bibitem [{\citenamefont {Dijkstra}(2013)}]{dijkstra2013}%
  \BibitemOpen
  \bibfield  {author} {\bibinfo {author} {\bibnamefont {Dijkstra},
  \bibfnamefont {H~A}}} (\bibinfo {year} {2013}),\ \href@noop {} {\emph
  {\bibinfo {title} {Nonlinear Climate Dynamics}}}\ (\bibinfo  {publisher}
  {Cambridge University Press},\ \bibinfo {address} {Cambridge,
  UK.})\BibitemShut {NoStop}%
\bibitem [{\citenamefont {Dijkstra}\ and\ \citenamefont
  {Burgers}(2002)}]{Dijkstra2002_ARFM}%
  \BibitemOpen
  \bibfield  {author} {\bibinfo {author} {\bibnamefont {Dijkstra},
  \bibfnamefont {H~A}}, \ and\ \bibinfo {author} {\bibfnamefont
  {G.}~\bibnamefont {Burgers}}} (\bibinfo {year} {2002}),\ \bibfield  {title}
  {\enquote {\bibinfo {title} {{Fluid mechanics of El Ni\~no variability}},}\
  }\href@noop {} {\bibfield  {journal} {\bibinfo  {journal} {Annu. Rev. Fluid
  Mech.}\ }\textbf {\bibinfo {volume} {34}},\ \bibinfo {pages}
  {531--558}}\BibitemShut {NoStop}%
\bibitem [{\citenamefont {Dijkstra}\ and\ \citenamefont {Ghil}(2005)}]{DG05}%
  \BibitemOpen
  \bibfield  {author} {\bibinfo {author} {\bibnamefont {Dijkstra},
  \bibfnamefont {H~A}}, \ and\ \bibinfo {author} {\bibfnamefont
  {M.}~\bibnamefont {Ghil}}} (\bibinfo {year} {2005}),\ \bibfield  {title}
  {\enquote {\bibinfo {title} {Low-frequency variability of the large-scale
  ocean circulation: A dynamical systems approach},}\ }\href {\doibase
  10.1029/2002RG000122} {\bibfield  {journal} {\bibinfo  {journal} {Reviews of
  Geophysics}\ }\textbf {\bibinfo {volume} {43}}~(\bibinfo {number} {3}),\
  \bibinfo {pages} {RG3002}}\BibitemShut {NoStop}%
\bibitem [{\citenamefont {Dijkstra}\ and\ \citenamefont
  {Katsman}(1997)}]{Dijkstra1997b}%
  \BibitemOpen
  \bibfield  {author} {\bibinfo {author} {\bibnamefont {Dijkstra},
  \bibfnamefont {H~A}}, \ and\ \bibinfo {author} {\bibfnamefont {C.~A.}\
  \bibnamefont {Katsman}}} (\bibinfo {year} {1997}),\ \bibfield  {title}
  {\enquote {\bibinfo {title} {Temporal variability of the wind-driven
  quasi-geostrophic double gyre ocean circulation: Basic bifurcation
  diagrams},}\ }\href@noop {} {\bibfield  {journal} {\bibinfo  {journal}
  {Geophys.\ Astrophys.\ Fluid Dyn.}\ }\textbf {\bibinfo {volume} {85}},\
  \bibinfo {pages} {195--232}}\BibitemShut {NoStop}%
\bibitem [{\citenamefont {Ditlevsen}\ and\ \citenamefont
  {Johnsen}(2010)}]{Ditlevsen2010}%
  \BibitemOpen
  \bibfield  {author} {\bibinfo {author} {\bibnamefont {Ditlevsen},
  \bibfnamefont {P~D}}, \ and\ \bibinfo {author} {\bibfnamefont {S.~J.}\
  \bibnamefont {Johnsen}}} (\bibinfo {year} {2010}),\ \bibfield  {title}
  {\enquote {\bibinfo {title} {Tipping points: {Early warning and wishful
  thinking}},}\ }\href {\doibase 10.1029/2010GL044486} {\bibfield  {journal}
  {\bibinfo  {journal} {Geophysical Research Letters}\ }\textbf {\bibinfo
  {volume} {37}}~(\bibinfo {number} {19}),\ 10.1029/2010GL044486}\BibitemShut
  {NoStop}%
\bibitem [{\citenamefont {Doblas-Reyes}\ \emph {et~al.}(2013)\citenamefont
  {Doblas-Reyes}, \citenamefont {Garcia-Serrano}, \citenamefont {Lienert},
  \citenamefont {Biescas},\ and\ \citenamefont {Rodrigues}}]{Doblas2013}%
  \BibitemOpen
  \bibfield  {author} {\bibinfo {author} {\bibnamefont {Doblas-Reyes},
  \bibfnamefont {Francisco~J}}, \bibinfo {author} {\bibfnamefont {Javier}\
  \bibnamefont {Garcia-Serrano}}, \bibinfo {author} {\bibfnamefont {Fabian}\
  \bibnamefont {Lienert}}, \bibinfo {author} {\bibfnamefont {Aida~Pinto}\
  \bibnamefont {Biescas}}, \ and\ \bibinfo {author} {\bibfnamefont {Luis
  R.~L.}\ \bibnamefont {Rodrigues}}} (\bibinfo {year} {2013}),\ \bibfield
  {title} {\enquote {\bibinfo {title} {Seasonal climate predictability and
  forecasting: status and prospects},}\ }\href {\doibase 10.1002/wcc.217}
  {\bibfield  {journal} {\bibinfo  {journal} {Wiley Interdisciplinary Reviews:
  Climate Change}\ }\textbf {\bibinfo {volume} {4}}~(\bibinfo {number} {4}),\
  \bibinfo {pages} {245--268}}\BibitemShut {NoStop}%
\bibitem [{\citenamefont {Dobrushin}(1970)}]{Dobrushin1970}%
  \BibitemOpen
  \bibfield  {author} {\bibinfo {author} {\bibnamefont {Dobrushin},
  \bibfnamefont {R}}} (\bibinfo {year} {1970}),\ \bibfield  {title} {\enquote
  {\bibinfo {title} {Prescribing a system of random variables by conditional
  distributions},}\ }\href {\doibase 10.1137/1115049} {\bibfield  {journal}
  {\bibinfo  {journal} {Theory of Probability \& Its Applications}\ }\textbf
  {\bibinfo {volume} {15}}~(\bibinfo {number} {3}),\ \bibinfo {pages}
  {458--486}}\BibitemShut {NoStop}%
\bibitem [{\citenamefont {Donges}\ \emph {et~al.}(2009)\citenamefont {Donges},
  \citenamefont {Zou}, \citenamefont {Marwan},\ and\ \citenamefont
  {Kurths}}]{Donges2009}%
  \BibitemOpen
  \bibfield  {author} {\bibinfo {author} {\bibnamefont {Donges}, \bibfnamefont
  {J~F}}, \bibinfo {author} {\bibfnamefont {Y.}~\bibnamefont {Zou}}, \bibinfo
  {author} {\bibfnamefont {N.}~\bibnamefont {Marwan}}, \ and\ \bibinfo {author}
  {\bibfnamefont {J.}~\bibnamefont {Kurths}}} (\bibinfo {year} {2009}),\
  \bibfield  {title} {\enquote {\bibinfo {title} {Complex networks in climate
  dynamics},}\ }\href {\doibase 10.1140/epjst/e2009-01098-2} {\bibfield
  {journal} {\bibinfo  {journal} {The European Physical Journal Special
  Topics}\ }\textbf {\bibinfo {volume} {174}}~(\bibinfo {number} {1}),\
  \bibinfo {pages} {157--179}}\BibitemShut {NoStop}%
\bibitem [{\citenamefont {Driver}(1977)}]{Driver.1977}%
  \BibitemOpen
  \bibfield  {author} {\bibinfo {author} {\bibnamefont {Driver}, \bibfnamefont
  {R~D}}} (\bibinfo {year} {1977}),\ \href@noop {} {\emph {\bibinfo {title}
  {Ordinary and Delay Differential Equations}}}\ (\bibinfo  {publisher}
  {Springer-Verlag, New York})\ \bibinfo {note} {{re-issued in 2012 by Springer
  Science \& Business Media}}\BibitemShut {NoStop}%
\bibitem [{\citenamefont {Dr{\'o}tos}\ \emph {et~al.}(2015)\citenamefont
  {Dr{\'o}tos}, \citenamefont {B{\'o}dai},\ and\ \citenamefont
  {T{\'e}l}}]{DBT15}%
  \BibitemOpen
  \bibfield  {author} {\bibinfo {author} {\bibnamefont {Dr{\'o}tos},
  \bibfnamefont {G}}, \bibinfo {author} {\bibfnamefont {T.}~\bibnamefont
  {B{\'o}dai}}, \ and\ \bibinfo {author} {\bibfnamefont {T.}~\bibnamefont
  {T{\'e}l}}} (\bibinfo {year} {2015}),\ \bibfield  {title} {\enquote {\bibinfo
  {title} {Probabilistic concepts in a changing climate: a snapshot attractor
  picture},}\ }\bibfield  {booktitle} {\emph {\bibinfo {booktitle} {Journal of
  Climate}},\ }\href {\doibase 10.1175/JCLI-D-14-00459.1} {\bibfield  {journal}
  {\bibinfo  {journal} {Journal of Climate}\
  }10.1175/JCLI-D-14-00459.1}\BibitemShut {NoStop}%
\bibitem [{\citenamefont {Duane}\ \emph {et~al.}(2017)\citenamefont {Duane},
  \citenamefont {Grabow}, \citenamefont {Selten},\ and\ \citenamefont
  {Ghil}}]{Duane.ea.2017}%
  \BibitemOpen
  \bibfield  {author} {\bibinfo {author} {\bibnamefont {Duane}, \bibfnamefont
  {G~S}}, \bibinfo {author} {\bibfnamefont {C.}~\bibnamefont {Grabow}},
  \bibinfo {author} {\bibfnamefont {F.}~\bibnamefont {Selten}}, \ and\ \bibinfo
  {author} {\bibfnamefont {M.}~\bibnamefont {Ghil}}} (\bibinfo {year} {2017}),\
  \bibfield  {title} {\enquote {\bibinfo {title} {Introduction to focus issue:
  {Synchronization in large networks and continuous media-data, models, and
  supermodels}},}\ }\href@noop {} {\bibfield  {journal} {\bibinfo  {journal}
  {Chaos}\ }\textbf {\bibinfo {volume} {27}},\ \bibinfo {pages}
  {126601}}\BibitemShut {NoStop}%
\bibitem [{\citenamefont {Duplessy}\ and\ \citenamefont
  {Shackleton}(1985)}]{Duplessy1985}%
  \BibitemOpen
  \bibfield  {author} {\bibinfo {author} {\bibnamefont {Duplessy},
  \bibfnamefont {J-C}}, \ and\ \bibinfo {author} {\bibfnamefont {N.~J.}\
  \bibnamefont {Shackleton}}} (\bibinfo {year} {1985}),\ \bibfield  {title}
  {\enquote {\bibinfo {title} {Response of global deep-water circulation to
  {Earth's climatic change 135,000-107,000 years ago}},}\ }\href@noop {}
  {\bibfield  {journal} {\bibinfo  {journal} {Nature}\ }\textbf {\bibinfo
  {volume} {316}},\ \bibinfo {pages} {500--507}}\BibitemShut {NoStop}%
\bibitem [{\citenamefont {E}\ and\ \citenamefont {Lu}(2011)}]{E.Lu.2011}%
  \BibitemOpen
  \bibfield  {author} {\bibinfo {author} {\bibnamefont {E}, \bibfnamefont {W}},
  \ and\ \bibinfo {author} {\bibfnamefont {J.}~\bibnamefont {Lu}}} (\bibinfo
  {year} {2011}),\ \bibfield  {title} {\enquote {\bibinfo {title} {Multiscale
  modeling},}\ }\href {\doibase 10.4249/scholarpedia.11527} {\bibfield
  {journal} {\bibinfo  {journal} {Scholarpedia}\ }\textbf {\bibinfo {volume}
  {6}}~(\bibinfo {number} {10}),\ \bibinfo {pages} {11527}},\ \bibinfo {note}
  {{revision \#91540}}\BibitemShut {NoStop}%
\bibitem [{\citenamefont {Eade}\ \emph {et~al.}(2014)\citenamefont {Eade},
  \citenamefont {Smith}, \citenamefont {Scaife}, \citenamefont {Wallace},
  \citenamefont {Dunstone}, \citenamefont {Hermanson},\ and\ \citenamefont
  {Robinson}}]{Eade2014}%
  \BibitemOpen
  \bibfield  {author} {\bibinfo {author} {\bibnamefont {Eade}, \bibfnamefont
  {R}}, \bibinfo {author} {\bibfnamefont {D.}~\bibnamefont {Smith}}, \bibinfo
  {author} {\bibfnamefont {A.}~\bibnamefont {Scaife}}, \bibinfo {author}
  {\bibfnamefont {E.}~\bibnamefont {Wallace}}, \bibinfo {author} {\bibfnamefont
  {N.}~\bibnamefont {Dunstone}}, \bibinfo {author} {\bibfnamefont
  {L.}~\bibnamefont {Hermanson}}, \ and\ \bibinfo {author} {\bibfnamefont
  {N.}~\bibnamefont {Robinson}}} (\bibinfo {year} {2014}),\ \bibfield  {title}
  {\enquote {\bibinfo {title} {Do seasonal-to-decadal climate predictions
  underestimate the predictability of the real world?}}\ }\href {\doibase
  10.1002/2014GL061146} {\bibfield  {journal} {\bibinfo  {journal} {Geophysical
  Research Letters}\ }\textbf {\bibinfo {volume} {41}}~(\bibinfo {number}
  {15}),\ \bibinfo {pages} {5620--5628}}\BibitemShut {NoStop}%
\bibitem [{\citenamefont {Eady}(1949)}]{Eady1949}%
  \BibitemOpen
  \bibfield  {author} {\bibinfo {author} {\bibnamefont {Eady}, \bibfnamefont
  {E~T}}} (\bibinfo {year} {1949}),\ \bibfield  {title} {\enquote {\bibinfo
  {title} {Long waves and cyclone waves},}\ }\href {\doibase
  10.1111/j.2153-3490.1949.tb01265.x} {\bibfield  {journal} {\bibinfo
  {journal} {Tellus}\ }\textbf {\bibinfo {volume} {1}}~(\bibinfo {number}
  {3}),\ \bibinfo {pages} {33--52}}\BibitemShut {NoStop}%
\bibitem [{\citenamefont {Eckmann}(1981)}]{Eckmann1981}%
  \BibitemOpen
  \bibfield  {author} {\bibinfo {author} {\bibnamefont {Eckmann}, \bibfnamefont
  {J~P}}} (\bibinfo {year} {1981}),\ \bibfield  {title} {\enquote {\bibinfo
  {title} {Roads to turbulence in disipative dynamical systems},}\ }\href@noop
  {} {\bibfield  {journal} {\bibinfo  {journal} {Rev. Mod. Physics}\ }\textbf
  {\bibinfo {volume} {53}},\ \bibinfo {pages} {643--654}}\BibitemShut {NoStop}%
\bibitem [{\citenamefont {Eckmann}\ and\ \citenamefont {Ruelle}(1985)}]{ER85}%
  \BibitemOpen
  \bibfield  {author} {\bibinfo {author} {\bibnamefont {Eckmann}, \bibfnamefont
  {J~P}}, \ and\ \bibinfo {author} {\bibfnamefont {D.}~\bibnamefont {Ruelle}}}
  (\bibinfo {year} {1985}),\ \bibfield  {title} {\enquote {\bibinfo {title}
  {Ergodic theory of chaos and strange attractors},}\ }\href {\doibase
  10.1103/RevModPhys.57.617} {\bibfield  {journal} {\bibinfo  {journal} {Rev.
  Mod. Phys.}\ }\textbf {\bibinfo {volume} {57}},\ \bibinfo {pages}
  {617--656}}\BibitemShut {NoStop}%
\bibitem [{\citenamefont {Edwards}(2010)}]{edwards10}%
  \BibitemOpen
  \bibfield  {author} {\bibinfo {author} {\bibnamefont {Edwards}, \bibfnamefont
  {P~N}}} (\bibinfo {year} {2010}),\ \href@noop {} {\emph {\bibinfo {title} {A
  Vast Machine: Computer Models, Climate Data, and the Politics of Global
  Warming}}}\ (\bibinfo  {publisher} {MIT Press},\ \bibinfo {address}
  {Cambridge})\BibitemShut {NoStop}%
\bibitem [{\citenamefont {Egger}(1978)}]{Egger.1978}%
  \BibitemOpen
  \bibfield  {author} {\bibinfo {author} {\bibnamefont {Egger}, \bibfnamefont
  {J}}} (\bibinfo {year} {1978}),\ \bibfield  {title} {\enquote {\bibinfo
  {title} {Dynamics of blocking highs},}\ }\href {\doibase
  10.1175/1520-0469(1978)035<1788:dobh>2.0.co;2} {\bibfield  {journal}
  {\bibinfo  {journal} {Journal of the Atmospheric Sciences}\ }\textbf
  {\bibinfo {volume} {35}}~(\bibinfo {number} {10}),\ \bibinfo {pages}
  {1788--1801}}\BibitemShut {NoStop}%
\bibitem [{\citenamefont {Einstein}(1905)}]{Einstein.1905}%
  \BibitemOpen
  \bibfield  {author} {\bibinfo {author} {\bibnamefont {Einstein},
  \bibfnamefont {A}}} (\bibinfo {year} {1905}),\ \bibfield  {title} {\enquote
  {\bibinfo {title} {{\"U}ber die von der molekularkinetischen {Theorie der
  W{\"a}rme geforderte Bewegung von in ruhenden Fl{\"u}ssigkeiten suspendierten
  Teilchen}},}\ }\href@noop {} {\bibfield  {journal} {\bibinfo  {journal}
  {Annalen der Physik}\ }\textbf {\bibinfo {volume} {322}}~(\bibinfo {number}
  {8}),\ \bibinfo {pages} {{549--560; reprinted in {\it Investigations on the
  Theory of the Brownian Movement, five articles by A. Einstein}, R. Furth
  (ed.) and A. D. Cowper (transl.), 1956, Dover Publ., New York, 122
  pp.}}}\BibitemShut {Stop}%
\bibitem [{\citenamefont {Emanuel}(1994)}]{Emanuel.1994}%
  \BibitemOpen
  \bibfield  {author} {\bibinfo {author} {\bibnamefont {Emanuel}, \bibfnamefont
  {K~A}}} (\bibinfo {year} {1994}),\ \href@noop {} {\emph {\bibinfo {title}
  {{Atmospheric Convection}}}}\ (\bibinfo  {publisher} {Oxford University Press
  on Demand})\BibitemShut {NoStop}%
\bibitem [{\citenamefont {Embrechts}\ \emph {et~al.}(1999)\citenamefont
  {Embrechts}, \citenamefont {Kl\"uppelberg},\ and\ \citenamefont
  {Mikosch}}]{EKM99}%
  \BibitemOpen
  \bibfield  {author} {\bibinfo {author} {\bibnamefont {Embrechts},
  \bibfnamefont {P}}, \bibinfo {author} {\bibfnamefont {C.}~\bibnamefont
  {Kl\"uppelberg}}, \ and\ \bibinfo {author} {\bibfnamefont {T.}~\bibnamefont
  {Mikosch}}} (\bibinfo {year} {1999}),\ \href@noop {} {\emph {\bibinfo {title}
  {Modelling Extremal Events for Insurance and Finance}}}\ (\bibinfo
  {publisher} {Springer-Verlag},\ \bibinfo {address} {New York})\BibitemShut
  {NoStop}%
\bibitem [{\citenamefont {Epstein}(1988)}]{Epstein.1988}%
  \BibitemOpen
  \bibfield  {author} {\bibinfo {author} {\bibnamefont {Epstein}, \bibfnamefont
  {E~S}}} (\bibinfo {year} {1988}),\ \bibfield  {title} {\enquote {\bibinfo
  {title} {Long-range weather prediction: {Limits} of predictability and
  beyond},}\ }\href@noop {} {\bibfield  {journal} {\bibinfo  {journal} {Weather
  and Forecasting}\ }\textbf {\bibinfo {volume} {3}},\ \bibinfo {pages}
  {69--75}}\BibitemShut {NoStop}%
\bibitem [{\citenamefont {Esper}\ \emph {et~al.}(2002)\citenamefont {Esper},
  \citenamefont {Cook},\ and\ \citenamefont {Schweingruber}}]{Esper.ea.02}%
  \BibitemOpen
  \bibfield  {author} {\bibinfo {author} {\bibnamefont {Esper}, \bibfnamefont
  {J}}, \bibinfo {author} {\bibfnamefont {E.~R.}\ \bibnamefont {Cook}}, \ and\
  \bibinfo {author} {\bibfnamefont {F.~H.}\ \bibnamefont {Schweingruber}}}
  (\bibinfo {year} {2002}),\ \bibfield  {title} {\enquote {\bibinfo {title}
  {Low-frequency signals in long tree-ring chronologies and the reconstruction
  of past temperature variability},}\ }\href@noop {} {\bibfield  {journal}
  {\bibinfo  {journal} {Science}\ }\textbf {\bibinfo {volume} {295}}~(\bibinfo
  {number} {5563}),\ \bibinfo {pages} {2250--2253}}\BibitemShut {NoStop}%
\bibitem [{\citenamefont {Eyring}\ \emph {et~al.}(2019)\citenamefont {Eyring},
  \citenamefont {Bock}, \citenamefont {Lauer}, \citenamefont {Righi},
  \citenamefont {Schlund}, \citenamefont {Andela}, \citenamefont {Arnone},
  \citenamefont {Bellprat}, \citenamefont {Br\"otz}, \citenamefont {Caron},
  \citenamefont {Carvalhais}, \citenamefont {Cionni}, \citenamefont {Cortesi},
  \citenamefont {Crezee}, \citenamefont {Davin}, \citenamefont {Davini},
  \citenamefont {Debeire}, \citenamefont {de~Mora}, \citenamefont {Deser},
  \citenamefont {Docquier}, \citenamefont {Earnshaw}, \citenamefont {Ehbrecht},
  \citenamefont {Gier}, \citenamefont {Gonzalez-Reviriego}, \citenamefont
  {Goodman}, \citenamefont {Hagemann}, \citenamefont {Hardiman}, \citenamefont
  {Hassler}, \citenamefont {Hunter}, \citenamefont {Kadow}, \citenamefont
  {Kindermann}, \citenamefont {Koirala}, \citenamefont {Koldunov},
  \citenamefont {Lejeune}, \citenamefont {Lembo}, \citenamefont {Lovato},
  \citenamefont {Lucarini}, \citenamefont {Massonnet}, \citenamefont
  {M\"uller}, \citenamefont {Pandde}, \citenamefont {P\'erez-Zan\'on},
  \citenamefont {Phillips}, \citenamefont {Predoi}, \citenamefont {Russell},
  \citenamefont {Sellar}, \citenamefont {Serva}, \citenamefont {Stacke},
  \citenamefont {Swaminathan}, \citenamefont {Torralba}, \citenamefont
  {Vegas-Regidor}, \citenamefont {von Hardenberg}, \citenamefont {Weigel},\
  and\ \citenamefont {Zimmermann}}]{Eyring2020}%
  \BibitemOpen
  \bibfield  {author} {\bibinfo {author} {\bibnamefont {Eyring}, \bibfnamefont
  {V}}, \bibinfo {author} {\bibfnamefont {L.}~\bibnamefont {Bock}}, \bibinfo
  {author} {\bibfnamefont {A.}~\bibnamefont {Lauer}}, \bibinfo {author}
  {\bibfnamefont {M.}~\bibnamefont {Righi}}, \bibinfo {author} {\bibfnamefont
  {M.}~\bibnamefont {Schlund}}, \bibinfo {author} {\bibfnamefont
  {B.}~\bibnamefont {Andela}}, \bibinfo {author} {\bibfnamefont
  {E.}~\bibnamefont {Arnone}}, \bibinfo {author} {\bibfnamefont
  {O.}~\bibnamefont {Bellprat}}, \bibinfo {author} {\bibfnamefont
  {B.}~\bibnamefont {Br\"otz}}, \bibinfo {author} {\bibfnamefont {L.-P.}\
  \bibnamefont {Caron}}, \bibinfo {author} {\bibfnamefont {N.}~\bibnamefont
  {Carvalhais}}, \bibinfo {author} {\bibfnamefont {I.}~\bibnamefont {Cionni}},
  \bibinfo {author} {\bibfnamefont {N.}~\bibnamefont {Cortesi}}, \bibinfo
  {author} {\bibfnamefont {B.}~\bibnamefont {Crezee}}, \bibinfo {author}
  {\bibfnamefont {E.}~\bibnamefont {Davin}}, \bibinfo {author} {\bibfnamefont
  {P.}~\bibnamefont {Davini}}, \bibinfo {author} {\bibfnamefont
  {K.}~\bibnamefont {Debeire}}, \bibinfo {author} {\bibfnamefont
  {L.}~\bibnamefont {de~Mora}}, \bibinfo {author} {\bibfnamefont
  {C.}~\bibnamefont {Deser}}, \bibinfo {author} {\bibfnamefont
  {D.}~\bibnamefont {Docquier}}, \bibinfo {author} {\bibfnamefont
  {P.}~\bibnamefont {Earnshaw}}, \bibinfo {author} {\bibfnamefont
  {C.}~\bibnamefont {Ehbrecht}}, \bibinfo {author} {\bibfnamefont {B.~K.}\
  \bibnamefont {Gier}}, \bibinfo {author} {\bibfnamefont {N.}~\bibnamefont
  {Gonzalez-Reviriego}}, \bibinfo {author} {\bibfnamefont {P.}~\bibnamefont
  {Goodman}}, \bibinfo {author} {\bibfnamefont {S.}~\bibnamefont {Hagemann}},
  \bibinfo {author} {\bibfnamefont {S.}~\bibnamefont {Hardiman}}, \bibinfo
  {author} {\bibfnamefont {B.}~\bibnamefont {Hassler}}, \bibinfo {author}
  {\bibfnamefont {A.}~\bibnamefont {Hunter}}, \bibinfo {author} {\bibfnamefont
  {C.}~\bibnamefont {Kadow}}, \bibinfo {author} {\bibfnamefont
  {S.}~\bibnamefont {Kindermann}}, \bibinfo {author} {\bibfnamefont
  {S.}~\bibnamefont {Koirala}}, \bibinfo {author} {\bibfnamefont {N.~V.}\
  \bibnamefont {Koldunov}}, \bibinfo {author} {\bibfnamefont {Q.}~\bibnamefont
  {Lejeune}}, \bibinfo {author} {\bibfnamefont {V.}~\bibnamefont {Lembo}},
  \bibinfo {author} {\bibfnamefont {T.}~\bibnamefont {Lovato}}, \bibinfo
  {author} {\bibfnamefont {V.}~\bibnamefont {Lucarini}}, \bibinfo {author}
  {\bibfnamefont {F.}~\bibnamefont {Massonnet}}, \bibinfo {author}
  {\bibfnamefont {B.}~\bibnamefont {M\"uller}}, \bibinfo {author}
  {\bibfnamefont {A.}~\bibnamefont {Pandde}}, \bibinfo {author} {\bibfnamefont
  {N.}~\bibnamefont {P\'erez-Zan\'on}}, \bibinfo {author} {\bibfnamefont
  {A.}~\bibnamefont {Phillips}}, \bibinfo {author} {\bibfnamefont
  {V.}~\bibnamefont {Predoi}}, \bibinfo {author} {\bibfnamefont
  {J.}~\bibnamefont {Russell}}, \bibinfo {author} {\bibfnamefont
  {A.}~\bibnamefont {Sellar}}, \bibinfo {author} {\bibfnamefont
  {F.}~\bibnamefont {Serva}}, \bibinfo {author} {\bibfnamefont
  {T.}~\bibnamefont {Stacke}}, \bibinfo {author} {\bibfnamefont
  {R.}~\bibnamefont {Swaminathan}}, \bibinfo {author} {\bibfnamefont
  {V.}~\bibnamefont {Torralba}}, \bibinfo {author} {\bibfnamefont
  {J.}~\bibnamefont {Vegas-Regidor}}, \bibinfo {author} {\bibfnamefont
  {J.}~\bibnamefont {von Hardenberg}}, \bibinfo {author} {\bibfnamefont
  {K.}~\bibnamefont {Weigel}}, \ and\ \bibinfo {author} {\bibfnamefont
  {K.}~\bibnamefont {Zimmermann}}} (\bibinfo {year} {2019}),\ \bibfield
  {title} {\enquote {\bibinfo {title} {Esmvaltool v2.0 -- extended set of
  large-scale diagnostics for quasi-operational and comprehensive evaluation of
  earth system models in cmip},}\ }\href {\doibase 10.5194/gmd-2019-291}
  {\bibfield  {journal} {\bibinfo  {journal} {Geoscientific Model Development
  Discussions}\ }\textbf {\bibinfo {volume} {2019}},\ \bibinfo {pages}
  {1--81}}\BibitemShut {NoStop}%
\bibitem [{\citenamefont {Eyring}\ \emph
  {et~al.}(2016{\natexlab{a}})\citenamefont {Eyring}, \citenamefont {Bony},
  \citenamefont {Meehl}, \citenamefont {Senior}, \citenamefont {Stevens},
  \citenamefont {Stouffer},\ and\ \citenamefont {Taylor}}]{Eyring2016}%
  \BibitemOpen
  \bibfield  {author} {\bibinfo {author} {\bibnamefont {Eyring}, \bibfnamefont
  {V}}, \bibinfo {author} {\bibfnamefont {S.}~\bibnamefont {Bony}}, \bibinfo
  {author} {\bibfnamefont {G.~A.}\ \bibnamefont {Meehl}}, \bibinfo {author}
  {\bibfnamefont {C.~A.}\ \bibnamefont {Senior}}, \bibinfo {author}
  {\bibfnamefont {B.}~\bibnamefont {Stevens}}, \bibinfo {author} {\bibfnamefont
  {R.~J.}\ \bibnamefont {Stouffer}}, \ and\ \bibinfo {author} {\bibfnamefont
  {K.~E.}\ \bibnamefont {Taylor}}} (\bibinfo {year} {2016}{\natexlab{a}}),\
  \bibfield  {title} {\enquote {\bibinfo {title} {Overview of the {Coupled
  Model Intercomparison Project Phase 6 (CMIP6) experimental design and
  organization}},}\ }\href {\doibase 10.5194/gmd-9-1937-2016} {\bibfield
  {journal} {\bibinfo  {journal} {Geoscientific Model Development}\ }\textbf
  {\bibinfo {volume} {9}}~(\bibinfo {number} {5}),\ \bibinfo {pages}
  {1937--1958}}\BibitemShut {NoStop}%
\bibitem [{\citenamefont {Eyring}\ \emph
  {et~al.}(2016{\natexlab{b}})\citenamefont {Eyring}, \citenamefont {Righi},
  \citenamefont {Lauer}, \citenamefont {Evaldsson}, \citenamefont {Wenzel},
  \citenamefont {Jones}, \citenamefont {Anav}, \citenamefont {Andrews},
  \citenamefont {Cionni}, \citenamefont {Davin}, \citenamefont {Deser},
  \citenamefont {Ehbrecht}, \citenamefont {Friedlingstein}, \citenamefont
  {Gleckler}, \citenamefont {Gottschaldt}, \citenamefont {Hagemann},
  \citenamefont {Juckes}, \citenamefont {Kindermann}, \citenamefont {Krasting},
  \citenamefont {Kunert}, \citenamefont {Levine}, \citenamefont {Loew},
  \citenamefont {M\"akel\"a}, \citenamefont {Martin}, \citenamefont {Mason},
  \citenamefont {Phillips}, \citenamefont {Read}, \citenamefont {Rio},
  \citenamefont {Roehrig}, \citenamefont {Senftleben}, \citenamefont {Sterl},
  \citenamefont {van Ulft}, \citenamefont {Walton}, \citenamefont {Wang},\ and\
  \citenamefont {Williams}}]{Eyring2016b}%
  \BibitemOpen
  \bibfield  {author} {\bibinfo {author} {\bibnamefont {Eyring}, \bibfnamefont
  {V}}, \bibinfo {author} {\bibfnamefont {M.}~\bibnamefont {Righi}}, \bibinfo
  {author} {\bibfnamefont {A.}~\bibnamefont {Lauer}}, \bibinfo {author}
  {\bibfnamefont {M.}~\bibnamefont {Evaldsson}}, \bibinfo {author}
  {\bibfnamefont {S.}~\bibnamefont {Wenzel}}, \bibinfo {author} {\bibfnamefont
  {C.}~\bibnamefont {Jones}}, \bibinfo {author} {\bibfnamefont
  {A.}~\bibnamefont {Anav}}, \bibinfo {author} {\bibfnamefont {O.}~\bibnamefont
  {Andrews}}, \bibinfo {author} {\bibfnamefont {I.}~\bibnamefont {Cionni}},
  \bibinfo {author} {\bibfnamefont {E.~L.}\ \bibnamefont {Davin}}, \bibinfo
  {author} {\bibfnamefont {C.}~\bibnamefont {Deser}}, \bibinfo {author}
  {\bibfnamefont {C.}~\bibnamefont {Ehbrecht}}, \bibinfo {author}
  {\bibfnamefont {P.}~\bibnamefont {Friedlingstein}}, \bibinfo {author}
  {\bibfnamefont {P.}~\bibnamefont {Gleckler}}, \bibinfo {author}
  {\bibfnamefont {K.-D.}\ \bibnamefont {Gottschaldt}}, \bibinfo {author}
  {\bibfnamefont {S.}~\bibnamefont {Hagemann}}, \bibinfo {author}
  {\bibfnamefont {M.}~\bibnamefont {Juckes}}, \bibinfo {author} {\bibfnamefont
  {S.}~\bibnamefont {Kindermann}}, \bibinfo {author} {\bibfnamefont
  {J.}~\bibnamefont {Krasting}}, \bibinfo {author} {\bibfnamefont
  {D.}~\bibnamefont {Kunert}}, \bibinfo {author} {\bibfnamefont
  {R.}~\bibnamefont {Levine}}, \bibinfo {author} {\bibfnamefont
  {A.}~\bibnamefont {Loew}}, \bibinfo {author} {\bibfnamefont {J.}~\bibnamefont
  {M\"akel\"a}}, \bibinfo {author} {\bibfnamefont {G.}~\bibnamefont {Martin}},
  \bibinfo {author} {\bibfnamefont {E.}~\bibnamefont {Mason}}, \bibinfo
  {author} {\bibfnamefont {A.~S.}\ \bibnamefont {Phillips}}, \bibinfo {author}
  {\bibfnamefont {S.}~\bibnamefont {Read}}, \bibinfo {author} {\bibfnamefont
  {C.}~\bibnamefont {Rio}}, \bibinfo {author} {\bibfnamefont {R.}~\bibnamefont
  {Roehrig}}, \bibinfo {author} {\bibfnamefont {D.}~\bibnamefont {Senftleben}},
  \bibinfo {author} {\bibfnamefont {A.}~\bibnamefont {Sterl}}, \bibinfo
  {author} {\bibfnamefont {L.~H.}\ \bibnamefont {van Ulft}}, \bibinfo {author}
  {\bibfnamefont {J.}~\bibnamefont {Walton}}, \bibinfo {author} {\bibfnamefont
  {S.}~\bibnamefont {Wang}}, \ and\ \bibinfo {author} {\bibfnamefont {K.~D.}\
  \bibnamefont {Williams}}} (\bibinfo {year} {2016}{\natexlab{b}}),\ \bibfield
  {title} {\enquote {\bibinfo {title} {{ESMValTool (v1.0) -- a community
  diagnostic and performance metrics tool for routine evaluation of Earth
  system models in CMIP}},}\ }\href {\doibase 10.5194/gmd-9-1747-2016}
  {\bibfield  {journal} {\bibinfo  {journal} {Geoscientific Model Development}\
  }\textbf {\bibinfo {volume} {9}}~(\bibinfo {number} {5}),\ \bibinfo {pages}
  {1747--1802}}\BibitemShut {NoStop}%
\bibitem [{\citenamefont {Faranda}\ \emph {et~al.}(2017)\citenamefont
  {Faranda}, \citenamefont {Messori},\ and\ \citenamefont
  {Yiou}}]{Faranda2017}%
  \BibitemOpen
  \bibfield  {author} {\bibinfo {author} {\bibnamefont {Faranda}, \bibfnamefont
  {D}}, \bibinfo {author} {\bibfnamefont {G.}~\bibnamefont {Messori}}, \ and\
  \bibinfo {author} {\bibfnamefont {P.}~\bibnamefont {Yiou}}} (\bibinfo {year}
  {2017}),\ \bibfield  {title} {\enquote {\bibinfo {title} {Dynamical proxies
  of {North Atlantic predictability and extremes}},}\ }\href {\doibase
  10.1038/srep41278} {\bibfield  {journal} {\bibinfo  {journal} {Scientific
  Reports}\ }\textbf {\bibinfo {volume} {7}},\ \bibinfo {pages} {41278 EP
  --}}\BibitemShut {NoStop}%
\bibitem [{\citenamefont {Farrell}\ and\ \citenamefont
  {Ioannou}(1996)}]{Farrell1996}%
  \BibitemOpen
  \bibfield  {author} {\bibinfo {author} {\bibnamefont {Farrell}, \bibfnamefont
  {B~F}}, \ and\ \bibinfo {author} {\bibfnamefont {P.~J.}\ \bibnamefont
  {Ioannou}}} (\bibinfo {year} {1996}),\ \bibfield  {title} {\enquote {\bibinfo
  {title} {{Generalized stability theory. I: Autonomous operators}},}\
  }\href@noop {} {\bibfield  {journal} {\bibinfo  {journal} {J.\ Atmos.\ Sci.}\
  }\textbf {\bibinfo {volume} {53}},\ \bibinfo {pages}
  {2025--2040}}\BibitemShut {NoStop}%
\bibitem [{\citenamefont {Feliks}\ \emph {et~al.}(2010)\citenamefont {Feliks},
  \citenamefont {Ghil},\ and\ \citenamefont {Robertson}}]{FGR2010}%
  \BibitemOpen
  \bibfield  {author} {\bibinfo {author} {\bibnamefont {Feliks}, \bibfnamefont
  {Y}}, \bibinfo {author} {\bibfnamefont {M.}~\bibnamefont {Ghil}}, \ and\
  \bibinfo {author} {\bibfnamefont {A.~W.}\ \bibnamefont {Robertson}}}
  (\bibinfo {year} {2010}),\ \bibfield  {title} {\enquote {\bibinfo {title}
  {Oscillatory climate modes in the {Eastern Mediterranean and their
  synchronization with the North Atlantic Oscillation}},}\ }\href@noop {}
  {\bibfield  {journal} {\bibinfo  {journal} {J. Climate}\ }\textbf {\bibinfo
  {volume} {23}},\ \bibinfo {pages} {4060--4079}}\BibitemShut {NoStop}%
\bibitem [{\citenamefont {Feliks}\ \emph {et~al.}(2011)\citenamefont {Feliks},
  \citenamefont {Ghil},\ and\ \citenamefont {Robertson}}]{FGR2011}%
  \BibitemOpen
  \bibfield  {author} {\bibinfo {author} {\bibnamefont {Feliks}, \bibfnamefont
  {Y}}, \bibinfo {author} {\bibfnamefont {M.}~\bibnamefont {Ghil}}, \ and\
  \bibinfo {author} {\bibfnamefont {A.~W.}\ \bibnamefont {Robertson}}}
  (\bibinfo {year} {2011}),\ \bibfield  {title} {\enquote {\bibinfo {title}
  {The atmospheric circulation over the {North Atlantic as induced by the SST
  field}},}\ }\href@noop {} {\bibfield  {journal} {\bibinfo  {journal} {J.
  Climate}\ }\textbf {\bibinfo {volume} {24}},\ \bibinfo {pages}
  {522--542}}\BibitemShut {NoStop}%
\bibitem [{\citenamefont {Feliks}\ \emph {et~al.}(2004)\citenamefont {Feliks},
  \citenamefont {Ghil},\ and\ \citenamefont {Simonnet}}]{FGS2004}%
  \BibitemOpen
  \bibfield  {author} {\bibinfo {author} {\bibnamefont {Feliks}, \bibfnamefont
  {Y}}, \bibinfo {author} {\bibfnamefont {M.}~\bibnamefont {Ghil}}, \ and\
  \bibinfo {author} {\bibfnamefont {E.}~\bibnamefont {Simonnet}}} (\bibinfo
  {year} {2004}),\ \bibfield  {title} {\enquote {\bibinfo {title}
  {Low-frequency variability in the mid-latitude atmosphere induced by an
  oceanic thermal front},}\ }\href@noop {} {\bibfield  {journal} {\bibinfo
  {journal} {J. Atmos. Sci.}\ }\textbf {\bibinfo {volume} {61}},\ \bibinfo
  {pages} {961--981}}\BibitemShut {NoStop}%
\bibitem [{\citenamefont {Feliks}\ \emph {et~al.}(2007)\citenamefont {Feliks},
  \citenamefont {Ghil},\ and\ \citenamefont {Simonnet}}]{FGS2007}%
  \BibitemOpen
  \bibfield  {author} {\bibinfo {author} {\bibnamefont {Feliks}, \bibfnamefont
  {Y}}, \bibinfo {author} {\bibfnamefont {M.}~\bibnamefont {Ghil}}, \ and\
  \bibinfo {author} {\bibfnamefont {E.}~\bibnamefont {Simonnet}}} (\bibinfo
  {year} {2007}),\ \bibfield  {title} {\enquote {\bibinfo {title}
  {Low-frequency variability in the mid-latitude baroclinic atmosphere induced
  by an oceanic thermal front},}\ }\href@noop {} {\bibfield  {journal}
  {\bibinfo  {journal} {{J. Atmos. Sci.}}\ }\textbf {\bibinfo {volume} {64}},\
  \bibinfo {pages} {97--116}}\BibitemShut {NoStop}%
\bibitem [{\citenamefont {Fenichel}(1979)}]{Fenichel.1979}%
  \BibitemOpen
  \bibfield  {author} {\bibinfo {author} {\bibnamefont {Fenichel},
  \bibfnamefont {N}}} (\bibinfo {year} {1979}),\ \bibfield  {title} {\enquote
  {\bibinfo {title} {Geometric singular perturbation theory for ordinary
  differential equations},}\ }\href@noop {} {\bibfield  {journal} {\bibinfo
  {journal} {{Journal of Differential Equations}}\ }\textbf {\bibinfo {volume}
  {31}}~(\bibinfo {number} {1}),\ \bibinfo {pages} {53--98}}\BibitemShut
  {NoStop}%
\bibitem [{\citenamefont {Ferranti}\ \emph {et~al.}(2015)\citenamefont
  {Ferranti}, \citenamefont {Corti},\ and\ \citenamefont
  {Janousek}}]{Ferranti2015}%
  \BibitemOpen
  \bibfield  {author} {\bibinfo {author} {\bibnamefont {Ferranti},
  \bibfnamefont {L}}, \bibinfo {author} {\bibfnamefont {S.}~\bibnamefont
  {Corti}}, \ and\ \bibinfo {author} {\bibfnamefont {M.}~\bibnamefont
  {Janousek}}} (\bibinfo {year} {2015}),\ \bibfield  {title} {\enquote
  {\bibinfo {title} {Flow-dependent verification of the {ECMWF ensemble over
  the Euro-Atlantic sector}},}\ }\href {\doibase 10.1002/qj.2411} {\bibfield
  {journal} {\bibinfo  {journal} {Quarterly Journal of the Royal Meteorological
  Society}\ }\textbf {\bibinfo {volume} {141}}~(\bibinfo {number} {688}),\
  \bibinfo {pages} {916--924}}\BibitemShut {NoStop}%
\bibitem [{\citenamefont {Feudel}\ \emph {et~al.}(2018)\citenamefont {Feudel},
  \citenamefont {Pisarchik},\ and\ \citenamefont {Showalter}}]{Feudel2018}%
  \BibitemOpen
  \bibfield  {author} {\bibinfo {author} {\bibnamefont {Feudel}, \bibfnamefont
  {U}}, \bibinfo {author} {\bibfnamefont {A.~N.}\ \bibnamefont {Pisarchik}}, \
  and\ \bibinfo {author} {\bibfnamefont {K.}~\bibnamefont {Showalter}}}
  (\bibinfo {year} {2018}),\ \bibfield  {title} {\enquote {\bibinfo {title}
  {Multistability and tipping: From mathematics and physics to climate and
  brain---minireview and preface to the focus issue},}\ }\href {\doibase
  10.1063/1.5027718} {\bibfield  {journal} {\bibinfo  {journal} {Chaos: An
  Interdisciplinary Journal of Nonlinear Science}\ }\textbf {\bibinfo {volume}
  {28}}~(\bibinfo {number} {3}),\ \bibinfo {pages} {033501}}\BibitemShut
  {NoStop}%
\bibitem [{\citenamefont {Fraedrich}(2012)}]{Fraedrich.e.2012}%
  \BibitemOpen
  \bibfield  {author} {\bibinfo {author} {\bibnamefont {Fraedrich},
  \bibfnamefont {K}}} (\bibinfo {year} {2012}),\ \bibfield  {title} {\enquote
  {\bibinfo {title} {A suite of user-friendly global climate models: hysteresis
  experiments},}\ }\href@noop {} {\bibfield  {journal} {\bibinfo  {journal}
  {The European Physical Journal Plus}\ }\textbf {\bibinfo {volume}
  {127}}~(\bibinfo {number} {5}),\ \bibinfo {pages} {53}}\BibitemShut {NoStop}%
\bibitem [{\citenamefont {Fraedrich}\ and\ \citenamefont
  {Bottger}(1978)}]{KF78}%
  \BibitemOpen
  \bibfield  {author} {\bibinfo {author} {\bibnamefont {Fraedrich},
  \bibfnamefont {K}}, \ and\ \bibinfo {author} {\bibfnamefont {H.}~\bibnamefont
  {Bottger}}} (\bibinfo {year} {1978}),\ \bibfield  {title} {\enquote {\bibinfo
  {title} {A wavenumber frequency analysis of the 500 mb geopotential at
  50$^\circ$ n},}\ }\href@noop {} {\bibfield  {journal} {\bibinfo  {journal}
  {J. Atmos. Sci.}\ }\textbf {\bibinfo {volume} {35}},\ \bibinfo {pages}
  {745--750}}\BibitemShut {NoStop}%
\bibitem [{\citenamefont {Fraedrich}\ \emph {et~al.}(2005)\citenamefont
  {Fraedrich}, \citenamefont {Jansen}, \citenamefont {Kirk}, \citenamefont
  {Luksch},\ and\ \citenamefont {Lunkeit}}]{Fraedrich.ea.2005}%
  \BibitemOpen
  \bibfield  {author} {\bibinfo {author} {\bibnamefont {Fraedrich},
  \bibfnamefont {K}}, \bibinfo {author} {\bibfnamefont {H.}~\bibnamefont
  {Jansen}}, \bibinfo {author} {\bibfnamefont {E.}~\bibnamefont {Kirk}},
  \bibinfo {author} {\bibfnamefont {U.}~\bibnamefont {Luksch}}, \ and\ \bibinfo
  {author} {\bibfnamefont {F.}~\bibnamefont {Lunkeit}}} (\bibinfo {year}
  {2005}),\ \bibfield  {title} {\enquote {\bibinfo {title} {The planet
  simulator: Towards a user friendly mode},}\ }\href@noop {} {\bibfield
  {journal} {\bibinfo  {journal} {Meteorologische Zeitschrift}\ }\textbf
  {\bibinfo {volume} {14}}~(\bibinfo {number} {3}),\ \bibinfo {pages}
  {299--304}}\BibitemShut {NoStop}%
\bibitem [{\citenamefont {Francis}\ and\ \citenamefont
  {Vavrus}(2012)}]{Francis.2012}%
  \BibitemOpen
  \bibfield  {author} {\bibinfo {author} {\bibnamefont {Francis}, \bibfnamefont
  {J~A}}, \ and\ \bibinfo {author} {\bibfnamefont {S.~J.}\ \bibnamefont
  {Vavrus}}} (\bibinfo {year} {2012}),\ \bibfield  {title} {\enquote {\bibinfo
  {title} {Evidence linking {Arctic amplification to extreme weather in
  mid-latitudes}},}\ }\href@noop {} {\bibfield  {journal} {\bibinfo  {journal}
  {Geophysical Research Letters}\ }\textbf {\bibinfo {volume} {39}}~(\bibinfo
  {number} {6})}\BibitemShut {NoStop}%
\bibitem [{\citenamefont {Franzke}\ \emph {et~al.}(2015)\citenamefont
  {Franzke}, \citenamefont {O'Kane}, \citenamefont {Berner}, \citenamefont
  {Williams},\ and\ \citenamefont {Lucarini}}]{Franzke.ea.2015}%
  \BibitemOpen
  \bibfield  {author} {\bibinfo {author} {\bibnamefont {Franzke}, \bibfnamefont
  {C~L~E}}, \bibinfo {author} {\bibfnamefont {T.~J.}\ \bibnamefont {O'Kane}},
  \bibinfo {author} {\bibfnamefont {J.}~\bibnamefont {Berner}}, \bibinfo
  {author} {\bibfnamefont {P.~D.}\ \bibnamefont {Williams}}, \ and\ \bibinfo
  {author} {\bibfnamefont {V.}~\bibnamefont {Lucarini}}} (\bibinfo {year}
  {2015}),\ \bibfield  {title} {\enquote {\bibinfo {title} {Stochastic climate
  theory and modeling},}\ }\href@noop {} {\bibfield  {journal} {\bibinfo
  {journal} {Wiley Interdisciplinary Reviews: Climate Change}\ }\textbf
  {\bibinfo {volume} {6}}~(\bibinfo {number} {1}),\ \bibinfo {pages}
  {63--78}}\BibitemShut {NoStop}%
\bibitem [{\citenamefont {Freidlin}\ and\ \citenamefont
  {Wentzell}(1984)}]{Freidlin1984}%
  \BibitemOpen
  \bibfield  {author} {\bibinfo {author} {\bibnamefont {Freidlin},
  \bibfnamefont {M~I}}, \ and\ \bibinfo {author} {\bibfnamefont {A.D.}\
  \bibnamefont {Wentzell}}} (\bibinfo {year} {1984}),\ \href@noop {} {\emph
  {\bibinfo {title} {Random Perturbations of Dynamical Systems}}}\ (\bibinfo
  {publisher} {Springer},\ \bibinfo {address} {New York})\BibitemShut {NoStop}%
\bibitem [{\citenamefont {Frisius}\ \emph {et~al.}(1998)\citenamefont
  {Frisius}, \citenamefont {Lunkeit}, \citenamefont {Fraedrich},\ and\
  \citenamefont {James}}]{Frisius1998}%
  \BibitemOpen
  \bibfield  {author} {\bibinfo {author} {\bibnamefont {Frisius}, \bibfnamefont
  {T}}, \bibinfo {author} {\bibfnamefont {F.}~\bibnamefont {Lunkeit}}, \bibinfo
  {author} {\bibfnamefont {K.}~\bibnamefont {Fraedrich}}, \ and\ \bibinfo
  {author} {\bibfnamefont {I.~N.}\ \bibnamefont {James}}} (\bibinfo {year}
  {1998}),\ \bibfield  {title} {\enquote {\bibinfo {title} {Storm-track
  organization and variability in a simplified atmospheric global circulation
  model},}\ }\href {\doibase 10.1002/qj.49712454802} {\bibfield  {journal}
  {\bibinfo  {journal} {Quarterly Journal of the Royal Meteorological Society}\
  }\textbf {\bibinfo {volume} {124}}~(\bibinfo {number} {548}),\ \bibinfo
  {pages} {1019--1043}}\BibitemShut {NoStop}%
\bibitem [{\citenamefont {G{\'{a}}lfi}\ \emph {et~al.}(2019)\citenamefont
  {G{\'{a}}lfi}, \citenamefont {Lucarini},\ and\ \citenamefont
  {Wouters}}]{Galfi2019}%
  \BibitemOpen
  \bibfield  {author} {\bibinfo {author} {\bibnamefont {G{\'{a}}lfi},
  \bibfnamefont {V~M}}, \bibinfo {author} {\bibfnamefont {V.}~\bibnamefont
  {Lucarini}}, \ and\ \bibinfo {author} {\bibfnamefont {J.}~\bibnamefont
  {Wouters}}} (\bibinfo {year} {2019}),\ \bibfield  {title} {\enquote {\bibinfo
  {title} {A large deviation theory-based analysis of heat waves and cold
  spells in a simplified model of the general circulation of the atmosphere},}\
  }\href {\doibase 10.1088/1742-5468/ab02e8} {\bibfield  {journal} {\bibinfo
  {journal} {Journal of Statistical Mechanics: Theory and Experiment}\ }\textbf
  {\bibinfo {volume} {2019}}~(\bibinfo {number} {3}),\ \bibinfo {pages}
  {033404}}\BibitemShut {NoStop}%
\bibitem [{\citenamefont {Gallavotti}\ and\ \citenamefont
  {Cohen}(1995)}]{gallavotti_dynamical_1995}%
  \BibitemOpen
  \bibfield  {author} {\bibinfo {author} {\bibnamefont {Gallavotti},
  \bibfnamefont {G}}, \ and\ \bibinfo {author} {\bibfnamefont {E.~G.~D.}\
  \bibnamefont {Cohen}}} (\bibinfo {year} {1995}),\ \bibfield  {title}
  {\enquote {\bibinfo {title} {Dynamical ensembles in stationary states},}\
  }\href@noop {} {\bibfield  {journal} {\bibinfo  {journal} {Journal of
  Statistical Physics}\ }\textbf {\bibinfo {volume} {80}}~(\bibinfo {number}
  {5-6}),\ \bibinfo {pages} {931--970}}\BibitemShut {NoStop}%
\bibitem [{\citenamefont {Ghil}(1976)}]{Ghil1976}%
  \BibitemOpen
  \bibfield  {author} {\bibinfo {author} {\bibnamefont {Ghil}, \bibfnamefont
  {M}}} (\bibinfo {year} {1976}),\ \bibfield  {title} {\enquote {\bibinfo
  {title} {{Climate stability for a Sellers-type model}},}\ }\href@noop {}
  {\bibfield  {journal} {\bibinfo  {journal} {J. Atmos. Sci.}\ }\textbf
  {\bibinfo {volume} {33}},\ \bibinfo {pages} {3--20}}\BibitemShut {NoStop}%
\bibitem [{\citenamefont {Ghil}(1989)}]{Ghil.89}%
  \BibitemOpen
  \bibfield  {author} {\bibinfo {author} {\bibnamefont {Ghil}, \bibfnamefont
  {M}}} (\bibinfo {year} {1989}),\ \bibfield  {title} {\enquote {\bibinfo
  {title} {Meteorological data assimilation for oceanographers. {Part I:
  Description and theoretical framework}},}\ }\href@noop {} {\bibfield
  {journal} {\bibinfo  {journal} {Dyn. Atmos. Oceans}\ }\textbf {\bibinfo
  {volume} {13}},\ \bibinfo {pages} {171--218}}\BibitemShut {NoStop}%
\bibitem [{\citenamefont {Ghil}(1994)}]{Ghil1994b}%
  \BibitemOpen
  \bibfield  {author} {\bibinfo {author} {\bibnamefont {Ghil}, \bibfnamefont
  {M}}} (\bibinfo {year} {1994}),\ \bibfield  {title} {\enquote {\bibinfo
  {title} {Cryothermodynamics: the chaotic dynamics of paleoclimate},}\
  }\href@noop {} {\bibfield  {journal} {\bibinfo  {journal} {{Physica D}}\
  }\textbf {\bibinfo {volume} {77}},\ \bibinfo {pages} {130--159}}\BibitemShut
  {NoStop}%
\bibitem [{\citenamefont {Ghil}(2001)}]{Ghil2001}%
  \BibitemOpen
  \bibfield  {author} {\bibinfo {author} {\bibnamefont {Ghil}, \bibfnamefont
  {M}}} (\bibinfo {year} {2001}),\ \bibfield  {title} {\enquote {\bibinfo
  {title} {{Hilbert problems for the geosciences in the 21st century}},}\
  }\href@noop {} {\bibfield  {journal} {\bibinfo  {journal} {Nonlin. Processes
  Geophys.}\ }\textbf {\bibinfo {volume} {8}},\ \bibinfo {pages}
  {211--222}}\BibitemShut {NoStop}%
\bibitem [{\citenamefont {Ghil}(2002)}]{Ghil2002}%
  \BibitemOpen
  \bibfield  {author} {\bibinfo {author} {\bibnamefont {Ghil}, \bibfnamefont
  {M}}} (\bibinfo {year} {2002}),\ \bibfield  {title} {\enquote {\bibinfo
  {title} {Natural climate variability},}\ }in\ \href@noop {} {\emph {\bibinfo
  {booktitle} {Encyclopedia of Global Environmental Change, Vol. 1}}},\
  \bibinfo {editor} {edited by\ \bibinfo {editor} {\bibfnamefont {T.~E.}\
  \bibnamefont {Munn}}, \bibinfo {editor} {\bibfnamefont {M.}~\bibnamefont
  {MacCracken}}, \ and\ \bibinfo {editor} {\bibfnamefont {J.}~\bibnamefont
  {Perry}}}\ (\bibinfo  {publisher} {J. Wiley and Sons},\ \bibinfo {address}
  {Chichester/New York})\ pp.\ \bibinfo {pages} {544--549}\BibitemShut
  {NoStop}%
\bibitem [{\citenamefont {Ghil}(2015)}]{Ghil2015}%
  \BibitemOpen
  \bibfield  {author} {\bibinfo {author} {\bibnamefont {Ghil}, \bibfnamefont
  {M}}} (\bibinfo {year} {2015}),\ \bibfield  {title} {\enquote {\bibinfo
  {title} {A mathematical theory of climate sensitivity or, {How to deal with
  both anthropogenic forcing and natural variability?}}}\ }in\ \href@noop {}
  {\emph {\bibinfo {booktitle} {{Climate Change : Multidecadal and Beyond}}}},\
  \bibinfo {editor} {edited by\ \bibinfo {editor} {\bibfnamefont {C.~P.}\
  \bibnamefont {Chang}}, \bibinfo {editor} {\bibfnamefont {M.}~\bibnamefont
  {Ghil}}, \bibinfo {editor} {\bibfnamefont {M.}~\bibnamefont {Latif}}, \ and\
  \bibinfo {editor} {\bibfnamefont {J.~M.}\ \bibnamefont {Wallace}}}\ (\bibinfo
   {publisher} {World Scientific Publishing Co./Imperial College Press})\ pp.\
  \bibinfo {pages} {31--51}\BibitemShut {NoStop}%
\bibitem [{\citenamefont {Ghil}(2017)}]{Ghil2016}%
  \BibitemOpen
  \bibfield  {author} {\bibinfo {author} {\bibnamefont {Ghil}, \bibfnamefont
  {M}}} (\bibinfo {year} {2017}),\ \bibfield  {title} {\enquote {\bibinfo
  {title} {The wind-driven ocean circulation: {Applying dynamical systems
  theory to a climate problem}},}\ }\href@noop {} {\bibfield  {journal}
  {\bibinfo  {journal} {Discr. Cont. Dyn. Syst.--A}\ }\textbf {\bibinfo
  {volume} {37}}~(\bibinfo {number} {1}),\ \bibinfo {pages}
  {189--228}}\BibitemShut {NoStop}%
\bibitem [{\citenamefont {Ghil}(2019)}]{Ghil.2019}%
  \BibitemOpen
  \bibfield  {author} {\bibinfo {author} {\bibnamefont {Ghil}, \bibfnamefont
  {M}}} (\bibinfo {year} {2019}),\ \bibfield  {title} {\enquote {\bibinfo
  {title} {A century of nonlinearity in the geosciences},}\ }\href {\doibase
  10.1029/2019EA000599} {\bibfield  {journal} {\bibinfo  {journal} {Earth and
  Space Science}\ }\textbf {\bibinfo {volume} {6}},\ \bibinfo {pages}
  {1007--1042}}\BibitemShut {NoStop}%
\bibitem [{\citenamefont {Ghil}\ \emph {et~al.}(2002)\citenamefont {Ghil},
  \citenamefont {Allen}, \citenamefont {Dettinger}, \citenamefont {Ide},
  \citenamefont {Kondrashov}, \citenamefont {Mann}, \citenamefont {Robertson},
  \citenamefont {Saunders}, \citenamefont {Tian}, \citenamefont {Varadi},\ and\
  \citenamefont {Yiou}}]{Ghil.SSA.2002}%
  \BibitemOpen
  \bibfield  {author} {\bibinfo {author} {\bibnamefont {Ghil}, \bibfnamefont
  {M}}, \bibinfo {author} {\bibfnamefont {M.~R.}\ \bibnamefont {Allen}},
  \bibinfo {author} {\bibfnamefont {M.~D.}\ \bibnamefont {Dettinger}}, \bibinfo
  {author} {\bibfnamefont {K.}~\bibnamefont {Ide}}, \bibinfo {author}
  {\bibfnamefont {D.}~\bibnamefont {Kondrashov}}, \bibinfo {author}
  {\bibfnamefont {M.~E.}\ \bibnamefont {Mann}}, \bibinfo {author}
  {\bibfnamefont {A.~W.}\ \bibnamefont {Robertson}}, \bibinfo {author}
  {\bibfnamefont {A.}~\bibnamefont {Saunders}}, \bibinfo {author}
  {\bibfnamefont {Y.}~\bibnamefont {Tian}}, \bibinfo {author} {\bibfnamefont
  {F.}~\bibnamefont {Varadi}}, \ and\ \bibinfo {author} {\bibfnamefont
  {P.}~\bibnamefont {Yiou}}} (\bibinfo {year} {2002}),\ \bibfield  {title}
  {\enquote {\bibinfo {title} {{Advanced spectral methods for climatic time
  series}},}\ }\href {\doibase 10.1029/2000RG000092} {\bibfield  {journal}
  {\bibinfo  {journal} {Rev. Geophys.}\ }\textbf {\bibinfo {volume}
  {40}}~(\bibinfo {number} {1}),\ \bibinfo {pages} {31--341}}\BibitemShut
  {NoStop}%
\bibitem [{\citenamefont {Ghil}\ \emph {et~al.}({2008})\citenamefont {Ghil},
  \citenamefont {Chekroun},\ and\ \citenamefont {Simonnet}}]{GCS08}%
  \BibitemOpen
  \bibfield  {author} {\bibinfo {author} {\bibnamefont {Ghil}, \bibfnamefont
  {M}}, \bibinfo {author} {\bibfnamefont {M.~D.}\ \bibnamefont {Chekroun}}, \
  and\ \bibinfo {author} {\bibfnamefont {E.}~\bibnamefont {Simonnet}}}
  (\bibinfo {year} {{2008}}),\ \bibfield  {title} {\enquote {\bibinfo {title}
  {Climate dynamics and fluid mechanics: Natural variability and related
  uncertainties},}\ }\href {\doibase 10.1016/j.physd.2008.03.036} {\bibfield
  {journal} {\bibinfo  {journal} {Physica D: Nonlinear Phenomena}\ }\textbf
  {\bibinfo {volume} {237}}~(\bibinfo {number} {14--17}),\ \bibinfo {pages}
  {2111--2126}}\BibitemShut {NoStop}%
\bibitem [{\citenamefont {Ghil}\ \emph {et~al.}(2015)\citenamefont {Ghil},
  \citenamefont {Chekroun},\ and\ \citenamefont {Stepan}}]{Ghil.ea.2015}%
  \BibitemOpen
  \bibfield  {author} {\bibinfo {author} {\bibnamefont {Ghil}, \bibfnamefont
  {M}}, \bibinfo {author} {\bibfnamefont {M.~D.}\ \bibnamefont {Chekroun}}, \
  and\ \bibinfo {author} {\bibfnamefont {G.}~\bibnamefont {Stepan}}} (\bibinfo
  {year} {2015}),\ \bibfield  {title} {\enquote {\bibinfo {title} {{A
  collection on 'Climate dynamics: multiple scales and memory effects,
  Introduction}},}\ }\href {\doibase 10.1098/rspa.2015.0097} {\bibfield
  {journal} {\bibinfo  {journal} {R. Soc. Proc. A}\ }\textbf {\bibinfo {volume}
  {471}},\ \bibinfo {pages} {20150097}}\BibitemShut {NoStop}%
\bibitem [{\citenamefont {Ghil}\ and\ \citenamefont
  {Childress}(1987)}]{Ghil1987}%
  \BibitemOpen
  \bibfield  {author} {\bibinfo {author} {\bibnamefont {Ghil}, \bibfnamefont
  {M}}, \ and\ \bibinfo {author} {\bibfnamefont {S.}~\bibnamefont {Childress}}}
  (\bibinfo {year} {1987}),\ \href@noop {} {\emph {\bibinfo {title} {{Topics in
  Geophysical Fluid Dynamics: Atmospheric Dynamics, Dynamo Theory, and Climate
  Dynamics}}}}\ (\bibinfo  {publisher} {Springer-Verlag},\ \bibinfo {address}
  {Berlin/Heidelberg/New York})\BibitemShut {NoStop}%
\bibitem [{\citenamefont {Ghil}\ \emph {et~al.}(2018)\citenamefont {Ghil},
  \citenamefont {Groth}, \citenamefont {Kondrashov},\ and\ \citenamefont
  {Robertson}}]{Ghil.ea.S2S}%
  \BibitemOpen
  \bibfield  {author} {\bibinfo {author} {\bibnamefont {Ghil}, \bibfnamefont
  {M}}, \bibinfo {author} {\bibfnamefont {A.}~\bibnamefont {Groth}}, \bibinfo
  {author} {\bibfnamefont {D.}~\bibnamefont {Kondrashov}}, \ and\ \bibinfo
  {author} {\bibfnamefont {A.~W.}\ \bibnamefont {Robertson}}} (\bibinfo {year}
  {2018}),\ \bibfield  {title} {\enquote {\bibinfo {title} {Extratropical
  sub-seasonal--to--seasonal oscillations and multiple regimes: {The dynamical
  systems view}},}\ }in\ \href@noop {} {\emph {\bibinfo {booktitle} {The Gap
  Between Weather and Climate Forecasting: Sub-Seasonal to Seasonal
  Prediction}}},\ \bibinfo {editor} {edited by\ \bibinfo {editor}
  {\bibfnamefont {A.~W.}\ \bibnamefont {Robertson}}\ and\ \bibinfo {editor}
  {\bibfnamefont {F.}~\bibnamefont {Vitart}}},\ Chap.~\bibinfo {chapter} {6}\
  (\bibinfo  {publisher} {Elsevier})\ \bibinfo {note} {in press}\BibitemShut
  {NoStop}%
\bibitem [{\citenamefont {Ghil}\ \emph {et~al.}(1979)\citenamefont {Ghil},
  \citenamefont {Halem},\ and\ \citenamefont {Atlas}}]{Ghil.ea.79}%
  \BibitemOpen
  \bibfield  {author} {\bibinfo {author} {\bibnamefont {Ghil}, \bibfnamefont
  {M}}, \bibinfo {author} {\bibfnamefont {M.}~\bibnamefont {Halem}}, \ and\
  \bibinfo {author} {\bibfnamefont {R.}~\bibnamefont {Atlas}}} (\bibinfo {year}
  {1979}),\ \bibfield  {title} {\enquote {\bibinfo {title} {Time-continuous
  assimilation of remote-sounding data and its effect on weather
  forecasting},}\ }\href@noop {} {\bibfield  {journal} {\bibinfo  {journal}
  {Mon. Wea. Rev.}\ }\textbf {\bibinfo {volume} {107}},\ \bibinfo {pages}
  {140--171}}\BibitemShut {NoStop}%
\bibitem [{\citenamefont {Ghil}\ \emph {et~al.}(1991)\citenamefont {Ghil},
  \citenamefont {Kimoto},\ and\ \citenamefont {Neelin}}]{Ghil1991b}%
  \BibitemOpen
  \bibfield  {author} {\bibinfo {author} {\bibnamefont {Ghil}, \bibfnamefont
  {M}}, \bibinfo {author} {\bibfnamefont {M.}~\bibnamefont {Kimoto}}, \ and\
  \bibinfo {author} {\bibfnamefont {J.~D.}\ \bibnamefont {Neelin}}} (\bibinfo
  {year} {1991}),\ \bibfield  {title} {\enquote {\bibinfo {title} {{Nonlinear
  dynamics and predictability in the atmospheric sciences}},}\ }\href@noop {}
  {\bibfield  {journal} {\bibinfo  {journal} {Rev. Geophys., Supplement (U.S.
  Nat'l Rept. to Int'l Union of Geodesy and Geophys. 1987--1990)}\ }\textbf
  {\bibinfo {volume} {29}},\ \bibinfo {pages} {46--55}}\BibitemShut {NoStop}%
\bibitem [{\citenamefont {Ghil}\ and\ \citenamefont
  {Malanotte-Rizzoli}(1991)}]{Ghil.Mal.1991}%
  \BibitemOpen
  \bibfield  {author} {\bibinfo {author} {\bibnamefont {Ghil}, \bibfnamefont
  {M}}, \ and\ \bibinfo {author} {\bibfnamefont {P.}~\bibnamefont
  {Malanotte-Rizzoli}}} (\bibinfo {year} {1991}),\ \bibfield  {title} {\enquote
  {\bibinfo {title} {Data assimilation in meteorology and oceanography},}\ }in\
  \href@noop {} {\emph {\bibinfo {booktitle} {Advances in Geophysics}}},\
  Vol.~\bibinfo {volume} {33}\ (\bibinfo  {publisher} {Academic Press},\
  \bibinfo {address} {New York})\ pp.\ \bibinfo {pages} {141--266}\BibitemShut
  {NoStop}%
\bibitem [{\citenamefont {Ghil}\ and\ \citenamefont {Mo}(1991)}]{Ghil1991}%
  \BibitemOpen
  \bibfield  {author} {\bibinfo {author} {\bibnamefont {Ghil}, \bibfnamefont
  {M}}, \ and\ \bibinfo {author} {\bibfnamefont {K.~C.}\ \bibnamefont {Mo}}}
  (\bibinfo {year} {1991}),\ \bibfield  {title} {\enquote {\bibinfo {title}
  {Intraseasonal oscillations in the global {atmosphere. Part 1: Northern
  Hemisphere} and tropics},}\ }\href@noop {} {\bibfield  {journal} {\bibinfo
  {journal} {J. Atmos. Sci.}\ }\textbf {\bibinfo {volume} {48}},\ \bibinfo
  {pages} {752--779}}\BibitemShut {NoStop}%
\bibitem [{\citenamefont {Ghil}\ \emph {et~al.}(1987)\citenamefont {Ghil},
  \citenamefont {Mulhaupt},\ and\ \citenamefont {Pestiaux}}]{Ghil.ea.1987}%
  \BibitemOpen
  \bibfield  {author} {\bibinfo {author} {\bibnamefont {Ghil}, \bibfnamefont
  {M}}, \bibinfo {author} {\bibfnamefont {A}~\bibnamefont {Mulhaupt}}, \ and\
  \bibinfo {author} {\bibfnamefont {P}~\bibnamefont {Pestiaux}}} (\bibinfo
  {year} {1987}),\ \bibfield  {title} {\enquote {\bibinfo {title} {{Deep water
  formation and Quaternary glaciations}},}\ }\href@noop {} {\bibfield
  {journal} {\bibinfo  {journal} {Climate Dynamics}\ }\textbf {\bibinfo
  {volume} {2}},\ \bibinfo {pages} {1--10}}\BibitemShut {NoStop}%
\bibitem [{\citenamefont {Ghil}\ and\ \citenamefont
  {Robertson}(2000)}]{Ghil2000}%
  \BibitemOpen
  \bibfield  {author} {\bibinfo {author} {\bibnamefont {Ghil}, \bibfnamefont
  {M}}, \ and\ \bibinfo {author} {\bibfnamefont {A.~W.}\ \bibnamefont
  {Robertson}}} (\bibinfo {year} {2000}),\ \bibfield  {title} {\enquote
  {\bibinfo {title} {{Solving problems with GCMs: General circulation models
  and their role in the climate modeling hierarchy}},}\ }in\ \href@noop {}
  {\emph {\bibinfo {booktitle} {{General Circulation Model Development: Past,
  Present and Future}}}},\ \bibinfo {editor} {edited by\ \bibinfo {editor}
  {\bibfnamefont {D.~A.}\ \bibnamefont {Randall}}}\ (\bibinfo  {publisher}
  {Academic Press},\ \bibinfo {address} {New York})\ pp.\ \bibinfo {pages}
  {285--325}\BibitemShut {NoStop}%
\bibitem [{\citenamefont {Ghil}\ and\ \citenamefont
  {Robertson}(2002)}]{Ghil.Rob.2002}%
  \BibitemOpen
  \bibfield  {author} {\bibinfo {author} {\bibnamefont {Ghil}, \bibfnamefont
  {M}}, \ and\ \bibinfo {author} {\bibfnamefont {A.~W.}\ \bibnamefont
  {Robertson}}} (\bibinfo {year} {2002}),\ \bibfield  {title} {\enquote
  {\bibinfo {title} {{"Waves" vs. "particles" in the atmosphere's phase space:
  A pathway to long-range forecasting?}}}\ }\href@noop {} {\bibfield  {journal}
  {\bibinfo  {journal} {Proc. Natl. Acad. Sci. USA}\ }\textbf {\bibinfo
  {volume} {99}},\ \bibinfo {pages} {2493--2500}}\BibitemShut {NoStop}%
\bibitem [{\citenamefont {Ghil}\ \emph {et~al.}(2011)\citenamefont {Ghil},
  \citenamefont {Yiou}, \citenamefont {Hallegatte}, \citenamefont {Malamud},
  \citenamefont {Naveau}, \citenamefont {Soloviev}, \citenamefont
  {Friederichs}, \citenamefont {Keilis-Borok}, \citenamefont {Kondrashov},
  \citenamefont {Kossobokov}, \citenamefont {Mestre}, \citenamefont {Nicolis},
  \citenamefont {Rust}, \citenamefont {Shebalin}, \citenamefont {Vrac},
  \citenamefont {Witt},\ and\ \citenamefont {Zaliapin}}]{ghil2011}%
  \BibitemOpen
  \bibfield  {author} {\bibinfo {author} {\bibnamefont {Ghil}, \bibfnamefont
  {M}}, \bibinfo {author} {\bibfnamefont {P.}~\bibnamefont {Yiou}}, \bibinfo
  {author} {\bibfnamefont {S.}~\bibnamefont {Hallegatte}}, \bibinfo {author}
  {\bibfnamefont {B.~D.}\ \bibnamefont {Malamud}}, \bibinfo {author}
  {\bibfnamefont {P.}~\bibnamefont {Naveau}}, \bibinfo {author} {\bibfnamefont
  {A.}~\bibnamefont {Soloviev}}, \bibinfo {author} {\bibfnamefont
  {P.}~\bibnamefont {Friederichs}}, \bibinfo {author} {\bibfnamefont
  {V.}~\bibnamefont {Keilis-Borok}}, \bibinfo {author} {\bibfnamefont
  {D.}~\bibnamefont {Kondrashov}}, \bibinfo {author} {\bibfnamefont
  {V.}~\bibnamefont {Kossobokov}}, \bibinfo {author} {\bibfnamefont
  {O.}~\bibnamefont {Mestre}}, \bibinfo {author} {\bibfnamefont
  {C.}~\bibnamefont {Nicolis}}, \bibinfo {author} {\bibfnamefont {H.~W.}\
  \bibnamefont {Rust}}, \bibinfo {author} {\bibfnamefont {P.}~\bibnamefont
  {Shebalin}}, \bibinfo {author} {\bibfnamefont {M.}~\bibnamefont {Vrac}},
  \bibinfo {author} {\bibfnamefont {A.}~\bibnamefont {Witt}}, \ and\ \bibinfo
  {author} {\bibfnamefont {I.}~\bibnamefont {Zaliapin}}} (\bibinfo {year}
  {2011}),\ \bibfield  {title} {\enquote {\bibinfo {title} {Extreme events:
  dynamics, statistics and prediction},}\ }\href {\doibase
  10.5194/npg-18-295-2011} {\bibfield  {journal} {\bibinfo  {journal}
  {Nonlinear Processes in Geophysics}\ }\textbf {\bibinfo {volume}
  {18}}~(\bibinfo {number} {3}),\ \bibinfo {pages} {295--350}}\BibitemShut
  {NoStop}%
\bibitem [{\citenamefont {Ghil}\ \emph {et~al.}({2008a})\citenamefont {Ghil},
  \citenamefont {Zaliapin},\ and\ \citenamefont {Thompson}}]{Ghil.ea.2008a}%
  \BibitemOpen
  \bibfield  {author} {\bibinfo {author} {\bibnamefont {Ghil}, \bibfnamefont
  {M}}, \bibinfo {author} {\bibfnamefont {I.}~\bibnamefont {Zaliapin}}, \ and\
  \bibinfo {author} {\bibfnamefont {S.}~\bibnamefont {Thompson}}} (\bibinfo
  {year} {{2008a}}),\ \bibfield  {title} {\enquote {\bibinfo {title} {{A delay
  differential model of ENSO variability: parametric instability and the
  distribution of extremes}},}\ }\href@noop {} {\bibfield  {journal} {\bibinfo
  {journal} {Nonlinear Processes in Geophysics}\ }\textbf {\bibinfo {volume}
  {15}}~(\bibinfo {number} {3}),\ \bibinfo {pages} {417--433}}\BibitemShut
  {NoStop}%
\bibitem [{\citenamefont {Gildor}\ and\ \citenamefont
  {Tziperman}(2001)}]{Gildor2001}%
  \BibitemOpen
  \bibfield  {author} {\bibinfo {author} {\bibnamefont {Gildor}, \bibfnamefont
  {H}}, \ and\ \bibinfo {author} {\bibfnamefont {E.}~\bibnamefont {Tziperman}}}
  (\bibinfo {year} {2001}),\ \bibfield  {title} {\enquote {\bibinfo {title} {{A
  sea ice climate switch mechanism for the 100-kyr glacial cycles}},}\
  }\href@noop {} {\bibfield  {journal} {\bibinfo  {journal} {J.\ Geophys.
  Res.}\ }\textbf {\bibinfo {volume} {106}},\ \bibinfo {pages}
  {9117--9133}}\BibitemShut {NoStop}%
\bibitem [{\citenamefont {Gill}(1982)}]{Gill1982}%
  \BibitemOpen
  \bibfield  {author} {\bibinfo {author} {\bibnamefont {Gill}, \bibfnamefont
  {A~E}}} (\bibinfo {year} {1982}),\ \href@noop {} {\emph {\bibinfo {title}
  {Atmosphere-Ocean Dynamics}}}\ (\bibinfo  {publisher} {Academic Press},\
  \bibinfo {address} {New York, U.S.A.})\BibitemShut {NoStop}%
\bibitem [{\citenamefont {Ginelli}\ \emph {et~al.}(2007)\citenamefont
  {Ginelli}, \citenamefont {Poggi}, \citenamefont {Turchi}, \citenamefont
  {Chat\'e}, \citenamefont {Livi},\ and\ \citenamefont {Politi}}]{GPTCLP07}%
  \BibitemOpen
  \bibfield  {author} {\bibinfo {author} {\bibnamefont {Ginelli}, \bibfnamefont
  {F}}, \bibinfo {author} {\bibfnamefont {P.}~\bibnamefont {Poggi}}, \bibinfo
  {author} {\bibfnamefont {A.}~\bibnamefont {Turchi}}, \bibinfo {author}
  {\bibfnamefont {H.}~\bibnamefont {Chat\'e}}, \bibinfo {author} {\bibfnamefont
  {R.}~\bibnamefont {Livi}}, \ and\ \bibinfo {author} {\bibfnamefont
  {A.}~\bibnamefont {Politi}}} (\bibinfo {year} {2007}),\ \bibfield  {title}
  {\enquote {\bibinfo {title} {Characterizing dynamics with covariant lyapunov
  vectors},}\ }\href {\doibase 10.1103/PhysRevLett.99.130601} {\bibfield
  {journal} {\bibinfo  {journal} {Phys. Rev. Lett.}\ }\textbf {\bibinfo
  {volume} {99}},\ \bibinfo {pages} {130601}}\BibitemShut {NoStop}%
\bibitem [{\citenamefont {Gladwell}(2000)}]{Gladwell00}%
  \BibitemOpen
  \bibfield  {author} {\bibinfo {author} {\bibnamefont {Gladwell},
  \bibfnamefont {M}}} (\bibinfo {year} {2000}),\ \href@noop {} {\emph {\bibinfo
  {title} {The Tipping Point: How Little Things Can Make a Big Difference}}}\
  (\bibinfo  {publisher} {Little Brown},\ \bibinfo {address}
  {Boston})\BibitemShut {NoStop}%
\bibitem [{\citenamefont {Gleckler}\ \emph {et~al.}(2016)\citenamefont
  {Gleckler}, \citenamefont {Doutriaux}, \citenamefont {Durack}, \citenamefont
  {Taylor}, \citenamefont {Zhang}, \citenamefont {Williams}, \citenamefont
  {Mason},\ and\ \citenamefont {Servonnat}}]{Gleckler.ea.2016}%
  \BibitemOpen
  \bibfield  {author} {\bibinfo {author} {\bibnamefont {Gleckler},
  \bibfnamefont {P~J}}, \bibinfo {author} {\bibfnamefont {C.}~\bibnamefont
  {Doutriaux}}, \bibinfo {author} {\bibfnamefont {P.~J.}\ \bibnamefont
  {Durack}}, \bibinfo {author} {\bibfnamefont {K.~E.}\ \bibnamefont {Taylor}},
  \bibinfo {author} {\bibfnamefont {Y.}~\bibnamefont {Zhang}}, \bibinfo
  {author} {\bibfnamefont {D.~N.}\ \bibnamefont {Williams}}, \bibinfo {author}
  {\bibfnamefont {E.}~\bibnamefont {Mason}}, \ and\ \bibinfo {author}
  {\bibfnamefont {J.}~\bibnamefont {Servonnat}}} (\bibinfo {year} {2016}),\
  \bibfield  {title} {\enquote {\bibinfo {title} {A more powerful reality test
  for climate models},}\ }\href@noop {} {\bibfield  {journal} {\bibinfo
  {journal} {Eos Trans. AGU}\ }\textbf {\bibinfo {volume} {97}}}\BibitemShut
  {NoStop}%
\bibitem [{\citenamefont {G{\'{o}}mez-Leal}\ \emph {et~al.}(2018)\citenamefont
  {G{\'{o}}mez-Leal}, \citenamefont {Kaltenegger}, \citenamefont {Lucarini},\
  and\ \citenamefont {Lunkeit}}]{Gomez_2018}%
  \BibitemOpen
  \bibfield  {author} {\bibinfo {author} {\bibnamefont {G{\'{o}}mez-Leal},
  \bibfnamefont {I}}, \bibinfo {author} {\bibfnamefont {L.}~\bibnamefont
  {Kaltenegger}}, \bibinfo {author} {\bibfnamefont {V.}~\bibnamefont
  {Lucarini}}, \ and\ \bibinfo {author} {\bibfnamefont {F.}~\bibnamefont
  {Lunkeit}}} (\bibinfo {year} {2018}),\ \bibfield  {title} {\enquote {\bibinfo
  {title} {Climate sensitivity to carbon dioxide and the moist greenhouse
  threshold of earth-like planets under an increasing solar forcing},}\ }\href
  {\doibase 10.3847/1538-4357/aaea5f} {\bibfield  {journal} {\bibinfo
  {journal} {The Astrophysical Journal}\ }\textbf {\bibinfo {volume}
  {869}}~(\bibinfo {number} {2}),\ \bibinfo {pages} {129}}\BibitemShut
  {NoStop}%
\bibitem [{\citenamefont {Goody}(2000)}]{Goody00}%
  \BibitemOpen
  \bibfield  {author} {\bibinfo {author} {\bibnamefont {Goody}, \bibfnamefont
  {R}}} (\bibinfo {year} {2000}),\ \bibfield  {title} {\enquote {\bibinfo
  {title} {Sources and sinks of climate entropy},}\ }\href@noop {} {\bibfield
  {journal} {\bibinfo  {journal} {Q. J. R. Meteorol. Soc.}\ }\textbf {\bibinfo
  {volume} {126}},\ \bibinfo {pages} {1953--1970}}\BibitemShut {NoStop}%
\bibitem [{\citenamefont {{Gozolchiani}}\ \emph {et~al.}(2011)\citenamefont
  {{Gozolchiani}}, \citenamefont {{Havlin}},\ and\ \citenamefont
  {{Yamasaki}}}]{Gozolchiani2011}%
  \BibitemOpen
  \bibfield  {author} {\bibinfo {author} {\bibnamefont {{Gozolchiani}},
  \bibfnamefont {A}}, \bibinfo {author} {\bibfnamefont {S.}~\bibnamefont
  {{Havlin}}}, \ and\ \bibinfo {author} {\bibfnamefont {K.}~\bibnamefont
  {{Yamasaki}}}} (\bibinfo {year} {2011}),\ \bibfield  {title} {\enquote
  {\bibinfo {title} {{Emergence of El Ni{\~n}o} as an autonomous component in
  the climate network},}\ }\href {\doibase 10.1103/PhysRevLett.107.148501}
  {\bibfield  {journal} {\bibinfo  {journal} {Physical Review Letters}\
  }\textbf {\bibinfo {volume} {107}}~(\bibinfo {number} {14}),\ \bibinfo {eid}
  {148501}},\ \Eprint {http://arxiv.org/abs/1010.2605} {arXiv:1010.2605
  [physics.ao-ph]} \BibitemShut {NoStop}%
\bibitem [{\citenamefont {Graham}\ \emph {et~al.}(1991)\citenamefont {Graham},
  \citenamefont {Hamm},\ and\ \citenamefont {T\'el}}]{Graham1991}%
  \BibitemOpen
  \bibfield  {author} {\bibinfo {author} {\bibnamefont {Graham}, \bibfnamefont
  {R}}, \bibinfo {author} {\bibfnamefont {A.}~\bibnamefont {Hamm}}, \ and\
  \bibinfo {author} {\bibfnamefont {T.}~\bibnamefont {T\'el}}} (\bibinfo {year}
  {1991}),\ \bibfield  {title} {\enquote {\bibinfo {title} {Nonequilibrium
  potentials for dynamical systems with fractal attractors or repellers},}\
  }\href {\doibase 10.1103/PhysRevLett.66.3089} {\bibfield  {journal} {\bibinfo
   {journal} {Phys. Rev. Lett.}\ }\textbf {\bibinfo {volume} {66}},\ \bibinfo
  {pages} {3089--3092}}\BibitemShut {NoStop}%
\bibitem [{\citenamefont {Grasman}(1987)}]{Grasman.1987}%
  \BibitemOpen
  \bibfield  {author} {\bibinfo {author} {\bibnamefont {Grasman}, \bibfnamefont
  {J}}} (\bibinfo {year} {1987}),\ \href@noop {} {\emph {\bibinfo {title}
  {{Asymptotic Methods for Relaxation Oscillations and Applications}}}}\
  (\bibinfo  {publisher} {Springer Science \& Business Media})\BibitemShut
  {NoStop}%
\bibitem [{\citenamefont {Grassberger}(1989)}]{Grassberger1989}%
  \BibitemOpen
  \bibfield  {author} {\bibinfo {author} {\bibnamefont {Grassberger},
  \bibfnamefont {P}}} (\bibinfo {year} {1989}),\ \bibfield  {title} {\enquote
  {\bibinfo {title} {Noise-induced escape from attractors},}\ }\href@noop {}
  {\bibfield  {journal} {\bibinfo  {journal} {Journal of Physics A:
  Mathematical and General}\ }\textbf {\bibinfo {volume} {22}}~(\bibinfo
  {number} {16}),\ \bibinfo {pages} {3283}}\BibitemShut {NoStop}%
\bibitem [{\citenamefont {Grebogi}\ \emph {et~al.}(1983)\citenamefont
  {Grebogi}, \citenamefont {Ott},\ and\ \citenamefont {Yorke}}]{Grebogi1983}%
  \BibitemOpen
  \bibfield  {author} {\bibinfo {author} {\bibnamefont {Grebogi}, \bibfnamefont
  {C}}, \bibinfo {author} {\bibfnamefont {E.}~\bibnamefont {Ott}}, \ and\
  \bibinfo {author} {\bibfnamefont {J.~A.}\ \bibnamefont {Yorke}}} (\bibinfo
  {year} {1983}),\ \bibfield  {title} {\enquote {\bibinfo {title} {Fractal
  basin boundaries, long-lived chaotic transients, and unstable-unstable pair
  bifurcation},}\ }\href {\doibase 10.1103/PhysRevLett.50.935} {\bibfield
  {journal} {\bibinfo  {journal} {Physical Review Letters}\ }\textbf {\bibinfo
  {volume} {50}},\ \bibinfo {pages} {935--938}}\BibitemShut {NoStop}%
\bibitem [{\citenamefont {Gritsun}\ \emph {et~al.}(2008)\citenamefont
  {Gritsun}, \citenamefont {Branstator},\ and\ \citenamefont
  {Majda}}]{gritsun2008b}%
  \BibitemOpen
  \bibfield  {author} {\bibinfo {author} {\bibnamefont {Gritsun}, \bibfnamefont
  {A}}, \bibinfo {author} {\bibfnamefont {G.}~\bibnamefont {Branstator}}, \
  and\ \bibinfo {author} {\bibfnamefont {A.~J.}\ \bibnamefont {Majda}}}
  (\bibinfo {year} {2008}),\ \bibfield  {title} {\enquote {\bibinfo {title}
  {Climate response of linear and quadratic functionals using the
  fluctuation-dissipation theorem},}\ }\href@noop {} {\bibfield  {journal}
  {\bibinfo  {journal} {J. Atmos. Sci.}\ }\textbf {\bibinfo {volume}
  {65}}}\BibitemShut {NoStop}%
\bibitem [{\citenamefont {Gritsun}\ and\ \citenamefont
  {Branstator}(2007)}]{gritsun2007}%
  \BibitemOpen
  \bibfield  {author} {\bibinfo {author} {\bibnamefont {Gritsun}, \bibfnamefont
  {Andrey}}, \ and\ \bibinfo {author} {\bibfnamefont {Grant}\ \bibnamefont
  {Branstator}}} (\bibinfo {year} {2007}),\ \bibfield  {title} {\enquote
  {\bibinfo {title} {Climate response using a three-dimensional operator based
  on the fluctuationÃ¯--dissipation theorem},}\ }\href {\doibase
  10.1175/JAS3943.1} {\bibfield  {journal} {\bibinfo  {journal} {Journal of the
  Atmospheric Sciences}\ }\textbf {\bibinfo {volume} {64}}~(\bibinfo {number}
  {7}),\ \bibinfo {pages} {2558--2575}}\BibitemShut {NoStop}%
\bibitem [{\citenamefont {Gritsun}\ and\ \citenamefont
  {Lucarini}(2017)}]{gritsun2017}%
  \BibitemOpen
  \bibfield  {author} {\bibinfo {author} {\bibnamefont {Gritsun}, \bibfnamefont
  {Andrey}}, \ and\ \bibinfo {author} {\bibfnamefont {Valerio}\ \bibnamefont
  {Lucarini}}} (\bibinfo {year} {2017}),\ \bibfield  {title} {\enquote
  {\bibinfo {title} {Fluctuations, response, and resonances in a simple
  atmospheric model},}\ }\href {\doibase 10.1016/j.physd.2017.02.015}
  {\bibfield  {journal} {\bibinfo  {journal} {Physica D: Nonlinear Phenomena}\
  }\textbf {\bibinfo {volume} {349}},\ \bibinfo {pages} {62--76}}\BibitemShut
  {NoStop}%
\bibitem [{\citenamefont {de~Groot}\ and\ \citenamefont
  {Mazur}(1984)}]{deGroot84}%
  \BibitemOpen
  \bibfield  {author} {\bibinfo {author} {\bibnamefont {de~Groot},
  \bibfnamefont {S~R}}, \ and\ \bibinfo {author} {\bibfnamefont
  {P}~\bibnamefont {Mazur}}} (\bibinfo {year} {1984}),\ \href@noop {} {\emph
  {\bibinfo {title} {Non-Equilibrium Thermodynamics}}}\ (\bibinfo  {publisher}
  {Dover})\BibitemShut {NoStop}%
\bibitem [{\citenamefont {Guckenheimer}\ and\ \citenamefont
  {Holmes}(1983)}]{Guckenheimer1983}%
  \BibitemOpen
  \bibfield  {author} {\bibinfo {author} {\bibnamefont {Guckenheimer},
  \bibfnamefont {J}}, \ and\ \bibinfo {author} {\bibfnamefont {P.}~\bibnamefont
  {Holmes}}} (\bibinfo {year} {1983}),\ \href@noop {} {\emph {\bibinfo {title}
  {{Nonlinear Oscillations, Dynamical Systems and Bifurcations of Vector
  Fields}}}},\ \bibinfo {edition} {2nd}\ ed.\ (\bibinfo  {publisher}
  {Springer-Verlag},\ \bibinfo {address} {Berlin/Heidelberg})\BibitemShut
  {NoStop}%
\bibitem [{\citenamefont {Guckenheimer}\ \emph {et~al.}(2003)\citenamefont
  {Guckenheimer}, \citenamefont {Hoffman},\ and\ \citenamefont
  {Weckesser}}]{Guck.ea.2003}%
  \BibitemOpen
  \bibfield  {author} {\bibinfo {author} {\bibnamefont {Guckenheimer},
  \bibfnamefont {John}}, \bibinfo {author} {\bibfnamefont {Kathleen}\
  \bibnamefont {Hoffman}}, \ and\ \bibinfo {author} {\bibfnamefont {Warren}\
  \bibnamefont {Weckesser}}} (\bibinfo {year} {2003}),\ \bibfield  {title}
  {\enquote {\bibinfo {title} {The forced van der {Pol equation I: The slow
  flow and its bifurcations}},}\ }\href@noop {} {\bibfield  {journal} {\bibinfo
   {journal} {SIAM Journal on Applied Dynamical Systems}\ }\textbf {\bibinfo
  {volume} {2}}~(\bibinfo {number} {1}),\ \bibinfo {pages} {1--35}}\BibitemShut
  {NoStop}%
\bibitem [{\citenamefont {Haines}(1994)}]{Haines1994}%
  \BibitemOpen
  \bibfield  {author} {\bibinfo {author} {\bibnamefont {Haines}, \bibfnamefont
  {K}}} (\bibinfo {year} {1994}),\ \bibfield  {title} {\enquote {\bibinfo
  {title} {{Low-frequency variability in atmospheric middle latitudes}},}\
  }\href@noop {} {\bibfield  {journal} {\bibinfo  {journal} {Surveys in
  Geophysics}\ }\textbf {\bibinfo {volume} {15}},\ \bibinfo {pages}
  {1--61}}\BibitemShut {NoStop}%
\bibitem [{\citenamefont {Hale}(1977)}]{Hale.1977}%
  \BibitemOpen
  \bibfield  {author} {\bibinfo {author} {\bibnamefont {Hale}, \bibfnamefont
  {J~K}}} (\bibinfo {year} {1977}),\ \href@noop {} {\emph {\bibinfo {title}
  {{Theory of Functional Differential Equations}}}}\ (\bibinfo  {publisher}
  {Springer-Verlag},\ \bibinfo {address} {New York})\BibitemShut {NoStop}%
\bibitem [{\citenamefont {Hamm}\ \emph {et~al.}(1994)\citenamefont {Hamm},
  \citenamefont {T\'el},\ and\ \citenamefont {Graham}}]{Hamm1994}%
  \BibitemOpen
  \bibfield  {author} {\bibinfo {author} {\bibnamefont {Hamm}, \bibfnamefont
  {A}}, \bibinfo {author} {\bibfnamefont {T.}~\bibnamefont {T\'el}}, \ and\
  \bibinfo {author} {\bibfnamefont {R.}~\bibnamefont {Graham}}} (\bibinfo
  {year} {1994}),\ \bibfield  {title} {\enquote {\bibinfo {title}
  {Noise-induced attractor explosions near tangent bifurcations},}\ }\href
  {\doibase 10.1016/0375-9601(94)90621-1} {\bibfield  {journal} {\bibinfo
  {journal} {Physics Letters A}\ }\textbf {\bibinfo {volume} {185}}~(\bibinfo
  {number} {3}),\ \bibinfo {pages} {313--320}}\BibitemShut {NoStop}%
\bibitem [{\citenamefont {Hanggi}(1986)}]{Hanggi1986}%
  \BibitemOpen
  \bibfield  {author} {\bibinfo {author} {\bibnamefont {Hanggi}, \bibfnamefont
  {P}}} (\bibinfo {year} {1986}),\ \bibfield  {title} {\enquote {\bibinfo
  {title} {Escape from a metastable state},}\ }\href {\doibase
  10.1007/BF01010843} {\bibfield  {journal} {\bibinfo  {journal} {Journal of
  Statistical Physics}\ }\textbf {\bibinfo {volume} {42}}~(\bibinfo {number}
  {1}),\ \bibinfo {pages} {105--148}}\BibitemShut {NoStop}%
\bibitem [{\citenamefont {Hannart}\ \emph {et~al.}({2016b})\citenamefont
  {Hannart}, \citenamefont {Carrassi}, \citenamefont {Bocquet}, \citenamefont
  {Ghil}, \citenamefont {Naveau}, \citenamefont {Pulido}, \citenamefont
  {Ruiz},\ and\ \citenamefont {Tandeo}}]{Hannart.ea.2016}%
  \BibitemOpen
  \bibfield  {author} {\bibinfo {author} {\bibnamefont {Hannart}, \bibfnamefont
  {A}}, \bibinfo {author} {\bibfnamefont {A.}~\bibnamefont {Carrassi}},
  \bibinfo {author} {\bibfnamefont {M.}~\bibnamefont {Bocquet}}, \bibinfo
  {author} {\bibfnamefont {M.}~\bibnamefont {Ghil}}, \bibinfo {author}
  {\bibfnamefont {P.}~\bibnamefont {Naveau}}, \bibinfo {author} {\bibfnamefont
  {M.}~\bibnamefont {Pulido}}, \bibinfo {author} {\bibfnamefont
  {J.}~\bibnamefont {Ruiz}}, \ and\ \bibinfo {author} {\bibfnamefont
  {P.}~\bibnamefont {Tandeo}}} (\bibinfo {year} {{2016b}}),\ \bibfield  {title}
  {\enquote {\bibinfo {title} {{DADA: data assimilation for the detection and
  attribution of weather and climate-related events}},}\ }\href@noop {}
  {\bibfield  {journal} {\bibinfo  {journal} {Climatic Change}\ }\textbf
  {\bibinfo {volume} {136}}~(\bibinfo {number} {2}),\ \bibinfo {pages}
  {155--174}}\BibitemShut {NoStop}%
\bibitem [{\citenamefont {Hannart}\ and\ \citenamefont
  {Naveau}(2018)}]{Hannart2018}%
  \BibitemOpen
  \bibfield  {author} {\bibinfo {author} {\bibnamefont {Hannart}, \bibfnamefont
  {A}}, \ and\ \bibinfo {author} {\bibfnamefont {P.}~\bibnamefont {Naveau}}}
  (\bibinfo {year} {2018}),\ \bibfield  {title} {\enquote {\bibinfo {title}
  {Probabilities of causation of climate changes},}\ }\href {\doibase
  10.1175/JCLI-D-17-0304.1} {\bibfield  {journal} {\bibinfo  {journal} {Journal
  of Climate}\ }\textbf {\bibinfo {volume} {31}}~(\bibinfo {number} {14}),\
  \bibinfo {pages} {5507--5524}},\ \Eprint
  {http://arxiv.org/abs/https://doi.org/10.1175/JCLI-D-17-0304.1}
  {https://doi.org/10.1175/JCLI-D-17-0304.1} \BibitemShut {NoStop}%
\bibitem [{\citenamefont {Hannart}\ \emph {et~al.}({2016a})\citenamefont
  {Hannart}, \citenamefont {Pearl}, \citenamefont {Otto}, \citenamefont
  {Naveau},\ and\ \citenamefont {Ghil}}]{Hannart2016}%
  \BibitemOpen
  \bibfield  {author} {\bibinfo {author} {\bibnamefont {Hannart}, \bibfnamefont
  {A}}, \bibinfo {author} {\bibfnamefont {J.}~\bibnamefont {Pearl}}, \bibinfo
  {author} {\bibfnamefont {F.~E.~L.}\ \bibnamefont {Otto}}, \bibinfo {author}
  {\bibfnamefont {P.}~\bibnamefont {Naveau}}, \ and\ \bibinfo {author}
  {\bibfnamefont {M.}~\bibnamefont {Ghil}}} (\bibinfo {year} {{2016a}}),\
  \bibfield  {title} {\enquote {\bibinfo {title} {Causal counterfactual theory
  for the attribution of weather and climate-related events},}\ }\href
  {\doibase 10.1175/BAMS-D-14-00034.1} {\bibfield  {journal} {\bibinfo
  {journal} {Bull. Amer. Meteorol. Soc.}\ }\textbf {\bibinfo {volume}
  {97}}~(\bibinfo {number} {1}),\ \bibinfo {pages} {99--110}}\BibitemShut
  {NoStop}%
\bibitem [{\citenamefont {Hassanzadeh}\ \emph {et~al.}(2014)\citenamefont
  {Hassanzadeh}, \citenamefont {Kuang},\ and\ \citenamefont
  {Farrell}}]{Hassanz.ea.2014}%
  \BibitemOpen
  \bibfield  {author} {\bibinfo {author} {\bibnamefont {Hassanzadeh},
  \bibfnamefont {P}}, \bibinfo {author} {\bibfnamefont {Z.}~\bibnamefont
  {Kuang}}, \ and\ \bibinfo {author} {\bibfnamefont {B.~F.}\ \bibnamefont
  {Farrell}}} (\bibinfo {year} {2014}),\ \bibfield  {title} {\enquote {\bibinfo
  {title} {Responses of midlatitude blocks and wave amplitude to changes in the
  meridional temperature gradient in an idealized dry {GCM}},}\ }\href@noop {}
  {\bibfield  {journal} {\bibinfo  {journal} {Geophysical Research Letters}\
  }\textbf {\bibinfo {volume} {41}}~(\bibinfo {number} {14}),\ \bibinfo {pages}
  {5223--5232}}\BibitemShut {NoStop}%
\bibitem [{\citenamefont {Hasselmann}(1976)}]{Hasselmann1976}%
  \BibitemOpen
  \bibfield  {author} {\bibinfo {author} {\bibnamefont {Hasselmann},
  \bibfnamefont {K}}} (\bibinfo {year} {1976}),\ \bibfield  {title} {\enquote
  {\bibinfo {title} {Stochastic climate models. {I: T}heory},}\ }\href@noop {}
  {\bibfield  {journal} {\bibinfo  {journal} {Tellus}\ }\textbf {\bibinfo
  {volume} {28}},\ \bibinfo {pages} {473--485}}\BibitemShut {NoStop}%
\bibitem [{\citenamefont {Hasselmann}\ \emph {et~al.}(1993)\citenamefont
  {Hasselmann}, \citenamefont {Sausen}, \citenamefont {Maier-Reimer},\ and\
  \citenamefont {Voss}}]{hasselmann.ea.1993}%
  \BibitemOpen
  \bibfield  {author} {\bibinfo {author} {\bibnamefont {Hasselmann},
  \bibfnamefont {K}}, \bibinfo {author} {\bibfnamefont {R.}~\bibnamefont
  {Sausen}}, \bibinfo {author} {\bibfnamefont {E.}~\bibnamefont
  {Maier-Reimer}}, \ and\ \bibinfo {author} {\bibfnamefont {R.}~\bibnamefont
  {Voss}}} (\bibinfo {year} {1993}),\ \bibfield  {title} {\enquote {\bibinfo
  {title} {On the cold start problem in transient simulations with coupled
  atmosphere-ocean models},}\ }\href {\doibase 10.1007/BF00210008} {\bibfield
  {journal} {\bibinfo  {journal} {Climate Dynamics}\ }\textbf {\bibinfo
  {volume} {9}}~(\bibinfo {number} {2}),\ \bibinfo {pages}
  {53--61}}\BibitemShut {NoStop}%
\bibitem [{\citenamefont {Hasson}\ \emph {et~al.}(2013)\citenamefont {Hasson},
  \citenamefont {Lucarini},\ and\ \citenamefont {Pascale}}]{Hasson13}%
  \BibitemOpen
  \bibfield  {author} {\bibinfo {author} {\bibnamefont {Hasson}, \bibfnamefont
  {S}}, \bibinfo {author} {\bibfnamefont {V.}~\bibnamefont {Lucarini}}, \ and\
  \bibinfo {author} {\bibfnamefont {S.}~\bibnamefont {Pascale}}} (\bibinfo
  {year} {2013}),\ \bibfield  {title} {\enquote {\bibinfo {title} {Hydrological
  cycle over {South and Southeast Asian river basins as simulated by
  PCMDI/CMIP3 experiments}},}\ }\href {\doibase 10.5194/esd-4-199-2013}
  {\bibfield  {journal} {\bibinfo  {journal} {Earth System Dynamics}\ }\textbf
  {\bibinfo {volume} {4}}~(\bibinfo {number} {2}),\ \bibinfo {pages}
  {199--217}}\BibitemShut {NoStop}%
\bibitem [{\citenamefont {Hawkins}\ \emph {et~al.}(2011)\citenamefont
  {Hawkins}, \citenamefont {Smith}, \citenamefont {Allison}, \citenamefont
  {Gregory}, \citenamefont {Woollings}, \citenamefont {Pohlmann},\ and\
  \citenamefont {Cuevas}}]{Hawkins2011}%
  \BibitemOpen
  \bibfield  {author} {\bibinfo {author} {\bibnamefont {Hawkins}, \bibfnamefont
  {E}}, \bibinfo {author} {\bibfnamefont {R.~S.}\ \bibnamefont {Smith}},
  \bibinfo {author} {\bibfnamefont {L.~C.}\ \bibnamefont {Allison}}, \bibinfo
  {author} {\bibfnamefont {J.~M.}\ \bibnamefont {Gregory}}, \bibinfo {author}
  {\bibfnamefont {T.~J.}\ \bibnamefont {Woollings}}, \bibinfo {author}
  {\bibfnamefont {H.}~\bibnamefont {Pohlmann}}, \ and\ \bibinfo {author}
  {\bibfnamefont {B.}~\bibnamefont {Cuevas}}} (\bibinfo {year} {2011}),\
  \bibfield  {title} {\enquote {\bibinfo {title} {Bistability of the atlantic
  overturning circulation in a global climate model and links to ocean
  freshwater transport},}\ }\href@noop {} {\bibfield  {journal} {\bibinfo
  {journal} {Geophysical Research Letters}\ }\textbf {\bibinfo {volume}
  {38}}~(\bibinfo {number} {1})},\ \bibinfo {note} {copyright - Copyright 2011
  by American Geophysical Union; Last updated - 2014-08-09}\BibitemShut
  {NoStop}%
\bibitem [{\citenamefont {Hayashi}(1971)}]{Haya71}%
  \BibitemOpen
  \bibfield  {author} {\bibinfo {author} {\bibnamefont {Hayashi}, \bibfnamefont
  {Y}}} (\bibinfo {year} {1971}),\ \bibfield  {title} {\enquote {\bibinfo
  {title} {A generalized method for resolving disturbances into progressive and
  retrogressive waves by space {Fourier} and time cross-spectral analysis},}\
  }\href@noop {} {\bibfield  {journal} {\bibinfo  {journal} {J. Meteorol. Soc.
  Jpn.}\ }\textbf {\bibinfo {volume} {49}},\ \bibinfo {pages}
  {125--128}}\BibitemShut {NoStop}%
\bibitem [{\citenamefont {Heinrich}(1988)}]{Heinrich1988}%
  \BibitemOpen
  \bibfield  {author} {\bibinfo {author} {\bibnamefont {Heinrich},
  \bibfnamefont {H}}} (\bibinfo {year} {1988}),\ \bibfield  {title} {\enquote
  {\bibinfo {title} {{Origin and Consequences of Cyclic Ice Rafting in the
  Northeast Atlantic Ocean during the Past 130,000 Years}},}\ }\href@noop {}
  {\bibfield  {journal} {\bibinfo  {journal} {Quaternary Research}\ }\textbf
  {\bibinfo {volume} {29}},\ \bibinfo {pages} {142--152}}\BibitemShut {NoStop}%
\bibitem [{\citenamefont {Held}({2001})}]{Held.transport.01}%
  \BibitemOpen
  \bibfield  {author} {\bibinfo {author} {\bibnamefont {Held}, \bibfnamefont
  {I~M}}} (\bibinfo {year} {{2001}}),\ \bibfield  {title} {\enquote {\bibinfo
  {title} {{The partitioning of the poleward energy transport between the
  tropical ocean and atmosphere}},}\ }\href {\doibase
  {10.1175/1520-0469(2001)058<0943:TPOTPE>2.0.CO;2}} {\bibfield  {journal}
  {\bibinfo  {journal} {J.\ Atmos.\ Sci.}\ }\textbf {\bibinfo {volume}
  {{58}}}~(\bibinfo {number} {{8}}),\ \bibinfo {pages} {943--948}}\BibitemShut
  {NoStop}%
\bibitem [{\citenamefont {Held}({2005})}]{Held.gap.05}%
  \BibitemOpen
  \bibfield  {author} {\bibinfo {author} {\bibnamefont {Held}, \bibfnamefont
  {I~M}}} (\bibinfo {year} {{2005}}),\ \bibfield  {title} {\enquote {\bibinfo
  {title} {{The gap between simulation and understanding in climate
  modeling}},}\ }\href@noop {} {\bibfield  {journal} {\bibinfo  {journal}
  {Bull. Am. Meteorol. Soc.}\ }\textbf {\bibinfo {volume} {{86}}},\ \bibinfo
  {pages} {{1609--1614}}}\BibitemShut {NoStop}%
\bibitem [{\citenamefont {Held}\ and\ \citenamefont {Suarez}(1974)}]{Held1974}%
  \BibitemOpen
  \bibfield  {author} {\bibinfo {author} {\bibnamefont {Held}, \bibfnamefont
  {I~M}}, \ and\ \bibinfo {author} {\bibfnamefont {M.~J.}\ \bibnamefont
  {Suarez}}} (\bibinfo {year} {1974}),\ \bibfield  {title} {\enquote {\bibinfo
  {title} {{Simple albedo feedback models of the ice caps}},}\ }\href@noop {}
  {\bibfield  {journal} {\bibinfo  {journal} {Tellus}\ }\textbf {\bibinfo
  {volume} {26}},\ \bibinfo {pages} {613--629}}\BibitemShut {NoStop}%
\bibitem [{\citenamefont {von~der Heydt}\ \emph {et~al.}(2016)\citenamefont
  {von~der Heydt}, \citenamefont {Dijkstra}, \citenamefont {van~de Wal},
  \citenamefont {Caballero}, \citenamefont {Crucifix}, \citenamefont {Foster},
  \citenamefont {Huber}, \citenamefont {K{\"o}hler}, \citenamefont {Rohling},
  \citenamefont {Valdes}, \citenamefont {Ashwin}, \citenamefont {Bathiany},
  \citenamefont {Berends}, \citenamefont {van Bree}, \citenamefont {Ditlevsen},
  \citenamefont {Ghil}, \citenamefont {Haywood}, \citenamefont {Katzav},
  \citenamefont {Lohmann}, \citenamefont {Lohmann}, \citenamefont {Lucarini},
  \citenamefont {Marzocchi}, \citenamefont {P{\"a}like}, \citenamefont
  {Baroni}, \citenamefont {Simon}, \citenamefont {Sluijs}, \citenamefont
  {Stap}, \citenamefont {Tantet}, \citenamefont {Viebahn},\ and\ \citenamefont
  {Ziegler}}]{vonderHeydt2016}%
  \BibitemOpen
  \bibfield  {author} {\bibinfo {author} {\bibnamefont {von~der Heydt},
  \bibfnamefont {Anna~S}}, \bibinfo {author} {\bibfnamefont {Henk~A.}\
  \bibnamefont {Dijkstra}}, \bibinfo {author} {\bibfnamefont {Roderik S.~W.}\
  \bibnamefont {van~de Wal}}, \bibinfo {author} {\bibfnamefont {Rodrigo}\
  \bibnamefont {Caballero}}, \bibinfo {author} {\bibfnamefont {Michel}\
  \bibnamefont {Crucifix}}, \bibinfo {author} {\bibfnamefont {Gavin~L.}\
  \bibnamefont {Foster}}, \bibinfo {author} {\bibfnamefont {Matthew}\
  \bibnamefont {Huber}}, \bibinfo {author} {\bibfnamefont {Peter}\ \bibnamefont
  {K{\"o}hler}}, \bibinfo {author} {\bibfnamefont {Eelco}\ \bibnamefont
  {Rohling}}, \bibinfo {author} {\bibfnamefont {Paul~J.}\ \bibnamefont
  {Valdes}}, \bibinfo {author} {\bibfnamefont {Peter}\ \bibnamefont {Ashwin}},
  \bibinfo {author} {\bibfnamefont {Sebastian}\ \bibnamefont {Bathiany}},
  \bibinfo {author} {\bibfnamefont {Tijn}\ \bibnamefont {Berends}}, \bibinfo
  {author} {\bibfnamefont {Loes G.~J.}\ \bibnamefont {van Bree}}, \bibinfo
  {author} {\bibfnamefont {Peter}\ \bibnamefont {Ditlevsen}}, \bibinfo {author}
  {\bibfnamefont {Michael}\ \bibnamefont {Ghil}}, \bibinfo {author}
  {\bibfnamefont {Alan~M.}\ \bibnamefont {Haywood}}, \bibinfo {author}
  {\bibfnamefont {Joel}\ \bibnamefont {Katzav}}, \bibinfo {author}
  {\bibfnamefont {Gerrit}\ \bibnamefont {Lohmann}}, \bibinfo {author}
  {\bibfnamefont {Johannes}\ \bibnamefont {Lohmann}}, \bibinfo {author}
  {\bibfnamefont {Valerio}\ \bibnamefont {Lucarini}}, \bibinfo {author}
  {\bibfnamefont {Alice}\ \bibnamefont {Marzocchi}}, \bibinfo {author}
  {\bibfnamefont {Heiko}\ \bibnamefont {P{\"a}like}}, \bibinfo {author}
  {\bibfnamefont {Itzel~Ruvalcaba}\ \bibnamefont {Baroni}}, \bibinfo {author}
  {\bibfnamefont {Dirk}\ \bibnamefont {Simon}}, \bibinfo {author}
  {\bibfnamefont {Appy}\ \bibnamefont {Sluijs}}, \bibinfo {author}
  {\bibfnamefont {Lennert~B.}\ \bibnamefont {Stap}}, \bibinfo {author}
  {\bibfnamefont {Alexis}\ \bibnamefont {Tantet}}, \bibinfo {author}
  {\bibfnamefont {Jan}\ \bibnamefont {Viebahn}}, \ and\ \bibinfo {author}
  {\bibfnamefont {Martin}\ \bibnamefont {Ziegler}}} (\bibinfo {year} {2016}),\
  \bibfield  {title} {\enquote {\bibinfo {title} {Lessons on climate
  sensitivity from past climate changes},}\ }\href {\doibase
  10.1007/s40641-016-0049-3} {\bibfield  {journal} {\bibinfo  {journal}
  {Current Climate Change Reports}\ }\textbf {\bibinfo {volume} {2}}~(\bibinfo
  {number} {4}),\ \bibinfo {pages} {148--158}}\BibitemShut {NoStop}%
\bibitem [{\citenamefont {Hochman}\ \emph {et~al.}(2019)\citenamefont
  {Hochman}, \citenamefont {Alpert}, \citenamefont {Harpaz}, \citenamefont
  {Saaroni},\ and\ \citenamefont {Messori}}]{Hochman2019}%
  \BibitemOpen
  \bibfield  {author} {\bibinfo {author} {\bibnamefont {Hochman}, \bibfnamefont
  {A}}, \bibinfo {author} {\bibfnamefont {P.}~\bibnamefont {Alpert}}, \bibinfo
  {author} {\bibfnamefont {T.}~\bibnamefont {Harpaz}}, \bibinfo {author}
  {\bibfnamefont {H.}~\bibnamefont {Saaroni}}, \ and\ \bibinfo {author}
  {\bibfnamefont {G.}~\bibnamefont {Messori}}} (\bibinfo {year} {2019}),\
  \bibfield  {title} {\enquote {\bibinfo {title} {A new dynamical systems
  perspective on atmospheric predictability: {Eastern Mediterranean weather
  regimes as a case study}},}\ }\href {\doibase 10.1126/sciadv.aau0936}
  {\bibfield  {journal} {\bibinfo  {journal} {Science Advances}\ }\textbf
  {\bibinfo {volume} {5}}~(\bibinfo {number} {6}),\
  10.1126/sciadv.aau0936}\BibitemShut {NoStop}%
\bibitem [{\citenamefont {Hoffman}\ \emph {et~al.}(2002)\citenamefont
  {Hoffman}, \citenamefont {Kaufman}, \citenamefont {Halverson},\ and\
  \citenamefont {Schrag}}]{Hoffman}%
  \BibitemOpen
  \bibfield  {author} {\bibinfo {author} {\bibnamefont {Hoffman}, \bibfnamefont
  {P~F}}, \bibinfo {author} {\bibfnamefont {A.~J.}\ \bibnamefont {Kaufman}},
  \bibinfo {author} {\bibfnamefont {G.~P.}\ \bibnamefont {Halverson}}, \ and\
  \bibinfo {author} {\bibfnamefont {D.~P.}\ \bibnamefont {Schrag}}} (\bibinfo
  {year} {2002}),\ \bibfield  {title} {\enquote {\bibinfo {title} {On the
  initiation of a snowball earth},}\ }\href@noop {} {\bibfield  {journal}
  {\bibinfo  {journal} {Science}\ }\textbf {\bibinfo {volume} {281}},\ \bibinfo
  {pages} {1342}}\BibitemShut {NoStop}%
\bibitem [{\citenamefont {Hoffman}\ and\ \citenamefont
  {Schrag}(2002)}]{HoffmanSchrag}%
  \BibitemOpen
  \bibfield  {author} {\bibinfo {author} {\bibnamefont {Hoffman}, \bibfnamefont
  {P~F}}, \ and\ \bibinfo {author} {\bibfnamefont {D.~P.}\ \bibnamefont
  {Schrag}}} (\bibinfo {year} {2002}),\ \bibfield  {title} {\enquote {\bibinfo
  {title} {The snowball earth hypothesis: testing the limits of global
  change},}\ }\href@noop {} {\bibfield  {journal} {\bibinfo  {journal} {Terra
  Nova}\ }\textbf {\bibinfo {volume} {14}},\ \bibinfo {pages}
  {129}}\BibitemShut {NoStop}%
\bibitem [{\citenamefont {Holland}\ \emph {et~al.}(2012)\citenamefont
  {Holland}, \citenamefont {Vitolo}, \citenamefont {Rabassa}, \citenamefont
  {Sterk},\ and\ \citenamefont {Broer}}]{HVR12}%
  \BibitemOpen
  \bibfield  {author} {\bibinfo {author} {\bibnamefont {Holland}, \bibfnamefont
  {Mark~P}}, \bibinfo {author} {\bibfnamefont {Renato}\ \bibnamefont {Vitolo}},
  \bibinfo {author} {\bibfnamefont {Pau}\ \bibnamefont {Rabassa}}, \bibinfo
  {author} {\bibfnamefont {Alef~E.}\ \bibnamefont {Sterk}}, \ and\ \bibinfo
  {author} {\bibfnamefont {Henk~W.}\ \bibnamefont {Broer}}} (\bibinfo {year}
  {2012}),\ \bibfield  {title} {\enquote {\bibinfo {title} {Extreme value laws
  in dynamical systems under physical observables},}\ }\href {\doibase
  10.1016/j.physd.2011.11.005} {\bibfield  {journal} {\bibinfo  {journal}
  {Physica D: Nonlinear Phenomena}\ }\textbf {\bibinfo {volume}
  {241}}~(\bibinfo {number} {5}),\ \bibinfo {pages} {497--513}}\BibitemShut
  {NoStop}%
\bibitem [{\citenamefont {Holton}\ and\ \citenamefont {Hakim}(2013)}]{Holton}%
  \BibitemOpen
  \bibfield  {author} {\bibinfo {author} {\bibnamefont {Holton}, \bibfnamefont
  {J~R}}, \ and\ \bibinfo {author} {\bibfnamefont {G.~J.}\ \bibnamefont
  {Hakim}}} (\bibinfo {year} {2013}),\ \href@noop {} {\emph {\bibinfo {title}
  {An Introduction to Dynamic Meteorology, 4th ed.}}}\ (\bibinfo  {publisher}
  {Academic Press},\ \bibinfo {address} {San Diego, CA})\BibitemShut {NoStop}%
\bibitem [{\citenamefont {{Hoskins}}\ \emph {et~al.}(1985)\citenamefont
  {{Hoskins}}, \citenamefont {{McIntyre}},\ and\ \citenamefont
  {{Robertson}}}]{Hoskins1985}%
  \BibitemOpen
  \bibfield  {author} {\bibinfo {author} {\bibnamefont {{Hoskins}},
  \bibfnamefont {B~J}}, \bibinfo {author} {\bibfnamefont {M.~E.}\ \bibnamefont
  {{McIntyre}}}, \ and\ \bibinfo {author} {\bibfnamefont {A.~W.}\ \bibnamefont
  {{Robertson}}}} (\bibinfo {year} {1985}),\ \bibfield  {title} {\enquote
  {\bibinfo {title} {{On the use and significance of isentropic potential
  vorticity maps}},}\ }\href {\doibase 10.1002/qj.49711147002} {\bibfield
  {journal} {\bibinfo  {journal} {Quarterly Journal of the Royal Meteorological
  Society}\ }\textbf {\bibinfo {volume} {111}},\ \bibinfo {pages}
  {877--946}}\BibitemShut {NoStop}%
\bibitem [{\citenamefont {Hu}\ \emph {et~al.}(2019)\citenamefont {Hu},
  \citenamefont {B\'odai},\ and\ \citenamefont {Lucarini}}]{Hu2019}%
  \BibitemOpen
  \bibfield  {author} {\bibinfo {author} {\bibnamefont {Hu}, \bibfnamefont
  {G}}, \bibinfo {author} {\bibfnamefont {T.}~\bibnamefont {B\'odai}}, \ and\
  \bibinfo {author} {\bibfnamefont {V.}~\bibnamefont {Lucarini}}} (\bibinfo
  {year} {2019}),\ \bibfield  {title} {\enquote {\bibinfo {title} {Effects of
  stochastic parametrization on extreme value statistics},}\ }\href {\doibase
  10.1063/1.5095756} {\bibfield  {journal} {\bibinfo  {journal} {Chaos: An
  Interdisciplinary Journal of Nonlinear Science}\ }\textbf {\bibinfo {volume}
  {29}}~(\bibinfo {number} {8}),\ \bibinfo {pages} {083102}},\ \Eprint
  {http://arxiv.org/abs/https://doi.org/10.1063/1.5095756}
  {https://doi.org/10.1063/1.5095756} \BibitemShut {NoStop}%
\bibitem [{\citenamefont {Huybers}(2005)}]{Huybers05}%
  \BibitemOpen
  \bibfield  {author} {\bibinfo {author} {\bibnamefont {Huybers}, \bibfnamefont
  {P}}} (\bibinfo {year} {2005}),\ \bibfield  {title} {\enquote {\bibinfo
  {title} {{Comment on "Hockey sticks, principal components, and spurious
  significance'' by S. McIntyre and R. McKitrick}},}\ }\href {\doibase
  10.1029/2005GL023395} {\bibfield  {journal} {\bibinfo  {journal} {Geophys.
  Res. Lett.}\ }\textbf {\bibinfo {volume} {32}}~(\bibinfo {number} {20}),\
  \bibinfo {pages} {{L20705}}}\BibitemShut {NoStop}%
\bibitem [{\citenamefont {Imbrie}\ and\ \citenamefont
  {Imbrie}(1986)}]{Imbrie1986}%
  \BibitemOpen
  \bibfield  {author} {\bibinfo {author} {\bibnamefont {Imbrie}, \bibfnamefont
  {J}}, \ and\ \bibinfo {author} {\bibfnamefont {K.~P.}\ \bibnamefont
  {Imbrie}}} (\bibinfo {year} {1986}),\ \href@noop {} {\emph {\bibinfo {title}
  {Ice Ages: Solving the Mystery}}}\ (\bibinfo  {publisher} {2nd Edn., Harvard
  Univ. Press},\ \bibinfo {address} {Cambridge, Mass.})\BibitemShut {NoStop}%
\bibitem [{\citenamefont {IPCC}(2001)}]{IPCC01}%
  \BibitemOpen
  \bibfield  {author} {\bibinfo {author} {\bibnamefont {IPCC},}} (\bibinfo
  {year} {2001}),\ \href@noop {} {\emph {\bibinfo {title} {{Climate Change
  2001: The Scientific Basis. Contribution of Working Group I to the Third
  Assessment Report of the Intergovernmental Panel on Climate Change}}}},\
  edited by\ \bibinfo {editor} {\bibnamefont {{J. T. Houghton et al.}}}\
  (\bibinfo  {publisher} {Cambridge University Press},\ \bibinfo {address}
  {Cambridge, UK})\BibitemShut {NoStop}%
\bibitem [{\citenamefont {{IPCC}}(2001)}]{ipcc2001}%
  \BibitemOpen
  \bibfield  {author} {\bibinfo {author} {\bibnamefont {{IPCC}},}} (\bibinfo
  {year} {2001}),\ \href@noop {} {\emph {\bibinfo {title} {Third Assessment
  Reportof the Intergovernmental Panel on Climate Change}}},\ edited by\
  \bibinfo {editor} {\bibfnamefont {Houghton}\ \bibnamefont {et~al.}}\
  (\bibinfo  {publisher} {Cambridge University Press})\BibitemShut {NoStop}%
\bibitem [{\citenamefont {IPCC}(2007)}]{IPCC07}%
  \BibitemOpen
  \bibfield  {author} {\bibinfo {author} {\bibnamefont {IPCC},}} (\bibinfo
  {year} {2007}),\ \href@noop {} {\emph {\bibinfo {title} {{Climate Change 2007
  - The Physical Science Basis: Working Group I Contribution to the Fourth
  Assessment Report of the IPCC}}}},\ edited by\ \bibinfo {editor}
  {\bibnamefont {{S. Solomon et al.}}}\ (\bibinfo  {publisher} {Cambridge
  University Press},\ \bibinfo {address} {Cambridge, UK and New York, NY,
  USA})\BibitemShut {NoStop}%
\bibitem [{\citenamefont {{IPCC}}(2012)}]{IPCC12}%
  \BibitemOpen
  \bibfield  {author} {\bibinfo {author} {\bibnamefont {{IPCC}},}} (\bibinfo
  {year} {2012}),\ \href@noop {} {\emph {\bibinfo {title} {{Managing the Risks
  of Extreme Events and Disasters to Advance Climate Change Adaptation. A
  Special Report of Working Groups I and II of the Intergovernmental Panel on
  Climate Change}}}},\ edited by\ \bibinfo {editor} {\bibfnamefont
  {C.B.~Field}\ \bibnamefont {et~al.}}\ (\bibinfo  {publisher} {Cambridge
  University Press},\ \bibinfo {address} {Cambridge, UK, and New York,
  USA})\BibitemShut {NoStop}%
\bibitem [{\citenamefont {IPCC}(2014a)}]{IPCC13}%
  \BibitemOpen
  \bibfield  {author} {\bibinfo {author} {\bibnamefont {IPCC},}} (\bibinfo
  {year} {2014a}),\ \href {\doibase 10.1017/cbo9781107415324} {\emph {\bibinfo
  {title} {{Climate Change 2013: The Physical Science Basis. Contribution of
  Working Group I to the Fifth Assessment Report of the Intergovernmental Panel
  on Climate Change}}}},\ edited by\ \bibinfo {editor} {\bibnamefont {{T.
  Stocker et al.}}}\ (\bibinfo  {publisher} {Cambridge University Press},\
  \bibinfo {address} {Cambridge})\BibitemShut {NoStop}%
\bibitem [{\citenamefont {IPCC}(2014b)}]{IPCC14b}%
  \BibitemOpen
  \bibfield  {author} {\bibinfo {author} {\bibnamefont {IPCC},}} (\bibinfo
  {year} {2014b}),\ \href@noop {} {\emph {\bibinfo {title} {{Climate Change
  2014. Impacts, Adaptation and Vulnerability. Contribution of Working Group II
  to the Fifth Assessment Report of the Intergovernmental Panel on Climate
  Change}}}},\ edited by\ \bibinfo {editor} {\bibnamefont {{C.B. Field et
  al.}}}\ (\bibinfo  {publisher} {Cambridge University Press},\ \bibinfo
  {address} {Cambridge})\BibitemShut {NoStop}%
\bibitem [{\citenamefont {IPCC}(2014c)}]{IPCC14c}%
  \BibitemOpen
  \bibfield  {author} {\bibinfo {author} {\bibnamefont {IPCC},}} (\bibinfo
  {year} {2014c}),\ \href@noop {} {\emph {\bibinfo {title} {{Climate Change
  2014. Mitigation of Climate Change. Contribution of Working Group III to the
  Fifth Assessment Report of the Intergovernmental Panel on Climate
  Change}}}},\ edited by\ \bibinfo {editor} {\bibnamefont {{O. Edenhofer et
  al.}}}\ (\bibinfo  {publisher} {Cambridge University Press},\ \bibinfo
  {address} {Cambridge})\BibitemShut {NoStop}%
\bibitem [{\citenamefont {Itoh}\ and\ \citenamefont
  {Kimoto}(1996)}]{Itoh.Kimoto.1996}%
  \BibitemOpen
  \bibfield  {author} {\bibinfo {author} {\bibnamefont {Itoh}, \bibfnamefont
  {H}}, \ and\ \bibinfo {author} {\bibfnamefont {M.}~\bibnamefont {Kimoto}}}
  (\bibinfo {year} {1996}),\ \bibfield  {title} {\enquote {\bibinfo {title}
  {Multiple attractors and chaotic itinerancy in a quasigeostrophic model with
  realistic topography: Implications for weather regimes and low-frequency
  variability},}\ }\href {\doibase
  10.1175/1520-0469(1996)053<2217:maacii>2.0.co;2} {\bibfield  {journal}
  {\bibinfo  {journal} {{J. Atmos. Sci.}}\ }\textbf {\bibinfo {volume}
  {53}}~(\bibinfo {number} {15}),\ \bibinfo {pages} {2217--2231}}\BibitemShut
  {NoStop}%
\bibitem [{\citenamefont {Itoh}\ and\ \citenamefont
  {Kimoto}(1997)}]{Itoh.Kimoto.1997}%
  \BibitemOpen
  \bibfield  {author} {\bibinfo {author} {\bibnamefont {Itoh}, \bibfnamefont
  {H}}, \ and\ \bibinfo {author} {\bibfnamefont {M.}~\bibnamefont {Kimoto}}}
  (\bibinfo {year} {1997}),\ \bibfield  {title} {\enquote {\bibinfo {title}
  {Chaotic itinerancy with preferred transition routes appearing in an
  atmospheric model},}\ }\href {\doibase 10.1016/s0167-2789(97)00064-x}
  {\bibfield  {journal} {\bibinfo  {journal} {Physica D}\ }\textbf {\bibinfo
  {volume} {109}}~(\bibinfo {number} {3-4}),\ \bibinfo {pages}
  {274--292}}\BibitemShut {NoStop}%
\bibitem [{\citenamefont {Jiang}\ \emph
  {et~al.}(1995{\natexlab{a}})\citenamefont {Jiang}, \citenamefont {Neelin},\
  and\ \citenamefont {Ghil}}]{JiangN1995}%
  \BibitemOpen
  \bibfield  {author} {\bibinfo {author} {\bibnamefont {Jiang}, \bibfnamefont
  {N}}, \bibinfo {author} {\bibfnamefont {J.~D.}\ \bibnamefont {Neelin}}, \
  and\ \bibinfo {author} {\bibfnamefont {M.}~\bibnamefont {Ghil}}} (\bibinfo
  {year} {1995}{\natexlab{a}}),\ \bibfield  {title} {\enquote {\bibinfo {title}
  {{ Quasi-quadrennial and quasi-biennial varibility in the equatorial Pacific.
  }},}\ }\href@noop {} {\bibfield  {journal} {\bibinfo  {journal} {Clim. Dyn.}\
  }\textbf {\bibinfo {volume} {12}},\ \bibinfo {pages} {101--112}}\BibitemShut
  {NoStop}%
\bibitem [{\citenamefont {Jiang}\ \emph
  {et~al.}(1995{\natexlab{b}})\citenamefont {Jiang}, \citenamefont {Jin},\ and\
  \citenamefont {Ghil}}]{Jiang1995}%
  \BibitemOpen
  \bibfield  {author} {\bibinfo {author} {\bibnamefont {Jiang}, \bibfnamefont
  {S}}, \bibinfo {author} {\bibfnamefont {F.-F.}\ \bibnamefont {Jin}}, \ and\
  \bibinfo {author} {\bibfnamefont {M.}~\bibnamefont {Ghil}}} (\bibinfo {year}
  {1995}{\natexlab{b}}),\ \bibfield  {title} {\enquote {\bibinfo {title}
  {Multiple equilibria and aperiodic solutions in a wind-driven double-gyre,
  shallow-water model},}\ }\href@noop {} {\bibfield  {journal} {\bibinfo
  {journal} {J.\ Phys.\ Oceanogr.}\ }\textbf {\bibinfo {volume} {25}},\
  \bibinfo {pages} {764--786}}\BibitemShut {NoStop}%
\bibitem [{\citenamefont {Jin}\ and\ \citenamefont
  {Ghil}(1990)}]{Jin.Ghil.1990}%
  \BibitemOpen
  \bibfield  {author} {\bibinfo {author} {\bibnamefont {Jin}, \bibfnamefont
  {F-F}}, \ and\ \bibinfo {author} {\bibfnamefont {M.}~\bibnamefont {Ghil}}}
  (\bibinfo {year} {1990}),\ \bibfield  {title} {\enquote {\bibinfo {title}
  {Intraseasonal oscillations in the extratropics: {Hopf} bifurcation and
  topographic instabilities},}\ }\href {\doibase
  10.1175/1520-0469(1990)047<3007:ioiteh>2.0.co;2} {\bibfield  {journal}
  {\bibinfo  {journal} {J. Atmos. Sci.}\ }\textbf {\bibinfo {volume}
  {47}}~(\bibinfo {number} {24}),\ \bibinfo {pages} {3007--3022}}\BibitemShut
  {NoStop}%
\bibitem [{\citenamefont {Jin}\ and\ \citenamefont {Neelin}(1993)}]{Jin1993p1}%
  \BibitemOpen
  \bibfield  {author} {\bibinfo {author} {\bibnamefont {Jin}, \bibfnamefont
  {F-F}}, \ and\ \bibinfo {author} {\bibfnamefont {J.~D.}\ \bibnamefont
  {Neelin}}} (\bibinfo {year} {1993}),\ \bibfield  {title} {\enquote {\bibinfo
  {title} {{Modes of interannual tropical ocean-atmosphere interaction - a
  unified view. I: Numerical results.}}}\ }\href@noop {} {\bibfield  {journal}
  {\bibinfo  {journal} {J.\ Atmos.\ Sci.}\ }\textbf {\bibinfo {volume} {50}},\
  \bibinfo {pages} {3477--3503}}\BibitemShut {NoStop}%
\bibitem [{\citenamefont {Jin}\ \emph {et~al.}(1994)\citenamefont {Jin},
  \citenamefont {Neelin},\ and\ \citenamefont {Ghil}}]{Jin1994}%
  \BibitemOpen
  \bibfield  {author} {\bibinfo {author} {\bibnamefont {Jin}, \bibfnamefont
  {F-F}}, \bibinfo {author} {\bibfnamefont {J.~D.}\ \bibnamefont {Neelin}}, \
  and\ \bibinfo {author} {\bibfnamefont {M.}~\bibnamefont {Ghil}}} (\bibinfo
  {year} {1994}),\ \bibfield  {title} {\enquote {\bibinfo {title} {{El Ni\~no
  on the devil's staircase: Annual subharmonic steps to chaos}},}\ }\href@noop
  {} {\bibfield  {journal} {\bibinfo  {journal} {Science}\ }\textbf {\bibinfo
  {volume} {264}},\ \bibinfo {pages} {70--72}}\BibitemShut {NoStop}%
\bibitem [{\citenamefont {Jin}\ \emph {et~al.}(1996)\citenamefont {Jin},
  \citenamefont {Neelin},\ and\ \citenamefont {Ghil}}]{Jin1996a}%
  \BibitemOpen
  \bibfield  {author} {\bibinfo {author} {\bibnamefont {Jin}, \bibfnamefont
  {F-F}}, \bibinfo {author} {\bibfnamefont {J.~D.}\ \bibnamefont {Neelin}}, \
  and\ \bibinfo {author} {\bibfnamefont {M.}~\bibnamefont {Ghil}}} (\bibinfo
  {year} {1996}),\ \bibfield  {title} {\enquote {\bibinfo {title} {{El
  Ni\~no/Southern Oscillation and the annual cycle: Subharmonic
  frequency-locking and aperiodicity}},}\ }\href@noop {} {\bibfield  {journal}
  {\bibinfo  {journal} {Physica D}\ }\textbf {\bibinfo {volume} {98}},\
  \bibinfo {pages} {442--465}}\BibitemShut {NoStop}%
\bibitem [{\citenamefont {Jones}(1995)}]{Jones.1995}%
  \BibitemOpen
  \bibfield  {author} {\bibinfo {author} {\bibnamefont {Jones}, \bibfnamefont
  {C~K R~T}}} (\bibinfo {year} {1995}),\ \bibfield  {title} {\enquote {\bibinfo
  {title} {{Geometric singular perturbation theory}},}\ }in\ \href@noop {}
  {\emph {\bibinfo {booktitle} {Dynamical Systems}}},\ \bibinfo {series}
  {Lecture Notes in Mathematics}, Vol.\ \bibinfo {volume} {1609},\ \bibinfo
  {editor} {edited by\ \bibinfo {editor} {\bibfnamefont {R.}~\bibnamefont
  {Johnson}}}\ (\bibinfo  {publisher} {Springer})\ pp.\ \bibinfo {pages}
  {44--118}\BibitemShut {NoStop}%
\bibitem [{\citenamefont {Jouzel}\ \emph {et~al.}(1991)\citenamefont {Jouzel},
  \citenamefont {Barkov}, \citenamefont {Barnola}, \citenamefont {Bender},
  \citenamefont {Chappellaz}, \citenamefont {Genthon}, \citenamefont
  {Kotlyakov}, \citenamefont {Lipenkov}, \citenamefont {Lorius}, \citenamefont
  {Petit}, \citenamefont {Raynaud}, \citenamefont {Raisbeck}, \citenamefont
  {Ritz}, \citenamefont {Sowers}, \citenamefont {Stievenard}, \citenamefont
  {Yiou},\ and\ \citenamefont {Yiou}}]{Jouzel1993}%
  \BibitemOpen
  \bibfield  {author} {\bibinfo {author} {\bibnamefont {Jouzel}, \bibfnamefont
  {J}}, \bibinfo {author} {\bibfnamefont {N.~I.}\ \bibnamefont {Barkov}},
  \bibinfo {author} {\bibfnamefont {J.~M.}\ \bibnamefont {Barnola}}, \bibinfo
  {author} {\bibfnamefont {M.}~\bibnamefont {Bender}}, \bibinfo {author}
  {\bibfnamefont {J.}~\bibnamefont {Chappellaz}}, \bibinfo {author}
  {\bibfnamefont {C.}~\bibnamefont {Genthon}}, \bibinfo {author} {\bibfnamefont
  {V.~M.}\ \bibnamefont {Kotlyakov}}, \bibinfo {author} {\bibfnamefont
  {V.}~\bibnamefont {Lipenkov}}, \bibinfo {author} {\bibfnamefont
  {C.}~\bibnamefont {Lorius}}, \bibinfo {author} {\bibfnamefont {J.~R.}\
  \bibnamefont {Petit}}, \bibinfo {author} {\bibfnamefont {D.}~\bibnamefont
  {Raynaud}}, \bibinfo {author} {\bibfnamefont {G.}~\bibnamefont {Raisbeck}},
  \bibinfo {author} {\bibfnamefont {C.}~\bibnamefont {Ritz}}, \bibinfo {author}
  {\bibfnamefont {T.}~\bibnamefont {Sowers}}, \bibinfo {author} {\bibfnamefont
  {M.}~\bibnamefont {Stievenard}}, \bibinfo {author} {\bibfnamefont
  {F.}~\bibnamefont {Yiou}}, \ and\ \bibinfo {author} {\bibfnamefont
  {P.}~\bibnamefont {Yiou}}} (\bibinfo {year} {1991}),\ \bibfield  {title}
  {\enquote {\bibinfo {title} {{Extending the Vostok ice-core record of
  paleoclimate to the penultimate glacial period.}}}\ }\href@noop {} {\bibfield
   {journal} {\bibinfo  {journal} {Nature}\ }\textbf {\bibinfo {volume}
  {364}},\ \bibinfo {pages} {407--412}}\BibitemShut {NoStop}%
\bibitem [{\citenamefont {Just}\ \emph {et~al.}(2001)\citenamefont {Just},
  \citenamefont {Kantz}, \citenamefont {R{\"o}denbeck},\ and\ \citenamefont
  {Helm}}]{Kantz.ea.2001}%
  \BibitemOpen
  \bibfield  {author} {\bibinfo {author} {\bibnamefont {Just}, \bibfnamefont
  {W}}, \bibinfo {author} {\bibfnamefont {H.}~\bibnamefont {Kantz}}, \bibinfo
  {author} {\bibfnamefont {C.}~\bibnamefont {R{\"o}denbeck}}, \ and\ \bibinfo
  {author} {\bibfnamefont {M.}~\bibnamefont {Helm}}} (\bibinfo {year} {2001}),\
  \bibfield  {title} {\enquote {\bibinfo {title} {Stochastic modelling:
  replacing fast degrees of freedom by noise},}\ }\href@noop {} {\bibfield
  {journal} {\bibinfo  {journal} {{Journal of Physics A: Mathematical and
  General}}\ }\textbf {\bibinfo {volume} {34}}~(\bibinfo {number} {15}),\
  \bibinfo {pages} {3199--3213}}\BibitemShut {NoStop}%
\bibitem [{\citenamefont {Kalnay}(2003)}]{kalnay2003}%
  \BibitemOpen
  \bibfield  {author} {\bibinfo {author} {\bibnamefont {Kalnay}, \bibfnamefont
  {E}}} (\bibinfo {year} {2003}),\ \href@noop {} {\emph {\bibinfo {title}
  {Atmospheric {Modeling, Data Assimilation and Predictability}}}}\ (\bibinfo
  {publisher} {Cambridge University Press},\ \bibinfo {address} {Cambridge,
  UK})\BibitemShut {NoStop}%
\bibitem [{\citenamefont {Karamperidou}\ and\ \citenamefont
  {Coauthors}(2015)}]{Kara.ea.15}%
  \BibitemOpen
  \bibfield  {author} {\bibinfo {author} {\bibnamefont {Karamperidou},
  \bibfnamefont {C}}, \ and\ \bibinfo {author} {\bibnamefont {Coauthors}}}
  (\bibinfo {year} {2015}),\ \bibfield  {title} {\enquote {\bibinfo {title}
  {The response of {ENSO flavors to mid-Holocene climate: Implications for
  proxy interpretation}},}\ }\href@noop {} {\bibfield  {journal} {\bibinfo
  {journal} {Paleoceanography}\ }\textbf {\bibinfo {volume} {30}},\ \bibinfo
  {pages} {527--547}}\BibitemShut {NoStop}%
\bibitem [{\citenamefont {Karspeck}\ \emph {et~al.}(2018)\citenamefont
  {Karspeck}, \citenamefont {Danabasoglu}, \citenamefont {Anderson},
  \citenamefont {Karol}, \citenamefont {Collins}, \citenamefont {Vertenstein},
  \citenamefont {Raeder}, \citenamefont {Hoar}, \citenamefont {Neale},
  \citenamefont {Edwards},\ and\ \citenamefont {Craig}}]{Karspeck2018}%
  \BibitemOpen
  \bibfield  {author} {\bibinfo {author} {\bibnamefont {Karspeck},
  \bibfnamefont {Alicia~R}}, \bibinfo {author} {\bibfnamefont {Gokhan}\
  \bibnamefont {Danabasoglu}}, \bibinfo {author} {\bibfnamefont {Jeffrey}\
  \bibnamefont {Anderson}}, \bibinfo {author} {\bibfnamefont {Svetlana}\
  \bibnamefont {Karol}}, \bibinfo {author} {\bibfnamefont {Nancy}\ \bibnamefont
  {Collins}}, \bibinfo {author} {\bibfnamefont {Mariana}\ \bibnamefont
  {Vertenstein}}, \bibinfo {author} {\bibfnamefont {Kevin}\ \bibnamefont
  {Raeder}}, \bibinfo {author} {\bibfnamefont {Tim}\ \bibnamefont {Hoar}},
  \bibinfo {author} {\bibfnamefont {Richard}\ \bibnamefont {Neale}}, \bibinfo
  {author} {\bibfnamefont {Jim}\ \bibnamefont {Edwards}}, \ and\ \bibinfo
  {author} {\bibfnamefont {Anthony}\ \bibnamefont {Craig}}} (\bibinfo {year}
  {2018}),\ \bibfield  {title} {\enquote {\bibinfo {title} {A global coupled
  ensemble data assimilation system using the community earth system model and
  the data assimilation research testbed},}\ }\href {\doibase 10.1002/qj.3308}
  {\bibfield  {journal} {\bibinfo  {journal} {Quarterly Journal of the Royal
  Meteorological Society}\ }\textbf {\bibinfo {volume} {144}}~(\bibinfo
  {number} {717}),\ \bibinfo {pages} {2404--2430}}\BibitemShut {NoStop}%
\bibitem [{\citenamefont {Katsman}\ \emph {et~al.}(1998)\citenamefont
  {Katsman}, \citenamefont {Dijkstra},\ and\ \citenamefont
  {Drijfhout}}]{Katsman1998}%
  \BibitemOpen
  \bibfield  {author} {\bibinfo {author} {\bibnamefont {Katsman}, \bibfnamefont
  {C~A}}, \bibinfo {author} {\bibfnamefont {H.~A.}\ \bibnamefont {Dijkstra}}, \
  and\ \bibinfo {author} {\bibfnamefont {S.~S.}\ \bibnamefont {Drijfhout}}}
  (\bibinfo {year} {1998}),\ \bibfield  {title} {\enquote {\bibinfo {title}
  {The rectification of the wind-driven circulation due to its
  instabilities},}\ }\href@noop {} {\bibfield  {journal} {\bibinfo  {journal}
  {J.\ Mar.\ Res.}\ }\textbf {\bibinfo {volume} {56}},\ \bibinfo {pages}
  {559--587}}\BibitemShut {NoStop}%
\bibitem [{\citenamefont {Katz}\ \emph {et~al.}(2002)\citenamefont {Katz},
  \citenamefont {Parlange},\ and\ \citenamefont {Naveau}}]{KPN02}%
  \BibitemOpen
  \bibfield  {author} {\bibinfo {author} {\bibnamefont {Katz}, \bibfnamefont
  {RW}}, \bibinfo {author} {\bibfnamefont {M.B.}\ \bibnamefont {Parlange}}, \
  and\ \bibinfo {author} {\bibfnamefont {P.}~\bibnamefont {Naveau}}} (\bibinfo
  {year} {2002}),\ \bibfield  {title} {\enquote {\bibinfo {title} {Statistics
  of extremes in hydrology},}\ }\href@noop {} {\bibfield  {journal} {\bibinfo
  {journal} {Advances in Water Resources}\ }\textbf {\bibinfo {volume} {25}},\
  \bibinfo {pages} {1287--1304}}\BibitemShut {NoStop}%
\bibitem [{\citenamefont {Kautz}(1987)}]{Kautz1987}%
  \BibitemOpen
  \bibfield  {author} {\bibinfo {author} {\bibnamefont {Kautz}, \bibfnamefont
  {RL}}} (\bibinfo {year} {1987}),\ \bibfield  {title} {\enquote {\bibinfo
  {title} {Activation energy for thermally induced escape from a basin of
  attraction},}\ }\href {\doibase 10.1016/0375-9601(87)90151-4} {\bibfield
  {journal} {\bibinfo  {journal} {Physics Letters A}\ }\textbf {\bibinfo
  {volume} {125}}~(\bibinfo {number} {6}),\ \bibinfo {pages}
  {315--319}}\BibitemShut {NoStop}%
\bibitem [{\citenamefont {Kennett}\ and\ \citenamefont
  {Stott}(1991)}]{Kennett.Stott.1991}%
  \BibitemOpen
  \bibfield  {author} {\bibinfo {author} {\bibnamefont {Kennett}, \bibfnamefont
  {J~P}}, \ and\ \bibinfo {author} {\bibfnamefont {L.~D.}\ \bibnamefont
  {Stott}}} (\bibinfo {year} {1991}),\ \bibfield  {title} {\enquote {\bibinfo
  {title} {Abrupt deep-sea warming, palaeoceanographic changes and benthic
  extinctions at the end of the {Palaeocene}},}\ }\href@noop {} {\bibfield
  {journal} {\bibinfo  {journal} {Nature}\ }\textbf {\bibinfo {volume} {353}},\
  \bibinfo {pages} {225--229}}\BibitemShut {NoStop}%
\bibitem [{\citenamefont {Kharin}\ \emph {et~al.}(2005)\citenamefont {Kharin},
  \citenamefont {Zwiers},\ and\ \citenamefont {Zhang}}]{KZZ05}%
  \BibitemOpen
  \bibfield  {author} {\bibinfo {author} {\bibnamefont {Kharin}, \bibfnamefont
  {{VV}}}, \bibinfo {author} {\bibfnamefont {{F.W. }}\ \bibnamefont {Zwiers}},
  \ and\ \bibinfo {author} {\bibfnamefont {X.}~\bibnamefont {Zhang}}} (\bibinfo
  {year} {2005}),\ \bibfield  {title} {\enquote {\bibinfo {title}
  {Intercomparison of near surface temperature and precipitation extremes in
  {AMIP-2} simulations, reanalyses and observations},}\ }\href@noop {}
  {\bibfield  {journal} {\bibinfo  {journal} {Journal of Climate}\ }\textbf
  {\bibinfo {volume} {18}},\ \bibinfo {pages} {5201--5223}}\BibitemShut
  {NoStop}%
\bibitem [{\citenamefont {Kim}\ and\ \citenamefont {Kim}(2013)}]{Kim13}%
  \BibitemOpen
  \bibfield  {author} {\bibinfo {author} {\bibnamefont {Kim}, \bibfnamefont
  {Y-H}}, \ and\ \bibinfo {author} {\bibfnamefont {M-H}\ \bibnamefont {Kim}}}
  (\bibinfo {year} {2013}),\ \bibfield  {title} {\enquote {\bibinfo {title}
  {Examination of the global {L}orenz energy cycle using {MERRA} and
  {NCEP}-reanalysis 2},}\ }\href@noop {} {\bibfield  {journal} {\bibinfo
  {journal} {Clim. Dyn.}\ }\textbf {\bibinfo {volume} {40}},\ \bibinfo {pages}
  {1499--1513}}\BibitemShut {NoStop}%
\bibitem [{\citenamefont {Kimoto}\ and\ \citenamefont
  {Ghil}({1993a})}]{Kimoto1993a}%
  \BibitemOpen
  \bibfield  {author} {\bibinfo {author} {\bibnamefont {Kimoto}, \bibfnamefont
  {M}}, \ and\ \bibinfo {author} {\bibfnamefont {M.}~\bibnamefont {Ghil}}}
  (\bibinfo {year} {{1993a}}),\ \bibfield  {title} {\enquote {\bibinfo {title}
  {{Multiple flow regimes in the Northern Hemisphere winter. Part I:
  Methodology and hemispheric regimes}},}\ }\href@noop {} {\bibfield  {journal}
  {\bibinfo  {journal} {{J. Atmos. Sci.}}\ }\textbf {\bibinfo {volume} {50}},\
  \bibinfo {pages} {2625--2643}}\BibitemShut {NoStop}%
\bibitem [{\citenamefont {Kimoto}\ and\ \citenamefont
  {Ghil}({1993b})}]{Kimoto1993b}%
  \BibitemOpen
  \bibfield  {author} {\bibinfo {author} {\bibnamefont {Kimoto}, \bibfnamefont
  {M}}, \ and\ \bibinfo {author} {\bibfnamefont {M.}~\bibnamefont {Ghil}}}
  (\bibinfo {year} {{1993b}}),\ \bibfield  {title} {\enquote {\bibinfo {title}
  {{Multiple flow regimes in the Northern Hemisphere winter. Part II: Sectorial
  regimes and preferred transitions}},}\ }\href@noop {} {\bibfield  {journal}
  {\bibinfo  {journal} {J.\ Atmos.\ Sci.}\ }\textbf {\bibinfo {volume} {50}},\
  \bibinfo {pages} {2645--2673}}\BibitemShut {NoStop}%
\bibitem [{\citenamefont {Kistler}\ and\ \citenamefont
  {Coauthors}(2001)}]{Kistler01}%
  \BibitemOpen
  \bibfield  {author} {\bibinfo {author} {\bibnamefont {Kistler}, \bibfnamefont
  {R}}, \ and\ \bibinfo {author} {\bibnamefont {Coauthors}}} (\bibinfo {year}
  {2001}),\ \bibfield  {title} {\enquote {\bibinfo {title} {{The NCEP-NCAR
  50-year reanalysis: Monthly means CD-ROM and documentation}},}\ }\href@noop
  {} {\bibfield  {journal} {\bibinfo  {journal} {Bull. Amer. Meteorol. Soc.}\
  }\textbf {\bibinfo {volume} {82}},\ \bibinfo {pages} {247--267}}\BibitemShut
  {NoStop}%
\bibitem [{\citenamefont {Kleidon}(2009)}]{Kleidon09}%
  \BibitemOpen
  \bibfield  {author} {\bibinfo {author} {\bibnamefont {Kleidon}, \bibfnamefont
  {A}}} (\bibinfo {year} {2009}),\ \bibfield  {title} {\enquote {\bibinfo
  {title} {Non-equilibrium thermodynamics and maximum entropy production},}\
  }\href@noop {} {\bibfield  {journal} {\bibinfo  {journal}
  {Naturwissenschaften}\ }\textbf {\bibinfo {volume} {96}},\ \bibinfo {pages}
  {653--677}}\BibitemShut {NoStop}%
\bibitem [{\citenamefont {Kleidon}(2010)}]{Kleidon10}%
  \BibitemOpen
  \bibfield  {author} {\bibinfo {author} {\bibnamefont {Kleidon}, \bibfnamefont
  {A}}} (\bibinfo {year} {2010}),\ \bibfield  {title} {\enquote {\bibinfo
  {title} {Life, hierarchy, and the thermodynamic machinery of planet
  {Earth}},}\ }\href {\doibase 10.1016/j.plrev.2010.10.002} {\bibfield
  {journal} {\bibinfo  {journal} {Physics Life Rev.}\ }\textbf {\bibinfo
  {volume} {7}}~(\bibinfo {number} {4}),\ \bibinfo {pages}
  {424--460}}\BibitemShut {NoStop}%
\bibitem [{\citenamefont {Kleidon}\ and\ \citenamefont
  {Lorenz}(2005)}]{Kleidon05}%
  \BibitemOpen
  \bibinfo {editor} {\bibnamefont {Kleidon}, \bibfnamefont {A}}, \ and\
  \bibinfo {editor} {\bibfnamefont {R.}~\bibnamefont {Lorenz}},\ Eds. (\bibinfo
  {year} {2005}),\ \href@noop {} {\emph {\bibinfo {title} {Non-equilibrium
  {Thermodynamics and the Production of Entropy}}}}\ (\bibinfo  {publisher}
  {Springer},\ \bibinfo {address} {Berlin})\BibitemShut {NoStop}%
\bibitem [{\citenamefont {Klein}(2010)}]{klein2010}%
  \BibitemOpen
  \bibfield  {author} {\bibinfo {author} {\bibnamefont {Klein}, \bibfnamefont
  {R}}} (\bibinfo {year} {2010}),\ \bibfield  {title} {\enquote {\bibinfo
  {title} {Scale-dependent models for atmospheric flows},}\ }\href {\doibase
  10.1146/annurev-fluid-121108-145537} {\bibfield  {journal} {\bibinfo
  {journal} {Annu. Rev. Fluid Mech.}\ }\textbf {\bibinfo {volume}
  {42}}~(\bibinfo {number} {1}),\ \bibinfo {pages} {249--274}}\BibitemShut
  {NoStop}%
\bibitem [{\citenamefont {Kloeden}\ and\ \citenamefont
  {Rasmussen}(2011)}]{KR11}%
  \BibitemOpen
  \bibfield  {author} {\bibinfo {author} {\bibnamefont {Kloeden}, \bibfnamefont
  {P~E}}, \ and\ \bibinfo {author} {\bibfnamefont {M.}~\bibnamefont
  {Rasmussen}}} (\bibinfo {year} {2011}),\ \href@noop {} {\emph {\bibinfo
  {title} {{Nonautonomous Dynamical Systems}}}}\ (\bibinfo  {publisher}
  {American Mathematical Society})\BibitemShut {NoStop}%
\bibitem [{\citenamefont {Kondrashov}\ \emph {et~al.}(2015)\citenamefont
  {Kondrashov}, \citenamefont {Chekroun},\ and\ \citenamefont
  {Ghil}}]{MSM2015}%
  \BibitemOpen
  \bibfield  {author} {\bibinfo {author} {\bibnamefont {Kondrashov},
  \bibfnamefont {D}}, \bibinfo {author} {\bibfnamefont {M.~D.}\ \bibnamefont
  {Chekroun}}, \ and\ \bibinfo {author} {\bibfnamefont {M.}~\bibnamefont
  {Ghil}}} (\bibinfo {year} {2015}),\ \bibfield  {title} {\enquote {\bibinfo
  {title} {Data-driven non-{Markovian closure models}},}\ }\href {\doibase
  10.1016/j.physd.2014.12.005} {\bibfield  {journal} {\bibinfo  {journal}
  {Physica D}\ }\textbf {\bibinfo {volume} {297}},\ \bibinfo {pages}
  {33--55}}\BibitemShut {NoStop}%
\bibitem [{\citenamefont {Kondrashov}\ \emph {et~al.}(2013)\citenamefont
  {Kondrashov}, \citenamefont {Chekroun}, \citenamefont {Robertson},\ and\
  \citenamefont {Ghil}}]{Kondras.MJO.2013}%
  \BibitemOpen
  \bibfield  {author} {\bibinfo {author} {\bibnamefont {Kondrashov},
  \bibfnamefont {D}}, \bibinfo {author} {\bibfnamefont {M.~D.}\ \bibnamefont
  {Chekroun}}, \bibinfo {author} {\bibfnamefont {A.~W.}\ \bibnamefont
  {Robertson}}, \ and\ \bibinfo {author} {\bibfnamefont {M.}~\bibnamefont
  {Ghil}}} (\bibinfo {year} {2013}),\ \bibfield  {title} {\enquote {\bibinfo
  {title} {Low-order stochastic model and "past-noise forecasting" of the
  {Madden-Julian Oscillation}},}\ }\href@noop {} {\bibfield  {journal}
  {\bibinfo  {journal} {Geophys. Res. Lett.}\ }\textbf {\bibinfo {volume}
  {40}},\ \bibinfo {pages} {5305--5310}}\BibitemShut {NoStop}%
\bibitem [{\citenamefont {Kondrashov}\ \emph
  {et~al.}(2005{\natexlab{a}})\citenamefont {Kondrashov}, \citenamefont
  {Feliks},\ and\ \citenamefont {Ghil}}]{Nile2005}%
  \BibitemOpen
  \bibfield  {author} {\bibinfo {author} {\bibnamefont {Kondrashov},
  \bibfnamefont {D}}, \bibinfo {author} {\bibfnamefont {Y.}~\bibnamefont
  {Feliks}}, \ and\ \bibinfo {author} {\bibfnamefont {M.}~\bibnamefont {Ghil}}}
  (\bibinfo {year} {2005}{\natexlab{a}}),\ \bibfield  {title} {\enquote
  {\bibinfo {title} {Oscillatory modes of extended {Nile River records (A.D.
  622--1922)}},}\ }\href {\doibase 10.1029/2004GL022156} {\bibfield  {journal}
  {\bibinfo  {journal} {Geophysical Research Letters}\ }\textbf {\bibinfo
  {volume} {32}},\ \bibinfo {pages} {L10702}}\BibitemShut {NoStop}%
\bibitem [{\citenamefont {Kondrashov}\ \emph {et~al.}(2004)\citenamefont
  {Kondrashov}, \citenamefont {Ide},\ and\ \citenamefont
  {Ghil}}]{Kondras.ea.2004}%
  \BibitemOpen
  \bibfield  {author} {\bibinfo {author} {\bibnamefont {Kondrashov},
  \bibfnamefont {D}}, \bibinfo {author} {\bibfnamefont {K.}~\bibnamefont
  {Ide}}, \ and\ \bibinfo {author} {\bibfnamefont {M.}~\bibnamefont {Ghil}}}
  (\bibinfo {year} {2004}),\ \bibfield  {title} {\enquote {\bibinfo {title}
  {Weather regimes and preferred transition paths in a three-level
  quasigeostrophic model},}\ }\href {\doibase
  10.1175/1520-0469(2004)061<0568:wraptp>2.0.co;2} {\bibfield  {journal}
  {\bibinfo  {journal} {J. Atmos. Sci.}\ }\textbf {\bibinfo {volume}
  {61}}~(\bibinfo {number} {5}),\ \bibinfo {pages} {568--587}}\BibitemShut
  {NoStop}%
\bibitem [{\citenamefont {Kondrashov}\ \emph {et~al.}(2006)\citenamefont
  {Kondrashov}, \citenamefont {Kravtsov},\ and\ \citenamefont
  {Ghil}}]{Kondrashov.Kravtsov.ea.2006}%
  \BibitemOpen
  \bibfield  {author} {\bibinfo {author} {\bibnamefont {Kondrashov},
  \bibfnamefont {D}}, \bibinfo {author} {\bibfnamefont {S.}~\bibnamefont
  {Kravtsov}}, \ and\ \bibinfo {author} {\bibfnamefont {M.}~\bibnamefont
  {Ghil}}} (\bibinfo {year} {2006}),\ \bibfield  {title} {\enquote {\bibinfo
  {title} {Empirical mode reduction in a model of extratropical low-frequency
  variability},}\ }\href {\doibase 10.1175/jas3719.1} {\bibfield  {journal}
  {\bibinfo  {journal} {{J. Atmos. Sci.}}\ }\textbf {\bibinfo {volume}
  {63}}~(\bibinfo {number} {7}),\ \bibinfo {pages} {1859--1877}}\BibitemShut
  {NoStop}%
\bibitem [{\citenamefont {Kondrashov}\ \emph
  {et~al.}(2005{\natexlab{b}})\citenamefont {Kondrashov}, \citenamefont
  {Kravtsov}, \citenamefont {Robertson},\ and\ \citenamefont
  {Ghil}}]{Kondrashov.Kravtsov.ea.2005}%
  \BibitemOpen
  \bibfield  {author} {\bibinfo {author} {\bibnamefont {Kondrashov},
  \bibfnamefont {D}}, \bibinfo {author} {\bibfnamefont {S.}~\bibnamefont
  {Kravtsov}}, \bibinfo {author} {\bibfnamefont {A.~W.}\ \bibnamefont
  {Robertson}}, \ and\ \bibinfo {author} {\bibfnamefont {M.}~\bibnamefont
  {Ghil}}} (\bibinfo {year} {2005}{\natexlab{b}}),\ \bibfield  {title}
  {\enquote {\bibinfo {title} {A hierarchy of data-based {ENSO} models},}\
  }\href {\doibase 10.1175/jcli3567.1} {\bibfield  {journal} {\bibinfo
  {journal} {J. Climate}\ }\textbf {\bibinfo {volume} {18}}~(\bibinfo {number}
  {21}),\ \bibinfo {pages} {4425--4444}}\BibitemShut {NoStop}%
\bibitem [{\citenamefont {Kondrashov}\ \emph {et~al.}(2007)\citenamefont
  {Kondrashov}, \citenamefont {Shen}, \citenamefont {Berk}, \citenamefont
  {DÃ¢ÂÂAndrea},\ and\ \citenamefont {Ghil}}]{Kondrashov.ea.2007}%
  \BibitemOpen
  \bibfield  {author} {\bibinfo {author} {\bibnamefont {Kondrashov},
  \bibfnamefont {D}}, \bibinfo {author} {\bibfnamefont {J.}~\bibnamefont
  {Shen}}, \bibinfo {author} {\bibfnamefont {R.}~\bibnamefont {Berk}}, \bibinfo
  {author} {\bibfnamefont {F.}~\bibnamefont {DÃ¢ÂÂAndrea}}, \ and\
  \bibinfo {author} {\bibfnamefont {M.}~\bibnamefont {Ghil}}} (\bibinfo {year}
  {2007}),\ \bibfield  {title} {\enquote {\bibinfo {title} {Predicting weather
  regime transitions in northern hemisphere datasets},}\ }\href@noop {}
  {\bibfield  {journal} {\bibinfo  {journal} {Climate dynamics}\ }\textbf
  {\bibinfo {volume} {29}}~(\bibinfo {number} {5}),\ \bibinfo {pages}
  {535--551}}\BibitemShut {NoStop}%
\bibitem [{\citenamefont {Kramers}(1940)}]{Kramers1940}%
  \BibitemOpen
  \bibfield  {author} {\bibinfo {author} {\bibnamefont {Kramers}, \bibfnamefont
  {HA}}} (\bibinfo {year} {1940}),\ \bibfield  {title} {\enquote {\bibinfo
  {title} {Brownian motion in a field of force and the diffusion model of
  chemical reactions},}\ }\href {\doibase 10.1016/S0031-8914(40)90098-2}
  {\bibfield  {journal} {\bibinfo  {journal} {Physica}\ }\textbf {\bibinfo
  {volume} {7}}~(\bibinfo {number} {4}),\ \bibinfo {pages}
  {284--304}}\BibitemShut {NoStop}%
\bibitem [{\citenamefont {Kraut}\ and\ \citenamefont
  {Feudel}(2002)}]{Kraut2002}%
  \BibitemOpen
  \bibfield  {author} {\bibinfo {author} {\bibnamefont {Kraut}, \bibfnamefont
  {S}}, \ and\ \bibinfo {author} {\bibfnamefont {U.}~\bibnamefont {Feudel}}}
  (\bibinfo {year} {2002}),\ \bibfield  {title} {\enquote {\bibinfo {title}
  {Multistability, noise, and attractor hopping: The crucial role of chaotic
  saddles},}\ }\href {\doibase 10.1103/PhysRevE.66.015207} {\bibfield
  {journal} {\bibinfo  {journal} {Phys. Rev. E}\ }\textbf {\bibinfo {volume}
  {66}},\ \bibinfo {pages} {015207}}\BibitemShut {NoStop}%
\bibitem [{\citenamefont {Kravtsov}\ \emph {et~al.}(2007)\citenamefont
  {Kravtsov}, \citenamefont {Berloff}, \citenamefont {Dewar}, \citenamefont
  {Ghil},\ and\ \citenamefont {McWilliams}}]{Kravtsov.ea.2007}%
  \BibitemOpen
  \bibfield  {author} {\bibinfo {author} {\bibnamefont {Kravtsov},
  \bibfnamefont {S}}, \bibinfo {author} {\bibfnamefont {P.}~\bibnamefont
  {Berloff}}, \bibinfo {author} {\bibfnamefont {W.~K.}\ \bibnamefont {Dewar}},
  \bibinfo {author} {\bibfnamefont {M.}~\bibnamefont {Ghil}}, \ and\ \bibinfo
  {author} {\bibfnamefont {J.~C.}\ \bibnamefont {McWilliams}}} (\bibinfo {year}
  {2007}),\ \bibfield  {title} {\enquote {\bibinfo {title} {Dynamical origin of
  low-frequency variability in a highly nonlinear mid-latitude coupled
  model},}\ }\href@noop {} {\bibfield  {journal} {\bibinfo  {journal} {J.
  Climate}\ }\textbf {\bibinfo {volume} {19}},\ \bibinfo {pages}
  {6391--6408}}\BibitemShut {NoStop}%
\bibitem [{\citenamefont {Kravtsov}\ \emph {et~al.}(2005)\citenamefont
  {Kravtsov}, \citenamefont {Kondrashov},\ and\ \citenamefont
  {Ghil}}]{KravtsovKondrashovGhil_JCL05}%
  \BibitemOpen
  \bibfield  {author} {\bibinfo {author} {\bibnamefont {Kravtsov},
  \bibfnamefont {S}}, \bibinfo {author} {\bibfnamefont {D.}~\bibnamefont
  {Kondrashov}}, \ and\ \bibinfo {author} {\bibfnamefont {M.}~\bibnamefont
  {Ghil}}} (\bibinfo {year} {2005}),\ \bibfield  {title} {\enquote {\bibinfo
  {title} {Multi-level regression modeling of nonlinear processes: {Derivation
  and applications to climatic variability}},}\ }\href {\doibase
  10.1175/JCLI3544.1} {\bibfield  {journal} {\bibinfo  {journal} {J. Climate}\
  }\textbf {\bibinfo {volume} {18}}~(\bibinfo {number} {21}),\ \bibinfo {pages}
  {4404--4424}}\BibitemShut {NoStop}%
\bibitem [{\citenamefont {Kravtsov}\ \emph {et~al.}(2009)\citenamefont
  {Kravtsov}, \citenamefont {Kondrashov},\ and\ \citenamefont
  {Ghil}}]{KravtsovGhilKondrashov_09}%
  \BibitemOpen
  \bibfield  {author} {\bibinfo {author} {\bibnamefont {Kravtsov},
  \bibfnamefont {S}}, \bibinfo {author} {\bibfnamefont {D.}~\bibnamefont
  {Kondrashov}}, \ and\ \bibinfo {author} {\bibfnamefont {M.}~\bibnamefont
  {Ghil}}} (\bibinfo {year} {2009}),\ \bibfield  {title} {\enquote {\bibinfo
  {title} {Empirical model reduction and the modeling hierarchy in climate
  dynamics and the geosciences},}\ }in\ \href@noop {} {\emph {\bibinfo
  {booktitle} {Stochastic Physics and Climate Modeling}}},\ \bibinfo {editor}
  {edited by\ \bibinfo {editor} {\bibfnamefont {T.~N.}\ \bibnamefont {Palmer}}\
  and\ \bibinfo {editor} {\bibfnamefont {P.}~\bibnamefont {Williams}}}\
  (\bibinfo  {publisher} {Cambridge University Press})\ pp.\ \bibinfo {pages}
  {35--72}\BibitemShut {NoStop}%
\bibitem [{\citenamefont {Kravtsov}\ \emph {et~al.}(2011)\citenamefont
  {Kravtsov}, \citenamefont {Kondrashov}, \citenamefont {Kamenkovich},\ and\
  \citenamefont {Ghil}}]{kravtsov2011empirical}%
  \BibitemOpen
  \bibfield  {author} {\bibinfo {author} {\bibnamefont {Kravtsov},
  \bibfnamefont {S}}, \bibinfo {author} {\bibfnamefont {D.}~\bibnamefont
  {Kondrashov}}, \bibinfo {author} {\bibfnamefont {Igor}\ \bibnamefont
  {Kamenkovich}}, \ and\ \bibinfo {author} {\bibfnamefont {M.}~\bibnamefont
  {Ghil}}} (\bibinfo {year} {2011}),\ \bibfield  {title} {\enquote {\bibinfo
  {title} {An empirical stochastic model of sea-surface temperatures and
  surface winds over the {Southern Ocean}},}\ }\href@noop {} {\bibfield
  {journal} {\bibinfo  {journal} {Ocean Science}\ }\textbf {\bibinfo {volume}
  {7}}~(\bibinfo {number} {6}),\ \bibinfo {pages} {755--770}}\BibitemShut
  {NoStop}%
\bibitem [{\citenamefont {Krishnamurti}\ \emph {et~al.}(1979)\citenamefont
  {Krishnamurti}, \citenamefont {Stefanova},\ and\ \citenamefont
  {Misra}}]{Krish.ea.1979}%
  \BibitemOpen
  \bibfield  {author} {\bibinfo {author} {\bibnamefont {Krishnamurti},
  \bibfnamefont {T~N}}, \bibinfo {author} {\bibfnamefont {L.}~\bibnamefont
  {Stefanova}}, \ and\ \bibinfo {author} {\bibfnamefont {V.}~\bibnamefont
  {Misra}}} (\bibinfo {year} {1979}),\ \href@noop {} {\emph {\bibinfo {title}
  {{Tropical Meteorology, An Introduction}}}}\ (\bibinfo  {publisher}
  {Springer})\BibitemShut {NoStop}%
\bibitem [{\citenamefont {Kubo}(1957)}]{K57}%
  \BibitemOpen
  \bibfield  {author} {\bibinfo {author} {\bibnamefont {Kubo}, \bibfnamefont
  {R}}} (\bibinfo {year} {1957}),\ \bibfield  {title} {\enquote {\bibinfo
  {title} {Statistical-mechanical theory of irreversible processes. i. general
  theory and simple applications to magnetic and conduction problems},}\ }\href
  {\doibase 10.1143/JPSJ.12.570} {\bibfield  {journal} {\bibinfo  {journal}
  {Journal of the Physical Society of Japan}\ }\textbf {\bibinfo {volume}
  {12}}~(\bibinfo {number} {6}),\ \bibinfo {pages} {570--586}}\BibitemShut
  {NoStop}%
\bibitem [{\citenamefont {Kubo}(1966)}]{Kubo.1966}%
  \BibitemOpen
  \bibfield  {author} {\bibinfo {author} {\bibnamefont {Kubo}, \bibfnamefont
  {R}}} (\bibinfo {year} {1966}),\ \bibfield  {title} {\enquote {\bibinfo
  {title} {The fluctuation-dissipation theorem},}\ }\href@noop {} {\bibfield
  {journal} {\bibinfo  {journal} {Reports on Progress in Physics}\ }\textbf
  {\bibinfo {volume} {29}}~(\bibinfo {number} {1}),\ \bibinfo {pages}
  {255--284}}\BibitemShut {NoStop}%
\bibitem [{\citenamefont {Kuhlbrodt}\ \emph {et~al.}(2007)\citenamefont
  {Kuhlbrodt}, \citenamefont {Griesel}, \citenamefont {Montoya}, \citenamefont
  {Levermann}, \citenamefont {Hofmann},\ and\ \citenamefont
  {Rahmstorf}}]{Kuhlbrodt2007}%
  \BibitemOpen
  \bibfield  {author} {\bibinfo {author} {\bibnamefont {Kuhlbrodt},
  \bibfnamefont {T}}, \bibinfo {author} {\bibfnamefont {A.}~\bibnamefont
  {Griesel}}, \bibinfo {author} {\bibfnamefont {M.}~\bibnamefont {Montoya}},
  \bibinfo {author} {\bibfnamefont {A.}~\bibnamefont {Levermann}}, \bibinfo
  {author} {\bibfnamefont {M.}~\bibnamefont {Hofmann}}, \ and\ \bibinfo
  {author} {\bibfnamefont {S.}~\bibnamefont {Rahmstorf}}} (\bibinfo {year}
  {2007}),\ \bibfield  {title} {\enquote {\bibinfo {title} {On the driving
  processes of the atlantic meridional overturning circulation},}\ }\href
  {\doibase 10.1029/2004RG000166} {\bibfield  {journal} {\bibinfo  {journal}
  {Reviews of Geophysics}\ }\textbf {\bibinfo {volume} {45}}~(\bibinfo {number}
  {2}),\ 10.1029/2004RG000166}\BibitemShut {NoStop}%
\bibitem [{\citenamefont {Kushnir}(1987)}]{Kushnir.1987}%
  \BibitemOpen
  \bibfield  {author} {\bibinfo {author} {\bibnamefont {Kushnir}, \bibfnamefont
  {Y}}} (\bibinfo {year} {1987}),\ \bibfield  {title} {\enquote {\bibinfo
  {title} {Retrograding wintertime low-frequency disturbances over the {North
  Pacific} ocean},}\ }\href {\doibase
  10.1175/1520-0469(1987)044<2727:rwlfdo>2.0.co;2} {\bibfield  {journal}
  {\bibinfo  {journal} {J. Atmos. Sci.}\ }\textbf {\bibinfo {volume}
  {44}}~(\bibinfo {number} {19}),\ \bibinfo {pages} {2727--2742}}\BibitemShut
  {NoStop}%
\bibitem [{\citenamefont {Kushnir}(1994)}]{Kushnir1994}%
  \BibitemOpen
  \bibfield  {author} {\bibinfo {author} {\bibnamefont {Kushnir}, \bibfnamefont
  {Y}}} (\bibinfo {year} {1994}),\ \bibfield  {title} {\enquote {\bibinfo
  {title} {{Interdecadal variations in North Atlantic sea surface temperature
  and associated atmospheric conditions}},}\ }\href@noop {} {\bibfield
  {journal} {\bibinfo  {journal} {J.\ Phys.\ Oceanogr.}\ }\textbf {\bibinfo
  {volume} {7}},\ \bibinfo {pages} {141--157}}\BibitemShut {NoStop}%
\bibitem [{\citenamefont {Lagerstrom}(1988)}]{Lagerstrom.1988}%
  \BibitemOpen
  \bibfield  {author} {\bibinfo {author} {\bibnamefont {Lagerstrom},
  \bibfnamefont {P~A}}} (\bibinfo {year} {1988}),\ \href@noop {} {\emph
  {\bibinfo {title} {{Matched Asymptotic Expansions: Ideas and Techniques}}}}\
  (\bibinfo  {publisher} {Springer Science \& Business Media})\BibitemShut
  {NoStop}%
\bibitem [{\citenamefont {Lai}\ and\ \citenamefont {T\'el}(2011)}]{LT:2011}%
  \BibitemOpen
  \bibfield  {author} {\bibinfo {author} {\bibnamefont {Lai}, \bibfnamefont
  {Y-C}}, \ and\ \bibinfo {author} {\bibfnamefont {T.}~\bibnamefont {T\'el}}}
  (\bibinfo {year} {2011}),\ \href@noop {} {\emph {\bibinfo {title} {Transient
  Chaos}}}\ (\bibinfo  {publisher} {Springer},\ \bibinfo {address} {New
  York})\BibitemShut {NoStop}%
\bibitem [{\citenamefont {Laloyaux}\ \emph {et~al.}(2018)\citenamefont
  {Laloyaux}, \citenamefont {de~Boisseson}, \citenamefont {Balmaseda},
  \citenamefont {Bidlot}, \citenamefont {Broennimann}, \citenamefont {Buizza},
  \citenamefont {Dalhgren}, \citenamefont {Dee}, \citenamefont {Haimberger},
  \citenamefont {Hersbach}, \citenamefont {Kosaka}, \citenamefont {Martin},
  \citenamefont {Poli}, \citenamefont {Rayner}, \citenamefont {Rustemeier},\
  and\ \citenamefont {Schepers}}]{Laloyaux2018}%
  \BibitemOpen
  \bibfield  {author} {\bibinfo {author} {\bibnamefont {Laloyaux},
  \bibfnamefont {P}}, \bibinfo {author} {\bibfnamefont {E.}~\bibnamefont
  {de~Boisseson}}, \bibinfo {author} {\bibfnamefont {M.}~\bibnamefont
  {Balmaseda}}, \bibinfo {author} {\bibfnamefont {J.-R.}\ \bibnamefont
  {Bidlot}}, \bibinfo {author} {\bibfnamefont {S.}~\bibnamefont {Broennimann}},
  \bibinfo {author} {\bibfnamefont {R.}~\bibnamefont {Buizza}}, \bibinfo
  {author} {\bibfnamefont {P.}~\bibnamefont {Dalhgren}}, \bibinfo {author}
  {\bibfnamefont {D.}~\bibnamefont {Dee}}, \bibinfo {author} {\bibfnamefont
  {L.}~\bibnamefont {Haimberger}}, \bibinfo {author} {\bibfnamefont
  {H.}~\bibnamefont {Hersbach}}, \bibinfo {author} {\bibfnamefont
  {Y.}~\bibnamefont {Kosaka}}, \bibinfo {author} {\bibfnamefont
  {M.}~\bibnamefont {Martin}}, \bibinfo {author} {\bibfnamefont
  {P.}~\bibnamefont {Poli}}, \bibinfo {author} {\bibfnamefont {N.}~\bibnamefont
  {Rayner}}, \bibinfo {author} {\bibfnamefont {E.}~\bibnamefont {Rustemeier}},
  \ and\ \bibinfo {author} {\bibfnamefont {D.}~\bibnamefont {Schepers}}}
  (\bibinfo {year} {2018}),\ \bibfield  {title} {\enquote {\bibinfo {title}
  {{CERA-20C: A} coupled reanalysis of the twentieth century},}\ }\href
  {\doibase 10.1029/2018MS001273} {\bibfield  {journal} {\bibinfo  {journal}
  {Journal of Advances in Modeling Earth Systems}\ }\textbf {\bibinfo {volume}
  {10}}~(\bibinfo {number} {5}),\ \bibinfo {pages} {1172--1195}}\BibitemShut
  {NoStop}%
\bibitem [{\citenamefont {Lamb}(1972)}]{Lamb1972}%
  \BibitemOpen
  \bibfield  {author} {\bibinfo {author} {\bibnamefont {Lamb}, \bibfnamefont
  {H~H}}} (\bibinfo {year} {1972}),\ \href@noop {} {\emph {\bibinfo {title}
  {Climate: {Present, Past and Future}}}}\ (\bibinfo  {publisher} {Methuen},\
  \bibinfo {address} {London})\BibitemShut {NoStop}%
\bibitem [{\citenamefont {Langen}\ and\ \citenamefont
  {Alexeev}(2005)}]{langen_estimating_2005}%
  \BibitemOpen
  \bibfield  {author} {\bibinfo {author} {\bibnamefont {Langen}, \bibfnamefont
  {P~L}}, \ and\ \bibinfo {author} {\bibfnamefont {V.~A.}\ \bibnamefont
  {Alexeev}}} (\bibinfo {year} {2005}),\ \bibfield  {title} {\enquote {\bibinfo
  {title} {Estimating 2 x {CO2} warming in an aquaplanet {GCM} using the
  fluctuation-dissipation theorem},}\ }\href@noop {} {\bibfield  {journal}
  {\bibinfo  {journal} {Geophysical Research Letters}\ }\textbf {\bibinfo
  {volume} {32}}~(\bibinfo {number} {23}),\ \bibinfo {pages}
  {L23708}}\BibitemShut {NoStop}%
\bibitem [{\citenamefont {Lawrence}\ \emph {et~al.}(2018)\citenamefont
  {Lawrence}, \citenamefont {Sch{\"a}fer}, \citenamefont {Muri}, \citenamefont
  {Scott}, \citenamefont {Oschlies}, \citenamefont {Vaughan}, \citenamefont
  {Boucher}, \citenamefont {Schmidt}, \citenamefont {Haywood},\ and\
  \citenamefont {Scheffran}}]{Lawrence2018}%
  \BibitemOpen
  \bibfield  {author} {\bibinfo {author} {\bibnamefont {Lawrence},
  \bibfnamefont {M~G}}, \bibinfo {author} {\bibfnamefont {S.}~\bibnamefont
  {Sch{\"a}fer}}, \bibinfo {author} {\bibfnamefont {H.}~\bibnamefont {Muri}},
  \bibinfo {author} {\bibfnamefont {V.}~\bibnamefont {Scott}}, \bibinfo
  {author} {\bibfnamefont {A.}~\bibnamefont {Oschlies}}, \bibinfo {author}
  {\bibfnamefont {N.~E.}\ \bibnamefont {Vaughan}}, \bibinfo {author}
  {\bibfnamefont {O.}~\bibnamefont {Boucher}}, \bibinfo {author} {\bibfnamefont
  {H.}~\bibnamefont {Schmidt}}, \bibinfo {author} {\bibfnamefont
  {J.}~\bibnamefont {Haywood}}, \ and\ \bibinfo {author} {\bibfnamefont
  {J.}~\bibnamefont {Scheffran}}} (\bibinfo {year} {2018}),\ \bibfield  {title}
  {\enquote {\bibinfo {title} {Evaluating climate geoengineering proposals in
  the context of the {Paris Agreement temperature goals}},}\ }\href {\doibase
  10.1038/s41467-018-05938-3} {\bibfield  {journal} {\bibinfo  {journal}
  {Nature Communications}\ }\textbf {\bibinfo {volume} {9}}~(\bibinfo {number}
  {1}),\ \bibinfo {pages} {3734}}\BibitemShut {NoStop}%
\bibitem [{\citenamefont {Ledrappier}\ and\ \citenamefont
  {Young}(1988)}]{LY.1988}%
  \BibitemOpen
  \bibfield  {author} {\bibinfo {author} {\bibnamefont {Ledrappier},
  \bibfnamefont {F}}, \ and\ \bibinfo {author} {\bibfnamefont {L.-S.}\
  \bibnamefont {Young}}} (\bibinfo {year} {1988}),\ \bibfield  {title}
  {\enquote {\bibinfo {title} {Entropy formula for random transformations},}\
  }\href@noop {} {\bibfield  {journal} {\bibinfo  {journal} {Prob. Theory
  Related Fields}\ }\textbf {\bibinfo {volume} {80}},\ \bibinfo {pages}
  {217--240}}\BibitemShut {NoStop}%
\bibitem [{\citenamefont {Lee}\ \emph {et~al.}(2009)\citenamefont {Lee},
  \citenamefont {Awaji}, \citenamefont {Balmaseda}, \citenamefont {Greiner},\
  and\ \citenamefont {Stammer}}]{Lee2009}%
  \BibitemOpen
  \bibfield  {author} {\bibinfo {author} {\bibnamefont {Lee}, \bibfnamefont
  {T}}, \bibinfo {author} {\bibfnamefont {T.}~\bibnamefont {Awaji}}, \bibinfo
  {author} {\bibfnamefont {M.~A.}\ \bibnamefont {Balmaseda}}, \bibinfo {author}
  {\bibfnamefont {E.}~\bibnamefont {Greiner}}, \ and\ \bibinfo {author}
  {\bibfnamefont {D.}~\bibnamefont {Stammer}}} (\bibinfo {year} {2009}),\
  \bibfield  {title} {\enquote {\bibinfo {title} {Ocean state estimation for
  climate research},}\ }\href {\doibase 10.5670/oceanog.2009.74} {\bibfield
  {journal} {\bibinfo  {journal} {Oceanography}\ }\textbf {\bibinfo {volume}
  {22}},\ 10.5670/oceanog.2009.74}\BibitemShut {NoStop}%
\bibitem [{\citenamefont {Legras}\ and\ \citenamefont
  {Ghil}(1985)}]{Legras1985}%
  \BibitemOpen
  \bibfield  {author} {\bibinfo {author} {\bibnamefont {Legras}, \bibfnamefont
  {B}}, \ and\ \bibinfo {author} {\bibfnamefont {M.}~\bibnamefont {Ghil}}}
  (\bibinfo {year} {1985}),\ \bibfield  {title} {\enquote {\bibinfo {title}
  {Persistent anomalies, blocking, and variations in atmospheric
  predictability},}\ }\href@noop {} {\bibfield  {journal} {\bibinfo  {journal}
  {J.\ Atmos.\ Sci.}\ }\textbf {\bibinfo {volume} {42}},\ \bibinfo {pages}
  {433--471}}\BibitemShut {NoStop}%
\bibitem [{\citenamefont {Leith}(1975)}]{Leith75}%
  \BibitemOpen
  \bibfield  {author} {\bibinfo {author} {\bibnamefont {Leith}, \bibfnamefont
  {C~E}}} (\bibinfo {year} {1975}),\ \bibfield  {title} {\enquote {\bibinfo
  {title} {Climate response and fluctuation dissipation},}\ }\bibfield
  {booktitle} {\emph {\bibinfo {booktitle} {Journal of the Atmospheric
  Sciences}},\ }\href {\doibase 10.1175/1520-0469(1975)032<2022:CRAFD>2.0.CO;2}
  {\bibfield  {journal} {\bibinfo  {journal} {Journal of the Atmospheric
  Sciences}\ }\textbf {\bibinfo {volume} {32}}~(\bibinfo {number} {10}),\
  \bibinfo {pages} {2022--2026}}\BibitemShut {NoStop}%
\bibitem [{\citenamefont {Leith}\ and\ \citenamefont
  {Kraichnan}(1972)}]{Leith.Kraichnan.1972}%
  \BibitemOpen
  \bibfield  {author} {\bibinfo {author} {\bibnamefont {Leith}, \bibfnamefont
  {C~E}}, \ and\ \bibinfo {author} {\bibfnamefont {R.~H.}\ \bibnamefont
  {Kraichnan}}} (\bibinfo {year} {1972}),\ \bibfield  {title} {\enquote
  {\bibinfo {title} {Predictability of turbulent flows},}\ }\href@noop {}
  {\bibfield  {journal} {\bibinfo  {journal} {J. Atmos. Sci.}\ }\textbf
  {\bibinfo {volume} {29}},\ \bibinfo {pages} {1041--1058}}\BibitemShut
  {NoStop}%
\bibitem [{\citenamefont {Lembo}\ \emph {et~al.}(2019)\citenamefont {Lembo},
  \citenamefont {Lunkeit},\ and\ \citenamefont {Lucarini}}]{Lembo2019}%
  \BibitemOpen
  \bibfield  {author} {\bibinfo {author} {\bibnamefont {Lembo}, \bibfnamefont
  {V}}, \bibinfo {author} {\bibfnamefont {F.}~\bibnamefont {Lunkeit}}, \ and\
  \bibinfo {author} {\bibfnamefont {V.}~\bibnamefont {Lucarini}}} (\bibinfo
  {year} {2019}),\ \bibfield  {title} {\enquote {\bibinfo {title} {Thediato
  (v1.0) -- a new diagnostic tool for water, energy and entropy budgets in
  climate models},}\ }\href {\doibase 10.5194/gmd-12-3805-2019} {\bibfield
  {journal} {\bibinfo  {journal} {Geoscientific Model Development}\ }\textbf
  {\bibinfo {volume} {12}}~(\bibinfo {number} {8}),\ \bibinfo {pages}
  {3805--3834}}\BibitemShut {NoStop}%
\bibitem [{\citenamefont {Lenton}\ \emph {et~al.}(2008)\citenamefont {Lenton},
  \citenamefont {Held}, \citenamefont {Kriegler}, \citenamefont {Hall},
  \citenamefont {Lucht}, \citenamefont {Rahmstorf},\ and\ \citenamefont
  {Schellnhuber}}]{Lenton.tip.08}%
  \BibitemOpen
  \bibfield  {author} {\bibinfo {author} {\bibnamefont {Lenton}, \bibfnamefont
  {T~M}}, \bibinfo {author} {\bibfnamefont {H.}~\bibnamefont {Held}}, \bibinfo
  {author} {\bibfnamefont {E.}~\bibnamefont {Kriegler}}, \bibinfo {author}
  {\bibfnamefont {J.~W.}\ \bibnamefont {Hall}}, \bibinfo {author}
  {\bibfnamefont {W.}~\bibnamefont {Lucht}}, \bibinfo {author} {\bibfnamefont
  {S.}~\bibnamefont {Rahmstorf}}, \ and\ \bibinfo {author} {\bibfnamefont
  {H.~J.}\ \bibnamefont {Schellnhuber}}} (\bibinfo {year} {2008}),\ \bibfield
  {title} {\enquote {\bibinfo {title} {Tipping elements in the {Earth's climate
  system}},}\ }\href@noop {} {\bibfield  {journal} {\bibinfo  {journal} {Proc.
  Natl. Acad. Sci. USA}\ }\textbf {\bibinfo {volume} {105}},\ \bibinfo {pages}
  {1786--1793}}\BibitemShut {NoStop}%
\bibitem [{\citenamefont {Li}\ \emph {et~al.}(1997)\citenamefont {Li},
  \citenamefont {Ide}, \citenamefont {Le~Treut},\ and\ \citenamefont
  {Ghil}}]{Li.ea.1997}%
  \BibitemOpen
  \bibfield  {author} {\bibinfo {author} {\bibnamefont {Li}, \bibfnamefont
  {Z-X}}, \bibinfo {author} {\bibfnamefont {K.}~\bibnamefont {Ide}}, \bibinfo
  {author} {\bibfnamefont {H.}~\bibnamefont {Le~Treut}}, \ and\ \bibinfo
  {author} {\bibfnamefont {M.}~\bibnamefont {Ghil}}} (\bibinfo {year} {1997}),\
  \bibfield  {title} {\enquote {\bibinfo {title} {Atmospheric radiative
  equilibria in a simple column model},}\ }\href@noop {} {\bibfield  {journal}
  {\bibinfo  {journal} {Climate Dyn.}\ }\textbf {\bibinfo {volume} {13}},\
  \bibinfo {pages} {429--440}}\BibitemShut {NoStop}%
\bibitem [{\citenamefont {Lindzen}(1986)}]{Lindzen.1986}%
  \BibitemOpen
  \bibfield  {author} {\bibinfo {author} {\bibnamefont {Lindzen}, \bibfnamefont
  {R~S}}} (\bibinfo {year} {1986}),\ \bibfield  {title} {\enquote {\bibinfo
  {title} {Stationary planetary waves, blocking, and interannual
  variability},}\ }in\ \href {\doibase 10.1016/s0065-2687(08)60042-4} {\emph
  {\bibinfo {booktitle} {Advances in Geophysics}}},\ Vol.~\bibinfo {volume}
  {29}\ (\bibinfo  {publisher} {Elsevier})\ pp.\ \bibinfo {pages}
  {251--273}\BibitemShut {NoStop}%
\bibitem [{\citenamefont {Liu}\ \emph {et~al.}(2012)\citenamefont {Liu},
  \citenamefont {Curry}, \citenamefont {Wang}, \citenamefont {Song},\ and\
  \citenamefont {Horton}}]{Liu.Curry.2012}%
  \BibitemOpen
  \bibfield  {author} {\bibinfo {author} {\bibnamefont {Liu}, \bibfnamefont
  {J}}, \bibinfo {author} {\bibfnamefont {J.~A.}\ \bibnamefont {Curry}},
  \bibinfo {author} {\bibfnamefont {H.}~\bibnamefont {Wang}}, \bibinfo {author}
  {\bibfnamefont {M.}~\bibnamefont {Song}}, \ and\ \bibinfo {author}
  {\bibfnamefont {R.~M.}\ \bibnamefont {Horton}}} (\bibinfo {year} {2012}),\
  \bibfield  {title} {\enquote {\bibinfo {title} {Impact of declining arctic
  sea ice on winter snowfall},}\ }\href@noop {} {\bibfield  {journal} {\bibinfo
   {journal} {Proceedings of the National Academy of Sciences}\ }\textbf
  {\bibinfo {volume} {109}}~(\bibinfo {number} {11}),\ \bibinfo {pages}
  {4074--4079}}\BibitemShut {NoStop}%
\bibitem [{\citenamefont {Lorenz}(1963)}]{Lorenz1963a}%
  \BibitemOpen
  \bibfield  {author} {\bibinfo {author} {\bibnamefont {Lorenz}, \bibfnamefont
  {E~N}}} (\bibinfo {year} {1963}),\ \bibfield  {title} {\enquote {\bibinfo
  {title} {Deterministic nonperiodic flow},}\ }\href@noop {} {\bibfield
  {journal} {\bibinfo  {journal} {J. Atmos. Sci.}\ }\textbf {\bibinfo {volume}
  {20}},\ \bibinfo {pages} {130--141}}\BibitemShut {NoStop}%
\bibitem [{\citenamefont {Lorenz}(1967)}]{Lor67}%
  \BibitemOpen
  \bibfield  {author} {\bibinfo {author} {\bibnamefont {Lorenz}, \bibfnamefont
  {E~N}}} (\bibinfo {year} {1967}),\ \href@noop {} {\emph {\bibinfo {title}
  {The {Nature and Theory of the General Circulation of the Atmosphere}}}}\
  (\bibinfo  {publisher} {World Meteorological Organization},\ \bibinfo
  {address} {Geneva, Switzerland})\BibitemShut {NoStop}%
\bibitem [{\citenamefont {Lorenz}({1969a})}]{Lorenz1969a}%
  \BibitemOpen
  \bibfield  {author} {\bibinfo {author} {\bibnamefont {Lorenz}, \bibfnamefont
  {E~N}}} (\bibinfo {year} {{1969a}}),\ \bibfield  {title} {\enquote {\bibinfo
  {title} {The predictability of a flow which possesses many scales of
  motion},}\ }\href@noop {} {\bibfield  {journal} {\bibinfo  {journal}
  {Tellus}\ }\textbf {\bibinfo {volume} {21}},\ \bibinfo {pages}
  {289--307}}\BibitemShut {NoStop}%
\bibitem [{\citenamefont {Lorenz}({1969b})}]{Lorenz1969b}%
  \BibitemOpen
  \bibfield  {author} {\bibinfo {author} {\bibnamefont {Lorenz}, \bibfnamefont
  {E~N}}} (\bibinfo {year} {{1969b}}),\ \bibfield  {title} {\enquote {\bibinfo
  {title} {Three approaches to atmospheric predictability},}\ }\href@noop {}
  {\bibfield  {journal} {\bibinfo  {journal} {Bull. Am. Meteorol. Soc.}\
  }\textbf {\bibinfo {volume} {50}},\ \bibinfo {pages} {345--349}}\BibitemShut
  {NoStop}%
\bibitem [{\citenamefont {Lorenz}(1976)}]{Lorenz76}%
  \BibitemOpen
  \bibfield  {author} {\bibinfo {author} {\bibnamefont {Lorenz}, \bibfnamefont
  {E~N}}} (\bibinfo {year} {1976}),\ \bibfield  {title} {\enquote {\bibinfo
  {title} {Nondeterministic theories of climatic change},}\ }\href {\doibase
  10.1016/0033-5894(76)90022-3} {\bibfield  {journal} {\bibinfo  {journal}
  {Quatern. Res.}\ }\textbf {\bibinfo {volume} {6}}~(\bibinfo {number} {4}),\
  \bibinfo {pages} {495--506}}\BibitemShut {NoStop}%
\bibitem [{\citenamefont {Lorenz}(1979)}]{lorenz79}%
  \BibitemOpen
  \bibfield  {author} {\bibinfo {author} {\bibnamefont {Lorenz}, \bibfnamefont
  {E~N}}} (\bibinfo {year} {1979}),\ \bibfield  {title} {\enquote {\bibinfo
  {title} {Forced and free variations of weather and climate},}\ }\href@noop {}
  {\bibfield  {journal} {\bibinfo  {journal} {J. Atmos. Sci}\ }\textbf
  {\bibinfo {volume} {36}},\ \bibinfo {pages} {1367--1376}}\BibitemShut
  {NoStop}%
\bibitem [{\citenamefont {Lorenz}(1984)}]{L84}%
  \BibitemOpen
  \bibfield  {author} {\bibinfo {author} {\bibnamefont {Lorenz}, \bibfnamefont
  {E~N}}} (\bibinfo {year} {1984}),\ \bibfield  {title} {\enquote {\bibinfo
  {title} {Irregularity: a fundamental property of the atmosphere},}\
  }\href@noop {} {\bibfield  {journal} {\bibinfo  {journal} {Tellus A}\
  }\textbf {\bibinfo {volume} {36}}~(\bibinfo {number} {2}),\ \bibinfo {pages}
  {98--110}}\BibitemShut {NoStop}%
\bibitem [{\citenamefont {Lorenz}(1996)}]{Lorenz_predictability_1996}%
  \BibitemOpen
  \bibfield  {author} {\bibinfo {author} {\bibnamefont {Lorenz}, \bibfnamefont
  {E~N}}} (\bibinfo {year} {1996}),\ \bibfield  {title} {\enquote {\bibinfo
  {title} {Predictability: a problem partly solved},}\ }in\ \href@noop {}
  {\emph {\bibinfo {booktitle} {Seminar on Predictability, 4--8 September
  1995}}}\ (\bibinfo  {publisher} {European Centre for Medium-range Weather
  Forecasts},\ \bibinfo {address} {Reading, UK})\ pp.\ \bibinfo {pages}
  {1--18}\BibitemShut {NoStop}%
\bibitem [{\citenamefont {Lorenz}(1955)}]{Lor55}%
  \BibitemOpen
  \bibfield  {author} {\bibinfo {author} {\bibnamefont {Lorenz}, \bibfnamefont
  {EN}}} (\bibinfo {year} {1955}),\ \bibfield  {title} {\enquote {\bibinfo
  {title} {Available potential energy and the maintenance of the general
  circulation},}\ }\href@noop {} {\bibfield  {journal} {\bibinfo  {journal}
  {Tellus}\ }\textbf {\bibinfo {volume} {7}},\ \bibinfo {pages}
  {157--167}}\BibitemShut {NoStop}%
\bibitem [{\citenamefont {Lu}\ \emph {et~al.}(2018)\citenamefont {Lu},
  \citenamefont {Fu}, \citenamefont {Hua}, \citenamefont {Yuan},\ and\
  \citenamefont {Chen}}]{Lu2018}%
  \BibitemOpen
  \bibfield  {author} {\bibinfo {author} {\bibnamefont {Lu}, \bibfnamefont
  {Z}}, \bibinfo {author} {\bibfnamefont {Z.}~\bibnamefont {Fu}}, \bibinfo
  {author} {\bibfnamefont {L.}~\bibnamefont {Hua}}, \bibinfo {author}
  {\bibfnamefont {N.}~\bibnamefont {Yuan}}, \ and\ \bibinfo {author}
  {\bibfnamefont {L.}~\bibnamefont {Chen}}} (\bibinfo {year} {2018}),\
  \bibfield  {title} {\enquote {\bibinfo {title} {Evaluation of {ENSO
  simulations in CMIP5 models: A new perspective based on percolation phase
  transition in complex networks}},}\ }\href {\doibase
  10.1038/s41598-018-33340-y} {\bibfield  {journal} {\bibinfo  {journal}
  {Scientific Reports}\ }\textbf {\bibinfo {volume} {8}}~(\bibinfo {number}
  {1}),\ \bibinfo {pages} {14912}}\BibitemShut {NoStop}%
\bibitem [{\citenamefont {Lucarini}(2002)}]{Lucarini2002}%
  \BibitemOpen
  \bibfield  {author} {\bibinfo {author} {\bibnamefont {Lucarini},
  \bibfnamefont {V}}} (\bibinfo {year} {2002}),\ \bibfield  {title} {\enquote
  {\bibinfo {title} {Towards a definition of climate science},}\ }\href
  {\doibase 10.1504/IJEP.2002.002336} {\bibfield  {journal} {\bibinfo
  {journal} {Intl. J. Environ. Pollution}\ }\textbf {\bibinfo {volume}
  {18}}~(\bibinfo {number} {5}),\ \bibinfo {pages} {413--422}}\BibitemShut
  {NoStop}%
\bibitem [{\citenamefont {Lucarini}(2008{\natexlab{a}})}]{lucarini08}%
  \BibitemOpen
  \bibfield  {author} {\bibinfo {author} {\bibnamefont {Lucarini},
  \bibfnamefont {V}}} (\bibinfo {year} {2008}{\natexlab{a}}),\ \bibfield
  {title} {\enquote {\bibinfo {title} {Response theory for equilibrium and
  non-equilibrium statistical mechanics: Causality and generalized
  kramers-kronig relations},}\ }\href {\doibase 10.1007/s10955-008-9498-y}
  {\bibfield  {journal} {\bibinfo  {journal} {Journal of Statistical Physics}\
  }\textbf {\bibinfo {volume} {131}},\ \bibinfo {pages} {543--558}}\BibitemShut
  {NoStop}%
\bibitem [{\citenamefont {Lucarini}(2008{\natexlab{b}})}]{Lucarini08enc}%
  \BibitemOpen
  \bibfield  {author} {\bibinfo {author} {\bibnamefont {Lucarini},
  \bibfnamefont {V}}} (\bibinfo {year} {2008}{\natexlab{b}}),\ \bibfield
  {title} {\enquote {\bibinfo {title} {Validation of climate models},}\ }in\
  \href@noop {} {\emph {\bibinfo {booktitle} {Encyclopedia of Global Warming
  and Climate Change}}},\ \bibinfo {editor} {edited by\ \bibinfo {editor}
  {\bibfnamefont {G.}~\bibnamefont {Philander}}}\ (\bibinfo  {publisher}
  {{SAGE}},\ \bibinfo {address} {Thousand Oaks, California})\ pp.\ \bibinfo
  {pages} {1053--1057}\BibitemShut {NoStop}%
\bibitem [{\citenamefont {Lucarini}(2009{\natexlab{a}})}]{L09}%
  \BibitemOpen
  \bibfield  {author} {\bibinfo {author} {\bibnamefont {Lucarini},
  \bibfnamefont {V}}} (\bibinfo {year} {2009}{\natexlab{a}}),\ \bibfield
  {title} {\enquote {\bibinfo {title} {Evidence of dispersion relations for the
  nonlinear response of the {Lorenz} 63 system},}\ }\href {\doibase
  10.1007/s10955-008-9675-z} {\bibfield  {journal} {\bibinfo  {journal}
  {Journal of Statistical Physics}\ }\textbf {\bibinfo {volume} {134}},\
  \bibinfo {pages} {381--400}}\BibitemShut {NoStop}%
\bibitem [{\citenamefont {Lucarini}(2009{\natexlab{b}})}]{Lucarini:2009_PRE}%
  \BibitemOpen
  \bibfield  {author} {\bibinfo {author} {\bibnamefont {Lucarini},
  \bibfnamefont {V}}} (\bibinfo {year} {2009}{\natexlab{b}}),\ \bibfield
  {title} {\enquote {\bibinfo {title} {Thermodynamic efficiency and entropy
  production in the climate system},}\ }\href {\doibase
  10.1103/PhysRevE.80.02118} {\bibfield  {journal} {\bibinfo  {journal} {Phys.
  Rev. E}\ }\textbf {\bibinfo {volume} {80}},\ \bibinfo {pages}
  {021118}}\BibitemShut {NoStop}%
\bibitem [{\citenamefont {Lucarini}(2012)}]{lucarini2012}%
  \BibitemOpen
  \bibfield  {author} {\bibinfo {author} {\bibnamefont {Lucarini},
  \bibfnamefont {V}}} (\bibinfo {year} {2012}),\ \bibfield  {title} {\enquote
  {\bibinfo {title} {Stochastic perturbations to dynamical systems: A response
  theory approach},}\ }\href {\doibase 10.1007/s10955-012-0422-0} {\bibfield
  {journal} {\bibinfo  {journal} {Journal of Statistical Physics}\ }\textbf
  {\bibinfo {volume} {146}}~(\bibinfo {number} {4}),\ \bibinfo {pages}
  {774--786}}\BibitemShut {NoStop}%
\bibitem [{\citenamefont {Lucarini}(2013)}]{lucarini_modelling_2013}%
  \BibitemOpen
  \bibfield  {author} {\bibinfo {author} {\bibnamefont {Lucarini},
  \bibfnamefont {V}}} (\bibinfo {year} {2013}),\ \bibfield  {title} {\enquote
  {\bibinfo {title} {Modeling complexity: the case of climate science},}\ }in\
  \href@noop {} {\emph {\bibinfo {booktitle} {Models, Simulations, and the
  Reduction of Complexity}}},\ \bibinfo {editor} {edited by\ \bibinfo {editor}
  {\bibfnamefont {U}~\bibnamefont {G\"{a}hde}}, \bibinfo {editor}
  {\bibfnamefont {S}~\bibnamefont {Hartmann}}, \ and\ \bibinfo {editor}
  {\bibfnamefont {JH}~\bibnamefont {Wolf}}}\ (\bibinfo  {publisher} {De
  Gruyter})\ pp.\ \bibinfo {pages} {229--254}\BibitemShut {NoStop}%
\bibitem [{\citenamefont {Lucarini}(2016)}]{Lucarini2016}%
  \BibitemOpen
  \bibfield  {author} {\bibinfo {author} {\bibnamefont {Lucarini},
  \bibfnamefont {V}}} (\bibinfo {year} {2016}),\ \bibfield  {title} {\enquote
  {\bibinfo {title} {Response operators for {Markov} processes in a finite
  state space: Radius of convergence and link to the response theory for {Axiom
  A} systems},}\ }\href {\doibase 10.1007/s10955-015-1409-4} {\bibfield
  {journal} {\bibinfo  {journal} {Journal of Statistical Physics}\ }\textbf
  {\bibinfo {volume} {162}}~(\bibinfo {number} {2}),\ \bibinfo {pages}
  {312--333}}\BibitemShut {NoStop}%
\bibitem [{\citenamefont {Lucarini}(2018)}]{Lucarini2018JSP}%
  \BibitemOpen
  \bibfield  {author} {\bibinfo {author} {\bibnamefont {Lucarini},
  \bibfnamefont {V}}} (\bibinfo {year} {2018}),\ \bibfield  {title} {\enquote
  {\bibinfo {title} {Revising and extending the linear response theory for
  statistical mechanical systems: Evaluating observables as predictors and
  predictands},}\ }\href {\doibase 10.1007/s10955-018-2151-5} {\bibfield
  {journal} {\bibinfo  {journal} {Journal of Statistical Physics}\ }\textbf
  {\bibinfo {volume} {173}}~(\bibinfo {number} {6}),\ \bibinfo {pages}
  {1698--1721}}\BibitemShut {NoStop}%
\bibitem [{\citenamefont {Lucarini}\ \emph {et~al.}(2014)\citenamefont
  {Lucarini}, \citenamefont {Blender}, \citenamefont {Herbert}, \citenamefont
  {Ragone}, \citenamefont {Pascale},\ and\ \citenamefont
  {Wouters}}]{Lucarini.ea.2014}%
  \BibitemOpen
  \bibfield  {author} {\bibinfo {author} {\bibnamefont {Lucarini},
  \bibfnamefont {V}}, \bibinfo {author} {\bibfnamefont {R.}~\bibnamefont
  {Blender}}, \bibinfo {author} {\bibfnamefont {C.}~\bibnamefont {Herbert}},
  \bibinfo {author} {\bibfnamefont {F.}~\bibnamefont {Ragone}}, \bibinfo
  {author} {\bibfnamefont {S.}~\bibnamefont {Pascale}}, \ and\ \bibinfo
  {author} {\bibfnamefont {J.}~\bibnamefont {Wouters}}} (\bibinfo {year}
  {2014}),\ \bibfield  {title} {\enquote {\bibinfo {title} {Mathematical and
  physical ideas for climate science},}\ }\href {\doibase 10.1002/2013RG000446}
  {\bibfield  {journal} {\bibinfo  {journal} {Rev. Geophys.}\ }\textbf
  {\bibinfo {volume} {52}}~(\bibinfo {number} {4}),\ \bibinfo {pages}
  {809--859}}\BibitemShut {NoStop}%
\bibitem [{\citenamefont {Lucarini}\ and\ \citenamefont
  {B\'odai}(2017)}]{Lucarini2017N}%
  \BibitemOpen
  \bibfield  {author} {\bibinfo {author} {\bibnamefont {Lucarini},
  \bibfnamefont {V}}, \ and\ \bibinfo {author} {\bibfnamefont {T.}~\bibnamefont
  {B\'odai}}} (\bibinfo {year} {2017}),\ \bibfield  {title} {\enquote {\bibinfo
  {title} {Edge states in the climate system: exploring global instabilities
  and critical transitions},}\ }\href@noop {} {\bibfield  {journal} {\bibinfo
  {journal} {Nonlinearity}\ }\textbf {\bibinfo {volume} {30}}~(\bibinfo
  {number} {7}),\ \bibinfo {pages} {R32}}\BibitemShut {NoStop}%
\bibitem [{\citenamefont {Lucarini}\ and\ \citenamefont
  {B\'odai}(2019a)}]{Lucarini2019}%
  \BibitemOpen
  \bibfield  {author} {\bibinfo {author} {\bibnamefont {Lucarini},
  \bibfnamefont {V}}, \ and\ \bibinfo {author} {\bibfnamefont {T.}~\bibnamefont
  {B\'odai}}} (\bibinfo {year} {2019a}),\ \bibfield  {title} {\enquote
  {\bibinfo {title} {Transitions across melancholia states in a climate model:
  Reconciling the deterministic and stochastic points of view},}\ }\href
  {\doibase 10.1103/PhysRevLett.122.158701} {\bibfield  {journal} {\bibinfo
  {journal} {Phys. Rev. Lett.}\ }\textbf {\bibinfo {volume} {122}},\ \bibinfo
  {pages} {158701}}\BibitemShut {NoStop}%
\bibitem [{\citenamefont {Lucarini}\ and\ \citenamefont
  {B\'odai}(2019b)}]{LucariniBodai2019arxiv}%
  \BibitemOpen
  \bibfield  {author} {\bibinfo {author} {\bibnamefont {Lucarini},
  \bibfnamefont {V}}, \ and\ \bibinfo {author} {\bibfnamefont {T.}~\bibnamefont
  {B\'odai}}} (\bibinfo {year} {2019b}),\ \bibfield  {title} {\enquote
  {\bibinfo {title} {{Global Stability Properties of the Climate: Melancholia
  States, Invariant Measures, and Phase Transitions}},}\ }\href@noop {}
  {\bibfield  {journal} {\bibinfo  {journal} {arXiv e-prints}\ ,\ \bibinfo
  {eid} {arXiv:1903.08348}}}\Eprint {http://arxiv.org/abs/1903.08348}
  {arXiv:1903.08348 [physics.ao-ph]} \BibitemShut {NoStop}%
\bibitem [{\citenamefont {Lucarini}\ \emph
  {et~al.}(2005{\natexlab{a}})\citenamefont {Lucarini}, \citenamefont
  {Calmanti},\ and\ \citenamefont {Artale}}]{Lucariniea2005}%
  \BibitemOpen
  \bibfield  {author} {\bibinfo {author} {\bibnamefont {Lucarini},
  \bibfnamefont {V}}, \bibinfo {author} {\bibfnamefont {S.}~\bibnamefont
  {Calmanti}}, \ and\ \bibinfo {author} {\bibfnamefont {V.}~\bibnamefont
  {Artale}}} (\bibinfo {year} {2005}{\natexlab{a}}),\ \bibfield  {title}
  {\enquote {\bibinfo {title} {Destabilization of the thermohaline circulation
  by transient changes in the hydrological cycle},}\ }\href {\doibase
  10.1007/s00382-004-0484-z} {\bibfield  {journal} {\bibinfo  {journal}
  {Climate Dynamics}\ }\textbf {\bibinfo {volume} {24}}~(\bibinfo {number}
  {2}),\ \bibinfo {pages} {253--262}}\BibitemShut {NoStop}%
\bibitem [{\citenamefont {Lucarini}\ \emph {et~al.}(2007)\citenamefont
  {Lucarini}, \citenamefont {Calmanti},\ and\ \citenamefont
  {Artale}}]{Lucariniea2007}%
  \BibitemOpen
  \bibfield  {author} {\bibinfo {author} {\bibnamefont {Lucarini},
  \bibfnamefont {V}}, \bibinfo {author} {\bibfnamefont {S.}~\bibnamefont
  {Calmanti}}, \ and\ \bibinfo {author} {\bibfnamefont {V.}~\bibnamefont
  {Artale}}} (\bibinfo {year} {2007}),\ \bibfield  {title} {\enquote {\bibinfo
  {title} {Experimental mathematics: Dependence of the stability properties of
  a two-dimensional model of the atlantic ocean circulation on the boundary
  conditions},}\ }\href {\doibase 10.1134/S1061920807020124} {\bibfield
  {journal} {\bibinfo  {journal} {Russian Journal of Mathematical Physics}\
  }\textbf {\bibinfo {volume} {14}}~(\bibinfo {number} {2}),\ \bibinfo {pages}
  {224--231}}\BibitemShut {NoStop}%
\bibitem [{\citenamefont {Lucarini}\ and\ \citenamefont
  {Colangeli}(2012)}]{LC12}%
  \BibitemOpen
  \bibfield  {author} {\bibinfo {author} {\bibnamefont {Lucarini},
  \bibfnamefont {V}}, \ and\ \bibinfo {author} {\bibfnamefont {M.}~\bibnamefont
  {Colangeli}}} (\bibinfo {year} {2012}),\ \bibfield  {title} {\enquote
  {\bibinfo {title} {Beyond the linear fluctuation-dissipation theorem: the
  role of causality},}\ }\href@noop {} {\bibfield  {journal} {\bibinfo
  {journal} {Journal of Statistical Mechanics: Theory and Experiment}\ }\textbf
  {\bibinfo {volume} {2012}}~(\bibinfo {number} {05}),\ \bibinfo {pages}
  {P05013}}\BibitemShut {NoStop}%
\bibitem [{\citenamefont {Lucarini}\ \emph {et~al.}(2016)\citenamefont
  {Lucarini}, \citenamefont {Faranda}, \citenamefont {de~Freitas},
  \citenamefont {de~Freitas}, \citenamefont {Holland}, \citenamefont {Kuna},
  \citenamefont {Nicol}, \citenamefont {Todd},\ and\ \citenamefont
  {Vaienti}}]{lucarini2016extremes}%
  \BibitemOpen
  \bibfield  {author} {\bibinfo {author} {\bibnamefont {Lucarini},
  \bibfnamefont {V}}, \bibinfo {author} {\bibfnamefont {D.}~\bibnamefont
  {Faranda}}, \bibinfo {author} {\bibfnamefont {A.~C. G. M.~M.}\ \bibnamefont
  {de~Freitas}}, \bibinfo {author} {\bibfnamefont {J.~M.~M.}\ \bibnamefont
  {de~Freitas}}, \bibinfo {author} {\bibfnamefont {M.}~\bibnamefont {Holland}},
  \bibinfo {author} {\bibfnamefont {T.}~\bibnamefont {Kuna}}, \bibinfo {author}
  {\bibfnamefont {M.}~\bibnamefont {Nicol}}, \bibinfo {author} {\bibfnamefont
  {M.}~\bibnamefont {Todd}}, \ and\ \bibinfo {author} {\bibfnamefont
  {S.}~\bibnamefont {Vaienti}}} (\bibinfo {year} {2016}),\ \href@noop {} {\emph
  {\bibinfo {title} {Extremes and Recurrence in Dynamical Systems}}}\ (\bibinfo
   {publisher} {Wiley})\BibitemShut {NoStop}%
\bibitem [{\citenamefont {Lucarini}\ \emph
  {et~al.}(2010{\natexlab{a}})\citenamefont {Lucarini}, \citenamefont
  {Fraedrich},\ and\ \citenamefont {Lunkeit}}]{Luchyst}%
  \BibitemOpen
  \bibfield  {author} {\bibinfo {author} {\bibnamefont {Lucarini},
  \bibfnamefont {V}}, \bibinfo {author} {\bibfnamefont {K.}~\bibnamefont
  {Fraedrich}}, \ and\ \bibinfo {author} {\bibfnamefont {F.}~\bibnamefont
  {Lunkeit}}} (\bibinfo {year} {2010}{\natexlab{a}}),\ \bibfield  {title}
  {\enquote {\bibinfo {title} {Thermodynamic analysis of snowball earth
  hysteresis experiment: Efficiency, entropy production, and
  irreversibility},}\ }\href@noop {} {\bibfield  {journal} {\bibinfo  {journal}
  {Q. J. Royal Met. Soc.}\ }\textbf {\bibinfo {volume} {136}},\ \bibinfo
  {pages} {2--11}}\BibitemShut {NoStop}%
\bibitem [{\citenamefont {Lucarini}\ \emph
  {et~al.}(2010{\natexlab{b}})\citenamefont {Lucarini}, \citenamefont
  {Fraedrich},\ and\ \citenamefont {Lunkeit}}]{Lucarini2010b}%
  \BibitemOpen
  \bibfield  {author} {\bibinfo {author} {\bibnamefont {Lucarini},
  \bibfnamefont {V}}, \bibinfo {author} {\bibfnamefont {K.}~\bibnamefont
  {Fraedrich}}, \ and\ \bibinfo {author} {\bibfnamefont {F.}~\bibnamefont
  {Lunkeit}}} (\bibinfo {year} {2010}{\natexlab{b}}),\ \bibfield  {title}
  {\enquote {\bibinfo {title} {Thermodynamics of climate change: generalized
  sensitivities},}\ }\href {\doibase 10.5194/acp-10-9729-2010} {\bibfield
  {journal} {\bibinfo  {journal} {Atmospheric Chemistry and Physics}\ }\textbf
  {\bibinfo {volume} {10}}~(\bibinfo {number} {20}),\ \bibinfo {pages}
  {9729--9737}}\BibitemShut {NoStop}%
\bibitem [{\citenamefont {Lucarini}\ and\ \citenamefont
  {Gritsun}(2020)}]{LucariniG2019}%
  \BibitemOpen
  \bibfield  {author} {\bibinfo {author} {\bibnamefont {Lucarini},
  \bibfnamefont {V}}, \ and\ \bibinfo {author} {\bibfnamefont {A.}~\bibnamefont
  {Gritsun}}} (\bibinfo {year} {2020}),\ \bibfield  {title} {\enquote {\bibinfo
  {title} {A new mathematical framework for atmospheric blocking events},}\
  }\href {\doibase 10.1007/s00382-019-05018-2} {\bibfield  {journal} {\bibinfo
  {journal} {Climate Dynamics}\ }\textbf {\bibinfo {volume} {54}}~(\bibinfo
  {number} {1}),\ \bibinfo {pages} {575--598}}\BibitemShut {NoStop}%
\bibitem [{\citenamefont {Lucarini}\ \emph {et~al.}(2014b)\citenamefont
  {Lucarini}, \citenamefont {Kuna}, \citenamefont {Faranda},\ and\
  \citenamefont {Wouters}}]{LKFW14}%
  \BibitemOpen
  \bibfield  {author} {\bibinfo {author} {\bibnamefont {Lucarini},
  \bibfnamefont {V}}, \bibinfo {author} {\bibfnamefont {T.}~\bibnamefont
  {Kuna}}, \bibinfo {author} {\bibfnamefont {D.}~\bibnamefont {Faranda}}, \
  and\ \bibinfo {author} {\bibfnamefont {J.}~\bibnamefont {Wouters}}} (\bibinfo
  {year} {2014b}),\ \bibfield  {title} {\enquote {\bibinfo {title} {Towards a
  general theory of extremes for observables of chaotic dynamical systems},}\
  }\href@noop {} {\bibfield  {journal} {\bibinfo  {journal} {Journal of
  Statistical Physics}\ }\textbf {\bibinfo {volume} {154}}}\BibitemShut
  {NoStop}%
\bibitem [{\citenamefont {Lucarini}\ \emph {et~al.}(2013)\citenamefont
  {Lucarini}, \citenamefont {Pascale}, \citenamefont {Boschi}, \citenamefont
  {Kirk},\ and\ \citenamefont {Iro}}]{Lucarini2013a}%
  \BibitemOpen
  \bibfield  {author} {\bibinfo {author} {\bibnamefont {Lucarini},
  \bibfnamefont {V}}, \bibinfo {author} {\bibfnamefont {S.}~\bibnamefont
  {Pascale}}, \bibinfo {author} {\bibfnamefont {V.}~\bibnamefont {Boschi}},
  \bibinfo {author} {\bibfnamefont {E.}~\bibnamefont {Kirk}}, \ and\ \bibinfo
  {author} {\bibfnamefont {N.}~\bibnamefont {Iro}}} (\bibinfo {year} {2013}),\
  \bibfield  {title} {\enquote {\bibinfo {title} {{Habitability and
  multistability in earth-like planets}},}\ }\href {\doibase
  10.1002/asna.201311903} {\bibfield  {journal} {\bibinfo  {journal}
  {Astronomische Nachrichten}\ }\textbf {\bibinfo {volume} {334}}~(\bibinfo
  {number} {6}),\ \bibinfo {pages} {576--588}}\BibitemShut {NoStop}%
\bibitem [{\citenamefont {Lucarini}\ and\ \citenamefont
  {Ragone}(2011)}]{LucariniRagone}%
  \BibitemOpen
  \bibfield  {author} {\bibinfo {author} {\bibnamefont {Lucarini},
  \bibfnamefont {V}}, \ and\ \bibinfo {author} {\bibfnamefont {F}~\bibnamefont
  {Ragone}}} (\bibinfo {year} {2011}),\ \bibfield  {title} {\enquote {\bibinfo
  {title} {Energetics of climate models: Net energy balance and meridional
  enthalpy transport},}\ }\href {\doibase 10.1029/2009RG000323} {\bibfield
  {journal} {\bibinfo  {journal} {Rev. Geophys.}\ }\textbf {\bibinfo {volume}
  {49}},\ \bibinfo {pages} {RG1001}}\BibitemShut {NoStop}%
\bibitem [{\citenamefont {Lucarini}\ \emph {et~al.}(2017)\citenamefont
  {Lucarini}, \citenamefont {Ragone},\ and\ \citenamefont
  {Lunkeit}}]{Lucarini2017}%
  \BibitemOpen
  \bibfield  {author} {\bibinfo {author} {\bibnamefont {Lucarini},
  \bibfnamefont {V}}, \bibinfo {author} {\bibfnamefont {F.}~\bibnamefont
  {Ragone}}, \ and\ \bibinfo {author} {\bibfnamefont {F.}~\bibnamefont
  {Lunkeit}}} (\bibinfo {year} {2017}),\ \bibfield  {title} {\enquote {\bibinfo
  {title} {Predicting climate change using response theory: Global averages and
  spatial patterns},}\ }\href {\doibase 10.1007/s10955-016-1506-z} {\bibfield
  {journal} {\bibinfo  {journal} {Journal of Statistical Physics}\ }\textbf
  {\bibinfo {volume} {166}}~(\bibinfo {number} {3}),\ \bibinfo {pages}
  {1036--1064}}\BibitemShut {NoStop}%
\bibitem [{\citenamefont {Lucarini}\ \emph
  {et~al.}(2005{\natexlab{b}})\citenamefont {Lucarini}, \citenamefont
  {Saarinen}, \citenamefont {Peiponen},\ and\ \citenamefont
  {Vartiainen}}]{lucarini2005}%
  \BibitemOpen
  \bibfield  {author} {\bibinfo {author} {\bibnamefont {Lucarini},
  \bibfnamefont {V}}, \bibinfo {author} {\bibfnamefont {J.~J.}\ \bibnamefont
  {Saarinen}}, \bibinfo {author} {\bibfnamefont {K.-E.}\ \bibnamefont
  {Peiponen}}, \ and\ \bibinfo {author} {\bibfnamefont {E.~M.}\ \bibnamefont
  {Vartiainen}}} (\bibinfo {year} {2005}{\natexlab{b}}),\ \href@noop {} {\emph
  {\bibinfo {title} {Kramers-Kronig relations in Optical Materials Research}}}\
  (\bibinfo  {publisher} {Springer},\ \bibinfo {address} {New
  York})\BibitemShut {NoStop}%
\bibitem [{\citenamefont {Lucarini}\ and\ \citenamefont
  {Sarno}(2011)}]{Lucarini2011}%
  \BibitemOpen
  \bibfield  {author} {\bibinfo {author} {\bibnamefont {Lucarini},
  \bibfnamefont {V}}, \ and\ \bibinfo {author} {\bibfnamefont {S.}~\bibnamefont
  {Sarno}}} (\bibinfo {year} {2011}),\ \bibfield  {title} {\enquote {\bibinfo
  {title} {A statistical mechanical approach for the computation of the
  climatic response to general forcings},}\ }\href@noop {} {\bibfield
  {journal} {\bibinfo  {journal} {Nonlin. Processes Geophys}\ }\textbf
  {\bibinfo {volume} {18}},\ \bibinfo {pages} {7--28}}\BibitemShut {NoStop}%
\bibitem [{\citenamefont {Lucarini}\ and\ \citenamefont
  {Stone}(2005)}]{LucariniStone2005}%
  \BibitemOpen
  \bibfield  {author} {\bibinfo {author} {\bibnamefont {Lucarini},
  \bibfnamefont {Valerio}}, \ and\ \bibinfo {author} {\bibfnamefont {Peter~H.}\
  \bibnamefont {Stone}}} (\bibinfo {year} {2005}),\ \bibfield  {title}
  {\enquote {\bibinfo {title} {Thermohaline circulation stability: A box model
  study. part i: Uncoupled model},}\ }\href {\doibase 10.1175/JCLI-3278.1}
  {\bibfield  {journal} {\bibinfo  {journal} {Journal of Climate}\ }\textbf
  {\bibinfo {volume} {18}}~(\bibinfo {number} {4}),\ \bibinfo {pages}
  {501--513}},\ \Eprint
  {http://arxiv.org/abs/https://doi.org/10.1175/JCLI-3278.1}
  {https://doi.org/10.1175/JCLI-3278.1} \BibitemShut {NoStop}%
\bibitem [{\citenamefont {Lynch}(2008)}]{Lynch08}%
  \BibitemOpen
  \bibfield  {author} {\bibinfo {author} {\bibnamefont {Lynch}, \bibfnamefont
  {P}}} (\bibinfo {year} {2008}),\ \bibfield  {title} {\enquote {\bibinfo
  {title} {{The ENIAC Forecasts: A Re-creation}},}\ }\href {\doibase
  10.1175/BAMS-89-1-45} {\bibfield  {journal} {\bibinfo  {journal} {Bull. Am.
  Meteorol. Soc.}\ }\textbf {\bibinfo {volume} {89}}~(\bibinfo {number} {1}),\
  \bibinfo {pages} {45--55}}\BibitemShut {NoStop}%
\bibitem [{\citenamefont {Madden}\ and\ \citenamefont
  {Julian}(1971)}]{Mad.Jul.71}%
  \BibitemOpen
  \bibfield  {author} {\bibinfo {author} {\bibnamefont {Madden}, \bibfnamefont
  {RA}}, \ and\ \bibinfo {author} {\bibfnamefont {P.R.}\ \bibnamefont
  {Julian}}} (\bibinfo {year} {1971}),\ \bibfield  {title} {\enquote {\bibinfo
  {title} {Detection of a 40-50 day oscillation in the zonal wind in the
  {Tropical Pacific}},}\ }\href@noop {} {\bibfield  {journal} {\bibinfo
  {journal} {J. Atmos. Sci.}\ }\textbf {\bibinfo {volume} {28}},\ \bibinfo
  {pages} {702--708}}\BibitemShut {NoStop}%
\bibitem [{\citenamefont {Majda}\ \emph {et~al.}(2009)\citenamefont {Majda},
  \citenamefont {Gershgorin},\ and\ \citenamefont {Yuan}}]{majda2010b}%
  \BibitemOpen
  \bibfield  {author} {\bibinfo {author} {\bibnamefont {Majda}, \bibfnamefont
  {A~J}}, \bibinfo {author} {\bibfnamefont {B.}~\bibnamefont {Gershgorin}}, \
  and\ \bibinfo {author} {\bibfnamefont {Y.}~\bibnamefont {Yuan}}} (\bibinfo
  {year} {2009}),\ \bibfield  {title} {\enquote {\bibinfo {title}
  {Low-frequency climate response and fluctuation--dissipation theorems: Theory
  and practice},}\ }\bibfield  {booktitle} {\emph {\bibinfo {booktitle}
  {Journal of the Atmospheric Sciences}},\ }\href@noop {} {\bibfield  {journal}
  {\bibinfo  {journal} {Journal of the Atmospheric Sciences}\ }\textbf
  {\bibinfo {volume} {67}}~(\bibinfo {number} {4}),\ \bibinfo {pages}
  {1186--1201}}\BibitemShut {NoStop}%
\bibitem [{\citenamefont {Majda}\ and\ \citenamefont
  {Wang}(2006)}]{MajdaWangBook}%
  \BibitemOpen
  \bibfield  {author} {\bibinfo {author} {\bibnamefont {Majda}, \bibfnamefont
  {A~J}}, \ and\ \bibinfo {author} {\bibfnamefont {X.}~\bibnamefont {Wang}}}
  (\bibinfo {year} {2006}),\ \href@noop {} {\emph {\bibinfo {title} {{Nonlinear
  Dynamics and Statistical Theories for Basic Geophysical Flows}}}}\ (\bibinfo
  {publisher} {Cambridge University Press},\ \bibinfo {address}
  {Cambridge})\BibitemShut {NoStop}%
\bibitem [{\citenamefont {Maloney}\ and\ \citenamefont
  {Hartmann}(2000)}]{Mal.Hart.00}%
  \BibitemOpen
  \bibfield  {author} {\bibinfo {author} {\bibnamefont {Maloney}, \bibfnamefont
  {E~D}}, \ and\ \bibinfo {author} {\bibfnamefont {D.~L.}\ \bibnamefont
  {Hartmann}}} (\bibinfo {year} {2000}),\ \bibfield  {title} {\enquote
  {\bibinfo {title} {{Modulation of hurricane activity in the Gulf of Mexico by
  the Madden-Julian oscillation}},}\ }\href@noop {} {\bibfield  {journal}
  {\bibinfo  {journal} {Science}\ }\textbf {\bibinfo {volume} {287}},\ \bibinfo
  {pages} {2002--2004}}\BibitemShut {NoStop}%
\bibitem [{\citenamefont {Manabe}\ and\ \citenamefont
  {Strickler}(1964)}]{Manabe1964}%
  \BibitemOpen
  \bibfield  {author} {\bibinfo {author} {\bibnamefont {Manabe}, \bibfnamefont
  {Syukuro}}, \ and\ \bibinfo {author} {\bibfnamefont {Robert~F.}\ \bibnamefont
  {Strickler}}} (\bibinfo {year} {1964}),\ \bibfield  {title} {\enquote
  {\bibinfo {title} {Thermal equilibrium of the atmosphere with a convective
  adjustment},}\ }\href {\doibase
  10.1175/1520-0469(1964)021<0361:TEOTAW>2.0.CO;2} {\bibfield  {journal}
  {\bibinfo  {journal} {Journal of the Atmospheric Sciences}\ }\textbf
  {\bibinfo {volume} {21}}~(\bibinfo {number} {4}),\ \bibinfo {pages}
  {361--385}}\BibitemShut {NoStop}%
\bibitem [{\citenamefont {Mann}\ \emph {et~al.}(1998)\citenamefont {Mann},
  \citenamefont {Bradley},\ and\ \citenamefont {Hughes}}]{Mann1998}%
  \BibitemOpen
  \bibfield  {author} {\bibinfo {author} {\bibnamefont {Mann}, \bibfnamefont
  {M~E}}, \bibinfo {author} {\bibfnamefont {R.~S.}\ \bibnamefont {Bradley}}, \
  and\ \bibinfo {author} {\bibfnamefont {M.~K.}\ \bibnamefont {Hughes}}}
  (\bibinfo {year} {1998}),\ \bibfield  {title} {\enquote {\bibinfo {title}
  {{Global-scale temperature patterns and climate forcing over the past six
  centuries}},}\ }\href@noop {} {\bibfield  {journal} {\bibinfo  {journal}
  {Nature}\ }\textbf {\bibinfo {volume} {392}},\ \bibinfo {pages}
  {779--787}}\BibitemShut {NoStop}%
\bibitem [{\citenamefont {Mann}\ \emph {et~al.}(1999)\citenamefont {Mann},
  \citenamefont {Bradley},\ and\ \citenamefont {Hughes}}]{mann99}%
  \BibitemOpen
  \bibfield  {author} {\bibinfo {author} {\bibnamefont {Mann}, \bibfnamefont
  {M~E}}, \bibinfo {author} {\bibfnamefont {R.~S.}\ \bibnamefont {Bradley}}, \
  and\ \bibinfo {author} {\bibfnamefont {M.~K.}\ \bibnamefont {Hughes}}}
  (\bibinfo {year} {1999}),\ \bibfield  {title} {\enquote {\bibinfo {title}
  {Northern {Hemisphere temperatures during the past millennium: Inferences,
  uncertainties, and limitations}},}\ }\href {\doibase 10.1029/1999GL900070}
  {\bibfield  {journal} {\bibinfo  {journal} {Geophys.~Res.~Lett.}\ }\textbf
  {\bibinfo {volume} {26}}~(\bibinfo {number} {6}),\ \bibinfo {pages}
  {759--762}}\BibitemShut {NoStop}%
\bibitem [{\citenamefont {Mann}\ \emph {et~al.}(2008)\citenamefont {Mann},
  \citenamefont {Zhang}, \citenamefont {Hughes}, \citenamefont {Bradley},
  \citenamefont {Miller},\ and\ \citenamefont {Rutherford}}]{mann08}%
  \BibitemOpen
  \bibfield  {author} {\bibinfo {author} {\bibnamefont {Mann}, \bibfnamefont
  {M~E}}, \bibinfo {author} {\bibfnamefont {Z.}~\bibnamefont {Zhang}}, \bibinfo
  {author} {\bibfnamefont {M.~K.}\ \bibnamefont {Hughes}}, \bibinfo {author}
  {\bibfnamefont {R.~S.}\ \bibnamefont {Bradley}}, \bibinfo {author}
  {\bibfnamefont {S.~K.}\ \bibnamefont {Miller}}, \ and\ \bibinfo {author}
  {\bibfnamefont {F.}~\bibnamefont {Rutherford}, \bibfnamefont {S.and~Ni}}}
  (\bibinfo {year} {2008}),\ \bibfield  {title} {\enquote {\bibinfo {title}
  {Proxy-based reconstructions of hemispheric and global surface temperature
  variations over the past two millennia},}\ }\href {\doibase
  10.1073/pnas.0805721105} {\bibfield  {journal} {\bibinfo  {journal} {Proc.
  Natl. Acad. Sci. USA}\ }\textbf {\bibinfo {volume} {105}}~(\bibinfo {number}
  {36}),\ \bibinfo {pages} {13252--13257}}\BibitemShut {NoStop}%
\bibitem [{\citenamefont {Marangio}\ \emph {et~al.}(2019)\citenamefont
  {Marangio}, \citenamefont {Sedro}, \citenamefont {Galatolo}, \citenamefont
  {Di~Garbo},\ and\ \citenamefont {Ghil}}]{Galatolo.ea.2019}%
  \BibitemOpen
  \bibfield  {author} {\bibinfo {author} {\bibnamefont {Marangio},
  \bibfnamefont {L}}, \bibinfo {author} {\bibfnamefont {J.}~\bibnamefont
  {Sedro}}, \bibinfo {author} {\bibfnamefont {S.}~\bibnamefont {Galatolo}},
  \bibinfo {author} {\bibfnamefont {A.}~\bibnamefont {Di~Garbo}}, \ and\
  \bibinfo {author} {\bibfnamefont {M.}~\bibnamefont {Ghil}}} (\bibinfo {year}
  {2019}),\ \bibfield  {title} {\enquote {\bibinfo {title} {Arnold maps with
  noise: {Differentiability and non-monotonicity of the rotation number}},}\
  }\href@noop {} {\bibinfo  {journal} {arXiv preprint arXiv:1904.11744}\
  }\BibitemShut {NoStop}%
\bibitem [{\citenamefont {Marconi}\ \emph {et~al.}(2008)\citenamefont
  {Marconi}, \citenamefont {Puglisi}, \citenamefont {Rondoni},\ and\
  \citenamefont {Vulpiani}}]{marconi2008}%
  \BibitemOpen
\bibfield  {journal} {  }\bibfield  {author} {\bibinfo {author} {\bibnamefont
  {Marconi}, \bibfnamefont {U~Marini~Bettolo}}, \bibinfo {author}
  {\bibfnamefont {A.}~\bibnamefont {Puglisi}}, \bibinfo {author} {\bibfnamefont
  {L.}~\bibnamefont {Rondoni}}, \ and\ \bibinfo {author} {\bibfnamefont
  {A.}~\bibnamefont {Vulpiani}}} (\bibinfo {year} {2008}),\ \bibfield  {title}
  {\enquote {\bibinfo {title} {Fluctuation-dissipation: Response theory in
  statistical physics},}\ }\href@noop {} {\bibfield  {journal} {\bibinfo
  {journal} {Phys. Rep.}\ }\textbf {\bibinfo {volume} {461}},\ \bibinfo {pages}
  {111}}\BibitemShut {NoStop}%
\bibitem [{\citenamefont {Marotzke}(2000)}]{Marotzke2000}%
  \BibitemOpen
  \bibfield  {author} {\bibinfo {author} {\bibnamefont {Marotzke},
  \bibfnamefont {J}}} (\bibinfo {year} {2000}),\ \bibfield  {title} {\enquote
  {\bibinfo {title} {{Abrupt climate change and thermohaline circulation:
  Mechanisms and predictability}},}\ }\href@noop {} {\bibfield  {journal}
  {\bibinfo  {journal} {Proc. Natl. Acad. Sci. USA}\ }\textbf {\bibinfo
  {volume} {97}},\ \bibinfo {pages} {1347--1350}}\BibitemShut {NoStop}%
\bibitem [{\citenamefont {Marques}\ \emph {et~al.}(2010)\citenamefont
  {Marques}, \citenamefont {Rocha},\ and\ \citenamefont
  {Corte-Real}}]{Marques10}%
  \BibitemOpen
  \bibfield  {author} {\bibinfo {author} {\bibnamefont {Marques}, \bibfnamefont
  {CAF}}, \bibinfo {author} {\bibfnamefont {A}~\bibnamefont {Rocha}}, \ and\
  \bibinfo {author} {\bibfnamefont {J}~\bibnamefont {Corte-Real}}} (\bibinfo
  {year} {2010}),\ \bibfield  {title} {\enquote {\bibinfo {title} {Comparative
  energetic of {ERA}-40, {JRA}-25 and {NCEP}-{R}2 reanalysis, in the wave
  number domai},}\ }\href@noop {} {\bibfield  {journal} {\bibinfo  {journal}
  {Dyn Atmos Oceans}\ }\textbf {\bibinfo {volume} {50}},\ \bibinfo {pages}
  {375--399}}\BibitemShut {NoStop}%
\bibitem [{\citenamefont {Marques}\ \emph {et~al.}(2009)\citenamefont
  {Marques}, \citenamefont {Rocha}, \citenamefont {Corte-Real}, \citenamefont
  {Castanheira}, \citenamefont {Ferreira},\ and\ \citenamefont
  {Melo-Goncalves}}]{Marques09}%
  \BibitemOpen
  \bibfield  {author} {\bibinfo {author} {\bibnamefont {Marques}, \bibfnamefont
  {CAF}}, \bibinfo {author} {\bibfnamefont {A}~\bibnamefont {Rocha}}, \bibinfo
  {author} {\bibfnamefont {J}~\bibnamefont {Corte-Real}}, \bibinfo {author}
  {\bibfnamefont {JM}~\bibnamefont {Castanheira}}, \bibinfo {author}
  {\bibfnamefont {J}~\bibnamefont {Ferreira}}, \ and\ \bibinfo {author}
  {\bibfnamefont {P}~\bibnamefont {Melo-Goncalves}}} (\bibinfo {year} {2009}),\
  \bibfield  {title} {\enquote {\bibinfo {title} {Global atmospheric energetic
  from {NCEP}-reanalysis 2 and {ECMWF}-{ERA}40 reanalysis},}\ }\href@noop {}
  {\bibfield  {journal} {\bibinfo  {journal} {Int. J. Climatol.}\ }\textbf
  {\bibinfo {volume} {29}},\ \bibinfo {pages} {159--174}}\BibitemShut {NoStop}%
\bibitem [{\citenamefont {Marshall}\ \emph {et~al.}(1997)\citenamefont
  {Marshall}, \citenamefont {Hill}, \citenamefont {Perelman},\ and\
  \citenamefont {Adcroft}}]{Marshall1997a}%
  \BibitemOpen
  \bibfield  {author} {\bibinfo {author} {\bibnamefont {Marshall},
  \bibfnamefont {J}}, \bibinfo {author} {\bibfnamefont {C.}~\bibnamefont
  {Hill}}, \bibinfo {author} {\bibfnamefont {L.}~\bibnamefont {Perelman}}, \
  and\ \bibinfo {author} {\bibfnamefont {A..}\ \bibnamefont {Adcroft}}}
  (\bibinfo {year} {1997}),\ \bibfield  {title} {\enquote {\bibinfo {title}
  {Hydrostatic, quasi-hydrostatic, and non-hydrostatic ocean modelling},}\
  }\href@noop {} {\bibfield  {journal} {\bibinfo  {journal} {J.\ Geophys.
  Res.}\ }\textbf {\bibinfo {volume} {102}},\ \bibinfo {pages}
  {5733--5752}}\BibitemShut {NoStop}%
\bibitem [{\citenamefont {Marshall}\ and\ \citenamefont
  {Molteni}(1993)}]{Mar.Mol.1993}%
  \BibitemOpen
  \bibfield  {author} {\bibinfo {author} {\bibnamefont {Marshall},
  \bibfnamefont {J}}, \ and\ \bibinfo {author} {\bibfnamefont {F.}~\bibnamefont
  {Molteni}}} (\bibinfo {year} {1993}),\ \bibfield  {title} {\enquote {\bibinfo
  {title} {Toward a dynamical understanding of atmospheric weather regimes},}\
  }\href@noop {} {\bibfield  {journal} {\bibinfo  {journal} {J. Atmos. Sci.}\
  }\textbf {\bibinfo {volume} {50}},\ \bibinfo {pages}
  {1993--2014}}\BibitemShut {NoStop}%
\bibitem [{\citenamefont {Martinson}\ \emph {et~al.}(1995)\citenamefont
  {Martinson}, \citenamefont {Bryan}, \citenamefont {Ghil}, \citenamefont
  {Hall}, \citenamefont {Karl}, \citenamefont {Sarachik}, \citenamefont
  {Sorooshian},\ and\ \citenamefont {Talley}}]{NRC1995}%
  \BibitemOpen
  \bibinfo {editor} {\bibnamefont {Martinson}, \bibfnamefont {D~G}}, \bibinfo
  {editor} {\bibfnamefont {K.}~\bibnamefont {Bryan}}, \bibinfo {editor}
  {\bibfnamefont {M.}~\bibnamefont {Ghil}}, \bibinfo {editor} {\bibfnamefont
  {M.~M.}\ \bibnamefont {Hall}}, \bibinfo {editor} {\bibfnamefont {T.~R.}\
  \bibnamefont {Karl}}, \bibinfo {editor} {\bibfnamefont {E.~S.}\ \bibnamefont
  {Sarachik}}, \bibinfo {editor} {\bibfnamefont {S.}~\bibnamefont
  {Sorooshian}}, \ and\ \bibinfo {editor} {\bibfnamefont {L.~D.}\ \bibnamefont
  {Talley}},\ Eds. (\bibinfo {year} {1995}),\ \href@noop {} {\emph {\bibinfo
  {title} {{Natural Climate Variability on Decade-to-Century Time Scales}}}}\
  (\bibinfo  {publisher} {National Academies Press},\ \bibinfo {address}
  {Washington, DC, U.S.A.})\BibitemShut {NoStop}%
\bibitem [{\citenamefont {McIntyre}\ and\ \citenamefont
  {McKitrick}(2005)}]{McMc05}%
  \BibitemOpen
  \bibfield  {author} {\bibinfo {author} {\bibnamefont {McIntyre},
  \bibfnamefont {S}}, \ and\ \bibinfo {author} {\bibfnamefont {R.}~\bibnamefont
  {McKitrick}}} (\bibinfo {year} {2005}),\ \bibfield  {title} {\enquote
  {\bibinfo {title} {Hockey sticks, principal components, and spurious
  significance},}\ }\href {\doibase 10.1029/2004GL021750} {\bibfield  {journal}
  {\bibinfo  {journal} {Geophys. Res. Lett.}\ }\textbf {\bibinfo {volume}
  {32}}~(\bibinfo {number} {3}),\ \bibinfo {pages} {n/a--n/a}}\BibitemShut
  {NoStop}%
\bibitem [{\citenamefont {McWilliams}(2006)}]{McW06}%
  \BibitemOpen
  \bibfield  {author} {\bibinfo {author} {\bibnamefont {McWilliams},
  \bibfnamefont {J~C}}} (\bibinfo {year} {2006}),\ \href@noop {} {\emph
  {\bibinfo {title} {Fundamentals of Geophysical Fluids}}}\ (\bibinfo
  {publisher} {Cambridge University Press},\ \bibinfo {address} {Cambridge, UK
  and New York, NY, USA})\BibitemShut {NoStop}%
\bibitem [{\citenamefont {McWilliams}(2019)}]{JCM.2019}%
  \BibitemOpen
  \bibfield  {author} {\bibinfo {author} {\bibnamefont {McWilliams},
  \bibfnamefont {J~C}}} (\bibinfo {year} {2019}),\ \bibfield  {title} {\enquote
  {\bibinfo {title} {A perspective on the legacy of {Edward Lorenz}},}\
  }\href@noop {} {\bibfield  {journal} {\bibinfo  {journal} {Earth and Space
  Sciences}\ }}\bibinfo {note} {{in press}}\BibitemShut {NoStop}%
\bibitem [{\citenamefont {Meacham}(2000)}]{Meacham2000}%
  \BibitemOpen
  \bibfield  {author} {\bibinfo {author} {\bibnamefont {Meacham}, \bibfnamefont
  {S~P}}} (\bibinfo {year} {2000}),\ \bibfield  {title} {\enquote {\bibinfo
  {title} {Low frequency variability of the wind-driven circulation},}\
  }\href@noop {} {\bibfield  {journal} {\bibinfo  {journal} {J.\ Phys.\
  Oceanogr.}\ }\textbf {\bibinfo {volume} {30}},\ \bibinfo {pages}
  {269--293}}\BibitemShut {NoStop}%
\bibitem [{\citenamefont {Meehl}\ \emph {et~al.}(2014)\citenamefont {Meehl},
  \citenamefont {Goddard}, \citenamefont {Boer}, \citenamefont {Burgman},
  \citenamefont {Branstator}, \citenamefont {Cassou}, \citenamefont {Corti},
  \citenamefont {Danabasoglu}, \citenamefont {Doblas-Reyes}, \citenamefont
  {Hawkins}, \citenamefont {Karspeck}, \citenamefont {Kimoto}, \citenamefont
  {Kumar}, \citenamefont {Matei}, \citenamefont {Mignot}, \citenamefont
  {Msadek}, \citenamefont {Navarra}, \citenamefont {Pohlmann}, \citenamefont
  {Rienecker}, \citenamefont {Rosati}, \citenamefont {Schneider}, \citenamefont
  {Smith}, \citenamefont {Sutton}, \citenamefont {Teng}, \citenamefont {van
  Oldenborgh}, \citenamefont {Vecchi},\ and\ \citenamefont
  {Yeager}}]{Meehl2014}%
  \BibitemOpen
  \bibfield  {author} {\bibinfo {author} {\bibnamefont {Meehl}, \bibfnamefont
  {G~A}}, \bibinfo {author} {\bibfnamefont {L.}~\bibnamefont {Goddard}},
  \bibinfo {author} {\bibfnamefont {G.}~\bibnamefont {Boer}}, \bibinfo {author}
  {\bibfnamefont {R.}~\bibnamefont {Burgman}}, \bibinfo {author} {\bibfnamefont
  {G.}~\bibnamefont {Branstator}}, \bibinfo {author} {\bibfnamefont
  {C.}~\bibnamefont {Cassou}}, \bibinfo {author} {\bibfnamefont
  {S.}~\bibnamefont {Corti}}, \bibinfo {author} {\bibfnamefont
  {G.}~\bibnamefont {Danabasoglu}}, \bibinfo {author} {\bibfnamefont
  {F.}~\bibnamefont {Doblas-Reyes}}, \bibinfo {author} {\bibfnamefont
  {E.}~\bibnamefont {Hawkins}}, \bibinfo {author} {\bibfnamefont
  {A.}~\bibnamefont {Karspeck}}, \bibinfo {author} {\bibfnamefont
  {M.}~\bibnamefont {Kimoto}}, \bibinfo {author} {\bibfnamefont
  {A.}~\bibnamefont {Kumar}}, \bibinfo {author} {\bibfnamefont
  {D.}~\bibnamefont {Matei}}, \bibinfo {author} {\bibfnamefont
  {J.}~\bibnamefont {Mignot}}, \bibinfo {author} {\bibfnamefont
  {R.}~\bibnamefont {Msadek}}, \bibinfo {author} {\bibfnamefont
  {A.}~\bibnamefont {Navarra}}, \bibinfo {author} {\bibfnamefont
  {H.}~\bibnamefont {Pohlmann}}, \bibinfo {author} {\bibfnamefont
  {M.}~\bibnamefont {Rienecker}}, \bibinfo {author} {\bibfnamefont
  {T.}~\bibnamefont {Rosati}}, \bibinfo {author} {\bibfnamefont
  {E.}~\bibnamefont {Schneider}}, \bibinfo {author} {\bibfnamefont
  {D.}~\bibnamefont {Smith}}, \bibinfo {author} {\bibfnamefont
  {R.}~\bibnamefont {Sutton}}, \bibinfo {author} {\bibfnamefont
  {H.}~\bibnamefont {Teng}}, \bibinfo {author} {\bibfnamefont {G.~J.}\
  \bibnamefont {van Oldenborgh}}, \bibinfo {author} {\bibfnamefont
  {G.}~\bibnamefont {Vecchi}}, \ and\ \bibinfo {author} {\bibfnamefont
  {S.}~\bibnamefont {Yeager}}} (\bibinfo {year} {2014}),\ \bibfield  {title}
  {\enquote {\bibinfo {title} {Decadal climate prediction: {An} update from the
  trenches},}\ }\href {\doibase 10.1175/BAMS-D-12-00241.1} {\bibfield
  {journal} {\bibinfo  {journal} {Bulletin of the American Meteorological
  Society}\ }\textbf {\bibinfo {volume} {95}}~(\bibinfo {number} {2}),\
  \bibinfo {pages} {243--267}}\BibitemShut {NoStop}%
\bibitem [{\citenamefont {Melbourne}\ and\ \citenamefont
  {Stuart}(2011)}]{Mel.Stuart.2011}%
  \BibitemOpen
  \bibfield  {author} {\bibinfo {author} {\bibnamefont {Melbourne},
  \bibfnamefont {I}}, \ and\ \bibinfo {author} {\bibfnamefont {A.~M.}\
  \bibnamefont {Stuart}}} (\bibinfo {year} {2011}),\ \bibfield  {title}
  {\enquote {\bibinfo {title} {A note on diffusion limits of chaotic
  skew-product flows},}\ }\href@noop {} {\bibfield  {journal} {\bibinfo
  {journal} {Nonlinearity}\ }\textbf {\bibinfo {volume} {24}}~(\bibinfo
  {number} {4}),\ \bibinfo {pages} {1361--1367}}\BibitemShut {NoStop}%
\bibitem [{\citenamefont {Members}\ \emph {et~al.}(2012)\citenamefont
  {Members}, \citenamefont {Rohling}, \citenamefont {Rohling}, \citenamefont
  {Sluijs}, \citenamefont {Dijkstra}, \citenamefont {K{\"o}hler}, \citenamefont
  {van~de Wal}, \citenamefont {von~der Heydt}, \citenamefont {Beerling},
  \citenamefont {Berger}, \citenamefont {Bijl}, \citenamefont {Crucifix},
  \citenamefont {DeConto}, \citenamefont {Drijfhout}, \citenamefont {Fedorov},
  \citenamefont {Foster}, \citenamefont {Ganopolski}, \citenamefont {Hansen},
  \citenamefont {H{\"o}nisch}, \citenamefont {Hooghiemstra}, \citenamefont
  {Huber}, \citenamefont {Huybers}, \citenamefont {Knutti}, \citenamefont
  {Lea}, \citenamefont {Lourens}, \citenamefont {Lunt}, \citenamefont
  {Masson-Delmotte}, \citenamefont {Medina-Elizalde}, \citenamefont
  {Otto-Bliesner}, \citenamefont {Pagani}, \citenamefont {P{\"a}like},
  \citenamefont {Renssen}, \citenamefont {Royer}, \citenamefont {Siddall},
  \citenamefont {Valdes}, \citenamefont {Zachos},\ and\ \citenamefont
  {Zeebe}}]{PALEOSENS}%
  \BibitemOpen
  \bibfield  {author} {\bibinfo {author} {\bibnamefont {Members}, \bibfnamefont
  {PALAEOSENS~Project}}, \bibinfo {author} {\bibfnamefont {E.~J.}\ \bibnamefont
  {Rohling}}, \bibinfo {author} {\bibfnamefont {E.~J.}\ \bibnamefont
  {Rohling}}, \bibinfo {author} {\bibfnamefont {A.}~\bibnamefont {Sluijs}},
  \bibinfo {author} {\bibfnamefont {H.~A.}\ \bibnamefont {Dijkstra}}, \bibinfo
  {author} {\bibfnamefont {P.}~\bibnamefont {K{\"o}hler}}, \bibinfo {author}
  {\bibfnamefont {R.~S.~W.}\ \bibnamefont {van~de Wal}}, \bibinfo {author}
  {\bibfnamefont {A.~S.}\ \bibnamefont {von~der Heydt}}, \bibinfo {author}
  {\bibfnamefont {D.~J.}\ \bibnamefont {Beerling}}, \bibinfo {author}
  {\bibfnamefont {A.}~\bibnamefont {Berger}}, \bibinfo {author} {\bibfnamefont
  {P.~K.}\ \bibnamefont {Bijl}}, \bibinfo {author} {\bibfnamefont
  {M.}~\bibnamefont {Crucifix}}, \bibinfo {author} {\bibfnamefont
  {R.}~\bibnamefont {DeConto}}, \bibinfo {author} {\bibfnamefont {S.~S.}\
  \bibnamefont {Drijfhout}}, \bibinfo {author} {\bibfnamefont {A.}~\bibnamefont
  {Fedorov}}, \bibinfo {author} {\bibfnamefont {G.~L.}\ \bibnamefont {Foster}},
  \bibinfo {author} {\bibfnamefont {A.}~\bibnamefont {Ganopolski}}, \bibinfo
  {author} {\bibfnamefont {J.}~\bibnamefont {Hansen}}, \bibinfo {author}
  {\bibfnamefont {B.}~\bibnamefont {H{\"o}nisch}}, \bibinfo {author}
  {\bibfnamefont {H.}~\bibnamefont {Hooghiemstra}}, \bibinfo {author}
  {\bibfnamefont {M.}~\bibnamefont {Huber}}, \bibinfo {author} {\bibfnamefont
  {P.}~\bibnamefont {Huybers}}, \bibinfo {author} {\bibfnamefont
  {R.}~\bibnamefont {Knutti}}, \bibinfo {author} {\bibfnamefont {D.~W.}\
  \bibnamefont {Lea}}, \bibinfo {author} {\bibfnamefont {L.~J.}\ \bibnamefont
  {Lourens}}, \bibinfo {author} {\bibfnamefont {D.}~\bibnamefont {Lunt}},
  \bibinfo {author} {\bibfnamefont {V.}~\bibnamefont {Masson-Delmotte}},
  \bibinfo {author} {\bibfnamefont {M.}~\bibnamefont {Medina-Elizalde}},
  \bibinfo {author} {\bibfnamefont {B.}~\bibnamefont {Otto-Bliesner}}, \bibinfo
  {author} {\bibfnamefont {M.}~\bibnamefont {Pagani}}, \bibinfo {author}
  {\bibfnamefont {H.}~\bibnamefont {P{\"a}like}}, \bibinfo {author}
  {\bibfnamefont {H.}~\bibnamefont {Renssen}}, \bibinfo {author} {\bibfnamefont
  {D.~L.}\ \bibnamefont {Royer}}, \bibinfo {author} {\bibfnamefont
  {M.}~\bibnamefont {Siddall}}, \bibinfo {author} {\bibfnamefont
  {P.}~\bibnamefont {Valdes}}, \bibinfo {author} {\bibfnamefont {J.~C.}\
  \bibnamefont {Zachos}}, \ and\ \bibinfo {author} {\bibfnamefont {R.~E.}\
  \bibnamefont {Zeebe}}} (\bibinfo {year} {2012}),\ \bibfield  {title}
  {\enquote {\bibinfo {title} {Making sense of palaeoclimate sensitivity},}\
  }\href {\doibase 10.1038/nature11574} {\bibfield  {journal} {\bibinfo
  {journal} {Nature}\ }\textbf {\bibinfo {volume} {491}},\ \bibinfo {pages}
  {683 EP --}}\BibitemShut {NoStop}%
\bibitem [{\citenamefont {Merkin}\ \emph {et~al.}(2016)\citenamefont {Merkin},
  \citenamefont {Kondrashov}, \citenamefont {Ghil},\ and\ \citenamefont
  {Anderson}}]{Merkin.ea.16}%
  \BibitemOpen
  \bibfield  {author} {\bibinfo {author} {\bibnamefont {Merkin}, \bibfnamefont
  {V~G}}, \bibinfo {author} {\bibfnamefont {D.}~\bibnamefont {Kondrashov}},
  \bibinfo {author} {\bibfnamefont {M.}~\bibnamefont {Ghil}}, \ and\ \bibinfo
  {author} {\bibfnamefont {B.~J.}\ \bibnamefont {Anderson}}} (\bibinfo {year}
  {2016}),\ \bibfield  {title} {\enquote {\bibinfo {title} {Data assimilation
  of low-altitude magnetic perturbations into a global magnetosphere model},}\
  }\href@noop {} {\bibfield  {journal} {\bibinfo  {journal} {Space Weather}\
  }\textbf {\bibinfo {volume} {14}},\ \bibinfo {pages} {165--184}}\BibitemShut
  {NoStop}%
\bibitem [{\citenamefont {Merryfield}\ \emph {et~al.}(2020)\citenamefont
  {Merryfield}, \citenamefont {Baehr}, \citenamefont {Batt{\'e}}, \citenamefont
  {Becker}, \citenamefont {Butler}, \citenamefont {Coelho}, \citenamefont
  {Danabasoglu}, \citenamefont {Dirmeyer}, \citenamefont {Doblas-Reyes},
  \citenamefont {Domeisen} \emph {et~al.}}]{Merryfield.ea.2020}%
  \BibitemOpen
  \bibfield  {author} {\bibinfo {author} {\bibnamefont {Merryfield},
  \bibfnamefont {William~J}}, \bibinfo {author} {\bibfnamefont {Johanna}\
  \bibnamefont {Baehr}}, \bibinfo {author} {\bibfnamefont {Lauriane}\
  \bibnamefont {Batt{\'e}}}, \bibinfo {author} {\bibfnamefont {Emily~J.}\
  \bibnamefont {Becker}}, \bibinfo {author} {\bibfnamefont {Amy~H.}\
  \bibnamefont {Butler}}, \bibinfo {author} {\bibfnamefont {Caio A.~S.}\
  \bibnamefont {Coelho}}, \bibinfo {author} {\bibfnamefont {Gokhan}\
  \bibnamefont {Danabasoglu}}, \bibinfo {author} {\bibfnamefont {Paul~A.}\
  \bibnamefont {Dirmeyer}}, \bibinfo {author} {\bibfnamefont {Francisco~J.}\
  \bibnamefont {Doblas-Reyes}}, \bibinfo {author} {\bibfnamefont {Daniela
  I.~V.}\ \bibnamefont {Domeisen}},  \emph {et~al.}} (\bibinfo {year} {2020}),\
  \bibfield  {title} {\enquote {\bibinfo {title} {Current and emerging
  developments in subseasonal to decadal prediction},}\ }\href@noop {}
  {\bibfield  {journal} {\bibinfo  {journal} {Bulletin of the American
  Meteorological Society}\ }~(\bibinfo {number} {2020})}\BibitemShut {NoStop}%
\bibitem [{\citenamefont {Mitchell}(1976)}]{Mitchell1976}%
  \BibitemOpen
  \bibfield  {author} {\bibinfo {author} {\bibnamefont {Mitchell},
  \bibfnamefont {J~M}}} (\bibinfo {year} {1976}),\ \bibfield  {title} {\enquote
  {\bibinfo {title} {{An overview of climate variability and its causal
  mechanisms}},}\ }\href@noop {} {\bibfield  {journal} {\bibinfo  {journal}
  {Quatern. Res.}\ }\textbf {\bibinfo {volume} {6}},\ \bibinfo {pages}
  {481--493}}\BibitemShut {NoStop}%
\bibitem [{\citenamefont {Mo}\ and\ \citenamefont {Ghil}(1987)}]{Mo.Ghil.1987}%
  \BibitemOpen
  \bibfield  {author} {\bibinfo {author} {\bibnamefont {Mo}, \bibfnamefont
  {KC}}, \ and\ \bibinfo {author} {\bibfnamefont {M.}~\bibnamefont {Ghil}}}
  (\bibinfo {year} {1987}),\ \bibfield  {title} {\enquote {\bibinfo {title}
  {Statistics and dynamics of persistent anomalies},}\ }\href {\doibase
  10.1175/1520-0469(1987)044<0877:sadopa>2.0.co;2} {\bibfield  {journal}
  {\bibinfo  {journal} {{J. Atmos. Sci.}}\ }\textbf {\bibinfo {volume}
  {44}}~(\bibinfo {number} {5}),\ \bibinfo {pages} {877--902}}\BibitemShut
  {NoStop}%
\bibitem [{\citenamefont {Molteni}(2002)}]{Molteni2002}%
  \BibitemOpen
  \bibfield  {author} {\bibinfo {author} {\bibnamefont {Molteni}, \bibfnamefont
  {F~M}}} (\bibinfo {year} {2002}),\ \bibfield  {title} {\enquote {\bibinfo
  {title} {Weather regimes and multiple equilibria},}\ }in\ \href@noop {}
  {\emph {\bibinfo {booktitle} {{Encyclopedia of Atmospheric Sciences}}}},\
  \bibinfo {editor} {edited by\ \bibinfo {editor} {\bibfnamefont {J.~R.}\
  \bibnamefont {Holton}}}\ (\bibinfo  {publisher} {Academic Press})\ pp.\
  \bibinfo {pages} {2577--2586}\BibitemShut {NoStop}%
\bibitem [{\citenamefont {Monge}(1781)}]{Monge1781}%
  \BibitemOpen
  \bibfield  {author} {\bibinfo {author} {\bibnamefont {Monge}, \bibfnamefont
  {G}}} (\bibinfo {year} {1781}),\ \bibfield  {title} {\enquote {\bibinfo
  {title} {M\'emoire sur la th\'eorie des d\'eblais et des remblais},}\
  }\href@noop {} {\bibfield  {journal} {\bibinfo  {journal} {Histoire de
  l'Acad\'emie Royale des Sciences}\ }\textbf {\bibinfo {volume} {1}},\
  \bibinfo {pages} {666--704}}\BibitemShut {NoStop}%
\bibitem [{\citenamefont {Mori}(1965)}]{mori_transport_1965}%
  \BibitemOpen
  \bibfield  {author} {\bibinfo {author} {\bibnamefont {Mori}, \bibfnamefont
  {H}}} (\bibinfo {year} {1965}),\ \bibfield  {title} {\enquote {\bibinfo
  {title} {Transport, collective motion, and {Brownian} motion},}\ }\href@noop
  {} {\bibfield  {journal} {\bibinfo  {journal} {Progress of Theoretical
  Physics}\ }\textbf {\bibinfo {volume} {33}}~(\bibinfo {number} {3}),\
  \bibinfo {pages} {423--455}}\BibitemShut {NoStop}%
\bibitem [{\citenamefont {Moron}\ \emph {et~al.}(1998)\citenamefont {Moron},
  \citenamefont {Vautard},\ and\ \citenamefont {Ghil}}]{Moron1998}%
  \BibitemOpen
  \bibfield  {author} {\bibinfo {author} {\bibnamefont {Moron}, \bibfnamefont
  {V}}, \bibinfo {author} {\bibfnamefont {R.}~\bibnamefont {Vautard}}, \ and\
  \bibinfo {author} {\bibfnamefont {M.}~\bibnamefont {Ghil}}} (\bibinfo {year}
  {1998}),\ \bibfield  {title} {\enquote {\bibinfo {title} {Trends,
  interdecadal and interannual oscillations in global sea-surace
  temperature},}\ }\href@noop {} {\bibfield  {journal} {\bibinfo  {journal}
  {Clim.\ Dyn.}\ }\textbf {\bibinfo {volume} {14}},\ \bibinfo {pages}
  {545--569}}\BibitemShut {NoStop}%
\bibitem [{\citenamefont {Mukhin}\ \emph {et~al.}({2015b})\citenamefont
  {Mukhin}, \citenamefont {Kondrashov}, \citenamefont {Loskutov}, \citenamefont
  {Gavrilov}, \citenamefont {Feigin},\ and\ \citenamefont
  {Ghil}}]{Mukhin2015b}%
  \BibitemOpen
  \bibfield  {author} {\bibinfo {author} {\bibnamefont {Mukhin}, \bibfnamefont
  {D}}, \bibinfo {author} {\bibfnamefont {D.}~\bibnamefont {Kondrashov}},
  \bibinfo {author} {\bibfnamefont {E.}~\bibnamefont {Loskutov}}, \bibinfo
  {author} {\bibfnamefont {A.}~\bibnamefont {Gavrilov}}, \bibinfo {author}
  {\bibfnamefont {A.}~\bibnamefont {Feigin}}, \ and\ \bibinfo {author}
  {\bibfnamefont {M.}~\bibnamefont {Ghil}}} (\bibinfo {year} {{2015b}}),\
  \bibfield  {title} {\enquote {\bibinfo {title} {Predicting critical
  transitions in {ENSO models. Part II: Spatially} dependent models},}\ }\href
  {\doibase 10.1175/JCLI-D-14-00240.1} {\bibfield  {journal} {\bibinfo
  {journal} {Journal of Climate}\ }\textbf {\bibinfo {volume} {28}}~(\bibinfo
  {number} {5}),\ \bibinfo {pages} {1962--1976}},\ \Eprint
  {http://arxiv.org/abs/https://doi.org/10.1175/JCLI-D-14-00240.1}
  {https://doi.org/10.1175/JCLI-D-14-00240.1} \BibitemShut {NoStop}%
\bibitem [{\citenamefont {Mukhin}\ \emph {et~al.}({2015a})\citenamefont
  {Mukhin}, \citenamefont {Loskutov}, \citenamefont {Mukhina}, \citenamefont
  {Feigin}, \citenamefont {Zaliapin},\ and\ \citenamefont
  {Ghil}}]{Mukhin2015a}%
  \BibitemOpen
  \bibfield  {author} {\bibinfo {author} {\bibnamefont {Mukhin}, \bibfnamefont
  {D}}, \bibinfo {author} {\bibfnamefont {E.}~\bibnamefont {Loskutov}},
  \bibinfo {author} {\bibfnamefont {A.}~\bibnamefont {Mukhina}}, \bibinfo
  {author} {\bibfnamefont {A.}~\bibnamefont {Feigin}}, \bibinfo {author}
  {\bibfnamefont {I.}~\bibnamefont {Zaliapin}}, \ and\ \bibinfo {author}
  {\bibfnamefont {M.}~\bibnamefont {Ghil}}} (\bibinfo {year} {{2015a}}),\
  \bibfield  {title} {\enquote {\bibinfo {title} {Predicting critical
  transitions in {ENSO Models. Part I: Methodology} and simple models with
  memory},}\ }\href {\doibase 10.1175/JCLI-D-14-00239.1} {\bibfield  {journal}
  {\bibinfo  {journal} {Journal of Climate}\ }\textbf {\bibinfo {volume}
  {28}}~(\bibinfo {number} {5}),\ \bibinfo {pages} {1940--1961}},\ \Eprint
  {http://arxiv.org/abs/https://doi.org/10.1175/JCLI-D-14-00239.1}
  {https://doi.org/10.1175/JCLI-D-14-00239.1} \BibitemShut {NoStop}%
\bibitem [{\citenamefont {Munk}\ and\ \citenamefont
  {Wunsch}(1982)}]{Munk.Wunsch.82}%
  \BibitemOpen
  \bibfield  {author} {\bibinfo {author} {\bibnamefont {Munk}, \bibfnamefont
  {W}}, \ and\ \bibinfo {author} {\bibfnamefont {C.}~\bibnamefont {Wunsch}}}
  (\bibinfo {year} {1982}),\ \bibfield  {title} {\enquote {\bibinfo {title}
  {Observing the ocean in the 1990s},}\ }\href@noop {} {\bibfield  {journal}
  {\bibinfo  {journal} {Philos. Trans. R. Soc. London Ser. A}\ }\textbf
  {\bibinfo {volume} {307}},\ \bibinfo {pages} {439--464}}\BibitemShut
  {NoStop}%
\bibitem [{\citenamefont {NAC}(1986)}]{Bretherton86}%
  \BibitemOpen
  \bibfield  {author} {\bibinfo {author} {\bibnamefont {NAC},}} (\bibinfo
  {year} {1986}),\ \href@noop {} {\emph {\bibinfo {title} {{\rm NASA Advisory
  Council:} Earth System Science Overview}}},\ edited by\ \bibinfo {editor}
  {\bibnamefont {{Bretherton, F., et al.}}}\ (\bibinfo  {publisher} {National
  Aeronautics and Space Administration},\ \bibinfo {address} {Washington,
  DC})\BibitemShut {NoStop}%
\bibitem [{\citenamefont {Nadiga}\ and\ \citenamefont
  {Luce}(2001)}]{Nadiga2001}%
  \BibitemOpen
  \bibfield  {author} {\bibinfo {author} {\bibnamefont {Nadiga}, \bibfnamefont
  {B~T}}, \ and\ \bibinfo {author} {\bibfnamefont {B.}~\bibnamefont {Luce}}}
  (\bibinfo {year} {2001}),\ \bibfield  {title} {\enquote {\bibinfo {title}
  {{Global bifurcation of Shil\~nikov type in a double-gyre model}},}\
  }\href@noop {} {\bibfield  {journal} {\bibinfo  {journal} {J.\ Phys.\
  Oceanogr.}\ }\textbf {\bibinfo {volume} {31}},\ \bibinfo {pages}
  {2669--2690}}\BibitemShut {NoStop}%
\bibitem [{\citenamefont {Namias}(1968)}]{Namias1968}%
  \BibitemOpen
  \bibfield  {author} {\bibinfo {author} {\bibnamefont {Namias}, \bibfnamefont
  {J}}} (\bibinfo {year} {1968}),\ \bibfield  {title} {\enquote {\bibinfo
  {title} {{Long-range weather forecasting: History, current status and
  outlook}},}\ }\href@noop {} {\bibfield  {journal} {\bibinfo  {journal} {Bull.
  Am. Meteorol. Soc.}\ }\textbf {\bibinfo {volume} {49}},\ \bibinfo {pages}
  {438--470}}\BibitemShut {NoStop}%
\bibitem [{\citenamefont {Neelin}\ \emph {et~al.}(1998)\citenamefont {Neelin},
  \citenamefont {Battisti}, \citenamefont {Hirst}, \citenamefont {Jin},
  \citenamefont {Wakata}, \citenamefont {Yamagata},\ and\ \citenamefont
  {Zebiak}}]{Neelin1998}%
  \BibitemOpen
  \bibfield  {author} {\bibinfo {author} {\bibnamefont {Neelin}, \bibfnamefont
  {J~D}}, \bibinfo {author} {\bibfnamefont {D.~S.}\ \bibnamefont {Battisti}},
  \bibinfo {author} {\bibfnamefont {A.~C.}\ \bibnamefont {Hirst}}, \bibinfo
  {author} {\bibfnamefont {F.-F.}\ \bibnamefont {Jin}}, \bibinfo {author}
  {\bibfnamefont {Y.}~\bibnamefont {Wakata}}, \bibinfo {author} {\bibfnamefont
  {T.}~\bibnamefont {Yamagata}}, \ and\ \bibinfo {author} {\bibfnamefont
  {S.~E.}\ \bibnamefont {Zebiak}}} (\bibinfo {year} {1998}),\ \bibfield
  {title} {\enquote {\bibinfo {title} {{ENSO Theory}},}\ }\href@noop {}
  {\bibfield  {journal} {\bibinfo  {journal} {J.\ Geophys. Res.}\ }\textbf
  {\bibinfo {volume} {103}},\ \bibinfo {pages} {14,261--14,290}}\BibitemShut
  {NoStop}%
\bibitem [{\citenamefont {Neelin}\ \emph {et~al.}(1994)\citenamefont {Neelin},
  \citenamefont {Latif},\ and\ \citenamefont {Jin}}]{Neelin1994}%
  \BibitemOpen
  \bibfield  {author} {\bibinfo {author} {\bibnamefont {Neelin}, \bibfnamefont
  {J~D}}, \bibinfo {author} {\bibfnamefont {M.}~\bibnamefont {Latif}}, \ and\
  \bibinfo {author} {\bibfnamefont {F.-F.}\ \bibnamefont {Jin}}} (\bibinfo
  {year} {1994}),\ \bibfield  {title} {\enquote {\bibinfo {title} {{Dynamics of
  coupled ocean-atmosphere models: The tropical problem}},}\ }\href@noop {}
  {\bibfield  {journal} {\bibinfo  {journal} {Ann. Rev. Fluid Mech.}\ }\textbf
  {\bibinfo {volume} {26}},\ \bibinfo {pages} {617--659}}\BibitemShut {NoStop}%
\bibitem [{\citenamefont {Newman}(2010)}]{newman_networks_2010}%
  \BibitemOpen
  \bibfield  {author} {\bibinfo {author} {\bibnamefont {Newman}, \bibfnamefont
  {Mark}}} (\bibinfo {year} {2010}),\ \href@noop {} {\emph {\bibinfo {title}
  {Networks: {An Introduction}}}}\ (\bibinfo  {publisher} {Oxford University
  Press},\ \bibinfo {address} {Oxford, U.K.})\BibitemShut {NoStop}%
\bibitem [{\citenamefont {North}(1975)}]{North1975}%
  \BibitemOpen
  \bibfield  {author} {\bibinfo {author} {\bibnamefont {North}, \bibfnamefont
  {G~R}}} (\bibinfo {year} {1975}),\ \bibfield  {title} {\enquote {\bibinfo
  {title} {Analytical solution to a simple climate model with diffusive heat
  transport},}\ }\href@noop {} {\bibfield  {journal} {\bibinfo  {journal} {J.\
  Atmos.\ Sci.}\ }\textbf {\bibinfo {volume} {32}},\ \bibinfo {pages}
  {1301--1307}}\BibitemShut {NoStop}%
\bibitem [{\citenamefont {North}\ \emph {et~al.}(1993)\citenamefont {North},
  \citenamefont {Bell},\ and\ \citenamefont {Hardin}}]{North1993}%
  \BibitemOpen
  \bibfield  {author} {\bibinfo {author} {\bibnamefont {North}, \bibfnamefont
  {G~R}}, \bibinfo {author} {\bibfnamefont {R.~E.}\ \bibnamefont {Bell}}, \
  and\ \bibinfo {author} {\bibfnamefont {J.~W.}\ \bibnamefont {Hardin}}}
  (\bibinfo {year} {1993}),\ \bibfield  {title} {\enquote {\bibinfo {title}
  {Fluctuation dissipation in a general circulation model},}\ }\href {\doibase
  10.1007/BF00209665} {\bibfield  {journal} {\bibinfo  {journal} {Climate
  Dynamics}\ }\textbf {\bibinfo {volume} {8}}~(\bibinfo {number} {6}),\
  \bibinfo {pages} {259--264}}\BibitemShut {NoStop}%
\bibitem [{\citenamefont {North}\ \emph {et~al.}(1981)\citenamefont {North},
  \citenamefont {Cahalan},\ and\ \citenamefont {Coakley}}]{North1981}%
  \BibitemOpen
  \bibfield  {author} {\bibinfo {author} {\bibnamefont {North}, \bibfnamefont
  {G~R}}, \bibinfo {author} {\bibfnamefont {R.~F.}\ \bibnamefont {Cahalan}}, \
  and\ \bibinfo {author} {\bibfnamefont {J.~A.}\ \bibnamefont {Coakley}}}
  (\bibinfo {year} {1981}),\ \bibfield  {title} {\enquote {\bibinfo {title}
  {Energy balance climate models},}\ }\href@noop {} {\bibfield  {journal}
  {\bibinfo  {journal} {Reviews of Geophysics and Space Physics}\ }\textbf
  {\bibinfo {volume} {19}},\ \bibinfo {pages} {19--121}}\BibitemShut {NoStop}%
\bibitem [{\citenamefont {North}\ \emph {et~al.}(1979)\citenamefont {North},
  \citenamefont {Howard}, \citenamefont {Pollard},\ and\ \citenamefont
  {Wielicki}}]{North.ea.1979}%
  \BibitemOpen
  \bibfield  {author} {\bibinfo {author} {\bibnamefont {North}, \bibfnamefont
  {G~R}}, \bibinfo {author} {\bibfnamefont {L.}~\bibnamefont {Howard}},
  \bibinfo {author} {\bibfnamefont {D.}~\bibnamefont {Pollard}}, \ and\
  \bibinfo {author} {\bibfnamefont {B.}~\bibnamefont {Wielicki}}} (\bibinfo
  {year} {1979}),\ \bibfield  {title} {\enquote {\bibinfo {title} {Variational
  formulation of {Budyko-Sellers climate models}},}\ }\href@noop {} {\bibfield
  {journal} {\bibinfo  {journal} {J. Atmos. Sci.}\ }\textbf {\bibinfo {volume}
  {36}},\ \bibinfo {pages} {255--259}}\BibitemShut {NoStop}%
\bibitem [{\citenamefont {{NRC}}(2006)}]{NRC.06}%
  \BibitemOpen
  \bibfield  {author} {\bibinfo {author} {\bibnamefont {{NRC}},}} (\bibinfo
  {year} {2006}),\ \href@noop {} {\emph {\bibinfo {title} {{{\rm National
  Research Council:} Surface Temperature Reconstructions for the Last 2000
  Years}}}}\ (\bibinfo  {publisher} {National Academies Press})\BibitemShut
  {NoStop}%
\bibitem [{\citenamefont {Onogi}\ and\ \citenamefont
  {Coauthors}(2007)}]{Onogi07}%
  \BibitemOpen
  \bibfield  {author} {\bibinfo {author} {\bibnamefont {Onogi}, \bibfnamefont
  {K}}, \ and\ \bibinfo {author} {\bibnamefont {Coauthors}}} (\bibinfo {year}
  {2007}),\ \bibfield  {title} {\enquote {\bibinfo {title} {The {JRA-25}
  reanalysis},}\ }\href@noop {} {\bibfield  {journal} {\bibinfo  {journal} {J.
  Meteor. Soc. Japan}\ }\textbf {\bibinfo {volume} {85}},\ \bibinfo {pages}
  {369--432}}\BibitemShut {NoStop}%
\bibitem [{\citenamefont {Onsager}(1931)}]{Onsager.1931}%
  \BibitemOpen
  \bibfield  {author} {\bibinfo {author} {\bibnamefont {Onsager}, \bibfnamefont
  {L}}} (\bibinfo {year} {1931}),\ \bibfield  {title} {\enquote {\bibinfo
  {title} {Reciprocal relations in irreversible processes. i.}}\ }\href@noop {}
  {\bibfield  {journal} {\bibinfo  {journal} {Physical Review}\ }\textbf
  {\bibinfo {volume} {37}}~(\bibinfo {number} {4}),\ \bibinfo {pages}
  {405--426}}\BibitemShut {NoStop}%
\bibitem [{\citenamefont {Ott}(2002)}]{Ott2002}%
  \BibitemOpen
  \bibfield  {author} {\bibinfo {author} {\bibnamefont {Ott}, \bibfnamefont
  {E}}} (\bibinfo {year} {2002}),\ \href@noop {} {\emph {\bibinfo {title}
  {Chaos in Dynamical Systems}}}\ (\bibinfo  {publisher} {Cambridge University
  Press})\BibitemShut {NoStop}%
\bibitem [{\citenamefont {Otto}\ \emph {et~al.}(2013)\citenamefont {Otto},
  \citenamefont {Otto}, \citenamefont {Boucher}, \citenamefont {Church},
  \citenamefont {Hegerl}, \citenamefont {Forster}, \citenamefont {Gillett},
  \citenamefont {Gregory}, \citenamefont {Johnson}, \citenamefont {Knutti},
  \citenamefont {Lewis}, \citenamefont {Lohmann}, \citenamefont {Marotzke},
  \citenamefont {Myhre}, \citenamefont {Shindell}, \citenamefont {Stevens},\
  and\ \citenamefont {Allen}}]{Otto2013}%
  \BibitemOpen
  \bibfield  {author} {\bibinfo {author} {\bibnamefont {Otto}, \bibfnamefont
  {A}}, \bibinfo {author} {\bibfnamefont {F.~E.~L.}\ \bibnamefont {Otto}},
  \bibinfo {author} {\bibfnamefont {O.}~\bibnamefont {Boucher}}, \bibinfo
  {author} {\bibfnamefont {J.}~\bibnamefont {Church}}, \bibinfo {author}
  {\bibfnamefont {G.}~\bibnamefont {Hegerl}}, \bibinfo {author} {\bibfnamefont
  {P.~M.}\ \bibnamefont {Forster}}, \bibinfo {author} {\bibfnamefont {N.~P.}\
  \bibnamefont {Gillett}}, \bibinfo {author} {\bibfnamefont {J.}~\bibnamefont
  {Gregory}}, \bibinfo {author} {\bibfnamefont {G.~C.}\ \bibnamefont
  {Johnson}}, \bibinfo {author} {\bibfnamefont {R.}~\bibnamefont {Knutti}},
  \bibinfo {author} {\bibfnamefont {N.}~\bibnamefont {Lewis}}, \bibinfo
  {author} {\bibfnamefont {U.}~\bibnamefont {Lohmann}}, \bibinfo {author}
  {\bibfnamefont {J.}~\bibnamefont {Marotzke}}, \bibinfo {author}
  {\bibfnamefont {G.}~\bibnamefont {Myhre}}, \bibinfo {author} {\bibfnamefont
  {D.}~\bibnamefont {Shindell}}, \bibinfo {author} {\bibfnamefont
  {B.}~\bibnamefont {Stevens}}, \ and\ \bibinfo {author} {\bibfnamefont
  {M.~R.}\ \bibnamefont {Allen}}} (\bibinfo {year} {2013}),\ \bibfield  {title}
  {\enquote {\bibinfo {title} {Energy budget constraints on climate
  response},}\ }\href {\doibase 10.1038/ngeo1836} {\bibfield  {journal}
  {\bibinfo  {journal} {Nature Geoscience}\ }\textbf {\bibinfo {volume} {6}},\
  \bibinfo {pages} {415--416}}\BibitemShut {NoStop}%
\bibitem [{\citenamefont {{PAGES}}(2013)}]{pages2013}%
  \BibitemOpen
  \bibfield  {author} {\bibinfo {author} {\bibnamefont {{PAGES}},}} (\bibinfo
  {year} {2013}),\ \bibfield  {title} {\enquote {\bibinfo {title} {{{\rm
  PAGES~2k~Consortium}}: Continental-scale temperature variability during the
  past two millennia},}\ }\href {\doibase 10.1038/ngeo1797} {\bibfield
  {journal} {\bibinfo  {journal} {Nature Geosci.}\ }\textbf {\bibinfo {volume}
  {6}}~(\bibinfo {number} {5}),\ \bibinfo {pages} {339--346}}\BibitemShut
  {NoStop}%
\bibitem [{\citenamefont {Palmer}(1951)}]{Palmer.1951}%
  \BibitemOpen
  \bibfield  {author} {\bibinfo {author} {\bibnamefont {Palmer}, \bibfnamefont
  {C~E}}} (\bibinfo {year} {1951}),\ \bibfield  {title} {\enquote {\bibinfo
  {title} {Tropical meteorology},}\ }in\ \href@noop {} {\emph {\bibinfo
  {booktitle} {Compendium of Meteorology}}}\ (\bibinfo  {publisher}
  {Springer})\ pp.\ \bibinfo {pages} {859--880}\BibitemShut {NoStop}%
\bibitem [{\citenamefont {Palmer}(2017)}]{Palmer2017}%
  \BibitemOpen
  \bibfield  {author} {\bibinfo {author} {\bibnamefont {Palmer}, \bibfnamefont
  {T~N}}} (\bibinfo {year} {2017}),\ \bibfield  {title} {\enquote {\bibinfo
  {title} {The primacy of doubt: {Evolution of numerical weather prediction
  from determinism to probability}},}\ }\href {\doibase 10.1002/2017MS000999}
  {\bibfield  {journal} {\bibinfo  {journal} {Journal of Advances in Modeling
  Earth Systems}\ }\textbf {\bibinfo {volume} {9}}~(\bibinfo {number} {2}),\
  \bibinfo {pages} {730--734}}\BibitemShut {NoStop}%
\bibitem [{\citenamefont {Palmer}\ \emph {et~al.}(2008)\citenamefont {Palmer},
  \citenamefont {Doblas-Reyes}, \citenamefont {Weisheimer},\ and\ \citenamefont
  {Rodwell}}]{palmer_2008}%
  \BibitemOpen
  \bibfield  {author} {\bibinfo {author} {\bibnamefont {Palmer}, \bibfnamefont
  {T~N}}, \bibinfo {author} {\bibfnamefont {F.~J.}\ \bibnamefont
  {Doblas-Reyes}}, \bibinfo {author} {\bibfnamefont {A.}~\bibnamefont
  {Weisheimer}}, \ and\ \bibinfo {author} {\bibfnamefont {M.~J.}\ \bibnamefont
  {Rodwell}}} (\bibinfo {year} {2008}),\ \bibfield  {title} {\enquote {\bibinfo
  {title} {Toward seamless prediction: {Calibration of climate change
  projections using seasonal forecasts}},}\ }\href@noop {} {\bibfield
  {journal} {\bibinfo  {journal} {Bull. Amer. Meteorol. Soc.}\ }\textbf
  {\bibinfo {volume} {89}},\ \bibinfo {pages} {459--470}}\BibitemShut {NoStop}%
\bibitem [{\citenamefont {Palmer}\ and\ \citenamefont
  {Williams}(2009)}]{palmer_stochastic_2009}%
  \BibitemOpen
  \bibinfo {editor} {\bibnamefont {Palmer}, \bibfnamefont {T~N}}, \ and\
  \bibinfo {editor} {\bibfnamefont {P.}~\bibnamefont {Williams}},\ Eds.
  (\bibinfo {year} {2009}),\ \href@noop {} {\emph {\bibinfo {title} {Stochastic
  Physics and Climate Modelling}}}\ (\bibinfo  {publisher} {Cambridge
  University Press},\ \bibinfo {address} {Cambridge})\BibitemShut {NoStop}%
\bibitem [{\citenamefont {Pan}\ and\ \citenamefont
  {Randall}(1998)}]{Pan.Randall.1998}%
  \BibitemOpen
  \bibfield  {author} {\bibinfo {author} {\bibnamefont {Pan}, \bibfnamefont
  {D-M}}, \ and\ \bibinfo {author} {\bibfnamefont {D.~A.}\ \bibnamefont
  {Randall}}} (\bibinfo {year} {1998}),\ \bibfield  {title} {\enquote {\bibinfo
  {title} {A cumulus parameterization with a prognostic closure},}\ }\href@noop
  {} {\bibfield  {journal} {\bibinfo  {journal} {Quarterly Journal of the Royal
  Meteorological Society}\ }\textbf {\bibinfo {volume} {124}}~(\bibinfo
  {number} {547}),\ \bibinfo {pages} {949--981}}\BibitemShut {NoStop}%
\bibitem [{\citenamefont {Papanicolaou}\ and\ \citenamefont
  {Kohler}(1974)}]{Papa.K.1974}%
  \BibitemOpen
  \bibfield  {author} {\bibinfo {author} {\bibnamefont {Papanicolaou},
  \bibfnamefont {G~C}}, \ and\ \bibinfo {author} {\bibfnamefont
  {W.}~\bibnamefont {Kohler}}} (\bibinfo {year} {1974}),\ \bibfield  {title}
  {\enquote {\bibinfo {title} {Asymptotic theory of mixing stochastic ordinary
  differential equations},}\ }\href@noop {} {\bibfield  {journal} {\bibinfo
  {journal} {{Communications on Pure and Applied Mathematics}}\ }\textbf
  {\bibinfo {volume} {27}}~(\bibinfo {number} {5}),\ \bibinfo {pages}
  {641--668}}\BibitemShut {NoStop}%
\bibitem [{\citenamefont {Parker}(2010)}]{Parker2010}%
  \BibitemOpen
  \bibfield  {author} {\bibinfo {author} {\bibnamefont {Parker}, \bibfnamefont
  {W~S}}} (\bibinfo {year} {2010}),\ \bibfield  {title} {\enquote {\bibinfo
  {title} {Whose probabilities? {Predicting} climate change with ensembles of
  models},}\ }\href {\doibase 10.1086/656815} {\bibfield  {journal} {\bibinfo
  {journal} {Philosophy of Science}\ }\textbf {\bibinfo {volume}
  {77}}~(\bibinfo {number} {5}),\ \bibinfo {pages} {985--997}}\BibitemShut
  {NoStop}%
\bibitem [{\citenamefont {Paterson}(1981)}]{Pat81}%
  \BibitemOpen
  \bibfield  {author} {\bibinfo {author} {\bibnamefont {Paterson},
  \bibfnamefont {W~S~B}}} (\bibinfo {year} {1981}),\ \href@noop {} {\emph
  {\bibinfo {title} {The Physics of Glaciers}}},\ \bibinfo {edition} {2nd}\
  ed.\ (\bibinfo  {publisher} {Pergamon Press},\ \bibinfo {address} {Oxford,
  UK})\BibitemShut {NoStop}%
\bibitem [{\citenamefont {Pauluis}\ and\ \citenamefont {Held}(2002)}]{Pauluis}%
  \BibitemOpen
  \bibfield  {author} {\bibinfo {author} {\bibnamefont {Pauluis}, \bibfnamefont
  {O}}, \ and\ \bibinfo {author} {\bibfnamefont {I~M}\ \bibnamefont {Held}}}
  (\bibinfo {year} {2002}),\ \bibfield  {title} {\enquote {\bibinfo {title}
  {Entropy budget of an atmosphere in radiative-convective equilibrium. {P}art
  {I}: {M}aximum work and frictional dissipation},}\ }\href@noop {} {\bibfield
  {journal} {\bibinfo  {journal} {J. Atmos. Sci.}\ }\textbf {\bibinfo {volume}
  {59}},\ \bibinfo {pages} {125--139}}\BibitemShut {NoStop}%
\bibitem [{\citenamefont {Pavliotis}(2014)}]{Pavliotis2014}%
  \BibitemOpen
  \bibfield  {author} {\bibinfo {author} {\bibnamefont {Pavliotis},
  \bibfnamefont {G~A}}} (\bibinfo {year} {2014}),\ \href@noop {} {\emph
  {\bibinfo {title} {{Stochastic Processes and Applications: Diffusion
  Processes, the Fokker-Planck and Langevin Equations}}}},\ Texts in Applied
  Mathematics\ (\bibinfo  {publisher} {Springer New York})\BibitemShut
  {NoStop}%
\bibitem [{\citenamefont {Pavliotis}\ and\ \citenamefont
  {Stuart}(2008)}]{Pavliotis2008}%
  \BibitemOpen
  \bibfield  {author} {\bibinfo {author} {\bibnamefont {Pavliotis},
  \bibfnamefont {G~A}}, \ and\ \bibinfo {author} {\bibfnamefont {A.~M.}\
  \bibnamefont {Stuart}}} (\bibinfo {year} {2008}),\ \href@noop {} {\emph
  {\bibinfo {title} {{Multiscale Methods}}}}\ (\bibinfo  {publisher}
  {Springer},\ \bibinfo {address} {New York, {NY}})\BibitemShut {NoStop}%
\bibitem [{\citenamefont {Pearl}(2009)}]{Pearl2009}%
  \BibitemOpen
  \bibfield  {author} {\bibinfo {author} {\bibnamefont {Pearl}, \bibfnamefont
  {J}}} (\bibinfo {year} {2009}),\ \href@noop {} {\emph {\bibinfo {title}
  {{Causality: Models, Reasoning, and Inference}}}}\ (\bibinfo  {publisher}
  {Cambridge University Press},\ \bibinfo {address} {Cambridge,
  U.K.})\BibitemShut {NoStop}%
\bibitem [{\citenamefont {Pedlosky}(1987)}]{Pedlosky1987}%
  \BibitemOpen
  \bibfield  {author} {\bibinfo {author} {\bibnamefont {Pedlosky},
  \bibfnamefont {J}}} (\bibinfo {year} {1987}),\ \href@noop {} {\emph {\bibinfo
  {title} {{Geophysical Fluid Dynamics}}}},\ \bibinfo {edition} {2nd}\ ed.\
  (\bibinfo  {publisher} {Springer-Verlag},\ \bibinfo {address} {New
  York})\BibitemShut {NoStop}%
\bibitem [{\citenamefont {Pedlosky}(1996)}]{Pedlosky1996}%
  \BibitemOpen
  \bibfield  {author} {\bibinfo {author} {\bibnamefont {Pedlosky},
  \bibfnamefont {J}}} (\bibinfo {year} {1996}),\ \href@noop {} {\emph {\bibinfo
  {title} {Ocean {C}irculation {T}heory}}}\ (\bibinfo  {publisher} {Springer},\
  \bibinfo {address} {New York})\BibitemShut {NoStop}%
\bibitem [{\citenamefont {Peixoto}\ and\ \citenamefont
  {Oort}(1992)}]{Peixoto1992}%
  \BibitemOpen
  \bibfield  {author} {\bibinfo {author} {\bibnamefont {Peixoto}, \bibfnamefont
  {J~P}}, \ and\ \bibinfo {author} {\bibfnamefont {A.~H.}\ \bibnamefont
  {Oort}}} (\bibinfo {year} {1992}),\ \href@noop {} {\emph {\bibinfo {title}
  {Physics of Climate}}}\ (\bibinfo  {publisher} {AIP Press},\ \bibinfo
  {address} {New York})\BibitemShut {NoStop}%
\bibitem [{\citenamefont {Penland}(1989)}]{Penland_MWR89}%
  \BibitemOpen
  \bibfield  {author} {\bibinfo {author} {\bibnamefont {Penland}, \bibfnamefont
  {C}}} (\bibinfo {year} {1989}),\ \bibfield  {title} {\enquote {\bibinfo
  {title} {Random forcing and forecasting using principal oscillation pattern
  analysis},}\ }\href {\doibase
  10.1175/1520-0493(1989)117<2165:rfafup>2.0.co;2} {\bibfield  {journal}
  {\bibinfo  {journal} {Mon. Wea. Rev.}\ }\textbf {\bibinfo {volume} {117}},\
  \bibinfo {pages} {2165--2185}}\BibitemShut {NoStop}%
\bibitem [{\citenamefont {Penland}(1996)}]{Penland_PD96}%
  \BibitemOpen
  \bibfield  {author} {\bibinfo {author} {\bibnamefont {Penland}, \bibfnamefont
  {C}}} (\bibinfo {year} {1996}),\ \bibfield  {title} {\enquote {\bibinfo
  {title} {A stochastic model of {IndoPacific sea surface temperature
  anomalies}},}\ }\href {\doibase 10.1016/0167-2789(96)00124-8} {\bibfield
  {journal} {\bibinfo  {journal} {Physica D}\ }\textbf {\bibinfo {volume}
  {98}},\ \bibinfo {pages} {534--558}}\BibitemShut {NoStop}%
\bibitem [{\citenamefont {Penland}\ and\ \citenamefont
  {Ghil}(1993)}]{PenlandGhil_MWR93}%
  \BibitemOpen
  \bibfield  {author} {\bibinfo {author} {\bibnamefont {Penland}, \bibfnamefont
  {C}}, \ and\ \bibinfo {author} {\bibfnamefont {M.}~\bibnamefont {Ghil}}}
  (\bibinfo {year} {1993}),\ \bibfield  {title} {\enquote {\bibinfo {title}
  {Forecasting {Northern H}emisphere 700-mb geopotential height anomalies using
  empirical normal modes},}\ }\href {\doibase
  10.1175/1520-0493(1993)121<2355:fnhmgh>2.0.co;2} {\bibfield  {journal}
  {\bibinfo  {journal} {Mon Weather Rev.}\ }\textbf {\bibinfo {volume} {121}},\
  \bibinfo {pages} {2355--2372}}\BibitemShut {NoStop}%
\bibitem [{\citenamefont {Penland}\ and\ \citenamefont
  {Sardeshmukh}(1995)}]{PenlandSardeshmukh_JCL95}%
  \BibitemOpen
  \bibfield  {author} {\bibinfo {author} {\bibnamefont {Penland}, \bibfnamefont
  {C}}, \ and\ \bibinfo {author} {\bibfnamefont {P.~D.}\ \bibnamefont
  {Sardeshmukh}}} (\bibinfo {year} {1995}),\ \bibfield  {title} {\enquote
  {\bibinfo {title} {The optimal growth of tropical sea surface temperature
  anomalies},}\ }\href {\doibase
  10.1175/1520-0442(1995)008<1999:togots>2.0.co;2} {\bibfield  {journal}
  {\bibinfo  {journal} {J. Climate}\ }\textbf {\bibinfo {volume} {8}},\
  \bibinfo {pages} {1999--2024}}\BibitemShut {NoStop}%
\bibitem [{\citenamefont {Penny}\ and\ \citenamefont
  {Hamill}(2017)}]{Penny2017}%
  \BibitemOpen
  \bibfield  {author} {\bibinfo {author} {\bibnamefont {Penny}, \bibfnamefont
  {S~G}}, \ and\ \bibinfo {author} {\bibfnamefont {T.~M.}\ \bibnamefont
  {Hamill}}} (\bibinfo {year} {2017}),\ \bibfield  {title} {\enquote {\bibinfo
  {title} {Coupled data assimilation for integrated earth system analysis and
  prediction},}\ }\href {\doibase 10.1175/BAMS-D-17-0036.1} {\bibfield
  {journal} {\bibinfo  {journal} {Bulletin of the American Meteorological
  Society}\ }\textbf {\bibinfo {volume} {98}}~(\bibinfo {number} {7}),\
  \bibinfo {pages} {ES169--ES172}}\BibitemShut {NoStop}%
\bibitem [{\citenamefont {Pfeffer}(1960)}]{Pfeffer.1960}%
  \BibitemOpen
  \bibinfo {editor} {\bibnamefont {Pfeffer}, \bibfnamefont {R~L}},\ Ed.
  (\bibinfo {year} {1960}),\ \href@noop {} {\emph {\bibinfo {title} {Dynamics
  of Climate}}}\ (\bibinfo  {publisher} {Pergamon Press})\BibitemShut {NoStop}%
\bibitem [{\citenamefont {Pfister}\ and\ \citenamefont
  {Stocker}(2017)}]{Pfister17}%
  \BibitemOpen
  \bibfield  {author} {\bibinfo {author} {\bibnamefont {Pfister}, \bibfnamefont
  {P~L}}, \ and\ \bibinfo {author} {\bibfnamefont {T.~F.}\ \bibnamefont
  {Stocker}}} (\bibinfo {year} {2017}),\ \bibfield  {title} {\enquote {\bibinfo
  {title} {State-dependence of the climate sensitivity in earth system models
  of intermediate complexity},}\ }\href {\doibase 10.1002/2017GL075457}
  {\bibfield  {journal} {\bibinfo  {journal} {Geophysical Research Letters}\
  }\textbf {\bibinfo {volume} {44}}~(\bibinfo {number} {20}),\ \bibinfo {pages}
  {10,643--10,653}}\BibitemShut {NoStop}%
\bibitem [{\citenamefont {Philander}(1990)}]{Philander1990}%
  \BibitemOpen
  \bibfield  {author} {\bibinfo {author} {\bibnamefont {Philander},
  \bibfnamefont {S~G~H}}} (\bibinfo {year} {1990}),\ \href@noop {} {\emph
  {\bibinfo {title} {{ El Ni\~no and the Southern Oscillation.}}}}\ (\bibinfo
  {publisher} {Academic Press},\ \bibinfo {address} {New York})\BibitemShut
  {NoStop}%
\bibitem [{\citenamefont {Pierini}\ \emph {et~al.}(2018)\citenamefont
  {Pierini}, \citenamefont {Chekroun},\ and\ \citenamefont
  {Ghil}}]{Pierini.ea.2018}%
  \BibitemOpen
  \bibfield  {author} {\bibinfo {author} {\bibnamefont {Pierini}, \bibfnamefont
  {S}}, \bibinfo {author} {\bibfnamefont {M.~D.}\ \bibnamefont {Chekroun}}, \
  and\ \bibinfo {author} {\bibfnamefont {M.}~\bibnamefont {Ghil}}} (\bibinfo
  {year} {2018}),\ \bibfield  {title} {\enquote {\bibinfo {title} {The onset of
  chaos in nonautonomous dissipative dynamical systems: {A low-order
  ocean-model case study}},}\ }\href {https://doi.org/10.5194/npg-25-671-2018}
  {\bibfield  {journal} {\bibinfo  {journal} {Nonlin. Processes Geophys.}\
  }\textbf {\bibinfo {volume} {25}},\ \bibinfo {pages} {671--692}}\BibitemShut
  {NoStop}%
\bibitem [{\citenamefont {Pierini}\ \emph {et~al.}(2016)\citenamefont
  {Pierini}, \citenamefont {Ghil},\ and\ \citenamefont
  {Chekroun}}]{Pierini2016}%
  \BibitemOpen
  \bibfield  {author} {\bibinfo {author} {\bibnamefont {Pierini}, \bibfnamefont
  {S}}, \bibinfo {author} {\bibfnamefont {M.}~\bibnamefont {Ghil}}, \ and\
  \bibinfo {author} {\bibfnamefont {M.~D.}\ \bibnamefont {Chekroun}}} (\bibinfo
  {year} {2016}),\ \bibfield  {title} {\enquote {\bibinfo {title} {Exploring
  the pullback attractors of a low-order quasigeostrophic ocean model: The
  deterministic case},}\ }\href {\doibase 10.1175/JCLI-D-15-0848.1} {\bibfield
  {journal} {\bibinfo  {journal} {Journal of Climate}\ }\textbf {\bibinfo
  {volume} {29}}~(\bibinfo {number} {11}),\ \bibinfo {pages}
  {4185--4202}}\BibitemShut {NoStop}%
\bibitem [{\citenamefont {Pierrehumbert}(2004)}]{Pierrehumbert.2004}%
  \BibitemOpen
  \bibfield  {author} {\bibinfo {author} {\bibnamefont {Pierrehumbert},
  \bibfnamefont {R~T}}} (\bibinfo {year} {2004}),\ \bibfield  {title} {\enquote
  {\bibinfo {title} {High levels of atmospheric carbon dioxide necessary for
  the termination of global glaciation},}\ }\href@noop {} {\bibfield  {journal}
  {\bibinfo  {journal} {Nature}\ }\textbf {\bibinfo {volume} {429}}~(\bibinfo
  {number} {6992}),\ \bibinfo {pages} {646}}\BibitemShut {NoStop}%
\bibitem [{\citenamefont {Plant}\ and\ \citenamefont
  {Yano}(2016)}]{plant2016parameterization}%
  \BibitemOpen
  \bibfield  {author} {\bibinfo {author} {\bibnamefont {Plant}, \bibfnamefont
  {RS}}, \ and\ \bibinfo {author} {\bibfnamefont {J.I.}\ \bibnamefont {Yano}}}
  (\bibinfo {year} {2016}),\ \href@noop {} {\emph {\bibinfo {title}
  {Parameterization of Atmospheric Convection}}},\ \bibinfo {series}
  {Parameterization of Atmospheric Convection}\ No.\ \bibinfo {number} {v. 1}\
  (\bibinfo  {publisher} {Imperial College Press})\BibitemShut {NoStop}%
\bibitem [{\citenamefont {Plaut}\ \emph {et~al.}(1995)\citenamefont {Plaut},
  \citenamefont {Ghil},\ and\ \citenamefont {Vautard}}]{Plaut1995}%
  \BibitemOpen
  \bibfield  {author} {\bibinfo {author} {\bibnamefont {Plaut}, \bibfnamefont
  {G}}, \bibinfo {author} {\bibfnamefont {M.}~\bibnamefont {Ghil}}, \ and\
  \bibinfo {author} {\bibfnamefont {R.}~\bibnamefont {Vautard}}} (\bibinfo
  {year} {1995}),\ \bibfield  {title} {\enquote {\bibinfo {title} {{Interannual
  and interdecadal variability in 335 years of Central England temperature}},}\
  }\href@noop {} {\bibfield  {journal} {\bibinfo  {journal} {Science}\ }\textbf
  {\bibinfo {volume} {268}},\ \bibinfo {pages} {710--713}}\BibitemShut
  {NoStop}%
\bibitem [{\citenamefont {Poincar\'e}(1902)}]{Poincare.1902}%
  \BibitemOpen
  \bibfield  {author} {\bibinfo {author} {\bibnamefont {Poincar\'e},
  \bibfnamefont {H}}} (\bibinfo {year} {1902}),\ \href@noop {} {\emph {\bibinfo
  {title} {La Science et l'Hypoth\`ese}}},\ \bibinfo {note} {translated in 1905
  into English as``Science and Hypothesis"; versions of this translation are
  (i) downloadable for free as Gutenberg Project EBook\#37157, published in
  2011, and (ii) online as a Wikisource book,
  \url{https://en.wikisource.org/wiki/Science_and_Hypothesis}, published in
  2017.}\BibitemShut {Stop}%
\bibitem [{\citenamefont {Poincar\'e}(1908)}]{Poincare.1908}%
  \BibitemOpen
  \bibfield  {author} {\bibinfo {author} {\bibnamefont {Poincar\'e},
  \bibfnamefont {H}}} (\bibinfo {year} {1908}),\ \href@noop {} {\emph {\bibinfo
  {title} {{Science et M\'ethode}}}}\ (\bibinfo  {publisher} {Ernest
  Flammarion},\ \bibinfo {address} {Paris})\BibitemShut {NoStop}%
\bibitem [{\citenamefont {Poli}\ \emph {et~al.}(2016)\citenamefont {Poli},
  \citenamefont {Hersbach}, \citenamefont {Dee}, \citenamefont {Berrisford},
  \citenamefont {Simmons}, \citenamefont {Vitart}, \citenamefont {Laloyaux},
  \citenamefont {Tan}, \citenamefont {Peubey}, \citenamefont {Th\'epaut},
  \citenamefont {Tr\'emolet}, \citenamefont {Holm}, \citenamefont {Bonavita},
  \citenamefont {Isaksen},\ and\ \citenamefont {Fisher}}]{Poli2016}%
  \BibitemOpen
  \bibfield  {author} {\bibinfo {author} {\bibnamefont {Poli}, \bibfnamefont
  {P}}, \bibinfo {author} {\bibfnamefont {H.}~\bibnamefont {Hersbach}},
  \bibinfo {author} {\bibfnamefont {D.~P.}\ \bibnamefont {Dee}}, \bibinfo
  {author} {\bibfnamefont {P.}~\bibnamefont {Berrisford}}, \bibinfo {author}
  {\bibfnamefont {A.~J.}\ \bibnamefont {Simmons}}, \bibinfo {author}
  {\bibfnamefont {F.}~\bibnamefont {Vitart}}, \bibinfo {author} {\bibfnamefont
  {P.}~\bibnamefont {Laloyaux}}, \bibinfo {author} {\bibfnamefont {D.~G.~H.}\
  \bibnamefont {Tan}}, \bibinfo {author} {\bibfnamefont {C.}~\bibnamefont
  {Peubey}}, \bibinfo {author} {\bibfnamefont {J.-N.}\ \bibnamefont
  {Th\'epaut}}, \bibinfo {author} {\bibfnamefont {Y.}~\bibnamefont
  {Tr\'emolet}}, \bibinfo {author} {\bibfnamefont {E.~V.}\ \bibnamefont
  {Holm}}, \bibinfo {author} {\bibfnamefont {M.}~\bibnamefont {Bonavita}},
  \bibinfo {author} {\bibfnamefont {L.}~\bibnamefont {Isaksen}}, \ and\
  \bibinfo {author} {\bibfnamefont {M.}~\bibnamefont {Fisher}}} (\bibinfo
  {year} {2016}),\ \bibfield  {title} {\enquote {\bibinfo {title} {Era-20c: An
  atmospheric reanalysis of the twentieth century},}\ }\href {\doibase
  10.1175/JCLI-D-15-0556.1} {\bibfield  {journal} {\bibinfo  {journal} {Journal
  of Climate}\ }\textbf {\bibinfo {volume} {29}}~(\bibinfo {number} {11}),\
  \bibinfo {pages} {4083--4097}}\BibitemShut {NoStop}%
\bibitem [{\citenamefont {Pollicott}(1985)}]{Pollicott1985}%
  \BibitemOpen
  \bibfield  {author} {\bibinfo {author} {\bibnamefont {Pollicott},
  \bibfnamefont {M}}} (\bibinfo {year} {1985}),\ \bibfield  {title} {\enquote
  {\bibinfo {title} {On the rate of mixing of {Axiom A} flows},}\ }\href
  {\doibase 10.1007/BF01388579} {\bibfield  {journal} {\bibinfo  {journal}
  {Inventiones Mathematicae}\ }\textbf {\bibinfo {volume} {81}}~(\bibinfo
  {number} {3}),\ \bibinfo {pages} {413--426}}\BibitemShut {NoStop}%
\bibitem [{\citenamefont {Pratt}(1976)}]{Pratt76}%
  \BibitemOpen
  \bibfield  {author} {\bibinfo {author} {\bibnamefont {Pratt}, \bibfnamefont
  {R~W}}} (\bibinfo {year} {1976}),\ \bibfield  {title} {\enquote {\bibinfo
  {title} {The interpretation of space-time spectral quantities},}\ }\href@noop
  {} {\bibfield  {journal} {\bibinfo  {journal} {J. Atmos. Sci.}\ }\textbf
  {\bibinfo {volume} {33}},\ \bibinfo {pages} {1060--1066}}\BibitemShut
  {NoStop}%
\bibitem [{\citenamefont {Preisendorfer}(1988)}]{Preisendorfer1988}%
  \BibitemOpen
  \bibfield  {author} {\bibinfo {author} {\bibnamefont {Preisendorfer},
  \bibfnamefont {R~W}}} (\bibinfo {year} {1988}),\ \href@noop {} {\emph
  {\bibinfo {title} {Principal Component Analysis in Meteorology and
  Oceanography}}}\ (\bibinfo  {publisher} {Elsevier},\ \bibinfo {address}
  {Amsterdam, The Netherlands})\BibitemShut {NoStop}%
\bibitem [{\citenamefont {Prigogine}(1961)}]{Prigogine61}%
  \BibitemOpen
  \bibfield  {author} {\bibinfo {author} {\bibnamefont {Prigogine},
  \bibfnamefont {I}}} (\bibinfo {year} {1961}),\ \href@noop {} {\emph {\bibinfo
  {title} {Thermodynamics of Irreversible Processes}}}\ (\bibinfo  {publisher}
  {Interscience},\ \bibinfo {address} {New York})\BibitemShut {NoStop}%
\bibitem [{\citenamefont {Proctor}\ \emph {et~al.}(2018)\citenamefont
  {Proctor}, \citenamefont {Hsiang}, \citenamefont {Burney}, \citenamefont
  {Burke},\ and\ \citenamefont {Schlenker}}]{Proctor2018}%
  \BibitemOpen
  \bibfield  {author} {\bibinfo {author} {\bibnamefont {Proctor}, \bibfnamefont
  {J}}, \bibinfo {author} {\bibfnamefont {S.}~\bibnamefont {Hsiang}}, \bibinfo
  {author} {\bibfnamefont {J.}~\bibnamefont {Burney}}, \bibinfo {author}
  {\bibfnamefont {M.}~\bibnamefont {Burke}}, \ and\ \bibinfo {author}
  {\bibfnamefont {W.}~\bibnamefont {Schlenker}}} (\bibinfo {year} {2018}),\
  \bibfield  {title} {\enquote {\bibinfo {title} {Estimating global
  agricultural effects of geoengineering using volcanic eruptions},}\ }\href
  {\doibase 10.1038/s41586-018-0417-3} {\bibfield  {journal} {\bibinfo
  {journal} {Nature}\ }\textbf {\bibinfo {volume} {560}}~(\bibinfo {number}
  {7719}),\ \bibinfo {pages} {480--483}}\BibitemShut {NoStop}%
\bibitem [{\citenamefont {Quon}\ and\ \citenamefont
  {Ghil}(1992)}]{Quon.Ghil.1992}%
  \BibitemOpen
  \bibfield  {author} {\bibinfo {author} {\bibnamefont {Quon}, \bibfnamefont
  {C}}, \ and\ \bibinfo {author} {\bibfnamefont {M.}~\bibnamefont {Ghil}}}
  (\bibinfo {year} {1992}),\ \bibfield  {title} {\enquote {\bibinfo {title}
  {Multiple equilibria in thermosolutal convection due to salt-flux boundary
  conditions},}\ }\href@noop {} {\bibfield  {journal} {\bibinfo  {journal} {J.
  Fluid Mech.}\ }\textbf {\bibinfo {volume} {245}},\ \bibinfo {pages}
  {449--484}}\BibitemShut {NoStop}%
\bibitem [{\citenamefont {Ragone}\ \emph {et~al.}(2016)\citenamefont {Ragone},
  \citenamefont {Lucarini},\ and\ \citenamefont {Lunkeit}}]{Ragone2016}%
  \BibitemOpen
  \bibfield  {author} {\bibinfo {author} {\bibnamefont {Ragone}, \bibfnamefont
  {F}}, \bibinfo {author} {\bibfnamefont {V.}~\bibnamefont {Lucarini}}, \ and\
  \bibinfo {author} {\bibfnamefont {F.}~\bibnamefont {Lunkeit}}} (\bibinfo
  {year} {2016}),\ \bibfield  {title} {\enquote {\bibinfo {title} {A new
  framework for climate sensitivity and prediction: a modelling perspective},}\
  }\href {\doibase 10.1007/s00382-015-2657-3} {\bibfield  {journal} {\bibinfo
  {journal} {Climate Dynamics}\ }\textbf {\bibinfo {volume} {46}}~(\bibinfo
  {number} {5}),\ \bibinfo {pages} {1459--1471}}\BibitemShut {NoStop}%
\bibitem [{\citenamefont {Ragone}\ \emph {et~al.}(2018)\citenamefont {Ragone},
  \citenamefont {Wouters},\ and\ \citenamefont {Bouchet}}]{Ragone2017}%
  \BibitemOpen
  \bibfield  {author} {\bibinfo {author} {\bibnamefont {Ragone}, \bibfnamefont
  {F}}, \bibinfo {author} {\bibfnamefont {J.}~\bibnamefont {Wouters}}, \ and\
  \bibinfo {author} {\bibfnamefont {F.}~\bibnamefont {Bouchet}}} (\bibinfo
  {year} {2018}),\ \bibfield  {title} {\enquote {\bibinfo {title} {Computation
  of extreme heat waves in climate models using a large deviation algorithm},}\
  }\href {\doibase 10.1073/pnas.1712645115} {\bibfield  {journal} {\bibinfo
  {journal} {Proceedings of the National Academy of Sciences}\ }\textbf
  {\bibinfo {volume} {115}}~(\bibinfo {number} {1}),\ \bibinfo {pages}
  {24--29}}\BibitemShut {NoStop}%
\bibitem [{\citenamefont {Rahmstorf}\ \emph {et~al.}(2005)\citenamefont
  {Rahmstorf}, \citenamefont {Crucifix}, \citenamefont {Ganopolski},
  \citenamefont {Goosse}, \citenamefont {Kamenkovich}, \citenamefont {Knutti},
  \citenamefont {Lohmann}, \citenamefont {Marsh}, \citenamefont {Mysak},
  \citenamefont {Wang},\ and\ \citenamefont {Weaver}}]{Rahmstorf2005}%
  \BibitemOpen
  \bibfield  {author} {\bibinfo {author} {\bibnamefont {Rahmstorf},
  \bibfnamefont {Stefan}}, \bibinfo {author} {\bibfnamefont {Michel}\
  \bibnamefont {Crucifix}}, \bibinfo {author} {\bibfnamefont {Andrey}\
  \bibnamefont {Ganopolski}}, \bibinfo {author} {\bibfnamefont {Hugues}\
  \bibnamefont {Goosse}}, \bibinfo {author} {\bibfnamefont {Igor}\ \bibnamefont
  {Kamenkovich}}, \bibinfo {author} {\bibfnamefont {Reto}\ \bibnamefont
  {Knutti}}, \bibinfo {author} {\bibfnamefont {Gerrit}\ \bibnamefont
  {Lohmann}}, \bibinfo {author} {\bibfnamefont {Robert}\ \bibnamefont {Marsh}},
  \bibinfo {author} {\bibfnamefont {Lawrence~A.}\ \bibnamefont {Mysak}},
  \bibinfo {author} {\bibfnamefont {Zhaomin}\ \bibnamefont {Wang}}, \ and\
  \bibinfo {author} {\bibfnamefont {Andrew~J.}\ \bibnamefont {Weaver}}}
  (\bibinfo {year} {2005}),\ \bibfield  {title} {\enquote {\bibinfo {title}
  {Thermohaline circulation hysteresis: A model intercomparison},}\ }\href
  {\doibase 10.1029/2005GL023655} {\bibfield  {journal} {\bibinfo  {journal}
  {Geophysical Research Letters}\ }\textbf {\bibinfo {volume} {32}}~(\bibinfo
  {number} {23}),\ 10.1029/2005GL023655}\BibitemShut {NoStop}%
\bibitem [{\citenamefont {Ramanathan}\ and\ \citenamefont
  {Coakley}(1978)}]{Ram.Coak.78}%
  \BibitemOpen
  \bibfield  {author} {\bibinfo {author} {\bibnamefont {Ramanathan},
  \bibfnamefont {V}}, \ and\ \bibinfo {author} {\bibfnamefont {J.~A.}\
  \bibnamefont {Coakley}}} (\bibinfo {year} {1978}),\ \bibfield  {title}
  {\enquote {\bibinfo {title} {Climate modeling through radiative-convective
  models},}\ }\href@noop {} {\bibfield  {journal} {\bibinfo  {journal} {Rev.
  Geophys.}\ }\textbf {\bibinfo {volume} {16}},\ \bibinfo {pages}
  {465--489}}\BibitemShut {NoStop}%
\bibitem [{\citenamefont {Randall}(2000)}]{Randall2000}%
  \BibitemOpen
  \bibinfo {editor} {\bibnamefont {Randall}, \bibfnamefont {D~A}},\ Ed.
  (\bibinfo {year} {2000}),\ \href@noop {} {\emph {\bibinfo {title} {{General
  Circulation Model Development: Past, Present and Future}}}}\ (\bibinfo
  {publisher} {Academic Press},\ \bibinfo {address} {New York})\BibitemShut
  {NoStop}%
\bibitem [{\citenamefont {Rasmusson}\ \emph {et~al.}(1990)\citenamefont
  {Rasmusson}, \citenamefont {Wang},\ and\ \citenamefont
  {Ropelewski}}]{Rasmusson1990}%
  \BibitemOpen
  \bibfield  {author} {\bibinfo {author} {\bibnamefont {Rasmusson},
  \bibfnamefont {E}}, \bibinfo {author} {\bibfnamefont {X.}~\bibnamefont
  {Wang}}, \ and\ \bibinfo {author} {\bibfnamefont {C.}~\bibnamefont
  {Ropelewski}}} (\bibinfo {year} {1990}),\ \bibfield  {title} {\enquote
  {\bibinfo {title} {{The biennial component of ENSO variability}},}\
  }\href@noop {} {\bibfield  {journal} {\bibinfo  {journal} {J. Marine
  Systems}\ }\textbf {\bibinfo {volume} {1}},\ \bibinfo {pages}
  {71--96}}\BibitemShut {NoStop}%
\bibitem [{\citenamefont {Richardson}(1922)}]{Richardson.1922}%
  \BibitemOpen
  \bibfield  {author} {\bibinfo {author} {\bibnamefont {Richardson},
  \bibfnamefont {L~F}}} (\bibinfo {year} {1922}),\ \href@noop {} {\emph
  {\bibinfo {title} {Weather Prediction by Numerical Process}}}\ (\bibinfo
  {publisher} {Cambridge University Press},\ \bibinfo {address} {Cambridge,
  UK})\BibitemShut {NoStop}%
\bibitem [{\citenamefont {Riehl}(1954)}]{Riehl.1954}%
  \BibitemOpen
  \bibfield  {author} {\bibinfo {author} {\bibnamefont {Riehl}, \bibfnamefont
  {H}}} (\bibinfo {year} {1954}),\ \href@noop {} {\emph {\bibinfo {title}
  {{Tropical Meteorology}}}}\ (\bibinfo  {publisher} {McGraw-Hill})\BibitemShut
  {NoStop}%
\bibitem [{\citenamefont {Robert}\ \emph {et~al.}(2000)\citenamefont {Robert},
  \citenamefont {Alligood}, \citenamefont {Ott},\ and\ \citenamefont
  {Yorke}}]{Robert2000}%
  \BibitemOpen
  \bibfield  {author} {\bibinfo {author} {\bibnamefont {Robert}, \bibfnamefont
  {C}}, \bibinfo {author} {\bibfnamefont {K.~T.}\ \bibnamefont {Alligood}},
  \bibinfo {author} {\bibfnamefont {E.}~\bibnamefont {Ott}}, \ and\ \bibinfo
  {author} {\bibfnamefont {J.~A.}\ \bibnamefont {Yorke}}} (\bibinfo {year}
  {2000}),\ \bibfield  {title} {\enquote {\bibinfo {title} {Explosions of
  chaotic sets},}\ }\href {\doibase 10.1016/S0167-2789(00)00074-9} {\bibfield
  {journal} {\bibinfo  {journal} {Physica D: Nonlinear Phenomena}\ }\textbf
  {\bibinfo {volume} {144}}~(\bibinfo {number} {1}),\ \bibinfo {pages}
  {44--61}}\BibitemShut {NoStop}%
\bibitem [{\citenamefont {Robert}\ and\ \citenamefont
  {Sommeria}(1991)}]{Robert1991}%
  \BibitemOpen
  \bibfield  {author} {\bibinfo {author} {\bibnamefont {Robert}, \bibfnamefont
  {R}}, \ and\ \bibinfo {author} {\bibfnamefont {J.}~\bibnamefont {Sommeria}}}
  (\bibinfo {year} {1991}),\ \bibfield  {title} {\enquote {\bibinfo {title}
  {Statistical equilibrium states for two-dimensional flows},}\ }\href
  {\doibase 10.1017/S0022112091003038} {\bibfield  {journal} {\bibinfo
  {journal} {J. Fluid Mech.}\ }\textbf {\bibinfo {volume} {229}},\ \bibinfo
  {pages} {291--310}}\BibitemShut {NoStop}%
\bibitem [{\citenamefont {Robertson}\ and\ \citenamefont
  {Vitart}(2018)}]{S2S.book}%
  \BibitemOpen
  \bibinfo {editor} {\bibnamefont {Robertson}, \bibfnamefont {A~W}}, \ and\
  \bibinfo {editor} {\bibfnamefont {F.}~\bibnamefont {Vitart}},\ Eds. (\bibinfo
  {year} {2018}),\ \href@noop {} {\emph {\bibinfo {title} {The Gap Between
  Weather and Climate Forecasting: Sub-Seasonal to Seasonal Prediction}}}\
  (\bibinfo  {publisher} {Elsevier},\ \bibinfo {address}
  {Amsterdam})\BibitemShut {NoStop}%
\bibitem [{\citenamefont {Robin}\ \emph {et~al.}(2017)\citenamefont {Robin},
  \citenamefont {Yiou},\ and\ \citenamefont {Naveau}}]{Robin2017}%
  \BibitemOpen
  \bibfield  {author} {\bibinfo {author} {\bibnamefont {Robin}, \bibfnamefont
  {Y}}, \bibinfo {author} {\bibfnamefont {P.}~\bibnamefont {Yiou}}, \ and\
  \bibinfo {author} {\bibfnamefont {P.}~\bibnamefont {Naveau}}} (\bibinfo
  {year} {2017}),\ \bibfield  {title} {\enquote {\bibinfo {title} {Detecting
  changes in forced climate attractors with wasserstein distance},}\ }\href
  {\doibase 10.5194/npg-24-393-2017} {\bibfield  {journal} {\bibinfo  {journal}
  {Nonlinear Processes in Geophysics}\ }\textbf {\bibinfo {volume}
  {24}}~(\bibinfo {number} {3}),\ \bibinfo {pages} {393--405}}\BibitemShut
  {NoStop}%
\bibitem [{\citenamefont {Robinson}(2010)}]{Rob.10}%
  \BibitemOpen
  \bibfield  {author} {\bibinfo {author} {\bibnamefont {Robinson},
  \bibfnamefont {I~S}}} (\bibinfo {year} {2010}),\ \href@noop {} {\emph
  {\bibinfo {title} {Discovering the Ocean From Space: The Unique Applications
  of Satellite Oceanography}}}\ (\bibinfo  {publisher} {Praxis Publishing},\
  \bibinfo {address} {UK})\BibitemShut {NoStop}%
\bibitem [{\citenamefont {Roe}\ and\ \citenamefont {Baker}(2007)}]{Roe2007}%
  \BibitemOpen
  \bibfield  {author} {\bibinfo {author} {\bibnamefont {Roe}, \bibfnamefont
  {G~H}}, \ and\ \bibinfo {author} {\bibfnamefont {M.~B.}\ \bibnamefont
  {Baker}}} (\bibinfo {year} {2007}),\ \bibfield  {title} {\enquote {\bibinfo
  {title} {Why is climate sensitivity so unpredictable?}}\ }\href {\doibase
  10.1126/science.1144735} {\bibfield  {journal} {\bibinfo  {journal}
  {Science}\ }\textbf {\bibinfo {volume} {318}}~(\bibinfo {number} {5850}),\
  \bibinfo {pages} {629--632}}\BibitemShut {NoStop}%
\bibitem [{\citenamefont {Rombouts}\ and\ \citenamefont
  {Ghil}(2015)}]{Rombouts.2015}%
  \BibitemOpen
  \bibfield  {author} {\bibinfo {author} {\bibnamefont {Rombouts},
  \bibfnamefont {J}}, \ and\ \bibinfo {author} {\bibfnamefont {M.}~\bibnamefont
  {Ghil}}} (\bibinfo {year} {2015}),\ \bibfield  {title} {\enquote {\bibinfo
  {title} {Oscillations in a simple climate--vegetation model},}\ }\href@noop
  {} {\bibfield  {journal} {\bibinfo  {journal} {Nonlinear Processes in
  Geophysics (Online)}\ }\textbf {\bibinfo {volume} {22}}~(\bibinfo {number}
  {3})}\BibitemShut {NoStop}%
\bibitem [{\citenamefont {Romeiras}\ \emph {et~al.}(1990)\citenamefont
  {Romeiras}, \citenamefont {Grebogi},\ and\ \citenamefont
  {Ott}}]{Ott.ea.1990}%
  \BibitemOpen
  \bibfield  {author} {\bibinfo {author} {\bibnamefont {Romeiras},
  \bibfnamefont {F~J}}, \bibinfo {author} {\bibfnamefont {C.}~\bibnamefont
  {Grebogi}}, \ and\ \bibinfo {author} {\bibfnamefont {E.}~\bibnamefont {Ott}}}
  (\bibinfo {year} {1990}),\ \bibfield  {title} {\enquote {\bibinfo {title}
  {Multifractal properties of snapshot attractors of random maps},}\
  }\href@noop {} {\bibfield  {journal} {\bibinfo  {journal} {Phys. Rev. A}\
  }\textbf {\bibinfo {volume} {41}}~(\bibinfo {number} {2}),\ \bibinfo {pages}
  {784}}\BibitemShut {NoStop}%
\bibitem [{\citenamefont {Rooth}(1982)}]{Rooth1982}%
  \BibitemOpen
  \bibfield  {author} {\bibinfo {author} {\bibnamefont {Rooth}, \bibfnamefont
  {C}}} (\bibinfo {year} {1982}),\ \bibfield  {title} {\enquote {\bibinfo
  {title} {Hydrology and ocean circulation},}\ }\href@noop {} {\bibfield
  {journal} {\bibinfo  {journal} {Progress in Oceanography}\ }\textbf {\bibinfo
  {volume} {11}},\ \bibinfo {pages} {131--149}}\BibitemShut {NoStop}%
\bibitem [{\citenamefont {Roques}\ \emph {et~al.}(2014)\citenamefont {Roques},
  \citenamefont {Chekroun}, \citenamefont {Cristofol}, \citenamefont
  {Soubeyrand},\ and\ \citenamefont {Ghil}}]{Roques.ea.14}%
  \BibitemOpen
  \bibfield  {author} {\bibinfo {author} {\bibnamefont {Roques}, \bibfnamefont
  {L}}, \bibinfo {author} {\bibfnamefont {M.~D.}\ \bibnamefont {Chekroun}},
  \bibinfo {author} {\bibfnamefont {M.}~\bibnamefont {Cristofol}}, \bibinfo
  {author} {\bibfnamefont {S.}~\bibnamefont {Soubeyrand}}, \ and\ \bibinfo
  {author} {\bibfnamefont {M.}~\bibnamefont {Ghil}}} (\bibinfo {year} {2014}),\
  \bibfield  {title} {\enquote {\bibinfo {title} {Parameter estimation for
  energy balance models with memory},}\ }\href@noop {} {\bibfield  {journal}
  {\bibinfo  {journal} {Proc R. Soc. A}\ }\textbf {\bibinfo {volume} {470}},\
  \bibinfo {pages} {20140349}}\BibitemShut {NoStop}%
\bibitem [{\citenamefont {Rossby}\ \emph {et~al.}(1939)\citenamefont {Rossby}
  \emph {et~al.}}]{Rossby.ea.1939}%
  \BibitemOpen
  \bibfield  {author} {\bibinfo {author} {\bibnamefont {Rossby}, \bibfnamefont
  {C-G}},  \emph {et~al.}} (\bibinfo {year} {1939}),\ \bibfield  {title}
  {\enquote {\bibinfo {title} {Relation between variations in the intensity of
  the zonal circulation of the atmosphere and the displacements of the
  semi-permanent centers of action},}\ }\href@noop {} {\bibfield  {journal}
  {\bibinfo  {journal} {J. Marine Res.}\ }\textbf {\bibinfo {volume}
  {2}}~(\bibinfo {number} {1}),\ \bibinfo {pages} {38--55}}\BibitemShut
  {NoStop}%
\bibitem [{\citenamefont {Rothman}\ \emph {et~al.}(2003)\citenamefont
  {Rothman}, \citenamefont {Hayes},\ and\ \citenamefont
  {Summons}}]{Rothman.ea.2003}%
  \BibitemOpen
  \bibfield  {author} {\bibinfo {author} {\bibnamefont {Rothman}, \bibfnamefont
  {D~H}}, \bibinfo {author} {\bibfnamefont {J.~M.}\ \bibnamefont {Hayes}}, \
  and\ \bibinfo {author} {\bibfnamefont {R.~E.}\ \bibnamefont {Summons}}}
  (\bibinfo {year} {2003}),\ \bibfield  {title} {\enquote {\bibinfo {title}
  {Dynamics of the {Neoproterozoic carbon cycle}},}\ }\href {\doibase
  10.1073/pnas.0832439100} {\bibfield  {journal} {\bibinfo  {journal}
  {Proceedings of the National Academy of Sciences}\ }\textbf {\bibinfo
  {volume} {100}}~(\bibinfo {number} {14}),\ \bibinfo {pages}
  {8124--8129}}\BibitemShut {NoStop}%
\bibitem [{\citenamefont {Ruelle}(1986)}]{Ruelle1986}%
  \BibitemOpen
  \bibfield  {author} {\bibinfo {author} {\bibnamefont {Ruelle}, \bibfnamefont
  {D}}} (\bibinfo {year} {1986}),\ \bibfield  {title} {\enquote {\bibinfo
  {title} {Resonances of chaotic dynamical systems},}\ }\href {\doibase
  10.1103/PhysRevLett.56.405} {\bibfield  {journal} {\bibinfo  {journal}
  {Physical Review Letters}\ }\textbf {\bibinfo {volume} {56}},\ \bibinfo
  {pages} {405--407}}\BibitemShut {NoStop}%
\bibitem [{\citenamefont {Ruelle}(1998)}]{Ruelle:1998}%
  \BibitemOpen
  \bibfield  {author} {\bibinfo {author} {\bibnamefont {Ruelle}, \bibfnamefont
  {D}}} (\bibinfo {year} {1998}),\ \bibfield  {title} {\enquote {\bibinfo
  {title} {General linear response formula in statistical mechanics, and the
  fluctuation-dissipation theorem far from equilibrium},}\ }\href@noop {}
  {\bibfield  {journal} {\bibinfo  {journal} {Phys. Lett. A}\ }\textbf
  {\bibinfo {volume} {245}},\ \bibinfo {pages} {220--224}}\BibitemShut
  {NoStop}%
\bibitem [{\citenamefont {Ruelle}(1999)}]{ruelle_smooth_1999}%
  \BibitemOpen
  \bibfield  {author} {\bibinfo {author} {\bibnamefont {Ruelle}, \bibfnamefont
  {D}}} (\bibinfo {year} {1999}),\ \bibfield  {title} {\enquote {\bibinfo
  {title} {Smooth dynamics and new theoretical ideas in nonequilibrium
  statistical mechanics},}\ }\href@noop {} {\bibfield  {journal} {\bibinfo
  {journal} {Journal of Statistical Physics}\ }\textbf {\bibinfo {volume}
  {95}}~(\bibinfo {number} {1})}\BibitemShut {NoStop}%
\bibitem [{\citenamefont {Ruelle}(2009)}]{ruelle2009}%
  \BibitemOpen
  \bibfield  {author} {\bibinfo {author} {\bibnamefont {Ruelle}, \bibfnamefont
  {D}}} (\bibinfo {year} {2009}),\ \bibfield  {title} {\enquote {\bibinfo
  {title} {A review of linear response theory for general differentiable
  dynamical systems},}\ }\href@noop {} {\bibfield  {journal} {\bibinfo
  {journal} {Nonlinearity}\ }\textbf {\bibinfo {volume} {22}},\ \bibinfo
  {pages} {855--870}}\BibitemShut {NoStop}%
\bibitem [{\citenamefont {Ruti}\ \emph {et~al.}(2006)\citenamefont {Ruti},
  \citenamefont {Lucarini}, \citenamefont {Dell'Aquila}, \citenamefont
  {Calmanti},\ and\ \citenamefont {Speranza}}]{Ruti2006}%
  \BibitemOpen
  \bibfield  {author} {\bibinfo {author} {\bibnamefont {Ruti}, \bibfnamefont
  {P~M}}, \bibinfo {author} {\bibfnamefont {V.}~\bibnamefont {Lucarini}},
  \bibinfo {author} {\bibfnamefont {A.}~\bibnamefont {Dell'Aquila}}, \bibinfo
  {author} {\bibfnamefont {S.}~\bibnamefont {Calmanti}}, \ and\ \bibinfo
  {author} {\bibfnamefont {A.}~\bibnamefont {Speranza}}} (\bibinfo {year}
  {2006}),\ \bibfield  {title} {\enquote {\bibinfo {title} {Does the
  subtropical jet catalyze the midlatitude atmospheric regimes?}}\ }\href
  {\doibase 10.1029/2005GL024620} {\bibfield  {journal} {\bibinfo  {journal}
  {Geophysical Research Letters}\ }\textbf {\bibinfo {volume} {33}}~(\bibinfo
  {number} {6}),\ 10.1029/2005GL024620},\ \Eprint
  {http://arxiv.org/abs/https://agupubs.onlinelibrary.wiley.com/doi/pdf/10.1029/2005GL024620}
  {https://agupubs.onlinelibrary.wiley.com/doi/pdf/10.1029/2005GL024620}
  \BibitemShut {NoStop}%
\bibitem [{\citenamefont {Salmon}(1998)}]{SalmonBook}%
  \BibitemOpen
  \bibfield  {author} {\bibinfo {author} {\bibnamefont {Salmon}, \bibfnamefont
  {Rick}}} (\bibinfo {year} {1998}),\ \href@noop {} {\emph {\bibinfo {title}
  {{Lectures on Geophysical Fluid Dynamics}}}}\ (\bibinfo  {publisher} {Oxford
  University Press},\ \bibinfo {address} {Oxford})\BibitemShut {NoStop}%
\bibitem [{\citenamefont {Saltzman}(2001)}]{saltzman_dynamical}%
  \BibitemOpen
  \bibfield  {author} {\bibinfo {author} {\bibnamefont {Saltzman},
  \bibfnamefont {B}}} (\bibinfo {year} {2001}),\ \href@noop {} {\emph {\bibinfo
  {title} {Dynamical Paleoclimatology: Generalized Theory of Global Climate
  Change}}}\ (\bibinfo  {publisher} {Academic Press New York},\ \bibinfo
  {address} {New York})\BibitemShut {NoStop}%
\bibitem [{\citenamefont {Sardeshmukh}\ and\ \citenamefont
  {Penland}(2015)}]{Sard.Pen.2015}%
  \BibitemOpen
  \bibfield  {author} {\bibinfo {author} {\bibnamefont {Sardeshmukh},
  \bibfnamefont {P~D}}, \ and\ \bibinfo {author} {\bibfnamefont
  {C.}~\bibnamefont {Penland}}} (\bibinfo {year} {2015}),\ \bibfield  {title}
  {\enquote {\bibinfo {title} {Understanding the distinctively skewed and heavy
  tailed character of atmospheric and oceanic probability distributions},}\
  }\href {\doibase 10.1063/1.4914169} {\bibfield  {journal} {\bibinfo
  {journal} {Chaos}\ }\textbf {\bibinfo {volume} {25}},\ \bibinfo {pages}
  {036410}}\BibitemShut {NoStop}%
\bibitem [{\citenamefont {Satoh}\ \emph {et~al.}(2018)\citenamefont {Satoh},
  \citenamefont {Noda}, \citenamefont {Seiki}, \citenamefont {Chen},
  \citenamefont {Kodama}, \citenamefont {Yamada}, \citenamefont {Kuba},\ and\
  \citenamefont {Sato}}]{Satoh2018}%
  \BibitemOpen
  \bibfield  {author} {\bibinfo {author} {\bibnamefont {Satoh}, \bibfnamefont
  {M}}, \bibinfo {author} {\bibfnamefont {A.~T.}\ \bibnamefont {Noda}},
  \bibinfo {author} {\bibfnamefont {T.}~\bibnamefont {Seiki}}, \bibinfo
  {author} {\bibfnamefont {Y.-W.}\ \bibnamefont {Chen}}, \bibinfo {author}
  {\bibfnamefont {C.}~\bibnamefont {Kodama}}, \bibinfo {author} {\bibfnamefont
  {Y.}~\bibnamefont {Yamada}}, \bibinfo {author} {\bibfnamefont
  {N.}~\bibnamefont {Kuba}}, \ and\ \bibinfo {author} {\bibfnamefont
  {Y.}~\bibnamefont {Sato}}} (\bibinfo {year} {2018}),\ \bibfield  {title}
  {\enquote {\bibinfo {title} {Toward reduction of the uncertainties in climate
  sensitivity due to cloud processes using a global non-hydrostatic atmospheric
  model},}\ }\href {\doibase 10.1186/s40645-018-0226-1} {\bibfield  {journal}
  {\bibinfo  {journal} {Progress in Earth and Planetary Science}\ }\textbf
  {\bibinfo {volume} {5}}~(\bibinfo {number} {1}),\ \bibinfo {pages}
  {67}}\BibitemShut {NoStop}%
\bibitem [{\citenamefont {Scheffer}\ \emph {et~al.}(2009)\citenamefont
  {Scheffer}, \citenamefont {Bascompte}, \citenamefont {Brock}, \citenamefont
  {Brovkin}, \citenamefont {Carpenter}, \citenamefont {Dakos}, \citenamefont
  {Held}, \citenamefont {Van~Nes}, \citenamefont {Rietkerk},\ and\
  \citenamefont {Sugihara}}]{scheffer2009early}%
  \BibitemOpen
  \bibfield  {author} {\bibinfo {author} {\bibnamefont {Scheffer},
  \bibfnamefont {M}}, \bibinfo {author} {\bibfnamefont {J.}~\bibnamefont
  {Bascompte}}, \bibinfo {author} {\bibfnamefont {W.~A.}\ \bibnamefont
  {Brock}}, \bibinfo {author} {\bibfnamefont {V.}~\bibnamefont {Brovkin}},
  \bibinfo {author} {\bibfnamefont {S.~R.}\ \bibnamefont {Carpenter}}, \bibinfo
  {author} {\bibfnamefont {V.}~\bibnamefont {Dakos}}, \bibinfo {author}
  {\bibfnamefont {H.}~\bibnamefont {Held}}, \bibinfo {author} {\bibfnamefont
  {Egbert~H.}\ \bibnamefont {Van~Nes}}, \bibinfo {author} {\bibfnamefont
  {M.}~\bibnamefont {Rietkerk}}, \ and\ \bibinfo {author} {\bibfnamefont
  {G.}~\bibnamefont {Sugihara}}} (\bibinfo {year} {2009}),\ \bibfield  {title}
  {\enquote {\bibinfo {title} {Early-warning signals for critical
  transitions},}\ }\href@noop {} {\bibfield  {journal} {\bibinfo  {journal}
  {Nature}\ }\textbf {\bibinfo {volume} {461}}~(\bibinfo {number} {7260}),\
  \bibinfo {pages} {53--59}}\BibitemShut {NoStop}%
\bibitem [{\citenamefont {Schneider}\ and\ \citenamefont
  {Dickinson}(1974)}]{Schneider1974}%
  \BibitemOpen
  \bibfield  {author} {\bibinfo {author} {\bibnamefont {Schneider},
  \bibfnamefont {S~H}}, \ and\ \bibinfo {author} {\bibfnamefont {R.~E.}\
  \bibnamefont {Dickinson}}} (\bibinfo {year} {1974}),\ \bibfield  {title}
  {\enquote {\bibinfo {title} {{Climate modelling}},}\ }\href@noop {}
  {\bibfield  {journal} {\bibinfo  {journal} {Rev. Geophys. Space Phys.}\
  }\textbf {\bibinfo {volume} {25}},\ \bibinfo {pages} {447--493}}\BibitemShut
  {NoStop}%
\bibitem [{\citenamefont {Schneider}\ \emph {et~al.}(2017)\citenamefont
  {Schneider}, \citenamefont {Teixeira}, \citenamefont {Bretherton},
  \citenamefont {Brient}, \citenamefont {Pressel}, \citenamefont {Sch{\"a}r},\
  and\ \citenamefont {Siebesma}}]{Schneider2017}%
  \BibitemOpen
  \bibfield  {author} {\bibinfo {author} {\bibnamefont {Schneider},
  \bibfnamefont {T}}, \bibinfo {author} {\bibfnamefont {J.}~\bibnamefont
  {Teixeira}}, \bibinfo {author} {\bibfnamefont {C.~S.}\ \bibnamefont
  {Bretherton}}, \bibinfo {author} {\bibfnamefont {F.}~\bibnamefont {Brient}},
  \bibinfo {author} {\bibfnamefont {K.~G.}\ \bibnamefont {Pressel}}, \bibinfo
  {author} {\bibfnamefont {C.}~\bibnamefont {Sch{\"a}r}}, \ and\ \bibinfo
  {author} {\bibfnamefont {A.~P.}\ \bibnamefont {Siebesma}}} (\bibinfo {year}
  {2017}),\ \bibfield  {title} {\enquote {\bibinfo {title} {Climate goals and
  computing the future of clouds},}\ }\href {\doibase 10.1038/nclimate3190}
  {\bibfield  {journal} {\bibinfo  {journal} {Nat Clim Chang}\ }\textbf
  {\bibinfo {volume} {7}},\ 10.1038/nclimate3190}\BibitemShut {NoStop}%
\bibitem [{\citenamefont {Schneider}\ \emph {et~al.}(2007)\citenamefont
  {Schneider}, \citenamefont {Eckhardt},\ and\ \citenamefont
  {Yorke}}]{Schneider2007}%
  \BibitemOpen
  \bibfield  {author} {\bibinfo {author} {\bibnamefont {Schneider},
  \bibfnamefont {T~M}}, \bibinfo {author} {\bibfnamefont {B.}~\bibnamefont
  {Eckhardt}}, \ and\ \bibinfo {author} {\bibfnamefont {J.~A.}\ \bibnamefont
  {Yorke}}} (\bibinfo {year} {2007}),\ \bibfield  {title} {\enquote {\bibinfo
  {title} {Turbulence transition and the edge of chaos in pipe flow},}\ }\href
  {\doibase 10.1103/PhysRevLett.99.034502} {\bibfield  {journal} {\bibinfo
  {journal} {Physical Review Letters}\ }\textbf {\bibinfo {volume} {99}},\
  \bibinfo {pages} {034502}}\BibitemShut {NoStop}%
\bibitem [{\citenamefont {Schneider}\ \emph {et~al.}(2019)\citenamefont
  {Schneider}, \citenamefont {Kaul},\ and\ \citenamefont
  {Pressel}}]{Schneider2019}%
  \BibitemOpen
  \bibfield  {author} {\bibinfo {author} {\bibnamefont {Schneider},
  \bibfnamefont {Tapio}}, \bibinfo {author} {\bibfnamefont {Colleen~M.}\
  \bibnamefont {Kaul}}, \ and\ \bibinfo {author} {\bibfnamefont {Kyle~G.}\
  \bibnamefont {Pressel}}} (\bibinfo {year} {2019}),\ \bibfield  {title}
  {\enquote {\bibinfo {title} {Possible climate transitions from breakup of
  stratocumulus decks under greenhouse warming},}\ }\href {\doibase
  10.1038/s41561-019-0310-1} {\bibfield  {journal} {\bibinfo  {journal} {Nature
  Geoscience}\ }\textbf {\bibinfo {volume} {12}}~(\bibinfo {number} {3}),\
  \bibinfo {pages} {163--167}}\BibitemShut {NoStop}%
\bibitem [{\citenamefont {Schubert}\ and\ \citenamefont
  {Lucarini}(2016)}]{Schubert2016}%
  \BibitemOpen
  \bibfield  {author} {\bibinfo {author} {\bibnamefont {Schubert},
  \bibfnamefont {Sebastian}}, \ and\ \bibinfo {author} {\bibfnamefont
  {Valerio}\ \bibnamefont {Lucarini}}} (\bibinfo {year} {2016}),\ \bibfield
  {title} {\enquote {\bibinfo {title} {Dynamical analysis of blocking events:
  spatial and temporal fluctuations of covariant lyapunov vectors},}\ }\href
  {\doibase 10.1002/qj.2808} {\bibfield  {journal} {\bibinfo  {journal}
  {Quarterly Journal of the Royal Meteorological Society}\ }\textbf {\bibinfo
  {volume} {142}}~(\bibinfo {number} {698}),\ \bibinfo {pages}
  {2143--2158}}\BibitemShut {NoStop}%
\bibitem [{\citenamefont {Scott}\ \emph {et~al.}(1999)\citenamefont {Scott},
  \citenamefont {Marotzke},\ and\ \citenamefont {Stone}}]{Scott1999}%
  \BibitemOpen
  \bibfield  {author} {\bibinfo {author} {\bibnamefont {Scott}, \bibfnamefont
  {J~R}}, \bibinfo {author} {\bibfnamefont {J.}~\bibnamefont {Marotzke}}, \
  and\ \bibinfo {author} {\bibfnamefont {P.~H.}\ \bibnamefont {Stone}}}
  (\bibinfo {year} {1999}),\ \bibfield  {title} {\enquote {\bibinfo {title}
  {Interhemispheric thermohaline circulation in a coupled box model},}\ }\href
  {\doibase 10.1175/1520-0485(1999)029<0351:ITCIAC>2.0.CO;2} {\bibfield
  {journal} {\bibinfo  {journal} {Journal of Physical Oceanography}\ }\textbf
  {\bibinfo {volume} {29}}~(\bibinfo {number} {3}),\ \bibinfo {pages}
  {351--365}}\BibitemShut {NoStop}%
\bibitem [{\citenamefont {Sell}(1967)}]{Sell.1967}%
  \BibitemOpen
  \bibfield  {author} {\bibinfo {author} {\bibnamefont {Sell}, \bibfnamefont
  {G~R}}} (\bibinfo {year} {1967}),\ \bibfield  {title} {\enquote {\bibinfo
  {title} {Nonautonomous differential equations and topological dynamics. {I}.
  {The basic theory}},}\ }\href@noop {} {\bibfield  {journal} {\bibinfo
  {journal} {Trans. Amer. Math. Soc.}\ }\textbf {\bibinfo {volume} {127}},\
  \bibinfo {pages} {241--262}}\BibitemShut {NoStop}%
\bibitem [{\citenamefont {Sell}(1971)}]{Sell.1971}%
  \BibitemOpen
  \bibfield  {author} {\bibinfo {author} {\bibnamefont {Sell}, \bibfnamefont
  {G~R}}} (\bibinfo {year} {1971}),\ \href@noop {} {\emph {\bibinfo {title}
  {{Topological Dynamics and Ordinary Differential Equations}}}}\ (\bibinfo
  {publisher} {Van Nostrand Reinhold})\BibitemShut {NoStop}%
\bibitem [{\citenamefont {Sellers}(1969)}]{Sellers}%
  \BibitemOpen
  \bibfield  {author} {\bibinfo {author} {\bibnamefont {Sellers}, \bibfnamefont
  {W~D}}} (\bibinfo {year} {1969}),\ \bibfield  {title} {\enquote {\bibinfo
  {title} {A global climatic model based on the energy balance of the earth
  atmosphere},}\ }\href@noop {} {\bibfield  {journal} {\bibinfo  {journal} {J.
  Appl. Meteorol.}\ }\textbf {\bibinfo {volume} {8}},\ \bibinfo {pages}
  {392--400}}\BibitemShut {NoStop}%
\bibitem [{\citenamefont {Sevellec}\ and\ \citenamefont
  {Fedorov}(2015)}]{Sevellec2015}%
  \BibitemOpen
  \bibfield  {author} {\bibinfo {author} {\bibnamefont {Sevellec},
  \bibfnamefont {F}}, \ and\ \bibinfo {author} {\bibfnamefont {A.~V.}\
  \bibnamefont {Fedorov}}} (\bibinfo {year} {2015}),\ \bibfield  {title}
  {\enquote {\bibinfo {title} {Unstable {AMOC during glacial intervals and
  millennial variability: The role of mean sea ice extent}},}\ }\href {\doibase
  10.1016/j.epsl.2015.07.022} {\bibfield  {journal} {\bibinfo  {journal} {Earth
  and Planetary Science Letters}\ }\textbf {\bibinfo {volume} {429}},\ \bibinfo
  {pages} {60--68}}\BibitemShut {NoStop}%
\bibitem [{\citenamefont {Sheremet}\ \emph {et~al.}(1997)\citenamefont
  {Sheremet}, \citenamefont {Ierley},\ and\ \citenamefont
  {Kamenkovich}}]{Sheremet1997}%
  \BibitemOpen
  \bibfield  {author} {\bibinfo {author} {\bibnamefont {Sheremet},
  \bibfnamefont {V~A}}, \bibinfo {author} {\bibfnamefont {G.~R.}\ \bibnamefont
  {Ierley}}, \ and\ \bibinfo {author} {\bibfnamefont {V.~M.}\ \bibnamefont
  {Kamenkovich}}} (\bibinfo {year} {1997}),\ \bibfield  {title} {\enquote
  {\bibinfo {title} {Eigenanalysis of the two-dimensional wind-driven ocean
  circulation problem},}\ }\href@noop {} {\bibfield  {journal} {\bibinfo
  {journal} {J.\ Mar.\ Res.}\ }\textbf {\bibinfo {volume} {55}},\ \bibinfo
  {pages} {57--92}}\BibitemShut {NoStop}%
\bibitem [{\citenamefont {Simonnet}(2005)}]{Simonnet2005}%
  \BibitemOpen
  \bibfield  {author} {\bibinfo {author} {\bibnamefont {Simonnet},
  \bibfnamefont {E}}} (\bibinfo {year} {2005}),\ \bibfield  {title} {\enquote
  {\bibinfo {title} {Quantization of the low-frequency variability of the
  double-gyre circulation},}\ }\href@noop {} {\bibfield  {journal} {\bibinfo
  {journal} {J. Phys. Oceanogr.}\ }\textbf {\bibinfo {volume} {35}},\ \bibinfo
  {pages} {2268--2290}}\BibitemShut {NoStop}%
\bibitem [{\citenamefont {Simonnet}\ and\ \citenamefont
  {Dijkstra}(2002)}]{Simonnet2002}%
  \BibitemOpen
  \bibfield  {author} {\bibinfo {author} {\bibnamefont {Simonnet},
  \bibfnamefont {E}}, \ and\ \bibinfo {author} {\bibfnamefont {H.~A.}\
  \bibnamefont {Dijkstra}}} (\bibinfo {year} {2002}),\ \bibfield  {title}
  {\enquote {\bibinfo {title} {Spontaneous generation of low-frequency modes of
  variability in the wind-driven ocean circulation},}\ }\href@noop {}
  {\bibfield  {journal} {\bibinfo  {journal} {J.\ Phys.\ Oceanogr.}\ }\textbf
  {\bibinfo {volume} {32}},\ \bibinfo {pages} {1747--1762}}\BibitemShut
  {NoStop}%
\bibitem [{\citenamefont {Simonnet}\ \emph {et~al.}(2005)\citenamefont
  {Simonnet}, \citenamefont {Ghil},\ and\ \citenamefont
  {Dijkstra}}]{Simonnet.ea.2005}%
  \BibitemOpen
  \bibfield  {author} {\bibinfo {author} {\bibnamefont {Simonnet},
  \bibfnamefont {E}}, \bibinfo {author} {\bibfnamefont {M.}~\bibnamefont
  {Ghil}}, \ and\ \bibinfo {author} {\bibfnamefont {H.~A.}\ \bibnamefont
  {Dijkstra}}} (\bibinfo {year} {2005}),\ \bibfield  {title} {\enquote
  {\bibinfo {title} {Homoclinc bifurcations in the quasi-geostrophic
  double-gyre circulation},}\ }\href@noop {} {\bibfield  {journal} {\bibinfo
  {journal} {J. Marine Res.}\ }\textbf {\bibinfo {volume} {63}},\ \bibinfo
  {pages} {931--956}}\BibitemShut {NoStop}%
\bibitem [{\citenamefont {Simonnet}\ \emph {et~al.}({2003a})\citenamefont
  {Simonnet}, \citenamefont {Ghil}, \citenamefont {Ide}, \citenamefont
  {Temam},\ and\ \citenamefont {Wang}}]{Simonnet2003a}%
  \BibitemOpen
  \bibfield  {author} {\bibinfo {author} {\bibnamefont {Simonnet},
  \bibfnamefont {E}}, \bibinfo {author} {\bibfnamefont {M.}~\bibnamefont
  {Ghil}}, \bibinfo {author} {\bibfnamefont {K.}~\bibnamefont {Ide}}, \bibinfo
  {author} {\bibfnamefont {R.}~\bibnamefont {Temam}}, \ and\ \bibinfo {author}
  {\bibfnamefont {S.}~\bibnamefont {Wang}}} (\bibinfo {year} {{2003a}}),\
  \bibfield  {title} {\enquote {\bibinfo {title} {Low-frequency variability in
  shallow-water models of the wind-driven ocean circulation. {P}art {I}:
  {S}teady-state solutions},}\ }\href@noop {} {\bibfield  {journal} {\bibinfo
  {journal} {J.\ Phys.\ Oceanogr.}\ }\textbf {\bibinfo {volume} {33}},\
  \bibinfo {pages} {712--728}}\BibitemShut {NoStop}%
\bibitem [{\citenamefont {Simonnet}\ \emph {et~al.}({2003b})\citenamefont
  {Simonnet}, \citenamefont {Ghil}, \citenamefont {Ide}, \citenamefont
  {Temam},\ and\ \citenamefont {Wang}}]{Simonnet2003b}%
  \BibitemOpen
  \bibfield  {author} {\bibinfo {author} {\bibnamefont {Simonnet},
  \bibfnamefont {E}}, \bibinfo {author} {\bibfnamefont {M.}~\bibnamefont
  {Ghil}}, \bibinfo {author} {\bibfnamefont {K.}~\bibnamefont {Ide}}, \bibinfo
  {author} {\bibfnamefont {R.}~\bibnamefont {Temam}}, \ and\ \bibinfo {author}
  {\bibfnamefont {S.}~\bibnamefont {Wang}}} (\bibinfo {year} {{2003b}}),\
  \bibfield  {title} {\enquote {\bibinfo {title} {Low-frequency variability in
  shallow-water models of the wind-driven ocean circulation. {P}art {II: Time
  dependent solutions}},}\ }\href@noop {} {\bibfield  {journal} {\bibinfo
  {journal} {J.\ Phys.\ Oceanogr.}\ }\textbf {\bibinfo {volume} {33}},\
  \bibinfo {pages} {729--752}}\BibitemShut {NoStop}%
\bibitem [{\citenamefont {Simonnet}\ \emph {et~al.}(1995)\citenamefont
  {Simonnet}, \citenamefont {Temam}, \citenamefont {Wang}, \citenamefont
  {Ghil},\ and\ \citenamefont {Ide}}]{SGITW1995}%
  \BibitemOpen
  \bibfield  {author} {\bibinfo {author} {\bibnamefont {Simonnet},
  \bibfnamefont {E}}, \bibinfo {author} {\bibfnamefont {R.}~\bibnamefont
  {Temam}}, \bibinfo {author} {\bibfnamefont {S.}~\bibnamefont {Wang}},
  \bibinfo {author} {\bibfnamefont {M.}~\bibnamefont {Ghil}}, \ and\ \bibinfo
  {author} {\bibfnamefont {K.}~\bibnamefont {Ide}}} (\bibinfo {year} {1995}),\
  \bibfield  {title} {\enquote {\bibinfo {title} {Successive bifurcations in a
  shallow-water ocean model},}\ }in\ \href@noop {} {\emph {\bibinfo {booktitle}
  {16\textsuperscript{th} Intl. Conf. Numerical Methods in Fluid Dynamics}}},\
  \bibinfo {series} {Lecture Notes in Physics}, Vol.\ \bibinfo {volume} {515}\
  (\bibinfo  {publisher} {Springer-Verlag})\ pp.\ \bibinfo {pages}
  {225--230}\BibitemShut {NoStop}%
\bibitem [{\citenamefont {Simonnet}\ \emph {et~al.}(2009)\citenamefont
  {Simonnet}, \citenamefont {Dijkstra},\ and\ \citenamefont
  {Ghil}}]{SDG.Hdbk.2009}%
  \BibitemOpen
  \bibfield  {author} {\bibinfo {author} {\bibnamefont {Simonnet},
  \bibfnamefont {S}}, \bibinfo {author} {\bibfnamefont {H.~A.}\ \bibnamefont
  {Dijkstra}}, \ and\ \bibinfo {author} {\bibfnamefont {M.}~\bibnamefont
  {Ghil}}} (\bibinfo {year} {2009}),\ \bibfield  {title} {\enquote {\bibinfo
  {title} {Bifurcation analysis of ocean, atmosphere and climate models},}\
  }in\ \href@noop {} {\emph {\bibinfo {booktitle} {Computational Methods for
  the Ocean and the Atmosphere}}},\ \bibinfo {editor} {edited by\ \bibinfo
  {editor} {\bibfnamefont {R.}~\bibnamefont {Temam}}\ and\ \bibinfo {editor}
  {\bibfnamefont {J.~J.}\ \bibnamefont {Tribbia}}}\ (\bibinfo  {publisher}
  {North-Holland},\ \bibinfo {address} {Amsterdam})\ pp.\ \bibinfo {pages}
  {187--229}\BibitemShut {NoStop}%
\bibitem [{\citenamefont {Skufca}\ \emph {et~al.}(2006)\citenamefont {Skufca},
  \citenamefont {Yorke},\ and\ \citenamefont {Eckhardt}}]{Skufca2006}%
  \BibitemOpen
  \bibfield  {author} {\bibinfo {author} {\bibnamefont {Skufca}, \bibfnamefont
  {J~D}}, \bibinfo {author} {\bibfnamefont {J.~A.}\ \bibnamefont {Yorke}}, \
  and\ \bibinfo {author} {\bibfnamefont {B.}~\bibnamefont {Eckhardt}}}
  (\bibinfo {year} {2006}),\ \bibfield  {title} {\enquote {\bibinfo {title}
  {Edge of chaos in a parallel shear flow},}\ }\href {\doibase
  10.1103/PhysRevLett.96.174101} {\bibfield  {journal} {\bibinfo  {journal}
  {Physical Review Letters}\ }\textbf {\bibinfo {volume} {96}},\ \bibinfo
  {pages} {174101}}\BibitemShut {NoStop}%
\bibitem [{\citenamefont {Slingo}\ and\ \citenamefont
  {Palmer}(2011)}]{Slingo2011}%
  \BibitemOpen
  \bibfield  {author} {\bibinfo {author} {\bibnamefont {Slingo}, \bibfnamefont
  {J}}, \ and\ \bibinfo {author} {\bibfnamefont {T.~N.}\ \bibnamefont
  {Palmer}}} (\bibinfo {year} {2011}),\ \bibfield  {title} {\enquote {\bibinfo
  {title} {Uncertainty in weather and climate prediction},}\ }\href {\doibase
  10.1098/rsta.2011.0161} {\bibfield  {journal} {\bibinfo  {journal}
  {Philosophical Transactions of the Royal Society A: Mathematical, Physical
  and Engineering Sciences}\ }\textbf {\bibinfo {volume} {369}}~(\bibinfo
  {number} {1956}),\ \bibinfo {pages} {4751--4767}}\BibitemShut {NoStop}%
\bibitem [{\citenamefont {Smith}\ and\ \citenamefont
  {Wagner}(2018)}]{Smith_2018}%
  \BibitemOpen
  \bibfield  {author} {\bibinfo {author} {\bibnamefont {Smith}, \bibfnamefont
  {W}}, \ and\ \bibinfo {author} {\bibfnamefont {G.}~\bibnamefont {Wagner}}}
  (\bibinfo {year} {2018}),\ \bibfield  {title} {\enquote {\bibinfo {title}
  {Stratospheric aerosol injection tactics and costs in the first 15 years of
  deployment},}\ }\href {\doibase 10.1088/1748-9326/aae98d} {\bibfield
  {journal} {\bibinfo  {journal} {Environmental Research Letters}\ }\textbf
  {\bibinfo {volume} {13}}~(\bibinfo {number} {12}),\ \bibinfo {pages}
  {124001}}\BibitemShut {NoStop}%
\bibitem [{\citenamefont {Smyth}\ \emph {et~al.}(1999)\citenamefont {Smyth},
  \citenamefont {Ide},\ and\ \citenamefont {Ghil}}]{Smyth1999}%
  \BibitemOpen
  \bibfield  {author} {\bibinfo {author} {\bibnamefont {Smyth}, \bibfnamefont
  {P}}, \bibinfo {author} {\bibfnamefont {K.}~\bibnamefont {Ide}}, \ and\
  \bibinfo {author} {\bibfnamefont {M.}~\bibnamefont {Ghil}}} (\bibinfo {year}
  {1999}),\ \bibfield  {title} {\enquote {\bibinfo {title} {Multiple regimes in
  {Northern Hemisphere height fields via mixture model clustering}},}\
  }\href@noop {} {\bibfield  {journal} {\bibinfo  {journal} {J. Atmos. Sci.}\
  }\textbf {\bibinfo {volume} {56}},\ \bibinfo {pages}
  {3704--3723}}\BibitemShut {NoStop}%
\bibitem [{\citenamefont {Speich}\ \emph
  {et~al.}(1995{\natexlab{a}})\citenamefont {Speich}, \citenamefont
  {Dijkstra},\ and\ \citenamefont {Ghil}}]{SDG1995}%
  \BibitemOpen
  \bibfield  {author} {\bibinfo {author} {\bibnamefont {Speich}, \bibfnamefont
  {S}}, \bibinfo {author} {\bibfnamefont {H.~A.}\ \bibnamefont {Dijkstra}}, \
  and\ \bibinfo {author} {\bibfnamefont {M.}~\bibnamefont {Ghil}}} (\bibinfo
  {year} {1995}{\natexlab{a}}),\ \bibfield  {title} {\enquote {\bibinfo {title}
  {Successive bifurcations in a shallow-water model applied to the wind-driven
  ocean circulation},}\ }\href@noop {} {\bibfield  {journal} {\bibinfo
  {journal} {Nonlin. Processes Geophys.}\ }\textbf {\bibinfo {volume}
  {2}}~(\bibinfo {number} {241--268})}\BibitemShut {NoStop}%
\bibitem [{\citenamefont {Speich}\ \emph
  {et~al.}(1995{\natexlab{b}})\citenamefont {Speich}, \citenamefont
  {Dijkstra},\ and\ \citenamefont {Ghil}}]{Speich1995}%
  \BibitemOpen
  \bibfield  {author} {\bibinfo {author} {\bibnamefont {Speich}, \bibfnamefont
  {S}}, \bibinfo {author} {\bibfnamefont {H.~A.}\ \bibnamefont {Dijkstra}}, \
  and\ \bibinfo {author} {\bibfnamefont {M.}~\bibnamefont {Ghil}}} (\bibinfo
  {year} {1995}{\natexlab{b}}),\ \bibfield  {title} {\enquote {\bibinfo {title}
  {Successive bifurcations of a shallow-water model with applications to the
  wind driven circulation},}\ }\href@noop {} {\bibfield  {journal} {\bibinfo
  {journal} {Nonlin. Proc. Geophys.}\ }\textbf {\bibinfo {volume} {2}},\
  \bibinfo {pages} {241--268}}\BibitemShut {NoStop}%
\bibitem [{\citenamefont {Speranza}(1983)}]{Speranza83}%
  \BibitemOpen
  \bibfield  {author} {\bibinfo {author} {\bibnamefont {Speranza},
  \bibfnamefont {A}}} (\bibinfo {year} {1983}),\ \bibfield  {title} {\enquote
  {\bibinfo {title} {Deterministic and statistical properties of the
  westerlies},}\ }\href@noop {} {\bibfield  {journal} {\bibinfo  {journal}
  {Paleogeophysics}\ }\textbf {\bibinfo {volume} {121}},\ \bibinfo {pages}
  {511--562}}\BibitemShut {NoStop}%
\bibitem [{\citenamefont {Stommel}(1961)}]{Stommel1961}%
  \BibitemOpen
  \bibfield  {author} {\bibinfo {author} {\bibnamefont {Stommel}, \bibfnamefont
  {H}}} (\bibinfo {year} {1961}),\ \bibfield  {title} {\enquote {\bibinfo
  {title} {Thermohaline convection with two stable regimes of flow},}\
  }\href@noop {} {\bibfield  {journal} {\bibinfo  {journal} {Tellus}\ }\textbf
  {\bibinfo {volume} {2}},\ \bibinfo {pages} {244--230}}\BibitemShut {NoStop}%
\bibitem [{\citenamefont {Stommel}(1963)}]{Stommel63}%
  \BibitemOpen
  \bibfield  {author} {\bibinfo {author} {\bibnamefont {Stommel}, \bibfnamefont
  {H}}} (\bibinfo {year} {1963}),\ \bibfield  {title} {\enquote {\bibinfo
  {title} {Varieties of oceanographic experience},}\ }\href {\doibase
  10.1126/science.139.3555.572} {\bibfield  {journal} {\bibinfo  {journal}
  {Science}\ }\textbf {\bibinfo {volume} {139}}~(\bibinfo {number} {3555}),\
  \bibinfo {pages} {572--576}}\BibitemShut {NoStop}%
\bibitem [{\citenamefont {Stommel}(1965)}]{Stommel1965}%
  \BibitemOpen
  \bibfield  {author} {\bibinfo {author} {\bibnamefont {Stommel}, \bibfnamefont
  {H}}} (\bibinfo {year} {1965}),\ \href@noop {} {\emph {\bibinfo {title} {{The
  Gulf Stream: A Physical and Dynamical Description}}}}\ (\bibinfo  {publisher}
  {University of California Press},\ \bibinfo {address} {Berkeley and Los
  Angeles, California, USA})\BibitemShut {NoStop}%
\bibitem [{\citenamefont {Sushama}\ \emph {et~al.}(2007)\citenamefont
  {Sushama}, \citenamefont {Ghil},\ and\ \citenamefont {Ide}}]{Laxmi.ea.2007}%
  \BibitemOpen
  \bibfield  {author} {\bibinfo {author} {\bibnamefont {Sushama}, \bibfnamefont
  {L}}, \bibinfo {author} {\bibfnamefont {M.}~\bibnamefont {Ghil}}, \ and\
  \bibinfo {author} {\bibfnamefont {K.}~\bibnamefont {Ide}}} (\bibinfo {year}
  {2007}),\ \bibfield  {title} {\enquote {\bibinfo {title} {Spatio-temporal
  variability in a mid-latitude ocean basin subject to periodic wind
  forcing},}\ }\href {\doibase 10.3137/ao.450404} {\bibfield  {journal}
  {\bibinfo  {journal} {Atmosphere-Ocean}\ }\textbf {\bibinfo {volume} {45}},\
  \bibinfo {pages} {227--250}}\BibitemShut {NoStop}%
\bibitem [{\citenamefont {Sverdrup}(1947)}]{Sverdrup1947}%
  \BibitemOpen
  \bibfield  {author} {\bibinfo {author} {\bibnamefont {Sverdrup},
  \bibfnamefont {H~U}}} (\bibinfo {year} {1947}),\ \bibfield  {title} {\enquote
  {\bibinfo {title} {Wind-driven currents in a baroclinic ocean with
  application to the equatorial current in the eastern {P}acific},}\
  }\href@noop {} {\bibfield  {journal} {\bibinfo  {journal} {Proc. Natl. Acad.
  Sci. Wash.}\ }\textbf {\bibinfo {volume} {33}},\ \bibinfo {pages}
  {318--326}}\BibitemShut {NoStop}%
\bibitem [{\citenamefont {Sverdrup}\ \emph {et~al.}(1946)\citenamefont
  {Sverdrup}, \citenamefont {Johnson},\ and\ \citenamefont
  {Fleming}}]{Sverdrup1946}%
  \BibitemOpen
  \bibfield  {author} {\bibinfo {author} {\bibnamefont {Sverdrup},
  \bibfnamefont {H~U}}, \bibinfo {author} {\bibfnamefont {M.~W.}\ \bibnamefont
  {Johnson}}, \ and\ \bibinfo {author} {\bibfnamefont {R.~H.}\ \bibnamefont
  {Fleming}}} (\bibinfo {year} {1946}),\ \href@noop {} {\emph {\bibinfo {title}
  {{The Oceans: Their Physics, Chemistry and General Biology}}}}\ (\bibinfo
  {publisher} {Prentice Hall})\BibitemShut {NoStop}%
\bibitem [{\citenamefont {Swinbank}\ \emph {et~al.}(2016)\citenamefont
  {Swinbank}, \citenamefont {Friederichs},\ and\ \citenamefont
  {Wahl}}]{Friedrichs2016}%
  \BibitemOpen
  \bibfield  {author} {\bibinfo {author} {\bibnamefont {Swinbank},
  \bibfnamefont {R}}, \bibinfo {author} {\bibfnamefont {P.}~\bibnamefont
  {Friederichs}}, \ and\ \bibinfo {author} {\bibfnamefont {S.}~\bibnamefont
  {Wahl}}} (\bibinfo {year} {2016}),\ \enquote {\bibinfo {title} {Forecasting
  high-impact weather using ensemble prediction systems},}\ in\ \href {\doibase
  10.1017/CBO9781107775541.008} {\emph {\bibinfo {booktitle} {Dynamics and
  Predictability of Large-Scale, High-Impact Weather and Climate Events}}},\
  \bibinfo {series and number} {Special Publications of the International Union
  of Geodesy and Geophysics},\ \bibinfo {editor} {edited by\ \bibinfo {editor}
  {\bibfnamefont {J.}~\bibnamefont {Li}}, \bibinfo {editor} {\bibfnamefont
  {R.}~\bibnamefont {Swinbank}}, \bibinfo {editor} {\bibfnamefont
  {R.}~\bibnamefont {Grotjahn}}, \ and\ \bibinfo {editor} {\bibfnamefont
  {H.}~\bibnamefont {Volkert}}}\ (\bibinfo  {publisher} {Cambridge University
  Press})\ pp.\ \bibinfo {pages} {95--112}\BibitemShut {NoStop}%
\bibitem [{\citenamefont {Tantet}\ \emph {et~al.}(2018)\citenamefont {Tantet},
  \citenamefont {Lucarini}, \citenamefont {Lunkeit},\ and\ \citenamefont
  {Dijkstra}}]{Tantet2018}%
  \BibitemOpen
  \bibfield  {author} {\bibinfo {author} {\bibnamefont {Tantet}, \bibfnamefont
  {A}}, \bibinfo {author} {\bibfnamefont {V.}~\bibnamefont {Lucarini}},
  \bibinfo {author} {\bibfnamefont {F.}~\bibnamefont {Lunkeit}}, \ and\
  \bibinfo {author} {\bibfnamefont {H.~A.}\ \bibnamefont {Dijkstra}}} (\bibinfo
  {year} {2018}),\ \bibfield  {title} {\enquote {\bibinfo {title} {Crisis of
  the chaotic attractor of a climate model: a transfer operator approach},}\
  }\href {\doibase 10.1088/1361-6544/aaaf42} {\bibfield  {journal} {\bibinfo
  {journal} {Nonlinearity}\ }\textbf {\bibinfo {volume} {31}}~(\bibinfo
  {number} {5}),\ \bibinfo {pages} {2221--2251}}\BibitemShut {NoStop}%
\bibitem [{\citenamefont {Taricco}\ \emph {et~al.}(2009)\citenamefont
  {Taricco}, \citenamefont {Ghil}, \citenamefont {Alessio},\ and\ \citenamefont
  {Vivaldo}}]{Taricco.ea.09}%
  \BibitemOpen
  \bibfield  {author} {\bibinfo {author} {\bibnamefont {Taricco}, \bibfnamefont
  {C}}, \bibinfo {author} {\bibfnamefont {M.}~\bibnamefont {Ghil}}, \bibinfo
  {author} {\bibfnamefont {S.}~\bibnamefont {Alessio}}, \ and\ \bibinfo
  {author} {\bibfnamefont {G.}~\bibnamefont {Vivaldo}}} (\bibinfo {year}
  {2009}),\ \bibfield  {title} {\enquote {\bibinfo {title} {Two millennia of
  climate variability in the {Central Mediterranean}},}\ }\href@noop {}
  {\bibfield  {journal} {\bibinfo  {journal} {Clim. Past}\ }\textbf {\bibinfo
  {volume} {5}},\ \bibinfo {pages} {171--181}}\BibitemShut {NoStop}%
\bibitem [{\citenamefont {Taylor}(1921)}]{Taylor.1921}%
  \BibitemOpen
  \bibfield  {author} {\bibinfo {author} {\bibnamefont {Taylor}, \bibfnamefont
  {G~I}}} (\bibinfo {year} {1921}),\ \bibfield  {title} {\enquote {\bibinfo
  {title} {Diffusion by continuous movements},}\ }\href@noop {} {\bibfield
  {journal} {\bibinfo  {journal} {{Proceedings of the London Mathematical
  Society, Series 2}}\ }\textbf {\bibinfo {volume} {20}},\ \bibinfo {pages}
  {196--211}}\BibitemShut {NoStop}%
\bibitem [{\citenamefont {Tebaldi}\ and\ \citenamefont
  {Knutti}(2007)}]{Tebaldi2007}%
  \BibitemOpen
  \bibfield  {author} {\bibinfo {author} {\bibnamefont {Tebaldi}, \bibfnamefont
  {C}}, \ and\ \bibinfo {author} {\bibfnamefont {R.}~\bibnamefont {Knutti}}}
  (\bibinfo {year} {2007}),\ \bibfield  {title} {\enquote {\bibinfo {title}
  {The use of the multi-model ensemble in probabilistic climate projections},}\
  }\href {\doibase 10.1098/rsta.2007.2076} {\bibfield  {journal} {\bibinfo
  {journal} {Philosophical Transactions of the Royal Society A: Mathematical,
  Physical and Engineering Sciences}\ }\textbf {\bibinfo {volume}
  {365}}~(\bibinfo {number} {1857}),\ \bibinfo {pages}
  {2053--2075}}\BibitemShut {NoStop}%
\bibitem [{\citenamefont {{Teisserenc de Bort}}(1881)}]{TeissBort1881}%
  \BibitemOpen
  \bibfield  {author} {\bibinfo {author} {\bibnamefont {{Teisserenc de Bort}},
  \bibfnamefont {L}}} (\bibinfo {year} {1881}),\ \bibfield  {title} {\enquote
  {\bibinfo {title} {Etude sur l'hiver de 1879--80 et recherches sur la
  position des centres d'action de l'atmosph\`ere dans les hivers anormaux},}\
  }\href@noop {} {\bibfield  {journal} {\bibinfo  {journal} {{Annales du Bureau
  Central M\'et\'eor. France}}\ }\textbf {\bibinfo {volume} {4}},\ \bibinfo
  {pages} {17--62}}\BibitemShut {NoStop}%
\bibitem [{\citenamefont {Temam}(1984)}]{Temam1984}%
  \BibitemOpen
  \bibfield  {author} {\bibinfo {author} {\bibnamefont {Temam}, \bibfnamefont
  {R}}} (\bibinfo {year} {1984}),\ \href@noop {} {\emph {\bibinfo {title}
  {Navier-Stokes Equations: Theory and Numerical Analysis}}}\ (\bibinfo
  {publisher} {North-Holland})\BibitemShut {NoStop}%
\bibitem [{\citenamefont {Temam}(1997)}]{Temam1997}%
  \BibitemOpen
  \bibfield  {author} {\bibinfo {author} {\bibnamefont {Temam}, \bibfnamefont
  {R}}} (\bibinfo {year} {1997}),\ \href@noop {} {\emph {\bibinfo {title}
  {Infinite-Dimensional Dynamical Systems in Mechanics and Physics}}},\
  \bibinfo {edition} {2nd}\ ed.\ (\bibinfo  {publisher} {Springer Nature},\
  \bibinfo {address} {New York})\BibitemShut {NoStop}%
\bibitem [{\citenamefont {Thompson}(1957)}]{Thompson.1957}%
  \BibitemOpen
  \bibfield  {author} {\bibinfo {author} {\bibnamefont {Thompson},
  \bibfnamefont {P~D}}} (\bibinfo {year} {1957}),\ \bibfield  {title} {\enquote
  {\bibinfo {title} {Uncertainty of initial state as a factor in the
  predictability of large scale atmospheric flow patterns},}\ }\href@noop {}
  {\bibfield  {journal} {\bibinfo  {journal} {Tellus}\ }\textbf {\bibinfo
  {volume} {9}},\ \bibinfo {pages} {275--295}}\BibitemShut {NoStop}%
\bibitem [{\citenamefont {Thual}\ and\ \citenamefont
  {McWilliams}(1992)}]{Thual1992}%
  \BibitemOpen
  \bibfield  {author} {\bibinfo {author} {\bibnamefont {Thual}, \bibfnamefont
  {O}}, \ and\ \bibinfo {author} {\bibfnamefont {J.~C.}\ \bibnamefont
  {McWilliams}}} (\bibinfo {year} {1992}),\ \bibfield  {title} {\enquote
  {\bibinfo {title} {The catastrophe structure of thermohaline convection in a
  two-dimensional fluid model and a comparison with low-order box models},}\
  }\href@noop {} {\bibfield  {journal} {\bibinfo  {journal} {Geophys.\
  Astrophys.\ Fluid Dyn.}\ }\textbf {\bibinfo {volume} {64}},\ \bibinfo {pages}
  {67--95}}\BibitemShut {NoStop}%
\bibitem [{\citenamefont {Titz}\ \emph {et~al.}(2002)\citenamefont {Titz},
  \citenamefont {Kuhlbrodt}, \citenamefont {Rahmstorf},\ and\ \citenamefont
  {Feudel}}]{Titz2001}%
  \BibitemOpen
  \bibfield  {author} {\bibinfo {author} {\bibnamefont {Titz}, \bibfnamefont
  {S}}, \bibinfo {author} {\bibfnamefont {T.}~\bibnamefont {Kuhlbrodt}},
  \bibinfo {author} {\bibfnamefont {S.}~\bibnamefont {Rahmstorf}}, \ and\
  \bibinfo {author} {\bibfnamefont {U.}~\bibnamefont {Feudel}}} (\bibinfo
  {year} {2002}),\ \bibfield  {title} {\enquote {\bibinfo {title} {On
  freshwater-dependent bifurcations in box models of the interhemispheric
  thermohaline circulation},}\ }\href {\doibase
  10.1034/j.1600-0870.2002.00303.x} {\bibfield  {journal} {\bibinfo  {journal}
  {Tellus A}\ }\textbf {\bibinfo {volume} {54}}~(\bibinfo {number} {1}),\
  \bibinfo {pages} {89--98}}\BibitemShut {NoStop}%
\bibitem [{\citenamefont {Torralba}\ \emph {et~al.}(2017)\citenamefont
  {Torralba}, \citenamefont {Doblas-Reyes}, \citenamefont {MacLeod},
  \citenamefont {Christel},\ and\ \citenamefont {Davis}}]{Torralba2017}%
  \BibitemOpen
  \bibfield  {author} {\bibinfo {author} {\bibnamefont {Torralba},
  \bibfnamefont {V}}, \bibinfo {author} {\bibfnamefont {F.~J.}\ \bibnamefont
  {Doblas-Reyes}}, \bibinfo {author} {\bibfnamefont {D.}~\bibnamefont
  {MacLeod}}, \bibinfo {author} {\bibfnamefont {I.}~\bibnamefont {Christel}}, \
  and\ \bibinfo {author} {\bibfnamefont {M.}~\bibnamefont {Davis}}} (\bibinfo
  {year} {2017}),\ \bibfield  {title} {\enquote {\bibinfo {title} {Seasonal
  climate prediction: A new source of information for the management of wind
  energy resources},}\ }\href {\doibase 10.1175/JAMC-D-16-0204.1} {\bibfield
  {journal} {\bibinfo  {journal} {Journal of Applied Meteorology and
  Climatology}\ }\textbf {\bibinfo {volume} {56}}~(\bibinfo {number} {5}),\
  \bibinfo {pages} {1231--1247}}\BibitemShut {NoStop}%
\bibitem [{\citenamefont {Touchette}(2009)}]{T09}%
  \BibitemOpen
  \bibfield  {author} {\bibinfo {author} {\bibnamefont {Touchette},
  \bibfnamefont {H}}} (\bibinfo {year} {2009}),\ \bibfield  {title} {\enquote
  {\bibinfo {title} {The large deviation approach to statistical mechanics},}\
  }\href@noop {} {\bibfield  {journal} {\bibinfo  {journal} {Physics Reports}\
  }\textbf {\bibinfo {volume} {478}},\ \bibinfo {pages} {1--69}}\BibitemShut
  {NoStop}%
\bibitem [{\citenamefont {Trefethen}\ \emph {et~al.}(1993)\citenamefont
  {Trefethen}, \citenamefont {Trefethen}, \citenamefont {Reddy},\ and\
  \citenamefont {Driscoll}}]{Trefethen1993}%
  \BibitemOpen
  \bibfield  {author} {\bibinfo {author} {\bibnamefont {Trefethen},
  \bibfnamefont {L~N}}, \bibinfo {author} {\bibfnamefont {A.~E.}\ \bibnamefont
  {Trefethen}}, \bibinfo {author} {\bibfnamefont {S.~C.}\ \bibnamefont
  {Reddy}}, \ and\ \bibinfo {author} {\bibfnamefont {T.~A.}\ \bibnamefont
  {Driscoll}}} (\bibinfo {year} {1993}),\ \bibfield  {title} {\enquote
  {\bibinfo {title} {Hydrodynamic stability without eigenvalues},}\ }\href@noop
  {} {\bibfield  {journal} {\bibinfo  {journal} {Science}\ }\textbf {\bibinfo
  {volume} {261}},\ \bibinfo {pages} {578--584}}\BibitemShut {NoStop}%
\bibitem [{\citenamefont {Trenberth}\ and\ \citenamefont
  {Caron}(2001)}]{TreCar}%
  \BibitemOpen
  \bibfield  {author} {\bibinfo {author} {\bibnamefont {Trenberth},
  \bibfnamefont {K~E}}, \ and\ \bibinfo {author} {\bibfnamefont {J~M}\
  \bibnamefont {Caron}}} (\bibinfo {year} {2001}),\ \bibfield  {title}
  {\enquote {\bibinfo {title} {Estimates of meridional atmosphere and ocean
  heat transports},}\ }\href@noop {} {\bibfield  {journal} {\bibinfo  {journal}
  {J. Clim}\ }\textbf {\bibinfo {volume} {14}},\ \bibinfo {pages}
  {3433--3443}}\BibitemShut {NoStop}%
\bibitem [{\citenamefont {Trenberth}\ \emph {et~al.}(2009)\citenamefont
  {Trenberth}, \citenamefont {Fasullo},\ and\ \citenamefont
  {Kiehl}}]{Trenberth2009}%
  \BibitemOpen
  \bibfield  {author} {\bibinfo {author} {\bibnamefont {Trenberth},
  \bibfnamefont {Kevin~E}}, \bibinfo {author} {\bibfnamefont {John~T.}\
  \bibnamefont {Fasullo}}, \ and\ \bibinfo {author} {\bibfnamefont {Jeffrey}\
  \bibnamefont {Kiehl}}} (\bibinfo {year} {2009}),\ \bibfield  {title}
  {\enquote {\bibinfo {title} {Earth's global energy budget},}\ }\href
  {\doibase 10.1175/2008BAMS2634.1} {\bibfield  {journal} {\bibinfo  {journal}
  {Bulletin of the American Meteorological Society}\ }\textbf {\bibinfo
  {volume} {90}}~(\bibinfo {number} {3}),\ \bibinfo {pages}
  {311--324}}\BibitemShut {NoStop}%
\bibitem [{\citenamefont {Trevisan}\ and\ \citenamefont
  {Buzzi}({1980})}]{Trevi.Buzzi.1980}%
  \BibitemOpen
  \bibfield  {author} {\bibinfo {author} {\bibnamefont {Trevisan},
  \bibfnamefont {A}}, \ and\ \bibinfo {author} {\bibfnamefont {A.}~\bibnamefont
  {Buzzi}}} (\bibinfo {year} {{1980}}),\ \bibfield  {title} {\enquote {\bibinfo
  {title} {Stationary response of barotropic weakly non-linear {Rossby} waves
  to quasi-resonant orographic forcing},}\ }\href {\doibase
  {10.1175/1520-0469(1980)037<0947:SROBWN>2.0.CO;2}} {\bibfield  {journal}
  {\bibinfo  {journal} {{J. Atmos. Sci.}}\ }\textbf {\bibinfo {volume}
  {{37}}}~(\bibinfo {number} {{5}}),\ \bibinfo {pages} {947--957}}\BibitemShut
  {NoStop}%
\bibitem [{\citenamefont {Tribbia}\ and\ \citenamefont
  {Anthes}(1987)}]{Tribbia.1987}%
  \BibitemOpen
  \bibfield  {author} {\bibinfo {author} {\bibnamefont {Tribbia}, \bibfnamefont
  {J~J}}, \ and\ \bibinfo {author} {\bibfnamefont {R.~A.}\ \bibnamefont
  {Anthes}}} (\bibinfo {year} {1987}),\ \bibfield  {title} {\enquote {\bibinfo
  {title} {Scientific basis of modern weather prediction},}\ }\href@noop {}
  {\bibfield  {journal} {\bibinfo  {journal} {Science}\ }\textbf {\bibinfo
  {volume} {237}},\ \bibinfo {pages} {493--499}}\BibitemShut {NoStop}%
\bibitem [{\citenamefont {{Tsonis}}\ and\ \citenamefont
  {{Roebber}}(2004)}]{Tsonis2004}%
  \BibitemOpen
  \bibfield  {author} {\bibinfo {author} {\bibnamefont {{Tsonis}},
  \bibfnamefont {A~A}}, \ and\ \bibinfo {author} {\bibfnamefont {P.~J.}\
  \bibnamefont {{Roebber}}}} (\bibinfo {year} {2004}),\ \bibfield  {title}
  {\enquote {\bibinfo {title} {{The architecture of the climate network}},}\
  }\href {\doibase 10.1016/j.physa.2003.10.045} {\bibfield  {journal} {\bibinfo
   {journal} {Physica A Statistical Mechanics and its Applications}\ }\textbf
  {\bibinfo {volume} {333}},\ \bibinfo {pages} {497--504}}\BibitemShut
  {NoStop}%
\bibitem [{\citenamefont {{Tsonis}}\ \emph {et~al.}(2006)\citenamefont
  {{Tsonis}}, \citenamefont {{Swanson}},\ and\ \citenamefont
  {{Roebber}}}]{Tsonis2006}%
  \BibitemOpen
  \bibfield  {author} {\bibinfo {author} {\bibnamefont {{Tsonis}},
  \bibfnamefont {A~A}}, \bibinfo {author} {\bibfnamefont {K.~L.}\ \bibnamefont
  {{Swanson}}}, \ and\ \bibinfo {author} {\bibfnamefont {P.~J.}\ \bibnamefont
  {{Roebber}}}} (\bibinfo {year} {2006}),\ \bibfield  {title} {\enquote
  {\bibinfo {title} {{What Do Networks Have to Do with Climate?}}}\ }\href
  {\doibase 10.1175/BAMS-87-5-585} {\bibfield  {journal} {\bibinfo  {journal}
  {Bulletin of the American Meteorological Society}\ }\textbf {\bibinfo
  {volume} {87}},\ \bibinfo {pages} {585--595}}\BibitemShut {NoStop}%
\bibitem [{\citenamefont {Turco}\ \emph {et~al.}(1983)\citenamefont {Turco},
  \citenamefont {Toon}, \citenamefont {Ackerman}, \citenamefont {Pollack},\
  and\ \citenamefont {Sagan}}]{TTAPS.1983}%
  \BibitemOpen
  \bibfield  {author} {\bibinfo {author} {\bibnamefont {Turco}, \bibfnamefont
  {RP}}, \bibinfo {author} {\bibfnamefont {O.B.}\ \bibnamefont {Toon}},
  \bibinfo {author} {\bibfnamefont {T.P.}\ \bibnamefont {Ackerman}}, \bibinfo
  {author} {\bibfnamefont {J.B.}\ \bibnamefont {Pollack}}, \ and\ \bibinfo
  {author} {\bibfnamefont {C.}~\bibnamefont {Sagan}}} (\bibinfo {year}
  {1983}),\ \bibfield  {title} {\enquote {\bibinfo {title} {Nuclear winter:
  {Global} consequences of multiple nuclear explosions},}\ }\href@noop {}
  {\bibfield  {journal} {\bibinfo  {journal} {Science}\ }\textbf {\bibinfo
  {volume} {2222}}~(\bibinfo {number} {4630}),\ \bibinfo {pages}
  {1283--1292}}\BibitemShut {NoStop}%
\bibitem [{\citenamefont {Turner}\ and\ \citenamefont
  {Annamalai}(2012)}]{Turner12}%
  \BibitemOpen
  \bibfield  {author} {\bibinfo {author} {\bibnamefont {Turner}, \bibfnamefont
  {AG}}, \ and\ \bibinfo {author} {\bibfnamefont {H.}~\bibnamefont
  {Annamalai}}} (\bibinfo {year} {2012}),\ \bibfield  {title} {\enquote
  {\bibinfo {title} {Climate change and the south asian summer monsoon},}\
  }\href {\doibase 10.1038/nclimate1495} {\bibfield  {journal} {\bibinfo
  {journal} {Nature Clim. Change}\ }\textbf {\bibinfo {volume} {2}},\
  10.1038/nclimate1495}\BibitemShut {NoStop}%
\bibitem [{\citenamefont {Tziperman}\ \emph
  {et~al.}(1994{\natexlab{a}})\citenamefont {Tziperman}, \citenamefont {Stone},
  \citenamefont {Cane},\ and\ \citenamefont {Jarosh}}]{Tziperman1994b}%
  \BibitemOpen
  \bibfield  {author} {\bibinfo {author} {\bibnamefont {Tziperman},
  \bibfnamefont {E}}, \bibinfo {author} {\bibfnamefont {L.}~\bibnamefont
  {Stone}}, \bibinfo {author} {\bibfnamefont {M.~A.}\ \bibnamefont {Cane}}, \
  and\ \bibinfo {author} {\bibfnamefont {H.}~\bibnamefont {Jarosh}}} (\bibinfo
  {year} {1994}{\natexlab{a}}),\ \bibfield  {title} {\enquote {\bibinfo {title}
  {{El Ni\~no chaos: overlapping of resonances between the seasonal cycle and
  the Pacific ocean-atmosphere oscillator}},}\ }\href@noop {} {\bibfield
  {journal} {\bibinfo  {journal} {Science}\ }\textbf {\bibinfo {volume}
  {264}},\ \bibinfo {pages} {72--74}}\BibitemShut {NoStop}%
\bibitem [{\citenamefont {Tziperman}\ \emph
  {et~al.}(1994{\natexlab{b}})\citenamefont {Tziperman}, \citenamefont {Stone},
  \citenamefont {Cane},\ and\ \citenamefont {Jarosh}}]{Tziperman.ea.1994}%
  \BibitemOpen
  \bibfield  {author} {\bibinfo {author} {\bibnamefont {Tziperman},
  \bibfnamefont {E}}, \bibinfo {author} {\bibfnamefont {L.}~\bibnamefont
  {Stone}}, \bibinfo {author} {\bibfnamefont {M.~A.}\ \bibnamefont {Cane}}, \
  and\ \bibinfo {author} {\bibfnamefont {H.}~\bibnamefont {Jarosh}}} (\bibinfo
  {year} {1994}{\natexlab{b}}),\ \bibfield  {title} {\enquote {\bibinfo {title}
  {El {N}i{\~n}o chaos: {O}verlapping of resonances between the seasonal cycle
  and the {P}acific ocean-atmosphere oscillator},}\ }\href@noop {} {\bibfield
  {journal} {\bibinfo  {journal} {Science}\ }\textbf {\bibinfo {volume}
  {264}}~(\bibinfo {number} {5155}),\ \bibinfo {pages} {72--74}}\BibitemShut
  {NoStop}%
\bibitem [{\citenamefont {Vallis}(2006)}]{Vallis_atmospheric_2006}%
  \BibitemOpen
  \bibfield  {author} {\bibinfo {author} {\bibnamefont {Vallis}, \bibfnamefont
  {G~K}}} (\bibinfo {year} {2006}),\ \href@noop {} {\emph {\bibinfo {title}
  {Atmospheric and Oceanic Fluid Dynamics: Fundamentals and Large-scale
  Circulation}}}\ (\bibinfo  {publisher} {Cambridge University Press},\
  \bibinfo {address} {Cambridge, UK})\BibitemShut {NoStop}%
\bibitem [{\citenamefont {Vannitsem}\ \emph {et~al.}(2015)\citenamefont
  {Vannitsem}, \citenamefont {Demaeyer}, \citenamefont {De~Cruz},\ and\
  \citenamefont {Ghil}}]{Vann.ea.2015}%
  \BibitemOpen
  \bibfield  {author} {\bibinfo {author} {\bibnamefont {Vannitsem},
  \bibfnamefont {S}}, \bibinfo {author} {\bibfnamefont {J.}~\bibnamefont
  {Demaeyer}}, \bibinfo {author} {\bibfnamefont {L.}~\bibnamefont {De~Cruz}}, \
  and\ \bibinfo {author} {\bibfnamefont {M.}~\bibnamefont {Ghil}}} (\bibinfo
  {year} {2015}),\ \bibfield  {title} {\enquote {\bibinfo {title}
  {Low-frequency variability and heat transport in a low-order nonlinear
  coupled ocean-atmosphere model},}\ }\href {\doibase
  10.1016/j.physd.2015.07.006} {\bibfield  {journal} {\bibinfo  {journal}
  {Physica D}\ }\textbf {\bibinfo {volume} {309}},\ \bibinfo {pages}
  {71--85}}\BibitemShut {NoStop}%
\bibitem [{\citenamefont {Vannitsem}\ and\ \citenamefont
  {Lucarini}(2016)}]{Vannitsem2016}%
  \BibitemOpen
  \bibfield  {author} {\bibinfo {author} {\bibnamefont {Vannitsem},
  \bibfnamefont {S}}, \ and\ \bibinfo {author} {\bibfnamefont {V.}~\bibnamefont
  {Lucarini}}} (\bibinfo {year} {2016}),\ \bibfield  {title} {\enquote
  {\bibinfo {title} {Statistical and dynamical properties of covariant lyapunov
  vectors in a coupled atmosphere-ocean model{\textemdash}multiscale effects,
  geometric degeneracy, and error dynamics},}\ }\href {\doibase
  10.1088/1751-8113/49/22/224001} {\bibfield  {journal} {\bibinfo  {journal}
  {Journal of Physics A: Mathematical and Theoretical}\ }\textbf {\bibinfo
  {volume} {49}}~(\bibinfo {number} {22}),\ \bibinfo {pages}
  {224001}}\BibitemShut {NoStop}%
\bibitem [{\citenamefont {Varadhan}(1966)}]{Varadhan.1966}%
  \BibitemOpen
  \bibfield  {author} {\bibinfo {author} {\bibnamefont {Varadhan},
  \bibfnamefont {S~R~S}}} (\bibinfo {year} {1966}),\ \bibfield  {title}
  {\enquote {\bibinfo {title} {Asymptotic probabilities and differential
  equations},}\ }\href@noop {} {\bibfield  {journal} {\bibinfo  {journal}
  {Communications on Pure and Applied Mathematics}\ }\textbf {\bibinfo {volume}
  {19}}~(\bibinfo {number} {3}),\ \bibinfo {pages} {261--286}}\BibitemShut
  {NoStop}%
\bibitem [{\citenamefont {Varadhan}(1984)}]{V84}%
  \BibitemOpen
  \bibfield  {author} {\bibinfo {author} {\bibnamefont {Varadhan},
  \bibfnamefont {SRS}}} (\bibinfo {year} {1984}),\ \href@noop {} {\emph
  {\bibinfo {title} {Large Deviations and Applications}}}\ (\bibinfo
  {publisher} {SIAM, Philadelphia})\BibitemShut {NoStop}%
\bibitem [{\citenamefont {Veronis}(1963)}]{Veronis1963}%
  \BibitemOpen
  \bibfield  {author} {\bibinfo {author} {\bibnamefont {Veronis}, \bibfnamefont
  {G}}} (\bibinfo {year} {1963}),\ \bibfield  {title} {\enquote {\bibinfo
  {title} {{An analysis of the wind-driven ocean circulation with a limited
  number of Fourier components}},}\ }\href@noop {} {\bibfield  {journal}
  {\bibinfo  {journal} {J.\ Atmos.\ Sci.}\ }\textbf {\bibinfo {volume} {20}},\
  \bibinfo {pages} {577--593}}\BibitemShut {NoStop}%
\bibitem [{\citenamefont {Villani}(2009)}]{Villani.2009}%
  \BibitemOpen
  \bibfield  {author} {\bibinfo {author} {\bibnamefont {Villani}, \bibfnamefont
  {C}}} (\bibinfo {year} {2009}),\ \href@noop {} {\emph {\bibinfo {title}
  {{Optimal Transport: Old and New}}}}\ (\bibinfo  {publisher} {{Springer
  Science \& Business Media}})\BibitemShut {NoStop}%
\bibitem [{\citenamefont {Vissio}\ and\ \citenamefont
  {Lucarini}({2018a})}]{Vissio2018a}%
  \BibitemOpen
  \bibfield  {author} {\bibinfo {author} {\bibnamefont {Vissio}, \bibfnamefont
  {G}}, \ and\ \bibinfo {author} {\bibfnamefont {V.}~\bibnamefont {Lucarini}}}
  (\bibinfo {year} {{2018a}}),\ \bibfield  {title} {\enquote {\bibinfo {title}
  {A proof of concept for scale-adaptive parametrizations: the case of the
  lorenz '96 model},}\ }\href {\doibase 10.1002/qj.3184} {\bibfield  {journal}
  {\bibinfo  {journal} {Quarterly Journal of the Royal Meteorological Society}\
  }\textbf {\bibinfo {volume} {144}}~(\bibinfo {number} {710}),\ \bibinfo
  {pages} {63--75}}\BibitemShut {NoStop}%
\bibitem [{\citenamefont {Vissio}\ and\ \citenamefont
  {Lucarini}({2018b})}]{Vissio2018}%
  \BibitemOpen
  \bibfield  {author} {\bibinfo {author} {\bibnamefont {Vissio}, \bibfnamefont
  {G}}, \ and\ \bibinfo {author} {\bibfnamefont {V.}~\bibnamefont {Lucarini}}}
  (\bibinfo {year} {{2018b}}),\ \bibfield  {title} {\enquote {\bibinfo {title}
  {Evaluating a stochastic parametrization for a fast--slow system using the
  wasserstein distance},}\ }\href {\doibase 10.5194/npg-25-413-2018} {\bibfield
   {journal} {\bibinfo  {journal} {Nonlinear Processes in Geophysics}\ }\textbf
  {\bibinfo {volume} {25}}~(\bibinfo {number} {2}),\ \bibinfo {pages}
  {413--427}}\BibitemShut {NoStop}%
\bibitem [{\citenamefont {Vollmer}\ \emph {et~al.}(2009)\citenamefont
  {Vollmer}, \citenamefont {Schneider},\ and\ \citenamefont
  {Eckhardt}}]{Vollmer2009}%
  \BibitemOpen
  \bibfield  {author} {\bibinfo {author} {\bibnamefont {Vollmer}, \bibfnamefont
  {J}}, \bibinfo {author} {\bibfnamefont {T.~M.}\ \bibnamefont {Schneider}}, \
  and\ \bibinfo {author} {\bibfnamefont {B.}~\bibnamefont {Eckhardt}}}
  (\bibinfo {year} {2009}),\ \bibfield  {title} {\enquote {\bibinfo {title}
  {Basin boundary, edge of chaos and edge state in a two-dimensional model},}\
  }\href@noop {} {\bibfield  {journal} {\bibinfo  {journal} {New Journal of
  Physics}\ }\textbf {\bibinfo {volume} {11}}~(\bibinfo {number} {1}),\
  \bibinfo {pages} {013040}}\BibitemShut {NoStop}%
\bibitem [{\citenamefont {Von~Neumann}(1960)}]{JvN.predict.1960}%
  \BibitemOpen
  \bibfield  {author} {\bibinfo {author} {\bibnamefont {Von~Neumann},
  \bibfnamefont {J}}} (\bibinfo {year} {1960}),\ \bibfield  {title} {\enquote
  {\bibinfo {title} {Some remarks on the problem of forecasting climatic
  fluctuations},}\ }in\ \href {\doibase 10.1016/b978-1-4831-9890-3.50009-8}
  {\emph {\bibinfo {booktitle} {Dynamics of Climate}}},\ \bibinfo {editor}
  {edited by\ \bibinfo {editor} {\bibfnamefont {R.~L.}\ \bibnamefont
  {Pfeffer}}}\ (\bibinfo  {publisher} {Pergamon Press})\ pp.\ \bibinfo {pages}
  {9--11}\BibitemShut {NoStop}%
\bibitem [{\citenamefont {Wallace}\ and\ \citenamefont
  {Gutzler}(1981)}]{Wallace.Gutzler.1981}%
  \BibitemOpen
  \bibfield  {author} {\bibinfo {author} {\bibnamefont {Wallace}, \bibfnamefont
  {J~M}}, \ and\ \bibinfo {author} {\bibfnamefont {D.~S.}\ \bibnamefont
  {Gutzler}}} (\bibinfo {year} {1981}),\ \bibfield  {title} {\enquote {\bibinfo
  {title} {{Teleconnections in the geopotential height field during the
  Northern Hemisphere winter}},}\ }\href@noop {} {\bibfield  {journal}
  {\bibinfo  {journal} {Mon. Wea. Rev.}\ }\textbf {\bibinfo {volume}
  {109}}~(\bibinfo {number} {784--812})}\BibitemShut {NoStop}%
\bibitem [{\citenamefont {Walsh}(2014)}]{Walsh.2014}%
  \BibitemOpen
  \bibfield  {author} {\bibinfo {author} {\bibnamefont {Walsh}, \bibfnamefont
  {J~E}}} (\bibinfo {year} {2014}),\ \bibfield  {title} {\enquote {\bibinfo
  {title} {Intensified warming of the {Arctic: Causes and impacts on middle
  latitudes}},}\ }\href@noop {} {\bibfield  {journal} {\bibinfo  {journal}
  {Global and Planetary Change}\ }\textbf {\bibinfo {volume} {117}},\ \bibinfo
  {pages} {52--63}}\BibitemShut {NoStop}%
\bibitem [{\citenamefont {Wang}(2013)}]{W13}%
  \BibitemOpen
  \bibfield  {author} {\bibinfo {author} {\bibnamefont {Wang}, \bibfnamefont
  {Q}}} (\bibinfo {year} {2013}),\ \bibfield  {title} {\enquote {\bibinfo
  {title} {Forward and adjoint sensitivity computation of chaotic dynamical
  systems},}\ }\href {\doibase 10.1016/j.jcp.2012.09.007} {\bibfield  {journal}
  {\bibinfo  {journal} {Journal of Computational Physics}\ }\textbf {\bibinfo
  {volume} {235}}~(\bibinfo {number} {0}),\ \bibinfo {pages}
  {1--13}}\BibitemShut {NoStop}%
\bibitem [{\citenamefont {{Wang}}\ \emph {et~al.}(2013)\citenamefont {{Wang}},
  \citenamefont {{Gozolchiani}}, \citenamefont {{Ashkenazy}}, \citenamefont
  {{Berezin}}, \citenamefont {{Guez}},\ and\ \citenamefont
  {{Havlin}}}]{Wang2014}%
  \BibitemOpen
  \bibfield  {author} {\bibinfo {author} {\bibnamefont {{Wang}}, \bibfnamefont
  {Y}}, \bibinfo {author} {\bibfnamefont {A.}~\bibnamefont {{Gozolchiani}}},
  \bibinfo {author} {\bibfnamefont {Y.}~\bibnamefont {{Ashkenazy}}}, \bibinfo
  {author} {\bibfnamefont {Y.}~\bibnamefont {{Berezin}}}, \bibinfo {author}
  {\bibfnamefont {O.}~\bibnamefont {{Guez}}}, \ and\ \bibinfo {author}
  {\bibfnamefont {S.}~\bibnamefont {{Havlin}}}} (\bibinfo {year} {2013}),\
  \bibfield  {title} {\enquote {\bibinfo {title} {{Dominant imprint of Rossby}
  waves in the climate network},}\ }\href {\doibase
  10.1103/PhysRevLett.111.138501} {\bibfield  {journal} {\bibinfo  {journal}
  {Physical Review Letters}\ }\textbf {\bibinfo {volume} {111}}~(\bibinfo
  {number} {13}),\ \bibinfo {eid} {138501}},\ \Eprint
  {http://arxiv.org/abs/1304.0946} {arXiv:1304.0946 [physics.ao-ph]}
  \BibitemShut {NoStop}%
\bibitem [{\citenamefont {Washington}\ and\ \citenamefont
  {Parkinson}(2005)}]{Washington2005}%
  \BibitemOpen
  \bibfield  {author} {\bibinfo {author} {\bibnamefont {Washington},
  \bibfnamefont {WM}}, \ and\ \bibinfo {author} {\bibfnamefont {C.L.}\
  \bibnamefont {Parkinson}}} (\bibinfo {year} {2005}),\ \href@noop {} {\emph
  {\bibinfo {title} {An Introduction to Three-dimensional Climate Modeling}}}\
  (\bibinfo  {publisher} {University Science Books},\ \bibinfo {address}
  {Sausalito})\BibitemShut {NoStop}%
\bibitem [{\citenamefont {Watson}\ and\ \citenamefont
  {Lovelock}(1983)}]{watson_biological_1983}%
  \BibitemOpen
  \bibfield  {author} {\bibinfo {author} {\bibnamefont {Watson}, \bibfnamefont
  {A~J}}, \ and\ \bibinfo {author} {\bibfnamefont {J.~E.}\ \bibnamefont
  {Lovelock}}} (\bibinfo {year} {1983}),\ \bibfield  {title} {\enquote
  {\bibinfo {title} {Biological homeostasis of the global environment: the
  parable of {Daisyworld}},}\ }\href {\doibase
  10.1111/j.1600-0889.1983.tb00031.x} {\bibfield  {journal} {\bibinfo
  {journal} {Tellus B}\ }\textbf {\bibinfo {volume} {35B}}~(\bibinfo {number}
  {4}),\ \bibinfo {pages} {284--289}}\BibitemShut {NoStop}%
\bibitem [{\citenamefont {Weeks}\ \emph {et~al.}(1997)\citenamefont {Weeks},
  \citenamefont {Tian}, \citenamefont {Urbach}, \citenamefont {Ide},
  \citenamefont {Swinney},\ and\ \citenamefont {Ghil}}]{Weeks.ea.1997}%
  \BibitemOpen
  \bibfield  {author} {\bibinfo {author} {\bibnamefont {Weeks}, \bibfnamefont
  {E~R}}, \bibinfo {author} {\bibfnamefont {Y.}~\bibnamefont {Tian}}, \bibinfo
  {author} {\bibfnamefont {J.~S.}\ \bibnamefont {Urbach}}, \bibinfo {author}
  {\bibfnamefont {K.}~\bibnamefont {Ide}}, \bibinfo {author} {\bibfnamefont
  {H.~L.}\ \bibnamefont {Swinney}}, \ and\ \bibinfo {author} {\bibfnamefont
  {M.}~\bibnamefont {Ghil}}} (\bibinfo {year} {1997}),\ \bibfield  {title}
  {\enquote {\bibinfo {title} {Transitions between blocked and zonal flows in a
  rotating annulus with topography},}\ }\href@noop {} {\bibinfo  {journal}
  {Science}\ ,\ \bibinfo {pages} {1598--1601}}\BibitemShut {NoStop}%
\bibitem [{\citenamefont {Wester}\ \emph {et~al.}(2019)\citenamefont {Wester},
  \citenamefont {Mishra}, \citenamefont {Mukherji},\ and\ \citenamefont
  {Shrestha}}]{HIMAP2019}%
  \BibitemOpen
\bibfield  {journal} {  }\bibinfo {editor} {\bibnamefont {Wester},
  \bibfnamefont {P}}, \bibinfo {editor} {\bibfnamefont {A.}~\bibnamefont
  {Mishra}}, \bibinfo {editor} {\bibfnamefont {A.}~\bibnamefont {Mukherji}}, \
  and\ \bibinfo {editor} {\bibfnamefont {A.~B.}\ \bibnamefont {Shrestha}},\
  Eds. (\bibinfo {year} {2019}),\ \href@noop {} {\emph {\bibinfo {title} {The
  Hindu Kush Himalaya Assessment - Mountains, Climate Change, Sustainability
  and People}}}\ (\bibinfo  {publisher} {Springer Nature Switzerland AG},\
  \bibinfo {address} {Cham})\BibitemShut {NoStop}%
\bibitem [{\citenamefont {Wetherald}\ and\ \citenamefont
  {Manabe}(1975)}]{Wether.Manabe.1975}%
  \BibitemOpen
  \bibfield  {author} {\bibinfo {author} {\bibnamefont {Wetherald},
  \bibfnamefont {R~T}}, \ and\ \bibinfo {author} {\bibfnamefont
  {S.}~\bibnamefont {Manabe}}} (\bibinfo {year} {1975}),\ \bibfield  {title}
  {\enquote {\bibinfo {title} {{The effect of changing the solar constant on
  the climate of a general circulation model}},}\ }\href@noop {} {\bibfield
  {journal} {\bibinfo  {journal} {J.\ Atmos.\ Sci.}\ }\textbf {\bibinfo
  {volume} {32}},\ \bibinfo {pages} {2044--2059}}\BibitemShut {NoStop}%
\bibitem [{\citenamefont {Wiener}(1949)}]{Wiener.49}%
  \BibitemOpen
  \bibfield  {author} {\bibinfo {author} {\bibnamefont {Wiener}, \bibfnamefont
  {N}}} (\bibinfo {year} {1949}),\ \href@noop {} {\emph {\bibinfo {title}
  {Extrapolation, Interpolation and Smoothing of Stationary Time Series, with
  Engineering Applications}}}\ (\bibinfo  {publisher} {MIT Press})\BibitemShut
  {NoStop}%
\bibitem [{\citenamefont {Wolf}\ \emph {et~al.}(2018)\citenamefont {Wolf},
  \citenamefont {Haqq-Misra},\ and\ \citenamefont {Toon}}]{Wolf2018}%
  \BibitemOpen
  \bibfield  {author} {\bibinfo {author} {\bibnamefont {Wolf}, \bibfnamefont
  {E~T}}, \bibinfo {author} {\bibfnamefont {J.}~\bibnamefont {Haqq-Misra}}, \
  and\ \bibinfo {author} {\bibfnamefont {O.~B.}\ \bibnamefont {Toon}}}
  (\bibinfo {year} {2018}),\ \bibfield  {title} {\enquote {\bibinfo {title}
  {Evaluating climate sensitivity to {CO2 across Earth's history}},}\ }\href
  {\doibase 10.1029/2018JD029262} {\bibfield  {journal} {\bibinfo  {journal}
  {Journal of Geophysical Research: Atmospheres}\ }\textbf {\bibinfo {volume}
  {123}}~(\bibinfo {number} {21}),\ \bibinfo {pages}
  {11,861--11,874}}\BibitemShut {NoStop}%
\bibitem [{\citenamefont {Woollings}\ \emph {et~al.}(2018)\citenamefont
  {Woollings}, \citenamefont {Barriopedro}, \citenamefont {Methven},
  \citenamefont {Son}, \citenamefont {Martius}, \citenamefont {Harvey},
  \citenamefont {Sillmann}, \citenamefont {Lupo},\ and\ \citenamefont
  {Seneviratne}}]{Woollings2018}%
  \BibitemOpen
  \bibfield  {author} {\bibinfo {author} {\bibnamefont {Woollings},
  \bibfnamefont {T}}, \bibinfo {author} {\bibfnamefont {D.}~\bibnamefont
  {Barriopedro}}, \bibinfo {author} {\bibfnamefont {J.}~\bibnamefont
  {Methven}}, \bibinfo {author} {\bibfnamefont {S.-W.}\ \bibnamefont {Son}},
  \bibinfo {author} {\bibfnamefont {O.}~\bibnamefont {Martius}}, \bibinfo
  {author} {\bibfnamefont {B.}~\bibnamefont {Harvey}}, \bibinfo {author}
  {\bibfnamefont {J.}~\bibnamefont {Sillmann}}, \bibinfo {author}
  {\bibfnamefont {A.~R.}\ \bibnamefont {Lupo}}, \ and\ \bibinfo {author}
  {\bibfnamefont {S.}~\bibnamefont {Seneviratne}}} (\bibinfo {year} {2018}),\
  \bibfield  {title} {\enquote {\bibinfo {title} {Blocking and its response to
  climate change},}\ }\href {\doibase 10.1007/s40641-018-0108-z} {\bibfield
  {journal} {\bibinfo  {journal} {Current Climate Change Reports}\ }\textbf
  {\bibinfo {volume} {4}}~(\bibinfo {number} {3}),\ \bibinfo {pages}
  {287--300}}\BibitemShut {NoStop}%
\bibitem [{\citenamefont {Wouters}\ and\ \citenamefont
  {Lucarini}(2012)}]{wouters_disentangling_2012}%
  \BibitemOpen
  \bibfield  {author} {\bibinfo {author} {\bibnamefont {Wouters}, \bibfnamefont
  {J}}, \ and\ \bibinfo {author} {\bibfnamefont {V.}~\bibnamefont {Lucarini}}}
  (\bibinfo {year} {2012}),\ \bibfield  {title} {\enquote {\bibinfo {title}
  {Disentangling multi-level systems: averaging, correlations and memory},}\
  }\href@noop {} {\bibfield  {journal} {\bibinfo  {journal} {Journal of
  Statistical Mechanics: Theory and Experiment}\ }\textbf {\bibinfo {volume}
  {2012}}~(\bibinfo {number} {03}),\ \bibinfo {pages} {P03003}}\BibitemShut
  {NoStop}%
\bibitem [{\citenamefont {Wouters}\ and\ \citenamefont
  {Lucarini}(2013)}]{wouters_multi-level_2013}%
  \BibitemOpen
  \bibfield  {author} {\bibinfo {author} {\bibnamefont {Wouters}, \bibfnamefont
  {J}}, \ and\ \bibinfo {author} {\bibfnamefont {V.}~\bibnamefont {Lucarini}}}
  (\bibinfo {year} {2013}),\ \bibfield  {title} {\enquote {\bibinfo {title}
  {Multi-level dynamical systems: {Connecting the Ruelle response theory and
  the Mori-Zwanzig approach}},}\ }\href {\doibase 10.1007/s10955-013-0726-8}
  {\bibfield  {journal} {\bibinfo  {journal} {Journal of Statistical Physics}\
  }\textbf {\bibinfo {volume} {151}}~(\bibinfo {number} {5}),\
  10.1007/s10955-013-0726-8}\BibitemShut {NoStop}%
\bibitem [{\citenamefont {Wunsch}(1999)}]{Wunsch1999}%
  \BibitemOpen
  \bibfield  {author} {\bibinfo {author} {\bibnamefont {Wunsch}, \bibfnamefont
  {C}}} (\bibinfo {year} {1999}),\ \bibfield  {title} {\enquote {\bibinfo
  {title} {The interpretation of short climate records, with comments on the
  {North Atlantic and Southern Oscillations}},}\ }\href@noop {} {\bibfield
  {journal} {\bibinfo  {journal} {Bull. Am. Meteorol. Soc.}\ }\textbf {\bibinfo
  {volume} {80}},\ \bibinfo {pages} {245--255}}\BibitemShut {NoStop}%
\bibitem [{\citenamefont {Wunsch}(2002)}]{Wunsch2002}%
  \BibitemOpen
  \bibfield  {author} {\bibinfo {author} {\bibnamefont {Wunsch}, \bibfnamefont
  {C}}} (\bibinfo {year} {2002}),\ \bibfield  {title} {\enquote {\bibinfo
  {title} {What is the thermohaline circulation?}}\ }\href@noop {} {\bibfield
  {journal} {\bibinfo  {journal} {Science}\ }\textbf {\bibinfo {volume}
  {298}},\ \bibinfo {pages} {1179--1180}}\BibitemShut {NoStop}%
\bibitem [{\citenamefont {Wunsch}(2013)}]{Wunsch2013}%
  \BibitemOpen
  \bibfield  {author} {\bibinfo {author} {\bibnamefont {Wunsch}, \bibfnamefont
  {C}}} (\bibinfo {year} {2013}),\ \enquote {\bibinfo {title} {The past and
  future ocean circulation from a contemporary perspective},}\ in\ \href
  {\doibase 10.1029/173GM06} {\emph {\bibinfo {booktitle} {Ocean Circulation:
  Mechanisms and Impacts--Past and Future Changes of Meridional Overturning}}}\
  (\bibinfo  {publisher} {American Geophysical Union (AGU)})\ pp.\ \bibinfo
  {pages} {53--74}\BibitemShut {NoStop}%
\bibitem [{\citenamefont {Yanai}(1975)}]{Yanai.1975}%
  \BibitemOpen
  \bibfield  {author} {\bibinfo {author} {\bibnamefont {Yanai}, \bibfnamefont
  {M}}} (\bibinfo {year} {1975}),\ \bibfield  {title} {\enquote {\bibinfo
  {title} {Tropical meteorology},}\ }\href@noop {} {\bibfield  {journal}
  {\bibinfo  {journal} {Reviews of Geophysics}\ }\textbf {\bibinfo {volume}
  {13}}~(\bibinfo {number} {3}),\ \bibinfo {pages} {685--710}}\BibitemShut
  {NoStop}%
\bibitem [{\citenamefont {Young}(2002)}]{young_what_2002}%
  \BibitemOpen
  \bibfield  {author} {\bibinfo {author} {\bibnamefont {Young}, \bibfnamefont
  {{L-S}}}} (\bibinfo {year} {2002}),\ \bibfield  {title} {\enquote {\bibinfo
  {title} {What are {SRB} measures, and which dynamical systems have them?}}\
  }\href@noop {} {\bibfield  {journal} {\bibinfo  {journal} {Journal of
  Statistical Physics}\ }\textbf {\bibinfo {volume} {108}},\ \bibinfo {pages}
  {733--754}}\BibitemShut {NoStop}%
\bibitem [{\citenamefont {Zaliapin}\ and\ \citenamefont
  {Ghil}(2010)}]{ghil2010}%
  \BibitemOpen
  \bibfield  {author} {\bibinfo {author} {\bibnamefont {Zaliapin},
  \bibfnamefont {I}}, \ and\ \bibinfo {author} {\bibfnamefont {M.}~\bibnamefont
  {Ghil}}} (\bibinfo {year} {2010}),\ \bibfield  {title} {\enquote {\bibinfo
  {title} {Another look at climate sensitivity},}\ }\href {\doibase
  10.5194/npg-17-113-2010} {\bibfield  {journal} {\bibinfo  {journal}
  {Nonlinear Processes in Geophysics}\ }\textbf {\bibinfo {volume}
  {17}}~(\bibinfo {number} {2}),\ \bibinfo {pages} {113--122}}\BibitemShut
  {NoStop}%
\bibitem [{\citenamefont {Zeng}\ and\ \citenamefont
  {Neelin}(2000)}]{zeng_role_2000}%
  \BibitemOpen
  \bibfield  {author} {\bibinfo {author} {\bibnamefont {Zeng}, \bibfnamefont
  {N}}, \ and\ \bibinfo {author} {\bibfnamefont {J.~D.}\ \bibnamefont
  {Neelin}}} (\bibinfo {year} {2000}),\ \bibfield  {title} {\enquote {\bibinfo
  {title} {The role of vegetation--climate interaction and interannual
  variability in shaping the {African} savanna},}\ }\href@noop {} {\bibfield
  {journal} {\bibinfo  {journal} {Journal of Climate}\ }\textbf {\bibinfo
  {volume} {13}}~(\bibinfo {number} {15}),\ \bibinfo {pages}
  {2665--2670}}\BibitemShut {NoStop}%
\bibitem [{\citenamefont {Zhang}(2005)}]{Zhang.2005}%
  \BibitemOpen
  \bibfield  {author} {\bibinfo {author} {\bibnamefont {Zhang}, \bibfnamefont
  {C}}} (\bibinfo {year} {2005}),\ \bibfield  {title} {\enquote {\bibinfo
  {title} {{Madden-Julian Oscillation}},}\ }\href@noop {} {\bibfield  {journal}
  {\bibinfo  {journal} {Rev. Geophys.}\ }\textbf {\bibinfo {volume} {43}},\
  \bibinfo {pages} {2004RG000158}}\BibitemShut {NoStop}%
\bibitem [{\citenamefont {Zilitinkevich}({1975})}]{Zil75}%
  \BibitemOpen
  \bibfield  {author} {\bibinfo {author} {\bibnamefont {Zilitinkevich},
  \bibfnamefont {S~S}}} (\bibinfo {year} {{1975}}),\ \bibfield  {title}
  {\enquote {\bibinfo {title} {Resistance laws and prediction equations for
  depth of planetary boundary-layer},}\ }\href {\doibase
  {10.1175/1520-0469(1975)032<0741:RLAPEF>2.0.CO;2}} {\bibfield  {journal}
  {\bibinfo  {journal} {J.\ Atmos.\ Sci.}\ }\textbf {\bibinfo {volume}
  {{32}}}~(\bibinfo {number} {{4}}),\ \bibinfo {pages} {741--752}}\BibitemShut
  {NoStop}%
\bibitem [{\citenamefont {Zwanzig}(1961)}]{zwanzig_memory_1961}%
  \BibitemOpen
  \bibfield  {author} {\bibinfo {author} {\bibnamefont {Zwanzig}, \bibfnamefont
  {R}}} (\bibinfo {year} {1961}),\ \bibfield  {title} {\enquote {\bibinfo
  {title} {Memory effects in irreversible thermodynamics},}\ }\href@noop {}
  {\bibfield  {journal} {\bibinfo  {journal} {Physical Review}\ }\textbf
  {\bibinfo {volume} {124}}~(\bibinfo {number} {4}),\ \bibinfo {pages}
  {983--992}}\BibitemShut {NoStop}%
\end{thebibliography}
%

\end{document}